\documentclass[12pt,superscriptaddress,letter]{revtex4}%
\usepackage{amssymb}
\usepackage{color}
\usepackage{graphicx}
\usepackage{dcolumn}
\usepackage{bm}
\usepackage[header,title,page,titletoc]{appendix}
\usepackage{amsmath}
\usepackage{amsfonts}%
\providecommand{\U}[1]{\protect \rule{.1in}{.1in}}
\begin{document}
\title{Quantum Physics -- A Theory of Dynamics for "\emph{Space}" on Space}
\author{Su-Peng Kou}
\thanks{Corresponding author}
\email{spkou@bnu.edu.cn}
\affiliation{Center for Advanced Quantum Studies, School of Physics and Astronomy, Beijing
Normal University, Beijing, 100875, China}

\begin{abstract}
Till now, the foundation of quantum physics is still mysterious. To explore
the mysteries in the foundation of quantum physics, people always take it for
granted that quantum processes must be some types of fields/objects on a rigid
space. In this paper, we give a new idea that the space is no more rigid and
the matter is the certain changing of "space" itself rather than extra things
on it. Based on this starting point, we develop a new framework based on
quantum and classical mechanics. Now, physical laws emerge from different
changings of regular changings on spacetime. Then, both quantum mechanics and
classical mechanics become phenomenological theories and are interpreted by
using the concepts of the microscopic properties of a single physical
framework. The expanding/contracting dynamics of "space" leads to quantum
physics. In addition, when we consider a physical variant with 2-th order
variability, quantum fields with gauge structure emerge. The 2-th order
variability is reduced into \textrm{U}$^{\mathrm{em}}$\textrm{(1)} local gauge
symmetry and $\mathrm{SU(N)}$ non-Abelian gauge symmetry. The corresponding
theory becomes QED$\mathrm{\times}$QCD. The belief of "symmetry induce
interaction" is now updates to "Higher order variability induce interaction".
After introducing chirality, a physical variant with 2-th order variability
becomes the true physical reality of our universe. The low energy effective
theory is just the Standard model -- an $\mathrm{SU}_{\mathrm{Strong}%
}\mathrm{(3)}\otimes \mathrm{SU}_{\mathrm{weak}}\mathrm{(2)}\otimes$%
\textrm{U}$_{\mathrm{Y}}$\textrm{(1)} gauge theory. An important progress is
about hidden topological structure for elementary particles of different
generations -- chiral para-statistics. Then, the mass spectra of different
elementary particles are determined by topological invariables from chiral
para-statistics. In particular, to obtain the entire mass spectra, we only
need to use only one free parameter. This progress on the foundation of
quantum physics will have a far-reaching impact on modern physics in the future.

\end{abstract}
\maketitle
\tableofcontents

\newpage

\newpage

\section{Introduction}

Physics involves the study of matter and its motion on spacetime. One goal of
physics is to understand the underlying physical reality and its rules.

According to Newton's theory, matter is a point-like object with mass that
obeys classical mechanics, between which there exists the inverse-square law
of universal gravitation interaction. Field is another important concept in
physics. By understanding electromagnetism, Maxwell introduced the concept of
the electromagnetic field and believed that the propagation of light required
a medium for the waves, named the luminiferous aether. Thomson came to the
idea that classical atoms were knots of swirling vortices in the luminiferous
aether. Chemical elements correspond to knots and links.

In the 20th century, revolution in physics occurred. The concepts in classical
physics have changed. Quantum physics including quantum mechanics and quantum
field theory describes the universe. In the framework of quantum field theory,
it is believed that the interaction comes from exchanging virtual bosons via
local gauge symmetry principles\cite{we,yang,1}. The \emph{Standard model} (an
$\mathrm{SU}_{\mathrm{Strong}}\mathrm{(3)}\otimes \mathrm{SU}_{\mathrm{weak}%
}\mathrm{(2)}\otimes$\textrm{U}$_{\mathrm{Y}}$\textrm{(1)} gauge theory with
Higgs mechanism) is a special type of quantum field theory that focuses on
three non-gravitational interactions: weak interaction, strong interaction,
and electromagnetic interaction\cite{stand,w1,bur,tong}. In quantum physics,
the matter becomes elementary particles in the Standard Model.

Although SM is a successful theory, people study the underlying physics of
quantum field theory. However, because SM is a complicated gauge theory with
more than $20$ free parameters, people try to develop a theory beyond SM.
Grand unified theory (GUT) is proposed to unite SM physics through a quantum
gauge theory with a larger gauge symmetry at higher energy (for example,
\textrm{SU(5)} gauge theory\cite{so}).

There are several approaches towards GUT. String theory is a possible theory
of GUT\cite{GSW,JPbook}. According to string theory, matter consists of
vibrating strings and different oscillatory patterns of strings become
different particles with different masses. In condensed matter physics, the
idea of our universe as an \textquotedblleft \emph{emergent}\textquotedblright%
\ phenomenon has become increasingly popular. In emergence approach, a deeper
and unified understanding of the universe is developed based on a complicated
many-body system. Different quantum fields correspond to different many-body
systems: the vacuum corresponds to the ground state and the elementary
particles correspond to the excitations of the systems\cite{wen}. In addition,
there exist many other proposals of GUT from different points of view.

In particular, I believe that a correct theory for GUT must provide a
satisfactory answer to the following three questions:

\begin{enumerate}
\item How to give an exact definition of "\emph{classical object}" and how to
give an exact definition of "\emph{quantum object}"? And "how to unify the two
types of objects into a single framework"? A true theory beyond quantum
mechanics and classical mechanics must provide a complete understanding of the
physical reality (quantum object or classical object) rather than merely
describe its motion. In quantum mechanics, measurement is quite different from
that in classical mechanics. In quantum measurement processes,
\emph{randomness} appears. Why? A satisfying GUT must provide a complete
understanding of the reason for randomness during quantum measurement.

\item It was known that the key point of quantum gauge fields is local gauge
symmetry. A satisfying GUT must provide a \emph{fully understanding of many
unsolved mysteries}, including quark confinement problem\cite{confinement},
Strong CP puzzle and the existence of axion\cite{axion}, Landau's pole
problem\cite{landau}. To solve above mysteries satisfactorily, a complete, new
theory beyond usual quantum gauge field theory must be developed, rather than
providing certain non-perturbative theories.

\item Today, the SM becomes a fundamental branch of physics that agrees very
well with experiments and provides an accurate description of the dynamic
behaviors of elementary particles. However, the SM is more like a
phenomenological model. There exists a lot of unsolved problems of the SM,
such as the trouble about mass spectra, neutrino
oscillation\cite{neu1,neu2,neu3}, the naturalness problem of Higgs
field\cite{higg1,higg2,higg4}, the trouble about dark matter\cite{dark
matter1, dark matter2, dark matter3} and dark energy\cite{dark energy}. A
satisfied GUT must provide a fully understanding of these unsolved mysteries.
\end{enumerate}

In this paper, to answer above these questions, we reexamine the entire
foundation of quantum field theory for the SM and find three \emph{hidden}
assumptions. These assumptions are commonly referred to as agreed upon by
people and are deeply hidden. The three assumptions are about "field", "gauge
symmetry/invariant" and "point-like structure" of elementary particles.

\emph{One} hidden assumption is about the "\emph{field}" that is a physical
object on rigid spacetime. In modern physics, all physical objects belong to
two different types -- field (or matter) and spacetime. People are familiar to
all kinds of physical processes of classical/quantum fields in a rigid space,
and take it for granted that all physical processes are similar to this. Then,
we may ask, \emph{is this really true? Can all physical processes of quantum
gauge fields be intrinsically described by the processes of extra objects on a
rigid space?}

The \emph{second} hidden assumption is about "\emph{symmetry/invariant}" for
quantum gauge field. It was known that the key point of the SM is local gauge
symmetry. Under a local Abelian/non-Abelian gauge transformation, the
Lagrangian $\mathcal{L}$ or action $\mathcal{S}$ is invariant. The belief of
"\emph{symmetry induce interaction}" is considered as a fundamental idea in
modern physics. \emph{Is it always like this?\emph{ Can we develop a complete
theory for quantum gauge fields beyond belief of "symmetry induce
interaction"?}}

The \emph{third} hidden assumption is about the \emph{point-like}
\emph{structure} of elementary particles. In traditional quantum mechanics (or
quantum field theory), the elementary particles (for example, an electron) are
an infinitesimal point. Strictly speaking, there is no evidence that the size
of them is finite yet. \emph{Do elementary particles have finite sizes? What's
the internal structure of them?}

In the following parts, we point out that the three hidden assumptions are all
misleading. In particular, the second hidden assumption is the crux of the
problem for quantum foundation. An inspiring idea is that \emph{the particle
is the basic block of spacetime and the spacetime is made of matter}.
Therefore, according to this idea, the matter is really certain "changing" of
\textquotedblleft spacetime\textquotedblright \ itself rather than extra things
on it. This is the \emph{new idea} for the foundation of quantum mechanics and
then it becomes the starting point of a new, complete theory.

This paper is organized as below. In Sec. II, to characterize the "space", we
generalize the usual classical "field" for "non-changing" to "variant" for
"changing" and develop a new mathematic theory -- variant theory. We then give
a new theoretical framework beyond quantum mechanics and classical mechanics.
Now, the physical reality becomes physical variants, a predecessor of our
spacetime. In Sec. III, we introduce a non-local theory beyond "field" (we
call it 2-nd order variant theory) by generalizing usual "gauge field" with
local gauge symmetry/invariant to non-local higher-order "space" ("2-nd order
variant", strictly speaking). Within the new theory, usual quantum gauge field
theory become a phenomenological theory and are interpreted by using the
concepts of the microscopic properties of a new physical framework. By using
the new theory, we give possible solution for the troubles about quantum gauge
fields, including quark confinement problem, Yang-Mill gap problem, Strong CP
Problem and the Landau's pole problem. In Sec. IV, a new approach beyond the
SM is proposed based on the theory of chiral variant. Now, the physical chiral
variant becomes fundamental physical object, of which elementary excitations
are gauge fields, fermionic particles and Higgs fields. The low energy
effective theory is just the Standard model -- an $\mathrm{SU}%
_{\mathrm{Strong}}\mathrm{(3)}\otimes \mathrm{SU}_{\mathrm{weak}}%
\mathrm{(2)}\otimes$\textrm{U}$_{\mathrm{Y}}$\textrm{(1)} gauge theory. An
important progress is about hidden topological structure for elementary
particles of different generations -- chiral para-statistics. Then, the mass
spectra of different elementary particles are determined by topological
invariables from chiral para-statistics. In particular, to obtain the entire
mass spectra, we only need to use one free parameter (see below discussion).
In Sec. V, finally, the conclusions are drawn.

\newpage

\section{Quantum Mechanics -- Mechanics of "Changings"}

For the objects in our usual world, people call them "classical" that are
accurately described by \emph{classical mechanics}. In classical theory, there
are two different types of physical reality -- \emph{matter} and
\emph{spacetime}. According to classical theory, matter is a point-like object
(or object composed of point-like objects) with mass that obeys classical
mechanics. In Newton's classical mechanics, the space (without considering
general relativity\cite{ein}) is rigid and regarded as an invariant background
or an invariant stage. Without considering interaction (a certain kind of
potential energy), a (static) classical object is stationary, constant,
non-changing structure and will not affect each other. We may call classical
objects (both spacetime and matter) "\emph{non-changing} structures" (or
\emph{non-operating} structures). In general, classical objects have
\emph{local} property that may be located in certain positions in spacetime.
From Hamilton's principle, the equations of motion are obtained by minimizing
the action of the classical system. In principle, after giving a starting
condition, the moving processes during time evolution could be predicted,
i.e., the positions, the velocities and the accelerations at certain time are
all known. We may call it "\emph{deterministic}". In a classical world, the
surveyors and instruments are classical. Now, the rulers and clocks are
"deterministic", "non-changing", and independent of the physical properties of
the measured object. As a result, in principle, the observers have the ability
to detect every detail of the measured object and obtain the complete
information of the measured object. In summary, "classical" means
"\emph{non-changing}" (or "\emph{non-operating}"), "\emph{locality}", and
"\emph{deterministic}".\emph{ }

Although in our usual world, classical mechanics is both natural to understand
and successful in characterizing different (classical) phenomena. However, in
a microcosmic world, the objects obey \emph{quantum mechanics} (also known as
quantum physics or quantum theory). In quantum mechanics, matter (or quantum
object) is a certain "\emph{changing}" (or "\emph{operating}") structure
rather than a "\emph{non-changing}" (or "\emph{non-operating}") one. Without
considering interaction, they also affect each other. For example, by
exchanging two electrons far away, an extra $\pi$ phase appears. We call the
quantum object "\emph{changing}" structure (or "\emph{operating}" structure).
Now, the space (without considering general relativity) is also assumed to be
an invariant background or an invariant stage. In quantum mechanics, the
motion of a \emph{quantum object} is fully described by certain wave functions
$\psi(x,t)$. Thus, the quantum objects will spread the whole spacetime and
show \emph{non-locality}. The Schr\"{o}dinger equation $i\hbar \frac
{d\psi(x,t)}{dt}=\hat{H}\psi(x,t)$ describes how wave functions evolve,
playing a role similar to Newton's second law in classical mechanics. Here,
the Hamiltonian $\hat{H}$ is a Hermitian operator and $\hslash$ is the Planck
constant. According to quantum mechanics, the energy is \emph{quantized} and
can only change by discrete amounts, i.e. $E=\hslash \omega$. In addition, due
to long-range \emph{entanglement}, the quantum states for many-body particles
show \emph{non-locality} again. During quantum measurement, the wave-function
describes random and indeterministic results and the predicted value of the
measurement is described by a probability distribution. We may call it
"\emph{randomness}".\ In summary, "quantum" means "\emph{changing}",
"\emph{non-locality}", and "\emph{randomness}".

Today, quantum mechanics has become a fundamental branch of physics that
agrees very well with experiments and provides an accurate description of the
dynamic behaviors of microcosmic objects. However, quantum mechanics is far
from being well understood. Einstein said, \textquotedblleft \textit{There is
no doubt that quantum mechanics has grasped the wonderful corner of truth...
But I don't believe that quantum mechanics is the starting point for finding
basic principles, just as people can't start from thermodynamics (or
statistical mechanics) to find the foundation of mechanics}.\textquotedblright%
\ The exploration of the underlying physics of quantum mechanics and the
development of a new quantum foundation has been going on since its
establishment\cite{jammer}. There are a lot of attempts\cite{cabello}, such as
De Broglie's pivot-wave theory\cite{de brogile}, the Bohmian hidden invariable
mechanics\cite{Bohm1}, the many-world theory\cite{many}, the Nelsonian
stochastic mechanics\cite{nelson}, and so on. These quantum interpretations
always try to provide an interpretation of quantum mechanics based on the
picture and description of our usual (classical) world. Obviously, these
theories are not fully satisfactory. Therefore, after one decade, the
exploration to develop a new foundation for quantum mechanics is still not successful.

A complete, new theory beyond both quantum mechanics and classical mechanics
must be developed, rather than providing certain interpretations of quantum
mechanics based on the description of our usual, classical world. The
situation is similar to the foundation of thermodynamics and the relationship
between thermodynamics and statistics mechanics. Classical thermodynamics
describes the thermodynamic systems at near-equilibrium by using macroscopic,
measurable properties, such as energy, work, and heat based on the laws of
thermodynamics. It was known that a microscopic interpretation of these
concepts was later given via statistical mechanics (statistical
thermodynamics) developed in the late 19th century and early 20th century.
Within statistical mechanics, classical thermodynamics has become
phenomenological theory and has been interpreted by using the concepts of the
microscopic interactions between individual states, i.e.,%
\begin{align*}
&  \text{Classical thermodynamics (a phenomenological theory)}\\
&  \Longrightarrow \text{Statistical mechanics (a microscopic theory).}%
\end{align*}
As a result, the macroscopic properties of material in classical
thermodynamics are explained as the microscopic properties of individual
particles and atoms as a natural result of statistics mechanics at the
microscopic level.

In particular, I believe that a correct theory beyond quantum mechanics must
provide a satisfactory answer to the following five questions:

\begin{enumerate}
\item How to understand "\emph{non-locality}" in wave function for a single
particle and that in quantum entanglement? A true theory beyond quantum
mechanics and classical mechanics must provide a complete understanding of the
"\emph{non-locality}" character of quantum mechanics;

\item How to understand "\emph{changing}" structure (or "\emph{operating}"
structure) for quantum objects in quantum mechanics? What exactly does
"\emph{changing}" here mean? A true theory beyond quantum mechanics and
classical mechanics must provide a complete understanding of the
"\emph{changing}" character of quantum mechanics;

\item What does $\hbar$ mean? And can $\hbar$ be changed? A true theory beyond
quantum mechanics and classical mechanics must provide a complete
understanding of the existence of $\hbar$ and give a possible way to
"\emph{change}" $\hbar$;

\item How to give an exact definition of "\emph{classical object}" and how to
give an exact definition of "\emph{quantum object}"? And "how to unify the two
types of objects into a single framework"? A true theory beyond quantum
mechanics and classical mechanics must provide a complete understanding of the
physical reality (quantum object or classical object) rather than merely
describe its motion;

\item In quantum mechanics, measurement is quite different from that in
classical mechanics. In quantum measurement processes, \emph{randomness}
appears. Why? A true theory beyond quantum mechanics and classical mechanics
must provide a complete understanding of the reason for randomness during
quantum measurement.
\end{enumerate}

To accurately and globally characterize "spacetime", we develop a new,
complete theory (we call it \emph{variant theory}) by generalizing the usual
local "field" to non-local "space" ("variant", strictly speaking). Within the
new theory, both quantum mechanics and classical mechanics become
\emph{phenomenological} theories and are interpreted by using the concepts of
the microscopic properties of a single physical framework, i.e.,%
\begin{align*}
&  \text{Quantum mechanics (a phenomenological theory)}\\
&  \Longrightarrow \text{One case of new mechanics}\\
&  \text{(a microscopic theory),}%
\end{align*}
and
\begin{align*}
&  \text{Classical mechanics (a phenomenological theory)}\\
&  \Longrightarrow \text{The other case of new mechanics }\\
&  \text{(a microscopic theory).}%
\end{align*}
In particular, with the new theory, we have the power to recover the intrinsic
"changing", and "non-local" structure of quantum mechanics.

\subsection{Variant theory -- mathematical foundation for "changings"}

Our classical world can be regarded as a \emph{"non-changing"} configuration
structure that is described by the usual classical "field" on Cartesian space.
In this section, we generalize the usual classical "field" to "space"
(strictly speaking, group-changing space). We call the new mathematic
structure to be \emph{variant theory}. In general, the usual classical field
(for example, $f(x)$) is suitable to characterize a system with a
"\emph{non-changing}" configuration structure, i.e.,
\[
\text{"Classical field on space": Non-changing structure;}%
\]
On the contrary, variant theory is suitable to characterize a system with a
"\emph{changing}" or "\emph{operating}" structure, i.e.,
\[
\text{"Space on space": Changing structure.}%
\]

\subsubsection{Review on classical fields of compact Lie group}

In general, classical "\emph{field}" is a certain extra "non-changing" object
on a rigid space that spreads throughout a large region of space, in which
each point has a physical quantity associated with it. Therefore, a group
field that describes a configuration for group elements becomes one of the
most important physical objects in modern physics. To characterize a classical
field, local functions are introduced and the set of numbers for the classical
field describes their definitive states.

\paragraph{Properties of fields of group \textrm{G}}

We take a classical field of a (compact) Lie group $\mathrm{G}$ (a special,
multi-component, scalar field) as an example to show its elementary
properties, including the \emph{object of study}, \emph{elements},
\emph{definition},\emph{ classification}, and,\emph{ changings}.

\begin{figure}[ptb]
\includegraphics[clip,width=0.72\textwidth]{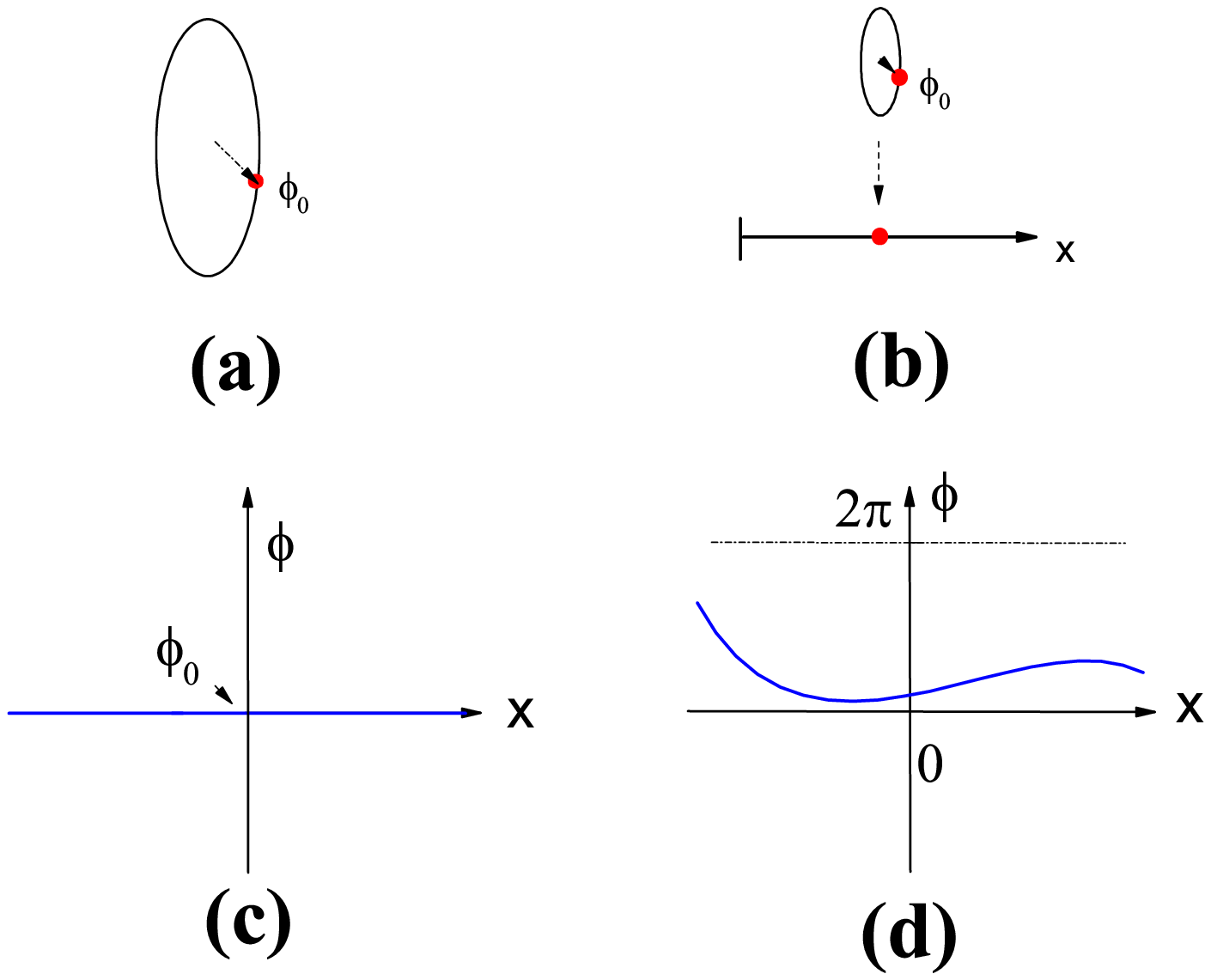}\caption{(Color online)
(a) An element of a compact \textrm{U(1)} group; (b) A mapping between the
elements for the compact \textrm{U(1)} group to the points in the
one-dimensional Cartesian space $C_{1}$. This is just the Geometry
representation of it; (c) An illustration of the reference for \textrm{U(1)}
group, $\phi(x)=\phi_{0}=0$; (d) The illustration of a general field of
compact \textrm{U(1)} group under analytic representation.}%
\end{figure}

For the field of a compact Lie group, the object of study is a \emph{group
space} on Cartesian space $\mathrm{C}_{d}$. For a (compact) Lie group
$\mathrm{G}$, $g$\ is the \emph{group element}. Fig.1(a) illustrates a group
element of a compact \textrm{U(1)}\ group. Then, all group elements make up
group space (or space of group elements $G$).

For example, for (non-Abelian) \textrm{SO(N)} group, the group element is
$g=e^{i\Theta}$ where $\Theta=\sum_{a=1}^{(n-1)n/2}\theta^{a}T^{a}$ and
$\theta^{a}$ are a set of $\frac{(N-1)N}{2}$ constant parameters, and $T^{a}$
are $\frac{(N-1)N}{2}$ matrices representing the generators of the Lie algebra
of $\mathrm{SO(N)}$. In general, we have spinor representation for
\textrm{SO(N)} group. By introducing Gamma matrices obeying Clifford algebra
$\Gamma^{a}$, $\{ \Gamma^{a},\Gamma^{b}\}=2\delta^{ab}$, the generators of the
Lie algebra of $\mathrm{SO(N)}$ become $-\frac{i}{4}[\Gamma^{a},\Gamma^{b}].$
For the case of $N=3$, both Gamma matrices and the generators for
$\mathrm{SO(3)}$ Lie group are Pauli matrices $\sigma^{x},$ $\sigma^{y},$
$\sigma^{z}$.

Field of the group \textrm{G} characterizes the dynamics of group space on
Cartesian space $\mathrm{C}_{d}.$ Then we give its definition.

\textit{Definition: A field of group }$\mathit{\mathrm{G}}$\textit{ is
described by the mapping between the group space (or the space of the group
elements }$\mathrm{G}$\textit{) and Cartesian space} $\mathrm{C}_{d}$.

According to the definition, a field of the group \textrm{G} is denoted by a
mapping between the group element and space point. In brief, the field $g(x)$
of the group \textrm{G} can be regarded as a \emph{point-to-point mapping}.
Fig.1(b) illustrates a point-to-point mapping between a group element of a
compact \textrm{U(1)}\ group and a point on a one-dimensional (1D) Cartesian space.

Next, we classify the field of the group \textrm{G}. Different fields are
classified by two values, one is about the group \textrm{G }that determines
the object of study, and the other is dimension number $d$ of Cartesian space
$\mathrm{C}_{d}.$

Finally, we address the issues of changings, the prelude for "classical
motion" in physics. The local changing of a field of the group \textrm{G
}comes from locally changing group elements by doing group operations.
Consequently, the field $g(x)$ of group \textrm{G} turns into another one
$g^{\prime}(x)$, i.e.,
\begin{equation}
g(x)\rightarrow g^{\prime}(x)\neq g(x).
\end{equation}
For example, for the group field of spins, the changings can be regarded as
different spin rotations on different positions. After doing spin rotations,
the original configuration of group elements turns into another.

\paragraph{Different representations}

There are different representations for a field of the group \textrm{G }from
different aspects, including \emph{algebra}, \emph{geometry}, and
\emph{algebra}, respectively. In general, to characterize the same field of
the group \textrm{G,} people can transform one representation to another.

\textit{Algebra representation}\textbf{:} In the algebra representation,
the field of group \textrm{G} is usually described by the function $g(x)$. The
set of numbers for $g(x)$ describes the definitive state of the study. In
other words, the element of a field is a point $g(x)$ which denotes an element
of a group. For the non-Abelian case, the function $g(x)$ becomes a matrix and
has $N$ variables ($N>1$), each of which corresponds to a group generator
$T^{a}$.

\textit{Algebra representation}: In general, to define a field $g(x),$ one
always chooses an initial one or its \emph{reference}. Difference group fields
are normalized by the reference, the relative deviation becomes the true
result. In Fig.1(c), we choose the reference for group $\mathrm{G}$ as a
constant group element $g_{0}(x)=1$ with fixed phase angle $\phi(x)=\phi_{0}%
.$\textit{ }In Fig.1(d), we show field $\varphi(x)$ of \textrm{U(1)} group
with the function's reference $\phi_{0}=0$. Hence, in algebra representation,
the group field is characterized by a group of (local) operations of the group
\textrm{G}. In a sentence, we can "\emph{generate}" a field of \textrm{G}
group $g(x)$ by a series of group operations on every position $x$ of
Cartesian space $\mathrm{C}_{d}$\textit{ }under a certain reference.

Then, $g(x)$ is obtained by an operation $\hat{U}(x)$ on the reference
$g_{0}(x)=1$. We call it a "local" operation. Here, the word "local" means
that the operations on different points (for example, $\hat{U}(x_{1})$,
$\hat{U}(x_{2}),$ $x_{1}\neq x_{2}$) are independent with each other, i.e,
\[
\lbrack \hat{U}(x_{1}),\hat{U}(x_{2})]\equiv0.
\]
The series of $\hat{U}(x)$ corresponds to the field of \textrm{G} group
$g(x).$ Now, the element of a field becomes a local operation $\hat{U}(x)$
that changes an element of the group. In the following parts, we call
operation $\hat{U}(x)$ with "$\wedge$" on $U$ to be \emph{group operation}.

\textit{Geometry representation}\textbf{:} Geometry representation provides an
alternative complete representation that gives a clear picture of the
"\emph{non-changing}" configurations of group fields. By using geometry
representation, people can plot a figure to characterize the group fields.

For a compact\textrm{ U(1)} group, the configuration of group elements is a
set of given phase angles $g(x)=e^{i\varphi(x)}$ on each position $x$ (see
Fig.1(b)); for the non-Abelian case, for example, a field of compact
\textrm{SU(2)} group,\ on each position $x$ a group element $g(x)=\exp
(i\sum_{a=1}^{3}\theta(x)^{a}\sigma^{a})$ corresponds to a point on Bloch
sphere. In general, on each point of Cartesian space, a point of a field of an
arbitrary compact group \textrm{G} corresponds to a point of a closed,
sphere-like super-manifold. This configuration structure of the group field
looks like a static picture. This is why we call a field of the group
\textrm{G} a "\emph{non-changing}" structure.

\subsubsection{Group-changing space -- object of study for variants}

To define a variant, we first introduce the object of study, which is a new
type of mathematic structure\emph{ }beyond group space -- \emph{group-changing
space}\textrm{. }

Before introducing group-changing space, we review the concept of the usual
\emph{Cartesian space}. (1D) Cartesian space is mathematic\ space described by
the coordinate $x$ that is a series of numbers arranged in size order (See
Fig.2(a)). Along the Cartesian space, the number changes correspondingly. The
element of 1D Cartesian space is infinitesimal line segments $\delta
x\rightarrow0$, rather than "point". A lot of infinitesimal line segments,
$\delta x$ make up a Cartesian space. For a Cartesian space with finite size
$L$, $L$ is a "topological" number. Using a similar idea, we could introduce
group-changing space $\mathrm{C}_{\mathrm{\tilde{G}},d}(\Delta \phi^{a})$ for
non-compact Lie group \textrm{\~{G}}.

\paragraph{Definition}

Then, we define group-changing space $\mathrm{C}_{\mathrm{\tilde{G}},d}%
(\Delta \phi^{a})$ for non-compact Lie group \textrm{\~{G}}. Here \textrm{G}
with "$\sim$" above means a non-compact Lie group.

\textit{Definition -- }$\mathit{d}$\textit{-dimensional group-changing space}
$\mathrm{C}_{\mathrm{\tilde{G}},d}(\Delta \phi^{a})$:\textit{\ For a
non-compact \textrm{\~{G}} Lie group, it has }$\mathit{N}$\textit{ generator.
The }$\mathit{d}$\textit{-dimensional group-changing space} $\mathrm{C}%
_{\mathrm{\tilde{G}},d}(\Delta \phi^{a})$\textit{\ of non-compact
\textrm{\~{G}} Lie group is described by }$\mathit{d}$\textit{ series of
numbers of group element }$\phi^{a}$\textit{ of a-th generator independently
in size order. }$\Delta \phi^{a}$\textit{ denotes the size of the
group-changing space along a-th direction that is a topological number. In
general, we have }$\mathit{N>d.}$

For example, 1D group-changing space $C_{\mathrm{\tilde{U}(1)},1}(\Delta \phi
)$\ of non-compact \textrm{\~{U}(1)} group is described by a series of numbers
of group element $\phi$ arranged in size order. $\Delta \phi$ denotes the total
size of the changing space that turns to infinity, i.e., $\Delta
\phi \rightarrow \infty$. "$1$" denotes dimension.

Clifford group-changing space ($d$-dimensional group-changing space of
non-compact $\mathrm{\tilde{S}\tilde{O}}$\textrm{(N)} group) is an interesting
higher dimensional group-changing space. For this case, besides the mutual
independence of different directions, there exists \emph{orthogonality},
i.e.,
\begin{equation}
\left \vert \mathbf{\phi}_{\mathrm{A}}-\mathbf{\phi}_{\mathrm{B}}\right \vert
^{2}=%
%TCIMACRO{\dsum \nolimits_{\mu}}%
%BeginExpansion
{\displaystyle \sum \nolimits_{\mu}}
%EndExpansion
(\phi_{\mathrm{A,}\mu}e^{\mu}-\phi_{\mathrm{B},\mu}e^{\mu})^{2}%
\end{equation}
where $\mathbf{\phi}_{\mathrm{A}}=%
%TCIMACRO{\dsum \nolimits_{\mu}}%
%BeginExpansion
{\displaystyle \sum \nolimits_{\mu}}
%EndExpansion
\phi_{\mathrm{A,}\mu}e^{\mu}$ and $\mathbf{\phi}_{\mathrm{B}}=%
%TCIMACRO{\dsum \nolimits_{\mu}}%
%BeginExpansion
{\displaystyle \sum \nolimits_{\mu}}
%EndExpansion
\phi_{\mathrm{B,}\mu}e^{\mu}$. Therefore, Clifford group-changing space is a
typical space obeying noncommutative geometry\cite{con}.

We point out that for a higher dimensional group-changing space $\mathrm{C}%
_{\mathrm{\tilde{G}},d}(\Delta \phi^{a})$ there exists global phase changing of
the system $\left \vert \Delta \phi^{\mu}(x)\right \vert =\sqrt{%
%TCIMACRO{\dsum \limits_{\mu}}%
%BeginExpansion
{\displaystyle \sum \limits_{\mu}}
%EndExpansion
(\Delta \phi^{\mu}(x))^{2}}$, the other is about $d-1$ internal relative angle
that is defined by $\mathrm{C}_{\mathrm{\tilde{G}},d}(\Delta \phi^{\mu
})/\mathrm{C}_{\mathrm{\tilde{U}(1)\in \tilde{G}},1}(\Delta \phi
_{\mathrm{global}})$, of which the degrees of freedom become compact. For
example, for a 2D group-changing space $\mathrm{C}_{\mathrm{\tilde{S}\tilde
{O}(2)},2}(\Delta \phi^{\mu}),$ except for the global phase changing of the
system, there exists an internal relative angle that rotates the original
group-changing space from one direction to the other.

\paragraph{Elements}

For a $d$-dimensional group-changing space $\mathrm{C}_{\mathrm{\tilde{G}}%
,d}(\Delta \phi^{a})$, the \emph{element} is an infinitesimal $d$-dimensional
\emph{group-changing} \emph{operation} $\delta \phi^{a}$ ($\delta \phi
^{a}\rightarrow0,$ $a=1,...,d$); or $\delta \phi^{a}$ is the \emph{piece} of
group-changing space $\mathrm{C}_{\mathrm{\tilde{G}},d}(\Delta \phi^{a})$.
Therefore, $\mathrm{C}_{\mathrm{\tilde{G}},d}(\Delta \phi^{a})$ is regarded as
a mathematical set of $n$\ infinitesimal changings of group element
($n\cdot \delta \phi \rightarrow \infty$). For a higher dimensional group-changing
space, along $a$-direction this group-changing space, the group element of
generator $T^{a}$ of \textrm{\~{G}} changes correspondingly. Therefore, the
group elements for different generator $T^{a}$ of \textrm{\~{G}} change
\emph{independently }(but not necessarily \emph{commutating})\emph{ }from each other.

\begin{figure}[ptb]
\includegraphics[clip,width=0.82\textwidth]{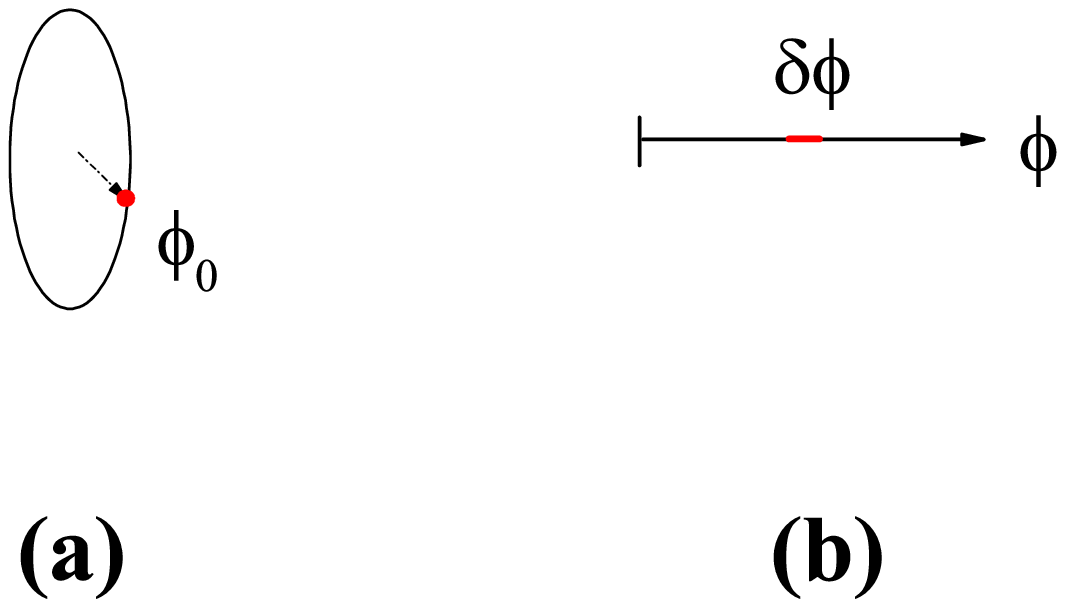}\caption{(Color online)
(a) An element $\phi_{0}$ of a compact \textrm{U(1)} group that denotes the
"non-changing" configuration of its field; (b) An element $\delta \phi$ (an
infinitesimal group-changing operation) of non-compact \textrm{\~{U}(1)} group
that denotes the "changing" configuration of a variant.}%
\end{figure}

For 1D group-changing space $\mathrm{C}_{\mathrm{\tilde{U}(1)},1}(\Delta \phi)$
for non-compact \textrm{\~{U}(1)} group, we have a series of infinitesimal
group-changing operations,
\begin{equation}
\prod_{i}(\tilde{U}(\delta \phi_{i}))
\end{equation}
with $%
%TCIMACRO{\dsum \nolimits_{i}}%
%BeginExpansion
{\displaystyle \sum \nolimits_{i}}
%EndExpansion
\delta \phi_{i}=\Delta \phi$. Here, $\tilde{U}(\delta \phi_{i})$\ with "$\sim$"
above it means an operation of contraction/expansion on group-changing space
that is different from group operation $\hat{U}(\delta \phi_{i})$. We can also
denote a $d$-dimensional group-changing space $\mathrm{C}_{\mathrm{\tilde{G}%
},d}(\Delta \phi^{a})$ for non-compact group \textrm{\~{G}} by a series of
infinitesimal operations of group-changing,
\begin{equation}
\prod_{i}(\tilde{U}(\delta \phi_{i}))=\prod_{i}(\prod_{a=1}^{d}(\tilde
{U}(\delta \phi_{i}^{a})))
\end{equation}
where $\tilde{U}(\delta \phi_{i})=\prod_{a=1}^{d}(\tilde{U}(\delta \phi_{i}%
^{a}))$ and $\tilde{U}(\delta \phi_{i}^{a})=e^{i((\delta \phi_{i}^{a}T^{a}%
)\cdot \hat{K}_{a})}$, $\hat{K}_{a}=-i\frac{d}{d\phi^{a}}.$\ Here, the i-th
operation $\hat{U}(\delta \phi_{i})$ generates an element of group-changing
that is infinitesimal group-changing operation with $d$ directions.

In particular, the operation $\tilde{U}(\delta \phi_{i})$ is a
"\emph{non-local}" operation that will change the size of the group-changing
space $\mathrm{C}_{\mathrm{\tilde{G}},d}(\Delta \phi^{a})$, i.e., $\Delta
\phi^{a}\rightarrow \Delta \phi^{a}\pm \delta \phi_{i}^{a}$. On the contrary, the
local group operation $\hat{U}(x_{i})=e^{\pm i\delta \phi_{i}^{a}T^{a}}$ will
never change the size of group-changing space. In the following part, we
call\ $\delta \phi^{a}$ that corresponds to $\tilde{U}(\delta \phi_{i}%
^{a})=e^{\pm i((\delta \phi_{i}^{a}T^{a})\cdot \hat{K}_{a})}$ ($\delta \phi
^{a}\rightarrow0$) to be \emph{group-changing element} for group-changing
space $\mathrm{C}_{\mathrm{\tilde{G}},d}(\Delta \phi^{a})$.

\paragraph{Classification of changings of group-changing space}

Then, we classify the changings of group-changing space. There are two types
of changings of the group-changing space $\mathrm{C}_{\mathrm{\tilde{G}}%
,d}(\Delta \phi^{a})$: one is topological, the other is non-topological. For
topological changings, there are globally \emph{expand} or \emph{contract.}
Under these changings, the sizes of a group-changing space $\mathrm{C}%
_{\mathrm{\tilde{G}},d}(\Delta \phi^{a})$ become different. For non-topological
changings, there are global shift and shape changings. Let's give more discussion.

\begin{figure}[ptb]
\includegraphics[clip,width=0.72\textwidth]{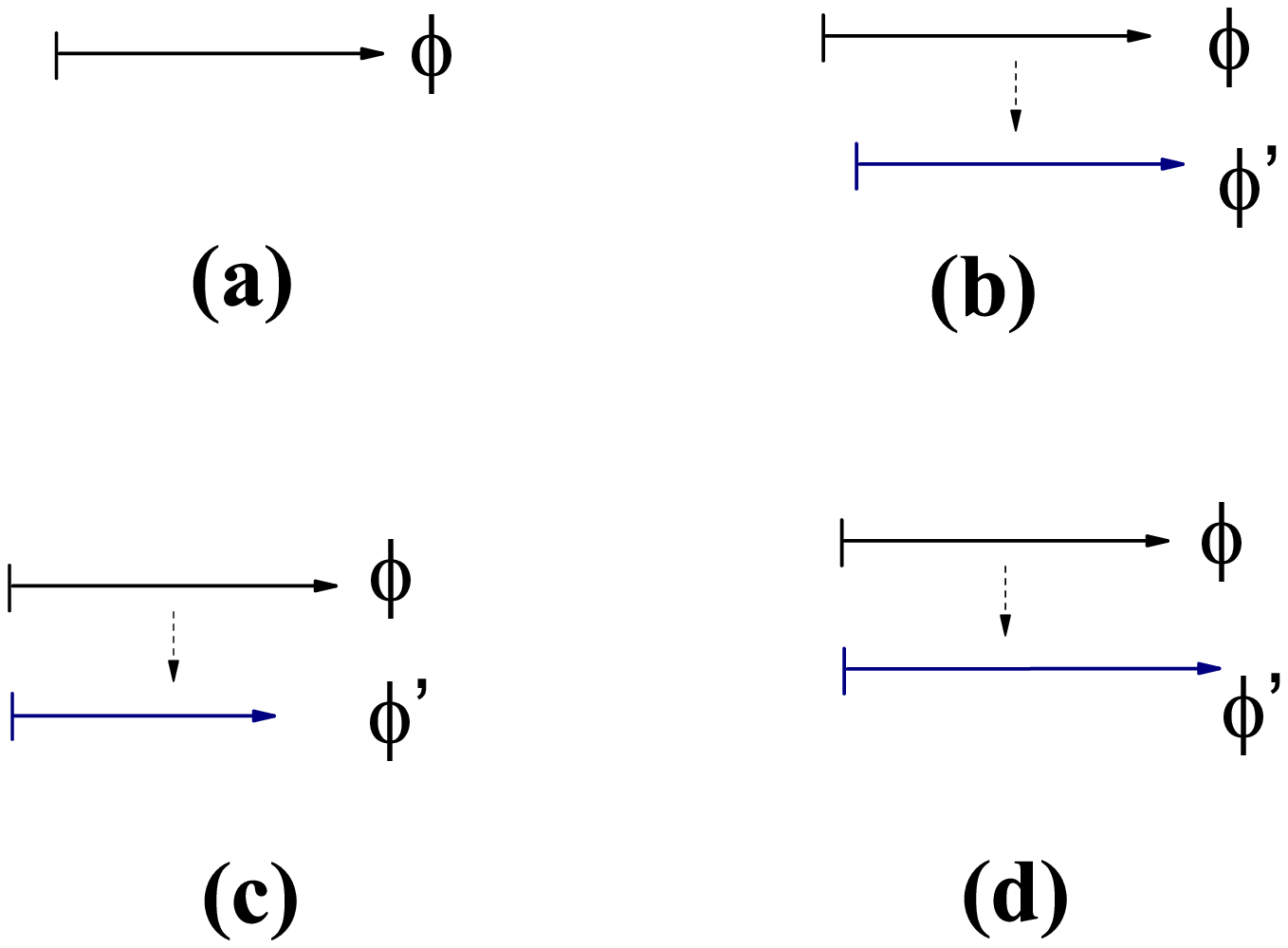}\caption{(Color online)
(a) A group-changing space of a non-compact \textrm{\~{U}(1)} group; (b)
Globally shift of the group-changing space; (c) and (d) denote the global
contraction and expansion of the group-changing space, respectively.}%
\end{figure}

Firstly, we consider a 1D group-changing space $\mathrm{C}_{\mathrm{\tilde{G}%
},1}(\Delta \phi^{a})$\ for $a$-th component of a non-compact \textrm{\~{G}}
Lie group. There are two types of changings:

1) Globally \emph{shift} of the 1D group-changing space $\mathrm{C}%
_{\mathrm{\tilde{G}},1}(\Delta \phi^{a})$ without changing its size: For a 1D
group-changing space $\mathrm{C}_{\mathrm{\tilde{G}},1}(\Delta \phi^{a}),$
under a globally shift $\phi_{0}^{a}$, the phase of it changes, $\phi
^{a}\rightarrow \phi^{a}+\phi_{0}^{a}$. The operation of such a globally shift
is denoted by $\hat{U}(\delta \phi^{a})=e^{i\phi_{0}^{a}T^{a}}$. Therefore,
$\phi_{0}^{a}$ plays the role of coordinate origin of group-changing space.
The size of it doesn't change and is still $\Delta \phi^{a}.$ See the
illustration in Fig.3(b);

2) Globally \emph{expand} or \emph{contract} with changing its size: Under
contraction/expansion, the original 1D group-changing space $\mathrm{C}%
_{\mathrm{\tilde{G}},1}(\Delta \phi^{a})$ turns into a new one $\mathrm{C}%
_{\mathrm{\tilde{G}},1}((\Delta \phi^{a})^{\prime})$. Under the changing of
"expansion", the total sizes of 1D group-changing space become larger, i.e.,
\[
\mathrm{C}_{\mathrm{\tilde{G}},1}(\Delta \phi^{a})\rightarrow \mathrm{C}%
_{\mathrm{\tilde{G}},1}((\Delta \phi^{a})^{\prime})\text{ with }(\Delta \phi
^{a})^{\prime}-\Delta \phi>0;
\]
Under the changing of "contraction", the total sizes $\Delta \phi$ of 1D
group-changing space become smaller, i.e.,
\[
\mathrm{C}_{\mathrm{\tilde{G}},1}(\Delta \phi^{a})\rightarrow \mathrm{C}%
_{\mathrm{\tilde{G}},1}((\Delta \phi^{a})^{\prime})\text{ with }(\Delta \phi
^{a})^{\prime}-\Delta \phi<0.
\]
The operation of contraction/expansion on 1D group-changing space
$\mathrm{C}_{\mathrm{\tilde{G}},1}(\Delta \phi^{a})$ is
\[
\tilde{U}(\delta \phi^{a})=e^{i((\delta \phi^{a}T^{a})\cdot \hat{K})}%
\]
where $\delta \phi^{a}=(\Delta \phi^{a})^{\prime}-\Delta \phi^{a}$ and $\hat
{K}=-i\frac{d}{d\phi^{a}}$ is its generator. See the illustration in Fig.3(c)
and Fig.3(d). In the following part, we point out that this type of changings
of group-changing space corresponds to the particle's generation and annihilation;

Using similar approach, we discuss the changings of a $d$-dimensional
group-changing space $\mathrm{C}_{\mathrm{\tilde{G}},d}(\Delta \phi^{a})$
($d>1$). There are four types of changings\ of the $d$-dimensional
group-changing space $\mathrm{C}_{\mathrm{\tilde{G}},d}(\Delta \phi^{a})$:

1) Globally \emph{shifting} $\mathrm{C}_{\mathrm{\tilde{G}},d}(\Delta \phi
^{a})$ along different directions without changing its size, i.e., $\phi
^{a}\rightarrow \phi^{a}+\phi_{0}^{a}$: The operation of such a globally shift
is $\hat{U}(\phi_{0}^{a})=e^{i\phi_{0}^{a}T^{a}}$. Now, the size of it is
still $\Delta \phi^{a}$;

2) Globally \emph{rotating} $\mathrm{C}_{\mathrm{\tilde{G}},d}(\Delta \phi
^{a})$ from $a$-direction to $b$-direction: The operation is $\hat{U}%
(\delta \varphi^{ab})=e^{\delta \varphi^{ab}T^{ab}}$ that changes $T^{a}$ to
$T^{b}$. The operation of globally rotation obeys rules of a compact Lie group;

3) Globally \emph{expanding} or \emph{contracting} $\mathrm{C}_{\mathrm{\tilde
{G}},d}(\Delta \phi^{a})$\ along $a$-th direction with changing its
corresponding size: The operation of contraction/expansion on group-changing
space is $\tilde{U}(\delta \phi^{a})=e^{i((\delta \phi^{a}T^{a})\cdot \hat{K}%
^{a})}$ where $\delta \phi^{a}=(\Delta \phi^{a})^{\prime}-\Delta \phi^{a}$ and
$\hat{K}^{a}=-i\frac{d}{d\phi^{a}}.$ For higher dimensional case, the group
elements for different generator $T^{a}$ of \textrm{\~{G}} are independently
(but not necessarily commutating) expanding or contracting\emph{ }from each other;

4) Locally rotating\emph{ }on Cartesian space $\mathrm{C}_{d}$: Locally
rotating of $\mathrm{C}_{\mathrm{\tilde{G}},d}(\Delta \phi^{a})$\ ($d>1$) leads
to the \emph{shape} of system locally changing.\emph{ }Due to noncommutative
character, the changings for $\mathrm{C}_{\mathrm{\tilde{G}},d}(\Delta \phi
^{a})$ from locally shape changing become very complex. This is related to
curved space and irrelevant to the issue of this paper. We don't discuss it.

In summary, different changings of a $d$-dimensional group-changing space
$\mathrm{C}_{\mathrm{\tilde{G}},d}(\Delta \phi^{a})$ can be characterized by
performing additional group-changing operations together with additional
possible group operations.

\subsubsection{Variant: fundamental concept, definition, classification, and
examples}

Variant describes a structure of "\emph{changings}". Here, the word
"\emph{changing}"\ means a space-like structure of a set of numbers' changing
on Cartesian space. Therefore, a variant is theory describing the space
dynamics rather than field dynamics on Cartesian space. In a word, we say that
"\emph{It describes space on space}".

\paragraph{Definition}

We firstly give a definition about a general variant (an object of
d-dimensional group-changing space \textrm{C}$_{\mathrm{\tilde{G}},d}$ on
d-dimensional (rigid) Cartesian space $\mathrm{C}_{d}$).

\textit{Definition -- Variant: A variant }$V_{\mathrm{\tilde{G},}d}[\Delta
\phi^{\mu},\Delta x^{\mu},k_{0}^{\mu}]$\textit{ is denoted by\ a mapping
between a d-dimensional group-changing space }$\mathrm{C}_{\mathrm{\tilde{G}%
,}d}$\textit{ with total size }$\Delta \phi^{\mu}$\textit{\ and
\textit{Cartesian }space }$\mathrm{C}_{d}$\textit{\ with total size }$\Delta
x^{\mu}$\textit{, i.e.,}%
\begin{align}
V_{\mathrm{\tilde{G},}d}[\Delta \phi^{\mu},\Delta x^{\mu},k_{0}^{\mu}]  &
:\mathrm{C}_{\mathrm{\tilde{G},}d}=\{ \delta \phi^{\mu}\} \nonumber \\
&  \Longleftrightarrow \mathrm{C}_{d}=\{ \delta x^{\mu}\}
\end{align}
\textit{where }$\Longleftrightarrow$\textit{\ denotes an ordered mapping under
fixed changing rate of integer multiple }$k_{0}$\textit{.\ In particular,
}$\delta \phi^{\mu}$\textit{ denotes group-changing element along }$\mu
$\textit{-th direction (or element of group-changing space along }$\mu$-th
direction\textit{) rather than group element (or element of group). }Here, the
total size $\Delta \phi^{\mu}$ of\textit{ }$\mathrm{C}_{\mathrm{\tilde{G}},d}%
$\textit{\ }can match the total size\textit{ }$\Delta x^{\mu}$\textit{
}of\textit{ }$\mathrm{C}_{d}$\textit{, }i.e.,\textit{ }$\Delta \phi^{\mu}%
=k_{0}^{\mu}\Delta x^{\mu}$ or not, i.e., $\Delta \phi^{\mu}\neq k_{0}^{\mu
}\Delta x^{\mu}$.

We then take 1D variant $V_{\mathrm{\tilde{U}(1),}1}[\Delta \phi,\Delta
x,k_{0}]$ as an example to show the mathematic structure of "\emph{space on
space}".

The 1D variant $V_{\mathrm{\tilde{U}(1),}1}[\Delta \phi,\Delta x,k_{0}]$ is a
mapping between 1D group-changing space $\mathrm{C}_{\mathrm{\tilde{U}(1)}%
,1}(\Delta \phi)$\textit{ }and 1D Cartesian space $\mathrm{C}_{1}$, i.e.,
\begin{align}
V_{\mathrm{\tilde{U}(1),}1}[\Delta \phi,\Delta x,k_{0}]  &  :\mathrm{C}%
_{\mathrm{\tilde{U}(1)},1}(\Delta \phi)=\{ \delta \phi \} \nonumber \\
&  \Longleftrightarrow \mathrm{C}_{1}=\{ \delta x\}
\end{align}
where $\Longleftrightarrow$\ denotes an ordered mapping under fixed changing
rate of integer multiple $k_{0}$.\textit{ }According to above definition,\ for
a 1D variant $V_{\mathrm{\tilde{U}(1),}1}[\Delta \phi,\Delta x,k_{0}],$ we have%
\begin{equation}
\delta \phi_{i}=k_{0}n_{i}\delta x_{i}%
\end{equation}
where $k_{0}$ is a constant real number and $n_{i}$ is an integer number.
$k_{0}n_{i}$ is changing rate for $i$-th space element, i.e., $k_{0}%
n_{i}=\delta \phi_{i}/\delta x_{i}$. Under the mapping, each of the
infinitesimal element of $\mathrm{C}_{\mathrm{\tilde{U}(1)},1}(\Delta \phi)$ is
marked by a given position $x_{i}$ in 1D Cartesian space $\mathrm{C}_{1},$
i.e., $\delta \phi_{i}\rightarrow \delta \phi_{i}(x_{i})$ or $n_{i}\rightarrow
n_{i}(x_{i})$. Therefore, for the 1D variant $\mathrm{C}_{\mathrm{\tilde
{U}(1)},1}(\Delta \phi)$, we have a series of numbers of infinitesimal elements
to record its information, i.e.,
\begin{align}
V_{\mathrm{\tilde{U}(1),}1}[\Delta \phi,\Delta x,k_{0}]  &  :\{n_{i}%
\} \nonumber \\
&  =(...n_{1},n_{2},n_{3},n_{4},n_{5},n_{6},...).
\end{align}
Different 1D variants $V_{\mathrm{\tilde{U}(1),}1}[\Delta \phi,\Delta x,k_{0}]$
are characterized by different distributions of $n_{i}$. As a result, in some
sense, a variant can be described by "\emph{function}" of $n_{i}.$

For higher dimensional variants, an infinitesimal element of group-changing
space has $d$ components. With fixing $k_{0}^{\mu}$, we have $d$ series of
numbers of infinitesimal elements, i.e.,
\begin{align}
V_{\mathrm{\tilde{G},}d}[\Delta \phi^{\mu},\Delta x^{\mu},k_{0}^{\mu}]  &
:\nonumber \\
\{n_{i}^{\mu}\}  &  =(...n_{1}^{\mu},n_{2}^{\mu},n_{3}^{\mu},n_{4}^{\mu}%
,n_{5}^{\mu},n_{6}^{\mu},...).
\end{align}

Therefore, according to above discussion, "field" of group $\mathrm{G}$ is a
group of group elements $\phi^{a}$ (or elements of local group operations
$\hat{U}(\delta \phi_{i}^{a})$) on Cartesian space; a variant
$V_{\mathrm{\tilde{G},}d}[\Delta \phi^{\mu},\Delta x^{\mu},k_{0}^{\mu}]$ is a
group of group-changing elements $\delta \phi^{a}$ (or elements of non-local
group-changing operations $\tilde{U}(\delta \phi_{i}^{a})$) on Cartesian space.

\paragraph{Classification of variants}

We classify the variant $V_{\mathrm{\tilde{G},}d}[\Delta \phi^{\mu},\Delta
x^{\mu},k_{0}^{\mu}]$ of non-compact Lie group \textrm{\~{G}}.

Different variants are classified by two values, one is about the non-compact
Lie group \textrm{\~{G} }that determines the whole structure, the other is
dimension number $d$ of Cartesian space $\mathrm{C}_{d}.$ In addition,
$\Delta \phi^{\mu},$ $\Delta x^{\mu},$ $k_{0}^{\mu}$ are meaningful.
$k_{0}^{\mu}$\ characterizes the changing rate along $\mu$-th spatial
direction. $\Delta \phi^{\mu}$ denotes the size of group changing space along
$\mu$-th spatial direction, $\Delta x^{\mu}$ denotes the size of $\mu$-th
spatial direction.

For example, if we consider the orthogonality of group-changing space, we have
$\mathrm{\tilde{S}\tilde{O}}$\textrm{(N)} variant (Clifford group-changing
space on $d$-dimensional Cartesian space $\mathrm{C}_{d}$). This variant is
very interesting due to its important role in quantum mechanics.

In the following parts, we introduce the concept of different variants, such
as uniform variants, perturbative uniform variants, "complementary pair" of
two variants. Uniform variants are simplest types of variants and perturbative
uniform variants are always generated by perturbatively changings on
corresponding uniform variants.

\paragraph{Examples}

\subparagraph{Uniform variant}

\begin{figure}[ptb]
\includegraphics[clip,width=0.9\textwidth]{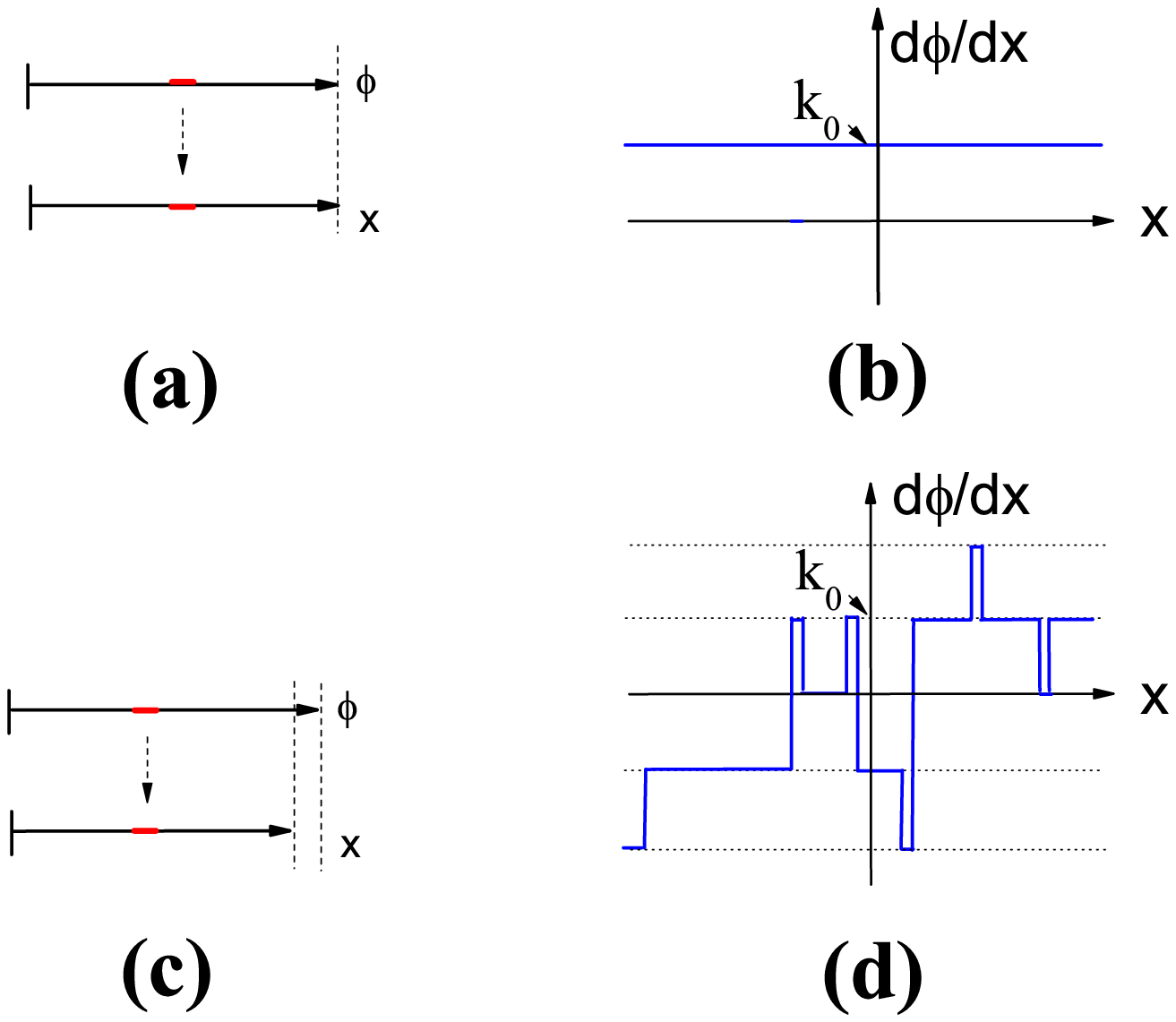}\caption{(Color online)
(a) A uniform variant of non-compact \textrm{\~{U}(1)} group that is a mapping
between the group-changing space to the one dimensional Cartesian space; (b)
The constant changing rate $d\phi/dx$ of a uniform variant of non-compact
\textrm{\~{U}(1)} group on the one dimensional Cartesian space. $k_{0}$
denotes the constant changing rate; (c) The mapping between the group-changing
space to one dimensional Cartesian space of a perturbative uniform variant of
non-compact \textrm{\~{U}(1)} group on the one dimensional Cartesian space;
(d) The changing rate of the non-uniform variant. }%
\end{figure}

Firstly, we discuss \emph{uniform variants}. The status of uniform variants in
variant theory is similar to the role of a constant group field in usual
mathematics. In the following parts, we abbreviate it by U-variant.

Then, we give the definition of a $d$-dimensional U-variant.

\textit{Definition -- d-dimensional U-variant }$V_{d}[\Delta \phi^{\mu},\Delta
x^{\mu},k_{0}^{\mu}]$\textit{ for group-changing space} $\mathrm{C}%
_{\mathrm{\tilde{G}},d}(\Delta \phi^{\mu})$\textit{ of non-compact Lie group}
\textrm{\~{G}} \textit{is defined by a perfect, ordered mapping between a
d-dimensional Clifford group-changing space} $\mathrm{C}_{\mathrm{\tilde{G}%
},d}(\Delta \phi^{\mu})$ \textit{and the d-dimensional Cartesian space
}$\mathrm{C}_{d}$\textit{, i.e., }%
\begin{equation}
V_{\mathrm{\tilde{G},}d}[\Delta \phi^{\mu},\Delta x^{\mu},k_{0}^{\mu}]:\{
\delta \phi^{\mu}\} \Leftrightarrow \{ \delta x^{\mu}\}.
\end{equation}
\textit{where }$\Leftrightarrow$\textit{\ denotes an ordered mapping under
fixed changing rate of integer }$k_{0}^{\mu},$\textit{\ and }$\mu$\textit{
labels the spatial direction. The adjective "perfect" means the total size
}$\Delta \phi^{\mu}$\textit{ of }$\mathrm{C}_{\mathrm{\tilde{G}},d}%
$\textit{\ exactly matches the total size }$\Delta x^{\mu}$\textit{ of
}$\mathrm{C}_{d}$\textit{, i.e., }$\Delta \phi^{\mu}=k_{0}^{\mu}\Delta x^{\mu}%
$. See the illustration in Fig.4(a).

For 1D U-variant $V_{\mathrm{\tilde{U}(1),}1}[\Delta \phi,\Delta x,k_{0}]$ of
non-compact $\mathrm{\tilde{U}(1)}$ Lie group, there exists only one type of
group-changing elements,
\begin{equation}
\delta \phi_{i}\equiv k_{0}\delta x_{i}%
\end{equation}
with fixed changing rate $\frac{d\phi}{dx}=k_{0}=\frac{\pi}{a}.$ Now, the
ordered mapping can be denoted by the series of same number "$1$", i.e.,
\begin{equation}
\{n_{i}\}=(...1,1,1,1,...).
\end{equation}
This number series indicates uniformity of a variant. Fig.4(b) illustrates of
a 1D U-variant\textit{ }$V_{\mathrm{\tilde{U}(1),}d}[\Delta \phi,\Delta
x,k_{0}]$ via $\phi$ and its changing rate $\frac{d\phi}{dx}$.

For higher dimensional U-variants, the $d$-dimensional infinitesimal element
of group-changing space is denoted by $d$ series of same number "$1$"
\begin{align}
&  V_{\mathrm{\tilde{G},}d}[\Delta \phi,\Delta x^{\mu},k_{0}^{\mu}]\nonumber \\
&  :\{n_{i}^{\mu}\}=\{%
\begin{array}
[c]{c}%
...1^{\mu},1^{\mu},1^{\mu},1^{\mu},...
\end{array}
\}.
\end{align}
Therefore,\ for a higher dimensional U-variant, the phase angles $\phi^{\mu}$
along different spatial directions belong to different group generators
$T^{\mu}$ of the non-compact Lie group $\mathrm{\tilde{G}}$.

\subparagraph{P-variant}

Another example is perturbative uniform variant. To obtain a perturbative
uniform variant, one can do \emph{perturbatively} changings on a uniform one.

We then give the definition of a perturbative uniform variant. In the
following parts, we abbreviate it by P-variant.

\textit{Definition -- d-dimensional P-variant }$V_{\mathrm{\tilde{G},}%
d}[\Delta \phi^{\mu},\Delta x^{\mu},k_{0}^{\mu}]$\textit{ for group-changing
space} $\mathrm{C}_{\mathrm{\tilde{G}},d}(\Delta \phi^{\mu})$ \textit{of
non-compact Lie group }\textrm{\~{G}} \textit{is defined by a quasi-perfect,
ordered mapping between a d-dimensional Clifford group-changing space}
$\mathrm{C}_{\mathrm{\tilde{G}},d}(\Delta \phi^{\mu})$ \textit{and the
d-dimensional Cartesian space }$\mathrm{C}_{d}$\textit{, i.e., }%
\begin{equation}
V_{\mathrm{\tilde{G},}d}[\Delta \phi^{\mu},\Delta x^{\mu},k_{0}^{\mu}]:\{
\delta \phi^{\mu}\} \Leftrightarrow \{ \delta x^{\mu}\}.
\end{equation}
\textit{where }$\Leftrightarrow$\textit{\ denotes an ordered mapping under
fixed changing rate of integer multiple }$k_{0}^{\mu},$\textit{\ and }$\mu
$\textit{ labels the spatial direction. The adjective "quasi-perfect" means
the total size }$\Delta \phi^{\mu}$\textit{ of }$\mathrm{C}_{\mathrm{\tilde{G}%
},d}$\textit{\ doesn't exactly match the total size }$\Delta x^{\mu}$\textit{
of }$\mathrm{C}_{d}$\textit{, i.e., }$\Delta \phi^{\mu}\neq k_{0}^{\mu}\Delta
x^{\mu},$\textit{ and} $\left \vert (\Delta \phi^{\mu}-k_{0}^{\mu}\Delta x^{\mu
})/\Delta \phi^{\mu}\right \vert \ll1$. See the illustration in Fig.5(c).
According to above mismatch condition $\Delta \phi^{\mu}\neq k_{0}^{\mu}\Delta
x^{\mu},$ and $\left \vert (\Delta \phi^{\mu}-k_{0}^{\mu}\Delta x^{\mu}%
)/\Delta \phi^{\mu}\right \vert \ll1$, for a P-variant, there must exist more
than one type of group-changing elements on it.

We take 1D P-variant $V_{\mathrm{\tilde{U}(1),}1}[\Delta \phi,\Delta x,k_{0}]$
of non-compact $\mathrm{\tilde{U}(1)}$ Lie group as an example to explain the concept.

We have
\begin{equation}
\{n_{i}\}=(...1,0,1,1,2,...0,1,1,1..).
\end{equation}
Here, "$0$" denotes a local contraction on group-changing space of the
original U-variant; "$2$", denotes local expansion on group-changing space of
the original U-variant. However, the word "perturbative" indicates that the
number of the group-changing elements "$1$" is much larger than all others,
"$0$", "$2$", ... See illustration in Fig.5(d).

For example, there are two types of P-variant -- one is about tiny contraction
on group-changing space of the original U-variant with only "$0$" and "$1$"
group-changing elements, i.e.,
\begin{equation}
\{n_{i}\}=(...1,0,1,1,...),
\end{equation}
the other is about tiny expansion on group-changing space of the original
variant with only "$1$" and "$2$" group-changing elements, i.e.,%
\begin{equation}
\{n_{i}\}=(...1,1,1,2,...).
\end{equation}
For both types of P-variant, there exist two kinds of group-changing elements
$\delta \phi^{A},$ $\delta \phi^{B}$ on d-dimensional Cartesian space
$\mathrm{C}_{d}$. The perturbative condition becomes
\begin{align}
\Delta \phi^{\mu}  &  =\sum \limits_{i}\delta \phi^{A}+\sum \limits_{j}\delta
\phi_{j}^{B},\nonumber \\
\left \vert \sum \limits_{i}\delta \phi^{A}\right \vert  &  \gg \left \vert
\sum \limits_{j}\delta \phi_{j}^{B}\right \vert .
\end{align}

One can see that the U-variants look like the ground states (vacuum), and the
P-variants look like the excited states in quantum physics.

\subparagraph{"Complementary pair" of two variants}

Finally, we introduce the concept of complementary of two variants.

\textit{Definition -- complementary of variants: For two} variants
$V_{\mathrm{\tilde{G},}d}[\Delta \phi^{\mu},\Delta x^{\mu},k_{0}^{\mu}]$ and
$V_{\mathrm{\tilde{G},}d}^{\prime}[\Delta \phi^{\mu},\Delta x^{\mu},k_{0}^{\mu
}],$\textit{ we call them complementary, if the series of numbers of
infinitesimal space elements of two variants }$\{n_{i}\}$\textit{ and
}$\{n_{i}^{\prime}\}$\textit{\ satisfy the following condition, }%
\[
\{n_{i}\}+\{n_{i}^{\prime}\}=\{n_{i}+n_{i}^{\prime}\}=\{1\} \mathit{.}%
\]

\begin{figure}[ptb]
\includegraphics[clip,width=0.8\textwidth]{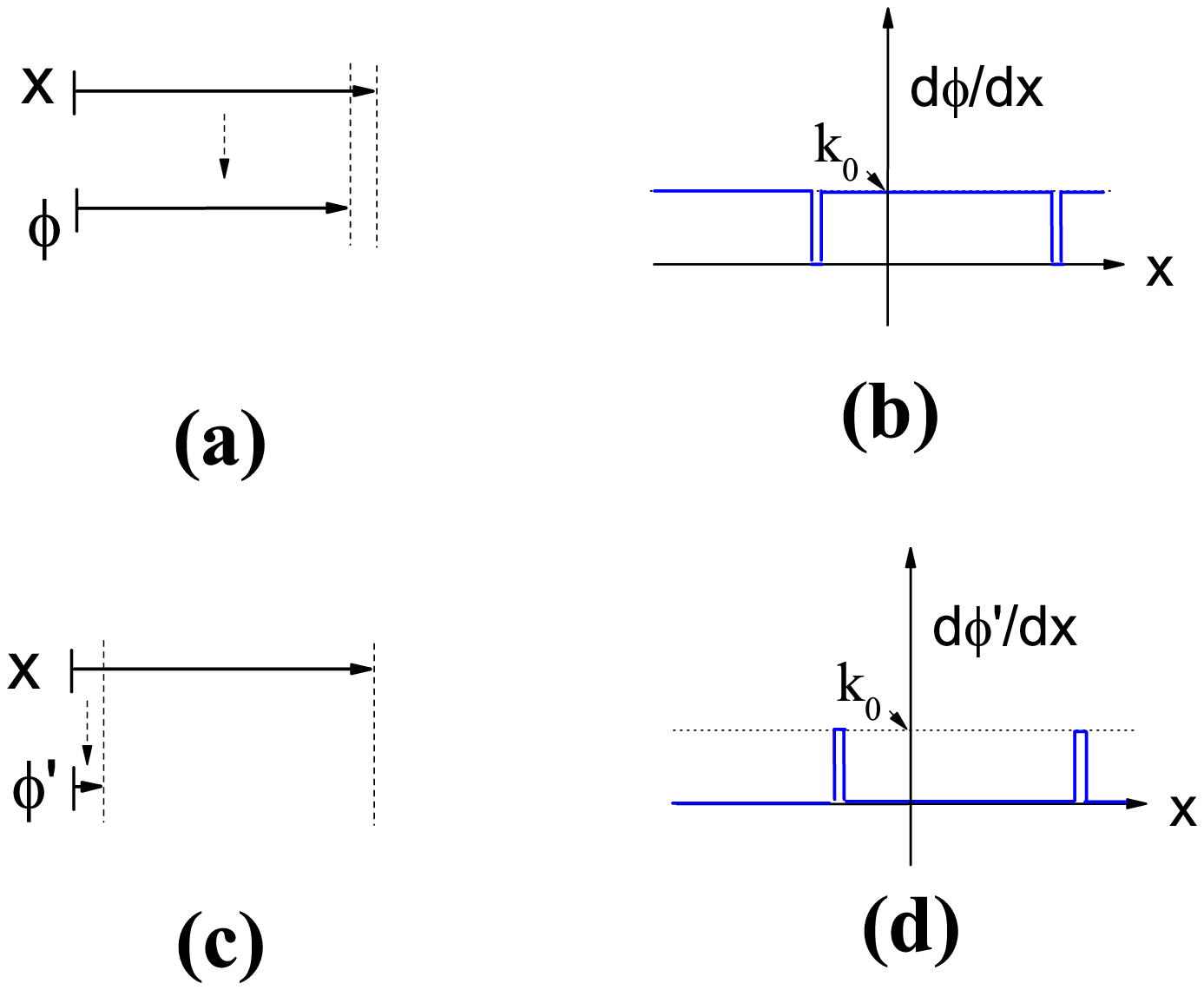}\caption{(Color online)
(a) A variant $V_{\mathrm{\tilde{U}(1),}1}[\Delta \phi,\Delta x,k_{0}]$ of
non-compact \textrm{\~{U}(1)} group that is a mapping between the
group-changing space to the one dimensional Cartesian space; (b) The changing
rate $d\phi/dx$ of $V_{\mathrm{\tilde{U}(1),}1}[\Delta \phi,\Delta x,k_{0}]$;
(c) The mapping of the complementary pair $V_{\mathrm{\tilde{U}(1),}1}%
^{\prime}[\Delta \phi,\Delta x,k_{0}]$ of $V_{\mathrm{\tilde{U}(1),}1}%
[\Delta \phi,\Delta x,k_{0}]$; (d) The changing rate $d\phi/dx$ of
$V_{\mathrm{\tilde{U}(1),}1}^{\prime}[\Delta \phi,\Delta x,k_{0}].$}%
\end{figure}

That means we can add two variants that satisfy complementary condition and
get a uniform one. Or, we can obtain a variant by choosing an U-variant
subtracting its complementary pair.

We also take 1D P-variants of the non-compact \textrm{\~{U}(1)} group as an
example to explain the concept of "complementary pair" of two variants.

The original U-variant for $V_{\mathrm{\tilde{U}(1),}1}[\Delta \phi,\Delta
x,k_{0}]$ is described by a series of number "$1$", $\{n_{i}%
\}=(...1,1,1,1,...)$. For a 1D P-variant $V_{\mathrm{\tilde{U}(1),}1}%
[\Delta \phi,\Delta x,k_{0}]$ that is described by a series of number "$1$" and
"$0$", $\{n_{i}\}=(...1,0,1,1,...),$ the complementary pair $V_{\mathrm{\tilde
{U}(1),}1}^{\prime}[\Delta \phi,\Delta x,k_{0}]$ is described by $\{n_{i}%
\}=(...0,1,0,0,...)$; For a 1D P-variant $V_{\mathrm{\tilde{U}(1),}1}%
[\Delta \phi,\Delta x,k_{0}]$ that is described by a series of number "$1$" and
"$2$", $\{n_{i}\}=(...1,1,1,2,...),$ the complementary pair $V_{\mathrm{\tilde
{U}(1),}1}^{\prime}[\Delta \phi,\Delta x,k_{0}]$ is described by $\{n_{i}%
\}=(...0,0,0,-1,...)$. Therefore, under varying reference from a natural
reference to an U-variant, $V_{\mathrm{\tilde{U}(1),}1}[\Delta \phi,\Delta
x,k_{0}]$ will change into its complementary pair $V_{\mathrm{\tilde{U}(1),}%
1}^{\prime}[\Delta \phi,\Delta x,k_{0}].$ See the illustration in Fig.5. In the
following parts, in certain cases, for simplicity, we use $V_{\mathrm{\tilde
{U}(1),}1}^{\prime}[\Delta \phi,\Delta x,k_{0}]$ to characterize
$V_{\mathrm{\tilde{U}(1),}1}[\Delta \phi,\Delta x,k_{0}]$.

\subsubsection{Significance of variant: higher-order variability}

It was known that a variant is a configuration of particular distribution of a
lot of group-changing elements $\delta \tilde{\phi}$. \emph{What's relationship
between a general variant and a usual field?}

On one hand, they are quite different. A variant is an ordered mapping between
group-changing space and Cartesian space; while a usual group field is a
(disordered) mapping between group space and Cartesian space. In usual group
field $g(x),$ the element object is "\emph{group element}"; while in a variant
$V_{\mathrm{\tilde{G},}d}[\Delta \phi^{\mu},\Delta x^{\mu},k_{0}^{\mu}]$ the
element object is "\emph{group-changing element}" $\delta \phi$. A variant
represents the object of nonlocal group operations; while the usual group
field $g(x)$ represents the object of local group operations. As a result, we
say that variants characterize "\emph{changing}" structure, while fields
characterize "\emph{non-changing}" structure.

On the other hand, a variant can be regarded as a \emph{special} group field
under \emph{global/local constraints}. The local constraint is about the fixed
changings of changing rate, i.e., $\frac{\delta \phi^{\mu}}{\delta x^{\mu}%
}=nk_{0}^{\mu}$ where $n$ is an integer number and $k_{0}^{\mu}$ is fixed.
This is certain "\emph{quantization condition}" enforced on a function. On the
contrary, for usual group field, $\frac{\delta \phi^{\mu}}{\delta x^{\mu}}$ can
arbitrarily change without additional condition. The other is about global
constraint with fixed size of the group-changing space, i.e., $\Delta \phi
^{\mu}$ are topological numbers. For usual group field, there is no such
constraint. To show the relationship between a general variant and an usual
field more clearly, we take 1D variant $V_{\mathrm{\tilde{U}(1),}1}[\Delta
\phi,\Delta x,k_{0}]$ as example that is one dimensional group-changing space
$\mathrm{C}_{\mathrm{\tilde{U}(1)},1}(\Delta \phi)$\textit{ }on Cartesian space
$\mathrm{C}_{1}$ with fixed changing rate of integer multiple $k_{0}$.\textit{
}Different 1D variants $V_{\mathrm{\tilde{U}(1),}1}[\Delta \phi,\Delta
x,k_{0}]$ are characterized by different distributions of $n_{i}$. As a
result, in some sense, a variant is a "function" of integer number $n_{i}.$ On
the contrary, for a usual group field described by function $g(x),$%
\emph{\ }$n_{i}$\emph{\ }is freely varied fractional/integer number.

\begin{figure}[ptb]
\includegraphics[clip,width=0.81\textwidth]{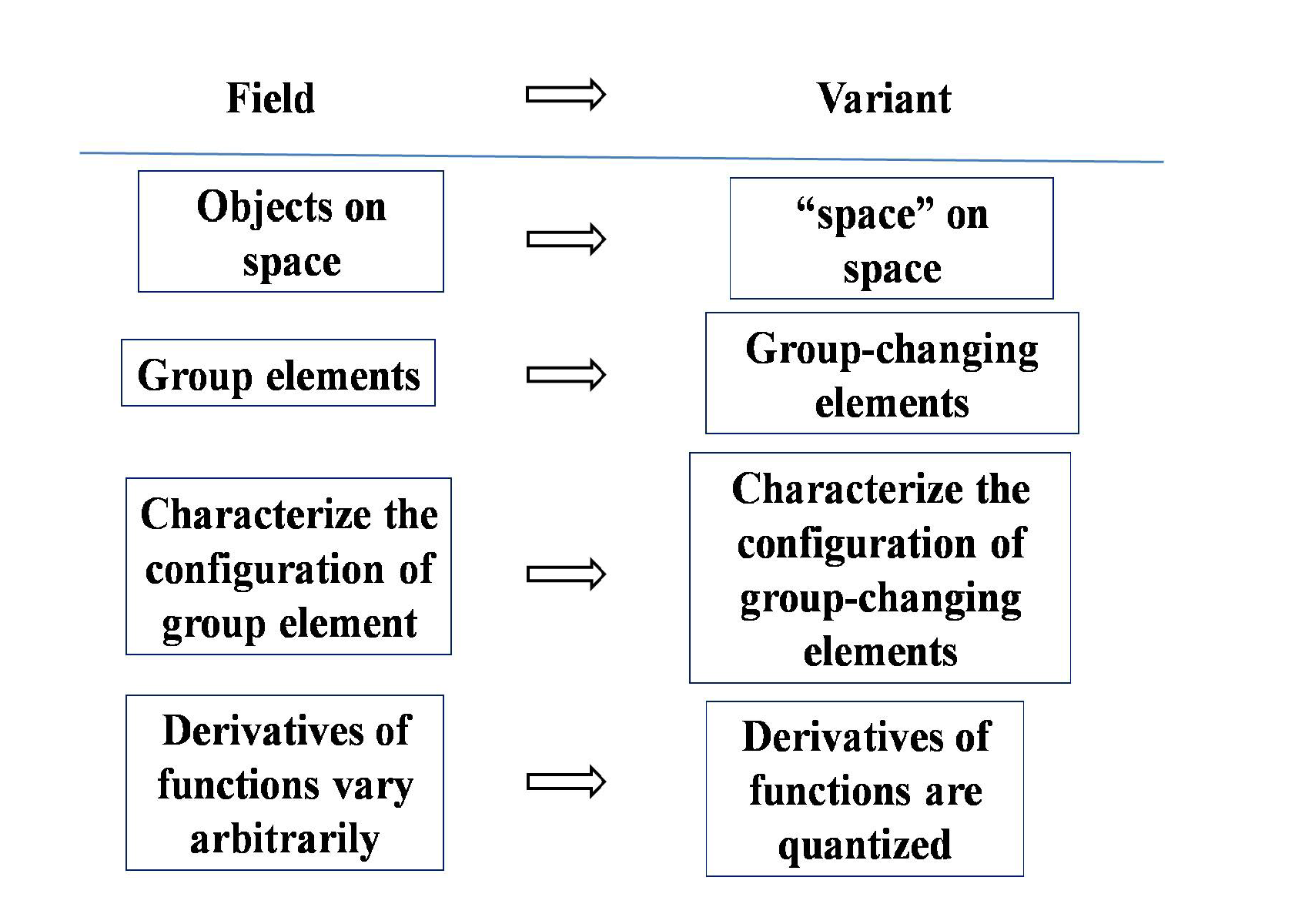}\caption{(Color online)
The comparison between usual field that is a function on space and variant
that is "space" on space.}%
\end{figure}

Therefore, in some sense, variant can be regarded as a special "field" under
constraint. Fig.6 is a table to show the difference between a field and a
variant (or a "space"). However, the significance of variant is the concept of
higher-order variability. Variant is a phenomenon of "\emph{changing}", i.e.,
an ordered space mapping between $\mathrm{C}_{\mathrm{\tilde{G}},d}(\Delta
\phi^{\mu})$ and\textit{ }$\mathrm{C}_{d}.$ To characterize this phenomenon,
we introduce the concept of "\emph{Higher-order} \emph{variability}".

In modern physics and modern mathematics, "\emph{symmetry}" or
"\emph{invariant}" is an important concept. For a usual group field with
uniform distribution $g(x)=g(x_{0})$ of (compact/non-compact) Lie group
\textrm{G} on $d$-dimensional Cartesian space\textit{ }$\mathrm{C}_{d}$ with
infinite size ($\Delta x^{\mu}\rightarrow \infty$), we have the following
correspondence,%
\begin{align}
\mathcal{T}(\delta x^{\mu})  &  =1,\\
\hat{U}(\delta \phi^{\mu}(x))  &  =\hat{U}(\delta \phi^{\mu}(x_{0}))
\end{align}
where $\mathcal{T}(\delta x^{\mu})$ is the spatial translation operation on
$\mathrm{C}_{d}$ along $x^{\mu}$-direction and $U(\delta \phi^{\mu})$ is local
group operation on the system that is space independent. We say that the
system has a global symmetry of (compact/non-compact) Lie group \textrm{G}.
That means when we do a global group operation, the system is globally rotating.

Another key point of this paper is to generalize "\emph{symmetry/invariant}"
of usual field to \emph{(higher-order)} \emph{variability. }This is a highly
non-trivial generalization.

For a U-variant with infinite size ($\Delta x\rightarrow \infty$), we have the
following relationship,%
\begin{equation}
\mathcal{T}(\delta x^{\mu})\leftrightarrow \hat{U}(\delta \phi^{\mu}%
)=e^{i\cdot \delta \phi^{\mu}T^{\mu}}%
\end{equation}
where $\mathcal{T}(\delta x^{\mu})$ is the spatial translation operation on
$\mathrm{C}_{d}$ along $x^{\mu}$-direction and $\hat{U}(\delta \phi^{\mu})$ is
shifting operation on group-changing space $\mathrm{C}_{\mathrm{\tilde{G}}%
,d}(\Delta \phi^{\mu})$, and $\delta \phi^{\mu}=k_{0}^{\mu}\delta x^{\mu}$. That
means when one translates along Cartesian space $\delta x^{\mu},$ the
corresponding shifting along group-changing space $\mathrm{C}_{\mathrm{\tilde
{G}},d}$ is $\delta \phi^{\mu}=k_{0}^{\mu}\delta x^{\mu}.$ We can regard usual
"\emph{symmetry/invariant}" to be \emph{zero-order variability} and
$V_{\mathrm{\tilde{G},}d}[\Delta \phi^{\mu},\Delta x^{\mu},k_{0}^{\mu}]$\ to be
a system with\emph{ 1-st order variability}! Then, variant has higher order
variability; while the variability of a usual group field is zero order. In
some sense, a variant can be regarded as \emph{"higher-order field"}.

In brief, the order of variability becomes a key value classifying the
complexity of mathematical systems. We point out that there may exist
mathematic objects with much higher-order variability, such as mathematic
objects of 2-nd order variability. In next part, we will show its highly
non-trivial application on quantum gauge theory.

\subsubsection{Classification of changings for variants}

The changings of variants are prelude of quantum motions in physics, by which
we could change one variant to another. For example, from point view of
changings of variant, each P-variant can be obtained by doing perturbative
changings on an U-variant.

Everyone is familiar with the changings of a usual group field. This is just
the changings of the its function, i.e., $g(x)\rightarrow g^{\prime}(x).$
However, for a variant, the situation becomes complex. There are two types of
changings of a variant $V_{\mathrm{\tilde{G},}d}[\Delta \phi^{\mu},\Delta
x^{\mu},k_{0}^{\mu}]$: one is topological, the other is non-topological. For
topological changings, the group-changing space of it is globally expand or
contract\emph{ }on Cartesian space $\mathrm{C}_{d}$. For non-topological
changings, there are global shift, local expansion/contraction and shape
changings. Let's give more detailed discussion.

We then classify the changings of variants. There are five types of
changings\ of the $d$-dimensional variant $V_{\mathrm{\tilde{G},}d}[\Delta
\phi^{\mu},\Delta x^{\mu},k_{0}^{\mu}]$:

1) Globally \emph{shifting} $\mathrm{C}_{\mathrm{\tilde{G}},d}(\Delta \phi
^{a})$ without changing its size on Cartesian space $\mathrm{C}_{d}$: The
operation of such a globally shift is $\hat{U}(\delta \phi^{a})=e^{i\delta
\phi^{a}T^{a}}$. Now, the size of it is still $\Delta \phi^{a}$. Under globally
shifting, the U-variant is invariant. Therefore, this is a symmetric operation
on a U-variant, such as the global phase symmetry. In the following part, we
point out that this type of time-dependent changings of a variant corresponds
to global phase rotation;

2) Globally \emph{rotating} $\mathrm{C}_{\mathrm{\tilde{G}},d}(\Delta \phi
^{a})$ from $a$-direction to $b$-direction on Cartesian space $\mathrm{C}_{d}%
$: The operation is $\hat{U}(\delta \varphi^{ab})=e^{\delta \varphi^{ab}T^{ab}}$
that rotates $T^{a}$ to $T^{b}$. The operation of globally rotation obeys a
compact Lie group and thus will not change the U-variant. Therefore, this is
also a symmetric operation on an U-variant. In the following part, we point
out that this type of time-dependent changings of a variant corresponds to
global rotation;

3) Globally \emph{expanding} or \emph{contracting} $\mathrm{C}_{\mathrm{\tilde
{G}},d}(\Delta \phi^{a})$\ with changing its corresponding size on Cartesian
space $\mathrm{C}_{d}$: The operation of contraction/expansion on
group-changing space is $\tilde{U}(\delta \phi^{a})=e^{i((\delta \phi^{a}%
T^{a})\cdot \hat{K}^{a})}$ where $\delta \phi^{a}=(\Delta \phi^{a})^{\prime
}-\Delta \phi^{a}$ and $\hat{K}^{a}=-i\frac{d}{d\phi^{a}}.$ This process
changes the size of the group-changing space. In the following part, we point
out that globally expansion/contraction of group-changing space in a variant
corresponds to the generation/annihilate of particles in quantum mechanics;

4) Locally rotating\emph{ }on Cartesian space $\mathrm{C}_{d}$: Locally
rotating of $\mathrm{C}_{\mathrm{\tilde{G}},d}(\Delta \phi^{a})$ on Cartesian
space $\mathrm{C}_{d}$\ ($d>1$) leads to the \emph{shape} of system locally
changing.\emph{ }This type of changings of a variant leads to a curving
spacetime and is irrelevant to the issue of this paper;

5) Locally \emph{expanding} or \emph{contracting} $\mathrm{C}_{\mathrm{\tilde
{G}},d}(\Delta \phi^{a})$\ without changing its corresponding size on Cartesian
space $\mathrm{C}_{d}$: The operation of contraction/expansion on
group-changing space becomes local. In the following part, we point out that
this type of time-dependent changings of a variant corresponds to the motion
of elementary particles in quantum mechanics with fixed particle's number;

In this part, we will focus on this globally/locally \emph{expanding} or
\emph{contracting} $\mathrm{C}_{\mathrm{\tilde{G}},d}(\Delta \phi^{a}%
)$\ with/without changing its corresponding size on Cartesian space
$\mathrm{C}_{d}$ in a P-variant.

\subsubsection{Representations for variants}

In this section, we discuss the representations for variants from 1D variants
to higher dimensional cases.

A variant is a mathematical object with 1-st order variability, of which the
representation is much more complex than usual classical field with 0-th order
variability. The reason comes from the existence of "\emph{projection}" on a
mathematical object with higher order variability. "Projection" is a procedure
reducing the variability order, for example, from variant with 1-st order
variability to a classical field with 0-th order variability. Therefore, to
characterize the "ability" about describing the highest order of variability,
different representations are classified by order of characterizing the
corresponding highest order of variability. In general, for variants with 1-st
order variability, there are two types of representations: one is 1-st order
without doing projection that is a complete, non-local description showing
"changing" structure, the other is 0-th order under knot projection that is an
incomplete, local description showing "non-changing" structure.

\paragraph{Representations for 1D variant}

We firstly study the representations for 1D variant of non-compact
\textrm{\~{U}(1)} group from U-variants to P-variants.

\subparagraph{Representations for 1D U-variant of non-compact
\textrm{\~{U}(1)} group}

1) 1-st order representations without projection

For a 1D U-variant of non-compact \textrm{\~{U}(1)} group,\textrm{ }there are
three kinds of 1-st order representations from different aspects, including
\emph{algebra}, \emph{geometry}, and \emph{algebra}, representations,
respectively. We also call them 1-st order algebra, geometry, and algebra,
representations. In general, people can transform one representation to
another to characterize the same U-variant.

\textit{1-st order algebra representation}: In 1-st order algebra
representation, the 1D U-variant is characterized by a series of (non-local)
group-changing elements of non-compact \textrm{\~{U}(1)} group according to
the definition of variants.

For an U-variant $V_{\mathrm{\tilde{U}(1),}1}[\Delta \phi,\Delta x,k_{0}]$
denoted by $\{n_{i}\}=(...1,1,1,1,...)$, there exists only one type of
group-changing elements with fixed changing rate $\frac{d\phi}{dx}=k_{0}%
=\frac{\pi}{a}.$ We can "\emph{generate}" the 1D variant by a series of
group-changing elements $\delta \phi_{i}(x_{i})$ on every position $x$ of
Cartesian space $\mathrm{C}_{1}$, i.e., $\tilde{U}(\delta \phi)=\prod_{i}%
\tilde{U}(\delta \phi_{i}(x_{i}))$ with\ $\tilde{U}(\delta \phi_{i}%
(x_{i}))=e^{i((\delta \phi_{i})\cdot \hat{K})}$ and $\hat{K}=-i\frac{d}{d\phi}%
.$\ Here, the i-th infinitesimal group-changing operation $\tilde{U}%
(\delta \phi_{i})$ generates a group-changing element on position $i$.

\begin{figure}[ptb]
\includegraphics[clip,width=0.92\textwidth]{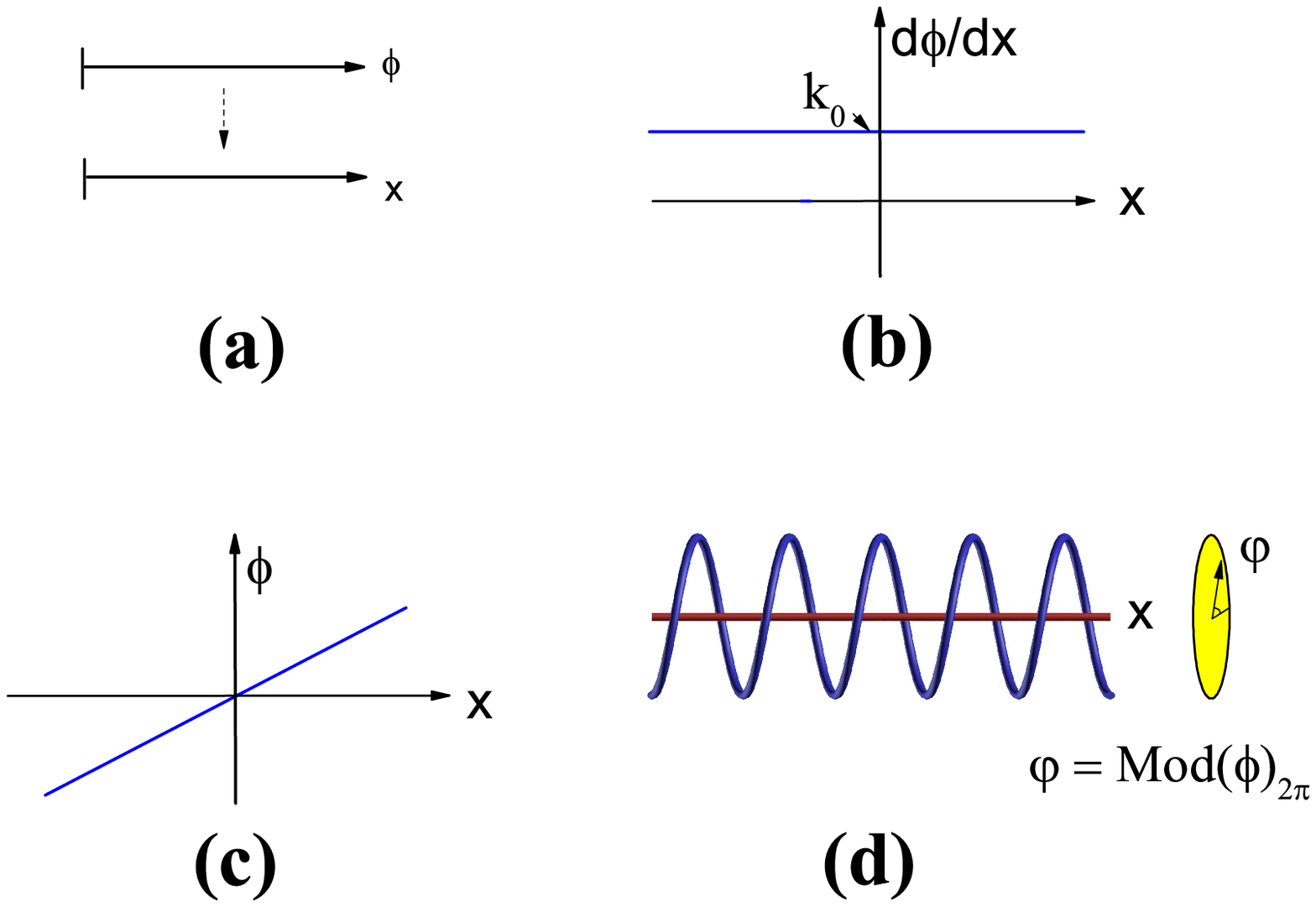}\caption{(Color online)
(a) A mapping of uniform variant of non-compact \textrm{\~{U}(1)} group; (b)
The changing rate $d\phi/dx$ of the one dimensional uniform variant; (c) 1-st
order algebra representation of the uniform distribution; (d) 1-st order
Geometry representation of the uniform variant. The uniform variant
corresponds to a spiral line on a cylinder with fixed radius that can be
regarded as a knot/link structure between the curved line and the straight
line at center.}%
\end{figure}

\textit{1-st order algebra representation}\textbf{:} In 1-st order algebra
representation, the 1D variant is usually described by a complex field
$\mathrm{z}=e^{i\phi(x)}$. To obtain its algebra representation, we must set
a \emph{reference}. In general, we have a\ natural choice, $\mathrm{z}%
_{0}=e^{i\phi_{0}}$. In the following parts, we call it "\emph{natural
reference}". We then do group-changing operation on natural reference and get
the 1-st order algebra representation of the corresponding variants.

The complex field $\mathrm{z}_{u}(x)$ for a U-variant is obtained by
\begin{equation}
\mathrm{z}_{u}(x)=\tilde{U}(\delta \phi)\mathrm{z}_{0}%
\end{equation}
where $\tilde{U}(\delta \phi)=\prod_{i}\tilde{U}(\delta \phi_{i}(x_{i}))$
denotes a series of group-changing operations\ with $\tilde{U}(\delta \phi
_{i}(x_{i}))=e^{i((\delta \phi_{i})\cdot \hat{K})}$ and $\hat{K}=-i\frac
{d}{d\phi}.$\ Here, the i-th group-changing operation $\tilde{U}(\delta
\phi_{i}(x))$ at $x$ generates a group-changing element.\ For the case of a
single group-changing element $\delta \phi_{i}(x_{i})$ on $\delta x_{i}$ at
$x_{i},$ the function is given by
\begin{equation}
\phi(x)=\left \{
\begin{array}
[c]{c}%
-\frac{\delta \phi_{i}}{2},\text{ }x\in(-\infty,x_{i}]\\
-\frac{\delta \phi_{i}}{2}+k_{0}x,\text{ }x\in(x_{i},x_{i}+\delta x_{i}]\\
\frac{\delta \phi_{i}}{2},\text{ }x\in(x_{i}+\delta x_{i},\infty)
\end{array}
\right \}  .
\end{equation}

Finally, under natural reference, a 1D U-variant $V_{\mathrm{\tilde{U}(1),}%
1}[\Delta \phi,\Delta x,k_{0}]$ can be described by a special complex field
$\mathrm{z}_{u}(x)$ in Cartesian space as
\[
\mathrm{z}_{u}(x)=\exp(i\phi(x))
\]
where $\phi(x)=\phi_{0}+k_{0}x$. See Fig.7 (c).

\textit{1-st order geometry representation}\textbf{:} For usual group field,
the geometry representation provides a clear picture for the configurations of
group elements in a variant. Therefore, by using 1-st order geometry
representation, we have a "\emph{changing}" picture for the configurations of
group-changing elements, of which the 1D variant shows the highly non-local
geometric structure -- \emph{knot/links}.

For the 1D variant $V_{\mathrm{\tilde{U}(1),}1}[\Delta \phi,\Delta x,k_{0}]$ of
non-compact \textrm{\~{U}(1)} group, we map the original complex field
$\mathrm{z}_{u}(x)=\exp(ik_{0}x+i\phi_{0})=\operatorname{Re}\xi
(x)+i\operatorname{Im}\eta(x)$ for a variant to a curved line $\{x,\xi
(x),\eta(x)\}$ in three dimensions. In Fig.7(d), an U-variant corresponds to a
spiral line on a cylinder with fixed radius that can be regarded as a
knot/link structure between the curved line of $\mathrm{z}_{u}(x)$ and the
straight line at center of $\mathrm{z}(x)=0$.

2) 0-th order representations under knot projection

In the above section, we introduce a 1-st order geometry representation for a
variant that can be regarded as knot/link. It is known that a knot/link can be
projected by counting the crossings (or zeros named in this paper) of the
corresponding lines. With the help of the knot projection (K-projection),
people can partially obtain the information of the variant.

We then introduce the\emph{ }K-projection of the curved line of 1D U-variant
along a given direction $\theta$ on the straight line at the center of
$\mathrm{z}(x)=0$ in 2D space $\{ \xi(x),\eta(x)\}$.

In mathematics, the K-projection is defined by
\begin{equation}
\hat{P}_{\theta}\left(
\begin{array}
[c]{c}%
\xi(x)\\
\eta(x)
\end{array}
\right)  =\left(
\begin{array}
[c]{c}%
\xi_{\theta}(x)\\
\left[  \eta_{\theta}(x)\right]  _{0}%
\end{array}
\right)
\end{equation}
where $\xi_{\theta}(x)$ is variable and $\left[  \eta_{\theta}(x)\right]
_{0}$ is constant. In the following parts we use $\hat{P}_{\theta}$ to denote
the projection operators. Because the projection direction out of the curved
line is characterized by an angle $\theta$ in $\{ \xi,\eta \}$ space, we have
\begin{equation}
\left(
\begin{array}
[c]{c}%
\xi_{\theta}\\
\eta_{\theta}%
\end{array}
\right)  =\left(
\begin{array}
[c]{cc}%
\cos \theta & \sin \theta \\
\sin \theta & -\cos \theta
\end{array}
\right)  \left(
\begin{array}
[c]{c}%
\xi \\
\eta
\end{array}
\right)
\end{equation}
where $\theta$ is angle \textrm{mod}($2\pi$), i.e. $\theta \operatorname{mod}%
2\pi=0.$ So the curved line of 1D variant is described by the function
\begin{equation}
\xi_{\theta}(x)=\xi(x)\cos \theta+\eta(x)\sin \theta.
\end{equation}
In the following parts, we call $\theta \in \lbrack0,2\pi)$ \emph{projection
angle}.

Under projection, each zero corresponds to a solution of the equation
\begin{equation}
\hat{P}_{\theta}[\mathrm{z}(x)]\equiv \xi_{\theta}(x)=0.\nonumber
\end{equation}
We call the equation \emph{zero-equation} and its solutions to be\emph{
zero-solution}. Now, a 1D U-variant becomes a 1D crystal of zeros (or 1D zero
lattice). The "changing" structure of phase factor disappears. Then, we also
call it "local", geometry representation.

Let us show the detailed results from K-projection.

For a 1D U-variant $V_{\mathrm{\tilde{U}(1)},1}(\Delta \phi,\Delta x,k_{0})$ of
non-compact \textrm{\~{U}(1)} group, from the its algebra representation
$\mathrm{z}_{u}(x)\sim e^{ik_{0}\cdot x}$, from zero-equation $\xi_{\theta
}(x)=0$ or $\cos(k_{0}x-\theta)=0,$ we get the zero-solutions are
\begin{equation}
x=l_{0}\cdot N/2+\frac{l_{0}}{2\pi}(\theta+\frac{\pi}{2})
\end{equation}
where $N$ is an integer number, and $l_{0}=2\pi/k_{0}$. The zero density
$\rho_{\text{\textrm{zero}}}$ is
\begin{equation}
\rho_{\text{\textrm{zero}}}=\frac{k_{0}}{\pi}.
\end{equation}
Fig.8 shows a 1D crystal of zeros for a U-variant (we also call it zero
lattice). One can see that each crossing corresponds to a zero.

\begin{figure}[ptb]
\includegraphics[clip,width=0.72\textwidth]{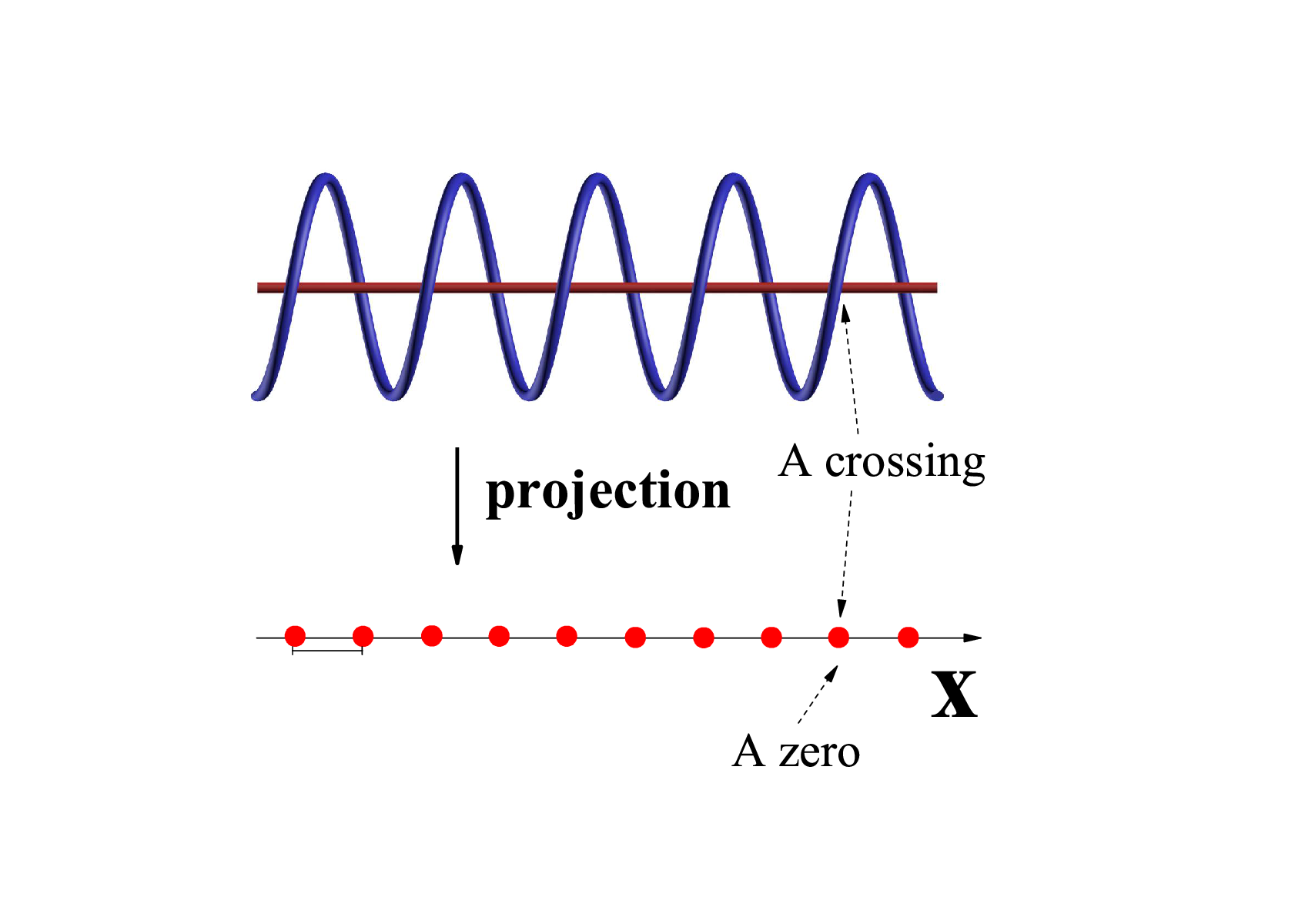}\caption{(Color online) A
1D crystal of zeros (or a zero lattice) for a projected uniform variant of
non-compact \textrm{\~{U}(1)} group. Each crossing corresponds to a zero.}%
\end{figure}

The zero lattice is "two-sublattice" with discrete spatial translation
symmetry. In other words, a unit cell with $2\pi$ phase change has two zeros.
The lattice distance is $l_{0}$. On the zero lattice, the group element is
projected to $\theta$ that is compact, i.e., $\theta \in \lbrack0,2\pi)$.
Consequently, after projection, the non-compact \textrm{\~{U}(1)} group of
$\phi(x)$ turns into a compact group on zero lattice of "two-sublattices",
i.e.,
\[
\phi(x)=2\pi N(x)+\theta.
\]
We then relabel the group-changing space $\mathrm{C}_{\mathrm{U(1)},1}%
(\Delta \phi)$ by two numbers ($N(x),\theta(N(x))$): $\theta(N(x))$ is compact
phase angle, the other is the integer winding number of unit cell of zero
lattice $N(x)$.

In summary, under K-projection, a U-variant turns into a uniform "crystal" of
zeros. In other words, a whole phase "\emph{changing}" structure is reduced
into a "\emph{non-changing}" configuration of points (zeroes) on space by
projected with fixed phase angle $\theta$. That means the projection is a
process to reduce an object with higher-order variability to a lower one. The
situation is similar to the measurement process in physics. To measure the
speed of a point-mass, one must determine the positions $x$ at given times
$t.$ This is a series "projection" processes that reduces a moving (or
"\emph{changing}") object to a static (or "non-changing") one. Therefore, we
may call the projection that reduces a object with higher-order variability to
a lower one to be "\emph{mathematic measurement}".

\subparagraph{1D P-variant of non-compact \textrm{\~{U}(1)} group}

In the above section, we provide different representations for 1D U-variant.
In this section, we discuss 1D P-variants. The approaches for 1D U-variants
can be easily generalized to the 1D P-variant of non-compact \textrm{\~{U}(1)}
group. For this case, $\mathrm{z}(x)$ is not uniform any more.

1) 1-st order representation without projection

In this section, we show the 1-st order representations for 1D P-variant
$V_{\mathrm{\tilde{U}(1),}1}[\Delta \phi,\Delta x,k_{0}]$. Without projection,
they are all representations that show complete information of the P-variant.

\textit{1-st order algebra representation}: In 1-st order algebra
representation, the 1D P-variant $V_{\mathrm{\tilde{U}(1),}1}[\Delta
\phi,\Delta x,k_{0}]$ is also characterized by a series of (non-local)
group-changing elements of non-compact \textrm{\~{U}(1)} group. In a sentence,
we can "\emph{generate}" the 1D P-variant by a series of group-changing
elements on every position $x$ of Cartesian space $\mathrm{C}_{1}$, i.e.,
\begin{equation}
\{n_{i}\}=(...1,0,1,1,2,...0,1,1,1..).
\end{equation}
The extra group-changing elements $0$ is denoted by $\tilde{U}(\delta \phi
_{i}(x_{i}))=e^{i((\delta \phi_{i})\cdot \hat{K})}$ with $\delta \phi_{i}=0$; the
extra group-changing elements $2$ is denoted by $\tilde{U}(\delta \phi
_{i}(x_{i}))=e^{i((\delta \phi_{i})\cdot \hat{K})}$ with $\delta \phi
_{i}\rightarrow2\delta \phi_{i}.$ However, the word "perturbative" indicates
that the number of the group-changing elements "$1$" is much larger than all
others, "$0$", "$2$", ...

\textit{1-st order algebra representation}\textbf{:} In 1-st order algebra
representation, the 1D P-variant $V_{\mathrm{\tilde{U}(1),}1}[\Delta
\phi,\Delta x,k_{0}]$ is usually described by a complex field $\mathrm{z}%
=e^{i\phi(x)}$. To obtain its algebra representation, we also set a natural
reference, $\mathrm{z}_{0}=e^{i\phi_{0}}$. We then do non-local group-changing
operation on $\mathrm{z}_{0}$ and get the non-local algebra representation
of the corresponding P-variants.

The complex field $\mathrm{z}_{p}(x)$ for an U-variant is obtained by
\begin{equation}
\mathrm{z}_{p}(x)=\tilde{U}(\delta \phi)\mathrm{z}_{0}%
\end{equation}
where $\tilde{U}(\delta \phi)=\prod_{i}\tilde{U}(\delta \phi_{i}(x_{i}))$
denotes a series of group-changing operations\ with $\tilde{U}(\delta \phi
_{i}(x_{i}))=e^{i((\delta \phi_{i})\cdot \hat{K})}$ and $\hat{K}=-i\frac
{d}{d\phi}.$\ Here, the i-th group-changing operation $\tilde{U}(\delta
\phi_{i}(x))$ at $x$ generates a group-changing element.

In addition, we have another approach to "generate" a P-variant by doing
non-local group-changing operation $\tilde{U}(\delta \phi^{B})$ of extra
group-changing elements $\delta \phi_{i}^{B}(x_{i})$ on U-variant. Now, the
P-variant is designed by adding a distribution of the extra group-changing
elements $\delta \phi_{i}^{B}(x_{i})$ on a 1D U-variant with a fixed total
phase changing $\Delta \phi^{B}=\sum \limits_{i}\delta \phi_{i}^{B}(x_{i}%
)\ll \Delta \phi$. Then, the original complex field $\mathrm{z}_{u}(x)$ for a
U-variant turns into another complex field $\mathrm{z}_{p}(x)$ for P-variant,
\begin{equation}
\mathrm{z}_{u}(x)\rightarrow \mathrm{z}_{p}(x)=\tilde{U}(\delta \phi
^{B})\mathrm{z}_{u}%
\end{equation}
where $\tilde{U}(\delta \phi^{B})=\prod_{i}\tilde{U}(\delta \phi_{i}^{B}%
(x_{i}))$ denotes a series of extra group-changing operations\ with $\tilde
{U}(\delta \phi_{i}^{B}(x_{i}))=e^{i((\delta \phi_{i}^{B})\cdot \hat{K})}$ and
$\hat{K}=-i\frac{d}{d\phi}.$\ Here, the i-th group-changing operation
$\tilde{U}(\delta \phi_{i}^{B}(x))$ at $x$ generates a group-changing element.\

Finally, we give the results. For P-variant with extra "$0$", i.e.,
$\{n_{i}\}=(...1,0,1,1,...),$ the extra group-changing elements $\delta
\phi_{i}^{B}(x_{i})$ on a 1D U-variant are denoted by $\tilde{U}(\delta
\phi_{i}^{B}(x_{i}))=e^{i((\delta \phi_{i}^{B})\cdot \hat{K})}$ and $\hat
{K}=-i\frac{d}{d\phi}$. Then, we get
\begin{equation}
\phi(x)=\left \{
\begin{array}
[c]{c}%
\phi_{0},\text{ }x\in(-\infty,\infty]
\end{array}
\right \}  ;
\end{equation}
for the case with extra "$2$", i.e, $\{n_{i}\}=(...1,1,1,2,...),$ the extra
group-changing elements $\delta \phi_{i}^{B}(x_{i})$ on a 1D U-variant are
denoted by $\tilde{U}(\delta \phi_{i}^{B}(x_{i}))=e^{i((\delta \phi_{i}%
^{B})\cdot \hat{K})}$ and $\hat{K}=-i\frac{d}{d\phi}$. Then, we get
\begin{equation}
\phi(x)=\left \{
\begin{array}
[c]{c}%
-\frac{\delta \phi_{i}}{2},\text{ }x\in(-\infty,x_{i}]\\
-\frac{\delta \phi_{i}}{2}+2k_{0}x,\text{ }x\in(x_{i},x_{i}+\delta x_{i}]\\
\frac{3\delta \phi_{i}}{2},\text{ }x\in(x_{i}+\delta x_{i},\infty)
\end{array}
\right \}  .
\end{equation}

\textit{1-st order geometry representation: }We discuss the 1-st order
geometry representation for P-variant.

For P-variant described by the complex field $\mathrm{z}_{p}(x)$, it also
corresponds to a curved line on a cylinder with fixed radius. As shown in
Fig.9, if we consider a 1D variant to be a continuous line with fixed radius
that is described by $\mathrm{z}(x)$, such a continuous line and the line of
its center\ can also be regarded as knot/link. See the illustration in Fig.9,
which is an illustration of "complementary pair" of two variants under 1-st
order geometry representation.

\begin{figure}[ptb]
\includegraphics[clip,width=0.72\textwidth]{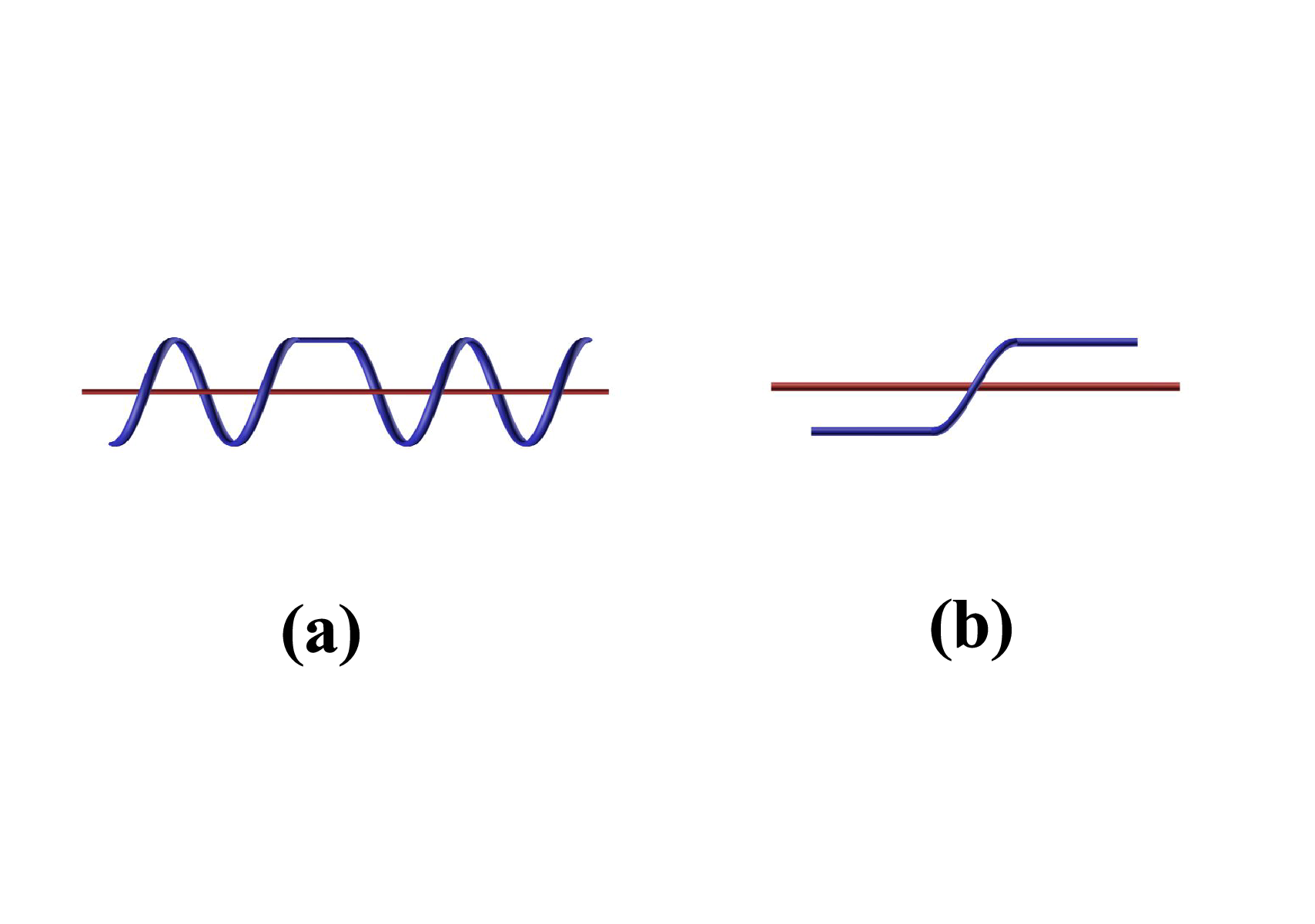}\caption{(Color online)
"Complementary pair" of two variants under 1-st order geometry representation,
(a), and (b).}%
\end{figure}

2) Hybrid-order representation under partial K-projection

For a P-variant, there exists a new type of representation -- Hybrid-order
representation under partial K-projection.

To get the Hybrid-order representation under partial K-projection, we consider
a P-variant $V_{\mathrm{\tilde{U}(1),}1}[\Delta \phi,\Delta x,k_{0}]$ as a
difference between an U-variant $V_{0,\mathrm{\tilde{U}(1),}1}[\Delta
\phi,\Delta x,k_{0}]$ and the partner $V_{\mathrm{\tilde{U}(1),}1}^{\prime
}[\Delta \phi,\Delta x,k_{0}]$ of its complementary\ pair. Then, we do
K-projection on the U-variant $V_{0,\mathrm{\tilde{U}(1),}1}[\Delta \phi,\Delta
x,k_{0}]$ and but no the partner $V_{\mathrm{\tilde{U}(1),}1}^{\prime}%
[\Delta \phi,\Delta x,k_{0}]$. Under K-projection, the U-variant
$V_{0,\mathrm{\tilde{U}(1),}1}[\Delta \phi,\Delta x,k_{0}]$ is reduced into a
uniform zero lattice. The extra group-changing elements are described by those
of the partner $V_{\mathrm{\tilde{U}(1),}1}^{\prime}[\Delta \phi,\Delta
x,k_{0}]$ on the uniform zero lattice $N^{\mu}$ and turns into a\emph{
}\textquotedblleft field\textquotedblright \ of compact $\mathrm{U(1)}$ group
on this discrete, rigid lattice.

\textit{Hybrid-level algebra representation:} In algebra representation of
Hybrid-order representation under partial K-projection, the 1D P-variant is
characterized by a series of (local) group operations of compact \textrm{U(1)} group.

Let us show the theory step by step.

The first step is to consider a P-variant $V_{\mathrm{\tilde{U}(1),}1}%
[\Delta \phi,\Delta x,k_{0}]$ as a difference between the original U-variant
$V_{0,\mathrm{\tilde{U}(1),}1}[\Delta \phi^{A},\Delta x,k_{0}]$ and the partner
of its complementary\ pair $V_{\mathrm{\tilde{U}(1),}1}^{\prime}[\Delta
\phi^{B},\Delta x,k_{0}]$, i.e., $V_{\mathrm{\tilde{U}(1),}1}[\Delta
\phi,\Delta x,k_{0}]=V_{0,\mathrm{\tilde{U}(1),}1}[\Delta \phi^{A},\Delta
x,k_{0}]-V_{\mathrm{\tilde{U}(1),}1}^{\prime}[\Delta \phi^{B},\Delta x,k_{0}]$
and $\Delta \phi=\Delta \phi^{A}-\Delta \phi^{B}$. For P-variant, the number of
extra group-changing elements is very small. Therefore, in continuum limit
$l_{0}\rightarrow0$, we can use $V_{\mathrm{\tilde{U}(1),}1}^{\prime}%
[\Delta \phi^{B},\Delta x,k_{0}]$ to characterize $V_{\mathrm{\tilde{U}(1),}%
1}[\Delta \phi,\Delta x,k_{0}]$. The zero solutions for the complementary pair
$V_{\mathrm{\tilde{U}(1),}1}^{\prime}[\Delta \phi^{B},\Delta x,k_{0}]$ of
$V_{\mathrm{\tilde{U}(1),}1}[\Delta \phi,\Delta x,k_{0}]$\ are "complementary",%
\begin{equation}
\{n_{i}\}+\{n_{i}\}^{\prime}=(...1,1,1,1,1,...1,1,1,1..).
\end{equation}
Here, $V_{\mathrm{\tilde{U}(1),}1}^{\prime}[\Delta \phi^{B},\Delta x,k_{0}]$ is
denoted by a dilute series of integers
\begin{equation}
\{n_{i}\}^{\prime}=(...0,-1,0,0,1,...-1,0,0,0..).
\end{equation}
As a result, without considering the contribution of background from
$V_{0,\mathrm{\tilde{U}(1),}1}[\Delta \phi^{A},\Delta x,k_{0}],$ one can
characterize the zero solutions of $V_{\mathrm{\tilde{U}(1),}1}[\Delta
\phi,\Delta x,k_{0}]$ by those of $V_{\mathrm{\tilde{U}(1),}1}^{\prime}%
[\Delta \phi^{B},\Delta x,k_{0}]$.

The second step is to do K-projection only on the original U-variant
$V_{0,\mathrm{\tilde{U}(1),}1}[\Delta \phi^{A},\Delta x,k_{0}]$, but not on
$V_{\mathrm{\tilde{U}(1),}1}^{\prime}[\Delta \phi^{B},\Delta x,k_{0}]$. This is
why we call it partial K-projection. After partial K-projection, the
group-changing space of the original U-variant $V_{0,\mathrm{\tilde{U}(1),}%
1}[\Delta \phi^{A},\Delta x,k_{0}]$ is projected to a uniform zero lattice. The
non-compact \textrm{\~{U}(1)} group of the original U-variant
$V_{0,\mathrm{\tilde{U}(1),}1}[\Delta \phi,\Delta x,k_{0}]$ turns into a
compact group on a zero lattice of "two-sublattice", i.e., $\phi(x)=2\pi
N(x)+\varphi(N(x)).$ We then relabel the group-changing space $\mathrm{C}%
_{\mathrm{U(1)},1}(\Delta \phi)$ by two numbers ($N(x),\varphi(N(x))$):
$\varphi(N(x))$ is compact phase angle, the other is the integer winding
number of unit cell of zero lattice. As a result, a variant that is a globally
changing structure, is cut into $N$ pieces, each of which is a zero and thus
turns into a locally changing structure with compact group structure;

The third step is to consider the extra group-changing elements of
$V_{\mathrm{\tilde{U}(1),}1}^{\prime}[\Delta \phi^{B},\Delta x,k_{0}]$ on the
uniform zero lattice $N(x)$. During this step, we assume that the zero lattice
is a rigid lattice and can be considered as the scale of Cartesian space
$\mathrm{C}_{d}$. The processes of the changing of original P-variant occur on
the rigid background of zero lattice. The situation is similar to the case of
atom lattices in solid physics, and the physical process of electron's moving
occurs on the rigid background of atom lattices.

The fourth step is to do \emph{compactification }for the extra group-changing
elements of $V_{\mathrm{\tilde{U}(1),}1}^{\prime}[\Delta \phi^{B},\Delta
x,k_{0}]$. On the zero lattice $N(x),$ to exactly determine an extra
group-changing element of $V_{\mathrm{\tilde{U}(1),}1}^{\prime}[\Delta \phi
^{B},\Delta x,k_{0}],$ one must know its position of the lattice site $N(x)$
together with its phase angle on this site $\varphi(N(x)).$ Here, the phase
angle is a compact field, i.e., $\varphi(N(x))=\varphi(N(x))\operatorname{mod}%
(2\pi)$. Fig.10 shows the compactification of a uniform variant of non-compact
\textrm{\~{U}(1)} group. Under compactification, a uniform variant of
non-compact \textrm{\~{U}(1)} group is reduced into a uniform field of compact
\textrm{U(1)} group on rigid zero lattice. In Fig.11, we show the
compactification of a perturbative uniform variant of non-compact
\textrm{\~{U}(1)} group. Now, perturbative uniform variant is a reduced into a
fluctuating field of compact \textrm{U(1)} group on zero lattice.
\begin{figure}[ptb]
\includegraphics[clip,width=0.72\textwidth]{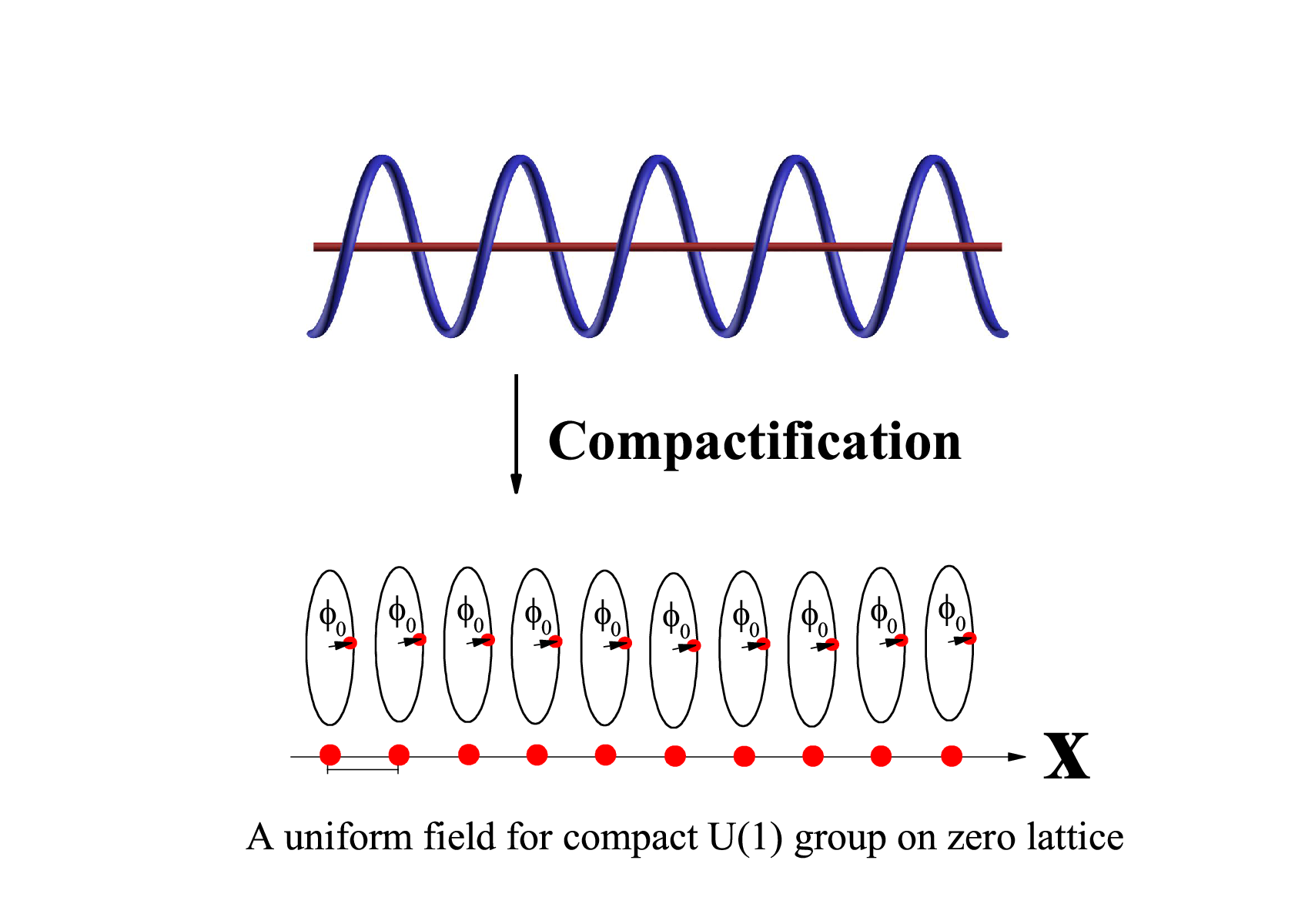}\caption{(Color online)
The compactification of a uniform variant of non-compact \textrm{\~{U}(1)}
group. Under compactification, a uniform variant of non-compact
\textrm{\~{U}(1)} group is reduced into a uniform field of compact
\textrm{U(1)} group on zero lattice. The distance between two zeroes is
$l_{0}/2$.}%
\end{figure}\begin{figure}[ptb]
\includegraphics[clip,width=0.92\textwidth]{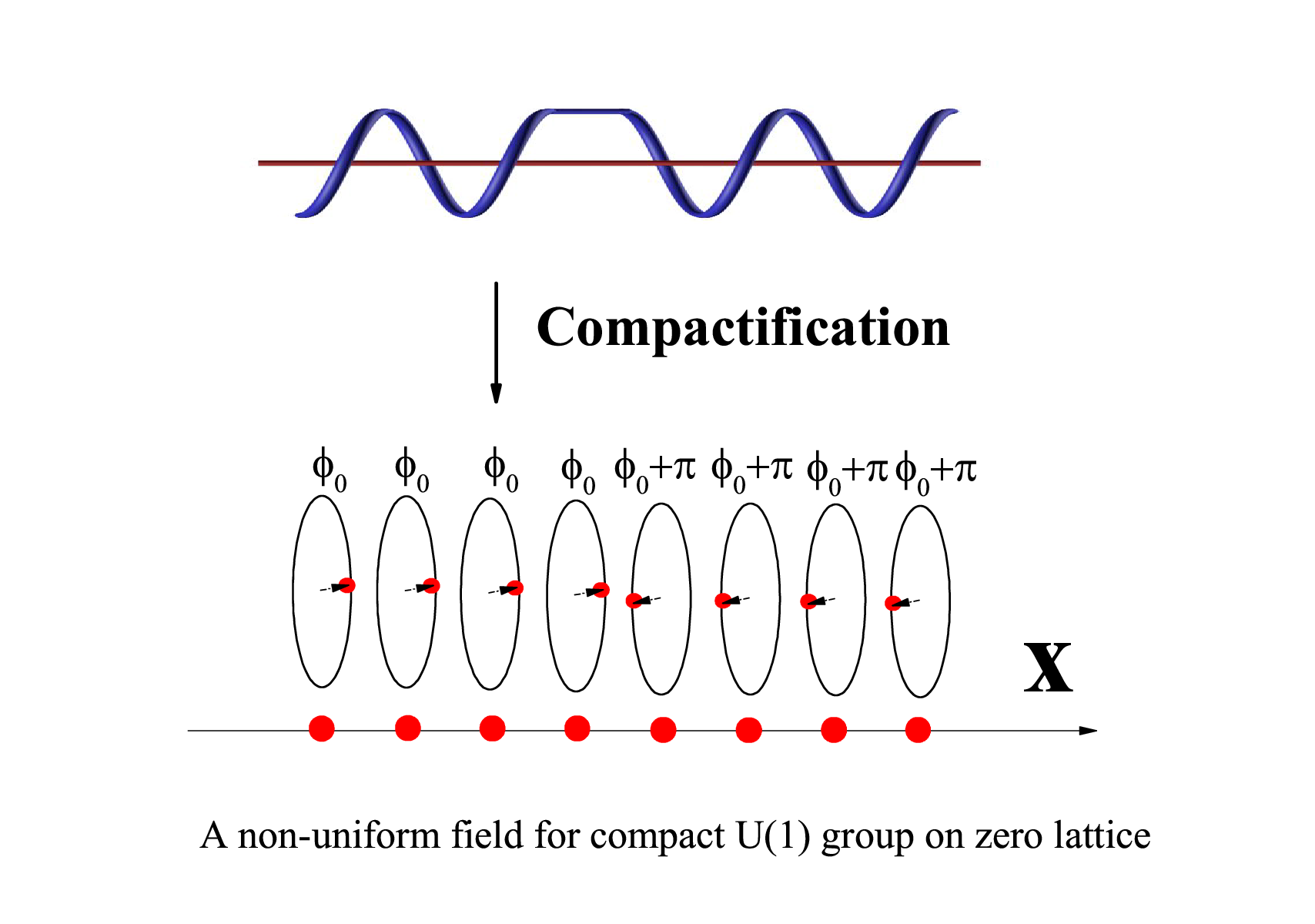}\caption{(Color online)
The compactification of a perturbative uniform variant of non-compact
\textrm{\~{U}(1)} group. Under compactification, a perturbative uniform
variant of non-compact \textrm{\~{U}(1)} group is reduced into a usual field
of compact \textrm{U(1)} group on zero lattice.}%
\end{figure}

The fifth step is to write down the local operation representation on zero
lattice. Now, the P-variant is designed by adding a distribution of the extra
group-changing elements $\delta \phi_{i}^{B}(x_{i})$ on the zero lattice with a
fixed total phase changing $\Delta \phi^{B}=\sum \limits_{i}\delta \phi_{i}%
^{B}(x_{i})\ll \Delta \phi$. Due to the compactification, the non-compact phase
angle $\phi$ turns into a compact one $\varphi.$ As a result, on the zero
lattice, the extra group-changing elements $\delta \phi_{i}^{B}(x_{i})$ of
$\tilde{U}(\delta \phi_{i}^{B}(x_{i}))$ are reduced to the group operations
$\hat{U}(\delta \varphi_{i}(N_{i}(x_{i})))$. Here, $\hat{U}(\delta \varphi
_{i}(N_{i}(x_{i})))$ is a local phase operation that changes phase angle from
$\varphi_{0}$ to $\varphi_{0}+\delta \varphi_{i}(N_{i}(x_{i})).$ Therefore, we
have a group of local phase operations on zero lattice. By using the usual
field of compact \textrm{U(1)} group, we can fully describe it.

Finally, by using algebra representation of Hybrid-order representation
under partial K-projection, a P-variant is reduced into a group of extra local
phase operations on zero lattice that is described by a field of compact
\textrm{U(1)} group. Each group-changing element $\tilde{U}(\delta \phi_{i}%
^{B}(x_{i}))$ is reduced into a group-operation element $\hat{U}(\delta
\varphi_{i}(N_{i}(x_{i})))$ with given compact phase $\varphi_{i}(N_{i}%
(x_{i}))$, i.e.,
\begin{align*}
\tilde{U}(\delta \phi_{i}^{B}(x_{i}))  &  \rightarrow \hat{U}(\delta \varphi
_{i}(N_{i}(x_{i}))),\\
\phi &  \rightarrow2\pi N_{i}(x_{i})+\varphi_{i}(N_{i}(x_{i})).
\end{align*}

In summary, under partial K-projection representation, the "group-changing
elements" $\tilde{U}(\delta \phi_{i}(x_{i}))$ turn into "group-operation
elements" $\hat{U}(\delta \varphi_{i}(x_{i}))$ and become extra objects on
Cartesian space. Therefore, one can say that under partial K-projection, the
non-local "\emph{global} \emph{changing}" structure of a P-variant is reduced
into a "\emph{non-changing}" structure (a fixed distribution of points on
space) together with a local "\emph{relative changing}" structure (a fixed
distribution of group-operation element $\hat{U}(\delta \varphi_{i}(N_{i}%
(x_{i})))$ on space).

In addition, we point out that by exchanging the two extra group-changing
elements on zero lattice, their total local phases change. This phenomenon
clearly reflects the characteristics of "\emph{changing}" structure. Let us
show the fact in detail.

We project the two extra group-changing elements $\tilde{U}(\delta \phi
_{1,.2}^{B}(x_{1,.2}))$ into $\hat{U}(\delta \varphi_{1,2}(N_{1,2}(x_{1,2}))).$
We then assume the phases of two elements to be $\varphi_{1}$ and $\varphi
_{2}$, respectively. Thus, the two extra group-operation elements are obtained
by the following function
\begin{equation}
\hat{U}(\delta \varphi_{1}(N_{1}),\varphi_{1})\hat{U}(\delta \varphi_{2}%
(N_{2}),\varphi_{2})e^{i\varphi_{1}+i\varphi_{2}}.
\end{equation}
After exchanging the two extra elements, we have an extra phase factor
\begin{align}
&  \hat{U}(\delta \varphi_{1}(N_{1}),\varphi_{1})\hat{U}(\delta \varphi
_{2}(N_{2}),\varphi_{2})e^{i\varphi_{1}+i\varphi_{2}}\nonumber \\
&  =\hat{U}(\delta \varphi_{2}(N_{2}),\varphi_{2})\hat{U}(\delta \varphi
_{1}(N_{1}),\varphi_{1})e^{i\varphi_{1}+i\varphi_{2}+i\Delta \varphi}%
\end{align}
where $\Delta \varphi=\frac{1}{\pi}\delta \varphi_{1}\cdot \delta \varphi_{2}.$

\textit{Hybrid-level algebra representation:} In algebra representation of
Hybrid-order representation under partial K-projection, the 1D P-variant is
characterized by a complex field $\mathrm{z}$ on uniform zero lattice $N(x)$,
i.e., $\mathrm{z}(n(x))=e^{i\varphi(N(x))}$. To obtain its algebra
representation, we also set a natural reference, $\mathrm{z}_{0}%
=e^{i\varphi_{0}}$. We then do local group operation on $\mathrm{z}_{0}$ and
get the local algebra representation of Hybrid-order representation under
partial K-projection for the corresponding P-variants.

Firstly, we consider the P-variant with a single extra group-changing element
$\delta \phi(x)$ for non-compact \textrm{\~{U}(1)} group as an example.

Now, we can label the extra group-changing element $\delta \phi(x)$ from
perturbation with two numbers, one is the position of the site of the original
uniform zero lattice $n(x)$, the other is the phase on this site $\varphi$.
Here, $\varphi$ is a compact phase angle for it, i.e, $\varphi=\phi
\operatorname{mod}(2\pi)$. We choose the uniform group configuration as
natural reference $\phi(x)=\phi_{0}$ and derive the local function
representation by do operation $\hat{U}(\delta \varphi(N(x),\varphi(x)))$ on a
natural reference. The extra group-changing element becomes an extra object on
a zero lattice and is characterized by compact Lie group \textrm{U(1). }

Thus, the variant with an extra group-changing element $\delta \phi(x)$ is
denoted by the following function
\begin{align}
\mathrm{z}  &  =\hat{U}(\delta \varphi(N(x),\varphi(x)))\mathrm{z}%
_{0}\nonumber \\
&  =\hat{U}(\delta \varphi(x))e^{i\varphi}\mathrm{z}_{0}%
\end{align}
where $\hat{U}(\delta \varphi(x))=e^{i((\delta \varphi(x))\cdot \hat{K})}$ is an
operator of compact \textrm{U(1)} group. Then, we get the local function
description of the extra group operation as
\begin{equation}
\varphi(x)=\left \{
\begin{array}
[c]{c}%
-\frac{\delta \varphi_{i}}{2},\text{ }x\in(-\infty,x_{i}]\\
-\frac{\delta \varphi_{i}}{2}+k_{0}x,\text{ }x\in(x_{i},x_{i}+\delta x_{i}]\\
\frac{\delta \varphi_{i}}{2},\text{ }x\in(x_{i}+\delta x_{i},\infty)
\end{array}
\right \}  .
\end{equation}
As a result, the group operator $\delta \phi(x)$ becomes "object" on discrete
lattice sites $n(x)$. In addition, because the lattice site has no size, on
such a lattice, the phase changing $\delta \varphi$ is only phase changing,
i.e., $\delta \varphi \neq0,$ $\delta x=0.$\

We then consider the case of many extra group-changing elements. Now, the
extra group-changing elements $\delta \phi_{i}(x_{i})$ are denoted by
($N_{i}(x_{i}),\varphi_{i}$). Here, $\varphi_{i}$ is compact phase angle for
it, i.e, $\varphi_{i}=\phi_{i}\operatorname{mod}(2\pi)$. Thus, the extra
group-operation elements are described by the following function
\begin{equation}
\mathrm{z}=\prod_{i}\hat{U}(\delta \varphi_{i}(x_{i}))e^{i%
%TCIMACRO{\dsum \nolimits_{i}}%
%BeginExpansion
{\displaystyle \sum \nolimits_{i}}
%EndExpansion
\varphi_{i}}\mathrm{z}_{0}%
\end{equation}
where $\hat{U}(\delta \varphi_{i}(x_{i}))\ $is group operation of compact
\textrm{U(1)} group.

To characterize the P-variant with many extra group-changing elements (or
group-operation elements), a key point is \emph{to classify it}. We point out
that the zero's number can be regarded as a topological invariant for
topological equivalence classes. Let us explain it.

For this 1D variant, the zero number is just \emph{crossing number}
$C(\mathrm{z}(x))$ of a knot/link $\mathrm{z}(x).$ It becomes a topological
invariant if it is the minimal number of crossings in all planar diagrams of
the knot/link. In particular, the crossing number $C(\mathrm{z}(x))$ is twice
of the \emph{linking number} for the knot/link that comes from entanglement
between the curve $\mathrm{z}(x)$ and the line of its center that is described
by $\mathrm{z}(x)=0$. Therefore, due to the topological character of zeroes,
the number of zeroes classifies the different topological equivalence classes
of P-variant. As a result, the systems with different number of zeroes belong
to different topological equivalence classes.

On the other hand, we point out that the existence of a zero is independent on
the directions of projection angle $\theta$. When one gets a zero-solution
along a given direction $\theta$, it will never split or disappear whatever
changing the projection direction, $\theta \rightarrow \theta^{\prime}$. This
fact indicates the conservativeness of a zero under projection and a zero is
elementary topological defect. This also indicates that the zero's number
could topological equivalence classes of P-variant.

For the case of a system with the extra zero, the total phase of
group-operation elements $\delta \phi_{i}(x_{i})$ are equal to $\pm \pi,$ i.e.,
$%
%TCIMACRO{\dsum \nolimits_{i}}%
%BeginExpansion
{\displaystyle \sum \nolimits_{i}}
%EndExpansion
\delta \phi_{i}(x_{i})=\pm \pi$. Thus, extra group-operation elements on zero
lattices are denoted by the following function
\begin{equation}
\mathrm{z}(x)=\prod_{i}\hat{U}(\delta \varphi_{i}(x_{i}))e^{i%
%TCIMACRO{\dsum \nolimits_{i}}%
%BeginExpansion
{\displaystyle \sum \nolimits_{i}}
%EndExpansion
\varphi_{i}}\mathrm{z}_{0}\nonumber
\end{equation}
where $\hat{U}(\delta \varphi_{i}(x_{i}))\ $is a group-operation element of
compact \textrm{U(1)} group.

The situation can be generalize to case of $N_{\mathrm{zero}}$ zeroes. For the
case of $N_{\mathrm{zero}}$ zeroes, the total phase of group-changing elements
$\delta \phi_{i}(x_{i})$ are $\pm N_{\mathrm{zero}}\pi,$ i.e., $%
%TCIMACRO{\dsum \nolimits_{i}}%
%BeginExpansion
{\displaystyle \sum \nolimits_{i}}
%EndExpansion
\delta \phi_{i}(x_{i})=\pm N_{\mathrm{zero}}\pi$. On the zero lattice, the
position of the group-changing element $\delta \phi_{i}(x_{i})$ is denoted by
($N_{i}(x_{i}),\varphi_{i}$). Here, $\varphi_{i}$ is the compact phase angle
for it, i.e, $\varphi_{i}=\phi_{i}\operatorname{mod}(2\pi)$. Thus, the extra
group-changing elements are denoted by the following function
\begin{equation}
\mathrm{z}(x)=\prod_{j}(\prod_{i}\hat{U}_{j}(\delta \varphi_{i}^{j}(x_{i}%
^{j})))e^{i%
%TCIMACRO{\dsum \nolimits_{j}}%
%BeginExpansion
{\displaystyle \sum \nolimits_{j}}
%EndExpansion
(%
%TCIMACRO{\dsum \nolimits_{i}}%
%BeginExpansion
{\displaystyle \sum \nolimits_{i}}
%EndExpansion
\varphi_{i}^{j})}\mathrm{z}_{0}\nonumber
\end{equation}
where the index $j$ denotes different zeroes and the index $i$ denotes
different group-changing elements of a given zero.

In summary, by using Hybrid-level algebra representation\textbf{ }under
partial K-projection representation, the "group-changing element" $\tilde
{U}(\delta \phi_{i}(x_{i}))$ is described by a field of compact $\mathrm{U(1)}$
group on zero lattice $\mathrm{z}(x)$ with fixed zeroes.

\textit{Hybrid-level geometry representation:} We discuss the geometry
representation of Hybrid-order representation under partial K-projection for P-variant.

From above discussion, by using algebra representation of Hybrid-order
representation under partial K-projection, the 1D P-variant is characterized
by a complex group field $\mathrm{z}=g(N(x))=e^{i\varphi(N(x))}$ of
compact\textrm{ U(1)} group on uniform zero lattice $N(x).$ It is known that
for a compact\textrm{ U(1)} group, the configuration of group elements is a
set of given phase angles $g(N(x))=e^{i\varphi(N(x))}$ on zero lattice. This
configuration structure of group field $e^{i\varphi(N(x))}$ finally becomes a
"\emph{non-changing}" structure.

3) 0-th order representation under fully K-projection

Next, we discuss the 0-th order representation under fully K-projection. There
are two types of 0-th order representations under different K-projections --
type-I and type-II.

To classify the difference of two types of 0-th order representations under
fully K-projections, we consider a P-variant $V_{\mathrm{\tilde{U}(1),}%
1}[\Delta \phi,\Delta x,k_{0}]$ as the difference between a U-variant
$V_{\mathrm{\tilde{U}(1),}1}[\Delta \phi^{A},\Delta x,k_{0}]$ and partner
$V_{\mathrm{\tilde{U}(1),}1}^{\prime}[\Delta \phi^{B},\Delta x,k_{0}]$ of its
complementary\ pair, i.e., $V_{\mathrm{\tilde{U}(1),}1}[\Delta \phi,\Delta
x,k_{0}]=V_{\mathrm{\tilde{U}(1),}1}[\Delta \phi^{A},\Delta x,k_{0}%
]-V_{\mathrm{\tilde{U}(1),}1}^{\prime}[\Delta \phi^{B},\Delta x,k_{0}]$. Then,
we separately do K-projections on the U-variant $V_{\mathrm{\tilde{U}(1),}%
1}[\Delta \phi,\Delta x,k_{0}]$ under projection angle $\theta_{0}$ and on the
partner $V_{\mathrm{\tilde{U}(1),}1}^{\prime}[\Delta \phi^{B},\Delta x,k_{0}]$
of its complementary\ pair under projection angle $\theta^{\prime}$. When the
projection angle $\theta_{0}$ for $V_{\mathrm{\tilde{U}(1),}1}[\Delta \phi
^{A},\Delta x,k_{0}]$ and the projection angle $\theta^{\prime}$ for
$V_{\mathrm{\tilde{U}(1),}1}^{\prime}[\Delta \phi^{B},\Delta x,k_{0}]$ are
equal, i.e., $\theta=\theta^{\prime},$ we have type-I fully K-projection; When
the projection angle $\theta_{0}$ for $V_{0,\mathrm{\tilde{U}(1),}1}%
[\Delta \phi^{A},\Delta x,k_{0}]$ and the projection angle $\theta^{\prime}$
for $V_{\mathrm{\tilde{U}(1),}1}^{\prime}[\Delta \phi^{B},\Delta x,k_{0}]$ are
different, i.e., $\theta \neq \theta^{\prime},$ we have type-II fully
K-projection under K-projection. For both cases, the U-variant
$V_{\mathrm{\tilde{U}(1),}1}[\Delta \phi^{A},\Delta x,k_{0}]$ is reduced into a
uniform zero lattice.

On the one hand, we study 0-th order representation under type-I fully
K-projection for a P-variant $V_{\mathrm{\tilde{U}(1),}1}[\Delta \phi
^{B},\Delta x,k_{0}]$ with $\theta=\theta^{\prime}.$

Now, for a P-variant under type-I fully K-projection, the extra group-changing
elements will lead to extra zero solutions. Consequently, we have a zero
lattice with defects.

For a P-variant $V_{\mathrm{\tilde{U}(1),}1}[\Delta \phi,\Delta x,k_{0}]$, due
to the mismatch condition $\Delta \phi \neq k_{0}\Delta x$ and $\left \vert
(\Delta \phi-k_{0}\Delta x)/\Delta \phi \right \vert \ll1$, we have a series of
numbers with small disorder,
\begin{equation}
\{n_{i}\}=(...1,0,1,1,2,...0,1,1,1..).
\end{equation}
Here, "$0$" means a local contraction on group-changing space of the original
variant; "$2$", means local expansion on group-changing space of the original
variant. When we do K-projection, the extra group-changing elements denoted by
"$0$" will not lead to zero solution, while the extra group-changing elements
denoted by "$2$" will lead to double zero solutions compared with the
group-changing element "$1$". Fig.12(a) is an illustration of a P-variant
under type-I fully K-projection. \begin{figure}[ptb]
\includegraphics[clip,width=0.82\textwidth]{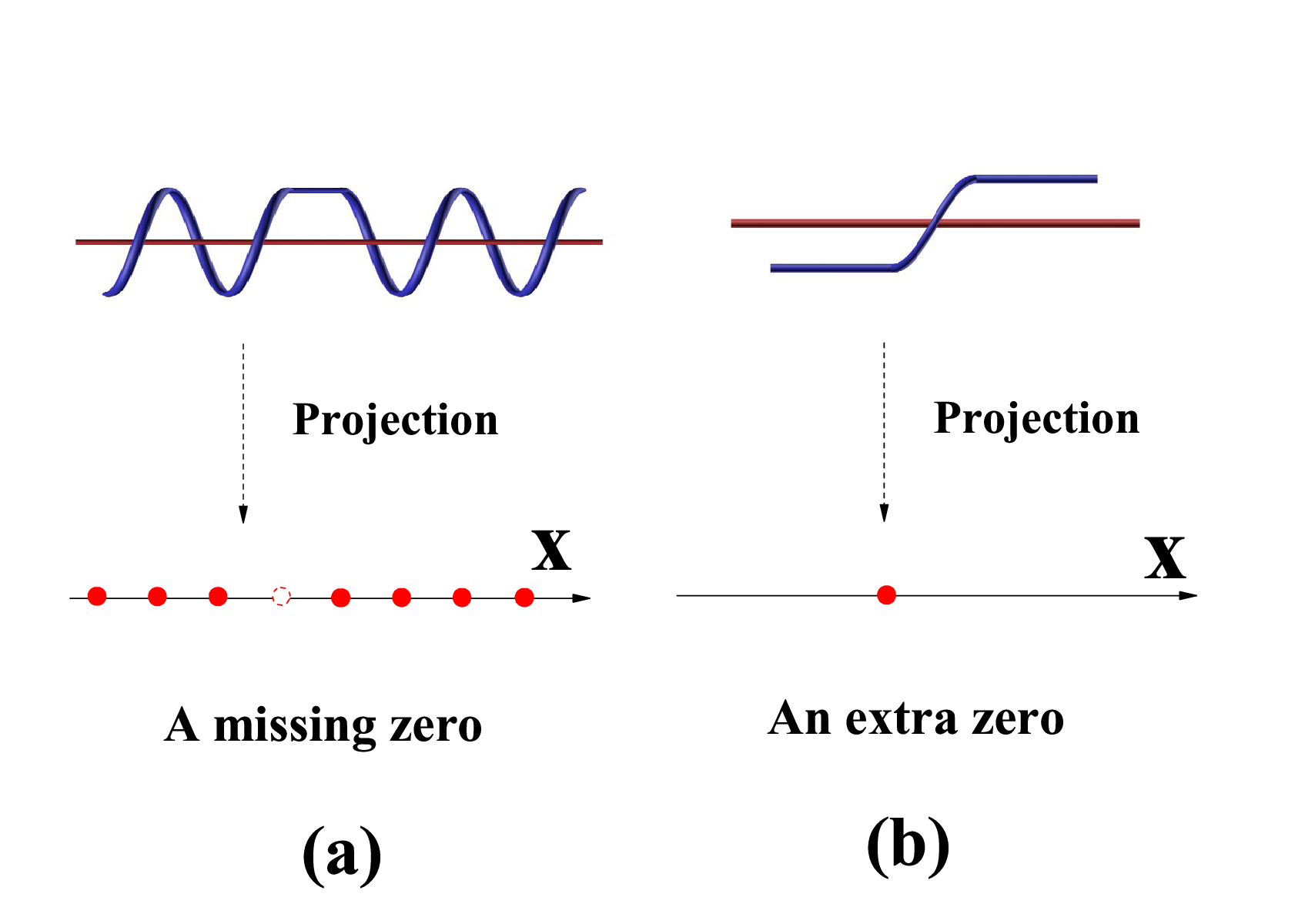}\caption{(Color online)
(a) and (b) show P-variant under type-I and type-II fully K-projection,
respectively. }%
\end{figure}

Next, we study the 0-th order representation under type-II fully K-projection
for a P-variant $V_{\mathrm{\tilde{U}(1),}1}[\Delta \phi,\Delta x,k_{0}]$ with
$\theta \neq \theta^{\prime}$.

Now, under a K-projection, the U-variant $V_{\mathrm{\tilde{U}(1),}1}%
[\Delta \phi^{A},\Delta x,k_{0}]$ is reduced into a uniform zero lattice; under
another K-projection, the partner $V_{\mathrm{\tilde{U}(1),}1}^{\prime}%
[\Delta \phi^{B},\Delta x,k_{0}]$ of its complementary\ pair is reduced into a
system with very small number of zeroes. Because we consider the uniform zero
lattice from U-variant $V_{\mathrm{\tilde{U}(1),}1}[\Delta \phi^{A},\Delta
x,k_{0}]$ as a rigid background, the dynamics of the original P-variant
$V_{\mathrm{\tilde{U}(1),}1}[\Delta \phi,\Delta x,k_{0}]$ is characterized by
the distribution of zeroes of $V_{\mathrm{\tilde{U}(1),}1}^{\prime}[\Delta
\phi^{B},\Delta x,k_{0}].$ Therefore, for a P-variant under type-II fully
K-projection, the extra group-changing elements will lead to very small number
of extra zeroes on a uniform zero lattice. Fig.12(b) is an illustration of a
P-variant under type-II fully K-projection.

In summary, under type-I fully K-projection, the whole "\emph{changing}"
structure of a variant is reduced into a "\emph{non-changing}" defective zero
lattice. $V_{\mathrm{\tilde{U}(1),}1}[\Delta \phi,\Delta x,k_{0}]$ becomes a
crystal of zeros with missing zeroes or extra zeroes; under type-II fully
K-projection, the whole "\emph{changing}" structure of a variant is reduced
into a "\emph{non-changing}" structure with very small number of zeroes
without considering the background of the uniform zero lattice.

\paragraph{Representations for higher dimensional variants}

In above section, we have discussed the representations for 1D variants
including 1D U-variants and 1D P-variant. In this section, we discuss the
cases for higher dimensional variants by focusing on the difference with 1D
cases. The key difference is, beside the usual (longitudinal) K-projection,
there exists transverse direction-projection for higher dimensional variants.

\subparagraph{Higher dimensional U-variants}

There are different representations for a higher dimensional U-variant of
non-compact \textrm{\~{G}} group\textrm{ }from different aspects, including
\emph{algebra}, \emph{geometry}, and \emph{algebra}, respectively.

\textit{1-st order algebra representation}: In 1-st order algebra
representation, the higher dimensional U-variant of non-compact \textrm{\~{G}}
group is characterized by a series of (non-local) group-changing elements of
non-compact \textrm{\~{G}} group.

For a higher dimensional U-variant $V_{\mathrm{\tilde{G},}d}[\Delta \phi^{\mu
},\Delta x^{\mu},k_{0}^{\mu}]$, there exists only one type of group-changing
elements with fixed changing rate $\frac{d\phi^{\mu}}{dx^{\mu}}=k_{0}^{\mu
}=\frac{\pi}{a^{\mu}}.$ The U-variant is designed by adding a uniform
distribution of the extra group-changing elements $\delta \phi_{i}^{\mu}(x)$,
which is described by a series of group-changing operations $\tilde{U}%
(\delta \phi)=\prod_{\mu}(\prod_{i}\tilde{U}(\delta \phi_{i}^{\mu}(x_{i})))$
with\ $\tilde{U}(\delta \phi_{i}^{\mu}(x))=e^{i((\delta \phi_{i}^{\mu}T^{\mu
})\cdot \hat{K}^{\mu})}$ and $\hat{K}^{\mu}=-i\frac{d}{d\phi^{\mu}}.$\ Here,
the i-th infinitesimal group-changing operation $\tilde{U}(\delta \phi_{i}%
^{\mu})$ generates a group-changing element on position $i$ with the condition
$\Delta \phi^{\mu}=k_{0}^{\mu}\Delta x^{\mu}$.

\textit{1-st order algebra representation:} In 1-st order algebra
representation, the higher dimensional U-variant $V_{\mathrm{\tilde{G},}%
d}[\Delta \phi^{\mu},\Delta x^{\mu},k_{0}^{\mu}]$ of non-compact \textrm{\~{G}}
group is usually described by a complex matrix $\mathrm{z}$. To obtain its
algebra representation, we must set a \emph{reference}. In general, we have
a natural choice, $\mathrm{z}_{0}=$ constant. We then do non-local
group-changing operation on the natural reference and get a non-local
algebra representation with corresponding variants, i.e.,
\begin{equation}
\mathrm{z}_{u}(x)=\tilde{U}(\delta \phi)\mathrm{z}_{0}%
\end{equation}
where $\tilde{U}(\delta \phi)=\prod_{\mu}(\prod_{i}\tilde{U}(\delta \phi
_{i}^{\mu}(x)))$.

\textit{Geometry representations under direction-projection}\textbf{:} Next,
we consider the 1-st order geometry representation of the higher dimensional
U-variant $V_{\mathrm{\tilde{G},}d}[\Delta \phi^{\mu},\Delta x^{\mu},k_{0}%
^{\mu}]$ of non-Abelian and non-compact Lie group $\mathrm{\tilde{G}}$. For a
variant in higher dimensions, we have a complex matrix,
\begin{equation}
\mathrm{z}_{u}(x)=\tilde{U}(\delta \phi)\mathrm{z}_{0}%
\end{equation}
where $\tilde{U}(\delta \phi)=\prod_{\mu}(\prod_{i}\tilde{U}(\delta \phi
_{i}^{\mu}(x)))$. Due to noncommutative structure along different spatial
directions, we cannot give an overall picture for the higher dimensional
variants. Instead, we can only characterize their changing structure along
given spatial direction by projecting the group-changing space, i.e.,
\[
\tilde{U}^{\mu}(\delta \phi)=\mathrm{Tr}(T^{\mu}\tilde{U}(\delta \phi)).
\]
In the following parts, we call the process of projection of a higher
dimensional group-changing space to 1D along its $\mu$-th spatial direction
"\emph{direction projection}" and abbreviate it to D-projection.

Therefore, for a $d$-dimensional variant, we have $d$ D-projected 1D variants,
each of which is described by a complex field of non-compact \textrm{\~{G}%
}$^{\mu}$ Abelian group
\begin{equation}
\mathrm{z}_{u}(x)=\tilde{U}^{\mu}(\delta \phi)\mathrm{z}_{0}%
\end{equation}
For each one, we can use the approach to 1D variant of non-compact
\textrm{\~{U}(1)} group.

As a result, under D-projection, by using the approach that is similar to 1D
variant, we also have a knot/link along this spatial direction, or $d$
different 1D knot/links in 3D space. This is a new type of knot/link in higher
dimensional space. We call it \emph{translation symmetry protected knot/links
in higher dimensional space}.

In addition to the 1-st order geometry representation, we discuss the 0-th
order geometry representation under both D-direction and K-projection.

By generalizing the K-projection to the $d$ 1D variants of non-compact Abelian
group \textrm{\~{G}}$^{\mu}$, for higher dimensional U-variant
$V_{\mathrm{\tilde{G},}d}[\Delta \phi^{\mu},\Delta x^{\mu},k_{0}^{\mu}]$, we
have $d$-dimensional zero lattice. Along $\mu$-th spatial direction of the
zero lattice, the lattice site is labeled by $N^{\mu}.$ Consequently, after
doing D-projection together with K-projection, the original non-compact
$\mathrm{\tilde{G}}$ group turns into a field of compact \textrm{G} group on
$d$-dimensional uniform zero lattice of "two-sublattice", i.e.,
\[
\phi^{\mu}(x)=2\pi N^{\mu}(x)+\varphi^{\mu}(x).\text{ }%
\]
We also can relabel the group-changing space $\mathrm{C}_{\mathrm{\tilde{G}%
},d}(\Delta \phi^{a})$ by $2d$ numbers ($N^{\mu}(x),\varphi^{\mu}(x)$):
$\varphi^{\mu}(x)$ is compact phase angle of $\mu$-th group generator of the
compact group \textrm{G}, the other is the integer winding number of unit cell
of zero lattice $N^{\mu}(x)\in \lbrack0,N^{\mu}]$. As a result, we organize the
$d$ compact phase angle $\varphi^{\mu}(x)$ into two groups: one is about
global phase changing $\left \vert \Delta \varphi^{\mu}(x)\right \vert =\sqrt{%
%TCIMACRO{\dsum \limits_{\mu}}%
%BeginExpansion
{\displaystyle \sum \limits_{\mu}}
%EndExpansion
(\Delta \varphi^{\mu}(x))^{2}}$, the other is about $d-1$ internal relative
compact angle.

In summary, for the case of $d$-dimensional U-variant, by using 1-st order
geometry representation under D-projection without K-projection, we have
translation symmetry protected knot/links in higher dimensional space; by
using 0-th order geometry representation under both K-projection and
D-projection, we have a $d$-dimensional uniform zero lattice.

\subparagraph{Higher dimensional P-variants}

For higher dimensional P-variants, there exist different representations
(algebra, geometry, and algebra representations) under different projections
(with/without K-projection, with/without D-projection). In this part, we only
discuss their Hybrid-level algebra representation under D-projection and
partial K-projection.

Along each spatial direction, under both D-projection and partial
K-projection, for higher dimensional U-variants we have a zero lattice
$N^{\mu}$ and a compact phase angle $\varphi^{\mu}$ of $\mu$-th group
generator. As a result, the position of group-changing space $\mathrm{C}%
_{\mathrm{\tilde{G}},d}$ is denoted by (discrete) coordinate $n^{\mu}$ and
compact\ phase angle $\varphi^{\mu}$.

Now, the extra group-changing elements on the U-variant turns into a\emph{
}\textquotedblleft field\textquotedblright \ of compact $\mathrm{G}$ group on
discrete lattice $N^{\mu}.$ The extra group-changing elements, $\delta \phi
_{i}^{\mu}(x_{i})$ along $\mu$-th direction are denoted by ($N_{i}^{\mu}%
(x_{i}),\varphi_{i}^{\mu}(N_{i}^{\mu}(x_{i}))$) where $\varphi_{i}^{\mu}%
(N_{i}^{\mu}(x_{i}))$ is compact phase angle for itself, i.e, $\varphi
_{i}^{\mu}(N_{i}^{\mu}(x_{i}))=\phi_{i}^{\mu}(N_{i}^{\mu}(x_{i}%
))\operatorname{mod}(2\pi)$. Thus, the extra group-changing elements are
denoted by the following function of matrix, i.e,
\begin{equation}
\mathrm{z}=\prod_{\mu}(\prod_{i}\hat{U}^{\mu}(\delta \varphi_{i}^{\mu}%
(N_{i})))\mathrm{z}_{0}^{\prime}%
\end{equation}
where $\hat{U}^{\mu}(\delta \varphi_{i}^{\mu}(N_{i}))\ $is an operation element
of $\mu$-th generator for compact \textrm{G} group.

For the case of $N_{\mathrm{zero}}$ zeroes, the total phase of group-changing
elements along arbitrary direction $\delta \phi_{i}^{\mu}(x_{i})$ is $\pm
N_{\mathrm{zero}}\pi,$ i.e., $%
%TCIMACRO{\dsum \nolimits_{i}}%
%BeginExpansion
{\displaystyle \sum \nolimits_{i}}
%EndExpansion
\delta \phi_{i}^{\mu}(x_{i})=\pm N_{\mathrm{zero}}\pi$. On the zero lattice,
the position of each group-changing element $\delta \phi_{i}^{\mu}(x_{i})$ is
denoted by ($N^{\mu}(x_{i}),\varphi_{i}^{\mu}$). Here, $\varphi_{i}^{\mu}$ is
a compact phase angle for itself, i.e, $\varphi_{i}^{\mu}=\phi_{i}^{\mu
}\operatorname{mod}(2\pi)$. Thus, the extra group-changing elements are
denoted by the following function of matrix $\mathrm{z}(x)$,
\begin{equation}
\mathrm{z}(x)=\prod_{j}(\prod_{\mu}(\prod_{i}\hat{U}_{j}(\delta \varphi
_{i}^{\mu,j}(x_{i}^{\mu,j}))))\mathrm{z}_{0}\nonumber
\end{equation}
where the index $j$ labels different zeroes, the index $i$ labels different
group-changing elements of a given zero, and the index $\mu$ labels the group
generator along $\mu$-th spatial direction.

In summary, under D-projection and partial K-projection representation, the
"group-changing elements" are considered as extra objects in Cartesian space
and above complex function of matrix $\mathrm{z}(x)$ characterizes
$N_{\mathrm{zero}}$ zeroes. \

\subsubsection{Summary}

In this section, we develop a new mathematic theory for "\emph{changing}"
structure -- variant theory that can characterize the changings of certain
"\emph{spaces}" (group-changing spaces) on Cartesian space. Under special
projections (K-projection, or/and D-projection), a variant is reduced into a
special "non-changing" structure (rigid background of space, or local field
with compact group) in Cartesian space. Consequently, a variant is reduced
into a special field. This powerful mathematic theory can help us understand
the non-local structure of quantum mechanics.

\subsection{A new theoretical framework for physics -- "\textbf{All from
Changings}"}

\subsubsection{From tower of changings to the tower of physics}

In this section, we will develop a new theoretical framework of physics beyond
quantum mechanics and classical mechanics.

We point out that all issues about quantum mechanics and classical mechanics
are relevant to the "\emph{changings}". Different physical laws emerge from
the changings in different levels. Fig.13 is the illustration of "\emph{Tower
of physics}" that is really "\emph{Tower of changings}". The base of the tower
is the uniform physical variant that is a uniform \emph{changing} structure on
Cartesian space. In modern physics, it's always named as "\emph{vacuum}" or
"\emph{ground state}". We call it 0-th level physics structure. Above 0-th
level are the expansion or contraction types of "\emph{changings}" of the
vacuum, which is named "\emph{matter}" or topological excited states in modern
physics. We call it 1-st level physics structure. Above 1-st level are the
time-dependent "\emph{changings}" of the local expansion or local contraction
changings of vacuum, which is named "\emph{motion}" of matter in usual
physics. The equations of motion (Schr\"{o}dinger's equation or Newton's
equation) inevitably emerge under certain approximations. We call it 2-nd
level physics structure. See the illustration of the "Tower of changings" in
Fig.13 that is the key point of this paper. \begin{figure}[ptb]
\includegraphics[clip,width=0.9\textwidth]{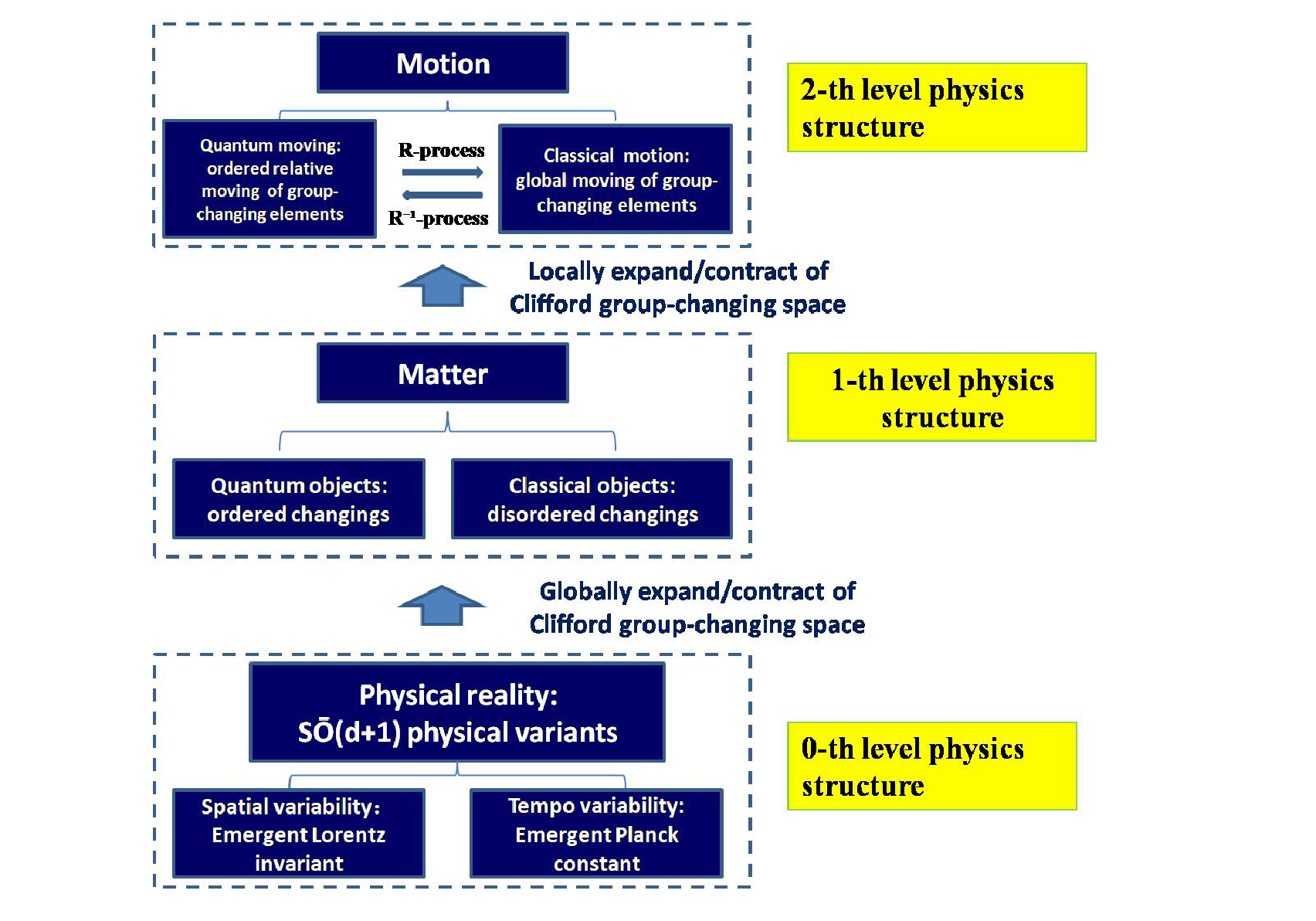}\caption{(Color online)
Tower of changings}%
\end{figure}

As a result, according to the tower of changings, we develop a new theoretical
framework of certain mechanics (the tower of physics) via three steps:

\begin{enumerate}
\item Step 1 is to develop theory about \emph{0-th level} physics structure by
giving the certain hypothesis about \emph{physical reality};

\item Step 2 is to develop theory about \emph{1-st level} physics structure by
giving the certain hypothesis about \emph{matter};

\item Step 3 is to develop theory about \emph{2-nd level} physics structure by
giving equation about the time-evolution of matter's \emph{motion}.
\end{enumerate}

From this spectacular scene about "\emph{changings}", we say that "\emph{All
from Changings}".

\subsubsection{$\mathrm{\tilde{S}\tilde{O}}$\textrm{(d+1)} physical variants:
concept and definition}

\emph{What's physical reality in a new theoretical framework beyond quantum
mechanics and classical mechanics?} The base of the tower of physics becomes
the key point to develop a new theoretical framework beyond quantum mechanics
and classical mechanics. In this paper, we point out that for quantum
mechanics and classical mechanics, the physical reality is ($d+1$) dimensional
$\mathrm{\tilde{S}\tilde{O}}$\textrm{(d+1)} \emph{physical variant}, a
predecessor of our spacetime and matter.

To get the correct type of variant of our universe, the following conditions
need to be met:

1) \emph{Variability condition}: This is just the assumption of "variants" for
our universe. We assume that along an arbitrary direction $(x,y,z,t)$ in
spacetime, the system must have 1-st order variability with fixed changing rate;

2) \emph{Symmetry condition}: We assume that the changing rate along different
directions of spacetime are same (by setting the light velocity $c$ to be $1$);

3) \emph{Orthogonal condition}: We assume another relationship of variability
for different directions -- the parallelogram rule, or $\left \vert
\mathbf{x}_{\mathrm{A}}-\mathbf{x}_{\mathrm{B}}\right \vert ^{2}=%
%TCIMACRO{\dsum \nolimits_{\mu}}%
%BeginExpansion
{\displaystyle \sum \nolimits_{\mu}}
%EndExpansion
(x_{\mathrm{A,}\mu}e^{\mu}-x_{\mathrm{B},\mu}e^{\mu})^{2}.$

To meet above conditions, our universe must be an $\mathrm{\tilde{S}\tilde{O}%
}$\textrm{(d+1)} physical variant that is a mapping between ($d+1$%
)-dimensional $\mathrm{\tilde{S}\tilde{O}}$\textrm{(d+1)} Clifford
group-changing space $\mathrm{C}_{\mathrm{\tilde{S}\tilde{O}(d+1)},d+1}%
(\Delta \phi^{\mu})$ and a rigid spacetime $\mathrm{C}_{d+1}(\Delta x^{\mu}).$
Here, $\mathrm{\tilde{S}\tilde{O}(d+1)}$ denotes an $\mathrm{\tilde{S}%
\tilde{O}}$\textrm{(d+1) }non-compact group and $\mu$ denotes an index for
arbitrary orthogonal direction of spacetime.

The following is the definition of ($d+1$)-dimensional Clifford group-changing
space $\mathrm{C}_{\mathrm{\tilde{S}\tilde{O}(d+1)},d+1}(\Delta \phi^{\mu})$:

\textit{Definition:\ A }($d+1$)\textit{-dimensional Clifford space}%
$\  \mathrm{C}_{\mathrm{\tilde{S}\tilde{O}(d+1)},d+1}(\Delta \phi^{\mu}%
)$\textit{ is described by }$d+1$\textit{ series of numbers of group elements
}$\phi^{\mu}$\textit{ arranged in size order with unit "vector" as }%
$(d+1)$-by-$(d+1)$\textbf{ }\textit{Gamma matrices }$\Gamma^{\mu}$\textit{
obeying Clifford algebra} $\{ \Gamma^{i},\Gamma^{j}\}=2\delta^{ij}$.\textit{
The total size along }$\mu$\textit{-direction of }$\mathrm{C}_{l,d+1}$\textit{
is }$\Delta \phi^{\mu}$\textit{.}

The ($d+1$)-dimensional Clifford group-changing space\textit{ }$\mathrm{C}%
_{\mathrm{\tilde{S}\tilde{O}(d+1)},d+1}(\Delta \phi^{\mu})$ has orthogonality.
A $d$-dimensional Clifford group-changing space $\mathrm{C}_{l,d+1}(\Delta
\phi^{\mu})$ obeys non-commutating geometry due to $\{ \Gamma^{\mu}%
,\Gamma^{\nu}\}=2\delta^{\mu \nu}$\cite{con}. Therefore, in ($d+1$)-dimensional
Clifford group-changing space $\mathrm{C}_{\mathrm{\tilde{S}\tilde{O}%
(d+1)},d+1}(\Delta \phi^{\mu})$, the parallelogram rule of vectors is similar
to Cartesian space's. For two vectors in $\mathrm{C}_{\mathrm{\tilde{S}%
\tilde{O}(d+1)},d+1}(\Delta \phi^{\mu})$, $\mathbf{\phi}_{\mathrm{A}}%
=\phi_{\mathrm{A,}\mu}e^{\mu}$ and $\mathbf{\phi}_{\mathrm{B}}=\phi
_{\mathrm{B},\mu}e^{\mu},$ the add and subtract rules become
\begin{equation}
\mathbf{\phi}_{\mathrm{A}}\pm \mathbf{\phi}_{\mathrm{B}}=%
%TCIMACRO{\dsum \nolimits_{\mu}}%
%BeginExpansion
{\displaystyle \sum \nolimits_{\mu}}
%EndExpansion
(\phi_{\mathrm{A,}\mu}e^{\mu}+\phi_{\mathrm{B},\mu}e^{\mu}).
\end{equation}
The distance between $\mathbf{\phi}_{\mathrm{A}}$ and $\mathbf{\phi
}_{\mathrm{B}}$ becomes
\begin{equation}
\left \vert \mathbf{\phi}_{\mathrm{A}}-\mathbf{\phi}_{\mathrm{B}}\right \vert
^{2}=%
%TCIMACRO{\dsum \nolimits_{\mu}}%
%BeginExpansion
{\displaystyle \sum \nolimits_{\mu}}
%EndExpansion
(\phi_{\mathrm{A,}\mu}e^{\mu}-\phi_{\mathrm{B},\mu}e^{\mu})^{2}.
\end{equation}
This leads to parallelogram rule in our spacetime.

Next, we give the definition of ($d+1$) dimensional $\mathrm{\tilde{S}%
\tilde{O}}$\textrm{(d+1)} physical variants:

\textit{Definition: }($d+1$)\textit{-dimensional} $\mathrm{\tilde{S}\tilde{O}%
}$\textrm{(d+1)} \textit{physical variant is a mapping between $\mathrm{\tilde
{S}\tilde{O}}$\textrm{(d+1)} Clifford group-changing space }$\mathrm{C}%
_{\mathrm{\tilde{S}\tilde{O}(d+1)},d+1}$\textit{\ and a rigid spacetime
}$\mathrm{C}_{d+1},$\textit{ i.e., }%
\begin{equation}
V_{\mathrm{\tilde{S}\tilde{O}(d+1)},d+1}[\Delta \phi^{\mu},\Delta x^{\mu}%
,k_{0}^{\mu}]:\{ \delta \phi^{\mu}\} \Leftrightarrow \{ \delta x^{\mu}\}
\end{equation}
\textit{where }$\Leftrightarrow$\textit{\ denotes an ordered mapping with
fixed changing rate of integer multiple }$k_{0}$ or $\omega_{0},$\textit{\ and
}$\mu$\textit{ labels the spatial direction}. \textit{In particular, we set
the light speed }$c=1,$\textit{ and have }$\omega_{0}=k_{0}$\textit{.}

Based on this Variant Hypothesis, we will develop a new, and complete
theoretical framework for quantum mechanics and classical mechanics step by step.

\subsubsection{Variant hypothesis of physical reality in our universe}

\paragraph{Variant hypothesis}

In this section, we develop theory about 0-th level physics structure based on
the Variant hypothesis about physical reality in our universe:

\textit{Variant Hypothesis of our universe: Physical reality in our universe
is a (}$d+1$\textit{)-dimensional }$\mathrm{\tilde{S}\tilde{O}}$\textrm{(d+1)}
\textit{physical variant }$V_{\mathrm{\tilde{S}\tilde{O}(d+1)},d+1}(\Delta
\phi^{\mu},\Delta x^{\mu},k_{0},\omega_{0})$ \textit{described by a mapping
between Clifford group-changing space }$\mathrm{C}_{\mathrm{\tilde{S}\tilde
{O}(d+1)},d+1}$\textit{\ and a rigid spacetime }$\mathrm{C}_{d+1}$\textit{.
Here, we have }$d=3$\textit{.}

\paragraph{Spatial/tempo variability of a uniform $\mathrm{\tilde{S}\tilde{O}%
}$\textrm{(d+1)} physical variants}

As the base of the tower, the uniform $\mathrm{\tilde{S}\tilde{O}}%
$\textrm{(d+1)}\textit{ }physical variant as a uniform changing structure in
Cartesian space becomes the starting point of the new theory. To accurately
characterize the physical variant, we consider its \emph{spatial/tempo
variability}, which characterizes its geometry/dynamic properties, respectively.

On the one hand, the geometry property is characterized by 1-st order
variability along an arbitrary spatial direction, i.e.,
\begin{align}
\mathcal{T}(\delta x^{i})  &  \leftrightarrow \hat{U}^{\mathrm{T}}(\delta
\phi^{i})=e^{i\cdot \delta \phi^{i}\Gamma^{i}},\text{ }\nonumber \\
i  &  =x_{1},x_{2},\text{...},x_{d},
\end{align}
where $\delta \phi^{i}=k_{0}\delta x^{i}$ and $\Gamma^{i}$ are the Gamma
matrices obeying Clifford algebra $\{ \Gamma^{i},\Gamma^{i}\}=2\delta^{ij}$.
Therefore, $\hat{U}^{\mathrm{T}}(\delta \phi^{i})$ is (spatial) translation
operation in Clifford group-changing space rather than the generator of a
(non-compact) $\mathrm{\tilde{S}\tilde{O}}$\textrm{(d)}\ group.

On the other hand, we consider the dynamic property that also can be
characterized by 1-st order variability along time direction, i.e.,
\begin{equation}
\mathcal{T}(\delta t)\leftrightarrow \hat{U}^{\mathrm{T}}(\delta \phi
^{t})=e^{i\cdot \delta \phi^{t}\Gamma^{t}},
\end{equation}
where $\delta \phi^{t}=\omega_{0}\delta t$ and $\Gamma^{t}$ is also Gamma
matrix anticommuting with $\Gamma^{i},$ $\{ \Gamma^{i},\Gamma^{t}%
\}=2\delta^{it}$. Therefore, $\hat{U}^{\mathrm{T}}(\delta \phi^{t})$ is (tempo)
translation operation in Clifford group-changing space. $\omega_{0}$ is an
"angular momentum" of the system in certain "extra" dimensions. In particular,
the system with 1-st order variability along time direction also indicates a
\emph{uniform motion} of the group-changing space along $\Gamma^{t}$ direction.

In addition, the uniform ($d+1$)-dimensional $\mathrm{\tilde{S}\tilde{O}}%
$\textrm{(d+1)} physical variants has a 1-st order rotation variability that
is defined by
\begin{equation}
\hat{U}^{\mathrm{R}}\leftrightarrow \hat{R}_{\mathrm{space}}%
\end{equation}
where $\hat{U}^{\mathrm{R}}$ is \textrm{SO(d+1)} rotation operator in Clifford
group-changing space $\hat{U}^{\mathrm{R}}\Gamma^{I}\mathbf{(}\hat
{U}^{\mathrm{R}})^{-1}=\Gamma^{I^{\prime}},$ and $\hat{R}_{\mathrm{space}}$ is
\textrm{SO(d+1)} rotation operator in Cartesian space, $\hat{R}%
_{\mathrm{space}}x^{I}\hat{R}_{\mathrm{space}}^{-1}=x^{I^{\prime}}.$ After
doing a global composite rotation operation $\hat{U}^{\mathrm{R}}\cdot \hat
{R}_{\mathrm{space}}$, the uniform ($d+1$)-dimensional $\mathrm{\tilde
{S}\tilde{O}}$\textrm{(d+1)} physical variant is invariant. This 1-st order
rotation variability will play important role in scattering processes.

\paragraph{Emergent physical laws from spatial/tempo variability}

Physical law always emerges from linearization on certain "\emph{uniform}
\emph{changing}" structures of a system. We take Hooke's law as an example to
illustrate the idea. When an object of solid materials is subjected to force,
there is a linear relationship between stress and strain (unit deformation) in
the material. The Hooke's law can be regarded as a law from linearization by
performing Taylor expansion around a smooth function. Then, we use similar
idea to study the dynamics of \textit{ }($d+1$)-dimensional $\mathrm{\tilde
{S}\tilde{O}}$\textrm{(d+1)} physical variants\textit{ }$V_{\mathrm{\tilde
{S}\tilde{O}(d+1)},d+1}(\Delta \phi^{\mu},\Delta x^{\mu},k_{0},\omega_{0})$.

We give a Variant Hypothesis of physical reality in our universe. Therefore,
our world comes from a variant with 1-st order spatial-tempo variability. Such
a spatial-tempo variability indicates a \emph{uniform}, \emph{holistic}
universe. In particular, we point out that there emerge remarkable physical
laws (Lorentz invariant, and quantization condition) from a system with 1-st
order spatial-tempo variability.

\textit{Emergent Lorentz invariant}\textbf{:}\emph{ }On the one hand, we
consider the emergent physical law from 1-st order spatial variability. From
above variant hypothesis of our universe, we have a fixed spatial changing
rate for vacuum, i.e., $k_{0}\neq0$. The direct physical consequence of this
fact is \emph{linear dispersion relation} and \emph{emergent Lorentz
invariant}. In general, we may assume the dispersion of the system is a smooth
function, such as $\omega(k).$ Here, $\omega(k)$ is \emph{uniform motion} of
pure phase changing without Gamma matrix. Near $k=k_{0},$ with linearization
at $k=k_{0}$, we have $\omega-\omega_{0}=c(k-k_{0})$. Consequently, an
effective "light" velocity can be got, i.e., $c=\frac{\partial \omega}{\partial
k}\mid_{k=k_{0}}$. Therefore, we can change light velocity $c$ by tuning
$k_{0}$.

\textit{Emergent Planck constant:}\textbf{ }On the other hand, we consider the
emergent physical law from 1-st order tempo variability (or a uniform motion
of the group-changing space along $\Gamma^{t}$ direction). Because the
momentum of the physical variants $V_{\mathrm{\tilde{S}\tilde{O}(d+1)},d+1}$
has a uniform distribution, the energy density $\rho_{E}=\frac{\Delta
E}{\Delta V}$ is constant. We assume that $\rho_{E}(\omega_{0})$ is also a
smooth function of $\omega_{0}$. Then, we have
\begin{equation}
\rho_{E}(\omega_{0}+\delta \omega)=\rho_{E}(\omega_{0})+\frac{\delta \rho_{E}%
}{\delta \omega}\mid_{\omega=\omega_{0}}\delta \omega+...
\end{equation}
where $\frac{\delta \rho_{E}}{\delta \omega}\mid_{\omega=\omega_{0}}=\rho
_{J}^{E}=\rho_{J}$ is called the density of (effective) "angular momentum". In
the following parts, we will point out that the "angular momentum" $\rho_{J}$
of an elementary particles is just Planck constant $\hbar$ and the
quantization condition in quantum mechanics comes from the linearization of
energy density $\rho_{E}$ via $\omega$ near $\omega_{0}$.

On the other hand, because the momentum of the physical variants
$V_{\mathrm{\tilde{S}\tilde{O}(d+1)},d+1}$ has a uniform distribution, the
momentum density $\rho_{p_{i}}=\frac{\Delta p_{i}}{\Delta V}$ is constant.
Then, we also assume that $\rho_{p_{i}}(k_{0})$ is also a smooth function of
$k_{0}$. Then, we have
\begin{equation}
\rho_{p_{i}}(k_{0}+\delta k_{i})=\rho_{E}(k_{0})+\frac{\delta \rho_{E}}{\delta
k_{i}}\mid_{k_{i}=k_{0}}\delta k_{i}+...
\end{equation}
where $\frac{\delta \rho_{E}}{\delta k_{i}}\mid_{k_{i}=k_{0}}=\rho_{J}^{p_{i}}$
is called the density of (effective) "angular momentum". In this paper, we
assume the following equivalent relationship exists along spatial direction
and tempo direction, i.e.,
\begin{equation}
\frac{\delta \rho_{E}}{\delta k_{i}}\mid_{k_{i}=k_{0}}=\frac{\delta \rho_{E}%
}{\delta \omega}\mid_{\omega=\omega_{0}}=\rho_{J}.
\end{equation}
This is consistent with the symmetry condition for different directions of spacetime.

\subsubsection{Classification of matter}

In this section, we develop theory about 1-st level physics structure by
classifying the types of matter that correspond to different types of
topological changings of $\mathrm{\tilde{S}\tilde{O}}$\textrm{(d+1)} physical
variants $V_{\mathrm{\tilde{S}\tilde{O}(d+1)},d+1}(\Delta \phi^{\mu},\Delta
x^{\mu},k_{0},\omega_{0})$.

Matter is about globally \emph{expanding} or \emph{contracting} $\mathrm{C}%
_{\mathrm{\tilde{S}\tilde{O}(d+1)},d+1}$ group-changing space\ with changing
its corresponding size in rigid space $\mathrm{C}_{d+1}$. The generation or
annihilation operations of matter is defined by the operator of
contraction/expansion of $\mathrm{C}_{\mathrm{\tilde{S}\tilde{O}(d+1)},d+1}$
group-changing space in Cartesian space\textit{ }$\mathrm{C}_{d}$, i.e.,
$\tilde{U}(\delta \phi^{a})=e^{i((\delta \phi^{a})\cdot \hat{K}^{a})}$ where
$\delta \phi^{a}=(\Delta \phi^{a})^{\prime}-\Delta \phi^{a}$ and $\hat{K}%
^{a}=-i\frac{d}{d\phi^{a}}$ ($a=x,y,z,t$).

In general, there are extra \emph{two} types of the perturbation on a variant:
one is \emph{ordered type}, in which the information of the local changings of
a physical variant is complete. Although the physical variant may be uniform
or not, we can completely characterize the whole system of variant; the other
is \emph{disordered type}, in which the information of changing of a physical
variant is unknown and we cannot completely characterize the whole system anymore.

To clarify different matters more clearly, we define \emph{ordered/disordered
P-variants}.

\textit{Definition -- Ordered/disordered perturbative variant }%
$V_{\mathrm{\tilde{S}\tilde{O}(d+1)},d+1}$\textit{: }$V_{\mathrm{\tilde
{S}\tilde{O}(d+1)},d+1}$ \textit{is defined by an ordered/disordered mapping
between the (}$d+1$\textit{)-dimensional Clifford group-changing space
}$\mathrm{C}_{\mathrm{\tilde{S}\tilde{O}(d+1)},d+1}$\textit{ and the (}%
$d+1$\textit{)-dimensional Cartesian space }$\mathrm{C}_{d}$\textit{, i.e., }%
\begin{align}
V_{\mathrm{\tilde{S}\tilde{O}(d+1)},d+1}  &  :\{ \delta \phi^{\mu,A},\delta
\phi^{\mu,B}\} \in \mathrm{C}_{\mathrm{\tilde{S}\tilde{O}(d+1)},d+1}\nonumber \\
&  \Leftrightarrow^{\mathrm{ordered/disorder}}\{ \delta x^{\mu}\}
\in \mathrm{C}_{d+1}.
\end{align}
\textit{where }$\Leftrightarrow^{\mathrm{ordered/disorder}}$\textit{\ denotes
an ordered/disordered mapping under fixed changing rate of integer
multiple.\ The total size }$\Delta \phi^{\mu,B}$\textit{ is much smaller than
that of }$\Delta \phi^{\mu,A}$\textit{. }In other words, the disordered
P-variant $V_{\mathrm{\tilde{S}\tilde{O}(d+1)},d+1}$ has a random distribution
of group-changing elements $\delta \phi_{j}^{\mu,B}$ that is named
\emph{classical object (or classical matter)}; while, the ordered P-variant
$V_{\mathrm{\tilde{S}\tilde{O}(d+1)},d+1}$ has a known (not random)
distribution of group-changing elements $\delta \phi_{j}^{\mu,B}$ that is named
\emph{quantum object (or quantum matter)}. Without extra group-changing
elements $\delta \phi_{j}^{\mu,B},$ we have a vacuum with matter.

In the following parts, we will show that how matter plays role of the carrier
of movement -- the ordered type of matter corresponds to the case of quantum
objects and the disordered type of matter corresponds to the case of classical objects.

\subsubsection{Classification of motions}

In this section, we develop theory about 2-nd level physics structure by
classifying the types of motion that corresponds to different types of
time-dependent changings of $\mathrm{\tilde{S}\tilde{O}}$\textrm{(d+1)}
physical variants $V_{\mathrm{\tilde{S}\tilde{O}(d+1)},d+1}(\Delta \phi^{\mu
},\Delta x^{\mu},k_{0},\omega_{0})$.

In above section, we pointed out that globally expansion/contraction of
group-changing space corresponds to the generation/annihilate of particles in
quantum mechanics. In this part, we point out that locally
expansion/contraction of group-changing space corresponds to the motion of
particles in quantum mechanics with fixed particle's number. The classical
objects are a group of group-changing elements with random distribution and
the quantum objects are a group of group-changing elements with regular
(ordered) distribution. For these two types of matter, quantum objects or
classical objects, there are totally four types of processes (or motions) in
our world, \textrm{U}-process, \textrm{C}-process, \textrm{R}-process,
\textrm{R}$^{-1}$-process. See illustration in Fig.14.

\begin{figure}[ptb]
\includegraphics[clip,width=0.73\textwidth]{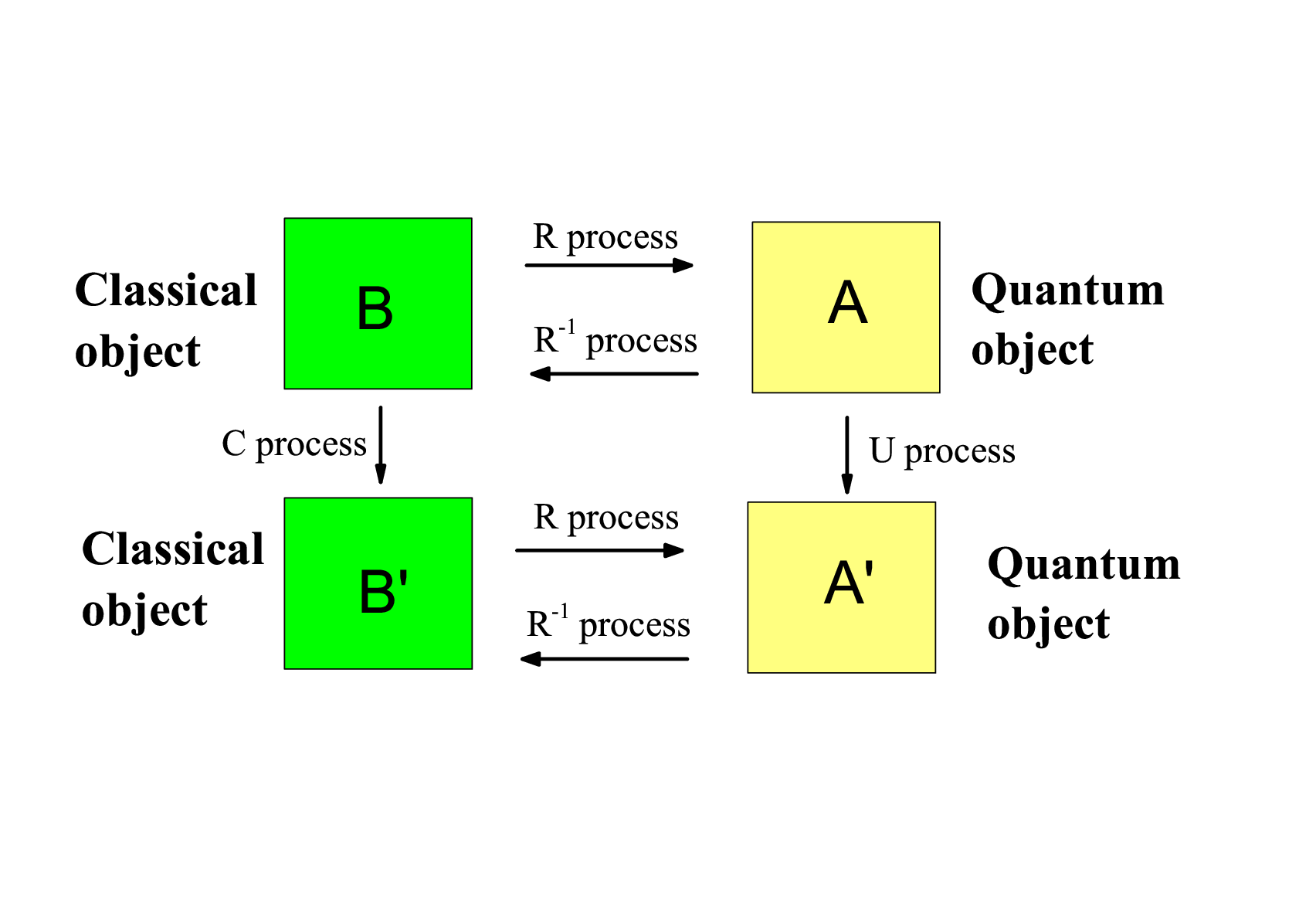}\caption{An illustration
of four types of processes between classical object and quantum object}%
\end{figure}

\textrm{U}-process denotes a quantum motion under unitary time evolutions,
that is characterized by Sch\"{o}rdinger equation. Now, the regular
distribution of the group-changing elements for a quantum object smoothly
changes. We may denote an \textrm{U}-process by
\begin{equation}
V_{1}\Longrightarrow V_{2}%
\end{equation}
where $V_{1}$ and $V_{2}$ are the initial ordered P-variant and final ordered
P-variant, respectively. Here, "$\Longrightarrow$" means time evolution.

\textrm{C}-process denotes a classical motion of time evolution in classical
mechanics, that is characterized by Newton equation. Now, the disordered
distribution of the group-changing elements (or classical object) globally
shift. We may denote a \textrm{C}-process by
\begin{equation}
\tilde{V}_{1}\Longrightarrow \tilde{V}_{2}%
\end{equation}
where $\tilde{V}_{1}$ and $\tilde{V}_{2}$ denote the initial P-variant and
final P-variant, respectively.

\textrm{R}-process denotes a process from a quantum object to a classical one,
that is characterized by Master equation. Now, a regular distribution of the
group-changing elements for a quantum object suddenly changes into a
disordered distribution of the group-changing elements for a classical object.
We may denote an \textrm{R}-process by
\begin{equation}
V_{1}\Longrightarrow \tilde{V}_{2}%
\end{equation}
where $V_{1}$ and $\tilde{V}_{2}$ denote the initial ordered P-variant and
final P-variant, respectively. In the following part, we point out that
quantum measurement is just a \textrm{R}-process from a quantum object to a
classical one.

\textrm{R}$^{-1}$-process denotes a process from a classical object to a
quantum one. Now, A disordered distribution of the group-changing elements for
a classical object changes into a regular distribution of the group-changing
elements for a quantum object. This is a process of preparing a quantum state.
We may denote a \textrm{R}$^{-1}$-process by
\begin{equation}
\tilde{V}_{1}\Longrightarrow V_{2}%
\end{equation}
where $\tilde{V}_{1}$ and $V_{2}$ denote the initial P-variant and final
ordered P-variant, respectively. In addition, we point out that the
preparation of quantum states is a \textrm{R}$^{-1}$-process from a classical
object to a quantum one.

In summary, there are four types of different processes, \textrm{U}-process,
\textrm{C}-process, \textrm{R}-process, \textrm{R}$^{-1}$-process.
\textrm{U/C}-process belongs to quantum/classical motion; \textrm{R/R}$^{-1}%
$-process belongs to the changings of types of matter's motions. In the
following parts, we will discuss them one by one in detail.

\subsection{Quantum mechanics: theory for quantum objects}

In above section, we assume that our universe is special variant --
\textbf{(}$3+1$\textbf{)}-dimensional\textit{ }$\mathrm{\tilde{S}\tilde
{O}(3+1)}$ physical variants\textit{ }$V_{\mathrm{\tilde{S}\tilde{O}%
(3+1)},3+1}(\Delta \phi^{\mu},\Delta x^{\mu},k_{0},\omega_{0}).$ In this part,
we focus on the issue of \emph{ordered P-variant} that comes from
\emph{ordered perturbation} on the physical variant. Now, in principle, we can
completely characterize the whole system. This new theoretical framework about
dynamics of physical variant becomes \emph{pre-quantum mechanics}. Under
partial K-projection, the pre-quantum mechanics is reduced to usual quantum
mechanics that people are very familiar with.

\subsubsection{Quantum elementary particle: Zero Hypothesis, topological
characteristics, dynamic property}

\paragraph{Mapping a uniform physical variant onto a many-particle system}

To develop a new, complete theoretical framework of quantum mechanics (we had
call it pre-quantum mechanics), an important question is "\emph{what is the
information unit of physical reality for quantum mechanics and what's
elementary particle (quantum object)?"} In this part, we will answer all these
questions and develop theory about 1-st level physics structure by mapping a
uniform physical variant onto a many-particle system.

A uniform ($3+1$)-dimensional $\mathrm{\tilde{S}\tilde{O}(3+1)}$\textit{
}physical variant $V_{\mathrm{\tilde{S}\tilde{O}(d+1)},d+1}[\Delta \phi^{\mu
},\Delta x^{\mu},k_{0}^{\mu}]$ is a mapping between\textit{ $\mathrm{\tilde
{S}\tilde{O}}$\textrm{(d+1)} }Clifford group-changing space\textit{
}$\mathrm{C}_{\mathrm{\tilde{S}\tilde{O}(d+1)},d+1}$\textit{\ }to a rigid
spacetime $\mathrm{C}_{d+1},$\textit{ }with size matching $\Delta \phi^{\mu
}=k_{0}\Delta x^{\mu}$. In particular, for this special U-variant, there
exists only one type of group-changing elements.

Firstly, we do D-projection. For uniform ($3+1$)-dimensional $\mathrm{\tilde
{S}\tilde{O}(3+1)}$\textit{ }physical variant $V_{\mathrm{\tilde{S}\tilde
{O}(d+1)},d+1}[\Delta \phi^{\mu},\Delta x^{\mu},k_{0}^{\mu}],$ under
D-projection, we can characterize its changing structure along $\mu$-th
direction by projecting the group-changing space, i.e.,
\[
\tilde{U}^{\mu}(\delta \phi)=\mathrm{Tr}(\Gamma^{\mu}\tilde{U}(\delta \phi)).
\]
Along $\mu$-th direction, we have a complex field of non-compact
\textrm{(}$\mathrm{\tilde{S}\tilde{O}(3+1)}$\textrm{)}$^{\mu}$ group
\begin{equation}
\mathrm{z}_{0}^{\mu}(x)=\tilde{U}^{\mu}(\delta \phi^{^{\mu}}(x^{^{\mu}%
})\mathrm{z}_{0}%
\end{equation}
where $\tilde{U}^{\mu}(\delta \phi^{^{\mu}}(x^{^{\mu}}))=\prod_{i}\tilde
{U}(\delta \phi_{i}^{\mu}(x_{i}))$ with\ $\tilde{U}(\delta \phi_{i}^{\mu
}(x))=e^{i((\delta \phi_{i}^{\mu})\cdot \hat{K}^{\mu})}$ and $\hat{K}^{\mu
}=-i\frac{d}{d\phi^{\mu}}.$ Here, \textrm{(}$\mathrm{\tilde{S}\tilde{O}(3+1)}%
$\textrm{)}$^{\mu}$ is an Abelian non-compact sub-group of its $\Gamma^{\mu}$ component.\

Then, we have a function of $V_{\mathrm{\tilde{S}\tilde{O}(3+1)},3+1}%
(\Delta \phi^{\mu},\Delta x^{\mu},k_{0},\omega_{0})$ along $\mu$-th direction,
i.e,
\[
\mathrm{z}_{p}^{\mu}(x^{^{\mu}})=\operatorname{Re}\xi(x^{^{\mu}}%
)+i\operatorname{Im}\eta(x^{^{\mu}})=e^{i\phi^{\mu}(x)}%
\]
where $\phi(x^{\mu})=\phi_{0}+k_{0}x^{^{\mu}}.$ As a result, under
D-projection, we have a translation symmetry protected knot/link given
spatial-tempo direction. This is new type of knot/links in higher dimensional
spacetime (not only higher dimensional space, but also time).

Finally, we do K-projection representation on the knot/link along a given
direction $\theta$ on $\{ \xi^{^{\mu}},\eta^{^{\mu}}\}$ 2D space. Under
K-projection along different directions on spacetime, the original uniform
physical variant is reduced into a (3+1)D uniform zero lattice.

Let us show the results in detail. According to zero-equation $\xi_{\theta
}(x^{^{\mu}})=0$ or $\cos(k_{0}x^{^{\mu}}-\theta)=0$ along $\mu$-th direction,
under D-projection and K-projection we get the zero-solutions to be
\begin{equation}
x^{^{\mu}}=l_{p}\cdot N^{^{\mu}}+\frac{l_{p}}{\pi}(\theta+\frac{\pi}{2})
\end{equation}
where $N^{^{\mu}}$ are integer numbers.

\paragraph{Zero Hypothesis of elementary particles}

Based on geometry representation under D-projection and K-projection, a
uniform physical variant is reduced into a uniform zero lattice. According to
above discussion, zero number is a \emph{topological} invariable that
characterizes different topological equivalence classes of the system. We
assume that each zero corresponds to an elementary particle and becomes the
changing unit (or information unit) for the system of "changings".

Then, to develop 1-st level physics structure, we give the second Hypothesis
for elementary particles in quantum physics.

\textit{Information Hypothesis of elementary particles: Elementary particle is
zero in a (}$d+1$\textit{)-dimensional }$\mathrm{\tilde{S}\tilde{O}}%
$\textrm{(d+1)} \textit{physical variants }$V_{\mathrm{\tilde{S}\tilde
{O}(d+1)},d+1}(\Delta \phi^{\mu},\Delta x^{\mu},k_{0},\omega_{0})$
\textit{under D-projection and K-projection.}

As a result, a uniform physical variant is mapped onto a many-particle system,
i.e.,
\[
\text{Uniform physical variant}\Longleftrightarrow \text{Many-particle system,
}%
\]
of which an elementary particle is mapped onto a zero that is the information
unit of the system, i.e.,%
\begin{align*}
\text{Information unit }  &  \Longleftrightarrow \text{Zero}\\
&  \Longleftrightarrow \text{Elementary particle. }%
\end{align*}
This fact also means that the spacetime is composed of elementary particles
and the block of space (or strictly speaking, spacetime) is an elementary particle!

\paragraph{Topological characteristics and dynamic property}

To develop a new, complete theoretical framework of quantum mechanics, another
important question is \emph{How does "quantization" appear in quantum
mechanics? and what does }$\hbar$\emph{ mean?} In this part, we will answer
all these questions and show the mechanism of quantization in quantum mechanics.

\begin{figure}[ptb]
\includegraphics[clip,width=1\textwidth]{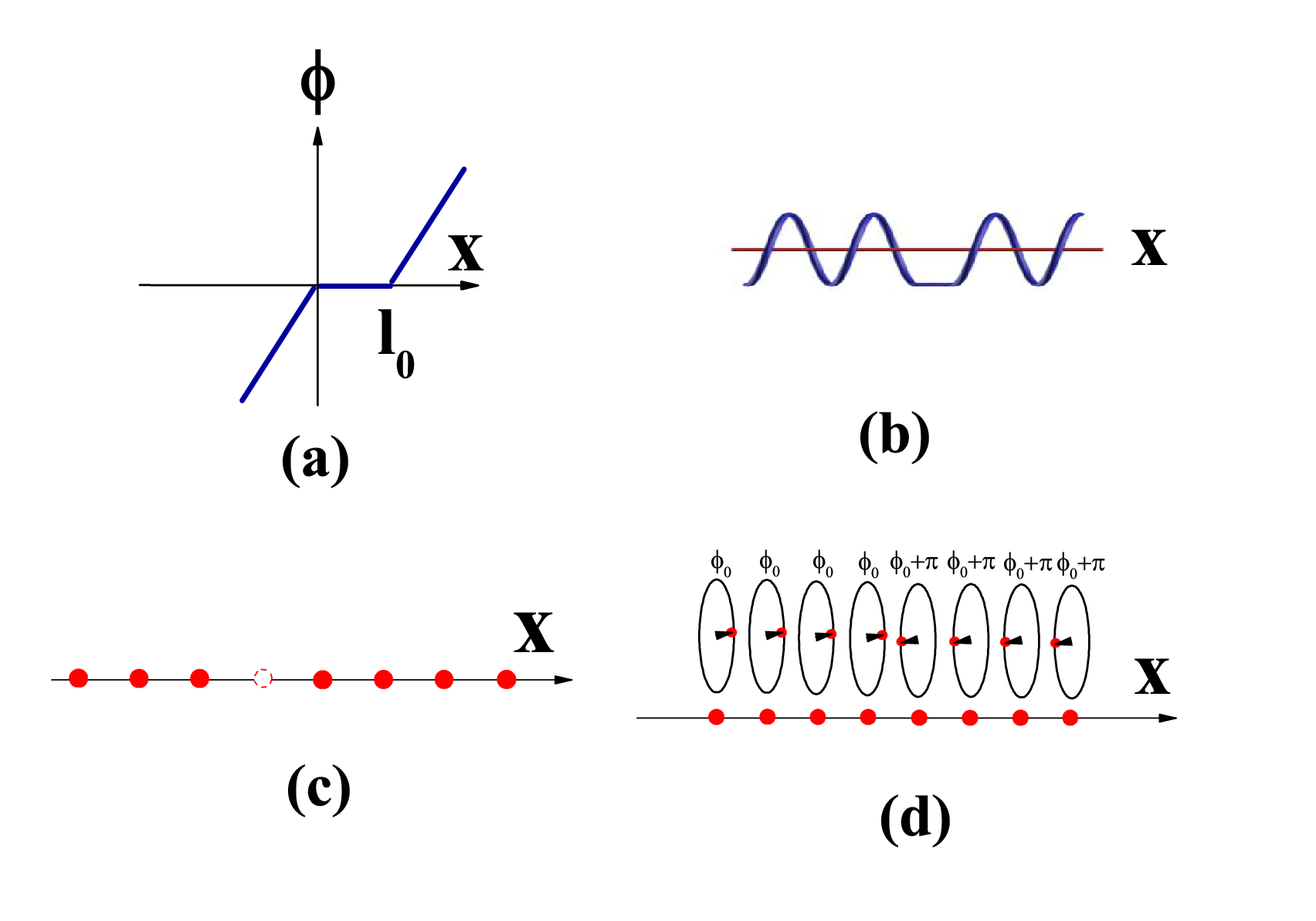}\caption{An illustration of
an elementary particle that is an additional zero with $\pi$-phase changing.
(a) 1-st order algebra representation; (b) 1-st order Geometry
representation; (c) 0-th order representation type-I fully K-projection; (d)
Hybrid-level Geometry representation.}%
\end{figure}

It was known that an elementary particle is changing unit (or information
unit) of the physical variants\textit{ }$V_{\mathrm{\tilde{S}\tilde{O}%
(d+1)},d+1}(\Delta \phi^{\mu},\Delta x^{\mu},k_{0},\omega_{0})$.

An important fact is the correspondence between a zero and $\pi$-phase
changing. Under K-projection along $\mu$-th direction, we have the zero
equation as
\[
\cos(\phi(x^{^{\mu}})-\theta)=0,
\]
of which the zero solution is given by $k_{0}(x^{^{\mu}}-x_{0}^{^{\mu}%
})-\theta=\pm \frac{\pi}{2}$. See the illustration of Fig.15.

As a result, when there exists an extra zero corresponding to an elementary,
the periodic boundary condition of systems along arbitrary direction is
changed into anti-periodic boundary condition, i.e.,
\begin{equation}
\frac{\mathrm{z}_{p}^{\mu}(x^{^{\mu}}\rightarrow-\infty)}{\mathrm{z}_{p}^{\mu
}(x^{^{\mu}}\rightarrow \infty)}=-\frac{\mathrm{z}_{0}(x^{^{\mu}}%
\rightarrow-\infty)}{\mathrm{z}_{0}(x^{^{\mu}}\rightarrow \infty)}%
\end{equation}
where $\mathrm{z}_{0}$ denotes a uniform physical variant.

For the uniform physical variant, each zero corresponds to an elementary
particle. Because the zero has uniform distribution, the size of the
elementary particle is $\pi/k_{0}=l_{0}/2$ where $l_{0}/2$\ is the minimum
distance between two zeroes. As a result, in $d$-dimensional space, the volume
of an elementary particle is given by $V_{F}=(\frac{l_{0}}{2})^{d}$. The exact
formula of the volume of an elementary particle will be calculated elsewhere.

The finite size of an elementary particle leads to fixed "angular momentum" to
it. It is known that the angular momentum of the physical variants has a
uniform distribution, or the angular momentum density $\rho_{J}$ is constant.
Then, for an elementary particle with fixed length $\frac{\pi}{k_{0}}%
=\frac{l_{0}}{2}$ along different spatial directions, the "angular momentum"
of it is a constant
\[
J_{F}=\rho_{J}\cdot(\frac{\pi}{k_{0}})^{d}=V_{F}\rho_{J}.
\]
$J_{F}$ plays the role of Planck constant $\hbar$ in quantum mechanics, i.e.,
\begin{align*}
&  \text{Fixed "angular momentum" }J\text{ }\\
&  \text{for an elementary particle}\\
&  \Longleftrightarrow \text{ Planck constant }\hbar \text{.}%
\end{align*}
Elsewhere, we point that $l_{0}=\frac{2\pi}{k_{0}}$ is twice of Planck lengths
$l_{p}$ (or $l_{0}/2=l_{p}$). The detailed calculations will be given elsewhere.

In summary, we point out that the quantization in quantum mechanics comes from
the \emph{topological characteristics} of elementary particle with fixed
"angular momentum", i.e.,
\begin{align*}
&  \text{Quantization in quantum mechanics}\\
&  \Longleftrightarrow \text{ Topological characteristics of an elementary }\\
&  \text{particle with fixed "angular momentum" }J_{F}\text{.}%
\end{align*}

In addition, we emphasize that because Planck constant $\hbar$ characterizes
the constant motion on Clifford group-changing space, and the changings of the
distribution of group-changing elements on Cartesian space\textit{
}$\mathrm{C}_{d+1}$ will never change its value, i.e., $\hbar=$
\textrm{constant}.

\subsubsection{Quantum motion of single elementary particle: definition and
representation}

To develop a new, complete theoretical framework for quantum mechanics, we
must answer the following questions "\emph{what is quantum motion?" }and\emph{
"How to characterize it?"} It was known that an elementary particle is
information unit of the physical variants that corresponds to a zero under
K-projection and D-projection. In this part, to develop theory about 2-nd
level physics structure, we focus on a system with an extra elementary
particle that corresponds to perturbatively expand or contract of the Clifford
group space $\mathrm{C}_{\mathrm{\tilde{S}\tilde{O}(d+1)},d+1}(\Delta \phi
^{\mu},\Delta x^{\mu},k_{0},\omega_{0})$ with an extra $\pi$ phase changings.
In particular, in this section, we point out that the time-dependent, local,
expand or contract changings for such a perturbative physical variant become
\emph{quantum motions.}

\paragraph{Definition of the states with an extra elementary particle}

To answer these two questions ("\emph{what is quantum motion?" }and\emph{ "How
to characterize it?"}), we firstly give an accurate definition on the states
with an extra elementary particle by defining the perturbative physical
variant on $V_{\mathrm{\tilde{S}\tilde{O}(3+1)},3+1}(\Delta \phi^{\mu},\Delta
x^{\mu},k_{0},\omega_{0})$. Such $\pi$-phase changing is reduced into a zero
under D-projection and K-projection.

\textit{Definition -- A perturbative physical variant with an extra elementary
particle }$V_{\mathrm{\tilde{S}\tilde{O}(3+1)},3+1}(\Delta \phi^{\mu}\pm
\pi,\Delta x^{\mu},k_{0},\omega_{0})$\textit{ is\ a mapping between a (d+1)
dimensional Clifford group-changing space} $\mathrm{C}_{\mathrm{\tilde
{S}\tilde{O}(3+1)},3+1}$\textit{ with total size }$\Delta \phi^{\mu}\pm \pi
$\textit{\ along }$\mu$\textit{-direction and \textit{Cartesian }space
}$\mathrm{C}_{d+1}$\textit{\ with total size }$\Delta x^{\mu}$\textit{, i.e.,}%
\begin{align}
V_{\mathrm{\tilde{S}\tilde{O}(3+1)},3+1}(\Delta \phi^{\mu}\pm \pi,\Delta x^{\mu
},k_{0},\omega_{0})  &  :\nonumber \\
\mathrm{C}_{\mathrm{\tilde{S}\tilde{O}(3+1)},3+1}(\Delta \phi^{\mu}\pm \pi)  &
=\{ \delta \phi^{\mu}\} \nonumber \\
&  \Longleftrightarrow \mathrm{C}_{d+1}=\{ \delta x^{\mu}\}
\end{align}
\textit{where }$\Longleftrightarrow$\textit{\ denotes an ordered mapping under
fixed changing rate of integer multiple }$k_{0}$\textit{ along spatial
direction and fixed changing rate of integer multiple }$\omega_{0}$\textit{
along time direction.}

As a result, the extra elementary particle is a $\pi$-phase changing of
Clifford group-changing space $\mathrm{C}_{\mathrm{\tilde{S}\tilde{O}%
(d+1)},d+1}(\Delta \phi^{\mu})$ on a uniform physical variant
$V_{\mathrm{\tilde{S}\tilde{O}(1+1)},1+1}(\Delta \phi^{\mu},\Delta x^{\mu
},k_{0},\omega_{0})$ along an arbitrary direction (including time direction).
If the total size of the Clifford group-changing space $V_{\mathrm{\tilde
{S}\tilde{O}(3+1)},3+1}$ along $\mu$-direction is $\Delta \phi^{\mu},$ when
there exists an extra elementary particle, the total size of $\mathrm{C}%
_{\mathrm{\tilde{S}\tilde{O}(3+1)},3+1}$ turns into $\Delta \phi^{\mu}\pm \pi.$

\paragraph{Quantum motion for an elementary particle}

An elementary particle is an extra group of group-changing elements with
totally $\pi$-phase changing, i.e., $%
%TCIMACRO{\dsum \limits_{i}}%
%BeginExpansion
{\displaystyle \sum \limits_{i}}
%EndExpansion
(\delta \phi_{i}^{\mu})=\pi.$ In a word, the generation/annihilation of an
elementary particle leads to local \emph{contraction/expansion} changing of
Clifford group-changing space on rigid spacetime from $\mathrm{C}%
_{\mathrm{\tilde{S}\tilde{O}(d+1)},d+1}(\Delta \phi^{\mu})$ to $\mathrm{C}%
_{\mathrm{\tilde{S}\tilde{O}(d+1)},d+1}(\Delta \phi^{\mu}\pm \pi)$. Such expand
or contract of the system indicates that an elementary particle can be
\emph{fragmented}. This fact looks strange. Let us explain it.\emph{ }

In Clifford group-changing space $\mathrm{C}_{\mathrm{\tilde{S}\tilde{O}%
(d+1)},d+1}(\Delta \phi^{\mu}),$ an elementary particle is always a whole and
cannot be divided. However, in Cartesian space\textit{ }$\mathrm{C}_{d+1},$ an
elementary particle can be divided into a group of group-changing elements.
The \emph{evolution of distribution of ordered group-changing elements} of an
elementary particle in Cartesian space $\mathrm{C}_{d+1}$ is quantum motion of
physical reality in quantum mechanics! Different distribution of
group-changing elements of the elementary particle are different states of
quantum motion of particles. Then, we answer the question about "\emph{what is
quantum motion"},%

\begin{align*}
&  \text{Quantum motion for particles}\Longleftrightarrow \text{Evolution of
}\\
&  \text{the distributions of ordered group-changing elements.}%
\end{align*}

\paragraph{Representation to characterize quantum motion}

We next try to answer the second question about \emph{"How to characterize
it". }There are different representations for the perturbative physical
variant $V_{\mathrm{\tilde{S}\tilde{O}(3+1)},3+1}(\Delta \phi^{\mu}\pm
\pi,\Delta x^{\mu},k_{0},\omega_{0})$ with an extra zero (or an extra
elementary particle) from different aspects, including \emph{algebra},
\emph{algebra}, and \emph{geometry}, under different projections, including
D-projection and (partial) K-projection.

\subparagraph{1-st order representation without K-projection}

\textit{1-st order algebra representation}: Now, the physical variants
$V_{\mathrm{\tilde{S}\tilde{O}(d+1)},d+1}(\Delta \phi^{\mu}\pm \pi,\Delta
x^{\mu},k_{0},\omega_{0})$ are approximatively represented by a mapping
between a Clifford group-changing space $\mathrm{C}_{\mathrm{\tilde{S}%
\tilde{O}(d+1)},d+1}$ with two types of space elements $\delta \phi^{A},$
$\delta \phi^{B}$ and the Cartesian space $\mathrm{C}_{d+1}$ with one type of
space elements $\delta x^{\mu}$, i.e.,
\begin{align}
V_{\mathrm{\tilde{S}\tilde{O}(d+1)},d+1}(\Delta \phi^{\mu}\pm \pi,\Delta x^{\mu
},k_{0},\omega_{0})  &  :\nonumber \\
\{ \delta \phi^{A},\delta \phi^{B}\}  &  \in \mathrm{C}_{\mathrm{\tilde{S}%
\tilde{O}(d+1)},d+1}\nonumber \\
&  \Leftrightarrow \{ \delta x\} \in \mathrm{C}_{d+1}%
\end{align}
As a result, $V_{\mathrm{\tilde{S}\tilde{O}(d+1)},d+1}(\Delta \phi^{\mu}\pm
\pi,\Delta x^{\mu},k_{0},\omega_{0})$ is determined by the distribution of the
space elements $\delta \phi^{B}$ on a uniform physical variants
$V_{\mathrm{\tilde{S}\tilde{O}(d+1)},d+1}(\Delta \phi^{\mu},\Delta x^{\mu
},k_{0},\omega_{0})$, of which the summation of total space elements along
arbitrary direction is $\pi$, i.e., $%
%TCIMACRO{\dsum \limits_{i}}%
%BeginExpansion
{\displaystyle \sum \limits_{i}}
%EndExpansion
(\delta \phi_{i}^{\mu,B})=\pi.$ Due to $\pi \ll \Delta \phi^{\mu}$, we can denote
it by the distribution of group-changing elements $\delta \phi^{\mu,B}$.

\textit{1-st order algebra representation}\textbf{:} In algebra
representation, the physical variant $V_{\mathrm{\tilde{S}\tilde{O}(3+1)}%
,3+1}(\Delta \phi^{\mu}\pm \pi,\Delta x^{\mu},k_{0},\omega_{0})$ is described by
a complex matrix $\mathrm{z}_{p}(x)$. By choosing natural reference, we get
the 1-st order algebra representation of the corresponding variants, i.e.,
$\mathrm{z}_{p}(x)=\tilde{U}(\delta \phi)\mathrm{z}_{0}$ where $\tilde
{U}(\delta \phi)=\prod_{\mu}(\prod_{i}\tilde{U}(\delta \phi_{i}^{\mu}(x)))$.

For a ($d+1$)-dimensional variant, we have ($d+1$) D-projected 1D variants,
each of which is described by a complex field of non-compact \textrm{\~{G}%
}$^{\mu}$ Abelian group
\begin{equation}
\mathrm{z}_{p}^{\mu}(x^{^{\mu}})=\tilde{U}^{\mu}(\delta \phi^{^{\mu}}(x^{^{\mu
}}))\mathrm{z}_{0}%
\end{equation}
where $\mu$ denotes an arbitrary direction in ($d+1$)d spacetime.

Then, we have two cases, $V_{\mathrm{\tilde{S}\tilde{O}(d+1)},d+1}(\Delta
\phi^{\mu}+\pi,\Delta x^{\mu},k_{0},\omega_{0})$ and $V_{\mathrm{\tilde
{S}\tilde{O}(d+1)},d+1}(\Delta \phi^{\mu}-\pi,\Delta x^{\mu},k_{0},\omega
_{0}).$\

For the case of $V_{\mathrm{\tilde{S}\tilde{O}(d+1)},d+1}(\Delta \phi^{\mu}%
+\pi,\Delta x^{\mu},k_{0},\omega_{0}),$ the function along $\mu$-th direction
$\mathrm{z}_{p}^{\mu}(x^{^{\mu}})$ is given by
\begin{align*}
\mathrm{z}_{p}^{\mu}(x^{^{\mu}})  &  =\operatorname{Re}\xi(x^{^{\mu}%
})+i\operatorname{Im}\eta(x^{^{\mu}})=e^{i\phi^{\mu}(x)}\\
&  =e^{i\phi^{A\mu}(x)}\text{ or }e^{i\phi^{B\mu}(x)}%
\end{align*}
where
\begin{equation}
\phi^{A,\mu}(x^{\mu})=\left \{
\begin{array}
[c]{c}%
\phi_{0}+k_{0}x^{^{\mu}},\text{ }x^{^{\mu}}\in(-\infty,x_{0}^{^{\mu}}]\\
\phi_{0}+k_{0}x^{^{\mu}},\text{ }x^{^{\mu}}\in(x_{0}^{^{\mu}},x_{0}^{^{\mu}%
}+\frac{\pi}{k_{0}}]\\
-\pi+\phi_{0}+k_{0}x^{^{\mu}},\text{ }x^{^{\mu}}\in(x_{0}^{^{\mu}}+\frac{\pi
}{k_{0}},\infty)
\end{array}
\right \}
\end{equation}
or
\begin{equation}
\phi^{B,\mu}(x^{\mu})=\left \{
\begin{array}
[c]{c}%
\phi_{0}+k_{0}x^{^{\mu}},\text{ }x^{^{\mu}}\in(-\infty,x_{0}^{^{\mu}}]\\
\phi_{0}+2k_{0}x^{^{\mu}},\text{ }x^{^{\mu}}\in(x_{0}^{^{\mu}},x_{0}^{^{\mu}%
}+\frac{\pi}{k_{0}}]\\
\pi+\phi_{0}+k_{0}x^{^{\mu}},\text{ }x^{^{\mu}}\in(x_{0}^{^{\mu}}+\frac{\pi
}{k_{0}},\infty)
\end{array}
\right \}  .
\end{equation}

For the case of $V_{\mathrm{\tilde{S}\tilde{O}(d+1)},d+1}(\Delta \phi^{\mu}%
-\pi,\Delta x^{\mu},k_{0},\omega_{0}),$ the function along $\mu$-th direction
$\mathrm{z}_{p}^{\mu}(x^{^{\mu}})$ is given by
\begin{align*}
\mathrm{z}_{p}^{\mu}(x^{^{\mu}})  &  =\operatorname{Re}\xi(x^{^{\mu}%
})+i\operatorname{Im}\eta(x^{^{\mu}})=e^{i\phi^{\mu}(x)}\\
&  =e^{i\phi^{A\mu}(x)}\text{ or }e^{i\phi^{B\mu}(x)}%
\end{align*}
where%
\begin{equation}
\phi^{\mu,A}(x^{\mu})=\left \{
\begin{array}
[c]{c}%
\phi_{0}+k_{0}x^{^{\mu}},\text{ }x^{^{\mu}}\in(-\infty,x_{0}^{^{\mu}}]\\
\phi_{0}+k_{0}x^{^{\mu}},\text{ }x^{^{\mu}}\in(x_{0}^{^{\mu}},x_{0}^{^{\mu}%
}+\frac{\pi}{k_{0}}]\\
\pi+\phi_{0}+k_{0}x^{^{\mu}},\text{ }x^{^{\mu}}\in(x_{0}^{^{\mu}}+\frac{\pi
}{k_{0}},\infty)
\end{array}
\right \}  .
\end{equation}
or
\begin{equation}
\phi^{\mu,B}(x^{\mu})=\left \{
\begin{array}
[c]{c}%
\phi_{0}+k_{0}x^{^{\mu}},\text{ }x^{^{\mu}}\in(-\infty,x_{0}^{^{\mu}}]\\
\phi_{0},\text{ }x^{^{\mu}}\in(x_{0}^{^{\mu}},x_{0}^{^{\mu}}+\frac{\pi}{k_{0}%
}]\\
-\pi+\phi_{0}+k_{0}x^{^{\mu}},\text{ }x^{^{\mu}}\in(x_{0}^{^{\mu}}+\frac{\pi
}{k_{0}},\infty)
\end{array}
\right \}
\end{equation}

\begin{figure}[ptb]
\includegraphics[clip,width=0.7\textwidth]{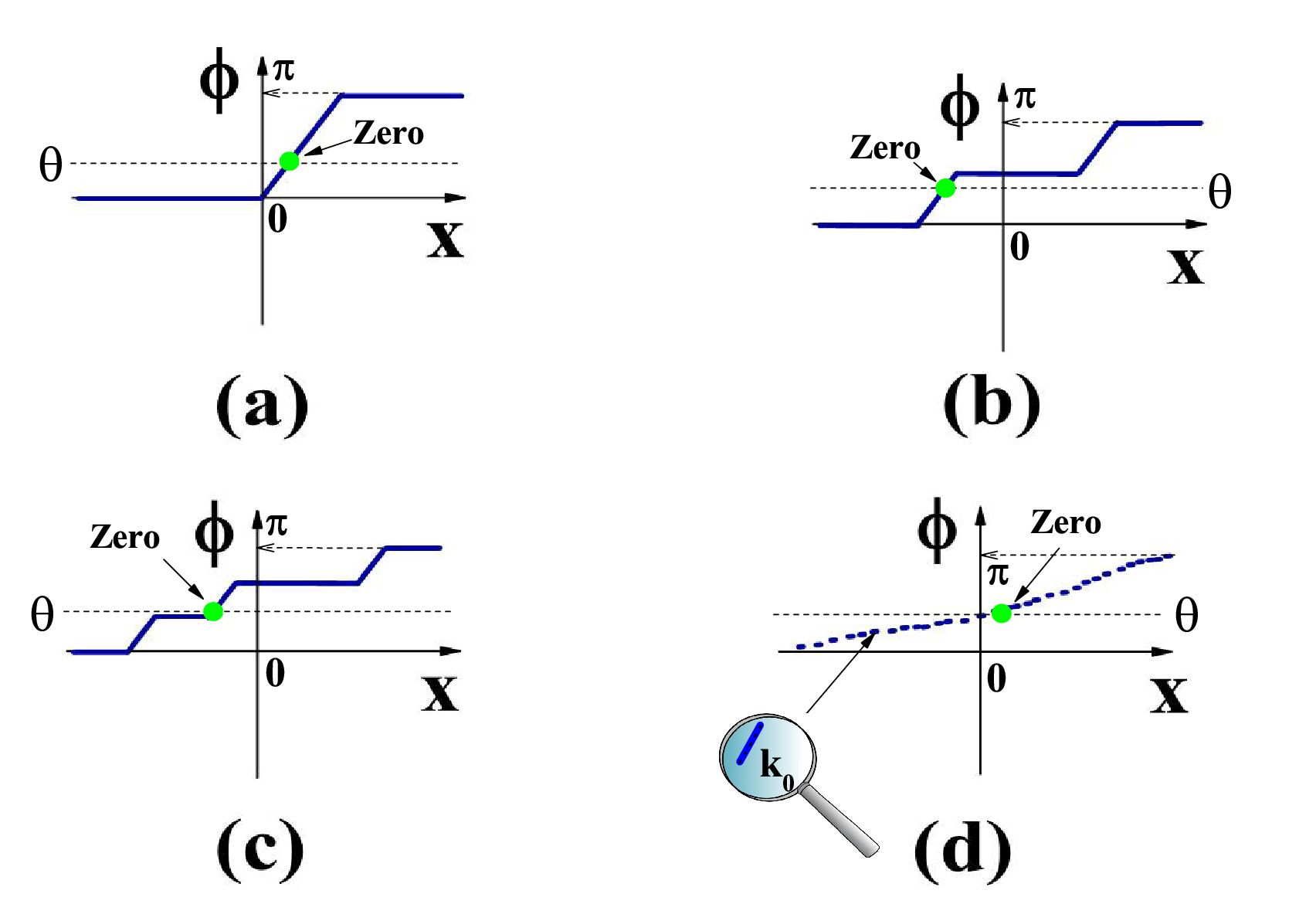}\caption{An illustration
of 1-st order algebra representation for the physical variant with an
additional elementary particle $V_{\mathrm{\tilde{S}\tilde{O}(3+1))}%
,3+1}(\Delta \phi^{\mu}\pm \pi,\Delta x^{\mu},k_{0},\omega_{0}).$ This figure
shows the phase $\phi$ of the complex matrix $\mathrm{z}_{p}(x)$. The green
spot denotes the position of the zero under projection. (a) An unified
elementary particle; (b) A fragmentized elementary particle that is split into
two pieces; (c) A fragmentized elementary particle that is split three pieces;
(d) A fragmentized elementary particle that is split into infinite pieces. The
blue points denote the changing pieces for the elementary particle with
$N\rightarrow \infty.$ }%
\end{figure}

In Fig.16, we show an illustration of 1-st order algebra representation for
the physical variant with an additional elementary particle $V_{\mathrm{\tilde
{U}(1)},1}(\Delta \phi+\pi,\Delta x^{\mu},k_{0},\omega_{0}).$ $\phi$ is the
phase of the complex matrix $\mathrm{z}_{p}(x)$.

\textbf{1-st order geometry representation under D-projection:} In 1-st order
geometry representation of $V_{\mathrm{\tilde{S}\tilde{O}(3+1)},3+1}%
(\Delta \phi^{\mu}\pm \pi,\Delta x^{\mu},k_{0},\omega_{0}),$ under D-projection,
we have knot/link structure along arbitrary spatial-tempo direction.

\subparagraph{Hybrid-order representations under partial K-projection --
quantum representation for an elementary particle}

Next, we introduce Hybrid-order representations under partial K-projection for
quantum motion of an elementary particle. If we only focus on the "changings"
of the physical variants $V_{\mathrm{\tilde{S}\tilde{O}(3+1)},3+1}(\Delta
\phi^{\mu},\Delta x^{\mu},k_{0},\omega_{0})$ rather than itself, we must
"hide" the whole uniform physical variants $V_{\mathrm{\tilde{S}\tilde
{O}(3+1)},3+1}(\Delta \phi^{\mu},\Delta x^{\mu},k_{0},\omega_{0})$ and project
it to a zero lattice. Such a zero lattice is then considered to be a rigid
spacetime. By using Hybrid-order representation under partial K-projection, we
locally characterize the information of the extra group-changing elements
$\delta \phi_{i}^{\mu,B}(x)$ on zero lattice by field of compact group. Such a
description of local field of compact group is just the usual quantum
representation for an elementary particle!

\textit{Algebra representation}: In algebra representation of Hybrid-order
representation under partial K-projection, the perturbative physical variant
with an extra elementary particle (or a zero) is characterized by a series of
(local) group operations of compact \textrm{SO(3+1)} group.

Let us show the theory step by step.

The first step is to consider the physical variant with an extra elementary
particle (or a zero) $V_{\mathrm{\tilde{S}\tilde{O}(3+1)},3+1}(\Delta \phi
^{\mu}\pm \pi,\Delta x^{\mu},k_{0},\omega_{0})$ as a summation of a U-variant
$V_{\mathrm{\tilde{S}\tilde{O}(3+1)},3+1}(\Delta \phi^{\mu},\Delta x^{\mu
},k_{0},\omega_{0})$ and the partner $V_{\mathrm{\tilde{S}\tilde{O}(3+1)}%
,3+1}^{\prime}(\pm \pi,\Delta x^{\mu},k_{0},\omega_{0})$ of its
complementary\ pair, i.e.,
\begin{align}
&  V_{\mathrm{\tilde{S}\tilde{O}(3+1)},3+1}(\Delta \phi^{\mu}\pm \pi,\Delta
x^{\mu},k_{0},\omega_{0})\nonumber \\
&  =V_{\mathrm{\tilde{S}\tilde{O}(3+1)},3+1}(\Delta \phi^{\mu},\Delta x^{\mu
},k_{0},\omega_{0})\\
&  -V_{\mathrm{\tilde{S}\tilde{O}(3+1)},3+1}^{\prime}(\pm \pi,\Delta x^{\mu
},k_{0},\omega_{0}).\nonumber
\end{align}
For P-variant, the number of extra group-changing elements is very small.
Therefore, we can use $V_{\mathrm{\tilde{S}\tilde{O}(3+1)},3+1}^{\prime}%
(\pm \pi,\Delta x^{\mu},k_{0},\omega_{0})$ to characterize $V_{\mathrm{\tilde
{S}\tilde{O}(3+1)},3+1}(\Delta \phi^{\mu}\pm \pi,\Delta x^{\mu},k_{0},\omega
_{0})$, of which $V_{\mathrm{\tilde{S}\tilde{O}(3+1)},3+1}^{\prime}(\pm
\pi,\Delta x^{\mu},k_{0},\omega_{0})$ and $V_{\mathrm{\tilde{S}\tilde{O}%
(3+1)},3+1}(\Delta \phi^{\mu}\pm \pi,\Delta x^{\mu},k_{0},\omega_{0})$ are
complementary pair.

The second step is to do K-projection on the U-variant $V_{\mathrm{\tilde
{S}\tilde{O}(3+1)},3+1}(\Delta \phi^{\mu},\Delta x^{\mu},k_{0},\omega_{0})$ but
not on its partner $V_{\mathrm{\tilde{S}\tilde{O}(3+1)},3+1}^{\prime}(\pm
\pi,\Delta x^{\mu},k_{0},\omega_{0})$. After partial K-projection, the
non-compact $\mathrm{\tilde{S}\tilde{O}(3+1)}$ group of the original U-variant
$V_{\mathrm{\tilde{S}\tilde{O}(3+1)},3+1}(\Delta \phi^{\mu},\Delta x^{\mu
},k_{0},\omega_{0})$ turns into a compact group on zero lattice of
"two-sublattice", i.e., $\phi^{\mu}(x)=2\pi N^{\mu}(x)+\varphi^{\mu}(x).$ We
then relabel the group-changing space by $2d=8$ numbers ($N^{\mu}%
(x),\varphi^{\mu}(x)$): $\varphi^{\mu}(x)$ denote compact phase angles,
$N^{\mu}(x)$ denote the integer winding numbers of unit cell of zero lattice
$N^{\mu}(x)$. Consequently, an U-variant is reduced into a uniform zero
lattcie and becomes a rigid background.

The third step is to consider the extra group-changing elements of
$V_{\mathrm{\tilde{S}\tilde{O}(3+1)},3+1}^{\prime}(\pm \pi,\Delta x^{\mu}%
,k_{0},\omega_{0})$ on the uniform zero lattice $N^{\mu}(x)$. During this
step, we assume that the zero lattice is rigid lattice and can be considered
as a \emph{background}. The processes of the changings of variant occur on the
rigid background of zero lattice.

The fourth step is to do \emph{compactification }for the extra group-changing
elements. On the zero lattice $N(x),$ to exactly determine an extra
group-changing element of $V_{\mathrm{\tilde{S}\tilde{O}(3+1)},3+1}^{\prime
}(\pm \pi,\Delta x^{\mu},k_{0},\omega_{0}),$ one must know its position of
lattice site $N^{\mu}(x)$ together with its phase angle on this site
$\varphi^{\mu}(x).$ Here, the phase angle is compact, i.e., $\varphi^{\mu
}(x)=\phi^{\mu}(x)\operatorname{mod}(2\pi)$.

The fifth step is to write down the local operation representation on uniform
zero lattice. Now, the P-variant is designed by adding a distribution of the
extra group-changing elements $\delta \phi_{i}^{\mu,B}(x_{i})$ on the zero
lattice with a fixed total phase changing $\Delta \phi^{\mu,B}=\sum
\limits_{i}\delta \phi_{i}^{\mu,B}(x_{i})\ll \Delta \phi^{\mu}$. The non-compact
phase angle $\phi^{\mu}$ turns into a compact one $\varphi^{\mu}$ due to the
compactification. As a result, on zero lattice, the extra group-changing
elements $\delta \phi_{i}^{\mu,B}(x_{i})$ of $\tilde{U}(\delta \phi_{i}^{\mu
,B}(x_{i}))$ is projected into group operation $\hat{U}(\delta \varphi_{i}%
^{\mu}(N_{i}^{\mu}(x_{i})))$. Here, $\hat{U}(\delta \varphi_{i}^{\mu}%
(N_{i}^{\mu}(x_{i})))$ is a local phase operation that changing phase angle
from $\varphi_{0}^{\mu}$ to $\varphi_{0}^{\mu}+\delta \varphi_{i}^{\mu}%
(N_{i}(x_{i})).$ Therefore, we have a certain distribution of local phase
operations on uniform zero lattice. By using the usual field of compact
\textrm{SO(3+1)} group, we can fully describe it.

Now, the $d+1$ compact phase angle $\varphi^{\mu}(x)$ can be reorganized into
two groups, one is global phase angle $\left \vert \varphi^{\mu}(x)\right \vert
=\sqrt{%
%TCIMACRO{\dsum \limits_{\mu}}%
%BeginExpansion
{\displaystyle \sum \limits_{\mu}}
%EndExpansion
(\varphi^{\mu}(x))^{2}}$ that denotes the size of the residue total phase
changing of the system, the other, $d$ phase angles denote \textrm{SO(d+1)}
rotation of the system. Therefore, we have a quantum field of compact
\textrm{U(1)}$\times$\textrm{SO(d+1)} group on ($d+1$)-dimensional zero
lattice. In continuum limit, a higher-dimensional P-variant $V_{\mathrm{\tilde
{S}\tilde{O}(d+1),}d+1}[\Delta \phi^{\mu},\Delta x^{\mu},k_{0}^{\mu}]$ is
characterized by a usual quantum field of compact \textrm{U(1)}$\times
$\textrm{SO(d+1)} group in quantum field theory.

Finally, by using algebra representation of Hybrid-order representation
under partial K-projection, a perturbative-uniform physical variant is reduced
into a group of extra local phase operations on zero lattice that is described
by a field of compact \textrm{U(1)}$\times$\textrm{SO(d+1)} group. Each
group-changing element $\tilde{U}(\delta \phi_{i}^{\mu,B}(x_{i}))$ is projected
into a group-operation element $\hat{U}(\delta \varphi_{i}^{\mu}(N_{i}^{\mu
}(x_{i})))$ with given compact phase $\varphi_{i}^{\mu}(N_{i}^{\mu}(x_{i}))$,
i.e.,
\begin{align*}
\tilde{U}(\delta \phi_{i}^{\mu,B}(x_{i}))  &  \rightarrow \hat{U}(\delta
\varphi_{i}^{\mu}(N_{i}^{\mu}(x_{i}))),\\
\phi^{\mu}  &  \rightarrow2\pi N_{i}^{\mu}(x_{i})+\varphi_{i}^{\mu}(N_{i}%
^{\mu}(x_{i})).
\end{align*}

In addition, we point out that total local phases can change $\pi$ by
exchanging the two zeroes on zero lattice.

\textit{Algebra representation}\textbf{:} In algebra representation of
Hybrid-order representation under partial K-projection, the perturbative
physical variant with an extra elementary particle (or a zero) is
characterized by a complex field $\mathrm{z}$ on uniform zero lattice $N^{\mu
}(x)$, i.e., $\mathrm{z}(N^{\mu}(x))=e^{i\varphi(N^{\mu}(x))}$. To obtain its
algebra representation, we also set a constant matrix as natural reference
$\mathrm{z}_{0}$. Then, we do local group operation on $\mathrm{z}_{0}$ and
get the local algebra representation of Hybrid-order representation under
partial K-projection for the corresponding P-variants.

Firstly, we consider the perturbative physical variant with an extra
elementary particle. Now, we can label the extra group-changing element
$\delta \phi^{\mu}(x)$ from perturbation with $2(d+1)$ numbers, $d+1$ is the
position of the site of the original uniform zero lattice $N^{\mu}(x)$, the
other $d+1$ is phase on this site $\varphi^{\mu}$. Here, $\varphi^{\mu}$ is a
compact phase angle for it, i.e, $\varphi^{\mu}=\phi^{\mu}\operatorname{mod}%
(2\pi)$. We choose the uniform group configuration as natural reference
$\phi(x)=\phi_{0}$ and derive the local function representation by doing
operation $\hat{U}(\delta \varphi^{\mu}(N^{\mu}(x),\varphi^{\mu}(x)))$ on a
natural reference. The extra group-changing element becomes extra object on
zero lattice.

Thus, the variant with an extra elementary particle is denoted by the
following function
\begin{equation}
\mathrm{z}=\hat{U}(\delta \varphi^{\mu}(N^{\mu}(x),\varphi^{\mu}(x)))\mathrm{z}%
_{0}%
\end{equation}
where $\hat{U}(\delta \varphi^{\mu}(N^{\mu}(x),\varphi^{\mu}(x)))$ is a usual
operator of compact \textrm{U(1)}$\times$\textrm{SO(d+1)} group. As a result,
the group operator $\delta \varphi^{\mu}(x)$ becomes "object" on discrete
lattice sites $N^{\mu}(x)$ without finite size.\ In particular, the total
phase changings of an elementary particle is $\pm \pi,$ i.e., $%
%TCIMACRO{\dsum \nolimits_{i}}%
%BeginExpansion
{\displaystyle \sum \nolimits_{i}}
%EndExpansion
\delta \varphi_{i}^{\mu}(N_{i}^{\mu})=\pm \pi$.

We point out that $\hat{U}(\delta \varphi^{\mu}(N^{\mu}(x),\varphi^{\mu}(x)))$
plays the role of creation/annihilation operator for an elementary
particle.\ Then we denote $\hat{U}(\delta \varphi^{\mu}(N^{\mu}(x),\varphi
^{\mu}(x)))$ to a creation operator $a^{\dagger}(N^{\mu})$ or annihilation
operator $a(N^{\mu})$ for an elementary particle. This will lead to the usual
quantum mechanics for an elementary particle.

For an arbitrary quantum state, a generalized function is defined by
\[
\mathrm{z}(n^{\mu})=%
%TCIMACRO{\dsum \nolimits_{\mu}}%
%BeginExpansion
{\displaystyle \sum \nolimits_{\mu}}
%EndExpansion%
%TCIMACRO{\dsum \nolimits_{k}}%
%BeginExpansion
{\displaystyle \sum \nolimits_{k}}
%EndExpansion
a_{k^{\mu}}^{\dagger}\exp(ik^{\mu}\cdot N^{\mu})
\]
where $a_{k^{\mu}}^{\dagger}$ is the amplitude of given plane wave $k^{\mu}$.

In long wave length limit, we replace the discrete numbers $N^{\mu}$ by
continuous coordinate $x^{\mu},$ and have
\begin{equation}
\mathrm{z}(N^{\mu})\rightarrow \mathrm{z}(x^{\mu}).
\end{equation}
As a result, a generalized function for quantum state is
\[
\mathrm{z}(x)\sim \int a_{k^{\mu}}^{\dagger}\exp(ik^{\mu}\cdot x^{\mu})dk.
\]
However, the information of the internal structure for an elementary particle
disappears. The size of an elementary particle on Cartesian space becomes
zero! Without information of $k_{0}$ in $a^{\dagger}$ or $a,$ the changing
rate $k_{0}$ of group-changing elements become hidden and people will never
know the changing rate $k_{0}$ of group-changing elements from the description
of generalized function.

Finally, we do \emph{normalization}$,$ $\mathrm{z}(x)\rightarrow \psi
(x)=C\cdot \mathrm{z}(x)$, and derive a usual "wave function" description for
quantum states of an elementary particle. Here the normalization factor $C$
guarantees\ that the total number of elementary particle is $1$.

In the end of this part, we discuss the physical meaning of wave functions.
For the sake of simplicity, we take 1D case of non-compact \textrm{\~{U}(1)}
as an example.

$\psi(x)$ is just the \emph{wave function} in usual quantum mechanics, denoted
as
\begin{equation}
\psi(x)=\sqrt{\Omega(x)}e^{i\varphi(x)},
\end{equation}
where the phase angle $\varphi(x)$ becomes the quantum phase angle of wave
function. An interesting fact is that the \emph{density} of group-changing
elements $\rho_{p}$ for an elementary particle is proportional to particle's
\emph{density} $\Omega(x)=\int \psi^{\ast}(x)\psi(x)dx$, which indicates
physical meaning of wave functions. We give a proof on the fact.

\textit{Proof:} The density of group-operation elements $\rho_{\mathrm{piece}%
}$ is defined by
\begin{align}
\rho_{\mathrm{piece}}  &  =\sum \limits_{i=1}^{N}\delta \varphi_{i}\nonumber \\
&  =C^{2}\int \mathrm{z}^{\ast}\hat{K}\mathrm{z}\text{ }d\varphi=\left \langle
\frac{\hat{K}}{\Delta V}\right \rangle
\end{align}
where $\hat{K}=-i\frac{d}{d\varphi}$.\ We can either label a group-changing
element according to its position $i_{x}$ on Cartesian space or label it
according to $\varphi_{i}$ on Clifford group-changing space. Here $\varphi
_{i}$ denotes ordering of $\varphi$ on Clifford group-changing space from
small to big and $i_{x}$ denotes a sorting of coordination $x$ with a given
order. Each $\delta \varphi_{i}$ corresponds to an $i_{x}.$ Then, we have
\begin{align}
\rho_{\mathrm{piece}}  &  =\left \langle \frac{\hat{K}}{\Delta V}\right \rangle
=C^{2}\int \mathrm{z}^{\ast}\hat{K}\mathrm{z}\text{ }d\varphi \\
&  =C^{2}\sum_{i_{\phi}}\left[  \mathrm{z}(x_{i_{\phi}})\right]  ^{\ast}%
\hat{K}\left[  \mathrm{z}(x_{i_{\phi}})\right] \nonumber \\
&  =C^{2}\sum_{i_{x}}\left[  \mathrm{z}(x_{i_{x}})\right]  ^{\ast}\hat
{K}\left[  \mathrm{z}(x_{i_{x}})\right] \nonumber \\
&  =C^{2}\left[  \mathrm{z}(x)\right]  ^{\ast}\hat{K}\left[  \mathrm{z}%
(x)\right]  dx\nonumber \\
&  =\frac{1}{\Delta V}\psi^{\ast}(x)(-i\frac{d}{d\varphi})\psi(x)dx\nonumber \\
&  =\psi^{\ast}(x)\psi(x)=\Omega(x).\nonumber
\end{align}
The result can be easily generalized to the case in high dimensions by
introducing global phase factor and internal relative phase factors and we
skipped the detailed discussion in this paper. According to above fact, we can
see that the essence of matter in a wave function is phase change. Finding
particles means finding changes. Therefore, in places with more changings,
there are more particles.

In summary, for quantum mechanics, it is wave functions that characterize the
distribution of extra group-changing elements of an extra elementary particle
on Cartesian space, i.e.,%
\begin{align}
&  \text{"Wave function" for quantum states }\\
&  \Longleftrightarrow \text{Algebra representation of Hybrid-order
}\nonumber \\
&  \text{representation under partial K-projection.}\nonumber
\end{align}

\textit{Hybrid-level geometry representation}\textbf{:} We discuss the
geometry representation of Hybrid-order representation under partial
K-projection and D-projection for the perturbative physical variant with an
extra elementary particle (or a zero).

From above discussion, by using algebra representation of Hybrid-order
representation under partial K-projection, the perturbative physical variant
with an extra elementary particle (or a zero) is characterized by a complex
group field
\begin{equation}
\mathrm{z}=\hat{U}(\delta \varphi^{\mu}(N^{\mu}(x),\varphi^{\mu}(x)))\mathrm{z}%
_{0}%
\end{equation}
where $\hat{U}(\delta \varphi^{\mu}(N^{\mu}(x),\varphi^{\mu}(x)))$ is an usual
operator of compact \textrm{U(1)}$\times$\textrm{SO(d+1)} group. Under
D-projection, it is reduced into Abelian sub-group along $\mu$-th direction,
i.e.,
\begin{equation}
\mathrm{z}^{\mu}=(\mathrm{Tr}(\Gamma^{\mu}\hat{U}(\delta \varphi^{\mu}(N^{\mu
}(x),\varphi^{\mu}(x)))))\mathrm{z}_{0}.
\end{equation}
The configuration of group elements is a set of given phase angles
$e^{i\varphi^{\mu}(N^{\mu}(x^{\mu}))\Gamma^{\mu}}$ on each position of zero
lattice. Finally, this configuration structure of group field $e^{i\varphi
^{\mu}(N^{\mu}(x^{\mu}))\Gamma^{\mu}}$ becomes a "\emph{non-changing}" structure.

In summary, we obtain wave function description in quantum mechanics by using
Hybrid-order representation under partial K-projection and D-projection.

\subparagraph{0-th order representations under fully K-projection}

Next, we do fully K-projection for the variant under D-projection. For the
function of $V_{\mathrm{\tilde{S}\tilde{O}(3+1)},3+1}(\Delta \phi^{\mu}\pm
\pi,\Delta x^{\mu},k_{0},\omega_{0})$ along $\mu$-th direction, there are two
types of 0-th order representations under different K-projections -- type-I
and type-II. Under fully K-projection, we have a zero lattice with defects.
Under two types of fully K-projections, the whole "\emph{changing}" structure
of a variant is reduced into two "\emph{non-changing}" structures.

To classify the difference of the two types of 0-th order representations
under fully K-projections, we consider $V_{\mathrm{\tilde{S}\tilde{O}%
(3+1)},3+1}(\Delta \phi^{\mu}\pm \pi,\Delta x^{\mu},k_{0},\omega_{0})$ as the
difference between a U-variant $V_{0,\mathrm{\tilde{S}\tilde{O}(3+1)}%
,3+1}(\Delta \phi^{\mu},\Delta x^{\mu},k_{0},\omega_{0})$ and the partner
$V_{\mathrm{\tilde{S}\tilde{O}(3+1)},3+1}^{\prime}(\pm \pi,\Delta x^{\mu}%
,k_{0},\omega_{0})$ of its complementary\ pair, i.e.,
\begin{align*}
&  V_{\mathrm{\tilde{S}\tilde{O}(3+1)},3+1}(\Delta \phi^{\mu}\pm \pi,\Delta
x^{\mu},k_{0},\omega_{0})\\
&  =V_{0,\mathrm{\tilde{S}\tilde{O}(3+1)},3+1}(\Delta \phi^{\mu},\Delta x^{\mu
},k_{0},\omega_{0})\\
&  -V_{\mathrm{\tilde{S}\tilde{O}(3+1)},3+1}^{\prime}(\pm \pi,\Delta x^{\mu
},k_{0},\omega_{0}).
\end{align*}

For 0-th order representation under type-I fully K-projection and D-projection
for $V_{\mathrm{\tilde{S}\tilde{O}(3+1)},3+1}(\Delta \phi^{\mu}\pm \pi,\Delta
x^{\mu},k_{0},\omega_{0})$\ with single projection angle $\theta,$ we have
defective zero lattice. On the other hand, For 0-th order representation under
type-II fully K-projection and D-projection for $V_{\mathrm{\tilde{S}\tilde
{O}(3+1)},3+1}(\pm \pi,\Delta x^{\mu},k_{0},\omega_{0}),$\ we do K-projection
on $V_{0,\mathrm{\tilde{S}\tilde{O}(3+1)},3+1}(\Delta \phi^{\mu},\Delta x^{\mu
},k_{0},\omega_{0})$ with single projection angle $\theta_{0}$ and another on
$V_{\mathrm{\tilde{S}\tilde{O}(3+1)},3+1}(\pm \pi,\Delta x^{\mu},k_{0}%
,\omega_{0})$ with projection angles $\theta$, respectively. Now, we have a
distribution of extra zero on a uniform zero lattice. Because we consider the
uniform zero lattice to be a rigid background, the original P-variant
$V_{\mathrm{\tilde{S}\tilde{O}(3+1)},3+1}(\Delta \phi^{\mu}\pm \pi,\Delta
x^{\mu},k_{0},\omega_{0})$ is characterized by the the distribution of zeroes
of $V_{\mathrm{\tilde{S}\tilde{O}(3+1)},3+1}(\pm \pi,\Delta x^{\mu}%
,k_{0},\omega_{0}).$

We finally compare the difference between a zero for an elementary particle
and a "point mass" in classical mechanics. In usual classical picture for our
world, an elementary particle is always regarded as a "point mass" and moves
on a rigid space. An interesting fact is that under D-projection and fully
K-projection, an elementary particle indeed turns into a "point mass".
Therefore, we call 0-th order representations under fully K-projection to be
"\emph{classical}" description.

When people try to understand quantum mechanics, they always insist on
"\emph{classical picture}". According to usual classical picture, they have a
hidden assumption -- "\emph{the elementary particle is an indivisible point on
a rigid space}", that looks like a classical mass point. The "classical"
picture leads to the existence of a lot of "misleading" confused
interpretations of quantum mechanics, such as hidden invariable
interpretation, many world interpretation, stochastic interpretation...

\paragraph{Summary}

Finally, the intrinsic relationship between different representations for
quantum mechanics becomes clear! According to it, the representation in usual
quantum mechanics is just "wave function" representation, that is Hybrid-order
representation under partial K-projection and D-projection.

\subsubsection{Quantum motion of many elementary particles: definition,
representation, Fermionic statistics, and quantum entanglement}

To develop a new, complete theoretical framework for quantum mechanics,
another important question is \emph{What is the relationship between different
information units of physical reality?} In this part, by taking a 1D physical
variant as an example, we will answer this question and show the description
of quantum states for two or more elementary particles. In addition, we show
the emergence of fermionic statistics and quantum entanglement.

\paragraph{Definition}

Firstly, we define a perturbative physical variant with many elementary
particles. Because each elementary particle has a $\pi$-phase changing on
Clifford group-changing space, a system with $N_{F}$ elementary particles have
an $N_{F}\pi$-phase changing on Clifford group-changing space.

\textit{Definition -- A perturbative physical variant with an extra elementary
particle }$V_{\mathrm{\tilde{S}\tilde{O}(3+1)},3+1}(\Delta \phi^{\mu}\pm
N_{F}\pi,\Delta x^{\mu},k_{0},\omega_{0})$\textit{ is\ a mapping between a
(d+1) dimensional Clifford group-changing space} $\mathrm{C}_{\mathrm{\tilde
{S}\tilde{O}(3+1)},3+1}$\textit{ with total size }$\Delta \phi^{\mu}\pm
N_{F}\pi$\textit{\ along }$\mu$\textit{-direction and \textit{Cartesian }space
}$\mathrm{C}_{d+1}$\textit{\ with total size }$\Delta x^{\mu}$\textit{, i.e.,}%
\begin{align}
V_{\mathrm{\tilde{S}\tilde{O}(3+1)},3+1}(\Delta \phi^{\mu}\pm N_{F}\pi,\Delta
x^{\mu},k_{0},\omega_{0})  &  :\nonumber \\
\mathrm{C}_{\mathrm{\tilde{S}\tilde{O}(3+1)},3+1}(\Delta \phi^{\mu}\pm N_{F}%
\pi)  &  =\{ \delta \phi^{\mu}\} \nonumber \\
&  \Longleftrightarrow \mathrm{C}_{d+1}=\{ \delta x^{\mu}\}
\end{align}
\textit{where }$\Longleftrightarrow$\textit{\ denotes an ordered mapping under
fixed changing rate of integer multiple }$k_{0}$\textit{ along spatial
direction and fixed changing rate of integer multiple }$\omega_{0}$\textit{
along time direction.}

\paragraph{Quantum motion for $N_{F}$ elementary particles}

For $V_{\mathrm{\tilde{S}\tilde{O}(d+1)},d+1}(\Delta \phi^{\mu}\pm N_{F}%
\pi,\Delta x^{\mu},k_{0},\omega_{0})$, $N_{F}$ elementary particle is an
$(N_{F}\pi)$-phase changing of Clifford group-changing space $\mathrm{C}%
_{\mathrm{\tilde{S}\tilde{O}(d+1)},d+1}(\Delta \phi^{\mu})$ along arbitrary
direction that leads to an extra group of group-changing elements, i.e., $%
%TCIMACRO{\dsum \limits_{i}}%
%BeginExpansion
{\displaystyle \sum \limits_{i}}
%EndExpansion
(\delta \phi_{i}^{\mu})=\pm N_{F}\pi.$ $N_{F}$ elementary particles lead to a
globally expand or contract of the system. And, locally expand or contract of
the system indicates that the quantum motion of $N_{F}$ elementary particles
is described by an ordered distribution of group-changing elements in
Cartesian space $\mathrm{C}_{d+1}$. Different distribution of group-changing
elements of the elementary particles are different states of quantum motion of particles.

\paragraph{Representations}

The (quantum) states for many elementary particles are determined by the
distribution of the extra group-changing elements $\delta \phi_{i}^{B}$ on a
uniform physical variant $V_{\mathrm{\tilde{S}\tilde{O}(3+1)},3+1}(\Delta
\phi^{\mu},\Delta x^{\mu},k_{0},\omega_{0})$, of which the summation of total
space elements is $N_{F}\pi$, i.e., $%
%TCIMACRO{\dsum \limits_{i}}%
%BeginExpansion
{\displaystyle \sum \limits_{i}}
%EndExpansion
(\delta \phi_{i}^{B})=N_{F}\pi.$ There are different representations for the
perturbative physical variant $V_{\mathrm{\tilde{S}\tilde{O}(3+1)},3+1}%
(\Delta \phi^{\mu}\pm N_{F}\pi,\Delta x^{\mu},k_{0},\omega_{0})$ with extra
$N_{F}$ zero under different aspects, including \emph{algebra},
\emph{algebra}, and \emph{geometry}, and under different projections,
including D-projection and (partial) K-projection.

In algebra representation of Hybrid-order representation under partial
K-projection, the perturbative physical variant with $N_{F}$ extra elementary
particles (or $N_{F}$ zeroes) is characterized by a complex field $\mathrm{z}$
on uniform zero lattice $N^{\mu}(x)$, i.e., $\mathrm{z}(N^{\mu}%
(x))=e^{i\varphi^{\mu}(N^{\mu}(x))}$. Therefore, we can obtain wave functions
for quantum many-particle states that characterize the distribution of extra
group-changing elements of extra $N_{F}$ elementary particles on Cartesian space.

\paragraph{Identical principle for elementary particles}

To distinguish the elementary particles, we consider two elementary particles.

In algebra representation of Hybrid-order representation under partial
K-projection, the functions of two particles for same quantum state possess
the same formula. For the group operators to generate elementary particles are
defined by $\hat{U}(\delta \phi^{B}(x))=\prod_{j=1}^{n}e^{i(\delta \phi_{i}%
^{B})\cdot \hat{K}_{j}}$ with $\hat{K}_{j}=-i\frac{d}{d\phi_{j}}$. Arbitrary
group-changing elements for the two particles are identical.\ So, the
elementary particles are \emph{identical} particle. This gives us identical
principle for elementary particles.

\paragraph{Fermionic statistics for elementary particles}

Next, we study the quantum statistics for elementary particles. Because an
elementary particle has a $\pi$-phase changing along arbitrary direction in
spacetime, when there exists an extra elementary particle, the periodic
boundary condition of systems along arbitrary direction is changed into
anti-periodic boundary condition. As a result, elementary particles
(topological defects of spacetime) obey fermionic statistics.

For two static particles, we have
\begin{equation}
\Psi(x,x^{\prime})\rightarrow \hat{U}(x^{\prime},t)\cdot \hat{U}(x,t)\mathrm{z}%
_{0}%
\end{equation}
where $\hat{U}(x^{\prime},t)$\ denotes the group operation of an elementary
particle with $\pi$-phase changing. After exchanging two particles, we get
\begin{align}
\Psi(x^{\prime},x)  &  \sim \lbrack \hat{U}(x^{\prime},t)\cdot \hat
{U}(x,t)]\mathrm{z}_{0}\\
&  \rightarrow \Psi(x,x^{\prime})=-[\hat{U}(x^{\prime},t)\cdot \hat
{U}(x,t)]\mathrm{z}_{0}\nonumber \\
&  \rightarrow-\Psi(x,x^{\prime}).\nonumber
\end{align}
See the illustration in Fig.17.

By using quantum description of "wave function" (or algebra representation
of Hybrid-order representation under partial K-projection), we introduce the
second quantization representation for fermionic particles by defining
fermionic operator $c^{\dagger}(x)$ as
\begin{equation}
\hat{U}(x^{\prime})\Longrightarrow c^{\dagger}(x).
\end{equation}
According to the fermionic statistics, there exists anti-commutation relation
\begin{equation}
\{c(x),\text{ }c^{\dagger}(x)\}=\delta(x-x^{\prime}).
\end{equation}

\begin{figure}[ptb]
\includegraphics[clip,width=0.5\textwidth]{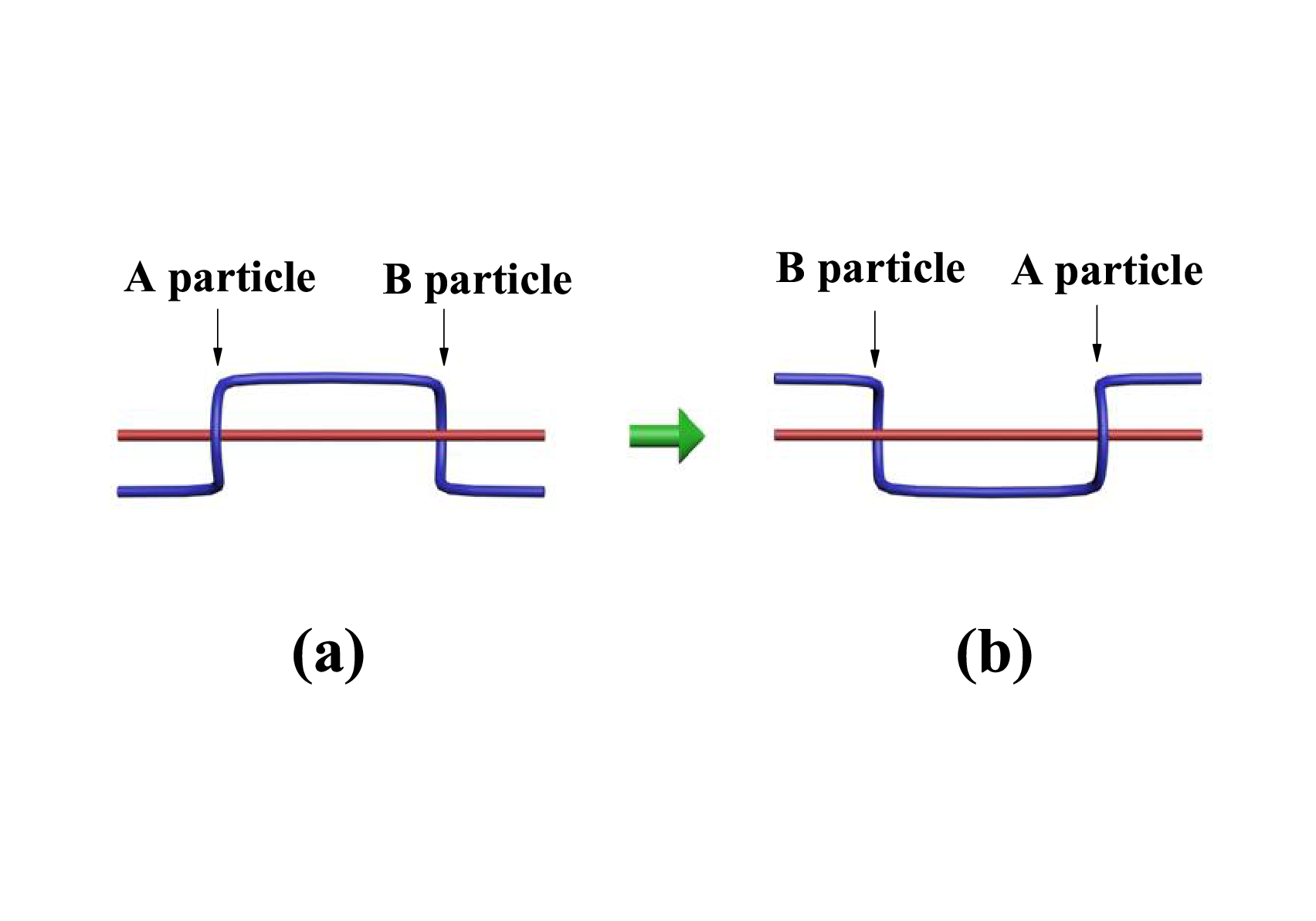}\caption{Under 1-st order
Geometry representation under D-projection, the exchanging two elementary
particles A and B with zero size on Cartesian space leads to $\pi$-phase
changing. Therefore, elementary particles obey fermionic statistics.}%
\end{figure}

In summary, the fermionic statistics comes from the algebraic relationship
between two elementary particles and indicates their \emph{non-local
}property, i.e.,
\begin{align}
&  \text{ Fermionic statistics for elementary particles}\\
&  \Longleftrightarrow \text{\ Algebraic relationship between}\nonumber \\
&  \text{ two information units of "variant". }\nonumber
\end{align}

\paragraph{Quantum entanglement}

Quantum entanglement is a physical phenomenon for many-body systems. According
to quantum entanglement, the quantum states of each particle cannot be
described independently of the others, even when the particles are separated
by a large distance. The starting point of quantum entanglement\cite{sch3} is
the Einstein-Podolsky-Rosen paradox \cite{epr} that revealed an unexpected
aspect of quantum physics which violates the main principle of special
relativity allowing information to be transmitted faster than light. In this
part, we show the approach to recover its "non-local" property from
representation without projection. This will help people to understand this
strange non-local phenomena in quantum mechanics clearly.

An entangled state for $N_{F}$-body quantum system comes from new type of
particles -- $N_{F}\pi$-particle that is a composite object with $N_{F}$ $\pi
$-phase changings. Such a composite object corresponds to $N_{F}$ zeroes.
Therefore, the quantum states for a $N_{F}\pi$-particle cannot be reduced into
a product state of the wave-function for $N_{F}$ particles and become entangled.

\begin{figure}[ptb]
\includegraphics[clip,width=0.7\textwidth]{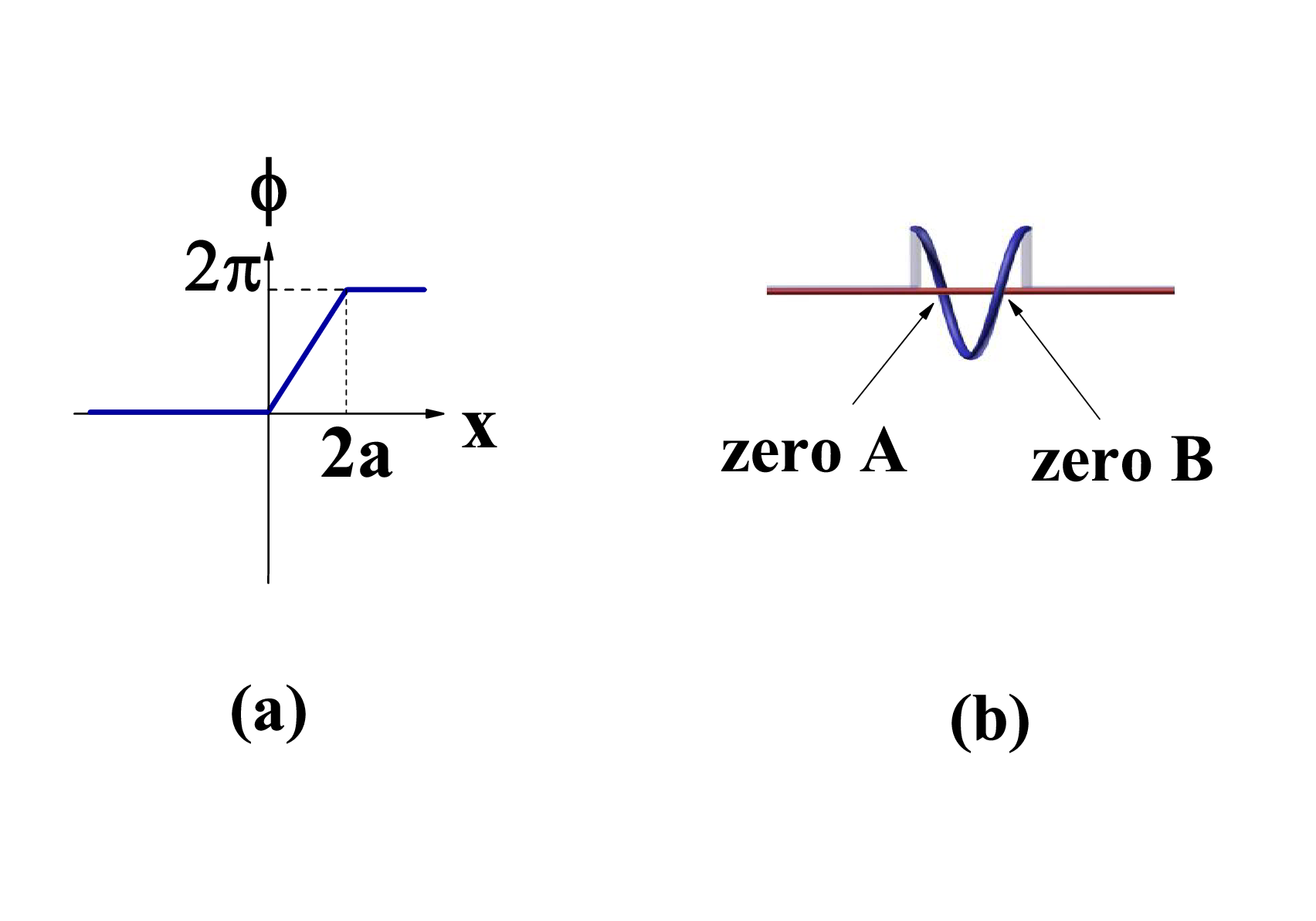}\caption{(a) The function
for a $2\pi$-particle with 2 \ particles under 1-st order algebra
representation under D-projection; (b) A picture for a unified $2\pi$-particle
with correlated 2 zeros under 1-st order Geometric representation and
D-projection. }%
\end{figure}

Because the non-local character of quantum entangled states for a composite
object can only be shown in 1-st order representations. Then we use the 1-st
order algebra representation under D-projection to show the detailed
structure of quantum entangled states.

By using 1-st order algebra representation under D-projection, the function
for a special $2\pi$-particle along $\mu$-th direction is given by
\begin{equation}
\mathrm{z}_{2\pi}=\exp[i\phi_{2\pi}(x)],
\end{equation}
with
\begin{equation}
\phi_{2\pi}(x)=\left \{
\begin{array}
[c]{c}%
\phi_{0},\text{ }x\in(-\infty,x_{0}]\\
\phi_{0}+k_{0}(x-x_{0}),\text{ }x\in(x_{0},x_{0}+2a]\\
\phi_{0}+2\pi,\text{ }x\in(x_{0}+2a,\infty)
\end{array}
\right \}
\end{equation}
where $\phi_{0}$ is constant. See the illustration of a $2\pi$-particle in
Fig.18. The group-changing elements for entangled two particles come from a
$2\pi$-particle rather than from two independent two $\pi$-particles.

The concept of composite particle can be generalized to the entangled states
for $N_{F}\pi$-particles. Under 1-st order algebra representation under
D-projection, the function for the composite $N_{F}\pi$-particle is given by
\begin{equation}
\mathrm{z}_{N_{F}\pi}=\exp[i\phi_{N_{F}\pi}(x,t)],
\end{equation}
with
\begin{equation}
\phi_{N_{F}\pi}(x)=\left \{
\begin{array}
[c]{c}%
\phi_{0},\text{ }x\in(-\infty,x_{0}]\\
\phi_{0}+k_{0}(x-x_{0}),\text{ }x\in(x_{0},x_{0}+na]\\
\phi_{0}+N_{F}\pi,\text{ }x\in(x_{0}+na,\infty)
\end{array}
\right \}  .
\end{equation}

The entangled states for $N_{F}$ elementary particles indicates a fact that
the $N_{F}\pi$-particles must be considered a unified object with $N_{F}\pi$
phase changing. Quantum entanglement comes from the coherent quantum motion
for $N_{F}\pi$-particles (that is a composite object of $N_{F}$ particles) and
indicates a \emph{hidden "space" structure} for quantum states of
multi-particles, i.e.,
\begin{align}
&  \text{Quantum entanglement}\\
&  \Longleftrightarrow \text{ Coherent quantum }\nonumber \\
&  \text{motion for }N_{F}\pi \text{-particles.}\nonumber
\end{align}

In addition, the quantum entangled states with $N_{F}\pi$-particle become very
strange by using 0-th order representation under type-II fully K-projection.
The coherent motion of $N_{F}\pi$-particles leads to correlated motion of
these $N_{F}$ "classical" objects. However, the correlation between the
$N_{F}$ zeroes is spooky, i.e, no matter how far apart they are connected each
other. To naturally understand this strange phenomenon, one must recover their
"non-local" character. After recovering its non-local character by using 1-st
order representations, we can completely predict the positions of the zeroes
according to wave functions. Then, this spooky phenomenon is no longer strange.

\subsubsection{Time-evolution of quantum states and the emergence of
Schr\"{o}dinger equation}

To uncover the underlying physics of quantum mechanics, an important question
is "\emph{what law does the time evolution of physical reality obey and what's
the corresponding equation?}" or\emph{ }"\emph{why the time-evolution of a
quantum states of an elementary particles obeys Schr\"{o}dinger equation?}" We
then discuss the time-evolution of a given state in a physical variant and try
to derive Schr\"{o}dinger equation.

\paragraph{Emergence of Schr\"{o}dinger equation}

When there exists an additional particle on uniform physical variant
$V_{\mathrm{\tilde{S}\tilde{O}(d+1)},d+1}(\Delta \phi^{\mu},\Delta x^{\mu
},k_{0},\omega_{0})$, the total energy of the system slightly changes
\begin{equation}
\mathrm{H}\rightarrow \mathrm{H}^{\prime}=\mathrm{H}+\Delta \mathrm{H}%
\end{equation}
where $\Delta \mathrm{H}\propto \Delta V$ is the volume changing, and $\Delta
V\ll V$. On the other hand, the energy of a particle is described by a
slightly changing of angular velocity on the system, $\omega_{0}%
\rightarrow \omega_{0}+\Delta \omega$. Because the system rotates globally with
very fast angular velocity, i.e., $\omega_{0}\gg \Delta \omega$, the energy
changing of a particle with fixed angular momentum $\hbar$ is obtained as
\begin{equation}
\Delta \mathrm{H}=J_{\mathrm{particle}}\cdot \Delta \omega=\hbar \cdot \Delta
\omega.
\end{equation}

Then, we choose the usual "wave function" representation (or Hybrid-order
representation under partial K-projection and D-projection). Under "wave
function" representation $\psi(\vec{x},t)=%
%TCIMACRO{\dsum \nolimits_{p}}%
%BeginExpansion
{\displaystyle \sum \nolimits_{p}}
%EndExpansion
c_{p}e^{-i\Delta \omega \cdot t+i\vec{k}\cdot \vec{x}}$, we have
\begin{align}
\left \langle \Delta \omega \right \rangle  &  =\int \psi^{\ast}(x,t)\Delta
\omega \psi(x,t)dV\nonumber \\
&  =\int[%
%TCIMACRO{\dsum \nolimits_{p}}%
%BeginExpansion
{\displaystyle \sum \nolimits_{p}}
%EndExpansion
c_{p}^{\ast}e^{i\Delta \omega \cdot t-i\vec{k}\cdot \vec{x}}](i\frac{\partial
}{\partial t})\nonumber \\
&  \times \lbrack%
%TCIMACRO{\dsum \nolimits_{p^{\prime}}}%
%BeginExpansion
{\displaystyle \sum \nolimits_{p^{\prime}}}
%EndExpansion
c_{p^{\prime}}e^{-i\Delta \omega \cdot t+i\vec{k}\cdot \vec{x}}]dV\\
&  =\int \psi^{\ast}(\vec{x},t)(i\frac{d}{dt})\psi(\vec{x},t)dV.\nonumber
\end{align}
These results ($E=\hbar \cdot \Delta \omega$ and $\Delta \omega \rightarrow
\hat{\omega}=i\frac{d}{dt}$) indicate that the energy becomes operator%
\begin{equation}
E\rightarrow \mathrm{\hat{H}}=\hbar \cdot i\frac{d}{dt}.
\end{equation}

As a result we derive the Schr\"{o}dinger equation for particles as%
\begin{equation}
i\hbar \frac{d\psi(\vec{x},t)}{dt}=\mathrm{\hat{H}}\psi(\vec{x},t)
\end{equation}
where $\mathrm{\hat{H}}$ is the Hamiltonian of elementary particles. For the
eigenstate with eigenvalue $E,$%
\begin{align}
\mathrm{\hat{H}}\psi(\vec{x},t)  &  =E\psi(\vec{x},t)\nonumber \\
&  =\hbar \cdot \Delta \omega \psi(\vec{x},t),
\end{align}
the wave-function becomes $\psi(\vec{x},t)=f(\vec{x})\exp(\frac{iEt}{\hbar})$
where $f(\vec{x})$ is spatial function.

In summary, Schr\"{o}dinger equation is an inevitable result of linearization
behavior of particle's energy around a periodically motion $\omega
_{0}\rightarrow \omega_{0}+\Delta \omega$. Therefore, the time-evolution of a
quantum state of an elementary particles obeys Schr\"{o}dinger equation,
i.e.,
\begin{align*}
&  \text{Schr\"{o}dingere quation }\\
&  \Longleftrightarrow \text{ An equation for perturbation on }\\
&  \text{periodical motion of group-changing space. }%
\end{align*}

\paragraph{Effective Hamiltonian for elementary particles}

In this section, we derive the effective Hamiltonian for elementary particles.

The effective Hamiltonian of the elementary particle is obtained as Dirac
model. In particular, there emerges another constant -- mass $m$ for
elementary particles.

We firstly define generation operator of elementary particle $c_{i}^{\dagger
}\left \vert 0\right \rangle =\left \vert i\right \rangle $ on (3+1)D uniform zero
lattice. We write down the hopping Hamiltonian. The hopping term between two
nearest neighbor sites $i$ and $j$ on (3+1)D uniform zero lattice becomes
\begin{equation}
\mathcal{H}_{\left \{  i,j\right \}  }=Jc_{i}^{\dagger}(t)\mathbf{T}_{\left \{
i,j\right \}  }c_{j}(t)
\end{equation}
where $\mathbf{T}_{\left \{  i,j\right \}  }$ is the transfer matrix between two
nearest neighbor sites $i$ and $j$ , $c_{i}(t)$ is the annihilation operator
of elementary particle at the site $i$. $J=\frac{c}{2l_{p}}$ is an effective
coupling constant between two nearest-neighbor sites. $l_{p}=l_{0}/2$ is
Planck length and $c$ is light speed. According to variability, $\left \vert
i\right \rangle =e^{il_{p}(\hat{k}^{\mu}\cdot \Gamma^{\mu})}\left \vert
j\right \rangle ,$ the transfer matrix $\mathbf{T}_{\left \{  i,j\right \}  }$
between $\left \vert i\right \rangle $ and $\left \vert j\right \rangle $ is
defined by
\[
\mathbf{T}_{\left \{  i,j\right \}  }=\left \langle i\mid j\right \rangle
=e^{il_{p}(\hat{k}^{\mu}\cdot \Gamma^{\mu})}.
\]
After considering the contribution of the terms from all sites, the effective
Hamiltonian is obtained as%
\begin{equation}
\mathcal{H}=%
%TCIMACRO{\dsum \limits_{\{i,j\}}}%
%BeginExpansion
{\displaystyle \sum \limits_{\{i,j\}}}
%EndExpansion
\mathcal{H}_{\left \{  i,j\right \}  }=J%
%TCIMACRO{\dsum \limits_{\{i,j\}}}%
%BeginExpansion
{\displaystyle \sum \limits_{\{i,j\}}}
%EndExpansion
c_{i}^{\dagger}\mathbf{T}_{\left \{  i,j\right \}  }c_{i+e^{I}}.
\end{equation}
See the illustration of 2D/3D zero lattices for fermionic elementary particles
in Fig.19.

\begin{figure}[ptb]
\includegraphics[clip,width=0.5\textwidth]{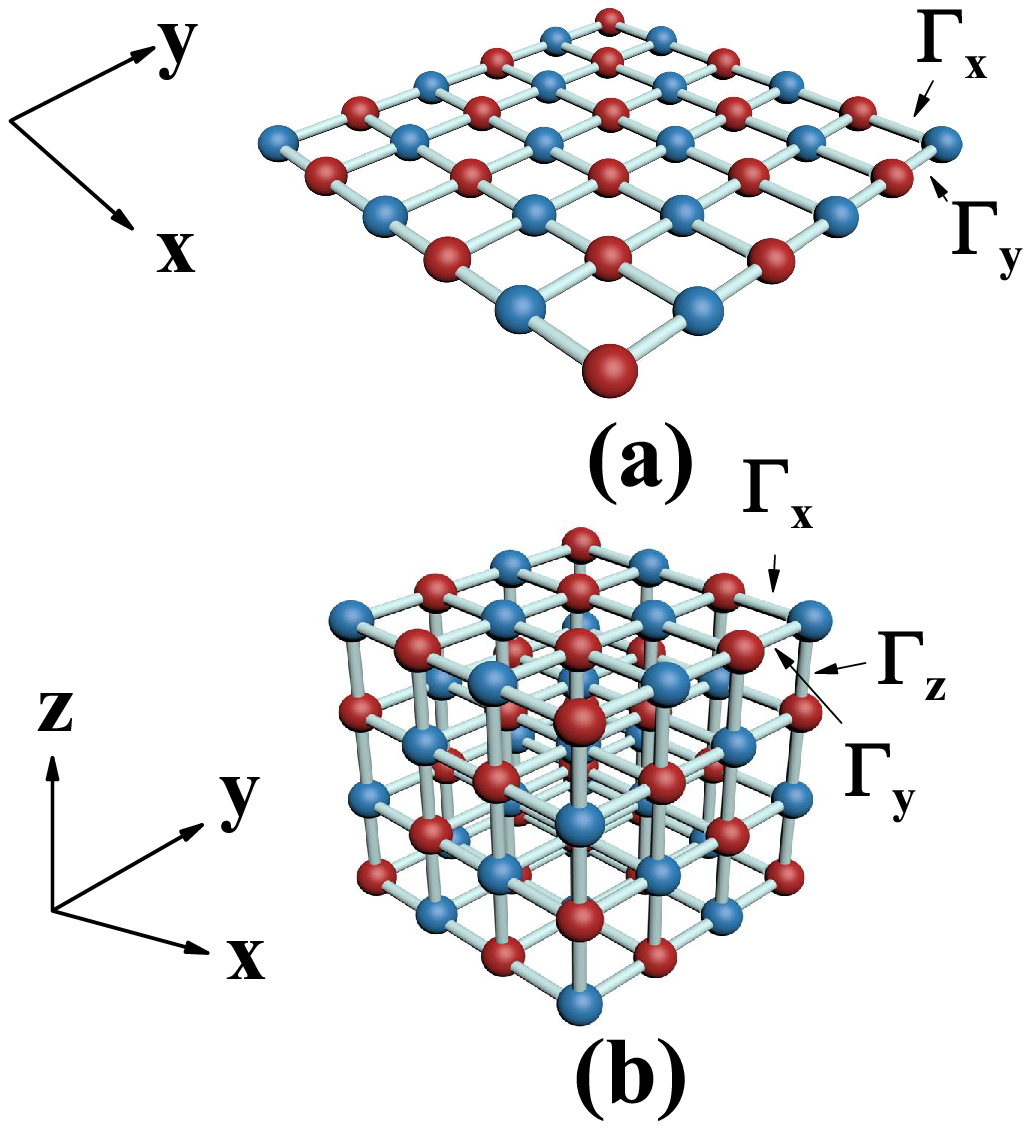}\caption{An illustration
of 2D/3D zero lattices for fermionic elementary particles}%
\end{figure}

In continuum limit, we have%
\begin{align}
\mathcal{H}  &  =J%
%TCIMACRO{\dsum \limits_{\{i,j\}}}%
%BeginExpansion
{\displaystyle \sum \limits_{\{i,j\}}}
%EndExpansion
c_{i}^{\dagger}(e^{il_{p}(\hat{k}^{\mu}\cdot \Gamma^{\mu})})c_{i+1}+h.c.\\
&  =2l_{p}J%
%TCIMACRO{\dsum \limits_{\mu}}%
%BeginExpansion
{\displaystyle \sum \limits_{\mu}}
%EndExpansion%
%TCIMACRO{\dsum \limits_{k^{\mu}}}%
%BeginExpansion
{\displaystyle \sum \limits_{k^{\mu}}}
%EndExpansion
c_{k^{\mu}}^{\dagger}[\cos(k^{\mu}\cdot \Gamma^{\mu})]c_{k^{\mu}}%
\end{align}
where the dispersion in continuum limit is
\begin{equation}
E_{k}\simeq \pm c\sqrt{[(\vec{k}-\vec{k}_{0})\cdot \vec{\Gamma}]^{2}%
+((\omega-\omega_{0})\cdot \Gamma^{t})^{2}},
\end{equation}
where $\vec{k}_{0}=\frac{1}{l_{p}}(\frac{\pi}{2},\frac{\pi}{2},\frac{\pi}%
{2}),$ and $\omega_{0}=\frac{\pi}{2}\frac{1}{l_{p}}c$.

We then re-write the effective Hamiltonian to be
\begin{equation}
\mathcal{H}=\int(\Psi^{\dagger}(\mathbf{x})\hat{H}\Psi(\mathbf{x}))d^{3}x
\end{equation}
where
\begin{equation}
\hat{H}=\vec{\Gamma}\cdot \Delta \vec{p}%
\end{equation}
with $\vec{\Gamma}=(\Gamma^{x},\Gamma^{y},\Gamma^{z})$ and
\begin{align}
\Gamma^{t}  &  =\tau^{z}\otimes \vec{1}\mathbf{,}\text{ }\Gamma^{x}=\tau
^{x}\otimes \sigma^{x},\\
\Gamma^{y}  &  =\tau^{x}\otimes \sigma^{y},\text{ }\Gamma^{z}=\tau^{x}%
\otimes \sigma^{z}.\nonumber
\end{align}
$\vec{p}=\hbar \Delta \vec{k}$ is the momentum operator. This is a model for
massless Dirac fermions.

To obtain the particle's mass, we must tune $\omega_{0}.$ If $\omega_{0}\neq
ck_{0}$, then the Dirac fermion have finite mass, i.e., $m=\hbar(\omega
_{0}-ck_{0})/c^{2}$. We then re-write the effective Hamiltonian to be
\begin{equation}
\mathcal{H}=\int(\Psi^{\dagger}(\mathbf{x})\hat{H}\Psi(\mathbf{x}))d^{3}x
\end{equation}
where
\begin{equation}
\hat{H}=\vec{\Gamma}\cdot \Delta \vec{p}+m\Gamma^{t}%
\end{equation}
with $\vec{\Gamma}=(\Gamma^{x},\Gamma^{y},\Gamma^{z})$ and
\begin{align}
\Gamma^{t}  &  =\tau^{z}\otimes \vec{1}\mathbf{,}\text{ }\Gamma^{x}=\tau
^{x}\otimes \sigma^{x},\\
\Gamma^{y}  &  =\tau^{x}\otimes \sigma^{y},\text{ }\Gamma^{z}=\tau^{x}%
\otimes \sigma^{z}.\nonumber
\end{align}

In future, when we consider more complex physical variants with 2-nd order
variability, there emerge gauge interactions. Then, we have alternative
Hamiltonian for matter.

\paragraph{Geometry representations for quantum motion of plane waves}

In usual classical mechanics, classical motion of a classical mass point is a
time-dependent shift on Cartesian space.\emph{ How about quantum motions}? In
this part, we try to give a picture for quantum motion of plane waves along
certain direction, $\psi(x,t)=Ce^{-i\Delta \omega \cdot t+ik\cdot x}$.

Firstly, we discuss the 1-st order geometry representations for quantum motion
of plane wave. In 1-st order representation, quantum motion describes an extra
uniformly shifting of extra group-changing elements on group-changing space
$\phi=t\cdot \Delta \omega$, of which the "velocity" is just $\Delta \omega$. On
Cartesian space, this is spiral motion by combining rotating in phase angle
$\varphi(t)=(t\cdot \Delta \omega)\operatorname{mod}(2\pi)$ and translating on
Cartesian space synchronously. The pitch on Cartesian space is $\frac{2\pi}%
{k}.$ The period of rotation motion of phase angle is $\frac{2\pi}%
{\Delta \omega}$.

Second, we discuss the geometry representations of "wave function"
representation for quantum motion of plane wave. This is hybrid-order
representation under partial K-projection and D-projection. After doing
K-projection, the non-compact phase angle becomes a compact one and the spiral
motion is reduced into a periodic rotation motion of phase angle without
shifting on Cartesian space. As a result, quantum motion is a periodical
motion of phase angle $\varphi(t)=(t\cdot \Delta \omega)\operatorname{mod}%
(2\pi)$.

Thirdly, we discuss the 0-th order geometry representations under type-II
fully projection for quantum motion of plane wave. Now, under fully
projection, the moving elementary particle is projected to a moving zero, of
which quantum motion describes a uniformly shifting of an extra zero on
Cartesian space. Let us show the details. We do projected representation along
$\theta$ direction on $\xi/\eta$-plane. If $\theta$ is fixed, the position of
zero solution becomes very \emph{strange}. During the time interval $\Delta
t=\frac{\pi}{\omega},$ the phase angle $\phi$ will be effectively changed
$\pi.$ Consequently, the zero will go through the whole system from one end to
the other during the time interval $\Delta t=\frac{\pi}{\omega}$. The speed of
zero's motion could turn to infinite. As a result, by using type-II fully
projection, the quantum motion is a periodical motion around the whole system
with period $\frac{\pi}{\omega}$ and speed $v_{eff}=\frac{\omega L}{\pi}$
where $L$ is size of the whole system along moving direction.

In summary, we have given geometric picture for quantum motions by using
different representations. Within higher order representation, the picture is
reasonable and becomes more non-local.

\paragraph{Path integral formulation for quantum mechanics}

Path integral formulation for quantum mechanics\cite{fey} is another
formulation describing the time-dependent evolution of the distribution of
group--changing elements.

We firstly take 1D case as an example to show its implication.

The probability amplitude $K(x^{\prime},t_{f};x,t_{i})$ for an elementary
particle from an initial position $x$ at time $t=t_{i}$ (that is described by
a state $\left \vert t_{i},x\right \rangle $) to position $x^{\prime}$ at a
later time $t=t_{f}$ ($\left \vert t_{f},x^{\prime}\right \rangle $) is obtained
as\cite{fey},
\begin{align}
K(x^{\prime},t_{f};x,t_{i})  &  =\langle t_{f},x^{\prime}\left \vert
t_{i},x\right \rangle =%
%TCIMACRO{\dsum \limits_{n}}%
%BeginExpansion
{\displaystyle \sum \limits_{n}}
%EndExpansion
e^{iS_{n}/\hbar}\nonumber \\
&  =\int \mathcal{D}\vec{p}(t)\mathcal{D}x(t)e^{iS/\hbar}%
\end{align}
where
\begin{align}
S  &  =%
%TCIMACRO{\dint }%
%BeginExpansion
{\displaystyle \int}
%EndExpansion
pdx-%
%TCIMACRO{\dint }%
%BeginExpansion
{\displaystyle \int}
%EndExpansion
E(p,x)dt\nonumber \\
&  =%
%TCIMACRO{\dint }%
%BeginExpansion
{\displaystyle \int}
%EndExpansion
p\dot{x}dt-%
%TCIMACRO{\dint }%
%BeginExpansion
{\displaystyle \int}
%EndExpansion
E(p,x)dt=%
%TCIMACRO{\dint }%
%BeginExpansion
{\displaystyle \int}
%EndExpansion
Ldt
\end{align}
and $L=p\dot{x}dt-E(p,x)$. Each group-changing element's path contributes
$e^{iS_{n}/\hbar}$ where $S_{n}$ is the $n$-th classical action for a
group-changing element. Therefore, in the path integral formulation, the
action is total phase changing from motion. $p$ and $E(p,x)$ play the roles of
phase changing rates along spatial and temporal directions, respectively. This
argument obviously provides the foundation of Canonical quantization.

Now, we consider the path integral formulation of multi-elementary particle.

The probability amplitude becomes a multi-variable function
\begin{align}
&  K(\vec{x}_{M}^{\prime},...,\vec{x}_{2}^{\prime},\vec{x}_{1}^{\prime}%
,t_{f};\vec{x}_{M},...,\vec{x}_{2},\vec{x}_{1},t_{i})\\
&  =\langle t_{f},\vec{x}_{M}^{\prime},...,\vec{x}_{2}^{\prime},\vec{x}%
_{1}^{\prime}\left \vert t_{i},\vec{x}_{M},...,\vec{x}_{2},\vec{x}%
_{1}\right \rangle \nonumber
\end{align}
where $x_{j}^{\prime}$ and $x_{j}$ denote the final position and initial
position of $j$-th elementary particle, respectively. For a multi-particle
system, quantum processes are described by
\begin{align}
&  K(\vec{x}_{M}^{\prime},...,\vec{x}_{2}^{\prime},\vec{x}_{1}^{\prime}%
,t_{f};\vec{x}_{M},...,\vec{x}_{2},\vec{x}_{1},t_{i})\nonumber \\
&  =\langle t_{f},\vec{x}_{M}^{\prime},...,\vec{x}_{2}^{\prime},\vec{x}%
_{1}^{\prime}\left \vert t_{i},\vec{x}_{M},...,\vec{x}_{2},\vec{x}%
_{1}\right \rangle \nonumber \\
&  =%
%TCIMACRO{\dprod \limits_{j}}%
%BeginExpansion
{\displaystyle \prod \limits_{j}}
%EndExpansion%
%TCIMACRO{\dsum \limits_{n}}%
%BeginExpansion
{\displaystyle \sum \limits_{n}}
%EndExpansion
e^{i\Delta \phi_{j,n}}=%
%TCIMACRO{\dprod \limits_{j}}%
%BeginExpansion
{\displaystyle \prod \limits_{j}}
%EndExpansion%
%TCIMACRO{\dsum \limits_{n}}%
%BeginExpansion
{\displaystyle \sum \limits_{n}}
%EndExpansion
e^{iS_{j,n}/\hbar}\nonumber \\
&  =%
%TCIMACRO{\dsum \limits_{n}}%
%BeginExpansion
{\displaystyle \sum \limits_{n}}
%EndExpansion
e^{i%
%TCIMACRO{\dsum \limits_{j}}%
%BeginExpansion
{\displaystyle \sum \limits_{j}}
%EndExpansion
S_{j,n}/\hbar}\nonumber \\
&  =%
%TCIMACRO{\dprod \limits_{p}}%
%BeginExpansion
{\displaystyle \prod \limits_{p}}
%EndExpansion
\psi_{p}^{\dagger}(\vec{x},t)\psi_{p}(\vec{x},t)e^{iS_{p}/\hbar}\nonumber \\
&  =\int \mathcal{D}\psi^{\dagger}(\vec{x},t)\mathcal{D}\psi(\vec
{x},t)e^{i\mathcal{S}/\hbar}%
\end{align}
where
\begin{equation}
\mathcal{S}=%
%TCIMACRO{\dsum \limits_{\omega,\vec{p}}}%
%BeginExpansion
{\displaystyle \sum \limits_{\omega,\vec{p}}}
%EndExpansion
S_{\omega,\vec{p}}=\int \mathcal{L}dtd^{3}x
\end{equation}
with
\begin{equation}
\mathcal{L}=i\psi^{\dagger}\partial_{t}\psi-\mathcal{\hat{H}}.
\end{equation}
The symbol $%
%TCIMACRO{\dsum \limits_{n}}%
%BeginExpansion
{\displaystyle \sum \limits_{n}}
%EndExpansion
$ denotes the summation of different group-changing elements and the symbol $%
%TCIMACRO{\dprod \limits_{j}}%
%BeginExpansion
{\displaystyle \prod \limits_{j}}
%EndExpansion
$ denotes the different elementary particles.

\subsubsection{New framework of quantum mechanics}

Quantum mechanics becomes \emph{phenomenological} theory and is interpreted by
using the concepts of the microscopic properties of physical variant. We
provide a new framework for quantum mechanics via the different levels of
physics structure:

\begin{enumerate}
\item Step 1 is to develop theory about \emph{0-th level} physics structure by
giving the Variant hypothesis. Such 0-th level physics structure is a physical
variant with 1-st order spatial-tempo variability;

\item Step 2 is to develop theory about \emph{1-st level} physics structure
(or \emph{matter}) by defining elementary particle (or the information unit of
physical reality). Under projection, each elementary particle corresponds to a
zero. Therefore, particles must be identical. The topological characteristics
of an elementary particle leads to the quantization of quantum mechanics. In
addition, under exchanging these particles, the wave functions for identical
particles are completely antisymmetric well;

\item Step 3 is to develop theory about \emph{2-nd level} physics structure
(or \emph{quantum motion}) by deriving the time-evolution of quantum states.
The quantum motion of physical reality in quantum mechanics corresponds to the
evolution of the distribution of the extra group-changing elements on a
uniform physical variant. Now, Schr\"{o}dinger equation is an inevitable
result of linearization behavior of particle's energy around a periodical
motion $\omega_{0}\rightarrow \omega_{0}+\Delta \omega$. This leads to the
development of dynamic theory for quantum mechanics.
\end{enumerate}

\paragraph{The explanation of fundamental principles in quantum mechanics}

There are several fundamental principles in quantum mechanics:
\emph{wave-particle duality} (objects exhibit both 'wave-like' behavior and
'particle-like' behavior), \emph{uncertainty principle} (attempting to measure
one attribute such as velocity or position may cause another attribute to
become less measurable), and \emph{superposition principle }(a wave-function
superimposes multiple co-existing states that have different probabilities).
Let us give an explanation on them based on variant theory.

\subparagraph{Complementarity principle}

In quantum mechanics, complementarity principle is fundamental proposed by
Born. From the point view of "space" dynamics, it comes from complementarity
property of elementary particles: On the one hand, an elementary particle is
information unit in group-changing space specifically a phase-changing of
$\Delta \phi^{\mu}=\pm \pi$ (or $\Delta \varphi^{\mu}=\pi$); On the other hand,
its quantum state has a given phase angle $\phi$ (or $\varphi$) that is
determined by wave function. One cannot exactly determine the phase angle of
an elementary particle by observing its phase-changing. We call this property
to be complementarity principle in quantum mechanics. We say that the
complementarity principle is related to the "\emph{changing}" characteristics
of quantum object in group-changing space.

\subparagraph{Wave--particle duality}

Wave--particle duality is the fact that elementary particles exhibit both
particle-like behavior and wave-like behavior. As Einstein wrote:
\textquotedblleft \textit{It seems as though we must use sometimes the one
theory and sometimes the other, while at times we may use either. We are faced
with a new kind of difficulty. We have two contradictory pictures of reality;
separately neither of them fully explains the phenomena of light, but together
they do}\textquotedblright.

Here, we point out that wave--particle duality of quantum particles is really
a duality between an information unit of group-changing space $\mathrm{C}%
_{\mathrm{\tilde{S}\tilde{O}(3+1)},3+1}$\ and its mapping to real space. On
the one hand, in group-changing space a particle is an information unit that
is a sharp, fixed topological phase-changing object and can never be divided
into two parts. Thus, it shows particle-like behavior; On the other hand,
after mapping to real space, it looks like a wave: the dynamic, smooth,
non-topological phase-changing shows wave-like behavior which is characterized
by wave-functions. This fact leads to \emph{particle-wave duality}.

In addition, we emphasize that an elementary particle is \emph{indivisible} in
Clifford group-changing space $\mathrm{C}_{\mathrm{\tilde{S}\tilde{O}%
(3+1)},3+1}$ of non-compact group. However, it is \emph{divisible} in
Cartesian space. Therefore, an elementary particle may spread the whole system
rather than localize a given point. The weight of finding a particle is
obvious proportional to local density of the group-changing elements. Although
the elementary particle can split and the size of it in $\mathrm{C}%
_{\mathrm{\tilde{S}\tilde{O}(3+1)},3+1}$ can never be changed, "angular
momentum" $\hbar$ is conserved after summarizing all elements.

\subparagraph{Uncertainty principle}

For quantum mechanics, the uncertainty principle is related to the
"\emph{fragmentation}" of an elementary particle in real space. Now, an
elementary particle may spread the whole system rather than localize a given
point. The weight of finding a particle is obvious proportional to local
density of the changing elements of it. The momentum denotes the spatial
distribution of group-changing elements; the energy denotes the temporal
distribution of group-changing elements. For example, a uniform distribution
of group-changing elements $\psi(x,t)\sim e^{-i\omega t+i\vec{k}\cdot \vec{x}}$
described by a wave-function of a plane wave has fixed projected momentum
$\vec{p}=\hbar \vec{k}$. For this case, we know momentum of the particle but it
has no given position. Another example is an elementary particle with unified
group-changing elements $\psi(x,t)\sim \delta(\vec{x}-\vec{x}_{0})$ which can
be regarded as a superposition state of $\psi(x,t)\sim%
%TCIMACRO{\dsum \limits_{k}}%
%BeginExpansion
{\displaystyle \sum \limits_{k}}
%EndExpansion
e^{-i\omega t+i\vec{k}\cdot \vec{x}}.$ For this case, we know the position of
the particle but it has no fixed momentum.

\paragraph{Incompleteness of quantum mechanics}

Einstein had questioned the completeness of quantum mechanics. In this
section, we address this issue. Before discussing the incompleteness of
quantum mechanics, we firstly show the relationship between two different
representations for quantum states, non-local, 1-st order representation
without projection and usual "wave function" representation for quantum states
(or Hybrid-order representation under partial K-projection and D-projection).
We try to recover the non-local character from quantum states in quantum
mechanics. This will help people to understand the non-local phenomena in
quantum mechanics and identify the incompleteness of quantum mechanics.

According to above discussion, wave function becomes a function describing the
distribution of extra group-changing elements. To recover the non-local
character for wave functions in quantum mechanics, there are following four steps:

Step 1 -- \textit{Describing wave function for quantum states by using (local)
algebra representation}: In (local) algebra representation, the wave function
$\psi(x)$ of the system is written into formula for extra elements on rigid
space,
\[
\psi(x)\rightarrow \mathrm{z}(x)=\hat{U}(\delta \varphi^{\mu}(N^{\mu}%
(x),\varphi^{\mu}(x)))\mathrm{z}_{0}%
\]
where $\hat{U}(\delta \varphi^{\mu}(N^{\mu}(x),\varphi^{\mu}(x)))$ denotes a
series of group operations\ with $e^{i((\delta \varphi_{i})\cdot \hat{K})}$ and
$\hat{K}=-i\frac{d}{d\varphi}.$

Step 2 -- \textit{Un-projection of uniform zero lattice}: To recover the fully
changing structure of the zero lattice, we try to un-project the uniform zero
lattice to a U-variant. The zero lattice is a rigid background. After
un-projecting, we have a uniform physical variant $V_{\mathrm{\tilde{S}%
\tilde{O}(3+1)},3+1}(\Delta \phi^{\mu},\Delta x^{\mu},k_{0},\omega_{0})$.
Without knowing the basic form of natural reference and the changing rate
$k_{0}$, we cannot completely do un-projection of a uniform zero lattice to a
uniform variant.

Step 3 -- \textit{Un-compactification of phase factor of quantum states}:
Next, we replace $\delta \varphi^{\mu}(n^{\mu}(x))$ and $\varphi^{\mu}(n^{\mu
}(x))$ of compact \textrm{SO(3+1)} Lie group by $\delta \phi^{\mu}(x)$ and
$\phi^{\mu}(x)$ of non-compact \textrm{\~{S}\~{O}(3+1)} Lie group
\[
\psi(x)\rightarrow \mathrm{z}(x)=\hat{U}(\delta \phi^{\mu}(x))\mathrm{z}_{0}%
\]
where $\hat{U}(\delta \phi^{\mu}(x))$ denotes a series of group
operations\ with $e^{i((\delta \phi)\cdot \hat{K})}$ and $\hat{K}=-i\frac
{d}{d\phi}.$ After un-compactification, we have the information of the partner
of its complementary pair $V_{\mathrm{\tilde{S}\tilde{O}(3+1)},3+1}^{\prime
}(\pm \pi,\Delta x^{\mu},k_{0},\omega_{0})$. In particular, after
non-compactification of phase factor of quantum states, the zero size of
group-operation elements of wave function $\delta \varphi_{i}(N_{i})$ with
$\delta x_{i}=0$ on Cartesian space is replaced by a finite size of them
$\delta \phi_{i}(x_{i})=k_{0}\delta x_{i}$. However, we point out that without
knowing the changing rate $k_{0}$, we cannot do un-compactification to a wave function.

Step 4 -- \textit{Combination of the two variants:} Finally, we combine
$V_{\mathrm{\tilde{S}\tilde{O}(3+1)},3+1}(\Delta \phi^{\mu},\Delta x^{\mu
},k_{0},\omega_{0})$ and $V_{\mathrm{\tilde{S}\tilde{O}(3+1)},3+1}^{\prime
}(\pm \pi,\Delta x^{\mu},k_{0},\omega_{0})$ into the original physical variant
$V_{\mathrm{\tilde{S}\tilde{O}(3+1)},3+1}(\Delta \phi^{\mu}\pm \pi,\Delta
x^{\mu},k_{0},\omega_{0}),$ i.e.,
\begin{align}
&  V_{\mathrm{\tilde{S}\tilde{O}(3+1)},3+1}(\Delta \phi^{\mu}\pm \pi,\Delta
x^{\mu},k_{0},\omega_{0})\nonumber \\
&  =V_{\mathrm{\tilde{S}\tilde{O}(3+1)},3+1}(\Delta \phi^{\mu},\Delta x^{\mu
},k_{0},\omega_{0})\\
&  -V_{\mathrm{\tilde{S}\tilde{O}(3+1)},3+1}^{\prime}(\pm \pi,\Delta x^{\mu
},k_{0},\omega_{0}).\nonumber
\end{align}

After doing above four steps, the non-local character of a wave function
$\psi(x)$ is recovered. However, without knowing additional information (for
example, the basic form of natural reference and the changing rate $k_{0}$),
we cannot recover the original physical variant. In usual quantum mechanics,
the information about $k_{0}$ loses. So, the size of an elementary particle is
believed to be zero on Cartesian space. In a word, \emph{it is the information
losing that leads to incompleteness of quantum mechanics!} Therefore, we
confirm the \emph{incompleteness} of quantum mechanics.\

In addition, we point out that the lower order of representations, the less
completeness of theories for quantum motions, i.e., 1-st order representation
is complete; hybrid-order representations (or quantum mechanics) are
incomplete; 0-th order representation (classical picture) is basically unable
to characterize the system.

\paragraph{Summary}

\begin{figure}[ptb]
\includegraphics[clip,width=0.7\textwidth]{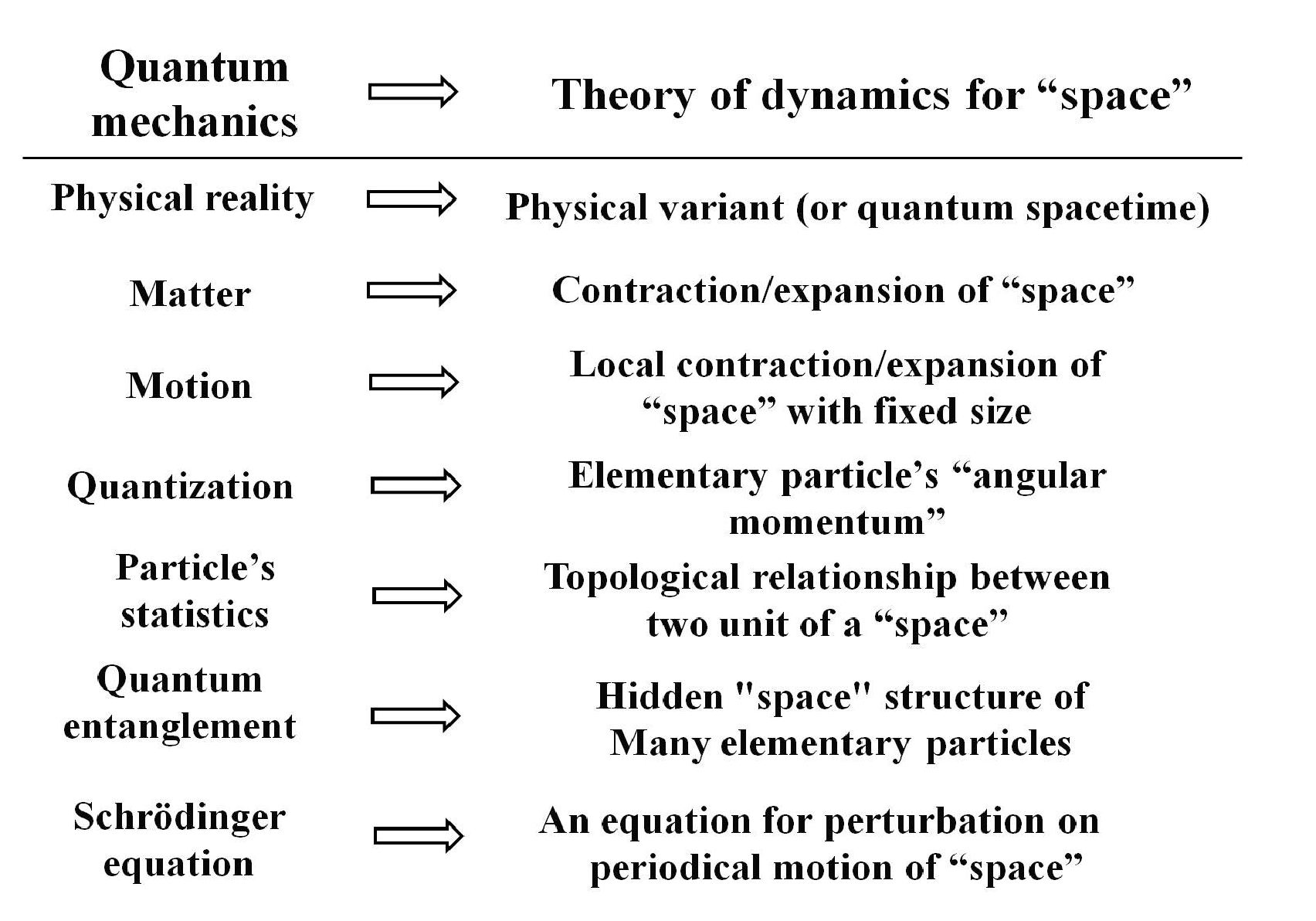}\caption{Quantum
mechanics is a phenomenological theory for ordered physical variant. This
table shows the corresponding between the concepts in quantum mechanics and
that in new theory.}%
\end{figure}

In this section, we try to develop a new theoretical framework beyond quantum
mechanics. Now, quantum mechanics emerges from regular changings on spacetime,
i.e.,
\begin{align*}
&  \text{Quantum mechanics (a phenomenological theory)}\\
&  \Longrightarrow \text{Theory for ordered physical variant }\\
&  \text{ }\  \text{\ }\  \  \text{(a microscopic theory).}%
\end{align*}
See Fig.20, in which we show the corresponding between the concepts in quantum
mechanics and that in new theory. Therefore, quantum mechanics partially
describes a "changing" structure of our world which endows the "non-local"
character of quantum physics.

In addition, we really recognize the "\emph{boundary}" of quantum mechanics --
When the matter is dense enough, the dilute approximation of group-changing
elements in P-variant begins to fail. The physical variant can no longer be
considered to be a P-variant. Now, quantum mechanics needs to use a non-local
representation, and the traditional Schrodinger equation from linearization
together with its wave function description is all no longer valid.

\subsection{Classical mechanics: theory for classical objects and classical
motion}

In above discussion, we show that quantum world really comes from an ordered
perturbative uniform physical variant. However, in our usual world, the
objects are "classical" that obey classical mechanics rather than quantum
mechanics. The formulas of classical mechanics deal with systems on rigid
spacetime having a finite number of degrees of freedom or infinitely
countable, for example, the mass point or rigid object. \emph{How to explain
this fact from the starting point of a physical variant? }In this part, we
will answer this question and develop a Monism theory for our world.

\subsubsection{Classical object: definition, representation, and
non-variability}

Before developing a Monism theory for our world, we must answer one more
fundamental question, i.e., \emph{"What is classical object?"} In classical
mechanics, one assumes that all objects are classical and consist of the mass
points. However, based on the physical reality of variant, the situation
becomes complex. In this part, we will show how classical objects emerge from
disordered perturbative uniform physical variant.

\paragraph{Definition}

Firstly, we define \emph{disordered variant}.

\textit{Definition: A disordered variant }$\tilde{V}_{\mathrm{\tilde{G}}%
,d}(\Delta \phi^{\mu},\Delta x^{\mu},k_{0}^{\mu},\omega_{0})$ \textit{is
defined by a disordered mapping between the d-dimensional Clifford
group-changing space }$\mathrm{\tilde{G}}$\textit{ and the d-dimensional
Cartesian space }$\mathrm{C}_{d}$\textit{, i.e., }%
\begin{align}
\tilde{V}_{\mathrm{\tilde{G}},d}^{D}(\Delta \phi^{\mu},\Delta x^{\mu}%
,k_{0}^{\mu},\omega_{0})  &  :\{ \delta \phi^{\mu}\} \in \mathrm{C}%
_{\mathrm{\tilde{G}},d}\nonumber \\
&  \Leftrightarrow^{\mathrm{disorder}}\{ \delta x^{\mu}\} \in \mathrm{C}_{d}.
\end{align}
\textit{where }$\Leftrightarrow^{\mathrm{disorder}}$\textit{\ denotes a
disordered mapping under fixed changing rate of integer multiple. "}$\sim
$\textit{" on\ }$\tilde{V}$\textit{ means disordered case.\ }In general, due
to disordered mapping, the group-changing elements on d-dimensional Cartesian
space\textit{ }$\mathrm{C}_{d}$ are all random.

Secondly, we define \emph{disordered\textit{-perturbative} variant}.

\textit{Definition -- }$\tilde{V}_{\mathrm{\tilde{G}},d}(\Delta \phi^{\mu
},\Delta x^{\mu},k_{0}^{\mu},\omega_{0})$ \textit{is a disordered-perturbative
uniform variant (DP-variants), if the partner} $(\tilde{V}_{\mathrm{\tilde{G}%
},d})^{\prime}(\Delta \phi^{\mu},\Delta x^{\mu},k_{0}^{\mu},\omega_{0})$
\textit{of its complementary\ pair} ($\tilde{V}_{\mathrm{\tilde{G}},d}%
(\Delta \phi^{\mu},\Delta x^{\mu},k_{0}^{\mu},\omega_{0})=V_{0,\mathrm{\tilde
{G}},d}(\Delta \phi^{\mu},\Delta x^{\mu},k_{0}^{\mu},\omega_{0})-(\tilde
{V}_{\mathrm{\tilde{G}},d})^{\prime}(\Delta \phi^{\mu},\Delta x^{\mu}%
,k_{0}^{\mu},\omega_{0})$) \textit{is a disordered variant. And, the number of
extra group-changing elements are tiny.}

Thirdly, we define \emph{locally-disordered\textit{-perturbative}
variant.}\textit{\ }

\textit{Definition -- }$\tilde{V}_{\mathrm{\tilde{G}},d}(\Delta \phi^{\mu
},\Delta x^{\mu},k_{0}^{\mu},\omega_{0})$ \textit{is
locally-disordered-perturbative uniform variant if the partner} $(\tilde
{V}_{\mathrm{\tilde{G}},d})^{\prime}(\Delta \phi^{\mu},\Delta x^{\mu}%
,k_{0}^{\mu},\omega_{0})$ \textit{of its complementary\ pair} ($\tilde
{V}_{\mathrm{\tilde{G}},d}(\Delta \phi^{\mu},\Delta x^{\mu},k_{0}^{\mu}%
,\omega_{0})=V_{0,\mathrm{\tilde{G}},d}(\Delta \phi^{\mu},\Delta x^{\mu}%
,k_{0}^{\mu},\omega_{0})-(\tilde{V}_{\mathrm{\tilde{G}},d})^{\prime}%
(\Delta \phi^{\mu},\Delta x^{\mu},k_{0}^{\mu},\omega_{0})$) \textit{is a
disordered variant, of which all group-changing elements }$\delta \phi^{\mu,B}%
$\textit{ have finite size in Cartesian space }$\mathrm{C}_{d}$\textit{ (for
example, }$\mathit{L}$\textit{). }

Fourthly , we define \emph{locally-disordered\textit{-perturbative} physical
variant.}

\textit{Definition -- }$\tilde{V}_{\mathrm{\tilde{S}\tilde{O}(d+1)}%
,d+1}(\Delta \phi^{\mu}\pm N_{F}\pi,\Delta x^{\mu},k_{0},\omega_{0})$\textit{
is a} \textit{locally-disordered-perturbative physical variant, if the
partner} $(\tilde{V}_{\mathrm{\tilde{S}\tilde{O}(d+1)},d+1})^{\prime}%
(\Delta \phi^{\mu},\Delta x^{\mu},k_{0}^{\mu},\omega_{0})$\textit{ of its
complementary\ pair is defined by a disordered variant with finite size,
\textit{mapping between the (d+1)-dimensional Clifford group-changing space
}$\mathrm{C}_{\mathrm{\tilde{S}\tilde{O}(d+1)},d+1}$.}

On the one hand, the locally-disordered-perturbative physical variant
$\tilde{V}_{\mathrm{\tilde{S}\tilde{O}(d+1)},d+1}(\Delta \phi^{\mu}\pm N_{F}%
\pi,\Delta x^{\mu},k_{0},\omega_{0})$ is a state with $N_{F}$ elementary
particles; On the other hand, it has a random distribution of extra
group-changing elements $\delta \phi_{j}^{\mu,B}$. It is obvious that it
doesn't describe a usual (pure) quantum state. Instead, it describes certain
mixed states.

Finally, we define the classical object by using the concept of
locally-disordered-perturbative physical variant $\tilde{V}_{\mathrm{\tilde
{S}\tilde{O}(d+1)},d+1}(\Delta \phi^{\mu}\pm N_{F}\pi,\Delta x^{\mu}%
,k_{0},\omega_{0}).$

\textit{Definition -- Classical object of }$N_{F}$\textit{ elementary
particles with finite size }$L$\textit{ in Cartesian space }$\mathrm{C}%
_{d+1}\ $\textit{is a locally-disordered-perturbative physical variant
}$\tilde{V}_{\mathrm{\tilde{S}\tilde{O}(d+1)},d+1}(\Delta \phi^{\mu}\pm
N_{F}\pi,\Delta x^{\mu},k_{0},\omega_{0},(x_{0}))$\textit{. Here, }$(x_{0}%
)$\textit{ denotes the mass center of }$N_{F}$\textit{ elementary particles
and is really collective coordinate for all group-changing elements. }

One important feature of classical object is fragmented. Or, a classical
object is a group of disordered group-changing elements rather than a rigid
mass point. The situation of "fragmented" is different from quantum objects.
The existence of a \emph{collective coordinate} $(x_{0})$ just means that all
these group-changing elements belong to the "\emph{same}" classical object. In
mathematic, we can define the collective coordinate to be the average position
of all group-changing elements; the condition of "\emph{finite size }$L$" of
all these group-changing elements denote a "\emph{locally}", rather a
"\emph{globally}" perturbation on the original uniform physical variant. In
addition, we point out that the size $L$ is only the size of wave packet, but
not the true size of an elementary particle that $l_{p}$ is much smaller than
$L$. The assumption of "randomness of group-changing elements" indicates
"\emph{Non-variability (non-changing structure) of classical objects}". This
is a key point, that will be discussed in following sections in detail.

See the illustration in Fig.21. Fig.21 shows the difference between a quantum
object and a classical object: for a quantum object, the group-changing
elements have ordered phase factor while those of classical objects have
disordered phase angle. In the magnifier of (a), we show that the
group-changing elements have finite size and the changing rate of
group-changing elements by using \textquotedblleft non-local" representation
is $k_{0}$; in the magnifier of (b), the group-changing elements have zero
size and the changing rate of group-changing elements by using
\textquotedblleft local" representation under K-projection is $0$ with a
random phase.

\begin{figure}[ptb]
\includegraphics[clip,width=0.7\textwidth]{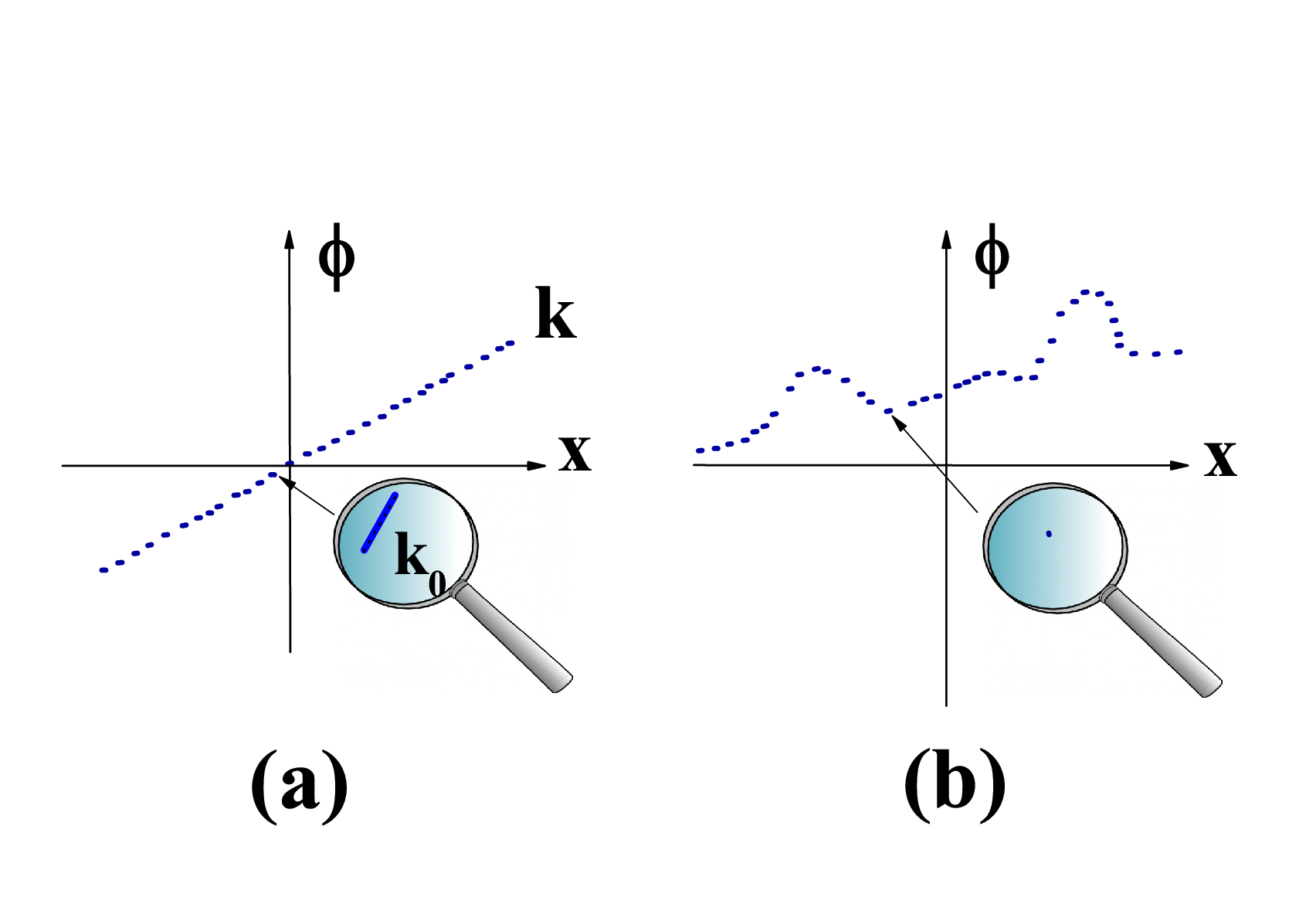}\caption{The comparison
between a quantum object (a) and a classical object (b): for a quantum object
(a), the changing pieces have ordered phase angle while the those of classical
object (b) has globally disordered phase angle. In the magnifier of (a), we
show that the changing rate for each piece is $k_{0}$; in the magnifier of
(b), the changing piece turns into a point with a random phase.}%
\end{figure}

\paragraph{Representation}

In this section, we show the property of classical object $\tilde
{V}_{\mathrm{\tilde{S}\tilde{O}(d+1)},d+1}(\Delta \phi^{\mu}\pm N_{F}\pi,\Delta
x^{\mu},k_{0},\omega_{0},(x_{0}))$ by using its 1-st order algebra representation.

We firstly unify the $N_{F}$ particles into a single object -- a composite
object of$\ N_{F}\pi$-particle. For example, by using 1-st order algebra
under D-projection, the function for a unified $N_{F}\pi$-particle is given
by
\begin{equation}
\mathrm{z}^{\mu}=\exp[i\phi^{\mu}(x)],
\end{equation}
with
\begin{equation}
\phi^{\mu}(x)=\left \{
\begin{array}
[c]{c}%
\phi_{0}^{\mu},\text{ }x^{\mu}\in(-\infty,x_{0}^{\mu}]\\
\phi_{0}^{\mu}+k_{0}(x^{\mu}-x_{0}^{\mu}),\text{ }x^{\mu}\in(x_{0},x_{0}^{\mu
}+na]\\
\phi_{0}^{\mu}+N_{F}\pi,\text{ }x^{\mu}\in(x_{0}^{\mu}+n^{\mu}a,\infty)
\end{array}
\right \}  .
\end{equation}
Then, we divide the unified Clifford group-changing space of $N_{F}\pi
$-particle into $n$ extra group-changing elements $(\hat{U}^{\mu}(x^{\mu
},t))^{j}$ ($n\rightarrow \infty$) in the region of $L$ ($L\gg \frac{N_{F}%
}{k_{0}}$).

Finally, we have 1-st order algebra under D-projection for a classical
object with a random distribution of all extra group-changing elements. Now,
$\phi^{\mu}(x)$ becomes a random number, i.e., $\phi^{\mu}(x)\in
\operatorname{rand}(0,k_{0}L\cdot2\pi)$ and function for variant is
\begin{equation}
\mathrm{z}^{\mu}(x^{\mu})=%
%TCIMACRO{\dprod \limits_{j=1}^{n}}%
%BeginExpansion
{\displaystyle \prod \limits_{j=1}^{n}}
%EndExpansion
(\hat{U}_{j}^{\mu}(x^{\mu},t)e^{i\phi_{j}^{\mu}(x)})\mathrm{z}_{0}\nonumber
\end{equation}
where $\phi^{\mu}(x)$ is random phase. As a result, the phase factor of
$\mathrm{z}^{\mu}(x^{\mu})$ is all random everywhere.

Next, we consider the hybrid-order representation under partial D-projection.
Now, the extra group-changing elements have random phase factors. Therefore,
the phase angle $\varphi(x)$ of wave function $\psi(x)$ is meaningless and one
has the information of particle's density $\Omega(x)=\int \psi^{\ast}%
(x)\psi(x)dx$ that is the density of extra group-changing elements.

Finally, we consider the 0-th order representation under type-II fully
D-projection. Now, the random distribution of extra group-changing elements
leads to random distribution of extra zeroes by considering random projection
angle $\theta$.

\paragraph{Non-variability (non-changing structure)}

According to above discussion, for a classical object there are two main
characteristics, \textquotedblleft \emph{random phase factor }$\phi^{\mu}(x)$",
and \textquotedblleft \emph{finite size L}". In particular, for a classical
object, there is an additional, important characteristics -- \textquotedblleft%
\emph{Non-variability (non-changing structure)}".

To define the characteristics of \textquotedblleft \emph{non-changing}", we do
an extra operation $\hat{U}(\delta \phi^{\prime \mu}(x))$ on it where $\hat
{U}(\delta \phi^{\prime \mu}(x))$ denotes a group-changing operation\ with
$e^{i((\delta \phi^{\prime \mu})\cdot \hat{K})}$ and $\hat{K}=-i\frac{d}%
{d\phi^{\mu}}.$ Under such a group-changing operation, the phase angles of all
group-changing elements shift $\delta \phi^{\prime \mu}$, i.e.,
\[
\phi^{\mu}(x)\rightarrow \phi^{\prime \mu}(x)=\phi^{\mu}(x)+\delta \phi
^{\prime \mu}%
\]
where $\phi^{\mu}(x)$ is a random number, i.e., $\phi^{\prime \mu}%
(x)\in \operatorname{rand}(0,k_{0}L\cdot2\pi).$ Therefore, $\phi^{\prime \mu
}(x)$ is also a random number, i.e., $\phi^{\mu}(x)\in \operatorname{rand}%
(0,k_{0}L\cdot2\pi).$ Due to randomness of phase factor, the state of the
classical object denoted by $\phi^{\mu}(x)$ is indistinguishable from that of
the classical object denoted by $\phi^{\prime \mu}(x)$.

As a result, when their phase factors become random, they will never change
each other. We say that they are same and the original classical object
doesn't change under the extra group-changing element. This is just the
so-called \textquotedblleft \emph{Non-variability (or non-changing) of
classical objects}".

\subsubsection{Classical motion}

According to above discussion, we explored the physics of quantum motion for
elementary particles, which is evolution of the distributions of extra
group-changing elements. Different distributions of group-changing elements of
the elementary particle represent different states of quantum motion of
particles. In particular, quantum motion describes the ordered relative motion
of group-changing elements of the elementary particle. See the illustration in
Fig.\ 22. In this part, we discuss classical motion for the elementary particles.

\begin{figure}[ptb]
\includegraphics[clip,width=0.7\textwidth]{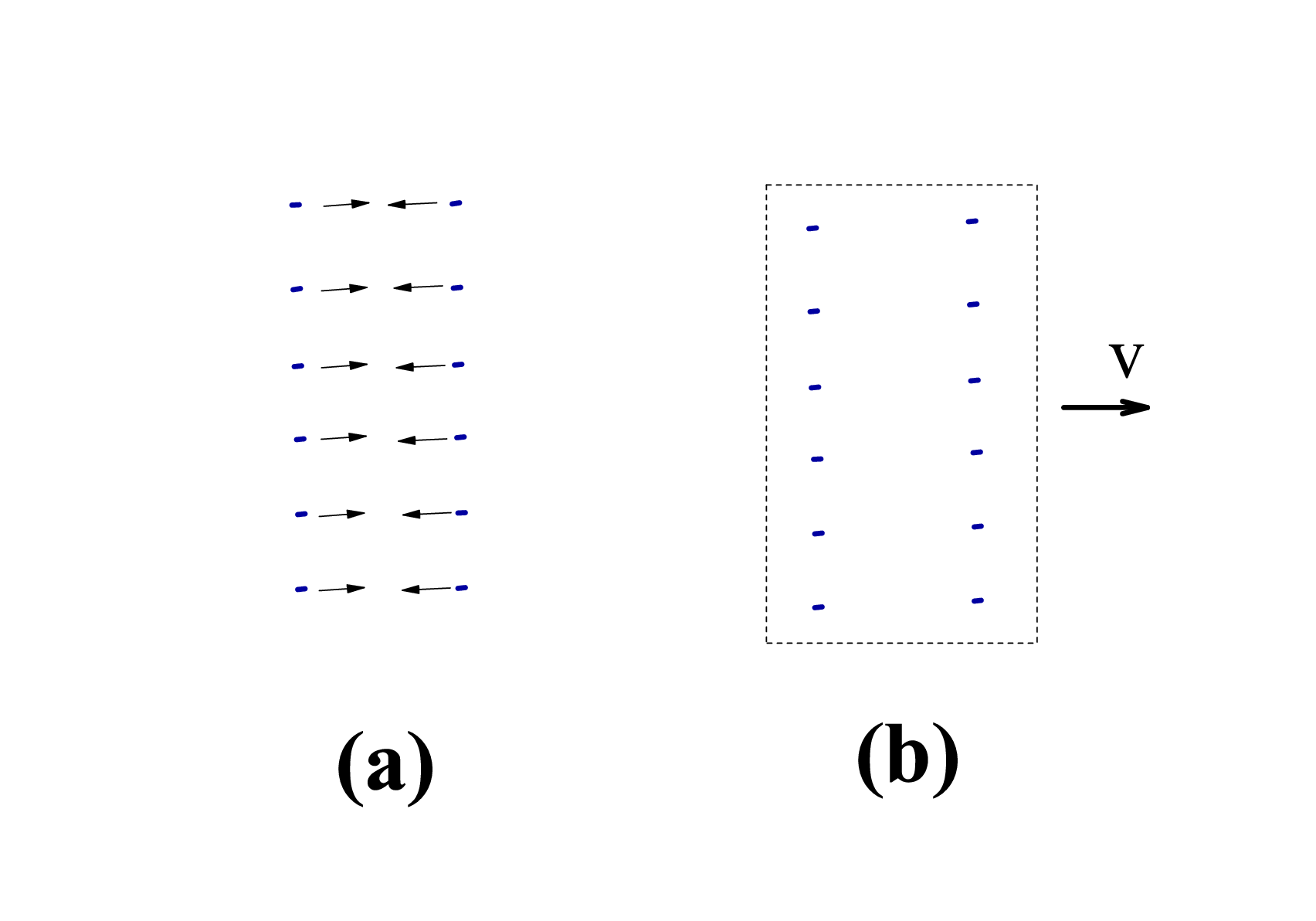}\caption{The comparison
between quantum motion (a) and classical motion (b): for quantum motion (a),
there are ordered relative motion for group-changing elements of the
elementary particle; while for classical motion (b), the group-changing
elements of elementary particles globally shift with fixed velocity $v$. }%
\end{figure}

\paragraph{Definition}

\textit{Definition: Classical motion is globally moving of extra
group-changing elements of given quantum/classical object, i.e., }%
\begin{align*}
\tilde{U}(\delta \phi^{B})  &  =\prod_{\mu}(\prod_{i}\tilde{U}(\delta \phi
_{i}^{B,\mu}(x_{i})))\rightarrow \\
\tilde{U}(\delta \phi^{B})  &  =\prod_{\mu}(\prod_{i}\tilde{U}(\delta \phi
_{i}^{B,\mu}(x_{i}(t))))
\end{align*}
\textit{where }$x_{i}(t)=x_{0,i}+\Delta x(t)$\textit{ denotes the position of
each group-changing element physical variant. Here, }$\Delta x(t)$\textit{ is
the time-dependent globally shift for all group-changing elements of given
quantum/classical object on rigid space. }

\subsubsection{Classical motion for quantum/classical objects}

From above definition, for a quantum/classical object with finite size $L,$ in
the long-wave length limit, $\Delta x(t)\gg L,$ we have classical motion
(globally shift) for both quantum objects and classical objects; in the
short-wave length limit, $L\gg \Delta x(t),$ we have quantum motion for quantum
objects and random motion for classical objects.

Let us give a brief proof on the classical motion of collective coordinate for
a quantum object. One can obtain an effective \textquotedblleft classical"
object from quantum one by K-projection with random projection angle $\theta$.
Consequently, the distribution of zeroes under random projection is very
similar to that of a classical object. Therefore, due to the inability to
distinguish their differences, the globally motion of the collective
coordinate must be described by a classical one.

\paragraph{Equation of motion}

According to above discussion, for a classical object, all group-changing
elements only have information of collective coordinates, that is the position
of the mass center $x_{0}(t).$ Then, to characterize the classical motion of
classical object in the long-wave length limit, $\Delta x(t)\gg L$, the
position and the global motion of it can be described by specifying the
Cartesian coordinates. We denote the position along the path of the moving
mass point in Cartesian coordinates to be $x_{0}(t).$ A series of positions
$x_{0}(t_{0}),$ $x_{0}(t_{1}),$ $x_{0}(t_{2}),$ ... is called the \emph{moving
path}. With the information of moving path, we can define the velocity
$v(t)=\frac{dx_{0}(t)}{dt}$ and acceleration $a(t)=\frac{dv(t)}{dt}$ for the
moving particle. In principle, the information for a moving classical object
is complete, i.e., we are able to know the position of mass center $x_{0}(t)$,
the velocity $v(t)$, the acceleration $a(t)$.

We then discuss the equation of classical motion for classical object in the
long-wave length limit, $\Delta x(t)\gg L$.

Because the phase factors for group-changing elements are random, relative
motion between them is meaningless. We only consider their global motion, or
globally shift on Cartesian space. When far away, the system can be regarded
as a mass point with total mass $M=N_{F}\cdot m$, where $m$ is the mass of
single particle. The motion of the \textquotedblleft classical" object can be
derived by considering Lorentz boost $x\rightarrow x^{\prime}=x-vt$. Under the
Lorentz boost, by using \textquotedblleft non-local" algebra representation
under D-projection, we have
\begin{align}
\mathrm{z}^{\mu}(x^{\mu})  &  =%
%TCIMACRO{\dprod \limits_{j=1}^{n}}%
%BeginExpansion
{\displaystyle \prod \limits_{j=1}^{n}}
%EndExpansion
(\hat{U}_{j}^{\mu}(x^{\mu})e^{i\phi_{j}^{\mu}(x^{\mu})})\mathrm{z}%
_{0}\nonumber \\
&  \rightarrow \mathrm{z}^{\mu}(x^{\mu}-vt)=%
%TCIMACRO{\dprod \limits_{j=1}^{n}}%
%BeginExpansion
{\displaystyle \prod \limits_{j=1}^{n}}
%EndExpansion
(\hat{U}_{j}^{\mu}(x^{\mu}-vt)e^{i\phi_{j}^{\mu}(x^{\mu}-vt)})\nonumber \\
&  \times \mathrm{z}_{0}%
\end{align}
Consequently, the group-changing elements have a global motion in Cartesian
space with velocity $v.$ Due to Lorentz invariant, we have the total energy
and total momentum to be $E=\sqrt{p^{2}+M^{2}}$ and $p=Mv$. These
relationships between $\vec{p},$ $E$, $\vec{v}$\ indicate a classical
mechanics for mass point at collective coordinates. Here, we set
\textquotedblleft speed of light" $c$ to be unity.

The equation for classical motion is just the Hamilton canonical equation that
is
\begin{align}
\frac{dx(t)}{dt}  &  =\frac{\partial H(x(t),p(t))}{\partial p(t)},\text{ }\\
-\frac{dp(t)}{dt}  &  =\frac{\partial H(x(t),p(t))}{\partial x(t)}.
\end{align}
Here, $H(x(t),p(t))=T+V$ is the Hamiltonian of the classical system. The
kinetic energy is $T=\sqrt{p^{2}+M^{2}}$, and the potential energy is zero,
$V=0$. From the above equations the Lagrangian of the system is defined to be
$L=T$ and the action is $S=\int Ldt.$ Hamilton's principle for the equations
of motion can be obtained by $\delta S=0,$ that is just the Euler-Lagrange
equations of motion
\begin{equation}
\frac{d}{dt}(\frac{\partial L}{\partial(\frac{d\vec{x}}{dt})})-\frac{\partial
L}{\partial \vec{x}}=0.
\end{equation}

In general, the validity of Euler-Lagrange equations in an arbitrary classical
system can be obtained by using path integral approach by setting the limit of
$\hbar \rightarrow0$. With the help of an assumption that the most probable
path is equivalent to the average value (collective coordinates mean the
average value), we can derive the Euler-Lagrange equations of motion in all
classical systems.

\paragraph{Summary}

In summary, we define classical motion that is global motion of extra
group-changing elements of given quantum/classical object. In other words,
classical motion is not the motion of classical objects. Instead, quantum
particles can also do classical motion! In following parts, we will focus on
the classical motion of classical objects in the long-wave length limit,
$\Delta x(t)\gg L$.

In addition, we address the \emph{completeness} of classical mechanics. In
principle, after giving a starting condition, the moving path could be
predicted, i.e., the position, the velocity and the acceleration at given time
are all known. However, the \textquotedblleft completeness" for classical
object of classical mechanics indicates the \textquotedblleft completeness"
for the information of collective coordinates of classical objects.

\subsubsection{New framework of classical mechanics}

I then provide a new framework for classical mechanics via the different
levels of physics structure:

\begin{enumerate}
\item Step 1 is to develop theory about \emph{0-th level} physics structure
(or \emph{physical reality}) by giving the Variant hypothesis. Such 0-th level
physics structure is a physical variant with 1-st order spatial-tempo
variability. The situation is similar to that in quantum mechanics;

\item Step 2 is to develop theory about \emph{1-st level} physics structure
(or \emph{matter}) by defining elementary particle (or the information unit of
physical reality). According to above discussion, a classical object is a
group of group-changing elements with random distribution. By focusing on its
collective coordinates, we regard a classical elementary particle as a mass
point, an infinite small object without size. We need to emphasize again that
the \textquotedblleft point" here has a finite size $L$, not an infinitesimal
point in mathematics;

\item Step 3 is to develop theory about \emph{2-nd level} physics structure
(or \emph{classical motion}). According to above discussion, for a classical
object, all group-changing elements only have collective information, i.e.,
the position of the mass center (or the collective coordinate) $x_{0}(t).$
Then, to characterize the classical motion of classical object in the
long-wave length limit, $\Delta x(t)\gg L$, the position and the global motion
of it can be described by specifying the Cartesian coordinates. In general,
the validity of Euler-Lagrange equations in an arbitrary classical system can
be obtained by using path integral approach by setting the limit of
$\hbar \rightarrow0$. With the help of an assumption that the most probable
path is equivalent to the average value (collective coordinates means the
average value), we can derive the Euler-Lagrange equations of motion in all
classical systems.
\end{enumerate}

\subsubsection{Summary}

In summary, based on the framework of physical variant, classical objects and
classical motion are both highly nontrivial --- they are all emergent
phenomena in long wave-length. Classical object is a \textquotedblleft
non-changing" object with disordered group-changing elements. Classical motion
describes globally motion of a quantum/classical object with
ordered/disordered group-changing elements. On the other hand, quantum object
is a \textquotedblleft changing" object with ordered group-changing elements.
Quantum motion describes the ordered relative motion between group-changing
elements of the elementary particles. As a result, classical motion describes
motion on a rigid spacetime; quantum motion describes locally expanding or
contracting group-changing space.

\subsection{Theory for quantum measurement}

In physics, measurement is a very important issue. People obtain the physical
properties of certain systems through experiments and then test the
rationality of physical laws. In particular, in quantum mechanics, measurement
is quite different from that in classical mechanics. Then, a question is
\textquotedblleft \emph{How to measure the motions for physical reality in
quantum mechanics}?" According to the Copenhagen interpretation, there exists
phenomenological \textquotedblleft wave-function collapse" during measurement
process. The wave-function collapse is random and indeterministic and the
predicted value of the measurement is described by a probability distribution.
In this part, we will answer above question and develop a systematic theory
about quantum measurement.

\subsubsection{Physical reality of measurement}

In this part, to develop a systematic theory about quantum measurement, we
must answer a more fundamental question, i.e., \textquotedblleft \emph{What is
physical reality during quantum measurement}?"

In classical mechanics, during the measurement process, we may assume that
there at least exist three physical objects --- measured object (classical
object A), the surveyors or instruments (classical object B), and rigid
spacetime. One describes measured object (classical object A) by the surveyors
or instruments (classical object B). For the observers \textrm{A}, the rulers
and clocks are independent of the physical properties of the measured object
\textrm{B}. The classical measurement process can be considered as a time
evolution of classical objects on rigid spacetime, i.e.,
\begin{equation}
\text{Classical measurement: }\tilde{V}_{A}\Longrightarrow \tilde{V}%
_{A^{\prime}}.
\end{equation}
During measurement, the classical object A changes to the classical object A'.

However, the situation for quantum measurement becomes complex. To develop a
measurement theory for quantum mechanics, we firstly study the physical
reality of the measurement processes in quantum mechanics. During the
measurement in quantum mechanics, people try to know the information of
quantum objects described by its wave function description $\psi(\vec{x},t).$
Therefore, during this process, we may assume that there at least exist three
physical objects --- measured object (quantum objects described by wave
function $\psi(\vec{x},t)$), the surveyors (instruments), and the fixed
spacetime. It is obvious that the surveyors (or the instruments) are large,
complex classical objects. This is a \emph{hidden} \emph{assumption} for
quantum measurement. In other words, the surveyors (or the instruments) are a
group of group-changing elements with random distribution.

\subsubsection{Quantum measurement: acquisition of global information via
indirect classical measurement}

In above parts, we show the physical reality of quantum measurement, based on
which we define the measurement processes in quantum mechanics.

\textit{Definition: The quantum measurement is a measurement of an unknown
quantum state from the changings of classical states of instruments B, i.e.,}%
\begin{equation}
\text{Quantum measurement: }\tilde{V}_{B}\Longrightarrow \tilde{V}_{B^{\prime}%
}.
\end{equation}
Therefore, from above definition, quantum measurement is \textquotedblleft%
\emph{indirect}" measurement.

Let us provide a detailed explanation. Before quantum measurement, we have
quantum object A (the original measured quantum object) and classical object B
(the original classical surveyors). During measurement, the quantum object A
changes to another (we denote it by A') and the classical object of instrument
B changes to another B'. One knows the total energy, total momentum (and other
global physical conserved quantities) of the quantum object A by checking the
difference between B and B'. In other words, the measurement process in
quantum mechanics is really a detection of the changings of the classical
object B between its original classical state and its final classical state on
rigid spacetime, i.e.,
\begin{align}
\text{Quantum measurement }  &  \rightarrow \text{Classical measurement:
}\nonumber \\
\tilde{V}_{B}  &  \Longrightarrow \tilde{V}_{B^{\prime}}.
\end{align}
This is actually very understandable: measurement is to see certain changings
of the instruments by surveyors. We then extrapolate the quantum states from
the changings of the instruments. See the illustration in Fig.\ 23.

In addition, we point out that quantum measurement is an \emph{event} from
quantum system to classical system. This issue will also be addressed elsewhere.

\begin{figure}[ptb]
\includegraphics[clip,width=0.7\textwidth]{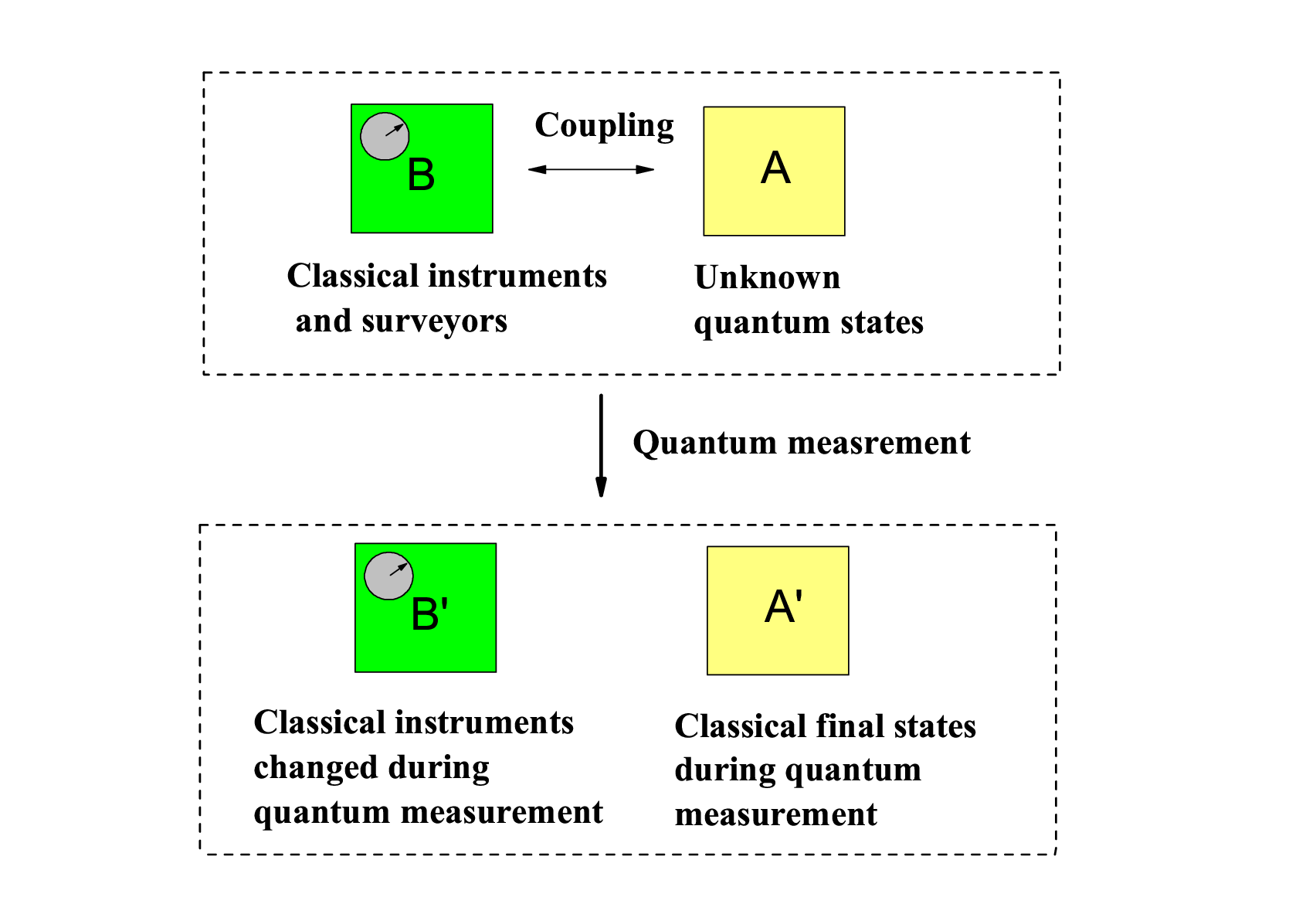}\caption{An illustration
of quantum measurement }%
\end{figure}

\subsubsection{Decoherence}

Although quantum measurement is an ``\emph{indirect}" measurement, people want
to know the final state of quantum object (or A'). We point out that A' is a
classical object on rigid spacetime after the measurement process, that is
denoted by the following process, i.e.,
\begin{equation}
\text{Quantum measurement: }V_{A^{\prime}}\Longrightarrow \tilde{V}_{A^{\prime
}}.
\end{equation}

Let us provide a detailed explanation. To obtain the global information of
quantum measured object (for example, the energy, or the momentum), one needs
to transfer the energy, or the momentum to classical surveyors. The more
complete the energy/momentum transfers, the more accurate the measurement
results. After energy/momentum transfers, the quantum states of quantum
objects undergo decoherence. As a result, the final state of the measured
quantum object is a static classical object that is a group of group-changing
elements with random distribution and without residue energy/momentum. Thus,
during quantum measurement there must exist a \textrm{R}-process that denotes
a process from a quantum object to a classical one. This is called decoherence
in traditional quantum physics.

It's already known that during quantum measurement, there exists
\emph{decoherence} for quantum objects. To accurately characterize the quantum
measurement processes and show decoherence, the theory is about the open
quantum mechanics and has already matured. By using the theory of open quantum
mechanics, one may consider the quantum measured objects to be a sub-system
coupling a thermal bath that is classical system. In principle, one can derive
the detailed results of the decoherence by solving the master equation.

In summary, from above discussion, after quantum measurement, people obtain
the global information of quantum measured objects and lose their internal
information at the same time. The physical reality of quantum measured objects
changes. Therefore, there indeed exists \textquotedblleft wave-function
collapse" during measurement process that corresponds to \textrm{R}-process
from a quantum object to a classical one.

\subsubsection{The probability in quantum measurement}

In quantum mechanics, \textrm{U}-process, a process of unitary time evolution
is deterministic and characterized by Sch\"{o}rdinger equation. However, the
situation becomes quite different during quantum measurement that corresponds
to random \textrm{R}-process. \emph{Why }\textrm{R}\emph{-process (or the
wave-function collapse) is random and indeterministic and the predicted value
of the measurement is described by a probability distribution?} Let us answer
this question.

Our starting point is the non-local representation of final measured state
that is a group of group-changing elements with random distribution. Then, we
introduce a new concept of \textquotedblleft \emph{quantum ensemble}":

\textit{Definition: A quantum ensemble is an ensemble of a lot of same final
measured states, of which all space-changing elements are identical and cannot
be distinguishable. }

\textit{Remark: Without additional internal information, due to
indistinguishability each space-changing element has the same probability
(that is }$\frac{1}{N}$\textit{) to find an elementary particle. }

Let us show the detail on the probability in quantum measurement.

Now, after quantum measurement, the original quantum object becomes
decoherence. We have a group of group-changing elements with random
distribution, each of which is $\frac{1}{N}$ particle. We consider a lot of
same final measured states (for example, $N_{F}$ particle, $N_{F}%
\rightarrow \infty$). This is a system with $N_{F}\times N$ identical
group-changing elements. Such a quantum ensemble is characterized by a group
of group-changing elements for $N_{F}$ elementary particle. Among $N_{F}\times
N$ group-changing elements, arbitrary $N$ group-changing elements correspond
to a particle. If the density of group-changing elements is $\rho
_{\mathrm{piece}},$ the density of group-changing elements $\frac{1}{N}%
\rho_{\mathrm{particle}}\ $becomes the probability to find a particle in a
given region $\psi^{\ast}(x,t)\psi(x,t)\Delta V$.

In addition, there exists mode selection effect in quantum measurement. For
example, we can observe the expected value along certain spin direction. This
corresponds to D-projection in different representations. Due to non
commutativity, we can control the group-changing elements of higher
dimensional variants to be ordered along one direction, but disordered along
another. This leads to the mode projection under quantum measurement.

We also discuss the relationship between quantum measurement and "math"
measurement by K-projection. One can obtain an effective \textquotedblleft
classical" object from quantum one by K-projection with random projection
angle $\theta$. As a result, the zeroes under random projection are very
similar to the zeroes of a classical elementary particle. The density of
group-changing elements $\frac{1}{N}\rho_{\mathrm{particle}}\ $is just the
probability to find a zero in a given region $\psi^{\ast}(x,t)\psi(x,t)\Delta
V.$

From aspect of quantum mechanics, the probability in quantum mechanics comes
from the measurement. During quantum measurement, quantum objects turn into
classical ones. Einstein had said, \textquotedblleft \textit{Quantum mechanics
is certainly imposing. But an inner voice tells me that it is not yet the real
thing. The theory says a lot, but does not really bring us any closer to the
secret of the `old one'. I, at any rate, am convinced that He does not throw
dice.}" Then, in principle, a classical observer will never obtain complete
information of a quantum object that is described by wave functions. The dice
is thrown by the \textquotedblleft classical" surveyors themselves (or
classical object B)!

\subsubsection{Application}

\paragraph{Double-slit experiment}

In this section, we provide an explanation of Feynman's gedanke double-slit
experiment with single electrons using a movable mask for closing or opening
one of the slits\cite{fey}.

Before measurement in double-slit experiment, the particle can be regarded as
a group of group-changing elements with regular distribution, that is
described by the wave-function. There is no classical path. There exists
particular interference pattern on the screen that agrees with the prediction
from quantum mechanics. If there exists an additional observer near one of a
slit, \textrm{R}-process occurs. The original quantum object changes into a
classical one that is a group of group-changing elements with random
distribution. Now, the result of measurement likes a classical result. As a
result, the phase coherence is destroyed and the interference disappears. See
the illustration in Fig.\ 24.

\begin{figure}[ptb]
\includegraphics[clip,width=0.8\textwidth]{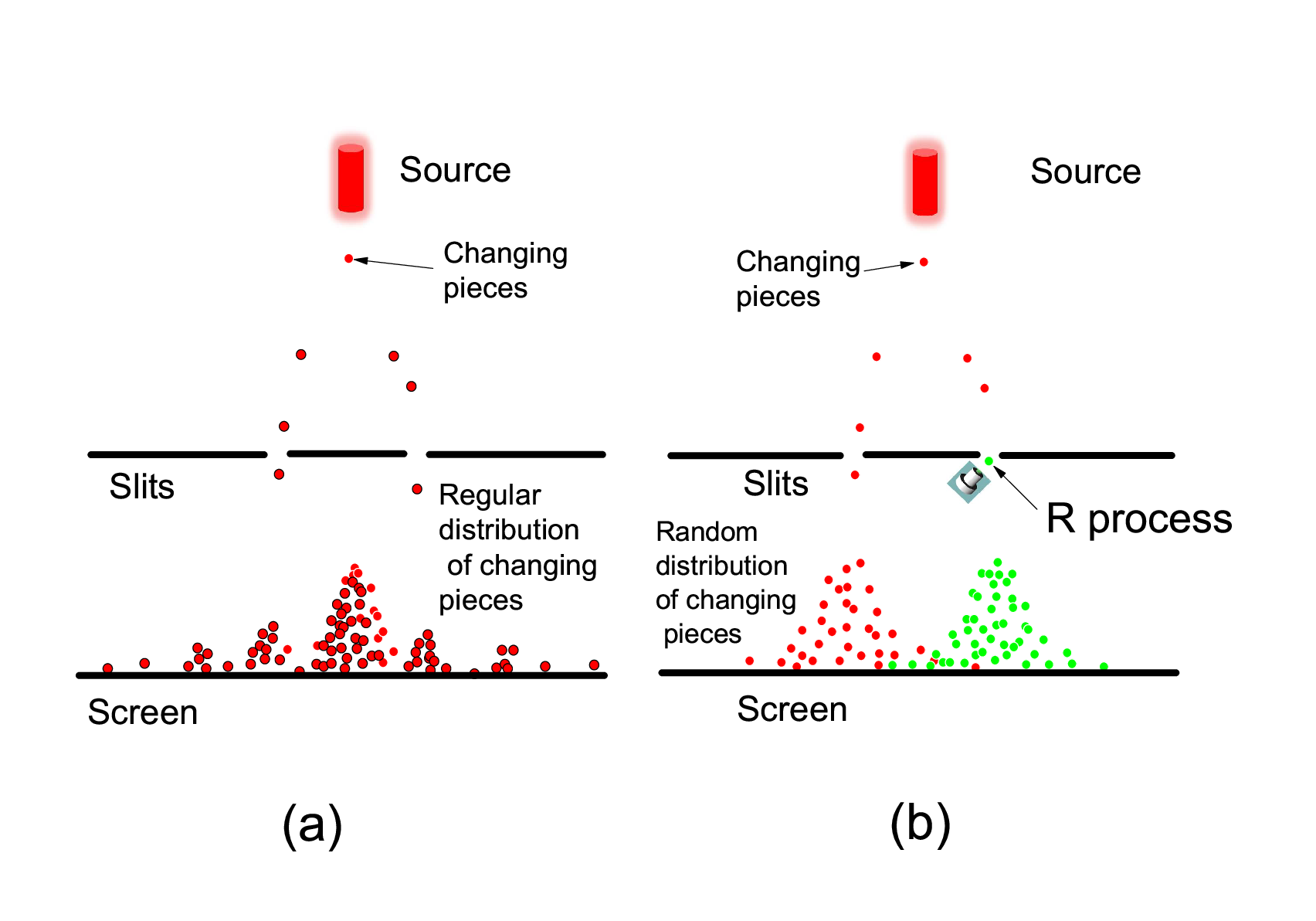}\caption{An illustration
of measurement-decoherence effect in double-slit experiment}%
\end{figure}

\paragraph{Schr\"{o}dinger's cat paradox}

Another famous puzzle of quantum foundation is the Schr\"{o}dinger's cat
paradox\cite{sch3}. In this part, we solve the paradox.

Firstly, we need to study the physical reality of this special process. In
particular, there at least exist five physical objects---the measured object
(quantum objects described by wave function $\psi(\vec{x},t)$), the instrument
to detect quantum states, the cat, the device for killing cats and the rigid
spacetime. It is obvious that the instrument to detect quantum states, the
cat, the device for killing cats are all large, complex classical objects. We
denote the measured object, the instrument to detect quantum states, the cat,
the device for killing cats as quantum object A, classical object B, C, D,
respectively. The key point is when the \textrm{R}-process (or decoherence, or
wave-function collapse) occurs, at which the quantum object turns into a
classical one. It is obvious the \textrm{R}-process occurs due to interaction
between quantum object A and classical object B. After the \textrm{R}-process,
the classical object B changes into classical object B'. After it, all
processes occur between classical object B, C, D that have nothing to do with
quantum measurement and will have no mystery.

Because the wave-function collapses at first step, all processes after it are
classical. As a result, there definitely doesn't exist a ``quantum state" of
dead cat and living cat.

\subsubsection{Summary}

In this part, we developed a systematic theory about quantum measurement. The
most amazing thing is the reversal of deterministic and stochastic characters!
People used to think that classical objects mean determinacy, and quantum
objects mean randomness. However, in this section, we point out that this
point of view is completely wrong --- classical objects mean randomness, and
quantum objects means determinacy. As a result, the probability in quantum
mechanics comes from the surveyors or instruments during quantum measurement.
In a word, it comes from \textrm{R}-process during quantum measurement. On the
other hand, if we consider \textrm{R}-process during quantum measurement as an
inversion of \textrm{R}$^{-1}$-process that is a process to prepare a quantum
state from a classical object. The situation is easily understood.

\subsection{Conclusion and discussion}

\subsubsection{Conclusion}

Finally, we give a summary. In this part, we developed a new framework on the
foundation of quantum mechanics and classical mechanics. Now, physical laws
emerge from different changes of regular changes on spacetime that is
characterized by 1-st order variability, i.e.,
\begin{equation}
\mathcal{T}(\delta x^{\mu})\leftrightarrow \hat{U}(\delta \phi^{\mu})=e^{i\cdot
k_{0}\delta x^{\mu}\Gamma^{\mu}}.
\end{equation}
Then, both quantum mechanics and classical mechanics become
\emph{phenomenological} theory and are interpreted by using the concepts of
the microscopic properties of a single physical framework, i.e.,%
\begin{align*}
&  \text{Quantum mechanics (a phenomenological theory)}\\
&  \Longrightarrow \text{Mechanics for ordered P-variant }\\
&  \text{(a microscopic theory),}%
\end{align*}
and
\begin{align*}
&  \text{Classical mechanics (a phenomenological theory)}\\
&  \Longrightarrow \text{Mechanics for disordered P-variant }\\
&  \text{(a microscopic theory).}%
\end{align*}

\subsubsection{Answers to five fundamental questions at beginning}

Consequently, with the help of physical variant, quantum mechanics is no more
a mystery, such as long range quantum entanglement, quantum non-locality,
wave--particle duality, the probability for quantum measurement, ...

Then we answer all the five fundamental questions at beginning:

1) How to understand "\emph{non-locality}" in wave function for single
particle and that in quantum entanglement?

\textbf{The answer:}

The non-locality in wave function for single particle and that in quantum
entanglement means that elementary particles are part of a "\emph{spacetime}",
i.e., particles expand and contract in group-changing space rather than being
extra objects on it. Therefore, the spacetime is composed of elementary
particles and the block of space (or strictly speaking, spacetime) is an
elementary particle. In a word, in quantum mechanics, "\emph{non-locality}"
means that all comes same regular changing structure (or physical variant);

2) and 3) How to understand "\emph{changing}" structure (or "\emph{operating}"
structure) for quantum objects in quantum mechanics? What exactly is
"\emph{changing}" here mean? How to give exact definition of "\emph{classical
object}" and how to give exact definition of "\emph{quantum object}"? And how
to unify the two types of objects into single framework?

\textbf{The answers:}

The "\emph{changing}" structure for quantum objects means that they are
ordered distribution of group-changing elements with ordered phases. Within
the framework of quantum mechanics, it indicates "\emph{operation}". On the
contrary, a classical object is a disordered distribution of group-changing
elements that is "\emph{non-changing}" and has no operator property. Thus, we
unify quantum objects and classical objects into single framework.

4) What does $\hbar$ mean? And can $\hbar$ be changed?

\textbf{The answer:}

The quantization in quantum mechanics comes from the fact of an elementary
particle with fixed "angular momentum" $J_{F}=(\frac{l_{p}}{2})^{d}\rho_{J}$.
Here, $\rho_{J}$ is the angular momentum density which is constant and
$(\frac{l_{p}}{2})^{d}$ is volume of an elementary particle. Therefore, to
alter $\hbar,$ one must change the volume of an elementary particle or modify
the angular momentum density $\rho_{J}.$ In our universe, it is impossible. In
a word, $\hbar$ means elementary particle's \emph{topological characteristics};

5) In quantum mechanics, measurement is quite different from that in classical
mechanics. In quantum measurement processes, \emph{randomness} appears. Why?

\textbf{The answer:}

The most amazing thing is the reversal of deterministic and stochastic
characters! People used to think that classical objects mean determinacy, and
quantum objects mean randomness. However, in this section, we point out that
this point of view is completely wrong -- classical objects mean randomness,
and quantum objects mean determinacy. As a result, the probability in quantum
mechanics comes from the surveyors or instruments during quantum measurement.
In a word, it comes from \textrm{R}-process during quantum measurement.

\subsubsection{Wholeness unification of quantum interpretations}

In this paper, we develop a new framework on the foundation of quantum
mechanics and classical mechanics rather than providing a new kind of
interpretation for quantum mechanics. Now, physical laws emerge from different
changings of our spacetime. Both quantum mechanics and classical mechanics
become phenomenological theory and are interpreted by using the concepts of
the microscopic properties of a single physical framework. In particular, the
expanding/contracting dynamics for "space" lead to quantum mechanics. We point
out that there are different representations, including \emph{algebra},
\emph{algebra}, and \emph{geometry} representations under different
projections, including D-projection and (partial) K-projection. The non-local
representation without projection is a complete description, and "wave
function" representation as a algebra representations under (partial)
K-projection is incomplete. However, although "wave function" representation
in the usual quantum mechanics is incomplete, it is good enough for
experiments. A question is "\emph{How about the local representation
projection under fully K-projection for quantum systems?}" The existence of
this representation leads to confusion on quantum foundation! Different
seemingly absurd interpretations of quantum mechanics originated from it, such
as hidden invariable interpretation, many world interpretation, stochastic
interpretation.... These interpretations of quantum mechanics have in common
is taking it for granted that elementary particles are indivisible mass point
on rigid spacetime. From the point of view of projection epistemology, we must
consider an \emph{un-projection process }to restore the original structure of
the system. Now, based on the "local" picture of quantum states under
K-projection, we have the same situation -- the elementary particle is indeed
an indivisible point with zero size on a rigid space.

In the end of this part, we will unify these different interpretations of
quantum mechanics\ within a single picture. See Fig.25.

\begin{figure}[ptb]
\includegraphics[clip,width=1\textwidth]{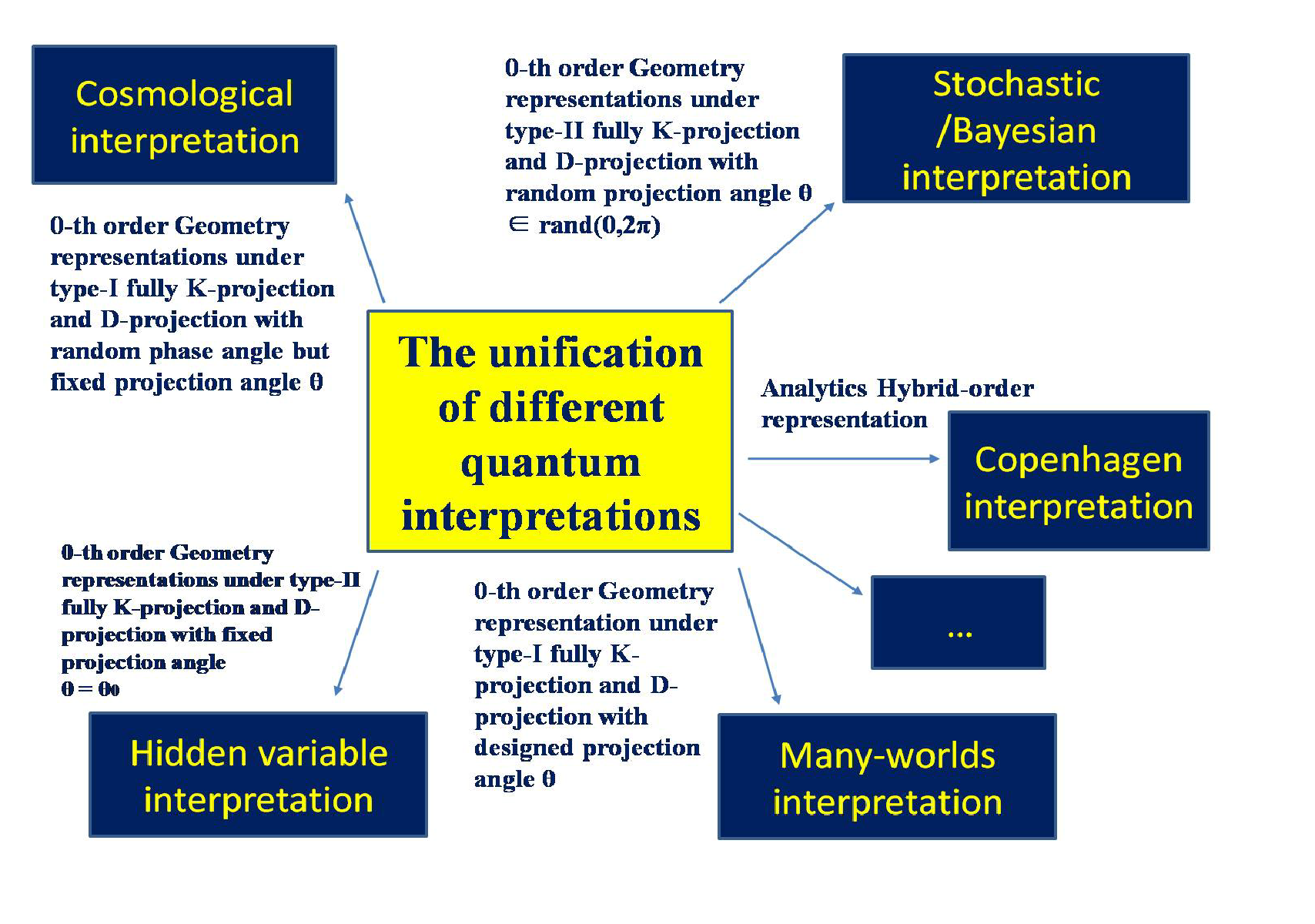}\caption{The unification of
different interpretations of quantum mechanics by different representations on
same physical variant}%
\end{figure}

\paragraph{Copenhagen interpretation}

The Copenhagen interpretation is a famous attempt to understand quantum
mechanics. N. Bohr, W. Heisenberg\cite{WH1}, M. Born\cite{Como} and others
provided a "phenomenal" interpretation on quantum mechanics. From the point of
phenomenology, it is successful. Today the Copenhagen interpretation is mostly
regarded as synonymous with indeterminism, Bohr's correspondence principle,
Born's statistical interpretation of the wave function, and Bohr's
complementarity interpretation of certain quantum phenomena. This is almost
the standard theory of quantum mechanics, an algebra representation of
Hybrid-order representation.

The nontrivial point is about quantum measurement that had been discussed in
above sections. According to the Copenhagen interpretation, there exists a
phenomenological "wave-function collapse" during measurement process. The
wave-function collapse is random and indeterministic and the predicted value
of the measurement is described by a probability. After quantum measurement,
people obtain the global information of quantum measured objects and lose its
internal information at the same time. The physical reality of quantum
measured objects changes. Therefore, there indeed exists "wave-function
collapse" during measurement process that corresponds to R-process that
denotes a process from a quantum object to a classical one.\

As a result, Copenhagen interpretation is a "field"\ representation for extra
elements of algebra representation of Hybrid-order representation. It is an
interpretation to explain physical experiments that focuses on "phenomena" of
physical object to explain "quantum motion" without pursuing the 0-th level
physics structure of our world.

\paragraph{Hidden variable interpretation}

Hidden variable theory, is a version of quantum theory discovered by Louis de
Broglie in 1927\cite{de brogile} and rediscovered by David Bohm in
1952\cite{Bohm1}. Hidden variables are variables unaccounted for in a
deterministic model of the quantum world. With the help of "hidden" variable,
the configuration of a system of particles evolves via a deterministic motion
choreographed by the wave function.

To develop a satisfactory theory for hidden variable interpretation, we use
0-th order representations under type-II fully K-projection and D-projection
on perturbative uniform physical variant $V_{\mathrm{\tilde{S}\tilde{O}%
(3+1)},3+1}(\pm \pi,\Delta x^{\mu},k_{0},\omega_{0})$.

The first step is to consider the physical variant with an extra elementary
particle (or a zero) $V_{\mathrm{\tilde{S}\tilde{O}(3+1)},3+1}(\Delta \phi
^{\mu}\pm \pi,\Delta x^{\mu},k_{0},\omega_{0})$ as a summation of an U-variant
$V_{\mathrm{\tilde{S}\tilde{O}(3+1)},3+1}(\Delta \phi^{\mu},\Delta x^{\mu
},k_{0},\omega_{0})$ and the partner $V_{\mathrm{\tilde{S}\tilde{O}(3+1)}%
,3+1}^{\prime}(\pm \pi,\Delta x^{\mu},k_{0},\omega_{0})$ of its
complementary\ pair, i.e.,
\begin{align}
&  V_{\mathrm{\tilde{S}\tilde{O}(3+1)},3+1}(\Delta \phi^{\mu}\pm \pi,\Delta
x^{\mu},k_{0},\omega_{0})\nonumber \\
&  =V_{\mathrm{\tilde{S}\tilde{O}(3+1)},3+1}(\Delta \phi^{\mu},\Delta x^{\mu
},k_{0},\omega_{0})\\
&  -V_{\mathrm{\tilde{S}\tilde{O}(3+1)},3+1}^{\prime}(\pm \pi,\Delta x^{\mu
},k_{0},\omega_{0}).\nonumber
\end{align}
Therefore, we can use $V_{\mathrm{\tilde{S}\tilde{O}(3+1)},3+1}^{\prime}%
(\pm \pi,\Delta x^{\mu},k_{0},\omega_{0})$ to fully characterize dynamics of
the extra elementary particle.

The second step is to do K-projection and D-projection on the partner
$V_{\mathrm{\tilde{S}\tilde{O}(3+1)},3+1}^{\prime}(\pm \pi,\Delta x^{\mu}%
,k_{0},\omega_{0})$. Then, under K-projection and D-projection, the extra
elementary particle is reduced to an extra zero of $V_{\mathrm{\tilde{S}%
\tilde{O}(3+1)},3+1}^{\prime}(\pm \pi,\Delta x^{\mu},k_{0},\omega_{0})$, of
which the position of the point $x$ is determined by zero-equation
$\xi_{\theta}(x^{\mu})=0.$

The third step is tracking the motion of the extra zero. By fixing the
projection angle $\theta$ to be a constant $\theta=\theta_{0}$, we derive
hidden variable interpretation for the extra elementary particle that
corresponds to a quantum state $\psi(x,t)$. For example, at $t=0,$ the
position of the zero with projection angle $\theta_{0}$ is at $x_{0}.$ Under
local geometry representation partial K-projection and D-projection, we
calculate the position of zero (or the elementary particle). At $t>0,$ we can
predict the position of zero from the same projection angle $\theta_{0}$.
During the time evolution, the zero's trajectory can be obtained.

As a result, we develop a theory with self-consistency for "hidden" variable
based on 0-th order representations under type-II fully K-projection and
D-projection on perturbative uniform physical variant $V_{\mathrm{\tilde
{S}\tilde{O}(3+1)},3+1}(\pm \pi,\Delta x^{\mu},k_{0},\omega_{0}).$ It is an
interpretation that focuses on "deterministic" of physical object to explain
"classical motion" without pursuing the 0-th level physics structure, and 1-st
level physics structure of our world. Now, it is the \emph{whole
"group-changing space"} that plays the role of "hidden" variable or guided
"wave". This is a non-local hidden-variable theory that does not violate
Bell's inequality.

\paragraph{Stochastic interpretation}

In quantum mechanics, the processes of measurement are stochastic. The
probability in quantum measurement is characterized by wave function. Then,
based on the assumption of indivisible mass point on rigid spacetime, several
physicists proposed stochastic interpretation, in which the evolution itself
can change in a random (or stochastic) way causing it to collapse all by
itself\cite{nelson}. Presumably this collapse process would occur very rapidly
for large (macroscopic) objects and slowly for subatomic particles.

To develop a theory for stochastic interpretation, we again use 0-th order
geometry representation under type-II K-projection and D-projection on partner
$V_{\mathrm{\tilde{S}\tilde{O}(3+1)},3+1}^{\prime}(\pm \pi,\Delta x^{\mu}%
,k_{0},\omega_{0})$ of its complementary\ pair. However, this time we randomly
do projection by considering the projection angle $\theta$ to be a random
number, i.e., $\theta \in \operatorname{rand}(0,2\pi)$. Consequently, the extra
zero that correspond to elementary particle moves randomly. In principle, the
information for unknown quantum states can be obtained. The situation is
similar to the approach of Monte Carlo to simulate a certain system and also
consistent with the Bayesian interpretation of quantum mechanics.

As a result, based on 0-th order representations under type-II fully
K-projection and D-projection on perturbative uniform physical variant
$V_{\mathrm{\tilde{S}\tilde{O}(3+1)},3+1}(\pm \pi,\Delta x^{\mu},k_{0}%
,\omega_{0})$, we develop a theory with self-consistency for stochastic
interpretation of quantum mechanics. It is an interpretation that focuses on
"stochastic" of physical object to explain "classical motion" without pursuing
the 0-th level physics structure, and 1-st level physics structure of our
world. However, we had recognized that the dice in quantum measurement comes
from the "classical" surveyors themselves rather than quantum states to be
measured. Therefore, strictly speaking, all stochastic interpretations
(including Nelsonian stochastic mechanics or Bayesian interpretation of
quantum mechanics) are very misleading.

\paragraph{Many-worlds interpretation}

The fundamental idea of the many-worlds interpretation, going back to Everett
1957\cite{many}, is that there are myriads of worlds in the Universe in
addition to the world we are aware of. Many-worlds interpretation is a certain
monism interpretation on quantum mechanics. Within Many-worlds interpretation,
every time a quantum experiment with different possible outcomes is performed,
all outcomes are obtained, each in a different newly created world, even if we
are only aware of the world with the outcome we have seen.

To develop a theory for Many-worlds interpretation, we use 0-th order geometry
representation under type-I fully K-projection and D-projection on the
perturbative uniform physical variant $V_{\mathrm{\tilde{S}\tilde{O}%
(3+1)},3+1}(\pm \pi,\Delta x^{\mu},k_{0},\omega_{0})$. Therefore, under type-I
fully K-projection and D-projection, for a physical variant, we have a
defective crystal of zeroes that corresponds to our world. If we regard the
projection of a crystal of zeroes projection as a true physical measurement,
we have a many-worlds interpretation! In particular, for each projection, the
projection angle $\theta$ is designed according to the measurement. Each newly
created world is generated by \emph{mathematical projection}. The information
of all these world "generation" can infinite approximation the truth.

Using similar ideas, we can also develop a theory for cosmological
interpretation of quantum mechanics.

As a result, based on 0-th order geometry representation under type-I fully
K-projection and D-projection on the perturbative uniform physical variant
$V_{\mathrm{\tilde{S}\tilde{O}(3+1)},3+1}(\pm \pi,\Delta x^{\mu},k_{0}%
,\omega_{0}),$ we give a theory with self-consistency on Many-Worlds
Interpretation for quantum mechanics. It is an interpretation that focuses on
the explanation of "stochastic" by "deterministic" and pursues the 0-th level
physics structure of our world.

\paragraph{Relational interpretation of quantum mechanics}

Relational quantum mechanics (RQM) is an interpretation of quantum mechanics
based on the idea that quantum states describe not an absolute property of a
system but rather a relationship between systems\cite{rov}. In other words,
RQM is about facts, not states. We point out that an absolute property of a
system is just \emph{"non-changing"} configuration structure in this paper; a
relationship between systems is just "\emph{changing}" ("\emph{operating}")
structure. As a result, RQM is an algebra Hybrid-order representation that
focuses on "\emph{changing}" ("\emph{operating}") of physical object to
explain "quantum motion" without pursuing the 0-th level physics structure of
our world.

\paragraph{The idea of "Implicate Order" of quantum mechanics}

In a book "\textit{Wholeness and the Implicate Order}"\cite{bohm}, D. Bohm
provides a deep idea of quantum mechanics -- "\emph{implicate order}". He said
that "\textit{Space is not empty. It is full, a plenum as opposed to a vacuum,
and is the ground for the existence of everything, including ourselves. The
universe is not separate from this cosmic sea of energy}.\textquotedblright%
\ The holo-movement is a key concept in David Bohm`s interpretation of quantum
mechanics and for his overall world-view. The holo-movement is the
\textquotedblleft fundamental ground of all matter.\textquotedblright \ It
brings together the holistic principle of \textquotedblleft undivided
wholeness\textquotedblright \ with the idea that everything is in a state of
process or becoming (or what he calls the \textquotedblleft universal flux").
For Bohm, wholeness is not a static oneness, but a dynamic wholeness-in-motion
in which everything moves together in an interconnected process.

Because "Implicate Order" of quantum mechanics is just a bold, radical idea
but not a systematic theory, we don't consider it as an interpretation for
quantum mechanics. However, it focuses on dynamic wholeness-in-motion of our
universe and try to pursuing the 0-th level physics structure of our world and
becomes valuable. In particular, the "Implicate Order" corresponds to higher
order variability. Physical laws indeed emerge from the uniform changing
structure (or physical variant) that can be regarded as "undivided wholeness".

\newpage

\section{Theory for Quantum Gauge Fields -- from Local Symmetry to
Higher-order Variability}

\subsection{Introduction}

Gauge field is an important concept in physics. To understand
electromagnetism, Maxwell introduced the concept of the electromagnetic field
and believed that the propagation of light required a medium for the waves,
named the luminiferous aether. In modern physics, the key point of
electromagnetic field is the existence of \emph{(local)\textrm{ }}%
\textrm{U}$^{\mathrm{em}}$\textrm{(1)}\emph{ gauge symmetry} that follows from
the redundance assumption that internal phase factors are not
measurable\cite{we}. Today, based on U$^{\mathrm{em}}$(1) gauge symmetry,
quantum electrodynamics (QED) becomes a fundamental theory of particle physics
that agrees very well with experiments and provides an accurate description of
quantum world. By generalizing Abelian, U(1) gauge symmetry to a non-Abelian
one, C.N. Yang and R.L. Mills proposed Yang-Mills field in 1954\cite{yang}.
With help of \textrm{SU(3)}\emph{ non-Abelian gauge symmetry}, Quantum
Chromodynamics (QCD) becomes a powerful theory that characterizes strong
interaction. According to QCD, quarks have three colors called red, blue and
green, respectively.

The success of gauge symmetry in developing QED and QCD has led to the belief
of "\emph{symmetry induce interaction}"\cite{1}. According to this belief,
when the gauge symmetry (a type of local "invariant") of a given quantum
system is acknowledged, the right formula of its Lagrangian $\mathcal{L}$ or
action $\mathcal{S}$ can be obtained straightforwardly, i.e.,%
\[
\text{Gauge/global symmetry}\rightarrow \text{Lagrangian }\mathcal{L}\text{ or
action }\mathcal{S}%
\]
In different physics fields, physical systems with gauge symmetry emerge, such
as emergent gauge fields in a special spacetime with partial compactification,
string-net condensation for quantum spin liquid in condensed matter
physics\cite{wen}, equivalent membrane in string theory...

However, there are several puzzles about gauge theories:

One is \emph{quark confinement} that refers to the confinement of quarks in
groups of two or three in a small region of space\cite{confinement}. If these
quarks stray too far from one another, the strong force pulls them back it. In
particular, the force is linear. Until the development of QCD, people try to
fully understand quark confinement. In the lattice gauge theory, the
confinement phenomenon is supported by numerical
calculations\cite{confinement1}. However, the clear physical mechanism has not
been provided.

In QCD, another trouble is \emph{Strong CP Problem}. For QCD, the theta
coefficient of theta term is unknown. People don't know why it is very closed
to zero (or the CP limit) and how to obtain its value. A possible solution is
axion that is quantum fluctuation of theta coefficient\cite{axion}. For a
dynamic theta coefficient, the lowest energy condition for ground state
determines a zero value after considering quantum tunneling effect via instantons.

In QED, a trouble is about Landau's pole, that is also called "\emph{Landau's
ghost}"\cite{landau}. Or, in QED, there exists ultraviolet divergence for the
coupling constant $\mathrm{e}$ of electromagnetic field. That means under
renormalization towards large energy, $\mathrm{e}$ turn into infinite, i.e.,
$\mathrm{e}(\Lambda \rightarrow \infty)\rightarrow \infty$.

To answer above questions satisfactorily, a complete, new theory beyond usual
quantum gauge field theory must be developed, rather than providing certain
non-perturbative theories. In addition, it must provide a \emph{fully
understanding the unsolved mysteries}, including quark confinement problem,
Yang-Mill gap problem, Strong CP puzzle and the existence of axion, Landau's
pole problem.

In the part, we develop a non-local theory beyond quantum field theory (we
call it \emph{(higher-order) variant theory}) by generalizing usual "gauge
field" with local gauge symmetry/invariant to non-local higher-order "space"
("higher-order variant", strictly speaking). Within the new theory, usual
quantum gauge field theory becomes \emph{phenomenological} theory and is
interpreted by using the concepts of the microscopic properties of a new
physical framework, i.e.,%
\begin{align*}
&  \text{Quantum gauge field (a phenomenological theory)}\\
&  \Longrightarrow \text{\emph{Higher-order physical variant }(a microscopic
theory).}%
\end{align*}

Therefore, in the part, we point out that the physical reality of quantum
gauge fields is really a higher order variant with an internal "group-changing
space" of particles. All physical processes of quantum gauge fields be
intrinsically described by the processes of the changings of "group-changing
space" inside elementary particles. Then, the "charge" is the topological unit
of the internal "group-changing space" and quantum gauge fields come from of
the collective fluctuations of the internal "group changing space". Abelian
gauge fields and non-Abelian gauge fields describe the global motion and
relative motion of internal degrees of freedom on particles, respectively.

The key point of the theory is \emph{higher-order variability }for the quantum
systems rather the gauge/global symmetry for its Lagrangian $\mathcal{L}$ or
action $\mathcal{S}$,%
\begin{align*}
&  \text{Local gauge symmetry (an extrinsic property)}\\
&  \Longrightarrow \text{Higher-order variability (an intrinsic property).}%
\end{align*}
Here, higher-order variability is a new concept that characterizes how the
internal struture of the given physical system changes. Now, the principle of
"symmetry induce interaction" is replaced by the principle of "variability
induce interaction". That means one can define a physical system\ with usual
gauge symmetry by using the concept of higher-order variability directly,
completely. Under certain projection, higher-order variability will be reduced
into usual symmetry/invariant that characterizes the system indirectly,
incompletely and shows the extrinsic property of the system.

According to variant theory, the quantum gauge fields of \textrm{SU(3)}
non-Abelian gauge symmetry in QCD and \textrm{U}$^{\mathrm{em}}$\textrm{(1)}
Abelian gauge symmetry in QED are not separate each other. Instead, they are
unified into single physics structure -- 2--th order variant, i.e.,
\begin{align*}
&  \text{QED + QCD (an extrinsic property)}\\
&  \Longrightarrow \text{A 2--th order physical variant }\\
&  \text{(an intrinsic property).}%
\end{align*}
QED and QCD describe the global motion and relative motion of a single
higher-order variant, respectively.

\subsection{Review on quantum gauge theory}

\subsubsection{Quantum field theory with global symmetry}

"Field" is one of the most important physical objects in modern physics. We
firstly take a "field" of a global, (compact) Lie group $\mathrm{G}%
=\mathrm{SU(2)}$ as an example to discuss this issue.

For (non-Abelian) \textrm{SU(2)} group, the group element is $g=e^{i\Theta}$
where $\Theta=\sum_{a=1}^{3}\theta^{a}T^{a}$ and $\theta^{a}$ are a set of $3$
constant parameters, and $T^{a}$ are $3$ $2\times2$ matrices representing the
generators of the Lie algebra of $\mathrm{SU(2),}$ i.e., $T^{a}$ are three
Pauli matrices $\sigma^{x},$ $\sigma^{y},$ $\sigma^{z}$. A field with
\emph{global} \textrm{SU(2)} symmetry is described by a \emph{point-to-point
mapping} between the group space (or the space of group elements) and
Cartesian space $\mathrm{C}_{d}$ (mathematic\ space described by the
coordinate $x$ that are a series of numbers arranged in size order). The field
is described by a usual (local) function of $2\times2$ matrices $g(x)$. In
general, to define a field of $g(x),$ one must choose an initial one, for
example $g_{0}$. We call it \emph{reference}. Difference functions are
normalized by the reference, the relative deviation becomes true result. The
Hamiltonian, Lagrangian, ground states of the system are all invariant under a
global $\mathrm{SU(2)}$ operation.

According to the principle of "symmetry induce interaction", the effective
Lagrangian $\mathcal{L}$ for a system with global $\mathrm{SU(2)}$ symmetry
and Lorentz invariant is always written as
\begin{equation}
\mathcal{L}_{\mathrm{Eff}}=\frac{1}{2\eta}[\partial_{\mu}g(x)\partial_{\mu
}g^{-1}(x)]+...
\end{equation}
where $\eta$ is coupling constant and $g(x)$ denotes the local direction of
group elements. The situation (like principle chiral model) can be generalized
to the cases with other continuous groups. The differential "$\partial_{\mu}$"
eliminates the effect of reference\ $g_{0}$ and then we have a model with
global $\mathrm{SU(2)}$ symmetry.

\subsubsection{Quantum field theory with local gauge symmetry}

Quantum field theory with local gauge symmetry is quite difference from those
with global one. In order to meet the condition of local gauge symmetry, the
Lagrangian $\mathcal{L}$ or action $\mathcal{S}$ becomes more complex.

We then review quantum gauge theories with the principle of "symmetry induce
interaction" together with the two assumptions ("field" assumption and
"symmetry/invariant" assumption). In general, there are two types of quantum
gauge fields, one with Abelian \textrm{U}$^{\mathrm{em}}$\textrm{(1)} local
gauge symmetry, the other with non-Abelian \textrm{G} local gauge symmetry.
For example, we have Abelian \textrm{U}$^{\mathrm{em}}$\textrm{(1)} local
gauge symmetry for electromagnetic interaction (or QED), non-Abelian
\textrm{SU(3)} local gauge symmetry for strong interaction (or QCD)\textrm{.}

\paragraph{QED}

QED is quantum \textrm{U}$^{\mathrm{em}}$\textrm{(1)} gauge field theory
coupled Dirac fermions. For QED with local \textrm{U}$^{\mathrm{em}}%
$\textrm{(1)} gauge symmetry, the Lagrangian $\mathcal{L}$ or action
$\mathcal{S}$ is invariant under the \emph{local} gauge transformation
$U_{\mathrm{U^{\mathrm{em}}(1)}}(\vec{x},t)=\exp(-i\varphi(\vec{x},t))$. As a
result, the Abelian gauge symmetry is represented by%
\begin{equation}
\psi^{\prime}\rightarrow U_{\mathrm{U^{\mathrm{em}}(1)}}(\vec{x},t)\psi,
\end{equation}
and
\begin{align}
A_{\mu}(\vec{x},t)  &  \rightarrow A_{\mu}(\vec{x},t)-i\frac{1}{\mathrm{e}%
}\left(  \partial_{\mu}U_{\mathrm{U^{\mathrm{em}}(1)}}(\vec{x},t)\right)
\left(  U_{\mathrm{U^{\mathrm{em}}(1)}}(\vec{x},t)\right)  ^{-1}\nonumber \\
&  =A_{\mu}(\vec{x},t)-\frac{1}{\mathrm{e}}\partial_{\mu}\varphi(\vec{x},t).
\end{align}
where \textrm{e} is electric coupling constant. In pure gauge condition, we
have $\mathrm{e}A_{\mu}(\vec{x},t)=-\partial_{\mu}\varphi(\vec{x},t)$. Now,
the strength of gauge fields becomes zero.

According to the belief of "\emph{symmetry induce interaction}" (or gauge
symmetry $\rightarrow$ Lagrangian $\mathcal{L}$), the action of QED is given
by $\mathcal{S}_{\mathrm{QED}}=\int \mathcal{L}_{\mathrm{QED}}d^{4}x$ where
\begin{equation}
\mathcal{L}_{\mathrm{QED}}=\mathrm{e}\bar{\psi}\gamma^{\mu}A_{\mu}\psi
-\frac{1}{4}F_{\mu \nu}F^{\mu \nu}+{\bar{\psi}}i\gamma^{\mu}\partial_{\mu}%
\psi+m{\bar{\psi}}\psi. \label{1}%
\end{equation}
The gauge field strength $F_{\mu \nu}$ is defined by $F_{\mu \nu}=\partial_{\mu
}A_{\nu}-\partial_{\nu}A_{\mu}.$ It is obvious that the action/Lagrangian is
invariant under gauge transformation. From Eq.\ref{1}, we can derive the
Maxwell equations
\begin{equation}
\partial_{\mu}F^{\mu \nu}=\mathrm{e}{j}_{(em)}^{\mu}%
\end{equation}
where ${j}_{(em)}^{\mu}=\bar{\psi}\gamma^{\mu}\psi$ is the electric current.

The two assumptions ("field" assumption and "symmetry/invariant" assumption)
indicate that $A_{\mu}(\vec{x},t)$ is a vector field on rigid spacetime and
describes local fluctuations of \textrm{U}$^{\mathrm{em}}$\textrm{(1)} field.

\paragraph{QCD}

QCD is quantum $\mathrm{SU(3)}\otimes$\textrm{U}$^{\mathrm{em}}$\textrm{(1)}
non-Abelian gauge field theory. Now, the idea of \textrm{U}$^{\mathrm{em}}%
$\textrm{(1)} Abelian gauge symmetry was generalized to non-Abelian
$\mathrm{SU(N)}$ gauge symmetry. Using the matrix representation of the
generators we can cast non-Abelian gauge fields $A_{\mu}^{a}(x)$ into an
$N\times N$ matrix: $\mathcal{A}_{\mu}(x)=\sum_{a=1}^{N^{2}-1}A_{\mu}%
^{a}(x)T^{a}$. We assume that the $\psi$'s transform linearly according to an
$N$-dimensional representation, $\Psi=\left(
\begin{array}
[c]{c}%
\psi^{1}\\
\vdots \\
\psi^{N}%
\end{array}
\right)  .$

For QCD with local gauge symmetry, the Lagrangian $\mathcal{L}$ or action
$\mathcal{S}$ is invariant under the non-Abelian gauge transformation
$U_{\mathrm{SU(N)}}(\vec{x},t)=e^{-i\Theta}$ where $\Theta=\sum_{a=1}%
^{N}\theta^{a}T^{a}$. where the $\theta^{a}$'s are a set of $N$ constant
parameters, and the $T^{a}$'s are $N$ $r\times r$ matrices representing the
$N$ generators of the Lie algebra of $\mathrm{SU(N)}$. They satisfy the
commutation rules:%
\begin{equation}
\left[  T^{a},T^{b}\right]  =if^{abc}T^{c} \label{commut}%
\end{equation}
where $f$'s are the structure constants of $\mathrm{SU(N)}$. Here, we have
$N=3$.

As a result, the non-Abelian gauge symmetry is represented by%
\begin{equation}
\psi^{\prime}\rightarrow U_{\mathrm{SU(N)}}(\vec{x},t)\psi
\end{equation}
and
\begin{align}
A_{\mu}(\vec{x},t)  &  \rightarrow U_{\mathrm{SU(N)}}(\vec{x},t)A_{\mu}%
(\vec{x},t)\left(  U_{\mathrm{SU(N)}}(\vec{x},t)\right)  ^{-1}\nonumber \\
&  -\frac{i}{g}\left(  \partial_{\mu}U_{\mathrm{SU(N)}}(\vec{x},t)\right)
\left(  U_{\mathrm{SU(N)}}(\vec{x},t)\right)  ^{-1}%
\end{align}
where $g$ an real constant. In pure gauge condition, we have $gA_{\mu}(\vec
{x},t)=-i\left(  \partial_{\mu}U_{\mathrm{SU(N)}}(\vec{x},t)\right)  \left(
U_{\mathrm{SU(N)}}(\vec{x},t)\right)  ^{-1}$. Now, the strength of gauge
fields becomes zero.

According to the belief of "\emph{symmetry induce interaction} (or gauge
symmetry $\rightarrow$ Lagrangian $\mathcal{L}$)", the action is given by
$\mathcal{S}=\int \mathcal{L}dtd^{3}x$ where
\begin{equation}
\mathcal{L}=-\frac{1}{2}TrF_{\mu \nu}F^{\mu \nu}+\mathcal{L}(\Psi,D\Psi)
\label{inv}%
\end{equation}
and%
\begin{equation}
F_{\mu \nu}=\partial_{\mu}A_{\nu}-\partial_{\nu}A_{\mu}-ig\left[  A_{\mu
},A_{\nu}\right]  . \label{curv}%
\end{equation}
The covariant derivatives of fermions is constructed as $D_{\mu}={\partial
}_{\mu}+igA_{\mu}$.

For QCD, the model is an $\mathrm{SU(3)}\otimes$\textrm{U}$^{\mathrm{em}}%
$\textrm{(1)} gauge theory, of which the Lagrangian is
\begin{align}
\mathcal{L}_{\mathrm{QCD}}  &  ={\bar{d}}i\gamma^{\mu}\partial_{\mu}%
d++{\bar{u}}i\gamma^{\mu}\partial_{\mu}u+m_{u}{\bar{u}}u+m_{d}{\bar{d}%
}d\nonumber \\
&  -\frac{1}{4}F_{\mu \nu}F^{\mu \nu}+\mathrm{e}{A}_{\mu}{j}_{(em)}^{\mu
}\nonumber \\
&  -\frac{1}{2}\mathrm{Tr}\mathcal{G}_{\mu \nu}\mathcal{G}^{\mu \nu
}+g\mathrm{Tr}J_{\mathrm{YM}}^{\mu}\mathcal{A}_{\mu}\nonumber
\end{align}
where $m_{u}$ and $m_{d}$ are masses for u-quark, d-quark, respectively. The
electric current is
\begin{equation}
{j}_{(em)}^{\mu}(x)=i\frac{2}{3}{\bar{u}}(x)\gamma^{\mu}u(x)-i\frac{1}{3}%
{\bar{d}}(x)\gamma^{\mu}d(x).\nonumber
\end{equation}
The $\mathrm{SU(3)}$ color current for strong interaction is
\begin{equation}
J_{\mathrm{YM}}^{a,\mu}(x)={\bar{u}}(x)i\gamma^{\mu}T^{a}u(x)+{\bar{d}%
}(x)i\gamma^{\mu}T^{a}d(x).
\end{equation}

The two assumptions ("field" assumption and "symmetry/invariant" assumption)
indicate that $A_{\mu}^{a}(x)$ is a multi-component vector field on rigid
spacetime and describes local fluctuations of gluon field. Now, \textrm{SU(3)}
non-Abelian gauge symmetry in QCD and \textrm{U}$^{\mathrm{em}}$\textrm{(1)}
Abelian gauge symmetry in QED are not independence each other, i.e., quarks
have both color charge and electric charge, i.e., the electric charge of
u-quark is $2/3$, the electric charge of d-quark is $-1/3$.

\subsection{Higher-order variant theory -- Mathematical tools for quantum
gauge fields}

Our classical world can be regarded as \emph{"non-changing"} structure that is
described by usual classical "field" on Cartesian space. However, in this
paper, we point out that quantum "field" is really a system with
"\emph{changing}" structure, i.e.,
\[
\text{Variant = "Space on space".}%
\]
We had called such a mathematic structure to be \emph{variant}. For quantum
gauge fields, the situation becomes more complex. We then generalize the usual
variant to \emph{higher-order variant}. In higher-order variant theory,
quantum gauge fields become one "\emph{changing}" structure on another
"\emph{changing}" structure, i.e.,
\begin{align*}
\text{Higher order variant }  &  \text{= "Space on space }\\
\text{on space" }  &  \text{= Space on variant.}%
\end{align*}

\subsubsection{Review on usual variant theory}

Before introducing higher-order variant, we briefly review the usual variant.
A variant describes "changing" structure, of which the element object is
"group-changing elements" $\delta \phi^{a}$. So, it is quite different from
usual fields $g(x)$ that characterize "non-changing" structure, of which the
element object is "group element". Variant describes a structure of
"changings". Here, the word "changing"\ means a space-like structure of a set
of number's changing on Cartesian space. Variant theory describes the space
dynamics rather than field dynamics on Cartesian space. In a word, we say "it
describes a space on the other space".

To define a variant, we firstly introduce the object of study --
group-changing space.

An arbitrary\textit{ }$d$-dimensional group-changing space $\mathrm{C}%
_{\mathrm{\tilde{G}},d}(\Delta \phi^{a})$\textit{\ }is described by a series of
numbers of group element $\phi^{a}$ of $a$-th generator independently in size
order along $a$-th direction. $\Delta \phi^{a}$ denotes the size of the
group-changing space along a-direction, a topological number. Here,
\textit{\textrm{\~{G}} }is a non-compact Lie group with\textit{ }$N$ generator
and $N<d$. \textrm{G} with "$\sim$" above it means a non-compact Lie group.

Let us take 1D group-changing space $C_{\mathrm{\tilde{U}(1)},1}(\Delta \phi
)$\ of non-compact \textrm{\~{U}(1)} group as an example to discuss.
$C_{\mathrm{\tilde{U}(1)},1}(\Delta \phi)$ is described by a series of numbers
of group elements $\phi$ arranged in size order. $\Delta \phi$ denotes the
total size of the changing space that turns to infinite, i.e., $\Delta
\phi \rightarrow \infty$. "$1$" denotes dimension.\quad On the other hand,
$\mathrm{C}_{\mathrm{\tilde{U}(1)},1}(\Delta \phi)$ can also be regarded as a
series of infinitesimal group-changing operations, $\prod_{i}(\tilde{U}%
(\delta \phi_{i}))$ with $%
%TCIMACRO{\dsum \nolimits_{i}}%
%BeginExpansion
{\displaystyle \sum \nolimits_{i}}
%EndExpansion
\delta \phi_{i}=\Delta \phi$. For a higher-dimensional case $\mathrm{C}%
_{\mathrm{\tilde{G}},d}(\Delta \phi^{a})$, along different directions (for
example, $a$-direction), the situation is similar to the 1D case by
considering a series of infinitesimal group-changing operations, $\prod
_{i}(\tilde{U}(\delta \phi_{i}^{a}))$ with $%
%TCIMACRO{\dsum \nolimits_{i}}%
%BeginExpansion
{\displaystyle \sum \nolimits_{i}}
%EndExpansion
\delta \phi_{i}^{a}=\Delta \phi^{a}.$

After introducing group-changing space $\mathrm{C}_{\mathrm{\tilde{G}}%
,d}(\Delta \phi^{a}),$ we define a variant. A variant $V_{\mathrm{\tilde{G},}%
d}[\Delta \phi^{\mu},\Delta x^{\mu},k_{0}^{\mu}]$ is denoted by\ a mapping
between a d-dimensional group-changing space $\mathrm{C}_{\mathrm{\tilde{G}%
,}d}$ with total size $\Delta \phi^{\mu}$\ and Cartesian space $\mathrm{C}_{d}%
$\ with total size $\Delta x^{\mu}$, i.e.,%
\begin{align}
V_{\mathrm{\tilde{G},}d}[\Delta \phi^{\mu},\Delta x^{\mu},k_{0}^{\mu}]  &
:\mathrm{C}_{\mathrm{\tilde{G},}d}=\{ \delta \phi^{\mu}\} \nonumber \\
&  \Longleftrightarrow \mathrm{C}_{d}=\{ \delta x^{\mu}\}
\end{align}
where $\Longleftrightarrow$\ denotes an ordered mapping under fixed changing
rate of integer multiple $k_{0}^{\mu}$.\ In particular, $\delta \phi^{\mu}$
denotes group-changing element along $\mu$-direction rather than
group-changing element.

Now, we take a 1D variant $V_{\mathrm{\tilde{U}(1),}1}[\Delta \phi,\Delta
x,k_{0}]$ as an example to show the concept.

$V_{\mathrm{\tilde{U}(1),}1}[\Delta \phi,\Delta x,k_{0}]$ is one dimensional
(1D) group-changing space $\mathrm{C}_{\mathrm{\tilde{U}(1)},1}(\Delta \phi
)$\textit{ }on Cartesian space $\mathrm{C}_{1}$, i.e.,
\begin{align*}
V_{\mathrm{\tilde{U}(1),}1}[\Delta \phi,\Delta x,k_{0}]  &  :\mathrm{C}%
_{\mathrm{\tilde{U}(1)},1}(\Delta \phi)=\{ \delta \phi \} \\
&  \Longleftrightarrow \mathrm{C}_{1}=\{ \delta x\}.
\end{align*}
\textit{ }According to above definition,\ for $V_{\mathrm{\tilde{U}(1),}%
1}[\Delta \phi,\Delta x,k_{0}],$ we have $\delta \phi_{i}=k_{0}n_{i}\delta
x_{i}$ where $k_{0}$ is a constant real number and $n_{i}$ is an integer
number. $k_{0}n_{i}$ is changing rate for $i$-th space element, i.e.,
$k_{0}n_{i}=\delta \phi_{i}/\delta x_{i}$. Therefore, for the 1D variant
$\mathrm{C}_{\mathrm{\tilde{U}(1)},1}(\Delta \phi)$, we have a series of
numbers of infinitesimal elements to record its information.\quad Different 1D
variants $V_{\mathrm{\tilde{U}(1),}1}[\Delta \phi,\Delta x,k_{0}]$ are
characterized by different distributions of $n_{i}$. As a result, in some
sense, a variant can be described by "\emph{function}" of $n_{i}.$ For a
higher-dimensional case $V_{\mathrm{\tilde{G},}d}[\Delta \phi^{\mu},\Delta
x^{\mu},k_{0}^{\mu}]$, along different directions (for example, $\mu
$-direction), the situation is similar to the 1D case by considering the
corresponding distributions of $n_{i}^{\mu}.${}

\subsubsection{Globally-mapping between different group-changing spaces}

Before introducing higher-order variant, we split the group-changing space
$\mathrm{C}_{\mathrm{\tilde{G}},d}(\Delta \phi^{\mu})$ of a variant
$V_{\mathrm{\tilde{G},}d}[\Delta \phi^{\mu},\Delta x^{\mu},k_{0}^{\mu}]$.

We point out that a group-changing space $\mathrm{C}_{\mathrm{\tilde{G}}%
,d}(\Delta \phi^{\mu})$ with 1-st order rotation variability can be split into
two kinds of group-changing subspace: one group-changing subspace
$\mathrm{C}_{\mathrm{\tilde{U}(1)\in \tilde{G}},1}(\Delta \phi_{\mathrm{global}%
})$ is about global phase changing of the system $\Delta \phi_{\mathrm{global}%
}=\sqrt{%
%TCIMACRO{\dsum \limits_{\mu}}%
%BeginExpansion
{\displaystyle \sum \limits_{\mu}}
%EndExpansion
(\Delta \phi^{\mu}(x))^{2}}$, the other is about $d-1$ internal relative
angles. For example, for a 1D group-changing space $\mathrm{C}_{\mathrm{\tilde
{U}(1)},1}(\Delta \phi),$ there doesn't exist internal relative angle and
exists only the global phase changing of the system. For a 2D group-changing
space $\mathrm{C}_{\mathrm{\tilde{S}\tilde{O}(2)},2}(\Delta \phi^{\mu}),$
except for the group-changing subspace $\mathrm{C}_{\mathrm{\tilde{U}%
(1)\in \tilde{S}\tilde{O}(2)},1}(\Delta \phi_{\mathrm{global}})$ for the global
phase changing of the system, there exists an internal relative angle that
rotates the original group-changing space from one direction to another.

Then, with the help of splitting of a group-changing space, we define a
globally-mapping between two group-changing spaces.

\textit{Definition -- Globally-mapping }$\mathrm{C}_{1,\mathrm{\tilde{G}}%
_{1},d_{1}}(\Delta \phi_{1}^{\mu})\Longleftrightarrow \mathrm{C}%
_{2,\mathrm{\tilde{G}}_{2},d_{2}}(\Delta \phi_{2}^{\mu})$\textit{ between two
group-changing spaces }$\mathrm{C}_{1,\mathrm{\tilde{G}}_{1},d_{1}}(\Delta
\phi_{1}^{\mu})$\textit{ and }$\mathrm{C}_{2,\mathrm{\tilde{G}}_{2},d_{2}%
}(\Delta \phi_{2}^{\mu})$\textit{ is a mapping between their group-changing
subspaces, }$\mathrm{C}_{1,\mathrm{\tilde{U}}_{1}\mathrm{(1)\in \tilde{G}}%
_{1},1}(\Delta \phi_{1,\mathrm{global}})$ and \textit{ }$\mathrm{C}%
_{2,\mathrm{\tilde{U}}_{2}\mathrm{(1)\in \tilde{G}}_{2},2}(\Delta
\phi_{2,\mathrm{global}})$\textit{, i.e.,}%
\begin{align*}
\mathrm{C}_{1,\mathrm{\tilde{G}}_{1},d_{1}}(\Delta \phi_{1}^{\mu})  &
\Longleftrightarrow \mathrm{C}_{2,\mathrm{\tilde{G}}_{2},d_{2}}(\Delta \phi
_{2}^{\mu})\\
&  \equiv \mathrm{C}_{1,\mathrm{\tilde{U}}_{1}\mathrm{(1)\in \tilde{G}}_{1}%
,1}(\Delta \phi_{1,\mathrm{global}})\\
&  \Longleftrightarrow \mathrm{C}_{2,\mathrm{\tilde{U}}_{2}\mathrm{(1)\in
\tilde{G}}_{2},1}(\Delta \phi_{2,\mathrm{global}}).
\end{align*}
\textit{In mathematic, it is defined by the mapping between global phase
changings}%
\begin{align*}
\mathrm{C}_{1,\mathrm{\tilde{U}}_{1}\mathrm{(1)\in \tilde{G}}_{1},1}(\Delta
\phi_{1,\mathrm{global}})  &  \Longleftrightarrow \mathrm{C}_{2,\mathrm{\tilde
{U}}_{2}\mathrm{(1)\in \tilde{G}}_{2},1}(\Delta \phi_{2,\mathrm{global}})\\
&  \equiv \delta \phi_{1,\mathrm{global}}=(\lambda^{\lbrack12]})^{-1}\delta
\phi_{2,\mathrm{global}}%
\end{align*}
\textit{ where }$\lambda^{\lbrack12]}$\textit{ is the ratio between the
changing rates of the global phases for two group-changing spaces, }%
$\delta \phi_{1,\mathrm{global}}=\sqrt{%
%TCIMACRO{\dsum \limits_{\mu}}%
%BeginExpansion
{\displaystyle \sum \limits_{\mu}}
%EndExpansion
(\delta \phi_{1}^{\mu}(x))^{2}}$\textit{ and }$\delta \phi_{2,\mathrm{global}%
}=\sqrt{%
%TCIMACRO{\dsum \limits_{\mu}}%
%BeginExpansion
{\displaystyle \sum \limits_{\mu}}
%EndExpansion
(\delta \phi_{2}^{\mu}(x))^{2}}$\textit{.} In this paper, we only focus on the
case of integer changing ratio $\lambda^{\lbrack12]},$ i.e., $\lambda
^{\lbrack12]}=1,2...n$.

\subsubsection{Higher-order variant: definition and classification}

A usual variant $V_{\mathrm{\tilde{G},}d}$ is a mapping between a
group-changing space $\mathrm{C}_{\mathrm{\tilde{G},}d}$ and Cartesian space
$\mathrm{C}_{d}.$ We may call it to be 1-st order variant and denote it by
$V_{\mathrm{\tilde{G},}d}$ that is a mapping between a group-changing space
$\mathrm{C}_{\mathrm{\tilde{G},}d}$ and Cartesian space $\mathrm{C}_{d}$.

Now, we generalized it to 2-nd order variant $V_{\mathrm{\tilde{G}}%
^{[2]}\mathrm{,\tilde{G}}^{[1]},d}^{[2]}$ by introducing higher-order mapping,
i.e., a mapping between a group-changing space $\mathrm{C}_{\mathrm{\tilde
{G}^{[2]},}d}^{[2]}$\ and another $\mathrm{C}_{\mathrm{\tilde{G}^{[1]},}%
d}^{[1]}$ that is defined on $d$-dimensional Cartesian space $\mathrm{C}_{d}$.
Because $\mathrm{C}_{\mathrm{\tilde{G}^{[1]},}d}^{[1]}$ and $\mathrm{C}%
_{\mathrm{\tilde{G}^{[2]},}d}^{[2]}$\ are not symmetric each other, we
introduce "level" to characterize them. We call $\mathrm{C}_{\mathrm{\tilde
{G}^{[1]},}d}^{[1]}$ level-1 group-changing space and $\mathrm{C}%
_{\mathrm{\tilde{G}^{[2]},}d}^{[2]}$ level-2 group-changing space, respectively.

\paragraph{Definition}

\textit{Definition -- A 2-nd order variant }$V_{\mathrm{\tilde{G}}%
^{[2]}\mathrm{,\tilde{G}}^{[1]},d}^{[2]}$\textit{\ is defined by\ a
higher-order mapping between }$\mathrm{C}_{\mathrm{\tilde{G}^{[1]},}d}^{[1]}$,
$\mathrm{C}_{\mathrm{\tilde{G}^{[2]},}d}^{[2]}$\textit{ and }$\mathrm{C}_{d},$%
\begin{equation}
V_{\mathrm{\tilde{G}}^{[2]}\mathrm{,\tilde{G}}^{[1]},d}^{[2]}:\mathrm{C}%
_{\mathrm{\tilde{G}^{[2]},}d}^{[2]}\Longleftrightarrow \mathrm{C}%
_{\mathrm{\tilde{G}^{[1]},}d}^{[1]}\Longleftrightarrow \mathrm{C}_{d},
\end{equation}
\textit{\ of which one is the mapping between }$\mathrm{C}_{\mathrm{\tilde
{G}^{[1]},}d}^{[1]}\ $\textit{and} \textit{Cartesian space} $\mathrm{C}_{d}%
$\textit{ with changing ratio }$k_{0}^{\mu}$ i.e.,
\[
\mathrm{C}_{\mathrm{\tilde{G}^{[1]},}d}^{[1]}\Longleftrightarrow \mathrm{C}%
_{d}.
\]
\textit{ The other is between }$\mathrm{C}_{\mathrm{\tilde{G}^{[2]},}d}^{[2]}%
$\textit{\ with total size }$(\Delta \phi^{\mu})^{[2]}$\textit{ and}
$\mathrm{C}_{\mathrm{\tilde{G}^{[1]},}d}^{[1]}$ \textit{with total
size\ }$(\Delta \phi^{\mu})^{[1]},$\textit{ i.e., }%
\begin{align*}
\mathrm{C}_{\mathrm{\tilde{G}^{[2]},}d}^{[2]}((\Delta \phi^{\mu})^{[2]})  &
\Longleftrightarrow \mathrm{C}_{\mathrm{\tilde{G}^{[1]},}d}^{[1]}((\Delta
\phi^{\mu})^{[1]})\\
&  \equiv \mathrm{C}_{\mathrm{\tilde{U}^{[2]}(1)\in \tilde{G}}_{2},1}%
((\Delta \phi_{\mathrm{global}})^{[2]})\\
&  \Longleftrightarrow \mathrm{C}_{\mathrm{\tilde{U}^{[1]}(1)\in \tilde{G}}%
_{1},1}((\Delta \phi_{\mathrm{global}})^{[1]})\\
&  \equiv \{ \delta \phi_{\mathrm{global}}^{[2]}\} \Leftrightarrow \{ \delta
\phi_{\mathrm{global}}^{[1]}\}
\end{align*}
\textit{where }$\Longleftrightarrow$\textit{\ denotes an ordered mapping under
fixed changing rate along different directions. Along }$\mu$\textit{-th
direction, the elements of two subgroup-changing spaces are }$\delta
\phi_{\mathrm{global}}^{[2]}=\sqrt{%
%TCIMACRO{\dsum \limits_{\mu}}%
%BeginExpansion
{\displaystyle \sum \limits_{\mu}}
%EndExpansion
((\delta \phi^{\mu})^{[2]})^{2}}$\textit{ and }$\delta \phi_{\mathrm{global}%
}^{[1]}=\sqrt{%
%TCIMACRO{\dsum \limits_{\mu}}%
%BeginExpansion
{\displaystyle \sum \limits_{\mu}}
%EndExpansion
((\delta \phi^{\mu})^{[1]})^{2}},$ \textit{respectively, The changing rate
between }$\mathrm{C}_{\mathrm{\tilde{G}^{[2]},}d}^{[2]}((\Delta \phi^{\mu
})^{[2]})\ $\textit{and} $\mathrm{C}_{\mathrm{\tilde{G}^{[1]},}d}%
^{[1]}((\Delta \phi^{\mu})^{[1]})$\textit{ is integer} multiple $(\lambda^{\mu
})^{[12]}.$\textit{ }

For example, for\ 1D 2-nd order variant for non-compact \textrm{\~{U}(1)}
group, the changing subspace of global phase is just the original
group-changing space. Then, 1D 2-nd order variant for non-compact
\textrm{\~{U}(1)} group is defined by a higher-order mapping between
$\mathrm{C}_{\phi^{\lbrack2]}}^{[2]},$ $\mathrm{C}_{\phi^{\lbrack1]}}^{[1]}$,
and $\mathrm{C}_{1}$, i.e.,%
\begin{equation}
V_{\mathrm{\tilde{U}^{[2]}(1),\tilde{U}^{[1]}(1),}1}^{[2]}:\mathrm{C}%
_{\phi^{\lbrack2]}}^{[2]}\Longleftrightarrow \mathrm{C}_{\phi^{\lbrack1]}%
}^{[1]}\Longleftrightarrow \mathrm{C}_{1}.
\end{equation}
Here, $\mathrm{C}_{\phi^{\lbrack2]}}^{[2]}$\textit{ }is described by a series
of numbers of group element $\phi^{\lbrack2]}\in \lbrack0,\Delta \phi
^{\lbrack2]}]\in$ A non-compact\  \textrm{U}$^{[2]}$\textrm{(1)} with a total
size of $\delta \phi^{\lbrack2]}\cdot N=\Delta \phi^{\lbrack2]}$ that is a fixed
number. For $\mathrm{C}_{\phi^{\lbrack2]}}^{[2]}$, the changing element is
$\delta \phi^{\lbrack2]}$ ($\delta \phi^{\lbrack2]}\rightarrow0$). As a result,
$\mathrm{C}_{\phi^{\lbrack2]}}^{[2]}$ is regarded as a mathematical set of $N$
infinitesimal changing of group element and $\delta \phi^{\lbrack2]}$ is the
piece of $\mathrm{C}_{\phi^{\lbrack2]}}^{[2]}$. For the case of
higher-dimensional 2-nd order variant, we focus on their global phases and get
similar results.

Using similar approach, we can define higher-order variants with order higher
than 2.

\textit{Definition -- A n-th order variant }$V_{\mathrm{\tilde{G}}%
^{[n]}\mathrm{,...,\tilde{G}}^{[2]}\mathrm{,\tilde{G}}^{[1]},d}^{[2]}%
$\textit{\ is defined by\ a higher-order mapping between }$\mathrm{C}%
_{\mathrm{\tilde{G}^{[n]},}d}^{[2]},$\textit{ ..., }$\mathrm{C}%
_{\mathrm{\tilde{G}^{[2]},}d}^{[2]},$\textit{ }$\mathrm{C}_{\mathrm{\tilde
{G}^{[1]},}d}^{[1]}$, \textit{and }$\mathrm{C}_{d},$%
\begin{align}
V_{\mathrm{\tilde{G}}^{[n]}\mathrm{,...,\tilde{G}}^{[2]}\mathrm{,\tilde{G}%
}^{[1]},d}^{[2]}  &  :\mathrm{C}_{\mathrm{\tilde{G}^{[n]},}d}^{[n]}%
\Longleftrightarrow...\nonumber \\
\mathrm{C}_{\mathrm{\tilde{G}^{[2]},}d}^{[2]}  &  \Longleftrightarrow
\mathrm{C}_{\mathrm{\tilde{G}^{[1]},}d}^{[1]}\Longleftrightarrow \mathrm{C}%
_{d}.
\end{align}

\paragraph{Classification of higher-order variant}

We classify the 2-nd order variant $V_{\mathrm{\tilde{G}}^{[1]},\mathrm{\tilde
{G}}^{[2]}\mathrm{,}d}^{[2]}$ of non-compact Lie groups \textrm{\~{G}}$^{[1]}$
and \textrm{\~{G}}$^{[2]}$. Different fields are classified by three values,
the level-1 non-compact Lie group $\mathrm{\tilde{G}}^{[1]}$ and the level-2
non-compact Lie group $\mathrm{\tilde{G}}^{[2]}$, the dimension number $d$ of
Cartesian space $\mathrm{C}_{d}.$

In general, if we consider the orthogonality, the level-1 group-changing space
$\mathrm{\tilde{G}}^{[1]}$ is always an $\mathrm{\tilde{S}\tilde{O}}%
$\textrm{(d)} Clifford group-changing space, i.e.,
\[
\mathrm{\tilde{G}}^{[1]}[\Delta \phi^{\mu},\Delta x^{\mu},k_{0}^{\mu
}]=\mathrm{\tilde{S}\tilde{O}(d)}[\Delta \phi^{\mu},\Delta x^{\mu},k_{0}^{\mu
}]
\]
where $k_{0}^{\mu}$\ characterizes the changing rate along $\mu$-th spatial
direction, $\Delta \phi^{\mu}$ denotes the size of group changing space along
$\mu$-th spatial direction, $\Delta x^{\mu}$ denotes the size of $\mu$-th
spatial direction. We may consider the simplest, symmetric case -- the
changing rates are same along different directions, $k_{0}^{\mu}\equiv k_{0}$.
In this paper, we focus on this case. After determine the type of level-1
group-changing space $\mathrm{\tilde{G}}^{[1]},$ a 2-nd order variant is
solely determined by $\mathrm{\tilde{G}}^{[2]}.$ In this paper, we focus on an
Abelian case of $\mathrm{\tilde{G}}^{[2]},$ i.e.,%
\[
\mathrm{\tilde{G}}^{[2]}\equiv \mathrm{\tilde{U}}^{[2]}\mathrm{(1)}.
\]

Therefore, in this paper, we only consider a special 2-nd order variant --
$V_{\mathrm{\tilde{U}}^{[2]}\mathrm{(1),\tilde{S}\tilde{O}^{[1]}\mathrm{(d)}%
,}d}^{[2]}$ that is a higher-order mapping between $\mathrm{C}_{\mathrm{\tilde
{U}}^{[2]}\mathrm{(1)}}^{[2]},$ $\mathrm{C}_{\mathrm{\tilde{S}\tilde{O}}%
^{[1]}\mathrm{(d),}d}^{[1]}$ and $\mathrm{C}_{d}$,%
\begin{equation}
V_{\mathrm{\tilde{U}}^{[2]}\mathrm{(1),\tilde{S}\tilde{O}^{[1]}\mathrm{(d)}%
,}d}^{[2]}:\mathrm{C}_{\mathrm{\tilde{U}}^{[2]}\mathrm{(1)}}^{[2]}%
\Longleftrightarrow \mathrm{C}_{\mathrm{\tilde{U}^{[2]}(1)\in \tilde{G}}_{1}%
,1}((\Delta \phi_{\mathrm{global}})^{[1]})\in \mathrm{C}_{\mathrm{\tilde
{S}\tilde{O}}^{[1]}\mathrm{(d),}d}^{[1]}\Longleftrightarrow \mathrm{C}_{d}.
\end{equation}

Finally, the type of a 2-nd order variant $V_{\mathrm{\tilde{U}(1),\tilde
{S}\tilde{O}^{[1]}\mathrm{(d)},}d}^{[2]}$ is fully determined by the ratio of
the changing rates of the two group-changing spaces
\begin{equation}
\lambda^{\lbrack12]}=\left \vert \frac{\delta \phi^{\lbrack2]}}{\delta
\phi_{\mathrm{global}}^{[1]}}\right \vert .
\end{equation}
In this paper, we only study the "\emph{commensurate}" case, i.e.,
$\lambda^{\lbrack12]}$ is an integer number $\lambda^{\lbrack12]}=1,2,3,$...
In the following parts, we show that the case of $\lambda^{\lbrack12]}=0$
corresponds to massive/massless Dirac model; the case of $\lambda^{\lbrack
12]}=1$ corresponds to a \textrm{QED} with a local \textrm{U}$^{\mathrm{em}}%
$\textrm{(1)} gauge symmetry; the case of $\lambda^{\lbrack12]}=2$ corresponds
to a quantum gauge theory with a local \textrm{SU(2)}$\times$\textrm{U}%
$^{\mathrm{em}}$\textrm{(1)} gauge symmetry; the case of $\lambda^{\lbrack
12]}=3$ corresponds to a \textrm{QCD}$\times$\textrm{QED} with a local
\textrm{SU(3)}$\times$\textrm{U}$^{\mathrm{em}}$\textrm{(1)} gauge symmetry...

In addition, there exist composite higher-order variants by coupling two or
more variants. This type of composite higher-order variants is relevant to the
quantum gauge field with "electromagnetic duality". In this paper, we don't
address this case.

\subsubsection{2-nd order uniform variants}

It was known that uniform variants play role of ground states. Then, in this
section, we discuss 2-nd order uniform variant (2-nd order U-variant).

\paragraph{Definition}

A (1-st order) U-variant $V_{d}^{[1]}[\Delta \phi^{\mu},\Delta x^{\mu}%
,k_{0}^{\mu}]$ for group-changing space $C_{\mathrm{\tilde{G}},d}(\Delta
\phi^{\mu})$ of non-compact Lie group \textrm{\~{G}} had been defined by a
perfect, ordered mapping between a d-dimensional Clifford group-changing space
$\mathrm{C}_{\mathrm{\tilde{G}},d}(\Delta \phi^{\mu})$ and the d-dimensional
Cartesian space $\mathrm{C}_{d}$, of which the total size $\Delta \phi^{\mu}$
of $\mathrm{C}_{\mathrm{\tilde{G}},d}$\ exactly matches the total size $\Delta
x^{\mu}$ of $\mathrm{C}_{d}$, i.e., $\Delta \phi^{\mu}=k_{0}^{\mu}\Delta
x^{\mu}$.

Then, we define 2-nd order U-variant $V_{\mathrm{\tilde{G}}^{[1]}%
,\mathrm{\tilde{G}}^{[2]}\mathrm{,}d}^{[2]}.$

\textit{Definition: A d-dimensional 2-nd order U-variant is a higher-order,
perfect ordered mapping between }$\mathrm{C}_{\mathrm{\tilde{G}^{[2]},}%
d}^{[2]}$, $\mathrm{C}_{\mathrm{\tilde{G}^{[1]},}d}^{[1]}$\textit{ and
}$\mathrm{C}_{d}$, \textit{of which the total sizes of them all match each
other, i.e., }$\Delta \phi_{\mathrm{global}}^{[1]}=\Delta \phi_{\mathrm{global}%
}^{[2]}$, $(\Delta \phi^{\mu})^{[1]}=\Delta x^{\mu}$.

Therefore, for the 2-nd order U-variant $V_{\mathrm{\tilde{G}}^{[1]}%
,\mathrm{\tilde{G}}^{[2]}\mathrm{,}d}^{[2]},$ we have two ordered number
series, one is $\{n_{i}^{[2]}\}=(...1,1,1,1,...)$ for level-2 group-changing
elements $\delta \phi_{i}^{[2]}(\phi^{\lbrack1]}(x_{i}))$ of $\mathrm{C}%
_{\mathrm{\tilde{G}^{[2]},}d}^{[2]}$ on $\mathrm{C}_{\mathrm{\tilde{G}^{[1]}%
,}d}^{[1]}$, the other $\{n_{i}^{[1]}\}=(...1,1,1,1,...)$ for level-1
group-changing elements $\delta \phi_{i}^{[1]}(x_{i})$ of $\mathrm{C}%
_{\mathrm{\tilde{G}^{[1]},}d}^{[1]}$ on Cartesian space.

\paragraph{Uniform knot/link under geometric representation}

In this section, we show that a 2-nd order U-variant $V_{\mathrm{\tilde{G}%
}^{[1]},\mathrm{\tilde{G}}^{[2]}\mathrm{,}d}^{[2]}$ becomes a uniform
knot/link under geometric representation. To show it clearer, we provide its
1-st order representation.

Firstly, we study a 1D 2-nd order U-variant $V_{\mathrm{\tilde{U}}%
^{[1]}\mathrm{(1),\tilde{U}}^{[2]}\mathrm{(1),}1}^{[2]}.$

On the one hand, under 1-st order representation, the level-1 group-changing
space $\mathrm{C}_{\mathrm{\tilde{U}}^{[1]}\mathrm{(1)}}^{[1]}$\ on 1D
Cartesian space is described by a complex field $\mathrm{z}^{[1]}%
=e^{i\phi^{\lbrack1]}(x)}$. The complex field $\mathrm{z}_{u}(x)$ for an
U-variant is obtained by $\mathrm{z}_{u}^{[1]}(x)=\tilde{U}^{[1]}(\delta
\phi^{\lbrack1]})\mathrm{z}_{0}^{[1]}$ where $\tilde{U}^{[1]}(\delta
\phi^{\lbrack1]})=\prod_{i}\tilde{U}^{[1]}(\delta \phi_{i}^{[1]}(x_{i}))$
denote a series of group-changing operations\ with $\tilde{U}(\delta \phi
_{i}^{[1]}(x_{i}))=e^{i((\delta \phi_{i}^{[1]})\cdot \hat{K}^{[1]})}$ and
$\hat{K}^{[1]}=-i\frac{d}{d\phi^{\lbrack1]}}.$\ Here, the i-th group-changing
operation $\tilde{U}^{[1]}(\delta \phi_{i}^{[1]}(x))$ at $x$ generates a
group-changing element.\ For the case of single group-changing element
$\delta \phi_{i}^{[1]}(x_{i})$ on $\delta x_{i}$ at $x_{i},$ the function is
given by
\begin{equation}
\phi^{\lbrack1]}(x)=\left \{
\begin{array}
[c]{c}%
-\frac{\delta \phi_{i}^{[1]}}{2},\text{ }x\in(-\infty,x_{i}]\\
-\frac{\delta \phi_{i}^{[1]}}{2}+k_{0}x,\text{ }x\in(x_{i},x_{i}+\delta
x_{i}]\\
\frac{\delta \phi_{i}^{[1]}}{2},\text{ }x\in(x_{i}+\delta x_{i},\infty)
\end{array}
\right \}  .
\end{equation}
Therefore, the level-1 group-changing space $\mathrm{C}_{\mathrm{\tilde{U}%
}^{[1]}\mathrm{(1)}}^{[1]}$\ is described by a complex field $\mathrm{z}%
_{u}^{[1]}(x)$ in Cartesian space as $\mathrm{z}_{u}^{[1]}(x)=\exp
(i\phi^{\lbrack1]}(x))$ where $\phi^{\lbrack1]}(x)=\phi_{0}^{[1]}+k_{0}x$.

We next discuss 1-st order representation of the level-2 group-changing space
$\mathrm{C}_{\mathrm{\tilde{U}}^{[2]}\mathrm{(1)}}^{[2]}$.

For the level-2 group-changing space $\mathrm{C}_{\mathrm{\tilde{U}}%
^{[2]}\mathrm{(1)}}^{[2]}$ on level-1 group-changing space $\mathrm{C}%
_{\mathrm{\tilde{U}}^{[1]}\mathrm{(1)}}^{[1]}$, we can use another complex
field $\mathrm{z}_{u}^{[2]}(\phi^{\lbrack1]})$ to characterize it,
\begin{equation}
\mathrm{z}_{u}^{[2]}(\phi^{\lbrack1]})=\tilde{U}(\delta \phi^{\lbrack
2]})\mathrm{z}_{0}(\phi^{\lbrack1]})
\end{equation}
where $\tilde{U}(\delta \phi^{\lbrack2]})=\prod_{i}\tilde{U}(\delta \phi
_{i}^{[2]}(\phi_{i}^{[1]}))$ denote a series of group-changing
operations\ with $\tilde{U}(\delta \phi_{i}^{[2]}(\phi_{i}^{[1]}))=e^{i((\delta
\phi_{i}^{[2]})\cdot \hat{K})}$ and $\hat{K}=-i\frac{d}{d\phi^{\lbrack2]}}%
.$\ Here, the i-th group-changing operation $\tilde{U}(\delta \phi_{i}%
^{[2]}(\phi^{\lbrack1]}))$ at $\phi^{\lbrack1]}$ generates a group-changing
element.\ For the case of single group-changing element $\delta \phi_{i}%
^{[2]}(\phi_{i}^{[1]})$ on $\delta \phi_{i}^{[1]}$ at $\phi_{i}^{[1]},$ the
function is given by
\begin{equation}
\phi^{\lbrack2]}(\phi^{\lbrack1]})=\left \{
\begin{array}
[c]{c}%
-\frac{\delta \phi_{i}^{[2]}}{2},\text{ }\phi^{\lbrack1]}\in(-\infty,\phi
_{i}^{[1]}]\\
-\frac{\delta \phi_{i}^{[2]}}{2}+\lambda^{\lbrack12]}\phi^{\lbrack1]},\text{
}\phi^{\lbrack1]}\in(\phi_{i}^{[1]},\phi_{i}^{[1]}+\delta \phi_{i}^{[1]}]\\
\frac{\delta \phi_{i}^{[2]}}{2},\text{ }\phi^{\lbrack1]}\in(\phi_{i}%
^{[1]}+\delta \phi_{i}^{[1]},\infty)
\end{array}
\right \}  .
\end{equation}
Therefore, for a 1D 2-nd order U-variant $V_{\mathrm{\tilde{U}}^{[1]}%
\mathrm{(1),\tilde{U}}^{[2]}\mathrm{(1),}1}^{[2]}$, the changings of
$\mathrm{C}_{\mathrm{\tilde{U}}^{[2]}\mathrm{(1)}}^{[2]}$ can be described by
another complex field $\mathrm{z}_{u}^{[2]}(\phi^{\lbrack1]})$ in Cartesian
space as $\mathrm{z}_{u}^{[2]}(\phi^{\lbrack1]})=\exp(i\lambda^{\lbrack
12]}\phi^{\lbrack1]}+i\phi_{0}^{[2]})$ where $\lambda^{\lbrack12]}$ is an
integer number.

Then, a 1D 2-nd order U-variant $V_{\mathrm{\tilde{U}(1),\tilde{G},}1}^{[2]}$
can be described by two special complex fields $\mathrm{z}_{u}^{[1]}(x)$ and
$\mathrm{z}_{u}^{[2]}(\phi^{\lbrack1]}(x))$, i.e,
\[
\mathrm{z}_{u}^{[1]}(x)=\exp(i\phi^{\lbrack1]}(x))=\exp(i\phi_{0}^{[1]}%
+ik_{0}x)
\]
and
\begin{align*}
\mathrm{z}_{u}^{[2]}(\phi^{\lbrack1]})  &  =\exp(i\phi^{\lbrack2]}%
(\phi^{\lbrack1]}))\\
&  =\exp(i\phi_{0}^{[2]}+i\lambda^{\lbrack12]}\phi^{\lbrack1]}).
\end{align*}

On the one hand, we can map $\mathrm{z}_{u}^{[1]}(x)=\exp(ik_{0}x+i\phi
_{0}^{[1]})=\operatorname{Re}\xi^{\lbrack1]}(x)+i\operatorname{Im}%
\eta^{\lbrack1]}(x)$ to a curved line $\{x,\xi^{\lbrack1]}(x),\eta^{\lbrack
1]}(x)\}$ in three dimensions. On the other hand, we map another complex field
$\mathrm{z}_{u}^{[2]}(\phi^{\lbrack1]})=\operatorname{Re}\xi^{\lbrack2]}%
(\phi^{\lbrack1]}(x))+i\operatorname{Im}\eta^{\lbrack2]}(\phi^{\lbrack1]}(x))$
for a variant to a curved line $\{ \phi^{\lbrack1]},\xi^{\lbrack2]}%
(\phi^{\lbrack1]}(x)),\eta^{\lbrack2]}(\phi^{\lbrack1]}(x))\}$ in three
dimensions as
\begin{align*}
\mathrm{z}_{u}^{[2]}(\phi^{\lbrack1]}(x))  &  =\epsilon \exp(i\phi^{\lbrack
2]}(\phi^{\lbrack1]}(x)))\\
&  =\epsilon \exp(i\phi_{0}^{[2]}+i\lambda^{\lbrack12]}\phi^{\lbrack1]}(x)).
\end{align*}
Here, $\epsilon$ is dimensionless value and the result is independent on it.
For simplicity, we set it to be small value, i.e., $\epsilon \rightarrow0$.
Finally, the 1D 2-nd order variant shows the highly non-local geometric
structure -- \emph{knot/links}. The knot/links consist of two lines, one is
about $\mathrm{z}_{u}^{[1]}(x)$, the other is about $\mathrm{z}_{u}^{[2]}%
(\phi^{\lbrack1]}(x))$. See the illsutration In Fig.26(b).

\begin{figure}[ptb]
\includegraphics[clip,width=0.7\textwidth]{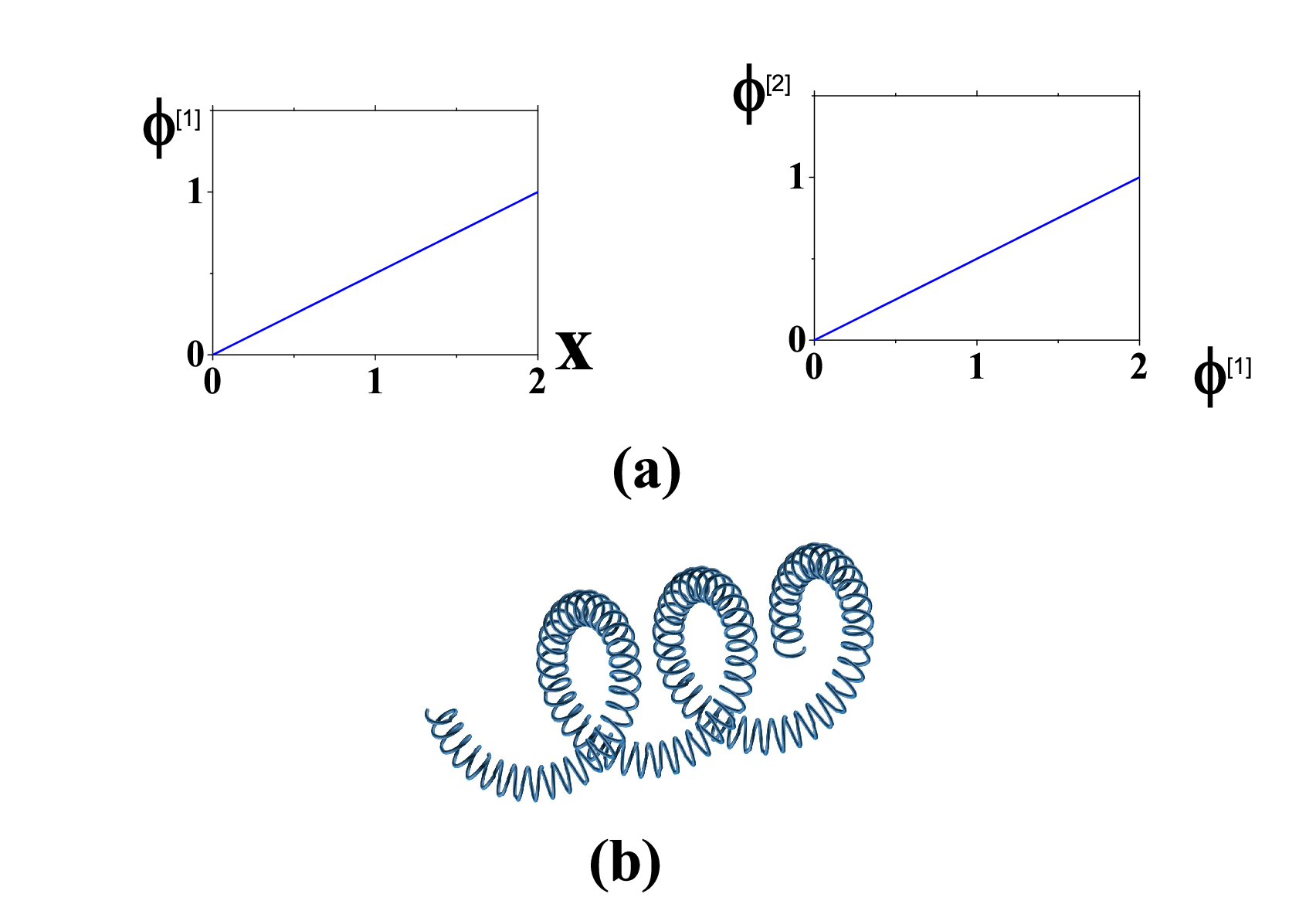}\caption{(Color online)
(a) An illustration of the changing rates of 2-nd order uniform variant of
non-compact \textrm{\~{U}(1)} group; (b) The illustration of 1-st order
geometry representation for level-2 group-changing space of the 2-nd order
uniform variant on Cartesian space. }%
\end{figure}

For a higher-dimensional 2-nd order U-variant, we have a complex uniform
knot/link structure as
\begin{align*}
\mathrm{z}_{u}^{[1]}(x^{\mu})  &  =\exp(i\phi^{\mu,[1]}(x^{\mu}))\\
&  =\exp(i\phi_{0}^{[1]}+ik_{0}^{\mu}x^{\mu})
\end{align*}
and
\begin{align*}
\mathrm{z}_{u}^{[2]}(\phi_{\mathrm{global}}^{[1]})  &  =\exp(i\phi
_{\mathrm{global}}^{[2]}(\phi_{\mathrm{global}}^{[1]}))\\
&  =\exp(i\phi_{0}^{[2]}+i\lambda^{\lbrack12]}\phi_{\mathrm{global}}^{[1]})
\end{align*}
where $\phi_{\mathrm{global}}^{[1]}=\sqrt{%
%TCIMACRO{\dsum \limits_{\mu}}%
%BeginExpansion
{\displaystyle \sum \limits_{\mu}}
%EndExpansion
((\phi^{\mu})^{[1]})^{2}}$\textit{ and }$\phi_{\mathrm{global}}^{[2]}=\sqrt{%
%TCIMACRO{\dsum \limits_{\mu}}%
%BeginExpansion
{\displaystyle \sum \limits_{\mu}}
%EndExpansion
((\phi^{\mu})^{[2]})^{2}}.$

\paragraph{Higher-order variability}

In modern physics and modern mathematics, "\emph{invariant/symmetry}" is an
important concept. However, invariant/symmetry is an indirect description on a
system and always characterizes the extrinsic property of a system. We had
proposed a new concept -- \emph{higher-order} \emph{variability }that is a
direct description on a system and characterizes the intrinsic property of a system.

Let firstly us review higher-order variability for 1-st order uniform
variants. For a (1-st order) uniform variant with infinite size ($\Delta
x\rightarrow \infty$), we have the following relationship,%
\begin{equation}
\mathcal{T}(\delta x^{\mu})\leftrightarrow \hat{U}(\delta \phi^{\mu}%
)=e^{i\cdot \delta \phi^{\mu}T^{\mu}}%
\end{equation}
where $\mathcal{T}(\delta x^{\mu})$ is the spatial translation operation on
$\mathrm{C}_{d}$ along $x^{\mu}$-direction and $\tilde{U}(\delta \phi^{\mu})$
is shift operation on $\mathrm{C}_{\mathrm{\tilde{G}},d}(\Delta \phi^{\mu})$,
and $\delta \phi^{\mu}=k_{0}^{\mu}\delta x^{\mu}$. That means when one
translate along Cartesian space $\delta x^{\mu},$ the corresponding shifting
along group-changing space $\mathrm{C}_{\mathrm{\tilde{G}},d}$ is $\delta
\phi^{\mu}=k_{0}^{\mu}\delta x^{\mu}.$ For simplicity, we can denote a system
with 1-st order variability by the following equation
\begin{equation}
\mathcal{T}\leftrightarrow \hat{U}.
\end{equation}

We called usual "\emph{symmetry}" or "\emph{invariant}" to be \emph{zero-order
variability} and $V_{\mathrm{\tilde{G},}d}^{[1]}[\Delta \phi^{\mu},\Delta
x^{\mu},k_{0}^{\mu}]$\ to be a system with\emph{ 1-st order variability}.
Therefore, the order of variability becomes a key value classifying the
complexity of mathematical systems. We point out that for 2-nd order U-variant
$V_{\mathrm{\tilde{G}}^{[1]},\mathrm{\tilde{G}}^{[2]}\mathrm{,}d}^{[2]}$ with
infinite size ($\Delta x^{\mu}\rightarrow \infty$), there exists 2-nd order
variability. Then, we give detailed discussion.

A 2-nd order variability for 2-nd order U-variant $V_{\mathrm{\tilde{G}}%
^{[1]},\mathrm{\tilde{G}}^{[2]}\mathrm{,}d}^{[2]}$ is described by the
following two equations%
\begin{align}
\hat{U}^{[1]}(\delta \phi_{\mathrm{global}}^{[1]})  &  \leftrightarrow \hat
{U}^{[2]}(\delta \phi_{\mathrm{global}}^{[2]})\nonumber \\
&  =\exp(i\lambda^{\lbrack12]}\delta \phi_{\mathrm{global}}^{[1]}),
\end{align}
and
\begin{align}
\mathcal{T}(\delta x^{\mu})  &  \leftrightarrow \hat{U}^{[1]}((\delta \phi^{\mu
})^{[1]})\nonumber \\
&  =\exp(i(T^{\mu})^{[1]}(\delta \phi^{\mu})^{[1]})\\
&  =\exp(i(T^{\mu})^{[1]}(k_{0}^{\mu}\delta x^{\mu})),\nonumber
\end{align}
where $\delta \phi_{\mathrm{global}}^{[1]}=\sqrt{%
%TCIMACRO{\dsum \limits_{\mu}}%
%BeginExpansion
{\displaystyle \sum \limits_{\mu}}
%EndExpansion
((\delta \phi^{\mu})^{[2]})^{2}}$\textit{ }and\textit{ }$\delta \phi
_{\mathrm{global}}^{[1]}=\sqrt{%
%TCIMACRO{\dsum \limits_{\mu}}%
%BeginExpansion
{\displaystyle \sum \limits_{\mu}}
%EndExpansion
((\delta \phi^{\mu})^{[1]})^{2}}.$ $\mathcal{T}(\delta x^{\mu})$ is the spatial
translation operation along $x^{\mu}$-direction, $\hat{U}^{[1]}((\delta
\phi^{\mu})^{[1]})=\exp(i(T^{\mu})^{[1]}(\delta \phi^{\mu})^{[1]})$ and
$\hat{U}^{[1]}(\delta \phi_{\mathrm{global}}^{[1]})$\ are shift operations on
$\mathrm{C}_{\mathrm{\tilde{G}}^{[1]}\mathrm{,}d}^{[1]}$, $\hat{U}%
^{[2]}(\delta \phi_{\mathrm{global}}^{[2]})$ is shift operation on
$\mathrm{C}_{\mathrm{\tilde{G}}^{[2]}\mathrm{,}d}^{[2]}$.

For example, for 1D 2-nd order U-variant $V_{\mathrm{\tilde{U}}^{[1]}%
\mathrm{(1),\tilde{U}}^{[2]}\mathrm{(1),}1}^{[2]},$ the 2-nd order variability
becomes simple, i.e.,
\begin{equation}
\hat{U}^{[1]}(\delta \phi^{\lbrack1]})\leftrightarrow \hat{U}^{[2]}(\delta
\phi^{\lbrack2]})=\exp(i\lambda^{\lbrack12]}\delta \phi^{\lbrack1]}),
\end{equation}
and
\begin{equation}
\mathcal{T}(\delta x)\leftrightarrow \hat{U}^{[1]}(\delta \phi^{\lbrack1]}%
)=\exp(i(k_{0}\delta x)).\nonumber
\end{equation}

Therefore, an operation $\hat{U}^{[1]}((\delta \phi_{\mathrm{global}}^{[1]}%
)$\ is always accompanied by an operation $\hat{U}^{[2]}(\delta \phi
_{\mathrm{global}}^{[2]})$; An operation $\mathcal{T}(\delta x^{\mu})$\ is
always accompanied by an operation $\hat{U}^{[1]}((\delta \phi_{\mathrm{global}%
}^{[1]}).$ For simplicity, we can denote a system with 2-nd order variability
by the following equation
\begin{equation}
\mathcal{T}\leftrightarrow \hat{U}^{[1]}\leftrightarrow \hat{U}^{[2]}.
\end{equation}
In addition, we may define a uniform variant with even higher-order
variability by
\begin{equation}
\mathcal{T}\leftrightarrow \hat{U}^{[1]}\leftrightarrow \hat{U}^{[2]}%
\rightarrow \hat{U}^{[3]}...
\end{equation}

\paragraph{Zero lattices under knot projection}

In above section, we show that under geometric representation, 1D 2-nd
U-variant $V_{0,\mathrm{\tilde{G}}^{[1]},\mathrm{\tilde{G}}^{[2]}\mathrm{,}%
d}^{[2]}$ can be regarded as a knot/link on 3D space. People had known that a
knot/link can be projected by counting the crossings (or zeroes named in this
paper) of the corresponding lines. With the help of the K-projection, people
can locally obtain the property of the variant. We then introduce the\emph{
}K-projection of the two curved lines of 1D 2-nd order U-variant.

Firstly, We show the knot projection for the 1D 2-nd order U-variant
$V_{0,\mathrm{\tilde{U}}^{[1]}\mathrm{(1),\tilde{U}}^{[2]}\mathrm{(1),}%
1}^{[2]}$.

On the one hand, we do K-projection on the group-changing space $\mathrm{C}%
_{\mathrm{\tilde{U}}^{[1]}\mathrm{(1)}}^{[1]}$ with the help of the complex
function $\mathrm{z}_{u}^{[1]}(x)=\exp(ik_{0}x+i\phi_{0})=\operatorname{Re}%
\xi^{\lbrack1]}(x)+i\operatorname{Im}\eta^{\lbrack1]}(x)$. In mathematics, the
K-projection is defined by
\begin{equation}
\hat{P}_{\theta^{\lbrack1]}}\left(
\begin{array}
[c]{c}%
\xi^{\lbrack1]}(x)\\
\eta^{\lbrack1]}(x)
\end{array}
\right)  =\left(
\begin{array}
[c]{c}%
\xi_{\theta^{\lbrack1]}}^{[1]}(x)\\
\left[  \eta_{\theta^{\lbrack1]}}^{[1]}(x)\right]  _{0}%
\end{array}
\right)
\end{equation}
where $\xi_{\theta^{\lbrack1]}}^{[1]}(x)$ is variable and $\left[
\eta_{\theta^{\lbrack1]}}^{[1]}(x)\right]  _{0}$ is constant. In the following
parts we use $\hat{P}_{\theta^{\lbrack1]}}$ to denote the projection
operators. Because the projection direction out of the curved line is
characterized by an angle $\theta^{\lbrack1]}$ in $\{ \xi^{\lbrack1]}%
,\eta^{\lbrack1]}\}$ space, we have
\begin{equation}
\left(
\begin{array}
[c]{c}%
\xi_{\theta}^{[1]}\\
\eta_{\theta}^{[1]}%
\end{array}
\right)  =\left(
\begin{array}
[c]{cc}%
\cos \theta^{\lbrack1]} & \sin \theta^{\lbrack1]}\\
\sin \theta^{\lbrack1]} & -\cos \theta^{\lbrack1]}%
\end{array}
\right)  \left(
\begin{array}
[c]{c}%
\xi^{\lbrack1]}\\
\eta^{\lbrack1]}%
\end{array}
\right)
\end{equation}
where $\theta^{\lbrack1]}$ is angle \textrm{mod}($2\pi$), i.e. $\theta
^{\lbrack1]}\operatorname{mod}2\pi=0.$ So the curved line of 1D variant is
described by the function
\begin{equation}
\xi_{\theta^{\lbrack1]}}^{[1]}(x)=\xi^{\lbrack1]}(x)\cos \theta^{\lbrack
1]}+\eta^{\lbrack1]}(x)\sin \theta^{\lbrack1]}.
\end{equation}
$\theta^{\lbrack1]}\in \lbrack0,2\pi)$ is projection angle.\begin{figure}[ptb]
\includegraphics[clip,width=0.7\textwidth]{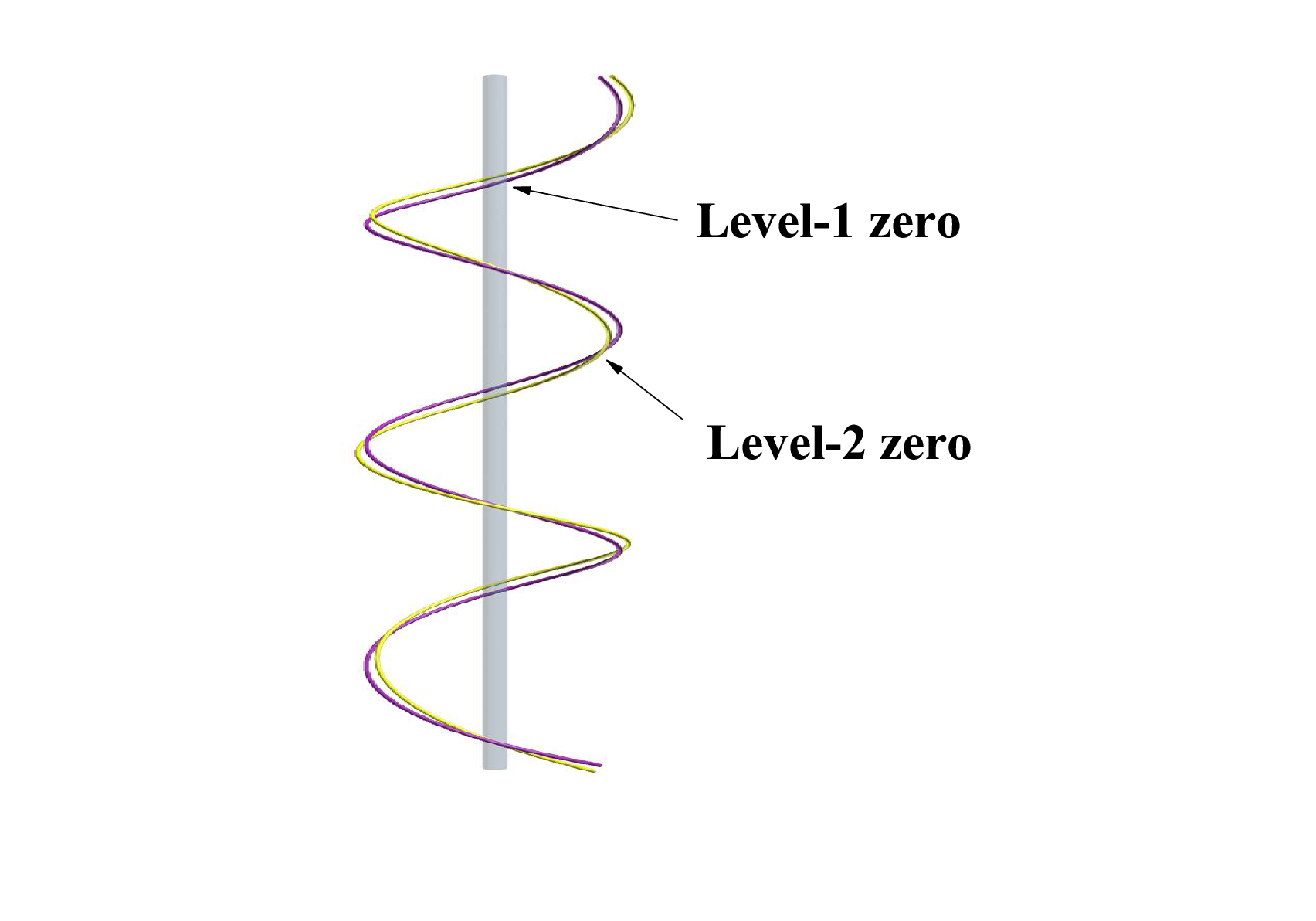}\caption{(Color online)
An illustration of higher-order variant under knot projection. There are two
types of zeroes (level-1 zero and level-2 zero) corresponding to two types of
crossings.}%
\end{figure}

Under projection, each zero corresponds to a solution of the equation
\begin{equation}
\hat{P}_{\theta^{\lbrack1]}}[\mathrm{z}_{u}^{[1]}(x)]\equiv \xi_{\theta
^{\lbrack1]}}^{[1]}(x)=0.\nonumber
\end{equation}

Now, a 1D U-variant becomes a 1D crystal of zeros (or 1D zero lattice). The
"changing" structure of phase factor disappears. From the zero-equation
$\xi_{\theta^{\lbrack1]}}^{[1]}(x)=0$ or $\cos(k_{0}x-\theta^{\lbrack1]})=0,$
we get the zero-solutions to be
\begin{equation}
x=l_{0}\cdot N^{[1]}/2+\frac{l_{0}}{2\pi}(\theta^{\lbrack1]}+\frac{\pi}{2})
\end{equation}
where $N$ is an integer number, and $l_{0}=2\pi/k_{0}$. The zero lattice is a
lattice of "two-sublattice" with discrete spatial translation symmetry. In
other words, a unit cell with $2\pi$ phase changing has two zeroes. The
lattice distance is $l_{0}$. Consequently, after projection, the non-compact
\textrm{\~{U}(1)} group of $\phi^{\lbrack1]}(x)$ turns into a compact group on
zero lattice of "two-sublattice", i.e.,
\[
\phi^{\lbrack1]}(x)=2\pi N^{[1]}(x)+\theta^{\lbrack1]}.
\]
We then relabel the group-changing space $\mathrm{C}_{\mathrm{\tilde{U}}%
^{[1]}\mathrm{(1)}}^{[1]}$ by two numbers ($N^{[1]}(x),\theta^{\lbrack1]}$):
$\theta^{\lbrack1]}$ is compact phase angle, the other is the integer winding
number of unit cell of zero lattice $N^{[1]}(x)$.

On the other hand, we show the K-projection on $\mathrm{C}_{\mathrm{\tilde{U}%
}^{[2]}\mathrm{(1)}}^{[2]}$ with the help of the complex function
$\mathrm{z}_{u}^{[2]}(\phi^{\lbrack1]})=\exp(i\phi_{0}^{[2]}+i\lambda
^{\lbrack12]}\phi^{\lbrack1]}(x))=\operatorname{Re}\xi^{\lbrack2]}%
(\phi^{\lbrack1]}(x))+i\operatorname{Im}\eta^{\lbrack2]}(\phi^{\lbrack1]}(x))$
on another three dimensions $\{ \phi^{\lbrack1]},\xi^{\lbrack2]}(\phi
^{\lbrack1]}),\eta^{\lbrack2]}(\phi^{\lbrack1]})\}$. Using similar approach,
under K-projection with angle $\theta^{\lbrack2]}$, we obtain zero function
\begin{equation}
\hat{P}_{\theta^{\lbrack2]}}[\mathrm{z}_{u}^{[2]}(\phi^{\lbrack1]})]\equiv
\xi^{\lbrack2]}(\phi^{\lbrack1]})=0.\nonumber
\end{equation}
The zero solution is
\begin{equation}
\phi^{\lbrack1]}=\frac{1}{\lambda^{\lbrack12]}}(\frac{\pi}{2}+2\pi
N^{[2]})+\theta^{\lbrack2]}.
\end{equation}
We have a level-2 zero lattice on level-1 phase changing space, of which each
site is denoted by an integer number $N^{[2]}$. Fig.27 shows a level-2 zero
and a level-1 zero for an 2-nd order U-variant under geometry representation.
One can see that each crossing corresponds to a zero.

Finally, under K-projection, we have two zero lattices, level-1 for
$\mathrm{C}_{\mathrm{\tilde{U}}^{[1]}\mathrm{(1)}}^{[1]}$ and level-2 for
$\mathrm{C}_{\mathrm{\tilde{U}}^{[2]}\mathrm{(1)}}^{[2]}.$\ It is obvious that
this is a composite zero lattice, of which each level-1 zero (or level-1
lattice site) of $\mathrm{C}_{\mathrm{\tilde{U}}^{[1]}\mathrm{(1)}}^{[1]}$
corresponds to $\lambda^{\lbrack12]}$ level-2 zeros (or level-2 lattice sites)
of $\mathrm{C}_{\mathrm{\tilde{U}}^{[2]}\mathrm{(1)}}^{[2]}.$

Next, we generalize above results to the cases for higher dimensional 2-nd
order U-variants.

For a higher-dimensional 2-nd order U-variant $V_{0,\mathrm{\tilde{G}}%
^{[1]},\mathrm{\tilde{G}}^{[2]}\mathrm{,}d}^{[2]}$, there also exist two types
of zero lattices, level-1 for $\mathrm{C}_{\mathrm{\tilde{G}}^{[1]}}^{[1]}$
and level-2 for $\mathrm{C}_{\mathrm{\tilde{G}}^{[2]}}^{[2]}$.

On the one hand, we discuss the level-1 zero lattice for $\mathrm{C}%
_{\mathrm{\tilde{G}}^{[1]}}^{[1]}.$ The situation is same to the case for a
higher-dimensional 1-st order U-variant $V_{\mathrm{\tilde{G}}^{[1]}%
\mathrm{,}d}^{[2]}.$ Now, we must do knot-projection on $\mathrm{C}%
_{\mathrm{\tilde{G}}^{[1]}}^{[1]}$ along its $\mu$-th spatial direction by
"\emph{direction projection}" (D-projection). Then, under D-projection, we
reduce the higher-dimensional 2-nd order U-variant $V_{\mathrm{\tilde{G}%
}^{[1]},\mathrm{\tilde{G}}^{[2]}\mathrm{,}d}^{[2]}$ to $d$ D-projected 1D 2-nd
order U-variants, each of which is described by a complex function
\begin{equation}
\mathrm{z}_{u}^{[1],\mu}(x)=\tilde{U}^{[1],\mu}((\delta \phi^{\mu}%
)^{[1]})\mathrm{z}_{0}%
\end{equation}
where $\tilde{U}^{[1],\mu}(\delta \phi^{\lbrack1]})=\mathrm{Tr}(T^{\mu}%
\tilde{U}^{[1]}(\delta \phi))$ and $\tilde{U}^{[1]}(\delta \phi)=\prod_{\mu
}(\prod_{i}\tilde{U}^{[1]}((\delta \phi_{i}^{\mu}(x))^{[1]}))$. As a result, we
have a $d$-dimensional level-1 zero lattice for K-projected $\mathrm{C}%
_{\mathrm{\tilde{G}}^{[1]}}^{[1]}$ that is denoted by $(N^{\mu})^{[1]}.$

On the other hand, we discuss level-2 zero lattice for $\mathrm{C}%
_{\mathrm{\tilde{G}}^{[2]}}^{[2]}.$ The situation is same to the case for a 1D
1-st order U-variant $V_{\mathrm{\tilde{U}}^{[1]}\mathrm{(1),\tilde{U}}%
^{[2]}\mathrm{(1),}1}^{[2]}$ where $\mathrm{\tilde{U}}^{[1]}\mathrm{(1)}$ and
$\mathrm{\tilde{U}}^{[2]}\mathrm{(1)}$ denotes group-changing subspace of
global changings for $\mathrm{\tilde{G}}^{[1]}$ and $\mathrm{\tilde{G}}%
^{[2]},$ respectively. Under K-projection, we have a level-2 zero lattice of
$\mathrm{C}_{\mathrm{\tilde{U}}^{[2]}\mathrm{(1)\in \tilde{G}}^{[2]}}^{[2]}%
.$\ It is obvious that each level-1 zero (or level-1 lattice site) of
$\mathrm{C}_{\mathrm{\tilde{U}}^{[1]}\mathrm{(1)\in \tilde{G}}^{[1]}}^{[1]}$
corresponds to $\lambda^{\lbrack12]}$ level-2 zeros (or level-2 lattice sites)
of $\mathrm{C}_{\mathrm{\tilde{U}}^{[2]}\mathrm{(1)\in \tilde{G}}^{[2]}}%
^{[2]}.$

Finally, under K-projection, we have a two levels of composite zero lattices
for a higher-dimensional 2-nd order U-variant $V_{0,\mathrm{\tilde{G}}%
^{[1]},\mathrm{\tilde{G}}^{[2]}\mathrm{,}d}^{[2]}$: the group-changing space
for $\mathrm{\tilde{G}}^{[1]}$ is reduced into a $d$-dimensional level-1 zero
lattice on Cartesian space, of which each zero corresponds to $\lambda
^{\lbrack12]}$ level-2 zeroes that is projected from $\mathrm{\tilde{G}}%
^{[2]}$.

\subsubsection{2-nd order perturbative variants}

In this part, we study the 2-nd order P-variants $V_{\mathrm{\tilde{G}}%
^{[1]},\mathrm{\tilde{G}}^{[2]}\mathrm{,}d}^{[2]}$ that can be regarded as a
variant by slightly perturbing a 2-nd order U-variant $V_{0,\mathrm{\tilde{G}%
}^{[1]},\mathrm{\tilde{G}}^{[2]}\mathrm{,}d}^{[2]}.$ Here, the subscript "$0$"
denotes a U-variant.

\textit{Definition -- d-dimensional 2-nd order P-variant }$V_{\mathrm{\tilde
{G}}^{[1]},\mathrm{\tilde{G}}^{[2]}\mathrm{,}d}^{[2]}$\textit{ is defined by
quasi-perfect, ordered mappings between }$\mathrm{C}_{\mathrm{\tilde{G}%
^{[2]},}d}^{[2]}$, $\mathrm{C}_{\mathrm{\tilde{G}^{[1]},}d}^{[1]}$\textit{ and
}$\mathrm{C}_{d}$\textit{, of which the total sizes of them don't exactly
match each other, i.e., }$\Delta \phi_{\mathrm{global}}^{[2]}\neq
\lambda^{\lbrack12]}\Delta \phi_{\mathrm{global}}^{[1]}$\textit{ with
}$\left \vert ((\Delta \phi_{\mathrm{global}}^{[2]}-\lambda^{\lbrack12]}%
\Delta \phi_{\mathrm{global}}^{[1]})/\Delta \phi_{\mathrm{global}}%
^{[2]}\right \vert \ll1$\textit{ and }$(\Delta \phi^{\mu})^{[1]}\neq k_{0}^{\mu
}\Delta x^{\mu}\ $with $\left \vert ((\Delta \phi^{\mu})^{[1]}-k_{0}^{\mu}\Delta
x^{\mu})/(\Delta \phi^{\mu})^{[1]}\right \vert \ll1$.

According to above mismatch conditions, for a 2-nd order P-variant, there must
exist more than one type of group-changing elements on either level-1 or
level-2 group-changing spaces.

In this part, we will focus on Hybrid-order representation under K-projection
for a 2-nd order P-variant.

According to above discussion, for a 1-st order P-variant, there must exist
more than one type of group-changing elements on the original U-variant
$V_{0,\mathrm{\tilde{G},}d}^{[1]}[\Delta \phi^{\mu},\Delta x^{\mu},k_{0}^{\mu
}]$. Then, we do K-projection on the original U-variant $V_{0,\mathrm{\tilde
{G},}d}^{[1]}[\Delta \phi^{\mu},\Delta x^{\mu},k_{0}^{\mu}]$ and obtain a
uniform zero lattice. If we consider the uniform zero lattice for
$V_{0,\mathrm{\tilde{G},}d}^{[1]}[\Delta \phi^{\mu},\Delta x^{\mu},k_{0}^{\mu
}]$ to be a rigid background, we have a usual quantum field with compact group
on the zero lattice for the 1-st order P-variant $V_{\mathrm{\tilde{G},}%
d}^{[1]}[\Delta \phi^{\mu},\Delta x^{\mu},k_{0}^{\mu}]$. Using similar
approach, we give the Hybrid-order representation under K-projection for a
2-nd order P-variant.

Firstly, we study the case for a 2-nd order P-variant as a 2-nd order
U-variant with an extra group-changing elements of $\mathrm{C}_{\mathrm{\tilde
{G}}^{[1]}}^{[1]}$ or those of $\mathrm{C}_{\mathrm{\tilde{G}}^{[2]}}^{[2]}$.
We can "generate" the 1D 2-nd order P-variant by do extra group-changing
operations $\delta \phi_{i}^{[1]}(x_{i})$ on the 2-nd order U-variant
$V_{0,\mathrm{\tilde{G}}^{[1]},\mathrm{\tilde{G}}^{[2]}\mathrm{,}d}^{[2]}$,
i.e., $\tilde{U}(\delta \phi_{i}^{[1]}(x_{i}))=e^{i((\delta \phi_{i}^{[1]}%
)\cdot \hat{K})}$ and $\hat{K}=-i\frac{d}{d\phi^{\lbrack1]}}.$\ Here, the i-th
infinitesimal group-changing operation $\tilde{U}(\delta \phi_{i}^{[1]})$
generates a group-changing element on position $i$. We then use similar
approach to "generate" the 1D 2-nd order P-variant by do an extra
group-changing elements $\delta \phi_{i}^{[2]}(\phi_{i}^{[1]})$ on its given
position $\phi_{i}^{[1]}$, i.e., $\tilde{U}(\delta \phi_{i}^{[2]}(\phi
_{i}^{[1]}))=e^{i((\delta \phi_{i}^{[2]})\cdot \hat{K})}$ and $\hat{K}%
=-i\frac{d}{d\phi^{\lbrack2]}}.$\ Here, the i-th infinitesimal group-changing
operation $\tilde{U}(\delta \phi_{i}^{[2]})$ generates a group-changing element
on position $i.$

Secondly, we do K-projection on the original U-variant $V_{0,\mathrm{\tilde
{G}}^{[1]},\mathrm{\tilde{G}}^{[2]}\mathrm{,}d}^{[2]}$. After K-projection,
$V_{0,\mathrm{\tilde{G}}^{[1]},\mathrm{\tilde{G}}^{[2]}\mathrm{,}d}^{[2]}$ is
projected to two levels of composite zero lattices denoted by $N^{[1]}(x)$ and
$N^{[2]}(\phi_{\mathrm{global}}^{[1]})$. As a result, $\mathrm{\tilde{G}%
}^{[1]}$ and $\mathrm{\tilde{G}}^{[2]}$ are projected to two compact groups on
zero lattices $(N^{\mu})^{[1]}(x^{\mu})$ and $N^{[2]}(\phi_{\mathrm{global}%
}^{[1]})$, respectively.

Thirdly, we put the extra group-changing elements $\delta \phi_{i}^{[1]}%
(x_{i})$ on uniform level-1 zero lattice and $\delta \phi_{i}^{[2]}(\phi
_{i}^{[1]})$ on the uniform level-2 zero lattice. During this step, we assume
that the two zero lattices are rigid lattices and can be considered as a
\emph{background}.

Fourthly, we do \emph{compactification }for the extra group-changing elements
of $\delta \phi_{i}^{[1]}(x_{i})$ or $\delta \phi_{i}^{[2]}(\phi
_{\mathrm{global}}^{[1]})$. Now, to exact determine an extra group-changing
element, one must know the corresponding positions of lattice site $(N^{\mu
})^{[1]}(x^{\mu})$ or $N^{[2]}(\phi_{\mathrm{global}}^{[1]})$ together with
phase angles $\varphi^{\lbrack1]}(x)$ or $\varphi^{\lbrack2]}(x)$ on
corresponding lattice sites. Here, the two phase angles are all compact,
i.e.,
\begin{align*}
\varphi^{\lbrack1]}((N^{\mu})^{[1]}(x^{\mu}))  &  =\phi^{\lbrack1]}((N^{\mu
})^{[1]}(x^{\mu}))\operatorname{mod}(2\pi),\\
\varphi^{\lbrack2]}(\phi_{\mathrm{global}}^{[1]}(N^{[2]}(\phi_{\mathrm{global}%
}^{[1]})))  &  =\phi^{\lbrack2]}(\phi_{\mathrm{global}}^{[1]}((N^{\mu}%
)^{[1]}(x^{\mu}))\\
&  \operatorname{mod}(2\pi).
\end{align*}
Consequently, on the uniform zero lattices of corresponding levels $(N^{\mu
})^{[1]}(x^{\mu})$ or $N^{[2]}(\phi_{\mathrm{global}}^{[1]}),$ the operation
for group-changing spaces are reduced into those for compact phase angles,
\begin{align*}
\tilde{U}_{i}^{[1]}(\delta \phi_{i}^{[1]}((N_{i}^{\mu})^{[1]}(x_{i}^{\mu})))
&  \rightarrow U_{i}^{[1]}(\delta \varphi_{i}^{[1]}((N_{i}^{\mu})^{[1]}%
(x_{i}^{\mu}))),\\
\tilde{U}_{i}^{[2]}(\delta \phi_{i}^{[2]}(N_{i}^{[2]}(\phi_{i,\mathrm{global}%
}^{[1]}))  &  \rightarrow U_{i}^{[2]}(\delta \varphi_{i}^{[2]}(N_{i}^{[2]}%
(\phi_{i,\mathrm{global}}^{[1]}))
\end{align*}
where $U_{i}^{[1]}(\delta \varphi_{i}^{[1]}((N_{i}^{\mu})^{[1]}(x_{i}^{\mu})))$
is a local i-th operation changing phase angle from $\varphi_{0}^{[1]}$ to
$\varphi_{0}^{[1]}+\delta \varphi_{i}^{[1]}((N_{i}^{\mu})^{[1]}(x_{i}^{\mu}))$
and $U_{i}^{[2]}(\delta \varphi_{i}^{[2]}(N_{i}^{[2]}(\phi_{i,\mathrm{global}%
}^{[1]}))$ is a local operation changing phase angle from $\varphi_{0}^{[2]}$
to $\varphi_{0}^{[2]}+\delta \varphi_{i}^{[2]}(N_{i}^{[2]}(\phi
_{i,\mathrm{global}}^{[1]})).$

Therefore, we can use a usual quantum field of compact \textrm{U(1)} group to
fully describe the effect of $U_{i}^{[1]}(\delta \varphi_{i}^{[1]},(N_{i}^{\mu
})^{[1]}(x_{i}^{\mu}))$ where $\delta \varphi_{i}^{[1]}$ denotes the value of
field of compact \textrm{U(1)} group and $(N_{i}^{\mu})^{[1]}(x_{i}^{\mu})$
denotes the position on $d$-dimensional Cartesian space $\mathrm{C}_{d}$. How
to charaterize the effect of $\hat{U}^{[2]}(\delta \varphi_{i}^{[2]}%
(N_{i}^{[2]}(\phi_{i,\mathrm{global}}^{[1]})))$? $\delta \varphi_{i}%
^{[2]}(N_{i}^{[2]}(\phi_{i,\mathrm{global}}^{[1]}))$ is an extra group
operation on the level-2 zero lattice $\phi_{i,\mathrm{global}}^{[1]}.$
Because we must characterize $\delta \varphi_{i}^{[2]}(\phi_{i,\mathrm{global}%
}^{[1]})$ on $d$-dimensional Cartesian space $\mathrm{C}_{d},$ the group
operation $\hat{U}^{[2]}(\delta \varphi_{i}^{[2]}(N_{i}^{[2]}(\phi
_{i,\mathrm{global}}^{[1]})))$ turns into $\hat{U}^{[2]}(\delta \varphi
_{i}^{[2]},N_{i}^{[2]},(N_{i}^{\mu})^{[1]}(x_{i}^{\mu})))$ where
$\delta \varphi_{i}^{[1]}$ denotes the value of field of compact \textrm{U(1)}
group, $(N_{i}^{\mu})^{[1]}(x_{i}^{\mu})$ denotes the position on
$d$-dimensional Cartesian space $\mathrm{C}_{d},$ and $N_{i}^{[2]}$ denotes
the position in level-2 group changing space. Therefore, to determine an extra
level-2 group-changing elements $\delta \varphi_{i}^{[2]},$ one must know the
phase angle $\delta \varphi_{i}^{[2]},$ the 2-nd order lattice site
$N_{i}^{[2]},$ the 1-st order lattice site $N_{i}^{[1]}(x_{i}).$ So, a usual
field of compact \textrm{U(1)} group cannot fully describe $\delta \varphi
_{i}^{[2]}(N_{i}^{[2]}(\phi_{i,\mathrm{global}}^{[1]}))$. Instead, except for
a local field of a compact \textrm{U(1)} group on level-1 lattice site
$N_{i}^{[1]}(x_{i}),$ we must introduce a \textrm{N}-component local field to
characterize the position of level-2 lattice site $N_{i}^{[2]}.$ Since
different level-2 lattice sites $N_{i}^{[2]}$ on a level-2 lattice site
$N_{i}^{[1]}(x_{i})$ are equivalent to each other, we introduce an
\textrm{N}-component local field obeying a symmetry of compact \textrm{SU(N)}
($\mathrm{N}=\lambda^{\lbrack12]}$) group. As a result, by using both a usual
field of compact \textrm{U(1)} group and a usual \textrm{N}-component field of
compact \textrm{SU(N)} ($\mathrm{N}=\lambda^{\lbrack12]}$) group, we fully
describe $\delta \varphi_{i}^{[2]}(N_{i}^{[2]}(\phi_{i,\mathrm{global}}%
^{[1]}))$. The results can be straightforwardly generalized to the cases of
many group-changing elements of an arbitrary 2-nd order P-variant.

Finally, by using Hybrid-order representation under K-projection for a 2-nd
order P-variant, in continuum limit, the extra level-1 group-changing elements
of $\mathrm{C}_{\mathrm{\tilde{G}}^{[1]}}^{[1]}$ are effectively characterized
by local field of a compact \textrm{U(1)} group, and the extra level-2
group-changing elements of $\mathrm{C}_{\mathrm{\tilde{G}}^{[2]}}^{[2]}$ are
effectively characterized by local field of a compact \textrm{U(1)} group and
a usual \textrm{N}-component field of compact \textrm{SU(N)} ($\mathrm{N}%
=\lambda^{\lbrack12]}$) group. Then, according to higher-order variability,
the two fields of compact \textrm{U(1)} groups for the two group-changing
spaces couple each other and cannot fluctuate individually. In the following
parts, we point out that this will lead to local gauge structures.

\subsubsection{Summary}

In this section, we develop a theory for higher-order variant
$V_{\mathrm{\tilde{G}}^{[1]},\mathrm{\tilde{G}}^{[2]}\mathrm{,}d}^{[2]}$.
Under a special projections (K-projection, or/and D-projection), a
higher-order variant is reduced into usual quantum fields with certain
symmetries. This powerful mathematic theory can help us understand the
non-local structure of quantum gauge field theories.

\subsection{A new theoretical framework for physics -- "\textbf{All from
Changings}"}

In this section, based on higher-order variant, we develop a new framework on
the foundation of quantum gauge field theory including QED and QCD. Different
physical laws emerge for the changings in different levels. We call it
"\emph{Tower of changings}". The base of the tower is 0-th level physics
structure that is the uniform physical variant named "\emph{vacuum}" or
"\emph{ground state}" in usual physics; above it is level-1 physics structure
that is the expansion and contraction types of "\emph{changings}" of the
vacuum, named "\emph{matter}" or physical excited states in usual physics;
above it is level-2 physics structure that is the time-dependent
"\emph{changings}" of the expansion and contraction changings of vacuum, named
"\emph{motion}" in usual physics.

\subsubsection{2--th order $\mathrm{\tilde{S}\tilde{O}}$\textrm{(d+1)}
physical variants: concept and definition}

\emph{What's physical reality in a new theoretical framework of quantum gauge
fields?} In this paper, we point out that for quantum gauge fields, the
physical reality is ($d+1$) dimensional 2-nd order $\mathrm{\tilde{S}\tilde
{O}}$\textrm{(d+1)} physical variant $V_{\mathrm{\tilde{U}}^{[2]}%
\mathrm{(1)},\mathrm{\tilde{S}\tilde{O}}^{[1]}\mathrm{(d+1)},d+1}^{[2]}$, a
predecessor of our spacetime, matter, and gauge fields.

Firstly, we give the definition of ($d+1$)-dimensional 2-nd order
$\mathrm{\tilde{S}\tilde{O}}$\textrm{(d+1)} physical variant
$V_{\mathrm{\tilde{U}}^{[2]}\mathrm{(1)},\mathrm{\tilde{S}\tilde{O}}%
^{[1]}\mathrm{(d+1)},d+1}^{[2]}$:

\textit{Definition -- }($d+1$)\textit{-dimensional} \textit{2-nd order
}$\mathrm{\tilde{S}\tilde{O}}$\textrm{(d+1)}\textit{ physical variants is
defined by a higher-order mapping between }$\mathrm{C}_{\mathrm{\tilde{U}%
}^{[2]}\mathrm{(1)}}^{[2]}$\textit{, $\mathrm{\tilde{S}\tilde{O}}%
$\textrm{(d+1)} Clifford group-changing space }$\mathrm{C}_{\mathrm{\tilde
{S}\tilde{O}(d+1)},d+1}^{[1]}$\textit{\ and a rigid spacetime }$\mathrm{C}%
_{d+1},$\textit{ i.e.,}%
\begin{align}
V_{\mathrm{\tilde{U}}^{[2]}\mathrm{(1)},\mathrm{\tilde{S}\tilde{O}}%
^{[1]}\mathrm{(d+1)},d+1}^{[2]}  &  :\mathrm{C}_{\mathrm{\tilde{U}}%
^{[2]}\mathrm{(1)}}^{[2]}\nonumber \\
&  \Longleftrightarrow \mathrm{C}_{\mathrm{\tilde{S}\tilde{O}}^{[1]}%
\mathrm{(d+1)},d+1}^{[1]}\nonumber \\
&  \Longleftrightarrow \mathrm{C}_{d+1}%
\end{align}
\textit{ where }$\Leftrightarrow$\textit{\ between }$\mathrm{C}%
_{\mathrm{\tilde{U}}^{[2]}\mathrm{(1)}}^{[2]}$ \textit{and }$\mathrm{C}%
_{\mathrm{\tilde{S}\tilde{O}}^{[1]}\mathrm{(d+1)},d+1}^{[1]}$\textit{ denotes
an ordered mapping under fixed ratio }$\lambda^{\lbrack12]}=\left \vert
\frac{\delta \phi^{\lbrack2]}}{\delta \phi_{\mathrm{global}}^{[1]}}\right \vert
$\textit{, }$\Leftrightarrow$\textit{\ between }$\mathrm{C}_{\mathrm{\tilde
{S}\tilde{O}}^{[1]}\mathrm{(d+1)},d+1}^{[1]}$\textit{ and} $\mathrm{C}_{d+1}%
$\textit{ denotes an ordered mapping under fixed changing rate of integer
multiple }$k_{0}$ or $\omega_{0}^{[1]},$\textit{\ and }$\mu$\textit{ labels
the spatial direction}.

Based on above this Hypothesis about 0-th level physics structure, we will
develop a new, complete theoretical framework for quantum gauge theory step by step.

\subsubsection{Higher-order Variability and its physical consequences}

Uniform 2-nd order $\mathrm{\tilde{S}\tilde{O}}$\textrm{(d+1)}\textit{
}physical variant $V_{0,\mathrm{\tilde{U}}^{[2]}\mathrm{(1)},\mathrm{\tilde
{S}\tilde{O}}^{[1]}\mathrm{(d+1)},d+1}^{[2]}$ is a complex uniform changing
structure on Cartesian space. To characterize the uniform 2-nd order physical
variant, we consider its higher-order variability of the two group-changing
spaces and then show its physical consequences.

Firstly, we discuss level-1 variability of level-1 group-changing space of a
uniform 2-nd order $\mathrm{\tilde{S}\tilde{O}}$\textrm{(d+1)}\textit{
}physical variant $V_{0,\mathrm{\tilde{U}}^{[2]}\mathrm{(1)},\mathrm{\tilde
{S}\tilde{O}}^{[1]}\mathrm{(d+1)},d+1}^{[2]}.$

\emph{Level-1 spatial variability}: There exists variability along an
arbitrary spatial direction of the level-1 group-changing space $\mathrm{C}%
_{\mathrm{\tilde{S}\tilde{O}}^{[1]}\mathrm{(d+1)},d+1}^{[1]}$, i.e.,
\begin{align}
\mathcal{T}(\delta x^{i})  &  \rightarrow \hat{U}^{[1]}((\delta \phi^{i}%
)^{[1]})=e^{i\cdot(\delta \phi^{i})^{[1]}\Gamma^{i}},\text{ }\nonumber \\
i  &  =x_{1},x_{2},\text{...},x_{d},
\end{align}
where $(\delta \phi^{i})^{[1]}=k_{0}\delta x^{i}$ and $\Gamma^{i}$ are the
Gamma matrices obeying Clifford algebra $\{ \Gamma^{i},\Gamma^{i}%
\}=2\delta^{ij}.$ This level-1 spatial variability is relevant to the Lorentz
invariance of quantum physics;

\emph{Level-1 tempo variability}: There exists a variability along time
direction of the level-1 group-changing space $\mathrm{C}_{\mathrm{\tilde
{S}\tilde{O}}^{[1]}\mathrm{(d+1)},d+1}^{[1]}$, i.e.,%

\begin{equation}
\mathcal{T}(\delta t)\rightarrow \hat{U}^{[1]}((\delta \phi^{i})^{[1]}%
)=e^{i\cdot(\delta \phi^{i})^{[1]}\Gamma^{t}},
\end{equation}
where $(\delta \phi^{i})^{[1]}=\omega_{0}^{[1]}\delta t$ and $\Gamma^{t}$ is
another Gamma matrix anticommuting with $\Gamma^{i},$ $\{ \Gamma^{i}%
,\Gamma^{t}\}=2\delta^{it}$. In particular, the system with 1-st order
variability along time direction also indicates a regular motion of the
level-1 group-changing space $\mathrm{C}_{\mathrm{\tilde{S}\tilde{O}}%
^{[1]}\mathrm{(d+1)},d+1}^{[1]}$ along $\Gamma^{t}$ direction. This level-1
tempo variability is relevant to the quantization in quantum mechanics;

\emph{Level-1 rotation variability}: There exists a rotation variability of
the level-1 group-changing space $\mathrm{C}_{\mathrm{\tilde{S}\tilde{O}%
}^{[1]}\mathrm{(d+1)},d+1}^{[1]}$ that is defined by
\begin{equation}
\hat{U}^{\mathrm{R}}\rightarrow \hat{R}_{\mathrm{space}}%
\end{equation}
where $\hat{U}^{\mathrm{R}}$ is \textrm{SO(d+1)} rotation operator on Clifford
group-changing space $\hat{U}^{\mathrm{R}}\Gamma^{I}\mathbf{(}\hat
{U}^{\mathrm{R}})^{-1}=\Gamma^{I^{\prime}},$ and $\hat{R}_{\mathrm{space}}$ is
\textrm{SO(d+1)} rotation operator on Cartesian space, $\hat{R}%
_{\mathrm{space}}x^{I}\hat{R}_{\mathrm{space}}^{-1}=x^{I^{\prime}}$. This
level-1 rotation variability is relevant to the spin dynamics and curved space
in modern physics.

Next, we discuss level-2 variability of level-2 group-changing space
$\mathrm{C}_{\mathrm{U(1)}^{[2]},1}^{[2]}.$

\emph{Level-2 "spatial" variability}: There exists a "spatial" variability of
the level-2 group-changing space $\mathrm{C}_{\mathrm{U(1)}^{[2]},1}^{[2]}$,
i.e.,%
\begin{equation}
\hat{U}^{[1]}(\delta \phi_{\mathrm{global}}^{[1]})\leftrightarrow \hat{U}%
^{[2]}(\delta \phi^{\lbrack2]})=\exp(i\lambda^{\lbrack12]}\delta \phi
_{\mathrm{global}}^{[1]}),
\end{equation}
where $\delta \phi_{\mathrm{global}}^{[1]}=\sqrt{%
%TCIMACRO{\dsum \limits_{\mu}}%
%BeginExpansion
{\displaystyle \sum \limits_{\mu}}
%EndExpansion
((\delta \phi^{\mu})^{[1]})^{2}}$. This level-2 "spatial" variability is
relevant to the gauge invariance of quantum physics;

\emph{Level-2 tempo variability}: There exists a tempo variability along time
direction of the level-2 group-changing space $\mathrm{C}_{\mathrm{U(1)}%
^{[2]},1}^{[2]}$, i.e.,%

\begin{equation}
\mathcal{T}(\delta t)\rightarrow \hat{U}^{[2]}(\delta \phi^{\lbrack
2]})=e^{i\delta \phi^{\lbrack2]}},
\end{equation}
where $\delta \phi^{\lbrack2]}=\omega_{0}^{[2]}\delta t$. In particular, the
system with level-2 tempo variability indicates a regular motion of the
level-2 group-changing space $\mathrm{C}_{\mathrm{U(1)}^{[2]},1}^{[2]}$. This
level-2 tempo variability is relevant to the coupling constant of quantum
gauge theory.

\subsubsection{Classification of matter}

Next, we develop theory about level-1 physics structure by classifying the
different types of matter that correspond to different types of globally
expand or contract the two group-changing spaces $\mathrm{C}_{\mathrm{\tilde
{S}\tilde{O}}^{[1]}\mathrm{(d+1)},d+1}^{[1]}$ or $\mathrm{C}_{\mathrm{U(1)}%
^{[2]},1}^{[2]}$ by changing their corresponding sizes. In general, different
types of matter are quantized -- the unit of $\mathrm{C}_{\mathrm{\tilde
{S}\tilde{O}}^{[1]}\mathrm{(d+1)},d+1}^{[1]}$ is fermionic elementary particle
and the unit of $\mathrm{C}_{\mathrm{U(1)}^{[2]},1}^{[2]}$ is color charge.

Firstly, we discuss the matter of level-1 group-changing space $\mathrm{C}%
_{\mathrm{\tilde{S}\tilde{O}}^{[1]}\mathrm{(d+1)},d+1}^{[1]}.$

Under D-projection, we have a translation symmetry protected knot/link along
spatial-tempo direction. This is new type of knot/links in higher dimensional
spacetime (not only higher dimensional space, but also time). Under
K-projection along different directions on spacetime, we have a (3+1)D uniform
zero lattice. The crossing number of a knot/link is an invariant that is the
minimal number of crossings in all planar diagrams of that knot. In
particular, the crossing number is double of linking number for two curved
lines. Therefore, due to the topological character of zeroes, the zero's
number classifies the different topological equivalence classes of level-1
group-changing space $\mathrm{C}_{\mathrm{\tilde{S}\tilde{O}}^{[1]}%
\mathrm{(d+1)},d+1}^{[1]}$ in a P-variant with many extra group-changing
elements. We point out that each zero corresponds to an elementary particle
and becomes the changing unit (or information unit) for of level-1
group-changing space $\mathrm{C}_{\mathrm{\tilde{S}\tilde{O}}^{[1]}%
\mathrm{(d+1)},d+1}^{[1]}$.

In addition, by introducing the group-changing subspace $\mathrm{C}%
_{\mathrm{\tilde{U}(1)\in \tilde{S}\tilde{O}}^{[1]}\mathrm{(d+1)},1}(\Delta
\phi_{\mathrm{global}}^{[1]})$ about global phase changing of the system
$\Delta \phi_{\mathrm{global}}^{[1]}=\sqrt{%
%TCIMACRO{\dsum \limits_{\mu}}%
%BeginExpansion
{\displaystyle \sum \limits_{\mu}}
%EndExpansion
(\Delta \phi^{\lbrack1],\mu}(x))^{2}},$ the zero of $\mathrm{C}_{\mathrm{\tilde
{U}(1)\in \tilde{S}\tilde{O}}^{[1]}\mathrm{(d+1)},1}(\Delta \phi
_{\mathrm{global}}^{[1]})$ characterizes changing unit (or information unit)
for of level-1 group-changing space $\mathrm{C}_{\mathrm{\tilde{S}\tilde{O}%
}^{[1]}\mathrm{(d+1)},d+1}^{[1]}.$ For the case of a system with extra $N_{F}$
zero, the total phase changings of all phase-changing elements $\delta
\phi_{\mathrm{global,}i}^{[1]}(x_{i})$ are equal to $\pm N_{F}\pi,$ i.e., $%
%TCIMACRO{\dsum \nolimits_{i}}%
%BeginExpansion
{\displaystyle \sum \nolimits_{i}}
%EndExpansion
\delta \phi_{\mathrm{global,}i}^{[1]}(x_{i})=\pm n^{[1]}\pi$.

Next, we discuss the matter of level-2 group-changing space $\mathrm{C}%
_{\mathrm{\tilde{U}(1)}^{[2]},1}^{[2]}.$

Under K-projection for $\mathrm{C}_{\mathrm{\tilde{U}}^{[2]}\mathrm{(1)}%
}^{[2]}$ on a level-1 zero of $\mathrm{C}_{\mathrm{\tilde{S}\tilde{O}}%
^{[1]}\mathrm{(d+1)},d+1}^{[1]}$, we have a level-2 zero lattice, of which the
lattice number is $\lambda^{\lbrack12]}$. For the case of a system with extra
$n^{[2]}$ level-2 zeroes, the total phase changings of phase-changing elements
$\delta \phi_{i}^{[2]}(x_{i})$ are equal to $\pm n^{[2]}\pi,$ i.e., $%
%TCIMACRO{\dsum \nolimits_{i}}%
%BeginExpansion
{\displaystyle \sum \nolimits_{i}}
%EndExpansion
\delta \phi_{i}^{[2]}(x_{i})=\pm n^{[2]}\pi$. Each level-2 zero corresponds to
a color charge of local field of \textrm{SU(}$\lambda^{\lbrack12]}$\textrm{)}
gauge symmetry.

We point out that the numbers $n^{[1]}$ and $n^{[2]}$ fully determine the
types of matter for a 2-nd order $\mathrm{\tilde{S}\tilde{O}}$\textrm{(d+1)}%
\textit{ }physical variant $V_{\mathrm{\tilde{U}}^{[2]}\mathrm{(1)}%
,\mathrm{\tilde{S}\tilde{O}}^{[1]}\mathrm{(d+1)},d+1}^{[2]}$. Due to the
level-2 "spatial" variability, the changings of the two group-changing spaces
lock, i.e.,
\begin{equation}
\lambda^{\lbrack12]}\delta \phi_{\mathrm{global}}^{[1]}\equiv \delta
\phi_{\mathrm{global}}^{[2]}=\delta \phi^{\lbrack2]}.
\end{equation}
Under compactification,\ we have a constraint on the changings on the compact
phase angles, i.e.,
\begin{equation}
\lambda^{\lbrack12]}\delta \varphi_{\mathrm{global}}^{[1]}\equiv \delta
\varphi^{\lbrack2]}%
\end{equation}
where $\delta \varphi_{\mathrm{global}}^{[1]}\in(0,2\pi]$, $\delta
\varphi^{\lbrack2]}\in(0,2\pi]$.

Therefore, for an elementary particle (an level-1 zero) with $n$ level-2
zeroes ($\lambda^{\lbrack12]}>n>0$), if the phase changing of the level-1 zero
(elementary particle) is $\delta \phi_{\mathrm{global}}^{[1]},$ the phase
changing of each level-2 zero is $\delta \phi_{\mathrm{global}}^{[1]}$ and the
total phase changing of the level-2 zeroes is $\delta \phi_{\mathrm{global}%
}^{[2]}=n^{[2]}\delta \phi_{\mathrm{global}}^{[1]}.$ For the case of
$n^{[2]}=\lambda^{\lbrack12]},$ we have $\delta \phi_{\mathrm{global}}%
^{[2]}=\lambda^{\lbrack12]}\delta \phi_{\mathrm{global}}^{[1]}$. We set an
electric charge for the case of $n^{[2]}=\lambda^{\lbrack12]}$ to be unit. A
level-2 zero has $1/\lambda^{\lbrack12]}$ electric charge and\ an elementary
particle (an level-1 zero) with $n^{[2]}$ level-2 zeroes ($\lambda
^{\lbrack12]}>n>0$) has $\frac{n^{[2]}}{\lambda^{\lbrack12]}}$ electric charge.

In summary, there are two types of matter of 2-nd order $\mathrm{\tilde
{S}\tilde{O}}$\textrm{(d+1)}\textit{ }physical variant $V_{\mathrm{\tilde{U}%
}^{[2]}\mathrm{(1)},\mathrm{\tilde{S}\tilde{O}}^{[1]}\mathrm{(d+1)},d+1}%
^{[2]}$ -- fermionic elementary particles corresponding to level-1 zeroes, and
color charges corresponding to level-2 zeroes, i.e.,
\begin{align*}
\text{Level-1 zero}  &  \rightarrow \text{A fermionic elementary particle,}\\
\text{Level-2 zero }  &  \rightarrow \text{A color charge with }\\
&  \frac{1}{\lambda^{\lbrack12]}}\text{ electric charge.}%
\end{align*}

\subsubsection{Classification of motions}

In this section, we develop theory about level-2 physics structure by
classifying the types of motion that corresponds to different types of
time-dependent changings of 2-nd order physical variants $V_{\mathrm{\tilde
{U}}^{[2]}\mathrm{(1)},\mathrm{\tilde{S}\tilde{O}}^{[1]}\mathrm{(d+1)}%
,d+1}^{[2]}$ without changings the size of the group-changing spaces. This is
a changings of mappings between two group-changing spaces $\mathrm{C}%
_{\mathrm{\tilde{U}}^{[2]}\mathrm{(1)}}^{[2]}$ and\textit{ }$\mathrm{C}%
_{\mathrm{\tilde{S}\tilde{O}}^{[1]}\mathrm{(d+1)},d+1}^{[1]}$. In particular,
for different types of motions, the size of the two group-changing spaces
$\mathrm{C}_{\mathrm{\tilde{U}}^{[2]}\mathrm{(1)}}^{[2]}$ and\textit{
}$\mathrm{C}_{\mathrm{\tilde{S}\tilde{O}}^{[1]}\mathrm{(d+1)},d+1}^{[1]}$ will
never change.

In above section, we pointed out that globally expand/contract of the two
group-changing spaces corresponds to the generation/annihilate of elementary
particles or color charges in quantum physics. In this part, we point out that
locally expand/contract of two group-changing space corresponds to the motion
of particles in quantum mechanics with fixed particle's number. There exist
two kinds of motions: one is the motion of matter that are extra zeroes (the
situation looks like the motion of itinerant electrons in solid physics), the
other is the collective motion of the two types of zero lattices without extra
zeroes (the situation looks like the collective motions of a lattice in solid
physics). Let us show the results.

Firstly, we discuss the motions of level-1 zeroes that corresponds to the
locally expand/contract of level-1 group-changing space $\mathrm{C}%
_{\mathrm{\tilde{S}\tilde{O}}^{[1]}\mathrm{(d+1)},d+1}^{[1]}$. There are two
kinds of motions, one is the motion of elementary particles, the other is the
collective motion of the level-1 zero lattice without defects.

On the one hand, the motion of elementary particles had been discussed in
detail in the paper of \cite{kou1}. In Cartesian space\textit{ }%
$\mathrm{C}_{d+1},$ an elementary particle can be divided into a group of
group-changing elements. The evolution of distribution of ordered level-1
group-changing elements of an elementary particle in Cartesian space
$\mathrm{C}_{d+1}$ are quantum motions in quantum mechanics! Different
distribution of group-changing elements of the elementary particle are
different states of quantum motions of particles. It is wave-function that
describes the different distribution of group-changing elements of the
elementary particle and characterize the changing of mapping between level-1
group-changing space $\mathrm{C}_{\mathrm{\tilde{S}\tilde{O}}^{[1]}%
\mathrm{(d+1)},d+1}^{[1]}$ and Cartesian space\textit{ }$\mathrm{C}_{d+1}$.

On the other hand, the collective motion of the level-1 zero lattice without
defects is gravitational wave. This issue is beyond the scope of this paper
and will be discussed in Ref\cite{kou1}.

Next, we discuss the motions of level-2 zeroes that corresponds to the locally
expand/contract of level-2 group-changing space $\mathrm{C}_{\mathrm{\tilde
{U}}^{[2]}\mathrm{(1)}}^{[2]}$. There are two kinds of motions, one is the
motion of extra level-2 zeroes (or color charges), the other is the collective
motions of the level-2 zero lattice without defects.

On the one hand, there exists the motion of extra level-2 zeroes (or color
charges) inside an elementary particle (a level-1 zero). The motion of level-2
zeroes is then described by "quantum mechanics" inside an elementary particle.
This is an effective model on a 1D lattice with $\lambda^{\lbrack12]}$ sites.
Then, the "wave-function" of level-2 zeroes describes the different
distributions of level-2 group-changing elements inside an elementary particle
and characterizes the changing of mapping between the two group-changing
spaces $\mathrm{C}_{\mathrm{\tilde{U}}^{[2]}\mathrm{(1)}}^{[2]}$ and
$\mathrm{C}_{\mathrm{\tilde{S}\tilde{O}}^{[1]}\mathrm{(d+1)},d+1}^{[1]}$.

On the other hand, there are two types of the collective motions of the
level-2 zero lattice, one is about the global motion of the level-2 zeroes
inside a level-1 zero, the other is about relative motion of the level-2
zeroes inside a level-1 zero. The first type and second type of collective
modes are similar to the acoustic phonons and optical phonons in composite
crystals, respectively. In particular, the collective motion of level-2
group-changing space is described by the gauge fields on a rigid
spacetime\textit{ }$\mathrm{C}_{d+1}:$ The global motion of the level-2 zeroes
corresponds to the fluctuations of quantum fields of \textrm{U}$^{\mathrm{em}%
}$\textrm{(1)} gauge symmetry, and the relative motion of the level-2 zeroes
corresponds to the fluctuations of quantum fields of \textrm{SU(}%
$\lambda^{\lbrack12]}$\textrm{).}

\subsubsection{Invariance from variability}

In above section, we point out that physical laws emerge from changings.
Different levels of changings have different physical laws. However, it was
known invariance/symmetry plays important role in physics. \emph{How
invariance/symmetry emerges from variability?} In this section, we will answer
this question and give an instructive discussion on invariance/symmetry in
modern physics.

According to above discussion, there are three levels of changings in physics.
We then discuss the invariances/symmetries for each levels of changings one by one.

\paragraph{Level-0 Invariant: The fixity of physical constants}

For the level-0 physics, we have a uniform 2-nd order physical variant
$V_{0,\mathrm{\tilde{U}}^{[2]}\mathrm{(1)},\mathrm{\tilde{S}\tilde{O}}%
^{[1]}\mathrm{(d+1)},d+1}^{[2]}$ with 2-nd order variability. The
\emph{fixity} of changing rates between different group-changing spaces is
known to be level-0 invariant. The \emph{fixity} indicates a invariant of
physical laws (Lorentz invariant, and quantization condition, gauge
interaction, Schr\"{o}dinger equation). The specific manifestation of
invariance is the fixity of physical constants, such as light speed $c$,
Planck constant $\hslash$, the coupling constant for electromagnetic
interaction $e$, the coupling constant for strong interaction, ... In other
words, all these physical constants don't change with time and place. So, we
say that such an invariance (or fixity) is protected by the 2-nd order
variability. This can be regarded as a higher-order generalization of
Noether's theorem.

Firstly, we consider light speed $c$ and the Lorentz invariant from spatial
variability of level-1 group-changing space $\mathrm{C}_{\mathrm{\tilde
{S}\tilde{O}}^{[1]}\mathrm{(d+1)},d+1}^{[1]}$. According to above discussion,
there exists a fixed spatial changing rate for level-1 group-changing space
$\mathrm{C}_{\mathrm{\tilde{S}\tilde{O}}^{[1]}\mathrm{(d+1)},d+1}^{[1]}$,
i.e., $k_{0}=(\delta \phi^{i})^{[1]}/\delta x^{i}\neq0$. The direct physical
consequence is linear dispersion relation and an emergent Lorentz invariant.
In general, we may assume the dispersion of the system is a smooth function,
such as $\omega^{\lbrack1]}(k).$ After linearization at $k=k_{0}$, we have
$\omega^{\lbrack1]}=\omega_{0}^{[1]}+c(k-k_{0})$. Consequently, an effective
"light" velocity, i.e., $c=\frac{\partial \omega^{\lbrack1]}}{\partial k}%
\mid_{k=k_{0}}$;

Seconly, we consider Planck constant $\hslash$ from tempo variability (or a
uniform motion) of level-1 group-changing space $\mathrm{C}_{\mathrm{\tilde
{S}\tilde{O}}^{[1]}\mathrm{(d+1)},d+1}^{[1]}$. Because the energy of the
physical variant always has a uniformly distribution, the energy density
$\rho_{E}=\frac{\Delta E}{\Delta V}$ must be constant. In addition, we assume
that $\rho_{E}(\omega_{0}^{[1]})$ is a smooth function of $\omega_{0}$. Then,
we have
\begin{align}
\rho_{E}(\omega_{0}^{[1]}+\delta \omega^{\lbrack1]})  &  =\rho_{E}(\omega
_{0}^{[1]})\nonumber \\
+\frac{\delta \rho_{E}}{\delta \omega^{\lbrack1]}}  &  \mid_{\omega^{\lbrack
1]}=\omega_{0}^{[1]}}\delta \omega^{\lbrack1]}+...
\end{align}
where $\frac{\delta \rho_{E}}{\delta \omega^{\lbrack1]}}\mid_{\omega^{\lbrack
1]}=\omega_{0}^{[1]}}=\rho_{J}^{E}$ is called the density of (effective)
"angular momentum". $\frac{\delta \rho_{E}}{\delta \omega}\mid_{\omega
^{\lbrack1]}=\omega_{0}^{[1]}}=\rho_{J}.$ The "angular momentum" $\rho_{J}$
for an elementary particle is just Planck constant $\hbar$ and the
quantization condition in quantum mechanics come from the linearization of
energy density $\rho_{E}$ via $\omega^{\lbrack1]}$ near $\omega_{0}^{[1]}$;

Thridly, we consider the coupling constant for electromagnetic interaction $e$
from tempo variability (or a uniform motion) of level-2 group-changing space
$\mathrm{C}_{\mathrm{U(1)}^{[2]},1}^{[2]}$. Because the energy of the physical
variant always has a uniformly distribution, the energy density $\rho
_{E}=\frac{\Delta E}{\Delta V}$ must be constant. In addition, we assume that
$\rho_{E}(\omega_{0}^{[2]})$ is a smooth function of $\omega_{0}^{[2]}$. Then,
we have
\begin{align}
\rho_{E}(\omega_{0}^{[2]}+\delta \omega^{\lbrack2]})  &  =\rho_{E}(\omega
_{0}^{[2]})\nonumber \\
+\frac{\delta \rho_{E}}{\delta \omega^{\lbrack2]}}  &  \mid_{\omega^{\lbrack
2]}=\omega_{0}^{[2]}}\delta \omega^{\lbrack2]}+...
\end{align}
where $\frac{\delta \rho_{E}}{\delta \omega^{\lbrack2]}}\mid_{\omega^{\lbrack
2]}=\omega_{0}^{[2]}}=\rho_{J}^{E}$ is called the density of (effective)
"angular momentum". $\frac{\delta \rho_{E}}{\delta \omega}\mid_{\omega
^{\lbrack2]}=\omega_{0}^{[2]}}=\rho_{J}.$ The "angular momentum" $\rho_{J}$
for level-2 zero determines the coupling constant for electromagnetic
interaction $\mathrm{e}.$ In other words, the existence of electromagnetic
interaction $\mathrm{e}$ comes from the linearization of energy density
$\rho_{E}$ via $\omega^{\lbrack2]}$ near $\omega_{0}^{[2]}$.

\paragraph{Level-1 Invariant: Topology stationarity}

Next, we discuss the invariant of level-1 physics structure for matter.

From point view of 2-nd order physical variant $V_{\mathrm{\tilde{U}}%
^{[2]}\mathrm{(1)},\mathrm{\tilde{S}\tilde{O}}^{[1]}\mathrm{(d+1)},d+1}^{[2]}%
$, the total sizes of the two group-changing spaces $\mathrm{C}%
_{\mathrm{\tilde{U}}^{[2]}\mathrm{(1)}}^{[2]}$ and $\mathrm{C}_{\mathrm{\tilde
{S}\tilde{O}}^{[1]}\mathrm{(d+1)},d+1}^{[1]}$ are all topological invariables.
Because the motions in quantum physics are changing the mapping between
$\mathrm{C}_{\mathrm{\tilde{U}}^{[2]}\mathrm{(1)}}^{[2]}$, $\mathrm{C}%
_{\mathrm{\tilde{S}\tilde{O}}^{[1]}\mathrm{(d+1)},d+1}^{[1]}$, and
$\mathrm{C}_{d+1},$ during the processes of motion, the total sizes of the two
group-changing spaces don't change obviously. Therefore, we call this type of
invariant to be topology stationarity.

It was known that matter corresponds to globally expand or contract of the two
group-changing spaces, $\mathrm{C}_{\mathrm{\tilde{S}\tilde{O}}^{[1]}%
\mathrm{(d+1)},d+1}^{[1]}$ or $\mathrm{C}_{\mathrm{U(1)}^{[2]},1}^{[2]}$ with
changing their corresponding sizes. For example, elementary particles with
different numbers of color charges are different types of matter. Therefore,
there exists an invariant for matter, i.e., the type of particles is never
changed. A moving electron cannot be changed into a quark without considering
interaction. We call such a invariance to be \emph{topology stationarity} of matter.

\paragraph{Level-2 Invariant: The symmetry of motion}

Finally, we discuss the invariant of level-2 physics structure for motions.

It was known that motion corresponds to locally expand or contract of the two
group-changing spaces, $\mathrm{C}_{\mathrm{\tilde{S}\tilde{O}}^{[1]}%
\mathrm{(d+1)},d+1}^{[1]}$ or $\mathrm{C}_{\mathrm{U(1)}^{[2]},1}^{[2]}$
without changing their corresponding sizes. Different states of motions
correspond to different mappings between $\mathrm{C}_{\mathrm{\tilde{U}}%
^{[2]}\mathrm{(1)}}^{[2]}$, $\mathrm{C}_{\mathrm{\tilde{S}\tilde{O}}%
^{[1]}\mathrm{(d+1)},d+1}^{[1]}$, and $\mathrm{C}_{d+1}$. If two states (or
different mappings between $\mathrm{C}_{\mathrm{\tilde{U}}^{[2]}\mathrm{(1)}%
}^{[2]}$, $\mathrm{C}_{\mathrm{\tilde{S}\tilde{O}}^{[1]}\mathrm{(d+1)}%
,d+1}^{[1]}$, and $\mathrm{C}_{d+1}$) have \emph{same energy}, we call such a
invariance to be \emph{symmetry} of motions.

Firstly, we discuss the symmetry for a level-1 group-changing space of a 2-nd
order physical U-variant that comes from invariant during a globally shifting
of the whole system.

According to the level-1 variability, under a globally shifting of the whole
system, we have
\begin{align}
\mathcal{T}(\delta x^{\mu})  &  \leftrightarrow \hat{U}^{[1]}((\delta \phi^{\mu
})^{[1]})=\exp(i(T^{\mu})^{[1]}(\delta \phi^{\mu})^{[1]})\\
&  =\exp(i(T^{\mu})^{[1]}(k_{0}^{\mu}\delta x^{\mu})).\nonumber
\end{align}
For simplify, we denote it by $\mathcal{T}\leftrightarrow \hat{U}^{[1]}.$ As a
result, the energy of whole system doesn't change.

We then do compactification on the system.

On the one hand, under compactification, the continuous translation operation
$\mathcal{T}(\delta x^{\mu})$ of the U-variant is reduced into a discrete
translation symmetry $T(\delta x^{\mu})$ on the level-1 zero lattice, i.e.,%
\[
\mathcal{T}(\delta x^{\mu})\leftrightarrow \tilde{T}(\delta N^{\mu,[1]}).
\]
For level-1 zero lattice, one lattice site is equivalence to another. Then,
for the uniform level-1 zero lattice, we have a reduced translation symmetry
denoted by the following equation
\[
\tilde{T}(\delta N^{\mu})\leftrightarrow1.
\]

On the other hand, under compactification, the operation $\hat{U}%
^{[1]}((\delta \phi^{\mu})^{[1]})$ of non-compact $\mathrm{\tilde{S}\tilde{O}%
}^{[1]}\mathrm{(d+1)}$ group is reduced to the operation $\hat{U}%
^{[1]}((\delta \varphi^{\mu})^{[1]})$ of compact $\mathrm{U}_{\mathrm{global}%
}^{[1]}\mathrm{(1)\otimes SO}^{[1]}$\textrm{(d+1)} group. On each lattice site
of level-1 zero lattice, we have an invariant under global phase operation
$\hat{U}_{\mathrm{U}_{\mathrm{global}}^{[1]}\mathrm{(1)}}^{[1]}(1)$ and global
compact \textrm{SO}$^{[1]}$\textrm{(d+1)} operation $\hat{U}_{\mathrm{SO}%
^{[1]}\mathrm{(d+1)}}^{[1]}$, i.e.,
\begin{equation}
\hat{U}^{\mu,[1]}\rightarrow \hat{U}_{\mathrm{U}_{\mathrm{global}}%
^{[1]}\mathrm{(1)}}^{[1]}(1)\times \hat{U}_{\mathrm{SO}^{[1]}\mathrm{(d+1)}%
}^{[1]}.
\end{equation}
For simplicity, we can denote it by the following equations
\begin{align}
U_{\mathrm{U}_{\mathrm{global}}^{[1]}\mathrm{(1)}}^{[1]}(1)  &
\leftrightarrow1\nonumber \\
\hat{U}_{\mathrm{SO}^{[1]}\mathrm{(d+1)}}^{[1]}  &  \leftrightarrow1.
\end{align}

Therefore, $\tilde{T}(\delta N^{\mu,[1]})\leftrightarrow1$ indicates that the
system is invariance under a global translation on level-1 zero lattice;
global phase symmetry $\hat{U}_{\mathrm{U}_{\mathrm{global}}^{[1]}%
\mathrm{(1)}}^{[1]}(\delta \phi_{\mathrm{global}}^{[1]})\leftrightarrow1$
indicates that the system is invariance under the global phase rotation on all
level-1 zero lattice; global $\mathrm{SO}^{[1]}\mathrm{(d+1)}$ rotation
symmetry $\hat{U}_{\mathrm{SO}^{[1]}\mathrm{(d+1)}}^{[1]}\leftrightarrow1$
indicates that the system is invariance under a global spacetime rotation.

As a result, we unify the global translation symmetry, global phase symmetry
and global rotation symmetry into level-1 variability!

Next, we discuss the symmetry for a level-2 group-changing space of a 2-nd
order physical U-variant.

First of all, this symmetry for a level-2 group-changing space of a 2-nd order
physical U-variant is a local symmetry. Here, the word "\emph{local}" means
the symmetry comes from the invariance for an operation on a level-1 zero,
rather a global operation on the whole system.

Under level-1 compactification $\hat{U}^{[1]}(\delta \phi_{\mathrm{global}%
}^{[1]})\rightarrow \hat{U}_{I}^{[1]}(\delta \varphi_{I,\mathrm{global}}^{[1]}%
)$, we consider the projected level-2 variability on a level-1 zero,
\begin{equation}
\hat{U}_{I}^{[1]}(\delta \varphi_{I,\mathrm{global}}^{[1]})\leftrightarrow
\hat{U}_{I}^{[2]}(\delta \phi_{I}^{[2]})
\end{equation}
where the index $I$ denotes $I$-th level-1 zero and $\varphi
_{I,\mathrm{global}}^{[1]}\in(0,2\pi]$. For simplify, we denote it by $\hat
{U}_{I}^{[1]}\leftrightarrow \hat{U}_{I}^{[2]}.$

According to the level-2 variability $\hat{U}_{I}^{[1]}\leftrightarrow \hat
{U}_{I}^{[2]}$, we do compactification and get a level-2 zero lattice with a
compact \textrm{U(1)} field.

On the one hand, under compactification, the continuous translation operation
$\hat{U}_{I}^{[1]}(\delta \varphi_{I,\mathrm{global}}^{[1]})$ is reduced into a
discrete translation operation $\hat{U}_{I}^{[1]}(\delta N_{I}^{[2]})$ on the
level-2 zero lattice, i.e.,%
\[
\hat{U}_{I}^{[1]}(\delta \varphi_{I,\mathrm{global}}^{[1]})\leftrightarrow
\hat{U}_{I}^{[1]}(\delta N_{I}^{[2]}).
\]
For level-2 zero lattice, one lattice site is equivalence to another. Then, we
have a reduced discrete $\mathrm{Z}_{N}$ ($N=\lambda^{\lbrack1]}$) symmetry
denoted by the following equation
\[
\hat{U}_{I}^{[1]}(\delta N_{I}^{[2]})\leftrightarrow1.
\]
This will lead to local \textrm{SU(N)} gauge symmetry.

On the other hand, under compactification, the operation $\hat{U}_{I}%
^{[2]}(\delta \phi_{I}^{[2]})$ of non-compact $\mathrm{\tilde{U}(1)}$ group is
reduced to the operation $\hat{U}_{I}^{[2]}(\delta \varphi_{I}^{[2]})$ for a
global compact \textrm{U}$^{[2]}$\textrm{(1)} group. On each lattice site of
level-2 zero lattice, we have an invariant under the global compact
\textrm{U}$^{[2]}$\textrm{(1)} group,
\[
\hat{U}_{I}^{[2]}(\delta \varphi^{\lbrack2]})\leftrightarrow1.
\]

However, due to $\hat{U}_{I}^{[1]}\leftrightarrow \hat{U}_{I}^{[2]},$ the
operation $\hat{U}_{I}^{[1]}(\delta \varphi_{I,\mathrm{global}}^{[1]})$ and the
operation $\hat{U}_{I}^{[2]}(\delta \varphi_{I}^{[2]})$ are not independent
each other. The symmetry from global compact \textrm{U}$_{\mathrm{global}%
}^{[1]}$\textrm{(1)} group and that from global compact \textrm{U}$^{[2]}%
$\textrm{(1)} group couple unify into a new local \textrm{U(1)} gauge
symmetry, i.e., $\hat{U}_{I}^{[2]}(\delta \varphi_{I}^{[2]})\leftrightarrow
\hat{U}_{I}^{[1]}(\delta \varphi_{I,\mathrm{global}}^{[1]}).$ This new local
\textrm{U(1)} gauge symmetry is just the \textrm{U}$^{\mathrm{em}}%
$\textrm{(1)} gauge symmetry for electromagnetic field, i.e.,
\[
\hat{U}_{I}^{[2]}(\delta \varphi^{\lbrack2]})=\hat{U}_{I}^{[1]}(\delta
\varphi_{\mathrm{global}}^{[1]})\leftrightarrow1.
\]

As a result, we unify the local \textrm{U}$^{\mathrm{em}}$\textrm{(1)} gauge
symmetry and the local \textrm{SU(N)} gauge symmetry into level-2 variability!

We give a brief summary.

On the one hand, the \emph{level-1} variability is reduced to different kinds
of \emph{global} symmetries: one is about spatial translation symmetry, the
other is about the rotating symmetry of compact \textrm{U}$_{\mathrm{global}%
}^{[1]}$\textrm{(1)} group and $\mathrm{SO}^{[1]}$\textrm{(d+1)} group.
Different conserved quantities are physical consequences of the level-1
variability of the original (uniform) 2-nd order variant: the energy/momentum
$E/p$ becomes a conserved quantity; the particle number $N$ becomes a
conserved quantity; the angular momentum becomes a conserved quantity.

On the other hand, the \emph{level-2} variability is reduced to different
kinds of \emph{gauge} symmetries: one is about local \textrm{U}$^{\mathrm{em}%
}$\textrm{(1)} gauge symmetry, the other is about local \textrm{SU(N)} gauge
symmetry. Therefore, with considering local \textrm{U}$^{\mathrm{em}}%
$\textrm{(1)} gauge symmetry, we have the theory for electromagnetic
interaction; with considering the local \textrm{SU(N)} gauge symmetry, we have
the theory for strong interaction. Different gauge theories are physical
consequences of the level-2 variability of the original (uniform) 2-nd order variant.

\paragraph{Summary -- Invariance as \emph{Shadow} of Variability}

In summary, we say that invariance can be regarded as \emph{shadow} of
variability: 0-level invariance (or fixity) determines the invariance of
physical laws with fixed physical constants; 1-level invariance (or topology
stationarity) determines the invariance of the matter under motion; 2-level
invariance (or symmetry) determines the invariance of motions. In addition,
the global symmetry comes from the level-1 variability and the local gauge
symmetry comes from the level-2 variability.

\subsection{N=0 case -- Dirac model}

In this section, we consider the $N=0$ case. Now, we have a 1-st order
physical variant.

\subsubsection{Physical variant}

Physical variant is a ($d+1$)-dimensional $\mathrm{\tilde{S}\tilde{O}}%
$\textrm{(d+1)} physical variants $V_{\mathrm{\tilde{S}\tilde{O}(d+1)}%
,d+1}(\Delta \phi^{\mu},\Delta x^{\mu},k_{0},\omega_{0})$ that is described by
a mapping between Clifford group-changing space $C_{\mathrm{\tilde{S}\tilde
{O}(d+1)},d+1}$\ to a rigid spacetime $\mathrm{C}_{d+1}$\textit{. }Here, we
have $d=3$. The 1-st order variability is
\begin{align}
\mathcal{T}(\delta x^{\mu})  &  \leftrightarrow \hat{U}((\delta \phi^{\mu
}))=\exp(i(T^{\mu}\delta \phi^{\mu}))\\
&  =\exp(i(T^{\mu}k_{0}^{\mu}\delta x^{\mu})).\nonumber
\end{align}
\begin{figure}[ptb]
\includegraphics[clip,width=0.7\textwidth]{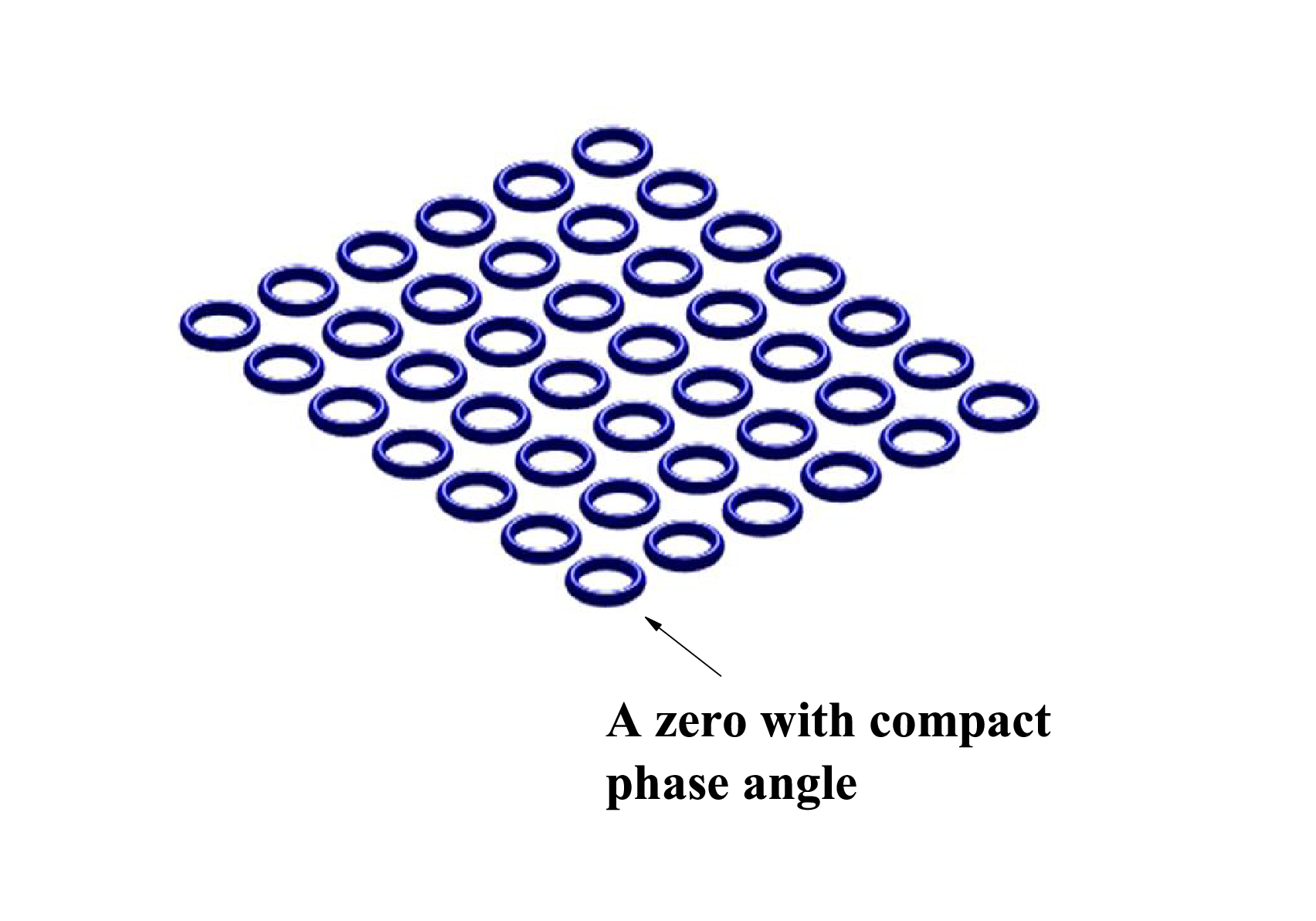}\caption{An illustration
of microscopic picture for Dirac model: a crystal of rings. Each ring
corresponds to a lattice site of the topological lattice with two zeroes (or
two elementary elementary particles). The symmetry of the rings indicates the
symmetry of the compact \textrm{U}$_{\mathrm{global}}$\textrm{(1)} group}%
\end{figure}

Under compactification, we have a topological lattice, of which there exist
compact \textrm{U}$_{\mathrm{global}}$\textrm{(1)} group and compact
\textrm{SO(d+1)} group on each lattice site. Because the compact
\textrm{SO(d+1)} group is relevant to curved spacetime, we focus on the
physical processes of compact \textrm{U}$_{\mathrm{global}}$\textrm{(1)}
group. As a result, the physical picture for ($d+1$)-dimensional
$\mathrm{\tilde{S}\tilde{O}}$\textrm{(d+1)} physical variants
$V_{\mathrm{\tilde{S}\tilde{O}(d+1)},d+1}(\Delta \phi^{\mu},\Delta x^{\mu
},k_{0},\omega_{0})$ with $N=0$ becomes a crystal of rings. See the
illustration in Fig.28. Each ring on each lattice site of the topological
lattice (with two zeroes) characterizes the $2\pi$ phase space $\varphi
_{\mathrm{global}}.$ The symmetry of the rings indicates the symmetry of the
compact \textrm{U}$_{\mathrm{global}}$\textrm{(1)} group. Sometimes, we may
use abbreviation $\varphi_{\mathrm{global}}$ by $\varphi$ and \textrm{U}%
$_{\mathrm{global}}$\textrm{(1)} group by \textrm{U(1)} group.

\subsubsection{Matter}

Matter comes from changings of group-changing space. Under K-projection, we
have the zero equation along $\mu$-th direction as $\cos(\phi(x^{^{\mu}%
})-\theta)=0,$ of which the zero solution is given by $k_{0}(x^{^{\mu}}%
-x_{0}^{^{\mu}})-\theta=\pm \frac{\pi}{2}$. An elementary particle is zero in a
($d+1$)-dimensional $\mathrm{\tilde{S}\tilde{O}}$\textrm{(d+1)} physical
variants $V_{\mathrm{\tilde{S}\tilde{O}(d+1)},d+1}(\Delta \phi^{\mu},\Delta
x^{\mu},k_{0},\omega_{0})$ under K-projection. Therefore, particle is basic
block of spacetime and the spacetime is really a multi-particle system and
made of matter. When there exists an additional zero corresponding to an
elementary, the periodic boundary condition of systems along arbitrary
direction is changed into anti-periodic boundary condition. That means the
elementary particle obeys fermionic statistics.

For an elementary particle of 1D system, the function is given by
$\mathrm{z}(x)=e^{i\phi(x)}$ where
\begin{equation}
\phi(x)=\left \{
\begin{array}
[c]{c}%
\phi_{0}\mp \frac{\pi}{2},\text{ }x\in(-\infty,x_{0}]\\
\phi_{0}\mp \frac{\pi}{2}\pm k_{0}(x-x_{0}),\text{ }x\in(x_{0},x_{0}+a]\\
\phi_{0}\pm \frac{\pi}{2},\text{ }x\in(x_{0}+a,\infty)
\end{array}
\right \}
\end{equation}
where $+$ ($\pm k_{0}(x-x_{0})$) denotes a clockwise winding and $-$ denotes a
counterclockwise winding. There is a linear relationship between $\phi(x)$ and
$x$ as $\phi(x)\propto x-x_{0}$ in the winding region of $x_{0}<x\leq x_{0}%
+a$. Thus, we obtain an anti-periodic boundary condition for the system,
\begin{equation}
\phi(x\rightarrow \infty)-\phi(x\rightarrow-\infty)=\pm \pi.
\end{equation}

Finally, we discuss the invariant of level-1 physics structure for matter.

It was known that matter corresponds to globally expand or contract of the
group-changing space, $\mathrm{C}_{\mathrm{\tilde{S}\tilde{O}(d+1)},d+1}$ with
$\pi$-phase changing along an arbitrary direction. Therefore, there exists an
invariant for matter, i.e., the statistics of an elementary particle is never
changed. We call such a invariance to be topology stationarity of matter. The
topology stationarity of matter comes from a fact that the total size of the
group-changing space $\mathrm{C}_{\mathrm{\tilde{S}\tilde{O}(d+1)},d+1}$ is a
topological invariable. During the processes of motion, the total size of the
group-changing space don't change.

\subsubsection{Motion}

In general, we can use quantum field theory to characterize the dynamics of
matter. To derive a quantum field theory for variant, we must do
compactification. Now, the ground state of the variant is uniform and its
dynamics can be regarded as that of a compact \textrm{U(1)} field. Then, we
firstly do compactification and derive its "field" description.

At first step, we do \emph{compactification}. After compactification, we have
a topological lattice, that is characterized by $\phi^{\mu}(x)=2\pi N^{\mu
}(x)+\varphi^{\mu}(x)$. We then relabel position in phase space by phase
angles of compact group $\varphi^{\mu}(x)$ and the winding numbers $N^{\mu
}(x).$ The lattice distance is $l_{0}$ on Cartesian space $\mathrm{C}_{3+1}$.
On each lattice site of the topological lattice, there exist compact
\textrm{U}$_{\mathrm{global}}$\textrm{(1)} group and compact \textrm{SO(d+1)} group.

In this paper, we focus on the changes from internal compact \textrm{U}%
$_{\mathrm{global}}$\textrm{(1)} group. The physical picture for
($d+1$)-dimensional $\mathrm{\tilde{S}\tilde{O}}$\textrm{(d+1)} physical
variants $V_{\mathrm{\tilde{S}\tilde{O}(d+1)},d+1}(\Delta \phi^{\mu},\Delta
x^{\mu},k_{0},\omega_{0})$ with $N=0$ becomes a crystal of rings. Each ring on
each lattice site of the topological lattice (with two zeroes) characterizes
the $2\pi$ phase space $\varphi_{\mathrm{global}}=\sqrt{%
%TCIMACRO{\dsum \limits_{\mu}}%
%BeginExpansion
{\displaystyle \sum \limits_{\mu}}
%EndExpansion
(\varphi^{\mu}(x))^{2}}.$ Sometimes, we may use abbreviation $\varphi
_{\mathrm{global}}$ by $\varphi$ and \textrm{U}$_{\mathrm{global}}%
$\textrm{(1)} group by \textrm{U(1)} group.

According to above discussion, it is $\psi(x,t)$ on each lattice site that
characterizes the dynamics of matter. In this part, the effective Hamiltonian
of the elementary particle is derived as Dirac model.

We firstly define generation operator of elementary particle $c_{i}^{\dagger
}\left \vert 0\right \rangle =\left \vert i\right \rangle $. We then write down
the hopping Hamiltonian. The hopping term between two nearest neighbor sites
$i$ and $j$ on topological lattice becomes
\begin{equation}
\mathcal{H}_{\left \{  i,j\right \}  }=Jc_{i}^{\dagger}(t)\mathbf{T}_{\left \{
i,j\right \}  }c_{j}(t)
\end{equation}
where $\mathbf{T}_{\left \{  i,j\right \}  }$ is the transfer matrix between two
nearest neighbor sites $i$ and $j$ and $c_{i}(t)$ is the annihilation operator
of elementary particle at the site $i$. $J=\frac{c}{2l_{p}}$ is an effective
coupling constant between two nearest-neighbor sites. $l_{p}=l_{0}/2$ is
Planck length and $c$ is light speed. According to hybrid symmetry,
$\left \vert i\right \rangle =e^{il_{p}(\hat{k}^{\mu}\cdot \Gamma^{\mu}%
)}\left \vert j\right \rangle ,$ the transfer matrix $\mathbf{T}_{\left \{
i,j\right \}  }$ between $\left \vert i\right \rangle $ and $\left \vert
j\right \rangle $ is defined by $\mathbf{T}_{\left \{  i,j\right \}  }%
=e^{il_{p}(\hat{k}^{\mu}\cdot \Gamma^{\mu})}.$ After considering the
contribution of the terms from all sites, the effective Hamiltonian is
obtained as%
\begin{equation}
\mathcal{H}=%
%TCIMACRO{\dsum \limits_{\{i,j\}}}%
%BeginExpansion
{\displaystyle \sum \limits_{\{i,j\}}}
%EndExpansion
\mathcal{H}_{\left \{  i,j\right \}  }=J%
%TCIMACRO{\dsum \limits_{\{i,j\}}}%
%BeginExpansion
{\displaystyle \sum \limits_{\{i,j\}}}
%EndExpansion
c_{i}^{\dagger}\mathbf{T}_{\left \{  i,j\right \}  }c_{j}+h.c..
\end{equation}

In continuum limit, we have%
\begin{align}
\mathcal{H}  &  =J%
%TCIMACRO{\dsum \limits_{\{i,j\}}}%
%BeginExpansion
{\displaystyle \sum \limits_{\{i,j\}}}
%EndExpansion
c_{i}^{\dagger}(e^{il_{p}(\hat{k}^{\mu}\cdot \Gamma^{\mu})})c_{i+1}+h.c.\\
&  =2l_{p}J%
%TCIMACRO{\dsum \limits_{\mu}}%
%BeginExpansion
{\displaystyle \sum \limits_{\mu}}
%EndExpansion%
%TCIMACRO{\dsum \limits_{k^{\mu}}}%
%BeginExpansion
{\displaystyle \sum \limits_{k^{\mu}}}
%EndExpansion
c_{k^{\mu}}^{\dagger}[\cos(k^{\mu}\cdot \Gamma^{\mu})]c_{k^{\mu}}%
\end{align}
where%
\begin{align}
\Gamma^{t}  &  =\tau^{z}\otimes \vec{1}\mathbf{,}\text{ }\Gamma^{x}=\tau
^{x}\otimes \sigma^{x},\\
\Gamma^{y}  &  =\tau^{x}\otimes \sigma^{y},\text{ }\Gamma^{z}=\tau^{x}%
\otimes \sigma^{z}.\nonumber
\end{align}
The dispersion in long wave length limit around $\omega=\omega_{0}=\frac{\pi
}{2}\frac{1}{l_{p}}c$ and $\vec{k}=\vec{k}_{0}$ is
\begin{equation}
E_{k}\simeq \pm c\sqrt{[(\vec{k}-\vec{k}_{0})\cdot \vec{\Gamma}]^{2}%
+((\omega-\omega_{0}^{\ast})\cdot \Gamma^{t})^{2}},
\end{equation}
where $\vec{k}_{0}=\frac{1}{l_{p}}(\frac{\pi}{2},\frac{\pi}{2},\frac{\pi}%
{2}).$ The mass of the Dirac particles is $m=\hbar(\omega_{0}^{\ast}%
-\omega_{0})/c^{2}$ where $\omega_{0}=ck_{0}$.

After doing \emph{continuation}, in long wave limit, we replace the discrete
numbers $N(x)$ by continuous coordinate $x=2l_{p}\cdot N(x)=l_{0}N(x),$ and
get a "field" description for the \textrm{U(1)} "field", $N(x)\rightarrow x.$

Finally, we find that under compactification the fluctuation of ($d+1$%
)-dimensional $\mathrm{\tilde{S}\tilde{O}}$\textrm{(d+1)} physical variants
$V_{\mathrm{\tilde{S}\tilde{O}(d+1)},d+1}(\Delta \phi^{\mu},\Delta x^{\mu
},k_{0},\omega_{0})$ becomes the field of massless/massive Dirac particles.

In paticular, the mass comes from extra changing rate along tempo direction
$m=\frac{\hbar}{c^{2}}\delta \omega=\frac{\hbar}{c^{2}}(\omega_{0}^{\ast
}-\omega_{0})$. Therefore, for a Dirac model with finite mass, the U-variant
is slightly changed along tempo direction, i.e.,
\begin{align}
&  V_{0,\mathrm{\tilde{S}\tilde{O}(d+1)},d+1}[\Delta \phi^{\mu},\Delta x^{\mu
},(k_{0}^{i},\omega_{0})]\nonumber \\
&  \rightarrow V_{0^{\prime},\mathrm{\tilde{S}\tilde{O}(d+1)},d+1}[\Delta
\phi^{\mu},\Delta x^{\mu},(k_{0}^{i},\omega_{0}^{\ast})].
\end{align}

\subsection{N=1 case -- QED}

In this section, we consider the $N=1$ case. Now, for the 2-nd order physical
variant, the corresponding 2-nd order variability is reduced into
\textrm{U}$^{\mathrm{em}}$\textrm{(1)} local gauge symmetry and the
corresponding quantum field theory becomes QED.

\subsubsection{Physical variant}

Now, we have a ($d+1$)-dimensional 2-nd order\textit{ }$\mathrm{\tilde
{S}\tilde{O}}$\textrm{(d+1)}, a higher-order mapping between\textit{
}$\mathrm{C}_{\mathrm{\tilde{U}}^{[2]}\mathrm{(1)}}^{[2]}$\textit{,
$\mathrm{\tilde{S}\tilde{O}}$}\textrm{(d+1)} Clifford group-changing
space\textit{ }$\mathrm{C}_{\mathrm{\tilde{S}\tilde{O}(d+1)},d+1}^{[1]}%
$\textit{\ }and a rigid spacetime\textit{ }$\mathrm{C}_{d+1}$\textit{.} The
ratio between the changing rates of the different levels is fixed to be $1$,
\[
\lambda^{\lbrack12]}=\left \vert \frac{\delta \phi^{\lbrack2]}}{\delta
\phi_{\mathrm{global}}^{[1]}}\right \vert \equiv1.
\]
The 2-nd order variability is
\begin{align}
\mathcal{T}(\delta x^{\mu})  &  \leftrightarrow \hat{U}^{[1]}((\delta
\phi^{\lbrack1]\mu}))\nonumber \\
&  =\exp(i(T^{\mu}\delta \phi^{\lbrack1]\mu}))\\
&  =\exp(i(T^{\mu}k_{0}^{\mu}\delta x^{\mu})),\nonumber
\end{align}
and%
\begin{equation}
\hat{U}_{I}^{[1]}(\delta \phi_{I,\mathrm{global}}^{[1]})\leftrightarrow \hat
{U}_{I}^{[2]}(\delta \phi_{I}^{[2]})
\end{equation}
For simplify, we denote it by $\mathcal{T}\leftrightarrow \hat{U}%
^{[1]}\leftrightarrow \hat{U}^{[1]}$ with $\lambda^{\lbrack12]}=1$.

Under K-projection, we have level-1 zero lattice $k_{0}(x^{^{\mu}}%
-x_{0}^{^{\mu}})-\theta=\pm \frac{\pi}{2}$ and no level-2 zero lattice. Now,
each level-1 (a fermionic elementary particle) zero corresponds to a level-2 zero.

Therefore, in Fig.29, we show a physical picture for the physical variant: On
each zero lattice, there exists two coupled rings. One ring denotes the
\textrm{U}$^{[2]}$\textrm{(1)} group space of $\varphi^{\lbrack2]};$ the other
denotes the \textrm{U}$^{[1]}$\textrm{(1)} group space of $\varphi^{\lbrack
1]}=\delta \varphi_{\mathrm{global}}^{[1]}$. The symmetries of rings indicate
the symmetrise of the two compact groups (\textrm{U}$^{[2]}$\textrm{(1)} and
\textrm{U}$^{[1]}$\textrm{(1)}). Due to the existence of the topological
constraint $\varphi^{\lbrack1]}=\delta \varphi_{\mathrm{global}}^{[1]}%
\equiv \delta \varphi^{\lbrack2]},$ the global compact \textrm{U}$^{[1]}%
$\textrm{(1)} group and global compact \textrm{U}$^{[2]}$\textrm{(1)} group
couple. To characterize this topological constraint, the two rings on a
lattice form a knot/link with unit linking number. Then, under
compactification, the physical picture for a 2-nd order physical variants
$V_{\mathrm{\tilde{U}}^{[2]}\mathrm{(1)},\mathrm{\tilde{S}\tilde{O}}%
^{[1]}\mathrm{(d+1)},d+1}^{[2]}$ with changing rate $\lambda^{\lbrack12]}=1$
becomes a crystal of knot/link!\begin{figure}[ptb]
\includegraphics[clip,width=0.7\textwidth]{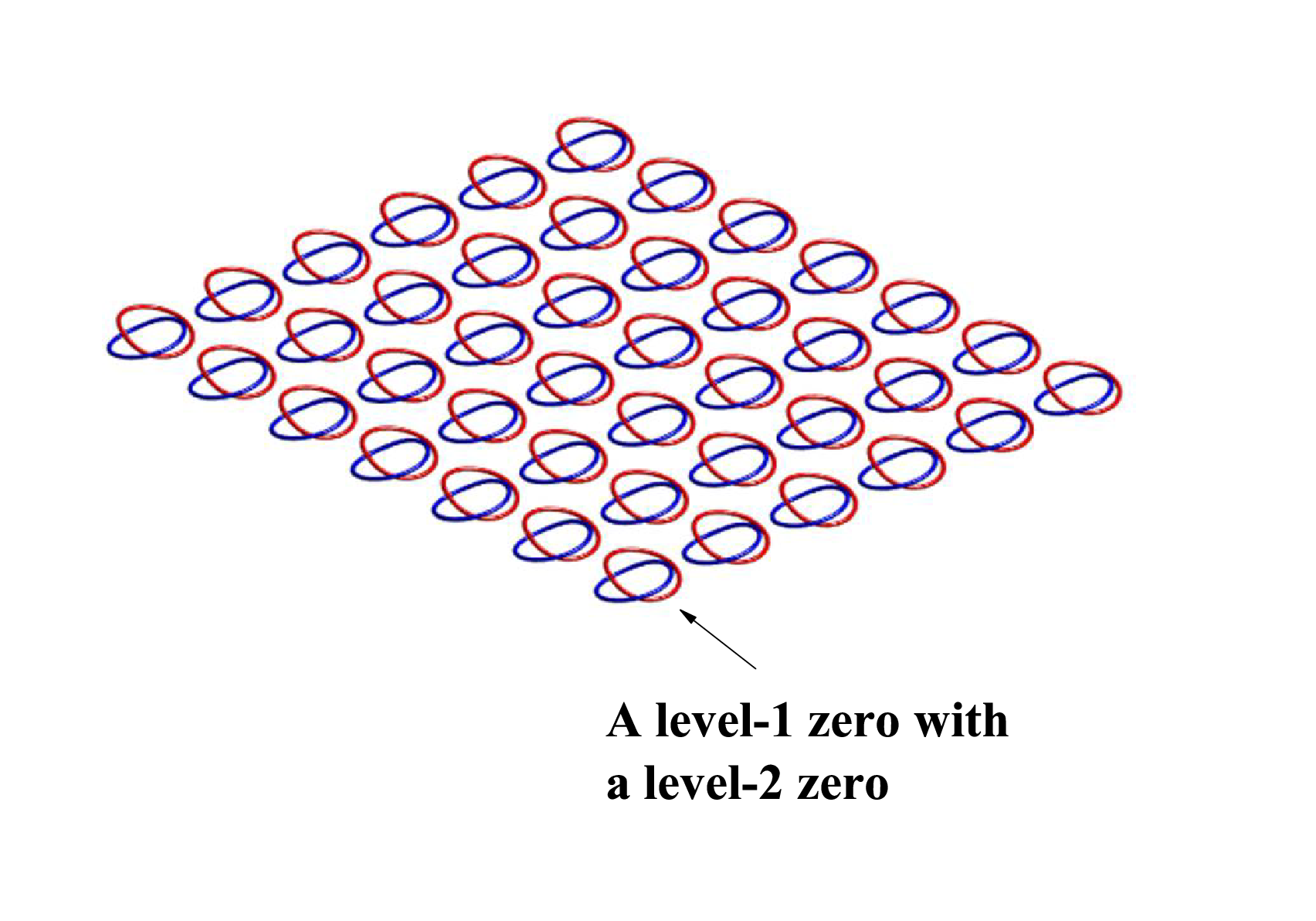}\caption{An illustration
of microscopic picture for QED: a crystal of two coupled rings (or
knot/links). Each ring corresponds to a lattice site of the topological
lattice with two (level-2 or level-1) zeroes. The symmetries of rings indicate
the symmetrise of the two compact groups (\textrm{U}$^{[2]}$\textrm{(1)} and
\textrm{U}$^{[1]}$\textrm{(1)}).}%
\end{figure}

\subsubsection{Matter}

Next, we discuss the matter for 2-nd order physical variants
$V_{\mathrm{\tilde{U}}^{[2]}\mathrm{(1)},\mathrm{\tilde{S}\tilde{O}}%
^{[1]}\mathrm{(d+1)},d+1}^{[2]}$ with changing rate $\lambda^{\lbrack12]}=1.$

Matter corresponds to globally expanding or contracting the two group-changing
spaces $\mathrm{C}_{\mathrm{\tilde{S}\tilde{O}}^{[1]}\mathrm{(d+1)},d+1}%
^{[1]}$ or $\mathrm{C}_{\mathrm{U(1)}^{[2]},1}^{[2]}$ with changing their
corresponding sizes. Therefore, for this system, the matter becomes the
elementary particle with a level-1 zero that is the unit of $\mathrm{C}%
_{\mathrm{\tilde{S}\tilde{O}}^{[1]}\mathrm{(d+1)},d+1}^{[1]}$ and a level-2
zero that is the unit of $\mathrm{C}_{\mathrm{U(1)}^{[2]},1}^{[2]}.$ A level-2
zero corresponds to an electric charge. The changing of both level-1 and
level-2 group-changing spaces are $\pi$. For this case, the color charge is meaningless.

Then, we show the detailed internal structure of an electric charge.

Each level-2 zero corresponds to a $\pi$-phase changing of level-2
group-changing space. Therefore, we have
\begin{equation}
\mathrm{z}^{[2]}(\phi^{\lbrack2]})\sim e^{i\phi^{\lbrack2]}(\phi
_{\mathrm{global}}^{[1]})}%
\end{equation}
where
\begin{equation}
\phi^{\lbrack2]}(\phi_{\mathrm{global}}^{[1]})=\left \{
\begin{array}
[c]{c}%
\phi_{0}^{[2]}\mp \frac{\pi}{2},\text{ }\phi_{\mathrm{global}}^{[1]}\in
(-\infty,\phi_{0}^{[2]}]\\
\mp \frac{\pi}{2}\pm \lambda^{\lbrack12]}(\phi_{\mathrm{global}}^{[1]}-\phi
_{0}^{[2]}),\text{ }\phi_{\mathrm{global}}^{[1]}\in(\phi_{0}^{[2]},\phi
_{0}^{[2]}+\pi/\lambda^{\lbrack12]}]\\
\pm \frac{\pi}{2},\text{ }\phi_{\mathrm{global}}^{[1]}\in(\pi/\lambda
^{\lbrack12]},\infty)
\end{array}
\right \}
\end{equation}
where $+$ denotes a clockwise winding and $-$ denotes a counterclockwise
winding. There is a linear relationship between $\phi^{\lbrack2]}$ and
$\phi_{\mathrm{global}}^{[1]}$ as $\phi^{\lbrack2]}\propto \lambda^{\lbrack
12]}(\phi_{\mathrm{global}}^{[1]}-\phi_{0}^{[2]})$ in the winding region of
$-\pi/2<\phi^{\lbrack2]}\leq \pi/2$. Thus, we obtain an anti-periodic boundary
condition on phase changing space,
\begin{equation}
\phi^{\lbrack2]}(\phi_{\mathrm{global}}^{[1]}\rightarrow \infty)-\phi
^{\lbrack2]}(\phi_{\mathrm{global}}^{[1]}\rightarrow-\infty)=\pm \pi.
\end{equation}

In addition, we discuss the invariant of level-1 physics structure for matter.

It was known that matter corresponds to globally expand or contract of the two
group-changing spaces, $\mathrm{C}_{\mathrm{\tilde{S}\tilde{O}}^{[1]}%
\mathrm{(d+1)},d+1}^{[1]}$ or $\mathrm{C}_{\mathrm{U(1)}^{[2]},1}^{[2]}$ with
$\pi$-phase changings. Under the changings, the matter becomes an elementary
particle with unit electric charge. The particle number is determined by the
number of level-1 zeros and electric charge number is determined by the number
of level-2 zeros.

Therefore, there exists an invariant for matter, i.e., the statistics of an
elementary particle and its electric charge are all invariant. We call such a
invariance to be topology stationarity of matter. The topology stationarity of
matter comes from a fact that the total sizes of the group-changing spaces
$\mathrm{C}_{\mathrm{\tilde{S}\tilde{O}}^{[1]}\mathrm{(d+1)},d+1}^{[1]}$ and
$\mathrm{C}_{\mathrm{U(1)}^{[2]},1}^{[2]}$ are all topological invariables.
During the processes of motion, the total size of the group-changing spaces
don't change.

\subsubsection{Motion}

There are two types motion -- one is about the motion for elementary
particles, the other is about the collective modes of the two zero lattices.
The collective modes of level-1 zero lattice are gravitational waves that are
described by Einstein's equation. The motion of the elementary particle is
described by Dirac model. So, we focus on the collective modes of the level-2
zero lattice and use quantum \textrm{U(1)} gauge field theory to characterize
its dynamics.

\paragraph{Internal quantum states and internal motion}

Firstly, we define the quantum states for the level-2 group-changing elements
of a level-1 zero.

Under compactification, the non-compact level-2 group-changing space turns
into the space of compact \textrm{U(1)} phase. We can "generate" an extra
level-2 group-changing elements $\delta \varphi_{I,i}^{[2]}(\varphi_{I,i}%
^{[1]})$ on a given position $\varphi_{I,i}^{[1]}$ by a level-2 group-changing
operation
\[
\tilde{U}(\delta \varphi_{I,i}^{[2]}(\varphi_{I,i}^{[1]}))=e^{i((\delta
\varphi_{I,i}^{[2]})\cdot \hat{K})}%
\]
where $\hat{K}=-i\frac{d}{d\varphi^{\lbrack2]}}$. Here, we have used
abbreviation $\varphi_{\mathrm{global}}^{[1]}$ by $\varphi^{\lbrack1]}$ and
\textrm{U}$_{\mathrm{global}}$\textrm{(1)} group by \textrm{U(1)} group. In
this part, we always use the abbreviation.

The motion of level-2 group-changing space comes from its local expansion and
contraction on level-1 zero lattice. Therefore, the motion for level-2
group-changing space is defined by the changings of the configuration of
level-2 phase factor $\varphi_{I}^{[2]}$ on different level-1 zeroes.

To define quantum states for the level-2 group-changing elements, we firstly
define the references of phases of both level-2 and level-1 group-changing
space on a given level-1 zero (for example, $I$-th), i.e., $\varphi
_{0,I}^{[1]}$ and $\varphi_{0,I}^{[2]}$. For different level-1 zeroes, the
reference of $\varphi_{0,I}^{[1]}$ (or the reference of $\varphi_{0,I}^{[2]}$)
may be differently defined. However, according to level-2 variability
$\tilde{U}_{I}^{[1]}(\delta \phi_{I}^{[1]})\leftrightarrow \hat{U}_{I}%
^{[2]}(\delta \phi_{I}^{[2]}),$ the reference of $\varphi_{0,I}^{[1]}$ and the
reference of $\varphi_{0,I}^{[2]}$ must be chosen \emph{synchronously}. Let us
explain the fact in detail: For a given level-1 zero, if one change the
reference of level-1 group-changing space from $\varphi_{0,I}^{[1]}$ to
$(\varphi_{0,I}^{[1]})^{\prime}$, the reference of level-2 group-changing
space $\varphi_{0,I}^{[2]}$ must be changed,
\[
\varphi_{0,I}^{[2]}\rightarrow(\varphi_{0,I}^{[2]})^{\prime}=\varphi
_{0,I}^{[2]}+(\varphi_{0,I}^{[1]})^{\prime}-(\varphi_{0,I}^{[1]})).
\]
Or, we have
\[
\delta \varphi_{0,I}^{[1]}=\delta \varphi_{0,I}^{[2]}%
\]
where $\delta \varphi_{0,I}^{[1]}=((\varphi_{0,I}^{[1]})^{\prime}-\varphi
_{0,I}^{[1]})$ and $\delta \varphi_{0,I}^{[2]}=((\varphi_{0,I}^{[2]})^{\prime
}-\varphi_{0,I}^{[2]})$. In this paper, we call $\delta \varphi_{0,I}%
^{[1]}=\delta \varphi_{0,I}^{[2]}$ to be (global) \emph{variability constraint}.

In the following parts, the locally changings of references will be named to
be gauge transformation for \emph{local }\textrm{U(1)}\emph{\ gauge symmetry}.
For simplicity, in continuum limit, we may denote the operation $\hat
{U}_{\mathrm{U(1)}}(x,t)$ of local\emph{\ }\textrm{U(1)} gauge symmetry by
synchronously changing of references of the \textrm{U(1)} fields of both
levels, i.e.,
\begin{equation}
\hat{U}_{\mathrm{U(1)}}(x,t)=\hat{U}(\delta \varphi_{0}^{[1]}(x))=\hat
{U}(\delta \varphi_{0}^{[2]}(x)).
\end{equation}
This local\emph{\ }\textrm{U(1)} gauge symmetry can be understood by locally
rotating a knot/link on a lattice site with changing the two rings together.

Finally, after defining the references $\varphi_{0,I}^{[1]}$ and
$\varphi_{0,I}^{[2]},$\ we can characterize different internal quantum states
by different distributions of extra level-2 group-changing elements $\tilde
{U}(\delta \varphi_{i}^{[2]})$.

\paragraph{Local \textrm{U(1)} gauge symmetry}

In modern physics, it is known that gauge symmetry appears as the redundancy
to define the particles. According to the variant theory, the redundancy comes
from the references $\delta \varphi_{0}^{[2]}(x)$ of level-2 group-changing
space (or the internal state of level-2 group-changing space of an elementary
particle). Because all physical processes are independent from the selection
of different references, $\delta \varphi_{0}^{[2]}(x)$ are not a physical
observable value. We may locally reorganize the particles by different ways
and get the same result. This becomes the underlying mechanism of the
\textrm{U}$^{\mathrm{em}}$\textrm{(1)} gauge symmetry from indistinguishable
phases inside the elementary particles.

Then, based on the local \textrm{U(1)} gauge transformation $\hat
{U}_{\mathrm{U(1)}}(x,t)$, we discuss the local \textrm{U(1)} gauge symmetry
that belongs to a 2-level invariance/symmetry for motion.

According to above discussion, due to level-2 variability, the changings of
references at position $x$ ($\varphi_{0}^{[1]}(x)$ and $\varphi_{0}^{[2]}(x)$)
must be synchronously changed, $\hat{U}(\delta \varphi_{0}^{[1]}(x))=\hat
{U}(\delta \varphi_{0}^{[2]}(x))$ ($\delta \varphi_{0}^{[1]}(x)=\delta
\varphi_{0}^{[2]}(x)$). However, this is a necessary condition, but not a
sufficient condition. To obey sufficient conditions, we must have
\textrm{U(1)} symmetries for both level-1 group-changing space and level-2
group-changing space on a lattice site of level-1 zero lattice. Let us check
whether there are the \textrm{U(1)} symmetries for both levels.

On the one hand, under compactification, the level-1 variability is reduced to
an invariant/symmetry of the \textrm{U(1)} group on the lattice site of
level-1 zero lattice.\ As a result, the system with constant $\delta
\varphi_{0}^{[1]}$ have same energy; On the other hand, under
compactification, on the lattice site of level-1 zero lattice, level-2
variability is reduced to an invariant/symmetry of the \textrm{U(1)}
group.\ As a result, the system with constant $\delta \varphi_{0}^{[2]}$ have
same energy.

In summary, under variability constraint $\delta \varphi_{0}^{[1]}%
(x)=\delta \varphi_{0}^{[2]}(x),$ the symmetry from \textrm{U}$^{[1]}%
$\textrm{(1)} group and that from \textrm{U}$^{[2]}$\textrm{(1)} group at
different lattice sites of level-1 zero lattice (or position $x$ in continuum
limit) unify into single local \textrm{U(1)} gauge symmetry. This is just the
\textrm{U}$^{\mathrm{em}}$\textrm{(1)} gauge symmetry for electromagnetic
field, i.e.,
\begin{align*}
&  \text{Local \textrm{U}}^{\mathrm{em}}\text{\textrm{(1)} gauge symmetry }\\
&  \text{=}\text{ Global \textrm{U}}^{[1]}\text{\textrm{(1)} group and Global
\textrm{U}}^{[2]}\text{\textrm{(1)} }\\
&  \text{group under variability constraint }%
\end{align*}

\paragraph{U$^{\mathrm{em}}$(1) gauge field}

In this section, we define \textrm{U}$^{\mathrm{em}}$\textrm{(1)} gauge field
that characterizes the quantum fluctuations of level-2 group-changing space.

In physics, the physical changings must be gauge invariant. Or, the physical
changings are independent on the different choosing references. Therefore, the
true physical changings that can be detected by experiments are changings of
the distribution of group-changing elements of level-2 group-changing space on
Cartesian space rather than on level-1 group-changing space. Then, \emph{how
to characterize the changings of the distribution of group-changing elements
of level-2 group-changing space on Cartesian space}? Or, \emph{what's the
physical consequence of the changings of the distribution of group-changing
elements of level-2 group-changing space on Cartesian space}? Let us anser
these questions.

Firstly, we define the \textrm{U}$^{\mathrm{em}}$\textrm{(1)} gauge field
$A_{I,I^{\prime}}$ to be a vector field on links of level-1 zero lattice that
characterizes the difference of phases $\varphi_{I}^{[2]}$ on lattice sites of
level-1 zero lattice, i.e.,
\begin{equation}
\mathrm{e}A_{I,I^{\prime}}=\varphi_{I}^{[2]}-\varphi_{I^{\prime}}^{[2]}.
\end{equation}
The \textrm{U}$^{\mathrm{em}}$\textrm{(1)} gauge field $A_{I,I^{\prime}}$
comes from the non-uniform distribution of level-2 group-changing elements on
different lattices of level-1 zero lattice. $\mathrm{e}$ is the coupling
constant that is calculated in the following parts.

To illustrate the local \textrm{U}$^{\mathrm{em}}$\textrm{(1)} gauge symmetry,
we do a local \textrm{U}$^{\mathrm{em}}$\textrm{(1)} gauge transformation
$U_{\text{\textrm{U}}^{\mathrm{em}}\text{\textrm{(1)}}}$ via changing the
references,%
\begin{equation}
\hat{U}_{\mathrm{U^{\mathrm{em}}(1)}}(x,t)=\hat{U}(\delta \varphi_{0}%
^{[1]}(x))=\hat{U}(\delta \varphi_{0}^{[2]}(x)).
\end{equation}
Under above local \textrm{U}$^{\mathrm{em}}$\textrm{(1)} gauge transformation
on site I of level-1 zero lattice, we have
\begin{equation}
\psi_{I}\rightarrow \psi_{I}^{\prime}=\hat{U}_{\mathrm{U^{\mathrm{em}}(1)}}%
\psi_{I}=e^{-i\varphi_{0,I}^{[1]}}\psi_{I},
\end{equation}
and
\begin{align}
A_{I,I^{\prime}}  &  \rightarrow A_{I,I^{\prime}}^{\prime}\nonumber \\
&  =A_{I,I^{\prime}}-\frac{1}{\mathrm{e}}(\varphi_{0,I}^{[2]}-\varphi
_{0,I^{\prime}}^{[2]}).
\end{align}

Next, we introduce \emph{loop current }to characterize the changings of the
distribution of group-changing elements of level-2 group-changing space on
Cartesian space,
\[
\Phi_{\mathcal{C}}^{[2]}=%
%TCIMACRO{\doint \limits_{\mathcal{C}}}%
%BeginExpansion
{\displaystyle \oint \limits_{\mathcal{C}}}
%EndExpansion
d\varphi^{\lbrack2]}.
\]
For example, if along a loop $\mathcal{C}$, the flux changes from
$\Phi_{\mathcal{C}}^{[2]}\ $to $(\Phi_{\mathcal{C}}^{[2]})^{\prime},$ the
level-2 group-changing space expand or contract. The changing of loop current
along $\mathcal{C}$ is $\delta \Phi_{\mathcal{C}}^{[2]}=(\Phi_{\mathcal{C}%
}^{[2]})^{\prime}-\Phi_{\mathcal{C}}^{[2]}.$

Now, we may use an elementary particle as a test object to detect the
existence of loop current along a given closed loop $\mathcal{C}$. Due to
variability constraint, if moving an elementary particle along $\mathcal{C},$
we have a finite phase changing of $\Delta \varphi^{\lbrack1]}=%
%TCIMACRO{\doint \limits_{\mathcal{C}}}%
%BeginExpansion
{\displaystyle \oint \limits_{\mathcal{C}}}
%EndExpansion
d\varphi^{\lbrack1]}=\Phi_{\mathcal{C}}^{[2]}=%
%TCIMACRO{\doint \limits_{\mathcal{C}}}%
%BeginExpansion
{\displaystyle \oint \limits_{\mathcal{C}}}
%EndExpansion
d\varphi^{\lbrack2]},$ there exists finite loop current along the loop.
Fig.30(a) is an illustration of 2D level-1 zero lattice with uniform global
loop current. Fig.30(b) is an illustration of a 3D level-1 zero with global
loop current on two plaquettes. \begin{figure}[ptb]
\includegraphics[clip,width=0.7\textwidth]{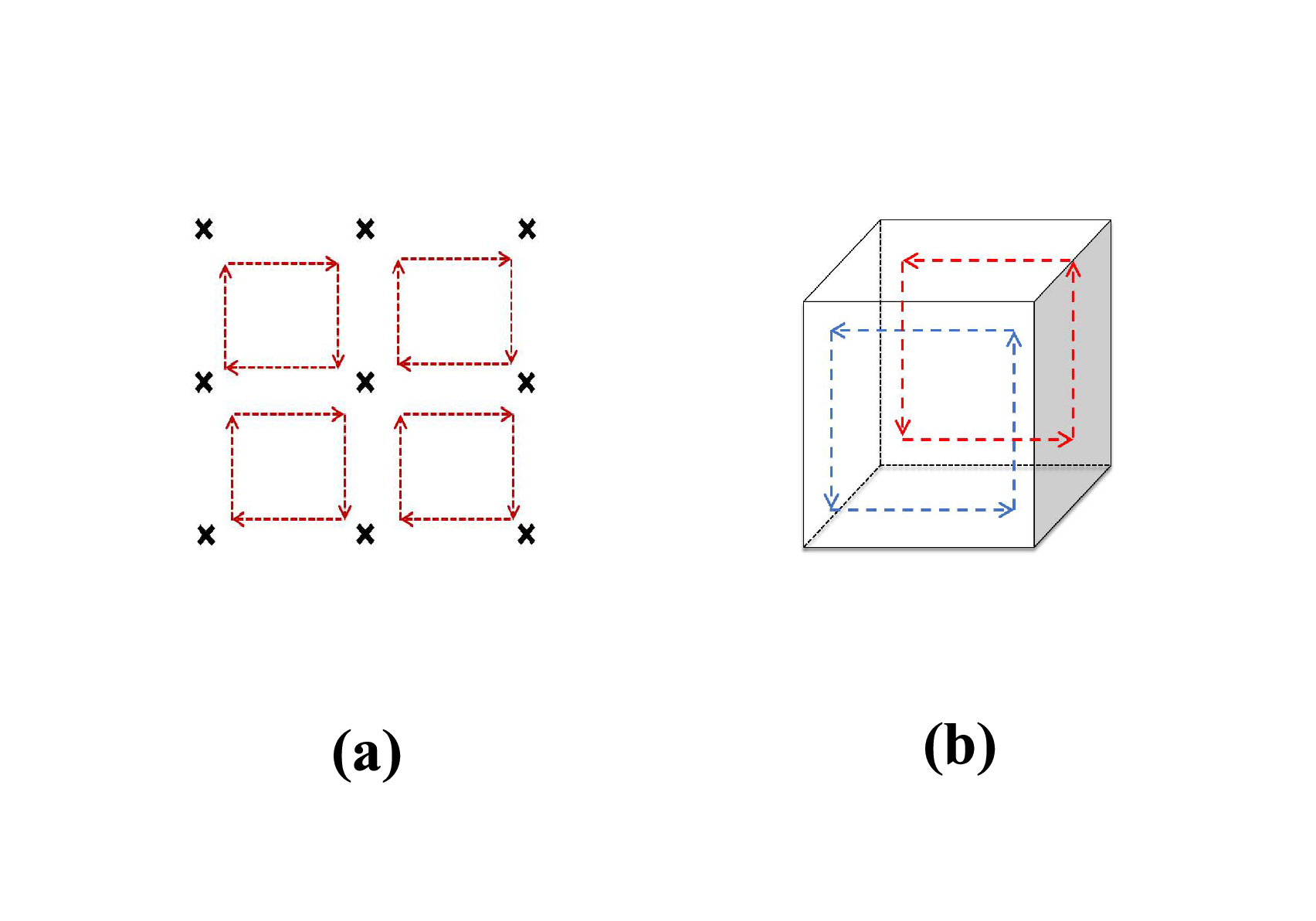}\caption{(a) An
illustration of 2D level-1 zero lattice with uniform global loop current. The
point denotes lattice sites of topological lattice; (b) An illustration of a
3D level-1 zero with global loop current on two plaquettes.}%
\end{figure}

To obtain the distribution of level-2 group-changing space on Cartesian space,
we must know the current along all loops on level-1 zero lattice. A simple
approach is to consider the current along all minimum loops that consist of
four links around four nearest neighbor lattice sites. We denote them by
$\Phi_{IJKL}^{[2]}$ on the plaquette of $IJKL$ lattice sites, i.e.,%
\begin{align}
\Phi_{\left \langle IJKL\right \rangle }^{[2]}  &  =\mathrm{e}A_{IJ}%
-\mathrm{e}A_{KL}\nonumber \\
&  =-(\varphi_{I}^{[2]}-\varphi_{J}^{[2]})+(\varphi_{K}^{[2]}-\varphi
_{L}^{[2]}).
\end{align}
The locally expanding and contracting of level-2 group-changing space are
described by
\[
\{ \Phi_{\left \langle IJKL\right \rangle }^{[2]}(I),\text{ }\left \langle
IJKL\right \rangle \in \mathrm{All}\}.
\]

In continuum limit, we have $\hat{U}_{I,\mathrm{U^{\mathrm{em}}(1)}%
}(t)\rightarrow \hat{U}_{\mathrm{U^{\mathrm{em}}(1)}}(x,t)$, $A_{I,I^{\prime}%
}\rightarrow A_{\mu}(x).$ $\Phi_{\left \langle IJKL\right \rangle }^{[2]}(I)$ is
reduced to the strength of gauge field $F_{\mu \nu},$%
\[
\Phi_{\left \langle IJKL\right \rangle }^{[2]}\sim F_{\mu \nu}.
\]
The $\mathrm{U^{\mathrm{em}}(1)}$ gauge symmetry is represented by%
\begin{equation}
\psi^{\prime}\rightarrow \hat{U}_{\mathrm{U^{\mathrm{em}}(1)}}(x,t)\psi
\end{equation}
and
\begin{align}
A_{\mu}(x,t)  &  \rightarrow A_{\mu}(x,t)\nonumber \\
&  -i\frac{1}{\mathrm{e}}(\partial_{\mu}\hat{U}_{\mathrm{U^{\mathrm{em}}(1)}%
}(x,t))(\hat{U}_{\mathrm{U^{\mathrm{em}}(1)}}(x,t))^{-1}\nonumber \\
&  =A_{\mu}(x,t)-\frac{1}{\mathrm{e}}\partial_{\mu}\varphi^{\lbrack
1]}(x,t)\nonumber \\
&  =A_{\mu}(x,t)-\frac{1}{\mathrm{e}}\partial_{\mu}\varphi^{\lbrack2]}(x,t).
\end{align}
One may easily check the local \textrm{U}$\mathrm{^{\mathrm{em}}}$\textrm{(1)}
gauge invariant for $\Phi_{\left \langle IJKL\right \rangle }^{[2]}$ or
$F_{\mu \nu}$, i.e.,%
\begin{equation}
F_{\mu \nu}\rightarrow F_{\mu \nu}=F_{\mu \nu}.
\end{equation}

\paragraph{Effective Hamiltonian for QED}

The key point for the theory of quantum gauge field is to derive the effective
model. To derive the effective model for quantum gauge fields, we consider two
types of motion of level-2 group-changing space: one for longitudinal motion,
the other for transverse motion.

Firstly, we discuss the longitudinal motion of level-2 group-changing space
that leads to the interaction term between the elementary particles and
\textrm{U}$\mathrm{^{\mathrm{em}}}$\textrm{(1)} gauge fields $A_{\mu}(x,t)$.
We consider phase angle of compact \textrm{U(1)} field for level-2
group-changing at $I$-th level-1 zero to $\varphi_{I}^{[2]}$. The longitudinal
motion for $\varphi_{I}^{[2]}$ will provide a contribute on the hopping term
for elementary particle,
\begin{equation}
J\psi_{I}^{\dagger}T_{I,I^{\prime}}\psi_{I^{\prime}}\rightarrow J\psi
_{I}^{\dagger}e^{-i\varphi_{I}^{[2]}}\cdot T_{I,I^{\prime}}\cdot
e^{i\varphi_{I^{\prime}}^{[2]}}\psi_{I^{\prime}}\nonumber
\end{equation}
where $T_{I,I^{\prime}}$ is transfer operator from $I$-site to $I^{\prime}%
$-site. By introducing a vector field $\mathrm{e}A_{I,I^{\prime}}=-\varphi
_{I}^{[2]}-\varphi_{I^{\prime}}^{[2]},$ the hopping term between level-1
zeroes becomes
\begin{equation}
J\psi_{I}^{\dagger}T_{I,I^{\prime}}e^{i\mathrm{e}A_{I,I^{\prime}}}%
\psi_{I^{\prime}}.
\end{equation}
Then, the total kinetic energy for longitudinal motion becomes
\begin{equation}
\mathcal{\hat{H}}=J%
%TCIMACRO{\dsum \nolimits_{\left\langle I,I^{\prime}\right\rangle }}%
%BeginExpansion
{\displaystyle \sum \nolimits_{\left \langle I,I^{\prime}\right \rangle }}
%EndExpansion
\psi_{I}^{\dagger}T_{I,I^{\prime}}e^{i\mathrm{e}A_{I,I^{\prime}}}%
\psi_{I^{\prime}}+h.c..
\end{equation}
Under local \textrm{U}$\mathrm{^{\mathrm{em}}}$\textrm{(1)}\ transformation,
after considering variability constraint $\varphi_{I}^{[1]}=\varphi_{I}^{[2]}%
$, the Hamiltonian doesn't change, $\mathcal{\hat{H}}=\mathcal{\hat{H}%
}^{\prime}.$

Secondly, we discuss the effective model for transverse motion of level-2
group-changing space.

The transverse motion of level-2 group-changing space comes from the
transverse motion of \textrm{U}$\mathrm{^{\mathrm{em}}}$\textrm{(1)} gauge
field $A_{I,I^{\prime}}.$ From the definition, we point out that loop current
$\Phi_{\left \langle IJKL\right \rangle }^{[2]}$ on the plaquette of $IJKL$
lattice sites characterizes the transverse motion of \textrm{U}%
$\mathrm{^{\mathrm{em}}}$\textrm{(1)} gauge field from $IJ$-link to $KL$-link,
i.e.,
\begin{align}
\Phi_{\left \langle IJKL\right \rangle }^{[2]}  &  =\mathrm{e}A_{IJ}%
-\mathrm{e}A_{KL}\nonumber \\
&  =-\varphi_{I}^{[2]}+\varphi_{J}^{[2]}+\varphi_{K}^{[2]}-\varphi_{L}^{[2]}.
\end{align}
The condition of compactification $\varphi_{I}^{[2]}=\phi_{I}^{[2]}%
\operatorname{mod}2\pi$ leads to a condition of compactification for
$\Phi_{\left \langle IJKL\right \rangle }^{[2]}$. So, the action $\mathcal{S}$
for $\Phi_{\left \langle IJKL\right \rangle }^{[2]}$ obeys the condition of
compactification
\[
\mathcal{S}(\Delta \Phi_{\left \langle ijkl\right \rangle }^{[2]})=\mathcal{S}%
(\Delta \Phi_{\left \langle ijkl\right \rangle }^{[2]}+2\pi).
\]
After considering the Hermitian condition, we have
\[
\mathcal{S}(\Delta \Phi_{\left \langle ijkl\right \rangle }^{[2]})\propto%
%TCIMACRO{\dsum \limits_{\left\langle ijkl\right\rangle }}%
%BeginExpansion
{\displaystyle \sum \limits_{\left \langle ijkl\right \rangle }}
%EndExpansion
e^{i(\Delta \Phi_{\left \langle ijkl\right \rangle }^{[2]}+\Phi_{0})}+h.c..
\]
After considering inverse symmetry of the level-1 zero lattice, we have
\[
\mathcal{S}(\Delta \Phi_{\left \langle ijkl\right \rangle }^{[2]})=\mathcal{S}%
(-\Delta \Phi_{\left \langle ijkl\right \rangle }^{[2]}).
\]
The action is obtained as
\[
\mathcal{S}(\Delta \Phi_{\left \langle ijkl\right \rangle }^{[2]})\propto%
%TCIMACRO{\dsum \limits_{\left\langle ijkl\right\rangle }}%
%BeginExpansion
{\displaystyle \sum \limits_{\left \langle ijkl\right \rangle }}
%EndExpansion
\cos(\Delta \Phi_{\left \langle ijkl\right \rangle }^{[2]}).
\]
In continuum limit, we derive
\[
\mathcal{S}(\Delta \Phi_{\left \langle ijkl\right \rangle }^{[2]})\simeq \frac
{1}{4}F_{\mu \nu}^{2}.
\]

Finally, we obtain the effective Lagrangian $\mathcal{L}_{\mathrm{EM}}$ for
QED,
\begin{align}
\mathcal{L}_{\mathrm{EM}}  &  \simeq{\bar{\Psi}}i\gamma^{\mu}\partial_{\mu
}\Psi+m{\bar{\Psi}}\Psi \nonumber \\
&  +{A}_{\mu}{j}_{(em)}^{\mu}-\frac{1}{4}F_{\mu \nu}F^{\mu \nu}%
\end{align}
where ${j}_{(em)}^{\mu}=i\mathrm{e}{\bar{\Psi}}\gamma^{\mu}\Psi$.

Due to variability constraint $\varphi_{I}^{[1]}=\varphi_{I}^{[2]}$, the
longitudinal degrees of freedom of gauge field is "eaten" by fermionic
particles. Then, local quantum fluctuations of gauge fields generate local
loop currents and carry energies and momenta. In addition, for a charged
elementary particle at $x$, by locally changing level-2 group changing space,
they will change the distribution of level-2 group-changing elements for
another charged elementary particle. As a result, there exists electromagnetic
interaction between different charged elementary particles.

\paragraph{Coupling constant}

Finally, we discuss the coupling constant $\mathrm{e}$ of QED.

For a 2-nd order physical variant, we always have finite tempo changing rate
for level-2 group-changing space, i.e., $\omega_{0}^{[2]}\neq0$. The direct
physical consequence of this fact leads to the finite density of effective
"angular momentum". Then, we consider the coupling constant for
electromagnetic interaction $\mathrm{e}$ from tempo variability (or a uniform
motion) of level-2 group-changing space $\mathrm{C}_{\mathrm{U(1)}^{[2]}%
,1}^{[2]}$. Because the energy of the physical variant always has a uniformly
distribution, the energy density $\rho_{E}=\frac{\Delta E}{\Delta V}$ must be
constant. In addition, we assume that $\rho_{E}(\omega_{0}^{[2]})$ is a smooth
function of $\omega_{0}^{[2]}$. Then, we have
\begin{align}
\rho_{E}(\omega_{0}^{[2]}+\delta \omega^{\lbrack2]})  &  =\rho_{E}(\omega
_{0}^{[2]})\nonumber \\
+\frac{\delta \rho_{E}}{\delta \omega^{\lbrack2]}}  &  \mid_{\omega^{\lbrack
2]}=\omega_{0}^{[2]}}\delta \omega^{\lbrack2]}+...
\end{align}
where $\frac{\delta \rho_{E}}{\delta \omega^{\lbrack2]}}\mid_{\omega^{\lbrack
2]}=\omega_{0}^{[2]}}=\rho_{J}^{[2],E}$ is called the density of (effective)
"angular momentum". $\frac{\delta \rho_{E}}{\delta \omega}\mid_{\omega
^{\lbrack2]}=\omega_{0}^{[2]}}=\rho_{J}^{[2]}.$ The "angular momentum"
$\rho_{J}^{[2]}$ for level-2 zero becomes the coupling constant for
electromagnetic interaction $\mathrm{e}$ from the linearization of energy
density $\rho_{E}$ via $\omega^{\lbrack2]}$ near $\omega_{0}^{[2]}$.

In particular, the effective Planck constant $\hbar^{\lbrack2]}$ for level-2
zeroes is different from usual Planck constant $\hbar=\hbar^{\lbrack1]}$ for
level-1 zeroes,%
\[
\hbar^{\lbrack2]}=\delta V\rho_{J}^{[2]}\lambda=\hbar \lambda
\]
where $\delta V$ is the effective volume in Cartesian space. $\lambda$ is a
dimensionless parameter as the ratio of Planck constants for the zeroes of
different levels, i.e., $\lambda=\frac{\hbar^{\lbrack2]}}{\hbar}$.

Now, we consider a pure gauge field $A_{\mu}(x,t)=-\frac{1}{\mathrm{e}%
}\partial_{\mu}\varphi^{\lbrack2]}(x,t)$ (\emph{not} ${F}_{\mu \nu}$) with
finite wave vector $\Delta k.$ The momentum is $\hbar^{\lbrack2]}\Delta k$.
Under a local gauge transformation, the function of gauge field becomes
spatial independent and instead, the wave function of elementary particles has
the finite momentum $\Delta p=\hbar \Delta k^{\prime}=\hbar^{\lbrack2]}\Delta
k=\frac{\mathrm{e}}{c}\partial_{t}\varphi^{\lbrack2]}(x,t).$ Therefore, the
electric charge is obtained as
\begin{equation}
\frac{\mathrm{e}}{c}=\hbar^{\lbrack2]}=\hbar \lambda,
\end{equation}
or
\begin{equation}
\mathrm{e}=\hbar \lambda c,
\end{equation}

Finally, we discuss the problem of \emph{Landua's ghost}.

In this paper, QED turns into lattice gauge theory that describes the
fluctuating loop currents on level-1 zero lattice. Due to existence of cutoff
from lattice distance and finite size of elementary particles, there will
never exist the ultraviolet divergence. Therefore, the Landua's ghost doesn't
exist at all.

\subsubsection{Summary}

In summary, the local \textrm{U}$\mathrm{^{\mathrm{em}}}$\textrm{(1)} gauge
symmetry comes from the 2-level invariance (or symmetry) that determines the
invariance of motions. The effective model is just QED. Under
compactification, the 2-nd order variability is reduced into \textrm{U}%
$\mathrm{^{\mathrm{em}}}$\textrm{(1)} gauge symmetry. Because one cannot
directly characterize phase changings of level-2 group-changing space, we use
the representation of loop currents on minimum plaquettes. Therefore, for
electromagnetic interaction, we provide a new picture: two elementary
particles exchange energy and momentum by changing level-2 group-changing
space. Now, the physical picture for QED becomes a crystal of links with unit
linking number, i.e.,
\[
\text{QED }\rightarrow \text{ A crystal of Links with }link=1\text{.}%
\]

\subsection{$N>1$ case: QED$\mathrm{\times}$QCD}

In this section, we consider the $N>1$ case. We find that the corresponding
2-nd order variability of the 2-nd order physical variant is reduced into
\textrm{U}$\mathrm{^{\mathrm{em}}}$\textrm{(1)} gauge symmetry and
$\mathrm{SU(\lambda}^{[12]}\mathrm{)}$ non-Abelian gauge symmetry. The
effective model is QED$\mathrm{\times}$QCD.

\subsubsection{Physical variant}

Now, we have a ($d+1$)-dimensional 2-nd order\textit{ }$\mathrm{\tilde
{S}\tilde{O}}$\textrm{(d+1) }physical variant that is a higher-order mapping
between\textit{ }$\mathrm{C}_{\mathrm{\tilde{U}}^{[2]}\mathrm{(1)}}^{[2]}%
$\textit{, }$\mathrm{\tilde{S}\tilde{O}}$\textrm{(d+1)} Clifford
group-changing space\textit{ }$\mathrm{C}_{\mathrm{\tilde{S}\tilde{O}%
(d+1)},d+1}^{[1]}$\textit{\ }and a rigid spacetime\textit{ }$\mathrm{C}_{d+1}%
$\textit{.} The 2-nd order variability is%
\begin{equation}
\mathcal{T}(\delta x^{\mu})\leftrightarrow \hat{U}^{[1]}(\delta \phi
^{\lbrack1]\mu})=\exp(i(T^{\mu}\delta \phi^{\lbrack1]\mu}))
\end{equation}
and%
\begin{equation}
\hat{U}^{[1]}(\delta \phi_{\mathrm{global}}^{[1]})\leftrightarrow \hat{U}%
^{[2]}(\delta \phi^{\lbrack2]})=\exp(i\lambda^{\lbrack12]}\delta \phi
_{\mathrm{global}}^{[1]}).
\end{equation}
In particular, the ratio between the changing rates of the two group-changing
space $\lambda^{\lbrack12]}=\left \vert \frac{\delta \phi^{\lbrack2]}}%
{\delta \phi_{\mathrm{global}}^{[1]}}\right \vert $ is an integer number larger
than $1$.

Under K-projection, we have level-1 zero lattice $k_{0}(x^{^{\mu}}%
-x_{0}^{^{\mu}})-\theta=\pm \frac{\pi}{2}$, on each of lattice site, there is a
level-2 zero lattice. Now, each level-1 zero corresponds to $\lambda
^{\lbrack12]}$ level-2 zero.

In Fig.31, we show a physical picture for physical variants: On each zero
lattice, there exists two coupled rings. One ring denotes the \textrm{U}%
$^{[2]}$\textrm{(1)} group space of $\varphi^{\lbrack2]};$ the other denotes
the \textrm{U}$^{[1]}$\textrm{(1)} group space of $\varphi^{\lbrack1]}$. The
symmetries of rings indicate the symmetrise of the two compact groups
(\textrm{U}$^{[2]}$\textrm{(1)} and \textrm{U}$^{[1]}$\textrm{(1)}). Due to
the existence of the topological constraint $\lambda^{\lbrack12]}\delta
\varphi_{\mathrm{global}}^{[1]}\equiv \delta \varphi^{\lbrack2]}$ ($\lambda
^{\lbrack12]}=3$)$,$ the global compact \textrm{U}$^{[1]}$\textrm{(1)} group
and global compact \textrm{U}$^{[2]}$\textrm{(1)} group couple. To
characterize this variability constraint, the two ring on a lattice form a
knot/link. Now, each knot/link corresponds to a level-1 zero with
$\lambda^{\lbrack12]}$ level-2 zero. The linking number is $\lambda
^{\lbrack12]}=3$.

In summary, under compactification, the physical picture for a 2-nd order
physical variant $V_{\mathrm{\tilde{U}}^{[2]}\mathrm{(1)},\mathrm{\tilde
{S}\tilde{O}}^{[1]}\mathrm{(d+1)},d+1}^{[2]}$ with $\lambda^{\lbrack12]}>1$
becomes a crystal of complex knot with linking number $\lambda^{\lbrack12]}%
$!\begin{figure}[ptb]
\includegraphics[clip,width=0.7\textwidth]{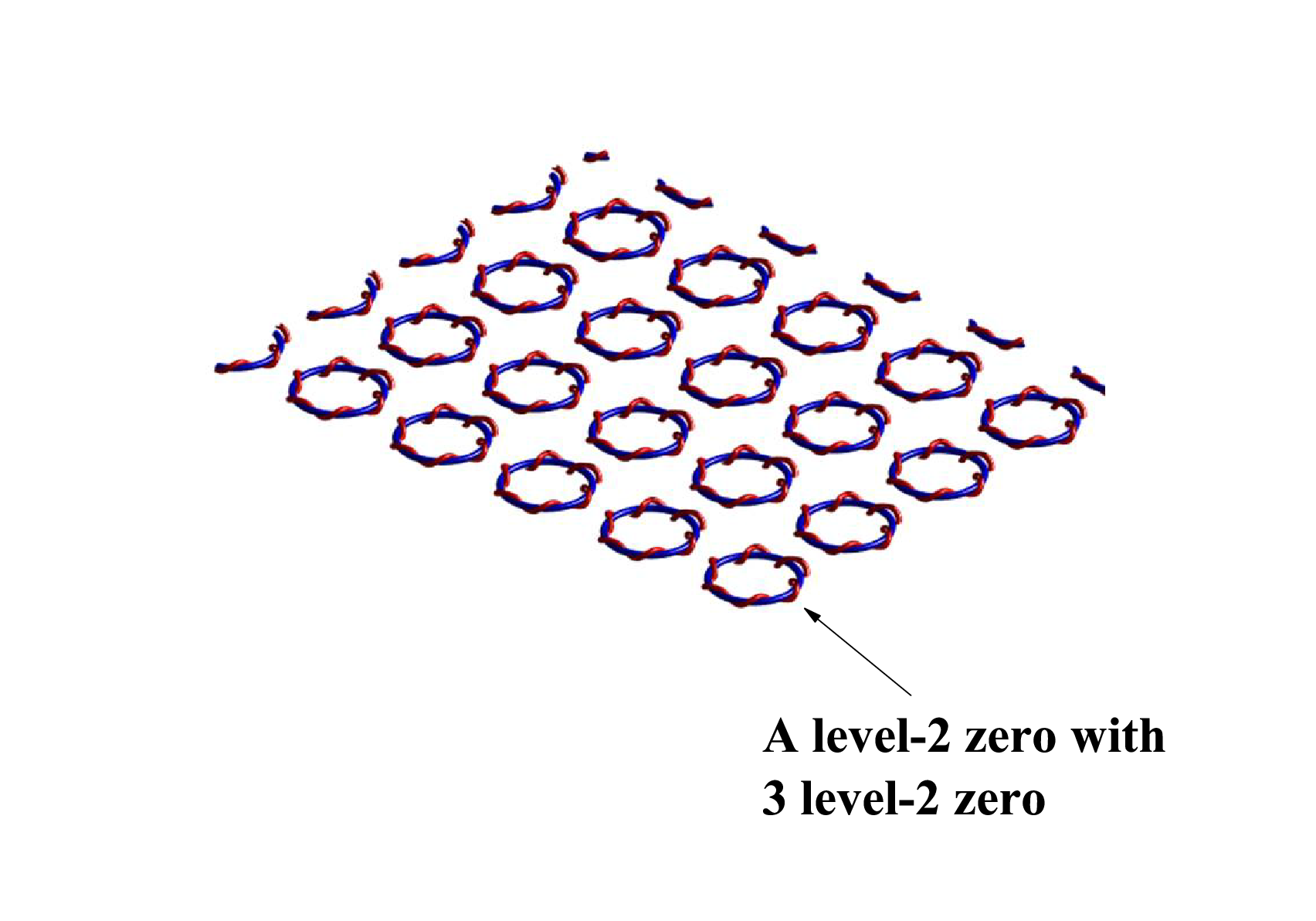}\caption{An illustration
of microscopic picture for QCD$\times$QED: a crystal of two coupled rings (or
knot/links). Each ring corresponds to a lattice site of the topological
lattice with two (level-2 or level-1) zeroes. The symmetries of rings indicate
the symmetrise of the two compact groups (\textrm{U}$^{[2]}$\textrm{(1)} and
\textrm{U}$^{[1]}$\textrm{(1)}). Each knot/link corresponds to a level-1 zero
with 3 level-2 zero. The linking number is 3.}%
\end{figure}

\subsubsection{Matter}

Next, we discuss the matter for 2-nd order physical variants
$V_{\mathrm{\tilde{U}}^{[2]}\mathrm{(1)},\mathrm{\tilde{S}\tilde{O}}%
^{[1]}\mathrm{(d+1)},d+1}^{[2]}$ with changing rate $\lambda^{\lbrack
12]}=\left \vert \frac{\delta \phi^{\lbrack2]}}{\delta \phi_{\mathrm{global}%
}^{[1]}}\right \vert >1.$

Matter corresponds to globally expand or contract the two group-changing
spaces $\mathrm{C}_{\mathrm{\tilde{S}\tilde{O}}^{[1]}\mathrm{(d+1)},d+1}%
^{[1]}$ or $\mathrm{C}_{\mathrm{U(1)}^{[2]},1}^{[2]}$ with changing their
corresponding sizes. Therefore, an object is classified by two integer numbers
$\{n^{[1]},n^{[2]}\}$ -- the number of level-1 zeroes $n^{[1]}$ and the number
of level-2 zeroes $n^{[2]}$.

Firstly, we point out that $n^{[1]}$ denotes the number of elementary
particles. We classify the types of elementary particles by its number of
level-2 zeroes $n^{[2]}$. For an elementary particle, $n^{[2]}$ determines
both color charge and electric charge. According to above discussion, each
level-2 zero has $1/\lambda^{\lbrack12]}$ electric charge. For an elementary
particle with $n^{[2]}$ level-2 zeroes, its electric charge is $\mathrm{e}%
=n^{[2]}/\lambda^{\lbrack12]}.$

For example, for the case $\lambda^{\lbrack12]}=2$, there are two types of
elementary particles: one is electron with $n^{[2]}=\lambda^{\lbrack12]}=2$,
of which the electric charge is unit, the other is quark with $n^{[2]}%
=\lambda^{\lbrack12]}=1$ and half electric charge, i.e.,
\begin{align*}
\{n^{[1]}  &  =1,n^{[2]}=2\}:\text{ Electron with }\\
&  1\text{ electric charge,}\\
\{n^{[1]}  &  =1,n^{[2]}=1\}:\text{ Quark with }\\
&  \frac{1}{2}\text{ electric charge.}%
\end{align*}

For the case $\lambda^{\lbrack12]}=3$, there are three types of elementary
particles:%
\begin{align*}
\{n^{[1]}  &  =1,n^{[2]}=3\}:\text{ Electron with }\\
&  1\text{ electric charge,}\\
\{n^{[1]}  &  =1,n^{[2]}=2\}:\text{ u-Quark with }\\
&  \frac{2}{3}\text{ electric charge,}\\
\{n^{[1]}  &  =1,n^{[2]}=1\}:\text{ d-Quark with }\\
&  \frac{1}{3}\text{ electric charge.}%
\end{align*}

For the case with higher $\lambda^{\lbrack12]},$ there are $\lambda
^{\lbrack12]}$ types of elementary particles that are labeled by
$\{n^{[1]}=1,n^{[2]}\}.$ Therefore, we have%
\begin{align*}
\{n^{[1]}  &  =1,n^{[2]}=\lambda^{\lbrack12]}\}:\text{ Electron with }\\
&  1\text{ electric charge,}\\
\{n^{[1]}  &  =1,n^{[2]}=\lambda^{\lbrack12]}-1\}:\text{ Quark with }\\
&  \frac{\lambda^{\lbrack12]}-1}{\lambda^{\lbrack12]}}\text{ electric
charge,}\\
\{n^{[1]}  &  =1,n^{[2]}=\lambda^{\lbrack12]}-2\}:\text{ Quark with }\\
&  \frac{\lambda^{\lbrack12]}-2}{\lambda^{\lbrack12]}}\text{ electric
charge,}\\
&  ...\\
\{n^{[1]}  &  =1,n^{[2]}=1\}:\text{ Quark with }\\
&  \frac{1}{\lambda^{\lbrack12]}}\text{ electric charge.}%
\end{align*}

In general, the quantum states of an elementary particle with $n^{[2]}$
level-2 zeroes are denoted by
\begin{equation}
\psi_{n^{[2]}}^{\dagger}(x,t)\left \vert \mathrm{vac}\right \rangle .
\end{equation}
With $n^{[2]}$ level-2 zeroes, we call the elementary particles to be
composite particles, the level-2 zero to be internal zero and the
corresponding particle to be parton. Thus, QED and QCD become two inseparable
parts of a single 2-nd order physical variant with $\lambda^{\lbrack12]}>1.$

Finally, we discuss the invariant of level-1 physics structure for matter. It
was known that matter corresponds to globally expand or contract of the two
group-changing spaces, $\mathrm{C}_{\mathrm{\tilde{S}\tilde{O}}^{[1]}%
\mathrm{(d+1)},d+1}^{[1]}$ or $\mathrm{C}_{\mathrm{U(1)}^{[2]},1}^{[2]}$. We
have an invariance named topology stationarity of matter. The topology
stationarity of matter comes from a fact that the total sizes of the
group-changing spaces $\mathrm{C}_{\mathrm{\tilde{S}\tilde{O}}^{[1]}%
\mathrm{(d+1)},d+1}^{[1]}$ and $\mathrm{C}_{\mathrm{U(1)}^{[2]},1}^{[2]}$ are
all topological invariables. During the processes of motion, the total size of
the group-changing spaces don't change.

\subsubsection{Motion}

In this section, we discuss the motion and the 2-level invariance (or
symmetry). There are two types of motions -- one is about the motions for
elementary particles, the other is about the collective motions of the two
levels of zero lattices.

It was known that the collective motion of level-1 zero lattice is
gravitational wave that is described by Einstein's equation, the motion of the
elementary particle is described by Dirac model. Here, we focus on the
collective motion of the level-2 zero lattice on Cartesian spacetime. In
general, we use quantum gauge field theory to characterize its dynamics.

\paragraph{Quantum states for motion}

\subparagraph{Quantum states for motion of level-2 zeroes}

Firstly, we discuss the quantum states for the level-2 group-changing elements
of a level-1 zero. So, we also call all these quantum states to be
\emph{internal }quantum states of the elementary particle that corresponds to
the level-1 zero.

Under compactification, the non-compact level-2 group-changing space turns
into level-2 zero lattice. On each site of the level-2 zero lattice (we denote
it by $I^{[2]}$), there is a local field of compact U(1) symmetry. We
"generate" an extra ($i$-th) level-2 group-changing elements $\varphi
_{I^{[2]},I^{[1]},i}^{[2]}$ on the position $\varphi_{I^{[2]},I^{[1]},i}%
^{[2]}$ of $I^{[2]}$-th level-2 zero and the position $\varphi_{I^{[1]}%
,i}^{[1]}$ of $I^{[1]}$-th level-1 zero by the level-2 group-changing
operation $\tilde{U}(\delta \varphi_{I^{[2]},I^{[1]},i}^{[2]}(\varphi
_{I^{[1]},i}^{[1]}))=e^{i((\delta \varphi_{I^{[2]},I^{[1]},i}^{[2]})\cdot
\hat{K})}$ and $\hat{K}=-i\frac{d}{d\varphi^{\lbrack2]}}.$\ Here, $I^{[2]}$
and $I^{[1]}$ label the level-2 zero and level-1 zero, respectively. We have
used abbreviation $\varphi_{\mathrm{global}}^{[1]}$ by $\varphi^{\lbrack1]}$
and \textrm{U}$_{\mathrm{global}}$\textrm{(1)} group by \textrm{U(1)} group.
In the following parts, we also use the abbreviation.

The motion of level-2 group-changing space comes from its local expansion and
contraction on different level-1 zeroes. If there exist $n^{[2]}$ level-2
zeroes, the total size of of all level-2 group-changing element is $\pm
n^{[2]}\pi$, i.e., $%
%TCIMACRO{\dsum }%
%BeginExpansion
{\displaystyle \sum}
%EndExpansion
\delta \varphi_{I^{[2]},I^{[1]}}^{[2]}=\pm n^{[2]}\pi$. We call a level-2 zero
to be parton inside an elementary particle (a level-1 zero). The local
expansion and contraction will shift the corresponding level-2 group-changing
space on two levels of zeroes, for example, changing level-2 phase factor
$\delta \varphi_{I^{[2]},I^{[1]}}^{[2]}$ on the position $\varphi_{I^{[1]}%
}^{[1]}$ on $I$-th level-1 zero, or changing the position of lattice sites of
level-2 group-changing space on $I$-th level-1 zero. Because there are total
$\lambda^{\lbrack12]}$ lattice sites for level-2 zeroes of a level-1 zero, we
have $\lambda^{\lbrack12]}$\ level-2 phase factors $\delta \varphi
_{I^{[2]},I^{[1]}}^{[2]}$ for a level-2 group-changing element of a level-1
zero. We define a global level-2 phase factor and $\lambda^{\lbrack12]}-1$
relative phase factors. The number $\lambda^{\lbrack12]}-1$ corresponds to the
number of Carton-Weyl base for \textrm{SU(N)} group.

To characterize these $\lambda^{\lbrack12]}$\ level-2 phase factors
$\delta \varphi_{I^{[2]},I^{[1]}}^{[2]}$, we must define $\lambda^{\lbrack12]}$
references, i.e., $\delta \varphi_{0,I^{[2]},I^{[1]}}^{[2]}.$ For the global
phase factor $\delta \varphi_{I^{[1]}}^{[2]}$, the reference is $\delta
\varphi_{0,I^{[1]}}^{[2]}$. According to level-2 variability, the changing of
reference $\delta \varphi_{0,I^{[1]}}^{[2]}$ for global phase factor is same to
the changing of the reference of level-1 global phase factor $\varphi
_{0,I}^{[1]}$%
\[
\delta \varphi_{0,I}^{[1]}=\lambda \delta \varphi_{0,I^{[1]}}^{[2]}%
\]
where $\delta \varphi_{0,I}^{[1]}=((\varphi_{0,I}^{[1]})^{\prime}-\varphi
_{0,I}^{[1]})$ and $\delta \varphi_{0,I^{[1]}}^{[2]}=((\varphi_{0,I^{[1]}%
}^{[2]})^{\prime}-\varphi_{0,I^{[1]}}^{[2]}).$ This leads to local
\textrm{U}$^{\mathrm{em}}$\textrm{(1)} gauge transformation. On the other
hand, there are $\lambda^{\lbrack12]}-1$ references for relative phase
factors. To set these references for $\lambda^{\lbrack12]}-1$ relative phase
factors, we define the reference state that is the representation of
compact\textrm{ SU(N)} group.

We consider the $I^{[1]}$-th level-1 zero with an extra level-2 zero on one
site of level-2 zero lattice, $1$, or $2$, ..., or $\lambda^{\lbrack12]}$. The
situation looks like an electron on a 1D crystal with $\lambda^{\lbrack12]}$
lattice sites. We may also use theory of solid state physics to characterize
the quantum states of them.

We denote its quantum states by a matrix, i.e.,
\begin{equation}
\left(
\begin{array}
[c]{c}%
\left \vert \psi_{1^{[2]},I^{[1]}}^{[2]}\right \rangle \\
\left \vert \psi_{2^{[2]},I^{[1]}}^{[2]}\right \rangle \\
...\\
\left \vert \psi_{(\lambda^{\lbrack12]})^{[2]},I^{[1]}}^{[2]}\right \rangle
\end{array}
\right)  .
\end{equation}
Here, $\left \vert \psi_{I^{[2]},I^{[1]}}^{[2]}\right \rangle $ denotes the
state of the level-2 zero on the $I^{[2]}$-th lattice site of level-2 zero
lattice inside $I^{[1]}$-th level-1 zero. Because there are $\lambda
^{\lbrack12]}$ lattice sites, the quantum states of $\left \vert \psi
_{I^{[2]},I^{[1]}}^{[2]}\right \rangle $ have $\lambda^{\lbrack12]}$
components. In addition, the quantum states of the level-2 zero on different
sites of the level-2 zero-lattice are orthogonal, i.e.,%
\begin{equation}
\langle \psi_{J^{[2]},I^{[1]}}^{[2]}\left \vert \psi_{I^{[2]},I^{[1]}}%
^{[2]}\right \rangle =\delta_{J^{[2]}I^{[2]}}.
\end{equation}
Hence, $\left \vert \psi_{1^{[2]},I^{[1]}}^{[2]}\right \rangle ,$ $\left \vert
\psi_{2^{[2]},I^{[1]}}^{[2]}\right \rangle ,$ ..., $\left \vert \psi
_{\lambda^{\lbrack12]},I^{[1]}}^{[2]}\right \rangle $ make up a complete basis.
In general, we can re-label the corresponding base of partons by a new one
$\left \vert \psi_{1^{[2]},I^{[1]}}^{[2]}\right \rangle ^{\prime},$ $\left \vert
\psi_{2^{[2]},I^{[1]}}^{[2]}\right \rangle ^{\prime},$ ..., $\left \vert
\psi_{\lambda^{\lbrack12]},I^{[1]}}^{[2]}\right \rangle ^{\prime}$%
.\textbf{\ }The relationship between the two basis is
\begin{equation}
\left(
\begin{array}
[c]{c}%
\left \vert \psi_{1^{[2]},I^{[1]}}^{[2]}\right \rangle ^{\prime}\\
\left \vert \psi_{2^{[2]},I^{[1]}}^{[2]}\right \rangle ^{\prime}\\
...\\
\left \vert \psi_{(\lambda^{\lbrack12]})^{[2]},I^{[1]}}^{[2]}\right \rangle
^{\prime}%
\end{array}
\right)  =\hat{U}_{\mathrm{SU(N)}}^{[2]}\left(  x,t\right)  \left(
\begin{array}
[c]{c}%
\left \vert \psi_{1^{[2]},I^{[1]}}^{[2]}\right \rangle \\
\left \vert \psi_{2^{[2]},I^{[1]}}^{[2]}\right \rangle \\
...\\
\left \vert \psi_{(\lambda^{\lbrack12]})^{[2]},I^{[1]}}^{[2]}\right \rangle
\end{array}
\right)
\end{equation}
where $\hat{U}_{\mathrm{SU(N)}}^{[2]}\left(  x,t\right)  =e^{-i\Theta \left(
\vec{x},t\right)  }$ is the matrix of the representation of $\mathrm{SU(N)}$
group. $\Theta \left(  x,t\right)  =\sum_{a=1}^{\mathrm{N}^{2}-1}\theta
^{a}\left(  x,t\right)  \tau^{a}$ and $\theta^{a}$ are a set of $\mathrm{N}%
^{2}-1$ constant parameters, and $\tau^{a}$ are $\mathrm{N}^{2}-1$
$\mathrm{N}\times \mathrm{N}$ matrices representing the $\mathrm{N}^{2}-1$
generators of the Lie algebra of $\mathrm{SU(N)}$\cite{yang}. The global phase
factor of $\left(
\begin{array}
[c]{c}%
\left \vert \psi_{1^{[2]},I^{[1]}}^{[2]}\right \rangle \\
\left \vert \psi_{2^{[2]},I^{[1]}}^{[2]}\right \rangle \\
...\\
\left \vert \psi_{(\lambda^{\lbrack12]})^{[2]},I^{[1]}}^{[2]}\right \rangle
\end{array}
\right)  $ is $\delta \varphi_{I^{[1]}}^{[2]},\ $of which the reference is
$\varphi_{0,I^{[1]}}^{[2]}.$\ The reference of relative phase angle can be
defined by a fixed group element of \textrm{SU(N)} group, i.e., $\hat
{U}_{\mathrm{SU(N)}}^{[2]}\left(  x,t\right)  =e^{-i\Theta_{0}\left(  \vec
{x},t\right)  }.$

\subparagraph{Quantum states for motion of level-1 zeroes}

In this part, we discuss the quantum states of a\ level-1 zero (or an
elementary particle) with $n^{[2]}$ level-1 zero (or $n^{[2]}$ partons).

Firstly, we consider the case of $n^{[2]}=1$. This is an elementary particle
with 1 parton that is a quark with a level-2 zero and $1/\lambda^{\lbrack12]}$
electric charge.

We use another $\lambda^{\lbrack12]}$-component field to describe the quantum
states of the elementary particle
\begin{equation}
\left(
\begin{array}
[c]{c}%
\left \vert \psi_{1^{[2]},I^{[1]}}^{[1]}\right \rangle \\
\left \vert \psi_{2^{[2]},I^{[1]}}^{[1]}\right \rangle \\
...\\
\left \vert \psi_{(\lambda^{\lbrack12]})^{[2]},I^{[1]}}^{[1]}\right \rangle
\end{array}
\right)  .
\end{equation}
$\left \vert \psi_{I^{[2]},I^{[1]}}^{[1]}\right \rangle $ denotes the quantum
state of its level-2 zero on the $I^{[2]}$-th lattice site of level-2 zero
lattice inside $I^{[1]}$-th level-1 zero. The changing of relative phase
factor of parton leads to corresponding changing of the relative phase factors
of elementary particle, i.e,
\begin{equation}
\left(
\begin{array}
[c]{c}%
\left \vert \psi_{1^{[2]},I^{[1]}}^{[2]}\right \rangle ^{\prime}\\
\left \vert \psi_{2^{[2]},I^{[1]}}^{[2]}\right \rangle ^{\prime}\\
...\\
\left \vert \psi_{(\lambda^{\lbrack12]})^{[2]},I^{[1]}}^{[2]}\right \rangle
^{\prime}%
\end{array}
\right)  =\hat{U}_{\mathrm{SU(N)}}^{[2]}\left(  x,t\right)  \left(
\begin{array}
[c]{c}%
\left \vert \psi_{1^{[2]},I^{[1]}}^{[2]}\right \rangle \\
\left \vert \psi_{2^{[2]},I^{[1]}}^{[2]}\right \rangle \\
...\\
\left \vert \psi_{(\lambda^{\lbrack12]})^{[2]},I^{[1]}}^{[2]}\right \rangle
\end{array}
\right)
\end{equation}
and
\begin{equation}
\left(
\begin{array}
[c]{c}%
\left \vert \psi_{1^{[2]},I^{[1]}}^{[1]}\right \rangle ^{\prime}\\
\left \vert \psi_{2^{[2]},I^{[1]}}^{[1]}\right \rangle ^{\prime}\\
...\\
\left \vert \psi_{(\lambda^{\lbrack12]})^{[2]},I^{[1]}}^{[1]}\right \rangle
^{\prime}%
\end{array}
\right)  =\hat{U}_{\mathrm{SU(N)}}^{[1]}\left(  x,t\right)  \left(
\begin{array}
[c]{c}%
\left \vert \psi_{1^{[2]},I^{[1]}}^{[1]}\right \rangle \\
\left \vert \psi_{2^{[2]},I^{[1]}}^{[1]}\right \rangle \\
...\\
\left \vert \psi_{(\lambda^{\lbrack12]})^{[2]},I^{[1]}}^{[1]}\right \rangle
\end{array}
\right)  .
\end{equation}
This provides a \emph{non-Abelian variability constraint}, i.e.,%
\[
U_{\mathrm{SU(N)}}^{[1]}\left(  x,t\right)  \equiv U_{\mathrm{SU(N)}}%
^{[2]}\left(  x,t\right)  .
\]
This non-Abelian variability constraint plays important role in non-Abelian
gauge symmetry for Yang-Mills field.

The situation can be straightforward generalized to the case of $n^{[2]}>1$.
Now, there are $C_{\lambda^{\lbrack12]}}^{n^{[2]}}$ internal quantum states of
the elementary particle. As a result, we use a $C_{\lambda^{\lbrack12]}%
}^{n^{[2]}}$-component field to characterize it.

\paragraph{Local \textrm{U}$^{\mathrm{em}}$\textrm{(1)}$\times$\textrm{SU(N)}
gauge symmetry}

\subparagraph{Global symmetry of partons}

We firstly discuss the symmetry for a parton (or a level-2 zero) of a level-1
zero, of which the wave function is $\left(
\begin{array}
[c]{c}%
\left \vert \psi_{1^{[2]},I^{[1]}}^{[2]}\right \rangle \\
\left \vert \psi_{2^{[2]},I^{[1]}}^{[2]}\right \rangle \\
...\\
\left \vert \psi_{(\lambda^{\lbrack12]})^{[2]},I^{[1]}}^{[2]}\right \rangle
\end{array}
\right)  $.

We point out that if all quantum states of a parton have same energy, there is
global symmetry; if different quantum states of a parton have different
energies, there isn't global symmetry. Then, to determine the global symmetry
for the partons, we calculate the effective model for them. Let us check it.

Firstly we check the discrete translation symmetry for the uniform level-2
zero lattice.

According to the level-2 variability $\hat{U}^{[1]}\leftrightarrow \hat
{U}^{[2]}$, for the uniform level-2 zero lattice, we have a reduced
translation symmetry denoted by the following equation $\hat{U}^{[2]}%
\leftrightarrow1.$ As a result, quantum states on different lattice sites of
level-2 zero lattice are equivalent each other.

We define generation operator $(c_{I^{[2]}}^{[2]})^{\dagger}$ of parton and
get $(c_{i^{I[2]}}^{[2]})^{\dagger}\left \vert 0\right \rangle =\left \vert
I^{[2]}\right \rangle $. We then write down the hopping Hamiltonian. The
hopping term between two nearest neighbor sites $I^{[2]}$ and $J^{[2]}$ on
topological lattice becomes
\begin{equation}
\mathcal{H}_{\left \{  i,j\right \}  }^{[2]}=\mathcal{J}^{[2]}(c_{I^{[2]}}%
^{[2]})^{\dagger}\mathbf{T}_{\left \{  I^{[2]},J^{[2]}\right \}  }c_{J^{[2]}%
}^{[2]}(t)
\end{equation}
where $\mathbf{T}_{\left \{  I^{[2]},J^{[2]}\right \}  }$ is the transfer matrix
between two nearest neighbor sites $I^{[2]}$ and $J^{[2]}$ and $c_{I^{[2]}%
}^{[2]}(t)$ is the annihilation operator of parton at the site $I^{[2]}$.
Because, there is no hopping term for a level-2 zero on a level-1 zero
lattice, $\mathcal{J}^{[2]}$ must be zero, i.e.,
\begin{equation}
\mathcal{J}^{[2]}\equiv0.
\end{equation}
In addition, there may exist the terms from on-site potential%
\begin{equation}
\mathcal{H}^{[2]}=%
%TCIMACRO{\dsum \limits_{I^{[2]}}}%
%BeginExpansion
{\displaystyle \sum \limits_{I^{[2]}}}
%EndExpansion
\mathcal{H}_{I^{[2]}}^{[2]}=V%
%TCIMACRO{\dsum \limits_{I^{[2]}}}%
%BeginExpansion
{\displaystyle \sum \limits_{I^{[2]}}}
%EndExpansion
(c_{I^{[2]}}^{[2]})^{\dagger}c_{I^{[2]}}^{[2]}+h.c..
\end{equation}
As a result, there is global \textrm{SU(N)} symmetry for a parton inside a
level-1 zero.

Secondly, we check the global \textrm{U}$^{[1]}$\textrm{(1)} symmetry for a
level-1 zero. Under compactification, the operation $\hat{U}^{[1]}$ is reduced
to that of a global \textrm{U}$^{[1]}$\textrm{(1)} group. Therefore, on each
lattice site, we have an invariant under the global compact \textrm{U}$^{[1]}%
$\textrm{(1)} group, i.e.,
\[
\hat{U}^{[1]}\leftrightarrow1.
\]
The quantum states of the parton with different global phase have same energy.
As a result, we have a global \textrm{U}$^{[1]}$\textrm{(1)} symmetry for a
parton inside a level-1 zero.

Thirdly, we check the global \textrm{U}$^{[2]}$\textrm{(1)} symmetry for a
parton of a level-1 zero. Under compactification, the operation $\hat{U}%
_{I}^{[2]}(\delta \phi_{I}^{[2]})$ is reduced to that of a global compact
\textrm{U}$^{[2]}$\textrm{(1)} group. On each lattice site of level-1 zero
lattice, we have an invariant under the global compact \textrm{U}$^{[2]}%
$\textrm{(1)} group, i.e., $\hat{U}_{I}^{[2]}(\delta \phi^{\lbrack
2]})\rightarrow \hat{U}_{I}^{[2]}(\delta \varphi^{\lbrack2]}).$ For simplicity,
we can denote it by the following equations $\hat{U}_{I}^{[2]}(\delta
\varphi^{\lbrack2]})\leftrightarrow1.$ Such an invariant under the global
\textrm{U(1)} group means that the system with different uniform phase angles
have same energy.

In summary, we have a global \textrm{SU(N)} symmetry for a level-2 zero, a
global \textrm{U}$^{[1]}$\textrm{(1)} symmetry for the level-1 zero, and a
global \textrm{U}$^{[2]}$\textrm{(1)} symmetry for a level-2 zero.

\subparagraph{Variability constraints}

There are two types of variability constraints - one is global variability
constraint, the other is relative variability constraint.

On the one hand, we discuss the global variability constraint. According to
above discussion, due to level-2 variability, the changings of references
$\varphi_{0,I}^{[1]}$ and $\varphi_{0,I}^{[2]}$ for the two group-changing
spaces must be synchronously, $\delta \varphi_{0,I}^{[1]}=\delta \varphi
_{0,I}^{[2]}$.

On the other hand, we discuss the relative variability constraint.

By trapping level-2 zeroes, there exist different internal states of
elementary particles. The internal states of the level-1 zero (or elementary
particle) are determined by the quantum states of the level-2 zeroes. We call
this phenomenon to be \emph{state nesting}. Now, the wave functions of quantum
states of the level-1 zero are functions of the wave functions of quantum
states of the level-2 zeroes, i.e.,%
\[
\psi_{1^{[2]},I^{[1]}}^{[1]}(\psi_{1^{[2]},I^{[1]}}^{[2]}).
\]
According to state nesting, we have relative variability constraint.

The changings of quantum states of level-2 zeroes are denoted by an operation
of SU(N) group on $\psi_{1^{[2]},I^{[1]}}^{[2]},$%
\[
\psi_{1^{[2]},I^{[1]}}^{[2]}\rightarrow(\psi_{1^{[2]},I^{[1]}}^{[2]})^{\prime
}=\hat{U}_{\mathrm{SU(N)}}^{[2]}\left(  x,t\right)  \psi_{1^{[2]},I^{[1]}%
}^{[2]}.
\]
Because the internal states of level-1 zero are determined by the quantum
states of level-2 zero, the changings of quantum states of level-2 zero lead
to the changings of internal states of level-1 zero, i.e.,
\begin{align*}
\psi_{1^{[2]},I^{[1]}}^{[1]}(\psi_{1^{[2]},I^{[1]}}^{[2]})  &  \rightarrow
(\psi_{1^{[2]},I^{[1]}}^{[1]}(\psi_{1^{[2]},I^{[1]}}^{[2]}))^{\prime}\\
&  =\psi_{1^{[2]},I^{[1]}}^{[1]}((\psi_{1^{[2]},I^{[1]}}^{[2]})^{\prime})\\
&  =\psi_{1^{[2]},I^{[1]}}^{[1]}(\hat{U}_{\mathrm{SU(N)}}^{[2]}\left(
x,t\right)  \psi_{1^{[2]},I^{[1]}}^{[2]})\\
&  =\hat{U}_{\mathrm{SU(N)}}^{[2]}\left(  x,t\right)  \psi_{1^{[2]},I^{[1]}%
}^{[1]}(\psi_{1^{[2]},I^{[1]}}^{[2]})\\
&  =\hat{U}_{\mathrm{SU(N)}}^{[1]}\left(  x,t\right)  \psi_{1^{[2]},I^{[1]}%
}^{[1]}(\psi_{1^{[2]},I^{[1]}}^{[2]}).
\end{align*}
Therefore, we have the relative variability constraint,
\[
\hat{U}_{\mathrm{SU(N)}}^{[2]}\left(  x,t\right)  \equiv \hat{U}%
_{\mathrm{SU(N)}}^{[1]}\left(  x,t\right)  .
\]

\subparagraph{Local gauge symmetry}

In this part, we discuss the local gauge symmetry. There are two types of
local gauge symmetries, one is Abelian, \textrm{U}$^{\mathrm{em}}$\textrm{(1)}
gauge symmetry for global motion of level-2 zeroes of a level-1 zero, the
other is non-Abelian, $\mathrm{SU(N)}$ gauge symmetry for relative motion of
level-2 zeroes of a level-1 zero.

Under compactification, the level-2 invariance is reduced to gauge invariant
under the global \textrm{U}$^{[2]}$\textrm{(1)} group and translation
invariant $\mathcal{T}^{[2]}$ on the level-2 zero lattice with $\lambda
^{\lbrack12]}$ lattice sites, i.e.,
\[
\hat{U}^{\mu}\rightarrow \hat{U}^{[2]}(\delta \varphi)\otimes \mathcal{T}^{[2]}.
\]
\ Because there doesn't exist hopping term for level-2 zero on the level-2
zero lattice, the translation invariant $\mathcal{T}^{[2]}$ on the level-2
zero lattice with $\lambda^{\lbrack12]}$ lattice sites turns into a global
\textrm{SU}$^{[2]}$\textrm{(}$\mathrm{\lambda^{\lbrack12]}}$\textrm{)}
symmetry for different quantum states of an extra level-2 zero. Therefore, the
symmetry for relative motion is reduced to a non-Abelian one,
\[
\mathcal{T}^{[2]}\rightarrow \hat{U}_{\mathrm{SU(\lambda^{\lbrack12]})}}%
^{[2]}.
\]

On the one hand, due to the existence of the\emph{ }global variability
constraint $\delta \varphi_{0,I}^{[1]}=\delta \varphi_{0,I}^{[2]},$ the symmetry
from global compact \textrm{U}$^{[1]}$\textrm{(1)} group and that from global
compact \textrm{U}$^{[2]}$\textrm{(1)} group couple and then unify into a
local \textrm{U}$^{\mathrm{em}}$\textrm{(1)} gauge symmetry, i.e., $\hat
{U}_{I}^{[2]}(\delta \varphi_{I}^{[2]})\leftrightarrow \hat{U}_{I}^{[1]}%
(\delta \varphi_{I}^{[1]}).$

On the other hand, we discuss the non-Ableian gauge symmetry for relative
motion. Due to the relative variability constraint from state nesting $\hat
{U}_{\mathrm{SU(N)}}^{[2]}\left(  x,t\right)  \equiv \hat{U}_{\mathrm{SU(N)}%
}^{[1]}\left(  x,t\right)  ,$ we have a local $\mathrm{SU(N)}$ symmetry that
denotes the indistinguishable internal quantum states of the elementary
particle,
\[
\hat{U}_{\mathrm{SU(N)}}(x,t)=\hat{U}_{\mathrm{SU(N)}}^{[2]}\left(
x,t\right)  \equiv \hat{U}_{\mathrm{SU(N)}}^{[1]}\left(  x,t\right)  .
\]
For simplicity, we have
\begin{align}
\psi_{1^{[2]},I^{[1]}}^{[1]}(\psi_{1^{[2]},I^{[1]}}^{[2]})  &  \rightarrow
(\psi_{1^{[2]},I^{[1]}}^{[1]}(\psi_{1^{[2]},I^{[1]}}^{[2]}))^{\prime
}\nonumber \\
&  =\hat{U}_{\mathrm{SU(N)}}(x,t)\psi_{1^{[2]},I^{[1]}}^{[1]}(\psi
_{1^{[2]},I^{[1]}}^{[2]}).
\end{align}

In summary, we have%
\begin{align*}
&  \text{Level-2 variability with }\lambda^{\lbrack12]}>1\text{ }\\
&  \rightarrow \text{ }\mathrm{U}^{\mathrm{em}}\mathrm{(1)}\text{ local gauge
symmetry }\\
&  \text{+ \textrm{SU(N)} local gauge symmetry, }%
\end{align*}%
\begin{align*}
&  \mathrm{U}^{\mathrm{em}}\mathrm{(1)}\text{ local gauge symmetry }\\
&  \text{=}\text{ Two global \textrm{U(1)} group with}\\
&  \text{ global variability constraint }%
\end{align*}
and
\begin{align*}
&  \text{Local \textrm{SU(N)} gauge symmetry }\\
&  \text{=}\text{ Two global \textrm{SU(N)} group with relative variability
}\\
&  \text{constraint due to state nesting effect. }%
\end{align*}

Let us give more discussion on this issue. In modern physics, it is known that
gauge symmetry appears as the redundancy to define the particles. According to
the variant theory, the redundancy of gauge symmetry comes from the redundancy
of references of level-2 group-changing space (or the internal state of an
elementary particle). Because all physical processes are independent from the
selection of different references, there exists redundancy to define
elementary particles. This leads to the underlying mechanism of the
\textrm{U(1)}$\times$SU(N) gauge symmetry from indistinguishable internal
state inside the elementary particles.

\paragraph{\textrm{U}$^{\mathrm{em}}$\textrm{(1)}$\times$\textrm{SU(N)} gauge
field and its physical picture}

In this section, we define \textrm{U}$^{\mathrm{em}}$\textrm{(1)}$\times
$\textrm{SU(N)} gauge field to characterize the fluctuations of level-2
group-changing space.

\subparagraph{\textrm{U}$^{\mathrm{em}}$\textrm{(1)} gauge field from global
collective motion of level-2 zeroes}

\textrm{U}$^{\mathrm{em}}$\textrm{(1)} gauge field characterizes the global
collective motion of level-2 zeroes. The situation is similar to the case of
the case of $\lambda^{\lbrack12]}=1.$ As a result, we introduce the vector
field $\mathrm{e}A_{I,I^{\prime}}=-\varphi_{I}^{[2]}+\varphi_{I^{\prime}%
}^{[2]}$ that plays the role of \textrm{U}$^{\mathrm{em}}$\textrm{(1)} gauge
field. To characterize the changings of level-2 group-changing space on
Cartesian space, we introduce the loop currents on level-1 zero lattice. A
simple approach is to consider the current around minimum loops that consist
of four links for four nearest neighbor lattice sites $\Phi_{\left \langle
IJKL\right \rangle }^{[2]}$ on the plaquette of $IJKL$ lattice site, i.e.,
$\Phi_{\left \langle IJKL\right \rangle }^{[2]}=\mathrm{e}A_{IJ}-\mathrm{e}%
A_{KL}=-\varphi_{I}^{[2]}+\varphi_{J}^{[2]}+\varphi_{K}^{[2]}-\varphi
_{L}^{[2]}.$ The quantum states for level-2 group-changing space on Cartesian
space are defined by $\{ \Phi_{\left \langle IJKL\right \rangle }^{[2]},$
$\left \langle IJKL\right \rangle \in \mathrm{All}\}.$ In continuum limit, we
have the \textrm{U}$^{\mathrm{em}}$\textrm{(1)} gauge field $A_{\mu}(x,t)$ and
the strength of gauge field $F_{\mu \nu}.$

However, the situation is more complex than the case of the case of
$\lambda^{\lbrack12]}=1$. For the elementary particles with partons (or extra
level-2 zeroes), they may have \emph{fractional }electric charges. Let us show detail.

It was known that due to the existence of the\emph{ }global variability
constraint $\delta \varphi_{0,I}^{[1]}=\delta \varphi_{0,I}^{[2]},$ the symmetry
from global compact \textrm{U}$^{[1]}$\textrm{(1)} group and that from global
compact \textrm{U}$^{[2]}$\textrm{(1)} group unify into a local \textrm{U}%
$^{\mathrm{em}}$\textrm{(1)} gauge symmetry, i.e., $U_{I}^{[2]}(\delta
\varphi_{I}^{[2]})\leftrightarrow U_{I}^{[1]}(\delta \varphi_{I}^{[1]}).$
However, when there exist partons, situation changes. The level-2 variability
$\hat{U}^{[2]}(\delta \phi^{\lbrack2]})=\exp(i\lambda^{\lbrack12]}\delta
\phi_{\mathrm{global}}^{[1]})$\ is changed into a new one, i.e.,%
\[
\hat{U}^{[2]}(\delta \phi^{\lbrack2]})=\exp(in^{[2]}\delta \phi_{\mathrm{global}%
}^{[1]}).
\]
As a result, the global variability constraint $\lambda^{\lbrack12]}%
\delta \varphi_{0,I}^{[1]}=\delta \varphi_{0,I}^{[2]}$ is changed into
\[
n^{[2]}\delta \varphi_{0I}^{[1]}=\delta \varphi_{0,I}^{[2]}.
\]
Consequently, the electric charge of the elementary particle changes from $1$
to\ $\frac{n^{[2]}}{\lambda^{\lbrack12]}}$. For example, for the case of
$\lambda^{\lbrack12]}=3,$ and $n^{[2]}=1,$\ with only one level-2 zero, the
elementary particle has $1/3$ electric charge.

\subparagraph{\textrm{SU(N)} gauge field from relative collective motion of
level-2 zeroes}

The \textrm{SU(N)} gauge field comes from the relative collective motion of
level-2 zero lattice. We then introduce the vector field $\mathcal{A}%
_{I,I^{\prime}}=%
%TCIMACRO{\dsum \limits_{a}}%
%BeginExpansion
{\displaystyle \sum \limits_{a}}
%EndExpansion
A_{I,I^{\prime}}T^{a}$ of $\lambda^{\lbrack12]}\times \lambda^{\lbrack12]}$
matrix that plays the role of \textrm{SU(N)} gauge field. Here, $T^{a}$ is
generate of \textrm{SU(N)} group along a-th direction.

\begin{figure}[ptb]
\includegraphics[clip,width=0.63\textwidth]{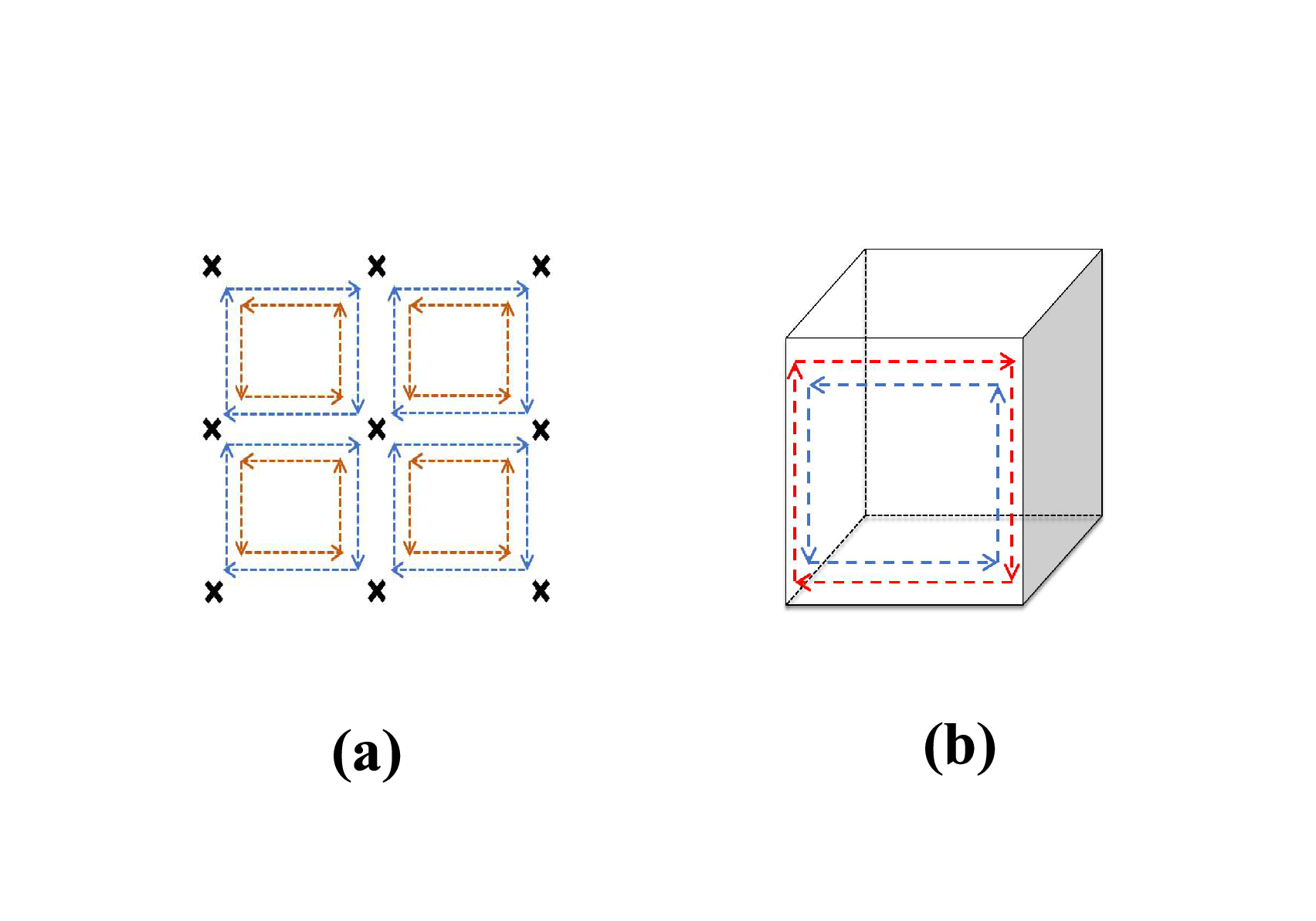}\caption{(a) An
illustration of 2D level-1 zero lattice with uniform relative loop current.
The point denotes lattice sites of topological lattice; (b) An illustration of
a 3D level-1 zero with relative loop current on a plaquette.}%
\end{figure}

Now, to characterize the relative collective motion of level-2 group-changing
space, we must know the colored loop currents $\Phi_{\left \langle
IJKL\right \rangle }^{[2]}$ on level-1 zero lattice. Fig.32(a) is an
illustration of 2D level-1 zero lattice with uniform relative loop current.
Fig.32(b) is an illustration of a 3D level-1 zero with relative loop current
on a plaquette.

A simple approach is to consider the colored loop currents along minimum loops
for four nearest neighbor lattice sites. We denote them by colored flux
$\Phi_{\left \langle IJKL\right \rangle }^{[2]}$ on the plaquette of $IJKL$
lattice site, i.e.,%
\begin{align}
\Phi_{\left \langle IJKL\right \rangle }^{[2]}  &  =%
%TCIMACRO{\dsum \limits_{a}}%
%BeginExpansion
{\displaystyle \sum \limits_{a}}
%EndExpansion
(\Phi_{\left \langle IJKL\right \rangle }^{a,[2]}T^{a})=g\mathcal{A}%
_{IJ}-g\mathcal{A}_{KL}\nonumber \\
&  =%
%TCIMACRO{\dsum \limits_{a}}%
%BeginExpansion
{\displaystyle \sum \limits_{a}}
%EndExpansion
(-\Delta \varphi_{I}^{a,[2]}+\Delta \varphi_{J}^{a,[2]}\nonumber \\
&  +\Delta \varphi_{K}^{a,[2]}-\Delta \varphi_{L}^{a,[2]})T^{a}%
\end{align}
where $g$ denotes the coupling constant for \textrm{SU(N)} gauge field.

The quantum states for fluctuations of level-2 group-changing space are
described by colored loop currents on the all minimum plaquetes $\left \langle
IJKL\right \rangle $,
\[
\{ \Phi_{\left \langle IJKL\right \rangle }^{[2]},\text{ }\left \langle
IJKL\right \rangle \in \mathrm{All}\}.
\]

\paragraph{Effective model}

In this section, we derive the effective model for the quantum gauge theory
with local \textrm{U}$^{\mathrm{em}}$\textrm{(1)}$\times$\textrm{SU(N)}
symmetry. The quantum theory for \textrm{U}$^{\mathrm{em}}$\textrm{(1)} gauge
fields $\mathcal{A}_{\mu}(x,t)$ is same to the case of $\lambda^{\lbrack
12]}=1$. So, we focus on the quantum gauge fields of \textrm{SU(N) }group.

In general, we discuss the effective model for two types of motion of quantum
gauge fields of \textrm{SU(N) }group, one for longitudinal motion, the other
for transverse motion.

On the one hand, we discuss the effective model for longitudinal motion.

Under changing the position of the internal zero inside an elementary
particle, the quantum state of the elementary particle at site $I^{[1]}$
changes correspondingly as%
\begin{equation}
\psi_{1^{[2]},I^{[1]}}^{[1]}\rightarrow(\psi_{1^{[2]},I^{[1]}}^{[1]})^{\prime
}=\hat{U}_{I^{[1]},\mathrm{SU}^{[1]}\mathrm{(N)}}\psi_{1^{[2]},I^{[1]}}^{[1]}.
\end{equation}
After considering the local changing of basis induced by $\hat{U}%
_{J^{[1]},\mathrm{SU(N)}}$, the local coupling between quantum states both
lattice sites changes, i.e.,
\begin{align}
\mathrm{J}(\psi_{1^{[2]},I^{[1]}}^{[1]})^{\dagger}T_{I^{[1]},J^{[1]}}%
\psi_{1^{[2]},J^{[1]}}^{[1]}  &  \rightarrow \mathrm{J}(\psi_{1^{[2]},I^{[1]}%
}^{[1]}\hat{U}_{I^{[1]},\mathrm{SU}^{[1]}\mathrm{(N)}})^{\dagger}\nonumber \\
&  \cdot T_{I^{[1]},J^{[1]}}\nonumber \\
&  \cdot(\hat{U}_{J^{[1]},\mathrm{SU}^{[1]}\mathrm{(N)}}\psi_{1^{[2]},J^{[1]}%
}^{[1]})
\end{align}
where $T_{I^{[1]},J^{[1]}}$ is translation operator from $I$-site to $J$-site.
We define a vector field $\mathcal{A}_{I^{[1]},J^{[1]}}$ to characterize the
local changing of basis%
\begin{equation}
e^{ig\mathcal{A}_{I^{[1]},J^{[1]}}}=(\hat{U}_{I^{[1]},\mathrm{SU}%
^{[1]}\mathrm{(N)}})^{-1}\hat{U}_{J^{[1]},\mathrm{SU}^{[1]}\mathrm{(N)}}%
\end{equation}
where $g$ is coupling constant of $\mathrm{SU(N)}$ non-Abelian gauge field. As
a result, the local coupling between two lattice sites becomes
\begin{equation}
\mathrm{J}(\psi_{1^{[2]},I^{[1]}}^{[1]})^{\dagger}e^{ig\mathcal{A}%
_{I^{[1]},J^{[1]}}}T_{I^{[1]},J^{[1]}}\psi_{1^{[2]},J^{[1]}}^{[1]}.
\end{equation}
The total kinetic energy is obtained as
\begin{align}
\mathcal{\hat{H}}  &  =\mathrm{J}%
%TCIMACRO{\dsum \nolimits_{\left\langle IJ\right\rangle }}%
%BeginExpansion
{\displaystyle \sum \nolimits_{\left \langle IJ\right \rangle }}
%EndExpansion
J(\psi_{1^{[2]},I^{[1]}}^{[1]})^{\dagger}e^{ig\mathcal{A}_{I^{[1]},J^{[1]}}%
}T_{I^{[1]},J^{[1]}}\psi_{1^{[2]},J^{[1]}}^{[1]}\nonumber \\
&  +h.c..
\end{align}

Because the changings of internal group space of level-2 zeroes are locked by
the changing of group space of level-1 zeroes by relative variability
constraint $U_{\mathrm{SU(N)}}^{[2]}\left(  x,t\right)  \equiv
U_{\mathrm{SU(N)}}^{[1]}\left(  x,t\right)  $, the longitudinal degrees of
freedom of non-Abelian gauge fields are always "eaten" by the elementary
particles (level-1 zeroes).

To illustrate the local $\mathrm{SU(N}$\textrm{)} gauge symmetry, we do a
local $\mathrm{SU(N}$\textrm{)} gauge transformation $U_{J^{[1]}%
,\mathrm{SU(N)}}$ that is the transformation of basis of internal quantum
states, i.e.,
\begin{equation}
\psi_{1^{[2]},J^{[1]}}^{[1]}\rightarrow(\psi_{1^{[2]},J^{[1]}}^{[1]})^{\prime
}=\hat{U}_{I^{[1]},\mathrm{SU}^{[1]}\mathrm{(N)}}\psi_{1^{[2]},J^{[1]}}^{[1]},
\end{equation}
and
\begin{align}
g\mathcal{A}_{I^{[1]},J^{[1]}}  &  \rightarrow g\mathcal{A}_{I^{[1]},J^{[1]}%
}^{\prime}\nonumber \\
&  =g\hat{U}_{I^{[1]},\mathrm{SU}^{[1]}\mathrm{(N)}}\mathcal{A}_{I^{[1]}%
,J^{[1]}}(\hat{U}_{J^{[1]},\mathrm{SU}^{[1]}\mathrm{(N)}})^{-1}\\
&  -i(\partial \hat{U}_{I^{[1]},\mathrm{SU}^{[1]}\mathrm{(N)}})(\hat
{U}_{I^{[1]},\mathrm{SU}^{[1]}\mathrm{(N)}})^{-1}.\nonumber
\end{align}
The total kinetic energy turns into
\begin{align}
\mathcal{\hat{H}}  &  \rightarrow \mathcal{\hat{H}}^{\prime}=J%
%TCIMACRO{\dsum \nolimits_{\left\langle \vec{j},\vec{j}^{\prime}\right\rangle
%}}%
%BeginExpansion
{\displaystyle \sum \nolimits_{\left \langle \vec{j},\vec{j}^{\prime
}\right \rangle }}
%EndExpansion
(\psi_{1^{[2]},I^{[1]}}^{^{\prime}[1]})^{\dagger}\nonumber \\
&  \times e^{ig\mathcal{A}_{I^{[1]},J^{[1]}}^{\prime}}T_{\vec{j},\vec
{j}^{\prime}}\psi_{1^{[2]},I^{[1]}}^{\prime \lbrack1]}+h.c..
\end{align}
The Hamiltonian doesn't change,%
\begin{equation}
\mathcal{\hat{H}}=\mathcal{\hat{H}}^{\prime}.
\end{equation}

On the other hand, we discuss the effective model for transverse motion.

Using the approach that is similar to the case of \textrm{U}$^{\mathrm{em}}%
$\textrm{(1)} gauge field, the action of \textrm{SU(N)} gauge field is
obtained as
\[
\mathcal{S}\propto \text{\textrm{Tr}}(%
%TCIMACRO{\dsum \limits_{a}}%
%BeginExpansion
{\displaystyle \sum \limits_{a}}
%EndExpansion
(%
%TCIMACRO{\dsum \limits_{\left\langle ijkl\right\rangle }}%
%BeginExpansion
{\displaystyle \sum \limits_{\left \langle ijkl\right \rangle }}
%EndExpansion
T^{a}\cos(\Phi_{\left \langle IJKL\right \rangle }^{a,[2]}))).
\]
The periodic relative position motion of level-2 zero lattices determines the
coupling constant $g$ of strong interaction. One can use the similar approach
of \textrm{U}$^{\mathrm{em}}$\textrm{(1)} gauge field to obtain it.

After doing \emph{continuation}, in long wave limit, we replace the discrete
numbers $X$ by continuous coordinate $x.$ Now, we have, $\hat{U}%
_{J,\mathrm{SU(N)}}(t)\rightarrow U_{\mathrm{SU(N)}}(\vec{x},t),$
$\mathcal{A}_{I,I^{\prime}}\rightarrow \mathcal{A}_{\mu}(x)\ $and
$\Phi_{\left \langle IJKL\right \rangle }^{[2]}\rightarrow \mathcal{G}_{\mu \nu
}(x)=%
%TCIMACRO{\dsum \limits_{a}}%
%BeginExpansion
{\displaystyle \sum \limits_{a}}
%EndExpansion
(T^{a}F_{\mu \nu}^{a})$. The Lagrangian of non-Abelian gauge field is obtained
as
\begin{equation}
\mathcal{L}_{\mathrm{YM}}(\mathrm{SU(N)})=-\frac{1}{2}\mathrm{Tr}\left(
\mathcal{G}_{\mu \nu}\mathcal{G}^{\mu \nu}\right)  +\mathrm{Tr}\left(
J_{\mathrm{YM}}^{\mu}\mathcal{A}_{\mu}\right)
\end{equation}
where $J_{\mathrm{YM}}^{\mu}=ig\bar{\Psi}\gamma^{\mu}\Psi.$

Now, the non-Abelian gauge symmetry is represented by%
\begin{equation}
\Psi=\hat{U}_{\mathrm{SU(N)}}(\vec{x},t)\Psi
\end{equation}
and%
\begin{align}
\mathcal{A}_{\mu}(\vec{x},t)  &  \rightarrow \hat{U}_{\mathrm{SU(N)}}(\vec
{x},t)\mathcal{A}_{\mu}(\vec{x},t)\left(  \hat{U}_{\mathrm{SU(N)}}(\vec
{x},t)\right)  ^{-1}\nonumber \\
&  -\frac{i}{g}\left(  \partial_{\mu}\hat{U}_{\mathrm{SU(N)}}(\vec
{x},t)\right)  \left(  \hat{U}_{\mathrm{SU(N)}}(\vec{x},t)\right)  ^{-1}.
\end{align}
The gauge strength is defined by $\mathcal{G}_{\mu \nu}$ as%
\begin{equation}
\mathcal{G}_{\mu \nu}=\partial_{\mu}\mathcal{A}_{\nu}-\partial_{\nu}%
\mathcal{A}_{\mu}-ig\left[  \mathcal{A}_{\mu},\mathcal{A}_{\nu}\right]
\end{equation}
or
\begin{align}
\mathcal{G}_{\mu \nu}  &  =G_{\mu \nu}^{a}t^{a}\text{, }\nonumber \\
G_{\mu \nu}^{a}  &  =\partial_{\mu}A_{\nu}^{a}-\partial_{\nu}A_{\mu}%
^{a}+gf^{abc}A_{\mu}^{b}A_{\nu}^{c}.
\end{align}

For QCD, we have $\lambda^{\lbrack12]}=3$. There are three sites of level-2
zero lattice for an elementary particle that is labeled by $1$, $2$, $3.$

For the case of $n^{[2]}=1$, there exists an level-2 zero inside the
elementary particle (or $d$-quark). Now, $d$-quark has three internal quantum
states described by $\left(
\begin{array}
[c]{c}%
d_{1}\left \vert \mathrm{vac}\right \rangle \\
d_{2}\left \vert \mathrm{vac}\right \rangle \\
d_{3}\left \vert \mathrm{vac}\right \rangle
\end{array}
\right)  .$ According to above discussion, there exists \textrm{SU(3)} local
non-Abelian gauge symmetry. When the position of the level-2 zero changes, the
quantum states of d-quarks is changed correspondingly, i.e.,
\begin{equation}
\left(
\begin{array}
[c]{c}%
d_{1}\left \vert \mathrm{vac}\right \rangle \\
d_{2}\left \vert \mathrm{vac}\right \rangle \\
d_{3}\left \vert \mathrm{vac}\right \rangle
\end{array}
\right)  \rightarrow \left(
\begin{array}
[c]{c}%
d_{1}^{\prime}\left \vert \mathrm{vac}\right \rangle \\
d_{2}^{\prime}\left \vert \mathrm{vac}\right \rangle \\
d_{3}^{\prime}\left \vert \mathrm{vac}\right \rangle
\end{array}
\right)  =U_{\mathrm{SU(3)}}\left(  \vec{x},t\right)  \left(
\begin{array}
[c]{c}%
d_{1}\left \vert \mathrm{vac}\right \rangle \\
d_{2}\left \vert \mathrm{vac}\right \rangle \\
d_{3}\left \vert \mathrm{vac}\right \rangle
\end{array}
\right)  .\nonumber
\end{equation}
Here $U_{\mathrm{SU(3)}}\left(  \vec{x},t\right)  =e^{-i\Theta \left(  \vec
{x},t\right)  }$ is the matrix of the representation of $\mathrm{SU(3)}$
group. $\Theta \left(  \vec{x},t\right)  =\sum_{a=1}^{8}\theta^{a}\left(
\vec{x},t\right)  \tau^{a}$ and $\theta^{a}$ are a set of $8$ constant
parameters, and $\tau^{a}$ are $8$ $3\times3$ matrices representing the $8$
generators of the Lie algebra of $\mathrm{SU(3)}$\cite{yang}.

For the case of $n^{[2]}=2$, there exists two level-2 zeroes inside the
elementary particle (or $u$-quark). Now, there exist a "hole" of the level-2
zero. $u$-quark also has three internal quantum states described by $\left(
\begin{array}
[c]{c}%
u_{1}\left \vert \mathrm{vac}\right \rangle \\
u_{2}\left \vert \mathrm{vac}\right \rangle \\
u_{3}\left \vert \mathrm{vac}\right \rangle
\end{array}
\right)  .$ According to above discussion, there exists \textrm{SU(3)} local
non-Abelian gauge symmetry for the $u$-quark. When the position of the "hole"
of level-2 zero changes, the quantum states of u-quarks is changed
correspondingly, i.e.,
\begin{equation}
\left(
\begin{array}
[c]{c}%
u_{1}\left \vert \mathrm{vac}\right \rangle \\
u_{2}\left \vert \mathrm{vac}\right \rangle \\
u_{3}\left \vert \mathrm{vac}\right \rangle
\end{array}
\right)  \rightarrow \left(
\begin{array}
[c]{c}%
u_{1}^{\prime}\left \vert \mathrm{vac}\right \rangle \\
u_{2}^{\prime}\left \vert \mathrm{vac}\right \rangle \\
u_{3}^{\prime}\left \vert \mathrm{vac}\right \rangle
\end{array}
\right)  =U_{\mathrm{SU(3)}}\left(  \vec{x},t\right)  \left(
\begin{array}
[c]{c}%
u_{1}\left \vert \mathrm{vac}\right \rangle \\
u_{2}\left \vert \mathrm{vac}\right \rangle \\
u_{3}\left \vert \mathrm{vac}\right \rangle
\end{array}
\right)  .\nonumber
\end{equation}

Finally, we give a brief summary.

An extra level-2 zero plays the role of source of $\mathrm{SU(N)}$ gauge field
and then carries color degree of freedom. There are $n^{[2]}$ level-2 zeroes
that determine the different quantum internal states of an elementary
particle. For the case of $\mathrm{SU(3)}$, there are three colors called red,
blue and green that correspond to the three positions $1$, $2$, $3$, of
level-2 zero lattice. The collective modes of $\mathrm{SU(3)}$ gauge field are
always called gluons that correspond to the fluctuations of the level-2
zeroes. Therefore, two colored particles interact by relatively shaking their
level-2 zero lattices. This provides a microscopic mechanism for the picture
of strong interaction.

\subsubsection{Summary}

In this section, we study the $N>1$ case and find that the corresponding 2-nd
order variability is reduced into \textrm{U}$^{\mathrm{em}}$\textrm{(1)} local
gauge symmetry and $\mathrm{SU(N)}$ non-Abelian gauge symmetry. The
corresponding theory becomes QED$\mathrm{\times}$QCD. Now, \textrm{SU(3)}
non-Abelian gauge symmetry in QCD and \textrm{U}$^{\mathrm{em}}$\textrm{(1)}
Abelian gauge symmetry in QED will never independence each other.

\subsection{Other issues relevant to Higher-order variants}

\subsubsection{Quark confinement}

It is known that quarks are always confined. The force between two quarks is
linear. Therefore, to free a quark, we must provide infinite energy. In this
section, we will show why there exists quark confinement in Yang-Mills theory.

Before discussing the quark confinement, we discuss the relationship between
geometric property of spacetime and internal dynamics for level-2 zeroes.

In above section, we discuss QCD and QED. The effective model of
QED$\mathrm{\times}$QCD becomes a quantum field theory of $\mathrm{SU(N)}%
\otimes$\textrm{U}$^{\mathrm{em}}$\textrm{(1)} gauge symmetry, of which the
effective Lagrangian is given by
\begin{align}
\mathcal{L}  &  ={\bar{e}}(x)i\gamma^{\mu}\partial_{\mu}e(x)+{\bar{\psi}%
}_{\mathrm{quark}}(x)i\gamma^{\mu}\partial_{\mu}\psi_{\mathrm{quark}}(x)\\
&  +m_{e}{\bar{e}}(x)e(x)+m_{\mathrm{quark}}{\bar{\psi}}_{\mathrm{quark}%
}(x)\psi_{\mathrm{quark}}(x)\nonumber \\
&  -\frac{1}{4}F_{\mu \nu}F^{\mu \nu}+\mathrm{e}{A}_{\mu}(x){j}_{(em)}^{\mu
}(x)\nonumber \\
&  -\frac{1}{2}\mathrm{Tr}\left(  \mathcal{G}_{\mu \nu}\mathcal{G}^{\mu \nu
}\right)  +\mathrm{Tr}\left(  J_{\mathrm{YM}}^{\mu}\mathcal{A}_{\mu}\right)
\nonumber
\end{align}
where $e(x)$ denotes annihilation operator of electrons and $\psi
_{\mathrm{quark}}(x)$ denotes annihilation operator for quarks.

\paragraph{Fractional particle number of level-1 zeroes induced by level-2
zeroes}

To understand quark confinement, the key point is to calculate the induced
particle number of level-2 zeroes for non-Abelian gauge fields.

Due to the variability constraint, each level-1 zero corresponds to
$\lambda^{\lbrack12]}$ level-2 zero, or each level-2 zero corresponds to
$\frac{1}{\lambda^{\lbrack12]}}$ level-1 zero from $\lambda^{\lbrack
12]}=\left \vert \frac{\delta \phi^{\lbrack2]}}{\delta \phi_{\mathrm{global}%
}^{[1]}}\right \vert .$ Therefore, the changing of a level-2 zero induces the
changing of the $\frac{1}{\lambda^{\lbrack12]}}$ level-1 zero. Hence, for a
quark with $n^{[2]}$ level-2 zeroes, induced particle number $\frac{n^{[2]}%
}{\lambda^{\lbrack12]}}$ is topological invariant that is same to its electric
charge
\[
N_{\text{\textrm{quark}}}^{[1]}=\frac{n^{[2]}}{\lambda^{\lbrack12]}}.
\]

Next, we calculate the induced particle number of level-2 zeroes from gluons.

To obtain the contribution from gluons, we calculate the level-2 "momentum"
(or level-2 "wave vector") from gluons. Here, the level-2 "wave vector" is
wave vector for "plane wave" on level-2 zero lattice of a level-1 zero. For
the uniform 2-nd order physical variant, the level-2 "momentum" or level-2
"wave vector" are good quantum number.

For a level-2 zero with "wave vector" $k^{[2]},$ due to the relation $\hat
{k}^{[2]}=i\frac{d}{d\varphi^{\lbrack1]}},$ the level-2 "wave vector"
$k^{[2]}$ becomes the particle number of level-1 zeroes! Then, we have the
\emph{relationship} between "momentum" of level-2 zero and particle number of
level-1 zeroes, i.e.,%
\[
\frac{1}{\pi}\hat{k}^{[2]}=N^{[1]}.
\]
The quantum states of level-2 zero with different level-2 "wave vector" have
the different induced particle number,%
\[
N_{\mathrm{gluon}}^{[1]}=\frac{1}{\pi}\Delta \hat{k}^{[2]}.
\]
Therefore, the changing of distributions of level-2 group-changing elements
for gluons induces the changing of particle number.

Let us calculate the level-2 "wave vectors" from gluons.

When there exists an extra level-2 zero, the number of residue level-2 zeroes
of the level-1 zero becomes $\lambda^{\lbrack12]}-1$. The phase-changing space
of the level-1 zero is $\Delta \varphi^{\lbrack1]}=\pi.$ To make the
phase-changing space to a system with periodic boundary condition, we consider
a lattice site of topological lattice by pairing a level-1 zero with its
conjugate pair that has a phase-changing space of the level-1 zero is
$\Delta \varphi^{\lbrack1]}=-\pi.$ Now, we have a pair of level-1 zeroes, of
which the phase-changing space is defined from $-\pi$ to $\pi$.

Then, the effective model for a level-2 zero on a lattice site of topological
lattice becomes that for a fermionic particle on 1D lattice with
$2\lambda^{\lbrack12]}$ lattice site. In particular, there are twisted
periodic boundary condition,
\begin{equation}
\psi_{I^{[2]}}^{[2]}=\psi_{I^{[2]}+2\lambda^{\lbrack12]}}^{[2]}e^{i\Delta
\varphi_{t}}%
\end{equation}
where $\Delta \varphi_{t}=(2\lambda^{\lbrack12]}-1)\pi$. For the case
$e^{i\hat{k}^{[2]}2\lambda^{\lbrack12]}}=e^{i\Delta \varphi_{t}},$ the level-2
"wave vector" is obtained as
\begin{align*}
\hat{k}_{\pm}^{[2]}  &  =\frac{2\pi M-2\lambda^{\lbrack12]}\pi+\pi}%
{2\lambda^{\lbrack12]}}\\
&  =(2M+1)\frac{\pi}{2\lambda^{\lbrack12]}}-\pi
\end{align*}
where $M\in \left[  0,2\lambda^{\lbrack12]}\right)  $. We then project out the
states with negative level-2 "wave vector" and focus on the positive ones.\

For example, for the case of $\lambda^{\lbrack12]}=2,$ $n^{[2]}=1,$ we have
$\hat{k}_{\pm}^{[2]}=\frac{\pi}{4}(2M+1)-\pi,$ ($M=0,1,2,3$) or $\hat{k}_{\pm
}^{[2]}=\pm \frac{3\pi}{4},$ $\pm \frac{\pi}{4}.$ After projecting out the
quantum states with negative phases, we have two quantum states with positive
level-2 "wave vector" $\hat{k}_{+}^{[2]}=\frac{3\pi}{4},$ $\frac{\pi}{4}$. The
corresponding induced particle number is $\frac{3}{4},$ $\frac{1}{4}.$ For the
case of $\lambda^{\lbrack12]}=3,$ $n^{[2]}=1,$ we have $\hat{k}_{\pm}%
^{[2]}=(2M+1)\frac{\pi}{6}-\pi,$ ($M=0,1,2,3,4,5$) or $\pm \frac{1}{6}\pi,$
$\pm \frac{1}{2}\pi,\  \pm \frac{5}{6}\pi$. After projecting out the quantum
states with negative phases, we have positive three quantum states with
level-2 "wave vector" $\hat{k}_{+}^{[2]}=\frac{1}{6}\pi,\frac{1}{2}\pi
,\frac{5}{6}\pi.$ The corresponding induced particle number is $\frac{1}{6},$
$\frac{1}{2},\  \frac{5}{6}.$

In general, the average wave vector for all quantum states of extra level-2
zeroes is defined by
\[
\left \langle \Delta \hat{k}^{[2]}\right \rangle =%
%TCIMACRO{\dsum \limits_{M}}%
%BeginExpansion
{\displaystyle \sum \limits_{M}}
%EndExpansion
(a_{M}\Delta k_{M}^{[2]})
\]
where $a_{M}$\ is the weight of the quantum states of $M$-th level-2 "wave
vector" $\Delta k_{M}^{[2]}$. $M$\ labels states with positive "wave vectors".
The corresponding induced particle number is
\begin{align*}
N_{\mathrm{gluon}}^{[1]}  &  =\frac{1}{\pi}\left \langle \Delta \hat{k}%
^{[2]}\right \rangle \\
&  =\frac{1}{\pi}%
%TCIMACRO{\dsum \limits_{M}}%
%BeginExpansion
{\displaystyle \sum \limits_{M}}
%EndExpansion
(a_{M}\Delta k_{M}^{[2]}).
\end{align*}
As a result, the average wave vector for gluons must finite, $\left \langle
\Delta \hat{k}^{[2]}\right \rangle =%
%TCIMACRO{\dsum \limits_{M}}%
%BeginExpansion
{\displaystyle \sum \limits_{M}}
%EndExpansion
(a_{M}\Delta k_{M}^{[2]})\neq0$ and the corresponding induced particle number
of level-1 zero $N_{\mathrm{gluon}}^{[1]}=\frac{1}{\pi}\left \langle \Delta
\hat{k}^{[2]}\right \rangle $ is always not zero.

\paragraph{Quark confinement as effect from topological defects on spacetime}

From above discussion, the total induced particle number are obtained as%
\[
N^{[1]}=N_{\text{\textrm{quark}}}^{[1]}+N_{\mathrm{gluon}}^{[1]}%
\]
where
\[
N_{\text{\textrm{quark}}}^{[1]}=\frac{n^{[2]}}{\lambda^{\lbrack12]}},
\]
and%
\begin{align*}
N_{\mathrm{gluon}}^{[1]}  &  =\frac{1}{\pi}\left \langle \Delta \hat{k}%
^{[2]}\right \rangle \\
&  =\frac{1}{\pi}%
%TCIMACRO{\dsum \limits_{M}}%
%BeginExpansion
{\displaystyle \sum \limits_{M}}
%EndExpansion
(a_{M}\Delta k_{M}^{[2]}).
\end{align*}

On the other hand, in Ref.\cite{kou1}. it was found that the
generation/annihilation of an elementary particle leads to
contraction/expansion $\pi$-phase changing of (level-1) group-changing space
along an arbitrary direction. This leads to unification of matter and
spacetime. The the changing of 3-volume of spacetime $\Delta V$ is determined
by its particle number%
\begin{equation}
N_{F}=N^{[1]}=-(4\pi l_{0}^{3})^{-1}\Delta V.
\end{equation}

As a result, according to the relationship between particle number and the
changing of 3-volume, $N^{[1]}=-(4\pi l_{0}^{3})^{-1}\Delta V,$ the induced
particle number in non-Abelian gauge fields leads to changing of 3-volume
changing, i.e.,
\begin{align*}
\Delta V  &  =-4\pi l_{0}^{3}\times N^{[1]}\\
&  =-4\pi l_{0}^{3}(N_{\text{\textrm{quark}}}^{[1]}+N_{\mathrm{gluon}}^{[1]}).
\end{align*}
Therefore, an important feature is the changings of quantum states of level-2
zeroes induce the changings of 3-volumes of spacetime!

The changing of 3-volume on quantum spacetime will disturb the zero lattice
and make a local frustration effect on it. See the illustration in Fig.9. The
extra 3-volume leads to frustration effect of spacetime. As a result, the
quark confinement becomes a physical consequence of the malposition of the
quantum spacetime (or level-1 zero lattice).

In addition, to quantitatively characterize the linear potential from
confinement, we consider the effect of the topological defects on quantum
spacetime. Another important feature is that the changings of quantum states
of level-2 zeroes change the fractional number of magnetic monopole of spacetime.

According to the relationship between changing of 3-volume and number of
magnetic monopole of spacetime $\Delta V=4\pi l_{0}^{3}q_{m}$ in
spacetime\cite{kou1}, the changing of 3-volumes leads to a fractional monopole
number,
\begin{align*}
\Delta q_{m}  &  =\frac{\Delta V}{4\pi l_{0}^{3}}\\
&  =-(N_{\text{\textrm{quark}}}^{[1]}+N_{\mathrm{gluon}}^{[1]}).
\end{align*}
\ Here, the number of magnetic monopole of spacetime is defined by
\begin{equation}
q_{m}=\frac{1}{4\pi}%
%TCIMACRO{\doint \nolimits_{\mathcal{S}}}%
%BeginExpansion
{\displaystyle \oint \nolimits_{\mathcal{S}}}
%EndExpansion
F_{\mathcal{S}}^{IJ},
\end{equation}
where $F_{\mathcal{S}}^{IJ}=dA_{\mathcal{S}}^{IJ}+A_{\mathcal{S}}^{Ic}\wedge
A_{\mathcal{S}}^{cJ}\equiv-A_{\mathcal{S}}^{I0}\wedge A_{\mathcal{S}}^{J0}.$
$A_{\mathcal{S}}^{IJ}$ are auxiliary gauge fields for quantum spacetime. See
the detailed discussion in Ref.\cite{kou1}.\

Now, the objects with fractional particle number induced by quarks or gluons
trap fractional number of magnetic monopole of spacetime. We consider a pair
of quark and anti-quark that are an object with positive fractional magnetic
monopole and the other with negative fractional magnetic monopole,
respectively. The total fractional particle number is zero, $N_{\mathrm{total}%
}^{[1]}=0$. So, we must have a string connecting the two ends of the pair of
the quark and anti-quark. In 3D case, the energy linearly increases as the
increasing the length of the string. As a result, a quark with fractional
fermion number or fractional magnetic charge must have infinite energy and
thus cannot exist. This explains the \emph{quark confinement} that can be
regarded as an effect of topological defect on spacetime.

\begin{figure}[ptb]
\includegraphics[clip,width=0.7\textwidth]{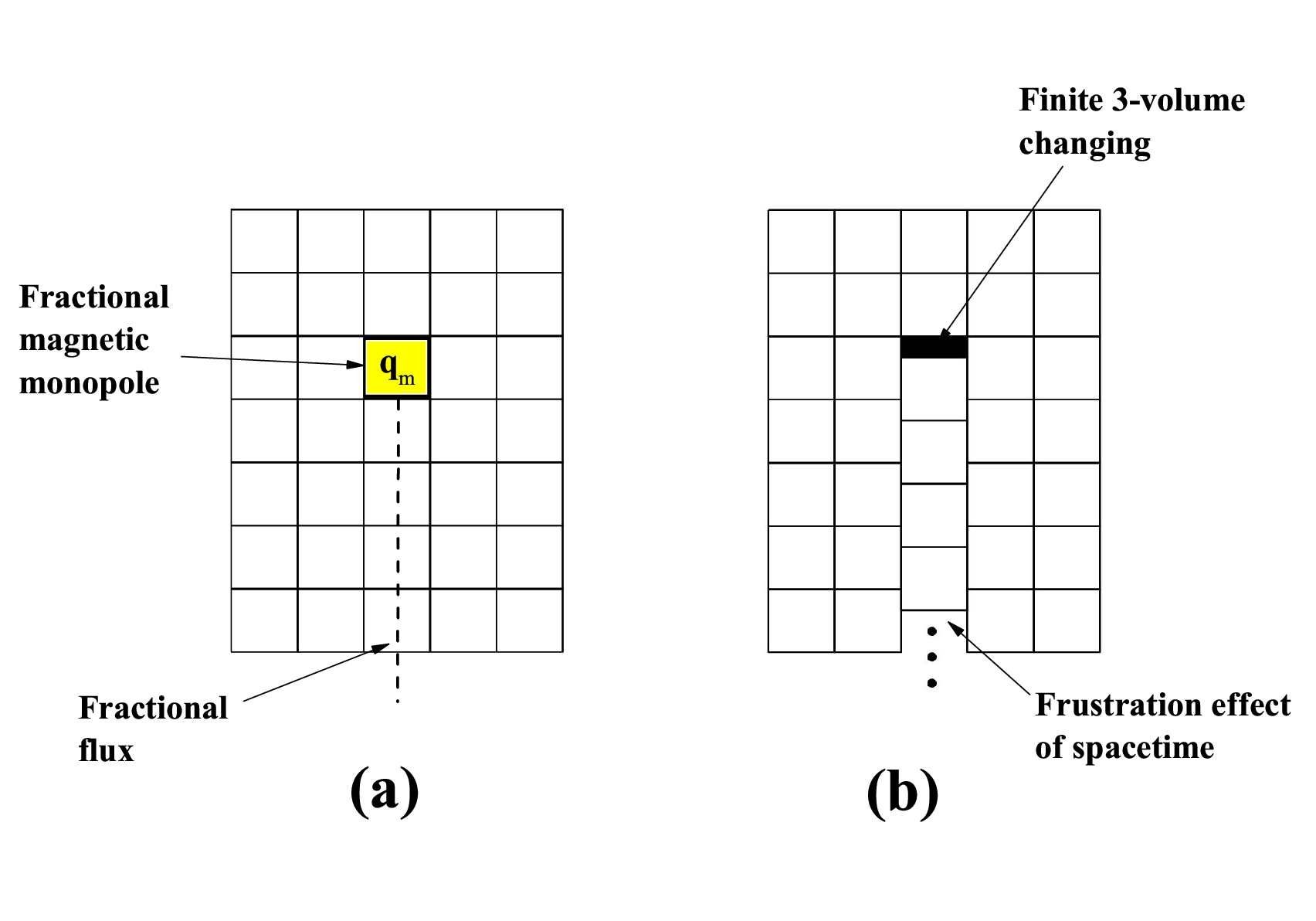}\caption{Quark
confinement as an effect of geometry frustration for quantum spacetime (or
zero lattice). (a) is an illustration of a zero lattice with a fractional
magnetic monopole, of which the fractional flux is trapped; (b) is an
illustration of a zero lattice with extra 3-volume that leads to frustration
effect on spacetime.}%
\end{figure}

See the illustration in Fig.9(a). Now, we have induced a pair of induced
fractional fluxes $\pm \Delta \phi$, between which the hopping terms change from
$J$\ to $Je^{i\Delta \phi}$. After considering the string connecting the pair
of induced fractional fluxes $\pm \Delta \phi,$ the energy $\Delta E$ of the
ground state linearly increases with $r$ (the distance between two induced
fractional fluxes $\pm \Delta \phi$), i.e.,
\[
\Delta E\sim \sigma r
\]
where the string tension $\sigma$ is estimated as%
\begin{align*}
\sigma &  \sim J\left \vert \sin \Delta \phi \right \vert =J\left \vert \sin(\pi
N^{[1]})\right \vert \\
&  =J\left \vert \sin(\pi(N_{\text{\textrm{quark}}}^{[1]}+N_{\mathrm{gluon}%
}^{[1]}))\right \vert .
\end{align*}
For 3D case, the situation is similar.

In summary, confinement becomes a problem of spacetime, or physical
consequence of topological defect with fractional monopole charge of quantum
spacetime, rather than just a problem of quantum field theory. In Fig.10, we
plot the triangular equivalence principle for quark confinement. This is an
intrinsic relationship between fractional particle particle, fractional
3-volume of quantum spacetime and fractional magnetic monopole.

It looks like that our results contradict those from lattice gauge theory, in
which there doesn't exist lattice distortion during confinement. We point the
lattice gauge theory can be a theory under kinetic representation with uniform
lattice and Gamma matrices. \begin{figure}[ptb]
\includegraphics[clip,width=0.7\textwidth]{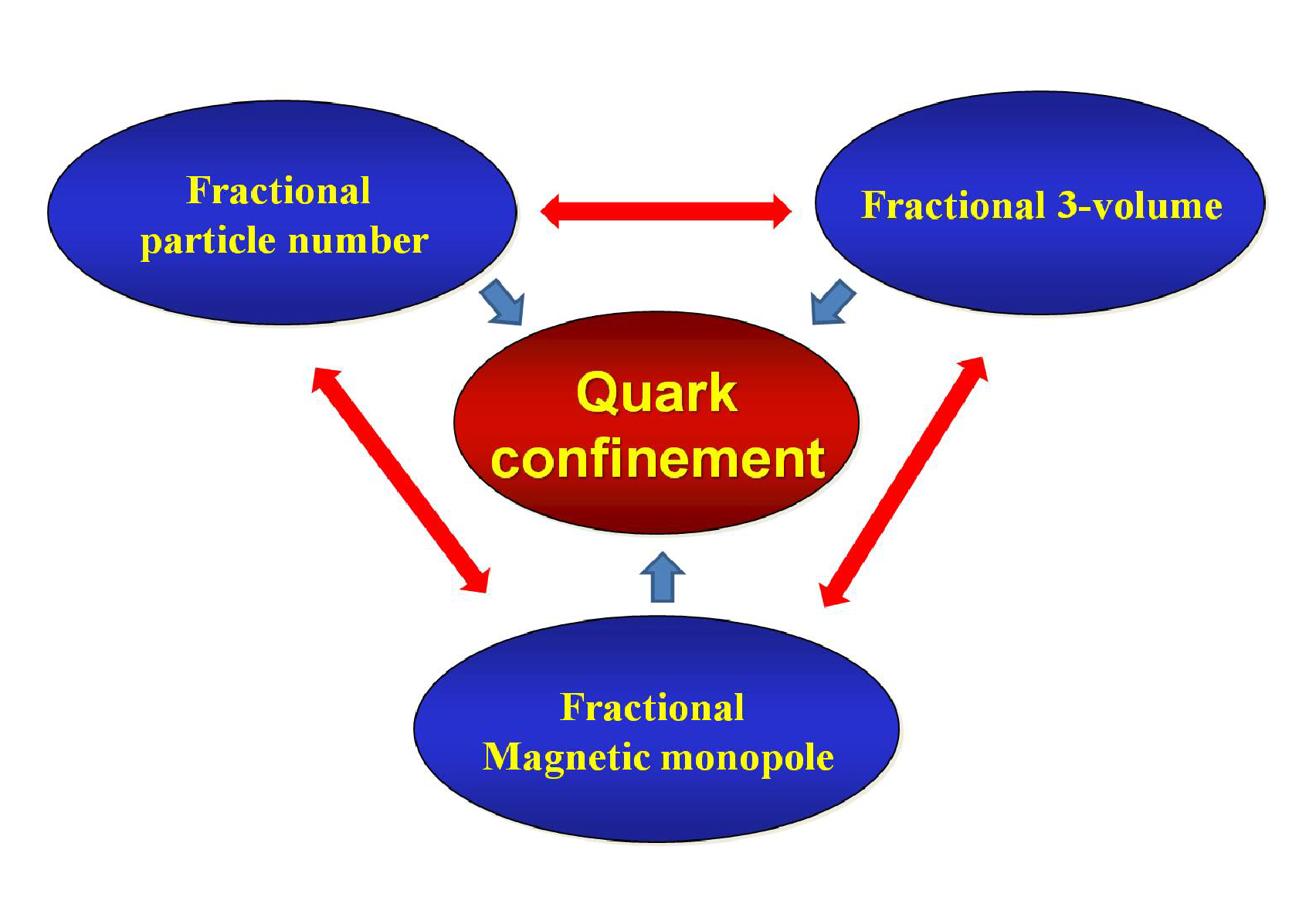}\caption{An
illustration of the triangular equivalence principle for quark confinement.
This is an intrinsic relationship between fractional particle particle,
fractional 3-volume of quantum spacetime and fractional magnetic monopole. }%
\end{figure}

\subsubsection{Effective gauge theory with internal dynamics and Kluza-Klein
mechanism}

\paragraph{Review on Kluza-Klein mechanism}

Kaluza-Klein theory is a five dimension (5D) extension of general relativity
which aim is to provide a unified description of gravitational and
electromagnetic interaction in a purely geometric description\cite{Kaluza}.
According to Kaluza-Klein theory, fifth dimension of the 5D spacetime becomes
a circus of length $L$, that is a compact space. Under dimensional reduction
along the fifth dimension, the particles have periodic motion and have finite
momentum. The non-trivial result obtained by the extra-dimensional
Kaluza-Klein theories relies on the emergence from the multi-dimensional
Einstein-Hilbert Lagrangian of the actions for electromagnetic field and
Yang-Mills field.

For example, we consider a dimensional reduction from 5D spacetime to 4D
spacetime.\ The compactness of the fifth dimension implies a quantization of
momentum along the fifth direction:
\begin{equation}
p_{5}=\frac{2\pi \hbar}{L}n\qquad n\in \mathbb{Z},
\end{equation}
where $L$ is the length of the circus describing the fifth dimension. The
electric charge \textrm{e} for the emergent electromagnetic field is obtained
as
\begin{equation}
\mathrm{e}=4\pi \frac{\hbar \sqrt{G}}{{L}c}n.
\end{equation}

In addition, for spacetime by doing compactification on extra higher
dimensional subspace, if the \emph{Holonomy} $\mathrm{Hol}(g)$ for sub-space
is $\mathrm{SU(N)}$, we have Yang-Mills fields with $\mathrm{SU(N)}$ gauge
symmetry. For example, if we consider a $6$ dimensional space rather than one,
we may get an effective \textrm{SU(2)} Yang-Mills field. If we consider a $7$
dimensional spacetime by doing compactification on extra three dimensional
space into Calabi-Yau manifold, we get an effective\textrm{ SU(3)} Yang-Mills field.

\paragraph{Effective gauge theory with internal dynamics}

Firstly, we consider a 2-nd order physical variant with dynamical level-2
group-changing space,
\begin{align*}
V_{\mathrm{\tilde{G}}_{\mathrm{dynamic}}^{[2]}\mathrm{,\tilde{S}\tilde
{O}^{[1]}(d+1)},d+1}^{[2]}  &  :\mathrm{C}_{\mathrm{\tilde{G}%
_{\mathrm{dynamic}}^{[2]}}}^{[2]}\\
&  \Longleftrightarrow \mathrm{C}_{\mathrm{\tilde{S}\tilde{O}^{[1]}(d+1),}%
d+1}^{[1]}\\
&  \Longleftrightarrow \mathrm{C}_{d+1}.
\end{align*}
\textit{\ }The elements of two group-changing spaces are $\delta
\phi_{\mathrm{global}}^{[2]}=\sqrt{%
%TCIMACRO{\dsum \limits_{\mu}}%
%BeginExpansion
{\displaystyle \sum \limits_{\mu}}
%EndExpansion
((\delta \phi^{\mu})^{[2]})^{2}}$ and $\delta \phi_{\mathrm{global}}^{[1]}%
=\sqrt{%
%TCIMACRO{\dsum \limits_{\mu}}%
%BeginExpansion
{\displaystyle \sum \limits_{\mu}}
%EndExpansion
((\delta \phi^{\mu})^{[1]})^{2}},$ respectively. The ratio between the changing
rates of the two group-changing spaces $\mathrm{C}_{\mathrm{\tilde
{G}_{\mathrm{dynamic}}^{[2]},}d}^{[2]}((\Delta \phi^{\mu})^{[2]})\ $and
$\mathrm{C}_{\mathrm{\tilde{S}\tilde{O}^{[1]}(d+1),}d+1}^{[1]}((\Delta
\phi^{\mu})^{[1]})$\textit{ }is integer multiple $(\lambda^{\mu})^{[12]}.$ In
particular, $\mathrm{C}_{\mathrm{\tilde{G}_{\mathrm{dynamic}}^{[2]}}}^{[2]}$
is an $\mathrm{\tilde{S}\tilde{O}^{[2]}}$\textrm{(b+1)} group-changing space
with dynamics\textit{,}
\[
\mathrm{\tilde{G}_{\mathrm{dynamic}}^{[2]}}\equiv \mathrm{\tilde{S}\tilde
{O}_{\mathrm{dynamic}}^{[2]}}\text{\textrm{(b+1)}}\mathrm{.}%
\]

Now, we have an $\mathrm{\tilde{S}\tilde{O}_{\mathrm{dynamic}}^{[2]}}%
$\textrm{(b+1)} group-changing space that is characterized by massless Dirac
partons. If there doesn't hopping of the Dirac partons, we have a theory
of\  \textrm{U(1)}$\times$\textrm{SU(}$\lambda^{\lbrack12]}$\textrm{)} gauge
fields that describes the dynamics of the non-Abelian $\mathrm{\tilde{S}%
\tilde{O}_{\mathrm{dynamic}}^{[2]}(b+1)}$ Clifford group space on Cartesian
space. Here, $\lambda^{\lbrack12]}$ is the number of level-2 zeroes of a
level-1 zero. In above sections, we had discussed this case in detail.
However, if the level-2 zero has its dynamics, the situation changes. Let us
on focus on this case.

If the level-2 zero has its dynamics, the energy degeneracy between different
quantum states for level-2 zeroes disappears. The local expansion and
contraction of the dynamical $\mathrm{\tilde{S}\tilde{O}_{\mathrm{dynamic}%
}^{[2]}(b+1)}$ Clifford group-changing space turns into gauge fields. For
example, for the case of $b=0,$ we have a dynamical \textrm{\={U}}%
$_{\mathrm{dynamic}}$\textrm{(1)} group-changing space, of which the low
energy theory is just \textrm{U}$^{\mathrm{em}}$\textrm{(1)} gauge field. In
the following parts of this section, we give detailed discussion on it.

\paragraph{Effective \textrm{U}$^{\mathrm{em}}$\textrm{(1)} gauge theory from
Kluza-Klein mechanism}

Firstly, we consider the Kaluza-Klein theory for a 5D 1-st order physical
variant that is defined as
\[
V_{\mathrm{\tilde{S}\tilde{O}(4+1)},4+1}:\mathrm{C}_{\mathrm{\tilde{S}%
\tilde{O}(4+1),}4+1}^{[1]}\Longleftrightarrow \mathrm{C}_{4+1}.
\]
Then, we do compactification on the fifth dimension. After the
compactification, we have an effective 2-nd order physical variant
$V_{\mathrm{\tilde{U}_{dynamic}^{[2]}(1),\tilde{S}\tilde{O}^{[1]}(3+1)}%
,3+1}^{[2]}$ with dynamical level-2 group-changing space $\mathrm{\tilde
{U}_{dynamic}^{[2]}(1)}.$

Now, the number of level-2 zeroes for a uniform physical variant is
\begin{equation}
\frac{L}{l_{p}}%
\end{equation}
where $l_{p}$ is distance of lattice unit of level-2 zero lattice and $L$ is
the length of the circus describing the fifth dimension. Due to the condition
of compactification, we must have the following constraint
\[
n^{[2]}\equiv1.
\]
That means for an elementary particle with a level-1 zero, the number of
level-2 zero is always $1$.

To deal with the dynamical level-2 group-changing space, we write down the
hopping Hamiltonian for the level-2 zero. The hopping term between two nearest
neighbor sites $I^{[2]}$ and $J^{[2]}$ on topological lattice becomes
\begin{equation}
\mathcal{H}_{\left \{  i,j\right \}  }^{[2]}=\mathcal{J}^{[2]}(c_{I^{[2]}}%
^{[2]})^{\dagger}\mathbf{T}_{\left \{  I^{[2]},J^{[2]}\right \}  }c_{J^{[2]}%
}^{[2]}(t)
\end{equation}
where $\mathbf{T}_{\left \{  I^{[2]},J^{[2]}\right \}  }$ is the transfer matrix
between two nearest neighbor sites $I^{[2]}$ and $J^{[2]}$ and $c_{I^{[2]}%
}^{[2]}(t)$ is the annihilation operator of level-2 zero at the site $I^{[2]}%
$. As a result, the total Hamiltonian is
\[
\mathcal{H}^{[2]}=\mathcal{J}^{[2]}%
%TCIMACRO{\dsum \limits_{\{I^{[2]},J^{[2]}\}}}%
%BeginExpansion
{\displaystyle \sum \limits_{\{I^{[2]},J^{[2]}\}}}
%EndExpansion
(c_{I^{[2]}}^{[2]})^{\dagger}\mathbf{T}_{\left \{  I^{[2]},J^{[2]}\right \}
}c_{J^{[2]}}^{[2]}(t).
\]
In the continuum limit, we have an effective 1D model of massless Dirac model
at half filling.

The quantum states for the extra level-2 zero are plane waves with fixed
momenta along the fifth direction,
\begin{equation}
p_{5}=\frac{2\pi \hbar}{L}n,\text{ }n\in \mathbb{Z}.
\end{equation}
The corresponding energy of the level-2 zero is
\begin{align*}
E_{n}  &  =cp_{5}\\
&  =\frac{2\pi \hbar c}{L}n.
\end{align*}

Let us show the local \textrm{U}$^{\mathrm{em}}$\textrm{(1) }gauge symmetry.
On one hand, we have the global \textrm{U}$^{[2]}$\textrm{(1)} symmetry for
all quantum plane waves $e^{ik_{5}x_{5}}=e^{i\frac{2\pi}{L}nx_{5}}$ by
changing the initial point of the level-2 zero lattice (or the circus along
the fifth dimension); on the other hand, there exist the global \textrm{U}%
$^{[1]}$\textrm{(1)} symmetry for the corresponding level-1 zero. After
considering the variability constraint between the phase changings,
$\delta \varphi^{\lbrack2]}=k_{5}\delta x_{5}=\delta \varphi^{\lbrack1]}$, we
have an effective local \textrm{U}$^{\mathrm{em}}$\textrm{(1) }gauge symmetry.

The gauge fluctuations come from locally fluctuations of the circus along the
fifth dimension. When the length of $L$ changes due to fluctuations of quantum
spacetime, $L\rightarrow L^{\prime}$, the energy changes correspondingly,
\begin{align*}
E_{n}  &  \rightarrow E_{n}^{\prime}=c(p_{5})^{\prime}\\
&  =\frac{2\pi \hbar c}{L^{\prime}}n.
\end{align*}
The energy different is $\frac{2\pi \hbar c}{L}\frac{\delta L}{L}n$. The phase
factor of level-2 zero changes correspondingly,
\begin{align*}
e^{ik_{5}x_{5}}  &  \rightarrow e^{ik_{5}^{\prime}x_{5}}=e^{i\frac{2\pi
}{L^{\prime}}nx_{5}}\\
&  =e^{i\frac{2\pi}{L}nx_{5}}e^{i\frac{2\pi}{L^{2}}n\delta Lx_{5}}.
\end{align*}
Because $\delta L$ may be locally changed, we have local gauge transformation.

We next calculate the coupling constant $\mathrm{e}$ that is ratio between the
effective Planch constant $\hbar^{\lbrack2]}$ and Planck constant $\hbar$.

Due to the time dependent phase factor $e^{i\omega_{n}t}$ of wave function of
the 1D system, the level-2 group-changing space has a global motion with
angular momentum $\omega_{0}^{[2]}=\frac{2\pi c}{L}.$ Therefore, after
considering the changing of energy, the density of angular momentum is
obtained as
\[
\rho_{\omega_{0}^{[2]}}=\frac{\delta E_{n}}{\delta \omega}\mid_{\omega
=\omega_{0}^{[2]}}=n\hbar.
\]
However, there are $\frac{L}{l_{p}}$ level-2 zeroes on the circus. For each
level-2 zero, we have effective Planck constant to be
\[
\hbar^{\lbrack2]}=\frac{l_{p}}{{L}}n\hbar
\]
that is an effective Planck constant by scaling the original one by a ratio
$\frac{nl_{p}}{{L}}$. As a result, we point out that, for the states with
$p_{5}=\frac{2\pi \hbar}{L}n$ ($n>1$), the effective Planck constant is
$\hbar^{\lbrack2]}=\frac{l_{p}}{{L}}\hbar n$ that depends on $n$.

Finally, the unit of coupling constant $\mathrm{e}$ is derived as
\begin{equation}
\mathrm{e}=c\hbar^{\lbrack2]}=\frac{l_{p}}{{L}}\hbar c.
\end{equation}
Different quantum states of different momenta $\frac{2\pi \hbar}{L}n$ have
different electric charge
\begin{equation}
\mathrm{e}^{\prime}=n\mathrm{e}=\frac{l_{p}}{{L}}n\hbar c.
\end{equation}

In addition, the dynamics of level-2 zero provides mass gap for level-1 zero.
The mass gap $m$ is $cp_{5}=\frac{2\pi}{L}n\hbar c$. So, we have a mass-charge
locking relationship between mass gap $m$ and electric charge $\mathrm{e}%
^{\prime}$,%
\[
m=cp_{5}\equiv \frac{2\pi}{l_{p}}\mathrm{e}^{\prime}.
\]
This large mass gap of elementary particles indicates the effective gauge
theory from Kluza-Klein mechanism will never be the correct theory in the
standard model.

\paragraph{Effective Yang-Mills theory from Kluza-Klein mechanism}

In this part, we discuss an effective Yang-Mills theory from Kluza-Klein compactification.

Firstly, we consider the Kaluza-Klein theory of a (3+1+M)D physical variant.
The original (3+1+M)D 1-st order physical variant is
\begin{equation}
V_{\mathrm{\tilde{S}\tilde{O}(3+1+M)},3+1+M}:\mathrm{C}_{\mathrm{\tilde
{S}\tilde{O}(3+1+M),}3+1+M}\Longleftrightarrow \mathrm{C}_{3+1+M}.
\end{equation}
Then, we do compactification on the extra $M$ dimensions. After the
compactification, we have an effective 2-nd order physical variant
$V_{\mathrm{\tilde{G},\tilde{S}\tilde{O}^{[1]}(5+1)},3+1}^{[2]}.$ Here, we
have $\mathrm{Hol}(g)=\mathrm{\tilde{G}}$. Due to the condition of
compactification, we also have the following constraint
\[
n^{[2]}\equiv1.
\]

If the level-2 group-changing space has no dynamics, we have a usual theory of
\textrm{U}$^{\mathrm{em}}$\textrm{(1)}$\times$\textrm{SU(}$\lambda
^{\lbrack12]}$\textrm{)} gauge field. However, now, the level-2 group-changing
space indeed has dynamics. We need to re-examine this issue. In the continuum
limit, we have an effective model of massless Dirac model at half filling on a
M dimensional sub-space with extra magnetic monopoles inside the center.

Now, the situation is similar to the case of a surface Dirac fermion on a
surface of M dimensional compact sub-space, of which extra magnetic monopoles
exist at center. The quantum states belong to different representations of
$\mathrm{Hol}(g)=\mathrm{\tilde{G}.}$ The energy levels $E_{n}$ is
proportional to $c/R$ where $R$ is its size. In particular, there may exist
degenerate quantum states with zero energy at "Fermi surface". In general, the
energy degeneracies comes from symmetry of M dimensional compact sub-space.
For example, for the case of $\mathrm{Hol}(g)=\mathrm{SU(N),}$ there may exist
$\mathrm{N}$-fold zero modes. If the M dimensional compact sub-space has
non-trivial topological structure, there exists the contribution of the energy
degeneracy from topology, i.e., $\mathrm{Index}(D)=n_{+}-n_{-}=\frac{1}%
{2}\left \vert \chi \right \vert $ where $\chi$ is its Eular number\cite{ka}.
Here, we don't consider the topological degeneracy and set $\chi=1$.

As a result, one can use the approach for the effective Yang-Mills theory of
local \textrm{SU(}$\mathrm{N}$\textrm{)} symmetry to deal with the low energy
degrees of freedom of the system. The "\textrm{N}" comes from the \textrm{N}
degeneracy of quantum states inside an elementary particles. The \textrm{N}
degenerate quantum states on the level-2 group-changing space play the role of
the \textrm{N} internal states of an elementary particles. Due to non-Abelian
variability constraint, in low energy limit, we have an effective local
\textrm{SU(}$\mathrm{N}$\textrm{)} symmetry rather than \textrm{U}%
$^{\mathrm{em}}$\textrm{(1)}$\times$\textrm{SU(}$\lambda^{\lbrack12]}%
$\textrm{)}. The quantum fluctuations of non-Abelian \textrm{SU(}$\mathrm{M}%
$\textrm{)} gauge fields (or gluons) come from the size changings of the
level-2 group-changing space that keep the shape invariant. So, we point out
that the symmetry-protected degeneracy of zero modes inside an elementary
particle plays the role of color number rather than the flavor number!

Then, we discuss the issues of confinement. \emph{Why people cannot understand
confinement with the help of effective SU(N) gauge theory from Kluza-Klein
mechanism?} Let us give an explanation.

The confinement is an effect of frustration effect of quantum spacetime by
fractional number of magnetic monopoles. To realize the confinement effect in
\textrm{SU(N)} gauge theory from Kluza-Klein mechanism, we point out that the
key point is the existence of the induced fractional particle number.

We consider massless Dirac model at half filling on M dimensional compact
sub-space with extra magnetic monopoles inside the center. The vacuum
expectation value of the fermion number $\langle N^{\mathrm{f}}\rangle$ is
related to the spectral asymmetry of the Dirac Hamiltonian
\begin{align}
\langle \Delta N_{F}\rangle &  =-\frac{1}{2}\int_{-\infty}^{\infty}dE\,
\frac{1}{\pi}\text{\textrm{Im} }\mathrm{Tr\,}(\frac{1}{H-E-i\epsilon})\, \\
&  \times \, \text{\textrm{sign}}(E)
\end{align}
where $H$ is the Hamiltonian of the fermionic elementary
particles\cite{NiemiSemenoff-1986}. And the fermionic number is also related
to the Atiah-Patodi-Singer invariant $\mathcal{\eta}_{_{H}}=-\frac{1}%
{2}\langle \Delta N_{F}\rangle$ which represents the difference between the
number of states with positive and negative energy. It represents a fact that
magnetic monopoles in extra dimensions induces fractional particle number%
\begin{equation}
\langle \Delta N^{[1]}\rangle.
\end{equation}
For the case of \textrm{SU(N)}, we \emph{assume} that the induced quantum
number is $\frac{1}{N}$.\textrm{ }Now, we have fractional number of extra
magnetic monopoles
\[
\Delta q_{m}=-\langle \Delta N^{[1]}\rangle.
\]
Based on the\emph{ assumption} of $\frac{1}{N}$ induced quantum number, the
quark confinement occurs. In addition, quantum fluctuations of the M
dimensional compact sub-space change its size. This effect will also lead to
induced fractional fermionic number. The result is then consistent to the case
of intrinsic Yang-Mills theory discussed above. As a result, we get a correct
effective Yang-Mills theory from Kluza-Klein compactification on a (3+1+M)
dimensional quantum spacetime.

In summary, based on Kluza-Klein compactification, the theory for physical
variants provides a deep insight on different branches of mathematics and
physics, including the APS index theorem for induced fractional fermion
number, topology of Calabi-Yau manifolds, non-perturbative theory for
Yang-Mills fields, ...

\subsubsection{Axion and Strong CP violate}

\paragraph{Axion and Strong CP violate}

The Standard Model may have a $\theta$ term for the ${\mathrm{SU}}(3)$ gluon
field,
\begin{equation}
\frac{\theta}{8\pi^{2}}\int \mathrm{tr}(\mathcal{G}\wedge \mathcal{G}^{\ast}).
\end{equation}
This is a CP-violating term. Due to the chiral anomaly, the value of $\theta$
may be shift by the quark mass matrix $m$. Therefore, the true value of
$\theta$ becomes ${\overline{\theta}}=\theta-\arg \det m.$ Consequently, a
CP-violating neutron electric dipole moment may exist. However, experiments
indicate a tiny ${\overline{\theta}}$, i.e., $|{\overline{\theta}}%
|\lesssim10^{-10}.$ \emph{Why is this number so small?} The puzzle of small
${\overline{\theta}}$ is known as the Strong CP Problem.

A possible solution to solve Strong CP Problem is the existence of an axion
field\cite{axion}. The reason comes from that QCD dynamics may generate a
potential for ${\overline{\theta}}$ which is minimized at the CP-preserving
value ${\overline{\theta}}=0$. Due to quantum tunneling effect from
instantons, one gets semiclassical contributions to an effective potential for
${\overline{\theta}}$ that is proportional to ${\mathrm{e}}^{-8\pi^{2}/g^{2}%
}\mathrm{\cos}\theta$.

In this section, we will answer additional three questions: 1) What is 2-nd
order physical variant for gauge fields with finite ${\overline{\theta}}$? 2)
What is the difference between our universe and that with finite
${\overline{\theta}}$? 3) why ${\overline{\theta}\equiv0}$ in our universe.

\paragraph{Physical variant for gauge field with finite theta term}

What is 2-nd order physical variant for gauge fields with finite
${\overline{\theta}?}$ In this section, we will answer this question. The
answer is ($d+2$)-dimensional 2-nd order $\mathrm{\tilde{S}\tilde{O}}%
$\textrm{(d+1)}\textit{ }physical variant $V_{\mathrm{\tilde{U}}%
^{[2]}\mathrm{(1)},\mathrm{\tilde{S}\tilde{O}}^{[1]}\mathrm{(d+2)},d+2}^{[2]}%
$, of which there exists a finite theta term..

The ($d+2$)-dimensional 2-nd order $\mathrm{\tilde{S}\tilde{O}}$%
\textrm{(d+1)}\textit{ }physical variant $V_{\mathrm{\tilde{U}}^{[2]}%
\mathrm{(1)},\mathrm{\tilde{S}\tilde{O}}^{[1]}\mathrm{(d+2)},d+2}^{[2]}$ is a
higher-order mapping between $\mathrm{C}_{\mathrm{\tilde{U}}^{[2]}%
\mathrm{(1)}}^{[2]}$\textit{, }\textrm{\~{S}\~{O}(d+1)}\textit{ }%
group-changing space\textit{ }$\mathrm{C}_{\mathrm{\tilde{S}\tilde{O}%
(d+2)},d+2}^{[1]}$\textit{\ }and a rigid spacetime\textit{ }$\mathrm{C}%
_{d+2},$\textit{ i.e.,}%
\begin{align}
V_{\mathrm{\tilde{U}}^{[2]}\mathrm{(1)},\mathrm{\tilde{S}\tilde{O}}%
^{[1]}\mathrm{(d+2)},d+2}^{[2]}  &  :\mathrm{C}_{\mathrm{\tilde{U}}%
^{[2]}\mathrm{(1)}}^{[2]}\nonumber \\
&  \Longleftrightarrow \mathrm{C}_{\mathrm{\tilde{S}\tilde{O}}^{[1]}%
\mathrm{(d+2)},d+2}^{[1]}\\
&  \Longleftrightarrow \mathrm{C}_{d+2}%
\end{align}
\textit{ }where $\Leftrightarrow$\ between $\mathrm{C}_{\mathrm{\tilde{U}%
}^{[2]}\mathrm{(1)}}^{[2]}$ and\textit{ }$\mathrm{C}_{\mathrm{\tilde{S}%
\tilde{O}}^{[1]}\mathrm{(d+2)},d+2}^{[1]}$\textit{ }denotes an ordered mapping
under fixed changing rate of $\lambda^{\lbrack12]}=\left \vert \frac{\delta
\phi^{\lbrack2]}}{\delta \phi_{\mathrm{global}}^{[1]}}\right \vert $,
$\Leftrightarrow$\ between\textit{ }$\mathrm{C}_{\mathrm{\tilde{S}\tilde{O}%
}^{[1]}\mathrm{(d+2)},d+2}^{[1]}$\textit{ }and $\mathrm{C}_{d+2}$ denotes an
ordered mapping under fixed changing rate.

In particular, there exist two tempo dimensions denoted by $t^{A}$\ and
$t^{B}$. The 1-st order variabilities along two tempo directions may be
difference,
\begin{align}
\mathcal{T}(\delta t^{A})  &  \rightarrow U^{\mathrm{T}^{A}}(\delta \phi
^{t^{A}})=e^{i\cdot \delta \phi^{t^{A}}\Gamma^{t^{A}}},\\
\mathcal{T}(\delta t^{B})  &  \rightarrow U^{\mathrm{T}^{B}}(\delta \phi
^{t^{B}})=e^{i\cdot \delta \phi^{t^{B}}\Gamma^{t^{B}}},
\end{align}
where $\delta \phi^{t^{A}}=\omega_{0}^{A}\delta t^{A},$ $\delta \phi^{t^{B}%
}=\omega_{0}^{B}\delta t^{B}$\ and $\Gamma^{t^{A}}$, $\Gamma^{t^{B}}$ are
Gamma matrix anticommuting with $\Gamma^{i},$ $\{ \Gamma^{i},\Gamma^{t^{A}%
}\}=2\delta^{it^{A}}$, $\{ \Gamma^{i},\Gamma^{t^{B}}\}=2\delta^{it^{B}}$, $\{
\Gamma^{t^{A}},\Gamma^{t^{B}}\}=0.$ The system with 1-st order variability
along two tempo directions indicates two orthogonal \emph{uniform motions} of
the group-changing spaces along $\Gamma^{t^{A}}$ direction and $\Gamma^{t^{B}%
}$ direction, respectively.

Now, there are three physical constants relevant the 1-st order tempo
variability, the Planck constant $\hslash$, the particle's mass $m$ and theta
angle $\theta$.

Firstly, because the energy of the physical variants has a uniformly
distribution, the energy density $\rho_{E}=\frac{\Delta E}{\Delta V}$ is
constant, we have the density of (effective) "angular momentum" as
$\frac{\delta \rho_{E}}{\delta \omega}\mid_{\omega=ck_{0}}=\rho_{J}^{E}$. The
"angular momentum" $\rho_{J}$ for an elementary particle (a level-1 zero) is
Planck constant $\hbar$. As a result, the quantization condition in quantum
mechanics come from the linearization of energy density $\rho_{E}$ via
$\omega$ near $\omega_{0}=\sqrt{(\omega_{0}^{A})^{2}+(\omega_{0}^{B})^{2}}$.

Secondly, the particle mass is defined by
\[
m=\hbar \sqrt{(\delta \omega_{0}^{A})^{2}+(\delta \omega_{0}^{B})^{2}}/c^{2}%
\]
where $\delta \omega_{0}^{A}=\omega_{0}^{A}-ck_{0}$ and $\delta \omega_{0}%
^{B}=\omega_{0}^{B}-ck_{0}$.

Thirdly, theta angle is determined by the ratio between $\delta \omega_{0}^{A}$
and $\delta \omega_{0}^{B},$
\[
\theta \equiv \arg(\frac{\delta \omega_{0}^{B}}{\delta \omega_{0}^{A}}).
\]
If $\frac{\delta \omega_{0}^{B}}{\delta \omega_{0}^{A}}$ fluctuates (or
$\delta \omega_{0}^{A}$ and $\delta \omega_{0}^{B}$ can fluctuate
independently), it leads to axion.

We derive the effective Hamiltonian for elementary particles (level-1 zeroes).
In the following parts, we set $\hbar=1$. We assume a tiny theta term as
$ck_{0}\gg \delta \omega_{0}^{A},$ $\left \vert \omega_{0}^{A}-ck_{0}\right \vert
\gg \delta \omega_{0}^{B}.$

We firstly write down the hopping Hamiltonian that is obtained as%
\begin{equation}
\mathcal{H}=%
%TCIMACRO{\dsum \limits_{\{i,j\}}}%
%BeginExpansion
{\displaystyle \sum \limits_{\{i,j\}}}
%EndExpansion
\mathcal{H}_{\left \{  i,j\right \}  }=J%
%TCIMACRO{\dsum \limits_{\{i,j\}}}%
%BeginExpansion
{\displaystyle \sum \limits_{\{i,j\}}}
%EndExpansion
c_{i}^{\dagger}\mathbf{T}_{\left \{  i,j\right \}  }c_{i+e^{I}}.
\end{equation}
In continuum limit, we have%
\begin{align}
\mathcal{H}  &  =J%
%TCIMACRO{\dsum \limits_{\{i,j\}}}%
%BeginExpansion
{\displaystyle \sum \limits_{\{i,j\}}}
%EndExpansion
c_{i}^{\dagger}(e^{il_{p}(\hat{k}^{\mu}\cdot \Gamma^{\mu})})c_{i+1}\\
&  =2l_{p}J%
%TCIMACRO{\dsum \limits_{\mu}}%
%BeginExpansion
{\displaystyle \sum \limits_{\mu}}
%EndExpansion%
%TCIMACRO{\dsum \limits_{k^{\mu}}}%
%BeginExpansion
{\displaystyle \sum \limits_{k^{\mu}}}
%EndExpansion
c_{k^{\mu}}^{\dagger}[\cos(k^{\mu}\cdot \Gamma^{\mu})]c_{k^{\mu}}%
\end{align}
where the dispersion in continuum limit is
\begin{align}
E_{k}  &  \simeq \pm c([(\vec{k}-\vec{k}_{0})\cdot \vec{\Gamma}]^{2}%
+((\omega_{0}^{A}-ck_{0})\cdot \Gamma^{t^{A}})^{2}\nonumber \\
&  +((\omega_{0}^{B}-ck_{0})\cdot \Gamma^{t^{B}})^{2})^{1/2},
\end{align}
where $\vec{k}_{0}=\frac{1}{l_{p}}(\frac{\pi}{2},\frac{\pi}{2},\frac{\pi}{2})$.

Then, the effective Hamiltonian is obtained as
\begin{equation}
\mathcal{H}=\int(\Psi^{\dagger}(\mathbf{x})\hat{H}\Psi(\mathbf{x}))d^{3}x
\end{equation}
where
\begin{equation}
\hat{H}=\vec{\Gamma}\cdot \Delta \vec{p}+m^{A}\Gamma^{t^{A}}+m^{B}\Gamma^{t^{B}}%
\end{equation}
with $m^{A}=\hbar \delta \omega_{0}^{A}/c^{2},$ $m^{B}=\hbar \delta \omega_{0}%
^{B}/c^{2}.$ This is a massive Dirac model with two mass terms. After consider
the dynamics for level-2 group-changing spaces, the Dirac fermions couple
\textrm{U}$^{\mathrm{em}}$\textrm{(1)}$\times$\textrm{SU(}$N$\textrm{)} gauge field.

Next, we can redefine the mass term and unify $m^{A}\Gamma^{t^{A}}+m^{B}%
\Gamma^{t^{B}}$ into one with corresponding Gamma matrix and have a gauge
field with theta terms. The two theta terms are obtained as%
\[
\mathcal{L}_{\theta,\mathrm{EM}}=\frac{\theta_{\mathrm{EM}}}{8\pi^{2}}%
\int \mathrm{tr}(F\wedge F),\text{ }%
\]
and
\[
\mathcal{L}_{\theta,\mathrm{YM}}=\frac{\theta_{\mathrm{YM}}}{8\pi^{2}}%
\int \mathrm{tr}(G\wedge G).
\]
where $\theta_{\mathrm{EM}}=\theta_{\mathrm{YM}}=\arg(\frac{\delta \omega
_{0}^{B}}{\delta \omega_{0}^{A}})$.

\paragraph{Difference between our universe and that with finite theta term}

Based on above discussion, we point out that the key point about the solution
for Strong CP Problem is to identify the correct type of physical variants for
our world, $V_{\mathrm{\tilde{U}}^{[2]}\mathrm{(1)},\mathrm{\tilde{S}\tilde
{O}}^{[1]}\mathrm{(d+1)},d+1}^{[2]}$ or $V_{\mathrm{\tilde{U}}^{[2]}%
\mathrm{(1)},\mathrm{\tilde{S}\tilde{O}}^{[1]}\mathrm{(d+2)},d+2}^{[2]}$. In
other words, the Strong CP Problem is relevant to the dynamics of spacetime
rather than a simple trouble of quantum field theory.

If our world is the physical variant $V_{\mathrm{\tilde{U}}^{[2]}%
\mathrm{(1)},\mathrm{\tilde{S}\tilde{O}}^{[1]}\mathrm{(d+1)},d+1}^{[2]},$
there exists only one tempo dimension. For the system with only one tempo
direction, there doesn't exist theta terms and axion field at all; If our
world is the physical variant $V_{\mathrm{\tilde{U}}^{[2]}\mathrm{(1)}%
,\mathrm{\tilde{S}\tilde{O}}^{[1]}\mathrm{(d+2)},d+2}^{[2]}$, we will have a
finite theta term $\theta_{\mathrm{YM}}=\arg(\frac{\delta \omega_{0}^{B}%
}{\delta \omega_{0}^{A}})\neq0$. The quantum phase fluctuations $\theta
\equiv \arg(\frac{\omega_{0}^{B}}{\omega_{0}^{A}})\ $becomes axion field.

\paragraph{Solution for Strong CP Problem}

Firstly, we provide the solution to Strong CP Problem.

Our world is the physical variant $V_{\mathrm{\tilde{U}}^{[2]}\mathrm{(1)}%
,\mathrm{\tilde{S}\tilde{O}}^{[1]}\mathrm{(d+1)},d+1}^{[2]},$ there exists
only one tempo dimension.

For a uniform physical variant, an elementary particle is a zero that is the
information unit of the system. Each elementary particle corresponds to a zero
with $\pi$-phase changing along arbitrary direction. Therefore, these
elementary particles becomes topological defect of quantum spacetime. Along
arbitrary direction including tempo direction, the size of an elementary
particle is $\pi/k_{0}=l_{p}$ where $l_{p}$\ is the minimum distance between
two zeroes. Now, the corresponding changing rate along the fifth direction is
zero, i.e., $\omega_{0}^{B}=0$. The size of an elementary particle along the
fifth direction is always zero. As a result, we always have $\delta \omega
_{0}^{B}=0$. That means there doesn't exist theta terms at all.

As a result, the Strong CP Problem in our universe is solved.

\subsection{Conclusion and discussion}

In the end of this part, we draw the conclusion. The\emph{ starting point} of
this theory is very simple -- a ($d+1$) dimensional 2-nd order $\mathrm{\tilde
{S}\tilde{O}}$\textrm{(d+1)} physical variant $V_{\mathrm{\tilde{U}}%
^{[2]}\mathrm{(1)},\mathrm{\tilde{S}\tilde{O}}^{[1]}\mathrm{(d+1)},d+1}^{[2]}$
with 2-nd order variability,%
\begin{equation}
\mathcal{T}(\delta x^{\mu})\leftrightarrow \hat{U}^{[1]}(\delta \phi^{\lbrack
1]})=e^{i\cdot k_{0}\delta x^{\mu}\Gamma^{\mu}}.
\end{equation}
and
\begin{equation}
\hat{U}^{[1]}(\delta \phi_{\mathrm{global}}^{[1]})\leftrightarrow \hat{U}%
^{[2]}(\delta \phi^{\lbrack2]})=\exp(i\lambda^{\lbrack12]}\delta \phi
_{\mathrm{global}}^{[1]}),
\end{equation}
where $\delta \phi_{\mathrm{global}}^{[1]}=\sqrt{%
%TCIMACRO{\dsum \limits_{\mu}}%
%BeginExpansion
{\displaystyle \sum \limits_{\mu}}
%EndExpansion
((\delta \phi^{\mu})^{[1]})^{2}}.$ The ratio of the changing rates of the two
group-changing spaces is $\lambda^{\lbrack12]}=\left \vert \frac{\delta
\phi^{\lbrack2]}}{\delta \phi_{\mathrm{global}}^{[1]}}\right \vert =3.$\ Based
on the simple starting point, we develop a theory for quantum gauge fields.

The 2-nd order variability is reduced into \textrm{U}$^{\mathrm{em}}%
$\textrm{(1)} local gauge symmetry and $\mathrm{SU(N)}$ non-Abelian gauge
symmetry. The corresponding theory becomes QED$\mathrm{\times}$QCD. There are
two types of the collective motions of the level-2 zero lattice, one is about
the global motion of the level-2 zeroes inside a level-1 zero, the other is
about relative motion of the level-2 zeroes inside a level-1 zero. The global
motion of the level-2 zeroes corresponds to the fluctuations of quantum fields
of \textrm{U}$^{\mathrm{em}}$\textrm{(1)} gauge symmetry, and the relative
motion of the level-2 zeroes corresponds to the quantum fields of
\textrm{SU(}$\mathrm{N}$\textrm{). } The belief of "symmetry induce
interaction" is now updates to \emph{\emph{"Higher order variability induce
interaction". }}See the logical structure of the quantum gauge theory in this
part in Fig.35.

\begin{figure}[ptb]
\includegraphics[clip,width=0.7\textwidth]{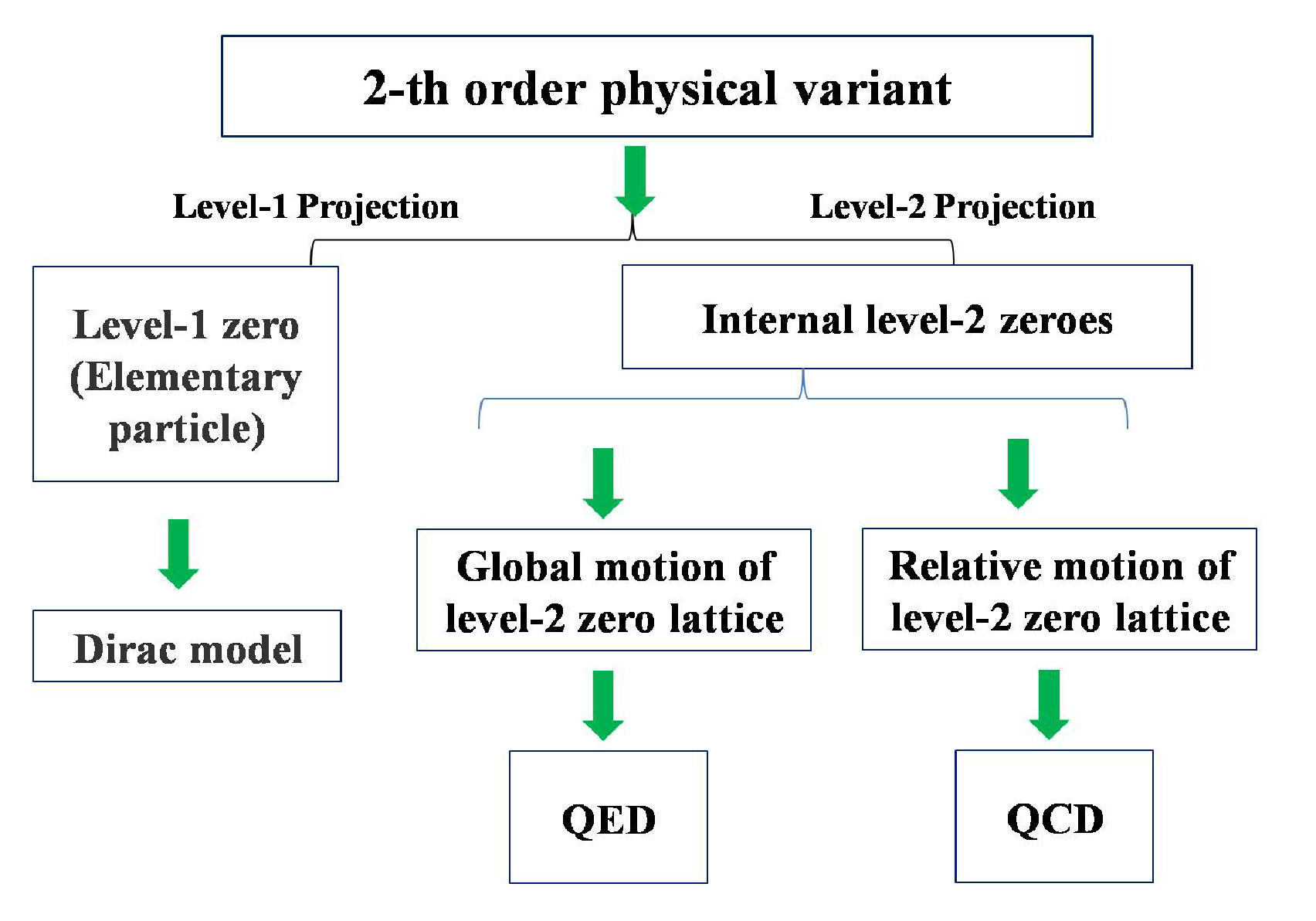}\caption{The logic
structure of the part of quantum gauge theory}%
\end{figure}In the end of this part, we answer all questions at beginning and
show how the troubles about quantum gauge fields disappear:

\emph{1. Problems for quark confinement}:

In particular, we point out that quark confinement is a problem of spacetime,
or physical consequence of topological defect with fractional monopole charge
of quantum spacetime, rather than just a field problem. This indicates the
incompleteness of quantum field theory.

Now, the objects with fractional particle number induced by quarks or gluons
trap fractional number of magnetic monopole of spacetime. As a result, a quark
with fractional fermion number or fractional magnetic charge must have
infinite energy and thus cannot exist. This explains the quark confinement.

\emph{2. Strong CP Problems}:

The key point about the solution for Strong CP Problem is to identify the
correct type of physical variants for our world, $V_{\mathrm{\tilde{U}}%
^{[2]}\mathrm{(1)},\mathrm{\tilde{S}\tilde{O}}^{[1]}\mathrm{(d+1)},d+1}^{[2]}$
or $V_{\mathrm{\tilde{U}}^{[2]}\mathrm{(1)},\mathrm{\tilde{S}\tilde{O}}%
^{[1]}\mathrm{(d+2)},d+2}^{[2]}$? Therefore, the Strong CP Problem is also
relevant to the dynamics of spacetime rather than a simple trouble of quantum
field theory. However, our world is the physical variant $V_{\mathrm{\tilde
{U}}^{[2]}\mathrm{(1)},\mathrm{\tilde{S}\tilde{O}}^{[1]}\mathrm{(d+1)}%
,d+1}^{[2]}$ with only one tempo dimension. There doesn't exist theta terms
and axion field at all.

\emph{3. Problem for Landau's ghost}:

According to above results, QED turns into lattice gauge theory that describes
the fluctuating loop currents on level-1 zero lattice. Due to existence of
cutoff from lattice distance and finite size of elementary particles, there
will never exist the ultraviolet divergence. Therefore, the Landua's ghost disappears.

In this part, we developed a theory for quantum gauge fields. However, there
are unsolved problems for Standard model, an \textrm{SU(3)}$_{\mathrm{C}%
}\times$\textrm{SU(2)}$_{\mathrm{L}}\times$\textrm{U(1)}$_{\mathrm{Y}}$ gauge
theory\cite{stand}. Therefore, in next part, we will apply the theory of
physical variant to \textrm{SU(2)}$_{\mathrm{L}}$ weak interaction and try to
learn the nature of Higgs mechanism and particularly, calculate the masses of
different elementary particles including neutrinos, leptons, quarks.

\newpage

\section{Microscopic Theory for the Standard Model -- from Variant to Chiral
Variant}

\subsection{Introduction}

Over the past several decades, the Standard Model (SM) for particle physics
was developed\cite{w1,bur,tong,stand}, of which the formulation has been
experimentally verified to a high degree of accuracy. According to the SM, the
properties of all known elementary particles are well described. Between
different elementary particles, there may exist different interactions by
exchanging gauge fields. When the Higgs field has finite expectation value,
the masses of elementary particles become finite and consequently the original
gauge symmetry may be broken.

Firstly, we discuss the properties of (fermionic) elementary particles. There
are three families of elementary particles. The first family includes the
electron, $e$-type neutrino, the up ($u$) and down ($d$) quarks; The second
family includes the $\mu$, $\mu$-type neutrino, the strange ($s$) and charm
($c$) quarks; The third family includes $\tau,$ $\tau$-type neutrino, the
bottom ($b$) and top ($t$) quarks. All elementary particles of these three
families obey fermionic statistics. For different families, the elementary
particles with same electric charges are identical except their masses.

According to the SM, there are three types of fundamental interactions, strong
interaction, weak interaction and electromagnetic interaction. All these three
types of fundamental interactions are mediated by the spin-1 fields that obey
bosonic statistics. The electromagnetic interaction that governs macroscopic
world is due to the exchange of the electromagnetic field (or photons), of
which the strength is characterized the fine structure constant. The weak
interaction is responsible for radioactive decay. Only left-handed fermions
feel weak interaction. This leads to parity breaking and thus the SM becomes
chirality\cite{ly}. The weak interaction is due to the exchange of the charged
\textrm{W}$^{\pm}$ and neutral \textrm{Z} bosons. The strong interaction
between quarks comes from exchanging of massless gluons. Due to the effect of
quark confinement, the energy for strong interaction between two colored
particles increases with increasing distance.

In particular, in the SM, the three fundamental interactions are described
by\ an \textrm{SU}$_{\mathrm{C}}$\textrm{(3)}$\times$\textrm{SU}%
$_{\mathrm{weak}}$\textrm{(2)}$\times$\textrm{U}$_{\mathrm{Y}}$\textrm{(1)}
gauge theory. The electromagnetic field is the\emph{ }gauge field with
\textrm{\textrm{U}$_{\mathrm{em}}$\textrm{(1)}} symmetry; the strong
interaction is characterized by \textrm{SU}$_{\mathrm{C}}$\textrm{(3)}
non-Abelian gauge theory; the weak interaction is described \textrm{SU}%
$_{\mathrm{weak}}$\textrm{(2)} non-Abelian gauge theory. Due to the existence
of \textrm{SU}$_{\mathrm{weak}}$\textrm{(2)} non-Abelian gauge symmetry, we
cannot distinguish a left-hand electron from a neutrino. In addition, in the
SM, the weak and electromagnetic interactions are unified into a single
electro-weak interaction with \textrm{SU}$_{\mathrm{weak}}$\textrm{(2)}%
$\times$\textrm{U}$_{\mathrm{Y}}$\textrm{(1)} gauge symmetry.

The Higgs field breaks the original \textrm{SU}$_{\mathrm{weak}}$%
\textrm{(2)}$\times$\textrm{U}$_{\mathrm{Y}}$\textrm{(1)} symmetries of the
electro-weak theory to the residual \textrm{\textrm{U}$_{\mathrm{em}}%
$\textrm{(1)}} gauge symmetry of QED that is mixed gauge symmetry of
\textrm{U}$_{\mathrm{Y}}$\textrm{(1)} gauge symmetry and diagonal component of
\textrm{SU}$_{\mathrm{weak}}$\textrm{(2)}gauge symmetry. Particles that
interact with the Higgs field have finite masses; Particles that do not
interact with the Higgs field - the photon, and gluons remain massless. As a
result, the fermionic elementary particles may have different masses. We call
it mass spectra of the SM. Then, one can distinguish a left-hand electron from
a neutrino when the Higgs field has finite expectation value.

The success of gauge symmetry in the SM has led to the belief of
"\emph{symmetry induce interaction}"\cite{we,yang}. According to this belief,
when the gauge symmetry (a type of local "invariant") of a given quantum
system is acknowledged, the right formula of its Lagrangian $\mathcal{L}$ or
action $\mathcal{S}$ can be obtained straightforwardly. Therefore, with the
help of the belief of "symmetry induce interaction", the SM became a unified
description of the strong, weak and electromagnetic forces in the language of
quantum gauge field theories.

Today, the SM becomes a fundamental branch of physics that agrees very well
with experiments and provides an accurate description of the dynamic behaviors
of elementary particles. However, the SM is more like a phenomenological
model. There exists a lot of unsolve problems of the SM. Then, I show five
unsolve problems:

1. The first trouble is about \emph{mass spectra. }There are many parameters
about masses of different elementary particles that people can't yet explain.
The masses of different elementary particles span at least $12$ orders of
magnitude. For example, the mass of top quarks is up to $173$\textrm{GeV} (the
energy scale of Higgs field), the mass of electrons is about $0.5$%
\textrm{MeV}, The mass of neutrino must be smaller than $0.15$\textrm{eV} (the
energy scale of dark energy). Why these particular masses? The answer is
obviously beyond the SM. So, people hope answer this question and calculate
the mass spectra by a deeper theory.

2. Another relevant puzzle is about \emph{neutrino oscillation}%
\cite{neu1,neu2,neu3}. According to the SM, the neutrino is Weyl fermion.
Therefore, its mass must be zero. However, the situation become complex after
the discovery of neutrino oscillation. Neutrino oscillation is a phenomenon
where neutrinos change from one type to another. To explain the neutrino
oscillation, a possible model comes from seesaw mechanism, that predicts the
existence of heavy right-hand neutrinos. Why neutrino becomes oscillating? Is
the seesaw mechanism right? Does heavy right-hand neutrino exist? The answer
about above question is all beyond the SM. So, again, a new theory beyond the
SM must be developed.

3. The third trouble is about \emph{Higgs field}. Higgs field is a scalar
field with spin-0\cite{higg1,higg2,higg4}. The energy scale of Higgs field
$10^{2}$\textrm{GeV} is much smaller than Planck energy $10^{19}$\textrm{GeV}.
Why? About Higgs field, the naturalness problem arise when we try to calculate
the Higgs potential in quantum field theory. The naturalness problem appears
if the cutoff is much larger than the energy scale of electro-weak
interaction, i.e., $\Lambda \gg10^{3}$\textrm{GeV}. This leads to unnatural
fine-tuning. A possible solution is supersymmetry which predicts new particles
and ensure that individual contributions come in pairs of almost equal size.
Another class of possible solution is technicolor. Now, there exists new
energy scale for cutoff close to $10^{3}$\textrm{GeV}. What's the underlying
physics of spontaneous symmetry breaking about Higgs mechanism?

4. The fourth trouble is \emph{Unification} of all different gauge fields into
single framework. The three fundamental interactions are described by\ an
\textrm{SU}$_{\mathrm{C}}$\textrm{(3)}$\times$\textrm{SU}$_{\mathrm{weak}}%
$\textrm{(2)}$\times$\textrm{U}$_{\mathrm{Y}}$\textrm{(1)} gauge theory. It
was known that weak interaction and electromagnetic interaction are unified
into a single framework - electro-weak interaction with \textrm{SU}%
$_{\mathrm{weak}}$\textrm{(2)}$\times$\textrm{U}$_{\mathrm{Y}}$\textrm{(1)}
gauge symmetry. Could strong interaction and electro-weak interaction be
unified a new framework? for example, \textrm{SU(5)}\cite{so} or
\textrm{SU(10)}?

5. The fourth trouble is about \emph{dark matter and dark energy. }It was
known that visible matter only makes up around $5\%$ of the universe\cite{dark
matter1, dark matter2, dark matter3}. The universe is dominated by dark energy
and dark matter that make up the other $27\%$ and $68\%$, respectively. What
are dark matter and dark energy? A possible solution for dark energy is the
vacuum energy (or the cosmological constant)\cite{dark energy}. However, the
vacuum energy density observed in cosmology is very small, for example,
$10^{-10}\mathrm{eV}^{4}.$ This huge difference between theory $10^{120}%
\mathrm{eV}^{4}$ and observation $10^{-10}\mathrm{eV}^{4}$ is a mystery. For
dark matter, the candidates may be weakly interacting massive particles (for
example, the super-partner in supersymmetry theory)\cite{wip1,wip2} or Axions.
However, in experiments, there don't exist any evidences for these particles.
What exactly are dark matter and dark energy?

To answer above questions satisfactorily, a complete, new theory beyond usual
quantum field theory of the SM must be developed, rather than providing
certain non-perturbative theories or considering a bigger group of gauge
fields. In addition, it must provide a \emph{fully understanding the unsolved
mysteries}, including mass spectra, neutrino oscillation problem, naturalness
problem, Higss mechanism, and the dark energy and dark matter problem.

In this part, for the SM, due to parity breaking and the chirality, the
higher-order physical variant for usual quantum gauge theories is generalized
to \emph{higher-order chiral physical variant}, of which the symmetry between
left-hand part right part is broken. The original physical variant is
characterized by left-hand sub-variant and right-hand sub-variant, or
equivalently relative sub-variant and global sub-variant. We found that the
detailed structure of \emph{relative sub-variant} determines the specific
properties of the SM, such as three flavors, mass spectra, neutrino
oscillations, Higgs field and dark energy and dark matter, ... So, different
gauge structures of the SM are unified into single physics structure -- 2-nd
order chiral physical variant, i.e.,
\begin{align}
&  \text{SM (an extrinsic property)}\\
&  \Longrightarrow \text{A 2-nd order chiral physical variant}\nonumber \\
&  \text{(an intrinsic property).}\nonumber
\end{align}

\subsection{Review on the Standard Model}

The Standard Model describes the physics of the building blocks of our world.
There are three parts in the total Lagrangian,
\begin{equation}
\mathcal{L}=\mathcal{L}_{\mathrm{QCD}}+\mathcal{L}_{\mathrm{EW}}%
+\mathcal{L}_{\mathrm{Higgs}}~.
\end{equation}
Let us give detailed discussion one by one.

\subsubsection{Sector of quantum Chromodynamics}

QCD is a non-Ableian \textrm{SU}$_{\mathrm{c}}$\textrm{(3)} gauge theory with
color triplet quark matter fields, i.e.,
\begin{equation}
\mathcal{L}_{\mathrm{QCD}}=-\frac{1}{4}\sum_{a=1}^{8}G^{a\mu \nu}G_{\mu \nu}%
^{a}+\sum_{j=1}^{n_{f}}\bar{q}_{j}(iD_{\mu}\gamma^{\mu}-m_{j})q_{j}%
\end{equation}
where $q_{j}$ are the quark fields (of $n_{f}=3$ different flavors) with mass
$m_{j}$, $\gamma^{\mu}$ are the Dirac matrices and $D_{\mu}=\partial_{\mu
}-ig_{s}\sum_{a}t^{a}G_{\mu}^{a}$ is the covariant derivative, and $g_{s}$ is
the gauge coupling for strong interaction. $A_{\mu}^{a}$ are the gluon fields
and $T^{a}$ are the $\mathrm{SU(3)}$ group generators in the triplet
representation of quarks (i.e. $T_{a}$ are $3\times3$ matrices) obeying the
commutation relations $[T^{a},T^{b}]=iC_{abc}T^{c}$ where $C_{abc}$ are the
complete antisymmetric structure constants of $\mathrm{SU(3)}$. The strength
of gauge fields is defined by
\begin{equation}
G_{\mu \nu}^{a}=\partial_{\mu}A_{\nu}^{a}-\partial_{\nu}A_{\mu}^{a}%
-g_{s}C_{abc}A_{\mu}^{b}A_{\nu}^{c}.
\end{equation}
As a result, there exists self-interaction for the gluons.

\subsubsection{Electro-weak sector}

$\mathcal{L}_{\mathrm{EW}}$ is the Yang--Mills Lagrangian for the gauge group
\textrm{SU}$_{\mathrm{weak}}$\textrm{(2)}$\times$\textrm{U}$_{\mathrm{Y}}%
$\textrm{(1)} with fermion matter fields that describes the electro-weak
interaction,
\begin{align}
\mathcal{L}_{\mathrm{EW}}  &  =-\frac{1}{4}\sum_{a=1}^{3}\mathcal{F}_{\mu \nu
}^{a}\mathcal{F}^{a\mu \nu}-\frac{1}{4}B_{\mu \nu}B^{\mu \nu}\nonumber \\
&  +\bar{\psi}_{L}i\gamma^{\mu}D_{\mu}\psi_{L}+\bar{\psi}_{R}i\gamma^{\mu
}D_{\mu}\psi_{R}%
\end{align}
where the \textrm{SU}$_{\mathrm{weak}}$\textrm{(2)}$\times$\textrm{U}%
$_{\mathrm{Y}}$\textrm{(1)} gauge fields are $W_{\mu}^{a}$ and $B_{\mu},$ the
corresponding strengths are
\begin{align}
B_{\mu \nu}  &  =\partial_{\mu}B_{\nu}-\partial_{\nu}B_{\mu},\quad \nonumber \\
\mathcal{F}_{\mu \nu}^{a}  &  =\partial_{\mu}W_{\nu}^{a}-\partial_{\nu}W_{\mu
}^{a}-g_{w}\epsilon_{abc}W_{\mu}^{B}W_{\nu}^{C}.
\end{align}

For the fermion fields, there are left-hand and right-hand components:
\begin{equation}
\psi_{L,R}=[(1\mp \gamma_{5})/2]\psi,\quad \bar{\psi}_{L,R}=\bar{\psi}%
[(1\pm \gamma_{5})/2].
\end{equation}
As a result, we have
\begin{align}
\bar{\psi}_{L}  &  =\psi_{L}^{\dag}\gamma_{0}=\psi^{\dag}[(1-\gamma
_{5})/2]\gamma_{0}\\
&  =\bar{\psi}[\gamma_{0}(1-\gamma_{5})/2]\gamma_{0}=\bar{\psi}[(1+\gamma
_{5})/2]~.\nonumber
\end{align}
The matrices $P_{\pm}=(1\pm \gamma_{5})/2$ are projectors. They satisfy the
relations
\begin{equation}
P_{\pm}P_{\pm}=P_{\pm},\text{ }P_{\pm}P_{\mp}=0,\text{ }P_{+}+P_{-}=1.
\end{equation}

For the SM, due to parity breaking and the chirality, $\psi_{L}$ and $\psi
_{R}$ behave differently under the gauge group, i.e., $\psi_{R}$ are singlets
and all $\psi_{L}$ are doublets. So, we have
\begin{equation}
D_{\mu}\psi_{L,R}=\left[  \partial_{\mu}+ig_{w}\sum_{A=1}^{3}T_{L,R}^{a}%
W_{\mu}^{a}+ig^{\prime}\frac{1}{2}Y_{L,R}B_{\mu}\right]  \psi_{L,R}~,
\end{equation}
where $T_{L,R}^{a}$ and $\frac{1}{2}Y_{L,R}$ are the \textrm{SU}%
$_{\mathrm{weak}}$\textrm{(2)} and \textrm{U}$_{\mathrm{Y}}$\textrm{(1)}
generators, respectively. The commutation relations of the $\mathrm{SU(2)}$
generators are given by
\begin{equation}
\lbrack T_{L}^{a},T_{L}^{b}]=i~\epsilon_{abc}T_{L}^{c},\quad \lbrack T_{R}%
^{a},T_{R}^{b}]=i\epsilon_{abc}T_{R}^{c}~.
\end{equation}
The electric charge generator $Q$ is given by
\begin{equation}
Q=T_{L}^{3}+\frac{1}{2}Y_{L}=T_{R}^{3}+\frac{1}{2}Y_{R}~.
\end{equation}

Then, the charged-current couplings are obtained as%
\begin{align}
V_{\bar{\psi}\psi W}  &  =g_{w}\bar{\psi}\gamma_{\mu}[(T_{L}^{+}/\sqrt
{2})(1-\gamma_{5})/2\nonumber \\
&  +(T_{R}^{+}/\sqrt{2})(1+\gamma_{5})/2]\psi W_{\mu}^{-}+\mathrm{h.c.}%
\end{align}
where $T^{\pm}=T^{1}\pm iT^{2}$ and $W^{\pm}=(W^{1}\pm iW^{2})/.$ The
neutral-current couplings are described by
\begin{equation}
\Gamma_{\bar{\psi}\psi Z}=g_{w}/(2\cos \theta_{W})\bar{\psi}\gamma_{\mu}%
[T_{R}^{3}(1+\gamma_{5})-2Q\sin^{2}\theta_{W}]\psi Z^{\mu},
\end{equation}
where $T_{L}^{3}=\pm \frac{1}{2}$. $\theta_{W}$ is named Weinberg angle that is
obtained by $\theta_{W}=\arctan \frac{g^{\prime}}{g_{w}}$. The relationships
between photon $A_{\mu}$ and $Z_{\mu}$ of the weak interaction and $B_{\mu}$
and $W_{\mu}^{3}$ are
\begin{align}
A_{\mu}  &  =\cos \theta_{W}B_{\mu}+\sin \theta_{W}W_{\mu}^{3},\nonumber \\
Z_{\mu}  &  =-\sin \theta_{W}B_{\mu}+\cos \theta_{W}W_{\mu}^{3}.
\end{align}

\subsubsection{Higgs sector}

We then discuss the Higgs sector.

After considering the Higgs condensation, fermionic elementary particles
together with $W^{\pm}$ and $Z$ masses become massive. The Lagrangian of Higgs
field $\Phi$ is described by
\begin{align}
\mathcal{L}_{\mathrm{Higgs}}  &  =(D_{\mu}\Phi)^{\dag}(D^{\mu}\Phi
)-V(\Phi^{\dag}\Phi)\nonumber \\
&  -\bar{\psi}_{L}(\Gamma \Phi)\psi_{R}-\bar{\psi}_{R}(\Gamma^{\dag}\Phi^{\dag
})\psi_{L}~,
\end{align}
where $\Phi$ denotes Higgs scalar fields that transform as a reducible
representation of the gauge group. $\Gamma$ determines the matrix of the
Yukawa couplings. The effective potential $V(\Phi^{\dag}\Phi)$ is usually
assumed to be
\begin{equation}
V(\Phi^{\dag}\Phi)=-\frac{1}{2}\mu^{2}\Phi^{\dag}\Phi+\frac{1}{4}\lambda
(\Phi^{\dag}\Phi)^{2}.
\end{equation}

When the spontaneous symmetry breaking occurs, we have the vacuum expectation
value (VEV) of $\Phi$, i.e.,
\begin{equation}
\langle0|\Phi(x)|0\rangle=\phi_{0}\not =0~.
\end{equation}
From the experimental result, we have $\phi_{0}=174.1~\mathrm{GeV}.$ After
considering the spontaneous symmetry breaking, the masses of elementary
particles are obtained from the Yukawa couplings by replacing $\Phi(x)$ by
$\phi_{0}$:
\begin{equation}
M=\bar{\psi}_{L}(\Gamma \cdot \phi_{0})\psi_{R}+\bar{\psi}_{R}(\Gamma^{\dag}%
\phi_{0}^{\ast})\psi_{L}.
\end{equation}
So, the masses of different elementary particles are proportional to $\phi
_{0}$. In general, the mass eigenstates and the weak eigenstates do not
coincide and a unitary transformation V connects the two sets that is called
the Cabibbo-Kobayashi-Maskawa (CKM) matrix.

In addition, for the case of $\phi_{0}\not =0$, the \textrm{SU}%
$_{\mathrm{weak}}$\textrm{(2)}$\times$\textrm{U}$_{\mathrm{Y}}$\textrm{(1)}
gauge symmetries of the electro-weak theory is broken to the residual
\textrm{\textrm{U}$_{\mathrm{em}}$\textrm{(1)}} gauge symmetry. These effects
are induced by the $(D_{\mu}\Phi)^{\dag}(D^{\mu}\Phi)$ term in $\mathcal{L}%
_{\mathrm{Higgs}}$. Under the condition of $Q|\phi_{0}\rangle=(T^{3}+\frac
{1}{2}Y)|\phi_{0}\rangle=0$, the photon remains massless. The $W^{\prime}$s
mass and the $Z^{\prime}$s mass are determined by the following terms,
\begin{equation}
m_{W}^{2}W_{\mu}^{+}W^{-\mu}=g_{w}^{2}|(T^{+}v/\sqrt{2})|^{2}W_{\mu}%
^{+}W^{-\mu}~, \label{53}%
\end{equation}
and
\begin{equation}
\frac{1}{2}m_{Z}^{2}Z_{\mu}Z^{\mu}=|[g_{w}\cos \theta_{W}T^{3}-g^{\prime}%
\sin \theta_{W}(Y/2)]\phi_{0}|^{2}Z_{\mu}Z^{\mu}. \label{54}%
\end{equation}
As a result, we have
\begin{equation}
m_{W}^{2}=1/2g_{w}^{2}\phi_{0}^{2},\quad m_{Z}^{2}=1/2g_{w}^{2}\phi_{0}%
^{2}/\cos^{2}\theta_{W}~. \label{58}%
\end{equation}

In the end, by fixing the value of $\lambda,$ the mass of Higgs field is
predicted by $m_{H}^{2}\sim \lambda(\phi_{0})^{2}$.

\subsubsection{8 key points for the SM}

In summary, we emphasize the following 8 key points for the SM:

\begin{enumerate}
\item \textit{Spacetime: }The spacetime is $(3+1)$-dimensional \emph{Minkowski
spacetime} with (global) symmetry of \emph{Poincar{\'{e}} group,}
\begin{equation}
G_{\text{gl}}=\mathbb{R}^{3,1}\rtimes \mathrm{O}(3,1)
\end{equation}
that is the semi-direct product of spacetime translations and the Lorentz group;

\item \textit{Gravity}: Gravity is described by \emph{general relativity}%
;\emph{ }

\item \textit{Validity of quantum field theory.} The theory of
\emph{relativistic quantum fields} is believed to the right approach to
describe the SM;

\item \textit{Elementary particles}: The (fermionic) elementary particles are
\emph{quarks} and \emph{leptons} with spin-1/2;

\item \textit{Gauge interactions: }The gauge interactions are determined by
following local gauge symmetry,
\begin{equation}
G_{\text{SM}}=\underbrace{\underbrace{\mathrm{SU}_{\mathrm{C}}\mathrm{(3)}%
}_{\text{color}}}_{\text{strong}}\times \underbrace{\underbrace{\mathrm{SU}%
_{\mathrm{weak}}\mathrm{(2)}}_{\text{left}}\times \underbrace{\mathrm{U}%
_{\mathrm{Y}}\mathrm{(1)}}_{\text{hypercharge}}}_{\text{electro-weak}}
\label{eq:GaugeGroupSM}%
\end{equation}
When the Higgs field has finite vacuum expectation value, the gauge group is
broken to a subgroup, i.e.,
\begin{equation}
\underbrace{\mathrm{SU}_{\mathrm{C}}\mathrm{(3)}}_{\text{QCD}}\times
\underbrace{\mathrm{U}_{\mathrm{em}}\mathrm{(1)}}_{\text{QED}}.
\end{equation}
Now, the corresponding gauge bosons of the broken group ($W^{\pm}$ and $Z^{0}%
$) become massive. While, the gluons and photons are still massless;

\item \textit{Three families}. For the quarks and leptons, there are $3$
families:
\begin{align*}
\text{Leptons}  &  \text{:}\quad%
\begin{pmatrix}
\nu_{e}\\
e
\end{pmatrix}
,%
\begin{pmatrix}
\nu_{\mu}\\
\mu
\end{pmatrix}
,%
\begin{pmatrix}
\nu_{\tau}\\
\tau
\end{pmatrix}
,\\
\text{Quarks}  &  \text{:}\quad%
\begin{pmatrix}
u\\
d
\end{pmatrix}
,%
\begin{pmatrix}
c\\
s
\end{pmatrix}
,%
\begin{pmatrix}
t\\
b
\end{pmatrix}
;
\end{align*}

\item \textit{Chirality}. For left- and right-handed fermions, the gauge
interactions are different. Left- and right-handed fermions transform in
different $\mathrm{SU}_{\mathrm{weak}}\mathrm{(2)}$-representations. For weak
interaction, \textrm{W} only couple left-handed fermions;

\item \textit{Charge quantization}. The electric charge is defined as
\begin{equation}
Q=T_{3}+Y
\end{equation}
where $T_{3}$ is the diagonal generator of $\mathrm{SU}_{\mathrm{weak}%
}\mathrm{(2)}$ and $Y$ is the $\mathrm{U}(1)_{Y}$ hyper-charge. For charged
leptons, the electric charge is quantized, i.e., $Q=T_{3}+Y=-e$. However, for
quarks, the electric charge becomes fractional, i.e., $+\frac{2}{3}e$ or
$-\frac{1}{3}e$.
\end{enumerate}

\subsection{Chiral variant theory -- Mathematical tools for the SM}

For quantum gauge fields, the usual variant is updated to \emph{higher-order
variant}. In higher-order variant theory, quantum gauge fields become one
"changing" structure on another "\emph{changing}" structure, i.e.,
\begin{equation}
\text{"Space on space on space" = Space on variant.}%
\end{equation}

For the SM, due to the parity breaking and its has non-trivial chirality. To
characterize the SM, we generalize the usual higher-order variant to
(higher-order) chiral variant. In this section, we develop the theory for
chiral variant.

\subsubsection{Review on higher-order variant theory}

To define a variant, we firstly introduce the object of study --
group-changing space. A $d$-dimensional group-changing space $\mathrm{C}%
_{\mathrm{\tilde{G}},d}(\Delta \phi^{a})$\textit{\ }is described by a series of
numbers of group element $\phi^{a}$ of $a$-th generator independently in size
order along $a$-th direction. $\Delta \phi^{a}$ denotes the size of the
group-changing space along a-direction, a topological number. Here,
\textit{\textrm{\~{G}} }is a non-compact Lie group with\textit{ }$N$ generator
and $N<d$. Here \textrm{G} with "$\sim$" above it means a non-compact Lie group.

A variant $V_{\mathrm{\tilde{G},}d}[\Delta \phi^{\mu},\Delta x^{\mu},k_{0}%
^{\mu}]$ is denoted by\ a mapping between a d-dimensional group-changing space
$\mathrm{C}_{\mathrm{\tilde{G},}d}$ with total size $\Delta \phi^{\mu}$\ and
Cartesian space $\mathrm{C}_{d}$\ with total size $\Delta x^{\mu}$, i.e.,%
\begin{align}
V_{\mathrm{\tilde{G},}d}[\Delta \phi^{\mu},\Delta x^{\mu},k_{0}^{\mu}]  &
:\mathrm{C}_{\mathrm{\tilde{G},}d}=\{ \delta \phi^{\mu}\} \nonumber \\
&  \Longleftrightarrow \mathrm{C}_{d}=\{ \delta x^{\mu}\}
\end{align}
where $\Longleftrightarrow$\ denotes an ordered mapping under fixed changing
rate of integer multiple $k_{0}^{\mu}$.\ In particular, $\delta \phi^{\mu}$
denotes group-changing element along $\mu$-direction rather than
group-changing element.

A uniform variant (U-variant) with infinite size ($\Delta x\rightarrow \infty$)
has 1-st order variability, i.e.,%
\begin{equation}
\mathcal{T}(\delta x^{\mu})\leftrightarrow \hat{U}(\delta \phi^{\mu}%
)=e^{i\cdot \delta \phi^{\mu}T^{\mu}}%
\end{equation}
where $\mathcal{T}(\delta x^{\mu})$ is the spatial translation operation on
$\mathrm{C}_{d}$ along $x^{\mu}$-direction and $\hat{U}(\delta \phi^{\mu})$ is
shift operation on group-changing space $\mathrm{C}_{\mathrm{\tilde{G}}%
,d}(\Delta \phi^{\mu})$, and $\delta \phi^{\mu}=k_{0}^{\mu}\delta x^{\mu}$. That
means when one translates along Cartesian space $\delta x^{\mu},$ the
corresponding shifting along group-changing space $\mathrm{C}_{\mathrm{\tilde
{G}},d}$ is $\delta \phi^{\mu}=k_{0}^{\mu}\delta x^{\mu}.$ For simplicity, we
can denote a system with 1-st order variability by the following equation
\begin{equation}
\mathcal{T}\leftrightarrow \hat{U}.
\end{equation}

To define higher-order variant, we split the group-changing space
$\mathrm{C}_{\mathrm{\tilde{G}},d}(\Delta \phi^{\mu})$ of a variant
$V_{\mathrm{\tilde{G},}d}[\Delta \phi^{\mu},\Delta x^{\mu},k_{0}^{\mu}]$ into
two kinds of group-changing subspace: one group-changing subspace
$\mathrm{C}_{\mathrm{\tilde{U}(1)\in \tilde{G}},1}(\Delta \phi_{\mathrm{global}%
})$ is about global phase changing of the system $\Delta \phi_{\mathrm{global}%
}=\sqrt{%
%TCIMACRO{\dsum \limits_{\mu}}%
%BeginExpansion
{\displaystyle \sum \limits_{\mu}}
%EndExpansion
(\Delta \phi^{\mu}(x))^{2}}$, the other is about $d-1$ internal relative angles.

Then, we define the mapping between global phase changings\textit{
}$\mathrm{C}_{1,\mathrm{\tilde{G}}_{1},d_{1}}(\Delta \phi_{1}^{\mu})$
and\textit{ }$\mathrm{C}_{2,\mathrm{\tilde{G}}_{2},d_{2}}(\Delta \phi_{2}^{\mu
})$\textit{,} i.e.,%
\begin{equation}
\mathrm{C}_{1,\mathrm{\tilde{U}}_{1}\mathrm{(1)\in \tilde{G}}_{1},1}(\Delta
\phi_{1,\mathrm{global}})\Longleftrightarrow \mathrm{C}_{2,\mathrm{\tilde{U}%
}_{2}\mathrm{(1)\in \tilde{G}}_{2},2}(\Delta \phi_{2,\mathrm{global}})
\end{equation}
or
\begin{equation}
\delta \phi_{1,\mathrm{global}}=\lambda^{\lbrack12]}\delta \phi
_{2,\mathrm{global}}%
\end{equation}
\textit{ }where $\lambda^{\lbrack12]}$ is the changing ratio between the
changings of the global phases for two group-changing spaces,\textit{ }%
$\delta \phi_{1,\mathrm{global}}=\sqrt{%
%TCIMACRO{\dsum \limits_{\mu}}%
%BeginExpansion
{\displaystyle \sum \limits_{\mu}}
%EndExpansion
(\delta \phi_{1}^{\mu}(x))^{2}}$\textit{ }and\textit{ }$\delta \phi
_{2,\mathrm{global}}=\sqrt{%
%TCIMACRO{\dsum \limits_{\mu}}%
%BeginExpansion
{\displaystyle \sum \limits_{\mu}}
%EndExpansion
(\delta \phi_{2}^{\mu}(x))^{2}}$\textit{.} In this paper, we only focus on the
case of integer changing ratio $\lambda^{\lbrack12]},$ i.e., $\lambda
^{\lbrack12]}=1,2...n$.

Next, we introduce 2-nd order variant $V_{\mathrm{\tilde{G}}^{[2]}%
\mathrm{,\tilde{G}}^{[1]},d}^{[2]}$ by introducing higher-order mapping, i.e.,
a mapping between a group-changing space $\mathrm{C}_{\mathrm{\tilde{G}%
^{[2]},}d}^{[2]}$\ and another $\mathrm{C}_{\mathrm{\tilde{G}^{[1]},}d}^{[1]}$
that is defined on $d$-dimensional Cartesian space $\mathrm{C}_{d}$. We call
$\mathrm{C}_{\mathrm{\tilde{G}^{[1]},}d}^{[1]}$ level-1 group-changing space
and $\mathrm{C}_{\mathrm{\tilde{G}^{[2]},}d}^{[2]}$ level-2group-changing
space, respectively.

In general, a 2-nd order variant $V_{\mathrm{\tilde{G}}^{[2]}\mathrm{,\tilde
{G}}^{[1]},d}^{[2]}$\ is defined by%
\begin{equation}
V_{\mathrm{\tilde{G}}^{[2]}\mathrm{,\tilde{G}}^{[1]},d}^{[2]}:\mathrm{C}%
_{\mathrm{\tilde{G}^{[2]},}d}^{[2]}\Longleftrightarrow \mathrm{C}%
_{\mathrm{\tilde{G}^{[1]},}d}^{[1]}\Longleftrightarrow \mathrm{C}_{d},
\end{equation}
\textit{\ }of which one is the mapping between\textit{ }$\mathrm{C}%
_{\mathrm{\tilde{G}^{[1]},}d}^{[1]}\ $and Cartesian space $\mathrm{C}_{d}%
$\textit{ }with changing ratio $k_{0}^{\mu}$ i.e.,
\begin{equation}
\mathrm{C}_{\mathrm{\tilde{G}^{[1]},}d}^{[1]}\Longleftrightarrow \mathrm{C}%
_{d}.
\end{equation}
The other is between $\mathrm{C}_{\mathrm{\tilde{G}^{[2]},}d}^{[2]}%
$\textit{\ }with total size $(\Delta \phi^{\mu})^{[2]}$\textit{ }and
$\mathrm{C}_{\mathrm{\tilde{G}^{[1]},}d}^{[1]}$ with total size\ $(\Delta
\phi^{\mu})^{[1]},$ i.e.,\textit{ }%
\begin{align}
\mathrm{C}_{\mathrm{\tilde{G}^{[2]},}d}^{[2]}((\Delta \phi^{\mu})^{[2]})  &
\Longleftrightarrow \mathrm{C}_{\mathrm{\tilde{G}^{[1]},}d}^{[1]}((\Delta
\phi^{\mu})^{[1]})\\
&  \equiv \mathrm{C}_{\mathrm{\tilde{U}^{[2]}(1)\in \tilde{G}}_{2},1}%
((\Delta \phi_{\mathrm{global}})^{[2]})\nonumber \\
&  \Longleftrightarrow \mathrm{C}_{\mathrm{\tilde{U}^{[1]}(1)\in \tilde{G}}%
_{1},1}((\Delta \phi_{\mathrm{global}})^{[1]})\nonumber \\
&  \equiv \{ \delta \phi_{\mathrm{global}}^{[2]}\} \Leftrightarrow \{ \delta
\phi_{\mathrm{global}}^{[1]}\} \nonumber
\end{align}
where $\Longleftrightarrow$\ denotes an ordered mapping under fixed changing
rate along different directions. Along $\mu$-th direction, the elements of two
subgroup-changing spaces are $\delta \phi_{\mathrm{global}}^{[2]}=\sqrt{%
%TCIMACRO{\dsum \limits_{\mu}}%
%BeginExpansion
{\displaystyle \sum \limits_{\mu}}
%EndExpansion
((\delta \phi^{\mu})^{[2]})^{2}}$ and $\delta \phi_{\mathrm{global}}^{[1]}%
=\sqrt{%
%TCIMACRO{\dsum \limits_{\mu}}%
%BeginExpansion
{\displaystyle \sum \limits_{\mu}}
%EndExpansion
((\delta \phi^{\mu})^{[1]})^{2}},$ respectively, The changing rate between
$C_{\mathrm{\tilde{G}^{[2]},}d}^{[2]}((\Delta \phi^{\mu})^{[2]})\ $and
$C_{\mathrm{\tilde{G}^{[1]},}d}^{[1]}((\Delta \phi^{\mu})^{[1]})$ is integer
multiple $(\lambda^{\mu})^{[12]}.$

Using similar approach, a n-th order variant $V_{\mathrm{\tilde{G}}%
^{[n]}\mathrm{,...,\tilde{G}}^{[2]}\mathrm{,\tilde{G}}^{[1]},d}^{[2]}$\ is
defined by\ a higher-order mapping between $\mathrm{C}_{\mathrm{\tilde
{G}^{[n]},}d}^{[2]},$\textit{ ..., }$\mathrm{C}_{\mathrm{\tilde{G}^{[2]},}%
d}^{[2]},$\textit{ }$\mathrm{C}_{\mathrm{\tilde{G}^{[1]},}d}^{[1]}$,
and\textit{ }$\mathrm{C}_{d},$%
\begin{align}
V_{\mathrm{\tilde{G}}^{[n]}\mathrm{,...,\tilde{G}}^{[2]}\mathrm{,\tilde{G}%
}^{[1]},d}^{[2]}  &  :\mathrm{C}_{\mathrm{\tilde{G}^{[2]},}d}^{[n]}%
\Longleftrightarrow...\nonumber \\
\mathrm{C}_{\mathrm{\tilde{G}^{[2]},}d}^{[2]}  &  \Longleftrightarrow
\mathrm{C}_{\mathrm{\tilde{G}^{[1]},}d}^{[1]}\Longleftrightarrow \mathrm{C}%
_{d}.
\end{align}

Therefore, the type of a 2-nd order variant $V_{\mathrm{\tilde{U}(1),\tilde
{S}\tilde{O}^{[1]}\mathrm{(d)},}d}^{[2]}$ is determined by the ratio between
changing rates of different levels,
\begin{equation}
\lambda^{\lbrack12]}=\left \vert \frac{\delta \phi_{\mathrm{global}}^{[2]}%
}{\delta \phi_{\mathrm{global}}^{[1]}}\right \vert .
\end{equation}
We point out that massive/massless Dirac model corresponds to the case of
$\lambda^{\lbrack21]}=0$; \textrm{QED} with a local \textrm{U}$_{\mathrm{em}}%
$\textrm{(1)} gauge symmetry corresponds to the case of $\lambda^{\lbrack
12]}=1$; a quantum gauge theory with a local \textrm{SU}$_{\mathrm{C}}%
$\textrm{(2)}$\times$\textrm{U}$_{\mathrm{em}}$\textrm{(1)} gauge symmetry
corresponds to the case of $\lambda^{\lbrack12]}=2$; a \textrm{QCD}$\times
$\textrm{QED} with a local \textrm{SU}$_{\mathrm{C}}$\textrm{(3)}$\times
$\textrm{U}$_{\mathrm{em}}$\textrm{(1)} gauge symmetry corresponds to the case
of $\lambda^{\lbrack12]}=3$.

For 2-nd order U-variant $V_{\mathrm{\tilde{G}}^{[1]},\mathrm{\tilde{G}}%
^{[2]}\mathrm{,}d}^{[2]}$ with infinite size ($\Delta x^{\mu}\rightarrow
\infty$), there exists 2-nd order variability.

A 2-nd order variability for 2-nd order U-variant $V_{\mathrm{\tilde{G}}%
^{[1]},\mathrm{\tilde{G}}^{[2]}\mathrm{,}d}^{[2]}$ is described by the
following two equations%
\begin{align}
\tilde{U}^{[1]}(\delta \phi_{\mathrm{global}}^{[1]})  &  \leftrightarrow \hat
{U}^{[2]}(\delta \phi_{\mathrm{global}}^{[2]})\nonumber \\
&  =\exp(i\lambda^{\lbrack12]}\delta \phi_{\mathrm{global}}^{[1]}),
\end{align}
and
\begin{align}
\mathcal{T}(\delta x^{\mu})  &  \leftrightarrow \hat{U}^{[1]}((\delta \phi^{\mu
})^{[1]})\nonumber \\
&  =\exp(i(T^{\mu})^{[1]}(\delta \phi^{\mu})^{[1]})\\
&  =\exp(i(T^{\mu})^{[1]}(k_{0}^{\mu}\delta x^{\mu})),\nonumber
\end{align}
where $\delta \phi_{\mathrm{global}}^{[1]}=\sqrt{%
%TCIMACRO{\dsum \limits_{\mu}}%
%BeginExpansion
{\displaystyle \sum \limits_{\mu}}
%EndExpansion
((\delta \phi^{\mu})^{[2]})^{2}}$\textit{ }and\textit{ }$\delta \phi
_{\mathrm{global}}^{[1]}=\sqrt{%
%TCIMACRO{\dsum \limits_{\mu}}%
%BeginExpansion
{\displaystyle \sum \limits_{\mu}}
%EndExpansion
((\delta \phi^{\mu})^{[1]})^{2}}.$ $\mathcal{T}(\delta x^{\mu})$ is the spatial
translation operation along $x^{\mu}$-direction, $\hat{U}^{[1]}((\delta
\phi^{\mu})^{[1]})=\exp(i(T^{\mu})^{[1]}(\delta \phi^{\mu})^{[1]})$ and
$\hat{U}^{[1]}(\delta \phi_{\mathrm{global}}^{[1]})$\ are shift operations on
$\mathrm{C}_{\mathrm{\tilde{G}}^{[1]}\mathrm{,}d}^{[1]}$, $\hat{U}%
^{[2]}(\delta \phi_{\mathrm{global}}^{[2]})$ is shift operation on
$\mathrm{C}_{\mathrm{\tilde{G}}^{[2]}\mathrm{,}d}^{[2]}$. For simplicity, we
can denote a system with 2-nd order variability by the following equation
\begin{equation}
\mathcal{T}\leftrightarrow \hat{U}^{[1]}\leftrightarrow \hat{U}^{[2]}.
\end{equation}

Under K-projection, we have a two levels of (composite) zero lattices for a
higher-dimensional 2-nd order U-variant $V_{0,\mathrm{\tilde{G}}%
^{[1]},\mathrm{\tilde{G}}^{[2]}\mathrm{,}d}^{[2]}$: the group-changing space
for $\mathrm{\tilde{G}}^{[1]}$ is reduced into a $d$-dimensional level-1 zero
lattice on Cartesian space, of which each zero corresponds to $\lambda
^{\lbrack12]}$ level-2 zeroes that is projected from $\mathrm{\tilde{G}}%
^{[1]}$.

Finally, we discuss 2-nd order P-variants $V_{\mathrm{\tilde{G}}%
^{[1]},\mathrm{\tilde{G}}^{[2]}\mathrm{,}d}^{[2]}$ that can be regarded as a
variant by slightly perturbing a 2-nd order U-variant $V_{0,\mathrm{\tilde{G}%
}^{[1]},\mathrm{\tilde{G}}^{[2]}\mathrm{,}d}^{[2]}.$ Here, the subscript "$0$"
denotes a U-variant. We then use the Hybrid-order representation under
K-projection for a 2-nd order P-variant. The extra level-1 group-changing
elements of $\mathrm{C}_{\mathrm{\tilde{G}}^{[1]}}^{[1]}$ are effectively
characterized by local field of a compact \textrm{U(1)} group, and the extra
level-2 group-changing elements of $\mathrm{C}_{\mathrm{\tilde{G}}^{[2]}%
}^{[2]}$ are effectively characterized by local field of a compact
\textrm{U(1)} group and a usual \textrm{N}-component field of compact
\textrm{SU(N)} ($\mathrm{N}=\lambda^{\lbrack12]}$) group. As a result, by
using both a usual field of compact \textrm{U(1)} group and a usual
\textrm{N}-component field of compact \textrm{SU(N)} ($\mathrm{N}%
=\lambda^{\lbrack12]}$) group, we fully describe $V_{\mathrm{\tilde{G}}%
^{[1]},\mathrm{\tilde{G}}^{[2]}\mathrm{,}d}^{[2]}$.

\subsubsection{Left/right-hand variants}

\emph{Chirality} is an asymmetric quality that is relevant in many fields of
science. The word chirality comes from the Greek, which means
\textquotedblleft \emph{hand}\textquotedblright. A chiral system is distinct
from its mirror image under mirror operation. For example, corkscrew is a
chiral object. One always has two different methods of use -- left-hand, or
right-hand.\ In this part, we introduce the concept of left/right-hand variants.

We firstly give a definition about a left/right-hand variant.

\textit{Definition -- A left/right-hand variant }$V_{L/R,\mathrm{\tilde{G}}%
,d}[\Delta \phi_{L/R}^{\mu},\Delta x^{\mu},k_{L/R,0}^{\mu}]$\textit{ is denoted
by\ a chiral mapping between a d-dimensional group-changing space }%
$\mathrm{C}_{L/R,\mathrm{\tilde{G},}d}$\textit{ and \textit{Cartesian }space
}$\mathrm{C}_{d}$\textit{\ with positive/negative changing rates, i.e.,}%
\begin{align}
V_{L/R,\mathrm{\tilde{G}},d}[\Delta \phi_{L/R}^{\mu},\Delta x^{\mu}%
,k_{L/R,0}^{\mu}]  &  :\mathrm{C}_{L/R,\mathrm{\tilde{G},}d}=\{ \delta
\phi_{L/R}^{\mu}\} \nonumber \\
&  \Longleftrightarrow^{L/R}\mathrm{C}_{d}=\{ \delta x^{\mu}\}
\end{align}
\textit{where }$\Longleftrightarrow^{L}$\textit{\ denotes a left-hand ordered
mapping under fixed changing rate of integer multiple, }$k_{L,0}^{\mu}%
>0$\textit{ and }$\Longleftrightarrow^{R}$\textit{\ denotes a right-hand
ordered mapping under fixed changing rate of integer multiple, }$k_{R,0}^{\mu
}<0.$\textit{\ }

We take 1D left/right-hand variant $V_{L/R,\mathrm{\tilde{U}(1),}1}[\Delta
\phi_{L/R},\Delta x,k_{L/R,0}]$ as example. $V_{L/R,\mathrm{\tilde{U}(1),}%
1}[\Delta \phi_{L/R},\Delta x,k_{L/R,0}]$ is described by a chiral mapping
between 1D group-changing space $\mathrm{C}_{L/R,\mathrm{\tilde{U}(1)}%
,1}(\Delta \phi_{L/R})$\textit{ }and 1D Cartesian space $\mathrm{C}_{1}$,
i.e.,
\begin{align}
V_{L/R,\mathrm{\tilde{U}(1),}1}[\Delta \phi_{L/R},\Delta x,k_{L/R,0}]  &
:\mathrm{C}_{L/R,\mathrm{\tilde{U}(1)},1}(\Delta \phi_{L/R})=\{ \delta
\phi_{L/R}\} \nonumber \\
&  \Longleftrightarrow^{L/R}\mathrm{C}_{1}=\{ \delta x\}
\end{align}
where $\Longleftrightarrow^{L/R}$\ denotes a chiral ordered mapping under
fixed changing rate of integer multiple $k_{L/R,0}$. For the left-hand case,
we have $k_{L,0}>0$; For the right-hand case, we have $k_{R,0}<0.$

Another example is three dimensional (3D) left/right-hand $\mathrm{\tilde
{S}\tilde{O}}$\textrm{(3)} variant $V_{L/R,\mathrm{\tilde{S}\tilde{O}(3)}%
,3}[\Delta \phi_{L/R}^{\mu},\Delta x^{\mu},k_{L/R,0}^{\mu}]$. A $d$-dimensional
left/right-hand $\mathrm{\tilde{S}\tilde{O}}$\textrm{(3)} variant is a chiral
mapping between Clifford group-changing space $\mathrm{C}_{L/R,\mathrm{\tilde
{S}\tilde{O}(3)},3}$\ and a rigid spacetime $\mathrm{C}_{3},$ i.e.,\textit{ }%
\begin{align}
V_{L/R,\mathrm{\tilde{S}\tilde{O}(3)},3}[\Delta \phi_{L/R}^{\mu},\Delta x^{\mu
},k_{L/R,0}^{\mu}]  &  :\nonumber \\
\mathrm{C}_{L/R,\mathrm{\tilde{S}\tilde{O}(3)},3}(\Delta \phi_{L/R}^{\mu})  &
=\{ \delta \phi_{L/R}^{\mu}\} \Leftrightarrow \mathrm{C}_{3}=\{ \delta x^{\mu}\}
\end{align}
where a Clifford group-changing space\textit{ }$\mathrm{C}_{L/R,\mathrm{\tilde
{S}\tilde{O}(3)},3}(\Delta \phi^{\mu})$\textit{\ }is described by $3$ series of
numbers of group elements $\phi^{\mu}$ arranged in size order with unit
"vector" as Pauli matrices $\sigma^{\mu}$ where $\mu=x,y,z$. $\Leftrightarrow
$\ denotes chiral ordered mapping with fixed changing rate of integer multiple
$k_{L/R,0}.$ For the left-hand case, we have $k_{L,0}^{\mu}>0$; For the
right-hand case, we have $k_{R,0}^{\mu}<0.$

Then, we focus on the left-hand uniform variant and show its properties.

We take 1D left-hand uniform variant of non-compact \textrm{\~{U}(1)} group as
an example. There are three kinds of 1-st order representations from different
aspects -- algebra, geometry, and algebra, representations.

In 1-st order algebra representation, the 1D left-hand uniform variant is
characterized by a series of (non-local) group-changing elements of
non-compact \textrm{\~{U}(1)} group. For either uniform left-hand variant,
there exists only one type of group-changing elements,
\begin{equation}
\delta \phi_{L,i}\equiv k_{L,0}\delta x_{i}%
\end{equation}
with fixed changing rate $\frac{d\phi_{L}}{dx}=k_{L,0}.$ Now, the ordered
mapping can be denoted by the series of same number "$1$", i.e.,
\begin{equation}
\{n_{L,i}\}=(...1,1,1,1,...).
\end{equation}
We can "\emph{generate}" the 1D uniform left-hand variant by a series of
group-changing elements $\delta \phi_{L,i}(x_{i})$ on every position $x$ of
Cartesian space $\mathrm{C}_{1}$, i.e., $\tilde{U}(\delta \phi_{L})=\prod
_{i}\tilde{U}(\delta \phi_{L,i}(x_{i}))$ with\ $\tilde{U}(\delta \phi
_{L,i}(x_{i}))=e^{i((\delta \phi_{L,i})\cdot \hat{K})}$ and $\hat{K}=-i\frac
{d}{d\phi_{L}}.$\ Here, the i-th infinitesimal group-changing operation
$\tilde{U}(\delta \phi_{L,i})$ generates a group-changing element on position
$i$.

In 1-st order algebra representation, the 1D uniform left-hand variant is
usually described by a complex field $\mathrm{z}_{L}=e^{i\phi_{L/R}(x)}$. We
then do group-changing operation on natural reference $\mathrm{z}_{0}%
=e^{i\phi_{0}}$ and get the algebra representation of the corresponding variants.

The complex field $\mathrm{z}_{L,u}(x)$ for a uniform left-hand variant is
obtained by
\begin{equation}
\mathrm{z}_{L,u}(x)=\tilde{U}(\delta \phi_{L})\mathrm{z}_{0}%
\end{equation}
where $\tilde{U}(\delta \phi_{L})=\prod_{i}\tilde{U}(\delta \phi_{L,i}(x_{i}))$
denotes a series of group-changing operations\ with $\tilde{U}(\delta
\phi_{L,i}(x_{i}))=e^{i((\delta \phi_{L,i})\cdot \hat{K})}$ and $\hat{K}%
=-i\frac{d}{d\phi_{L}}.$\ Here, the i-th group-changing operation $\tilde
{U}(\delta \phi_{L,i}(x))$ at $x$ generates a group-changing element.\ For the
case of a single group-changing element $\delta \phi_{L,i}(x_{i})$ on $\delta
x_{i}$ at $x_{i},$ the function is given by
\begin{equation}
\phi_{L}(x)=\left \{
\begin{array}
[c]{c}%
-\frac{\delta \phi_{L,i}}{2},\text{ }x\in(-\infty,x_{i}]\\
-\frac{\delta \phi_{L,i}}{2}+k_{L,0}x,\text{ }x\in(x_{i},x_{i}+\delta x_{i}]\\
\frac{\delta \phi_{L,i}}{2},\text{ }x\in(x_{i}+\delta x_{i},\infty)
\end{array}
\right \}  .
\end{equation}
So, a 1D uniform left-hand variant $V_{\mathrm{\tilde{U}(1),}1}^{L}[\Delta
\phi_{L},\Delta x,k_{L,0}]$ can be described by a special complex field
$\mathrm{z}_{L,u}(x)$ in Cartesian space as
\[
\mathrm{z}_{L,u}(x)=\exp(i\phi_{L}(x))
\]
where $\phi_{L}(x)=\phi_{0}+k_{L,0}x$. In particular, we have $k_{L,0}>0$.

For a uniform left-hand variant with infinite size ($\Delta x\rightarrow
\infty$), we have the following relationship,%
\begin{equation}
\mathcal{T}_{L}(\delta x)\leftrightarrow \hat{U}(\delta \phi_{L})=e^{i\cdot
\delta \phi_{L}}%
\end{equation}
where $\mathcal{T}_{L}(\delta x)$ is the spatial translation operation on
Cartesian space and $\hat{U}(\delta \phi_{L})$ is shifting operation on
group-changing space $\mathrm{C}_{\mathrm{\tilde{U}(1)},1}(\Delta \phi_{L})$,
and $\delta \phi_{L}=k_{L,0}\delta x$. That means when one translates along
Cartesian space $\delta x,$ the corresponding shifting along group-changing
space $\mathrm{C}_{L,\mathrm{\tilde{U}(1)},d}$ is $\delta \phi_{L}%
=k_{L,0}\delta x.$ For the left-hand case, due to $k_{L,0}>0,$ we have
$\delta \phi_{L}>0.$

For the 1D uniform left-hand variant $V_{\mathrm{\tilde{U}(1),}1}[\Delta
\phi,\Delta x,k_{0}]$ of non-compact \textrm{\~{U}(1)} group, we map the
original complex fields $\mathrm{z}_{L,u}(x)=\exp(ik_{L,0}x+i\phi_{0})=\xi
_{L}(x)+i\eta_{L}(x)$ to two curved lines $\{x,\xi_{L}(x),\eta_{L}(x)\}$ in
three dimensions. See the illustration of 1D uniform left-hand variant in
Fig.36 (a) and 1D uniform right-hand variant in Fig.36 (b), respectively. The
uniform left-hand variant corresponds to a spiral line on a cylinder with
fixed radius that can be regarded as a knot/link structure between the curved
line of $\mathrm{z}_{L,u}(x)$ and the straight line at center of
$\mathrm{z}(x)=0$. From this figure, one can obviously understand the physical
meaning of chirality.

\begin{figure}[ptb]
\includegraphics[clip,width=0.92\textwidth]{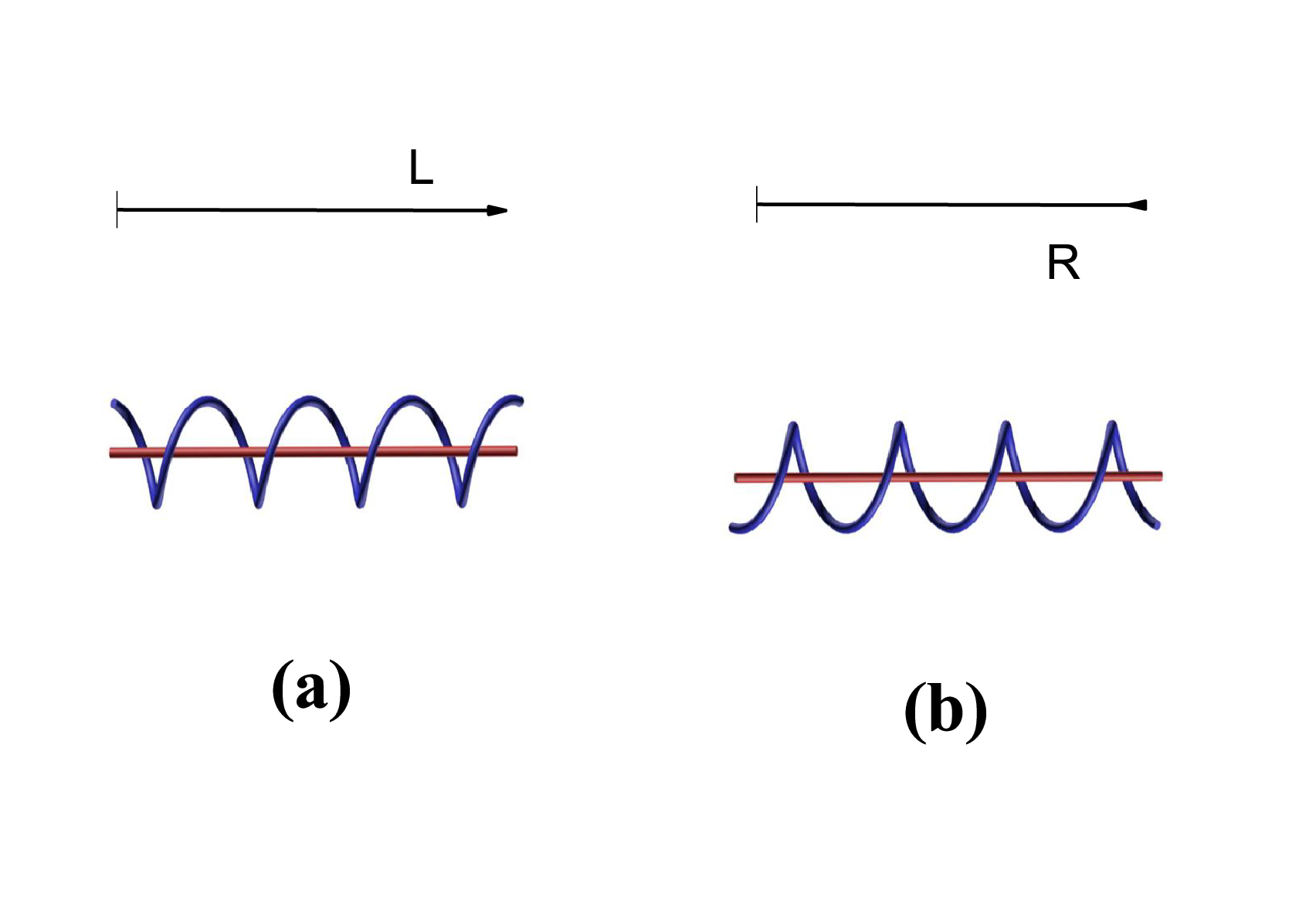}\caption{(Color online)
(a) Upper: the illustration of 1D uniform left-hand variant, of which the
length of the arrow denotes its changing rate and the direction of arrow
denotes chirality, Lower: 1-st order Geometry representation of the uniform
left-hand variant. The uniform variant corresponds to a spiral line on a
cylinder with fixed radius; (b) Upper: the illustration of 1D uniform
right-hand variant, Lower: 1-st order Geometry representation of the uniform
right-hand variant. }%
\end{figure}

We next do\emph{ }K-projection on uniform left-hand variant and get zero lattice.

In mathematics, the K-projection is defined by
\begin{equation}
\hat{P}_{\theta}\left(
\begin{array}
[c]{c}%
\xi_{L}(x)\\
\eta_{L}(x)
\end{array}
\right)  =\left(
\begin{array}
[c]{c}%
\xi_{L,\theta}(x)\\
\left[  \eta_{L,\theta}(x)\right]  _{0}%
\end{array}
\right)
\end{equation}
where $\xi_{L,\theta}(x)$ is variable and $\left[  \eta_{L,\theta}(x)\right]
_{0}$ is constant. In the following parts we use $\hat{P}_{\theta}$ to denote
the projection operators. Because the projection direction out of the curved
line is characterized by an angle $\theta$ in $\{ \xi_{L},\eta_{L}\}$ space,
we have
\begin{equation}
\left(
\begin{array}
[c]{c}%
\xi_{L,\theta}\\
\eta_{L,\theta}%
\end{array}
\right)  =\left(
\begin{array}
[c]{cc}%
\cos \theta & \sin \theta \\
\sin \theta & -\cos \theta
\end{array}
\right)  \left(
\begin{array}
[c]{c}%
\xi_{L}\\
\eta_{L}%
\end{array}
\right)
\end{equation}
where $\theta$ is angle \textrm{mod}($2\pi$), i.e. $\theta \operatorname{mod}%
2\pi=0.$ So the curved line of 1D variant is described by the function
\begin{equation}
\xi_{L,\theta}(x)=\xi_{L}(x)\cos \theta+\eta_{L}(x)\sin \theta.
\end{equation}

Under projection, each zero corresponds to a solution of the equation
\begin{equation}
\hat{P}_{\theta}[\mathrm{z}_{L,u}(x)]\equiv \xi_{L,\theta}(x)=0.\nonumber
\end{equation}
1D uniform left-hand variant becomes a 1D crystal of zeros (or 1D zero
lattice). For a 1D uniform left-hand variant $V_{\mathrm{\tilde{U}(1)}%
,1}^{L/R}(\Delta \phi_{L},\Delta x,k_{L,0})$ of non-compact \textrm{\~{U}(1)}
group, from the its algebra representation $\mathrm{z}_{L,u}(x)\sim
e^{ik_{L,0}\cdot x}$, from the zero-equation $\xi_{L,\theta}(x)=0$ or
$\cos(k_{L,0}x-\theta)=0,$ we get the zero-solutions to be
\begin{equation}
x_{L}=l_{L,0}\cdot N/2+\frac{l_{L,0}}{2\pi}(\theta+\frac{\pi}{2})
\end{equation}
where $N$ is an integer number, and $l_{L,0}=2\pi/k_{L,0}$. This is a 1D
crystal of zeros for a U-variant (we also call it zero lattice). Each crossing
corresponds to a zero. Under K-projection, the chirality of the variant disappears.

To obtain a perturbative left-hand uniform variant, one can do
\emph{perturbatively} changings on a uniform one. Now, there must exist more
than one type of group-changing elements on it. The number extra
group-changing elements is tiny. In general, we use "field" description to
characterize the distributions of the extra group-changing elements.

In the end, we discuss the property of a 1D right-hand variant of non-compact
\textrm{\~{U}(1)} group. For a right-hand variant, the changing rate becomes
negative, $k_{R,0}<0.$ So. we can use similar approach to describe the
right-hand variant and don't show the detailed discussion. In particular,
under geometric representation, a left-hand variant corresponds to spiral line
winding clockwise; a right-hand variant corresponds to spiral line winding counterclockwise.

\subsubsection{Chiral variant}

In this part, we introduce the concept of \emph{chiral variant}.

We consider a composite variant $V_{\mathrm{\tilde{G}},d}^{\mathrm{chiral}}$
with two sub-variants -- a left-hand sub-variant $V_{L,\mathrm{\tilde{G}}%
^{L},d}^{\mathrm{sub}}[\Delta \phi_{L}^{\mu},\Delta x^{\mu},k_{L,0}^{\mu}]$ and
a right-hand sub-variant $V_{R,\mathrm{\tilde{G}}^{R},d}^{\mathrm{sub}}%
[\Delta \phi_{R}^{\mu},\Delta x^{\mu},k_{R,0}^{\mu}]$. So, we denote the
variant by
\begin{equation}
V_{\mathrm{\tilde{G}},d}=\{V_{L,\mathrm{\tilde{G}}^{L},d}^{\mathrm{sub}%
},V_{R,\mathrm{\tilde{G}}^{R},d}^{\mathrm{sub}}\}.
\end{equation}
Here \textrm{G} with "$\sim$" above it means a non-compact Lie group. $L/R$
denotes left-hand or right-hand.

The \emph{chirality} of variants $V_{\mathrm{\tilde{G}},d}^{\mathrm{chiral}}$
is classified by the symmetries of left/right-hand sub-variants. If
$k_{L,0}^{\mu}=-k_{R,0}^{\mu},$ the chirality of the variant
$V_{\mathrm{\tilde{G}},d}$ is symmetric; If $k_{L,0}^{\mu}\neq-k_{R,0}^{\mu},$
the chirality of the variant is asymmetric. See the illustration in Fig.37.

\begin{figure}[ptb]
\includegraphics[clip,width=0.92\textwidth]{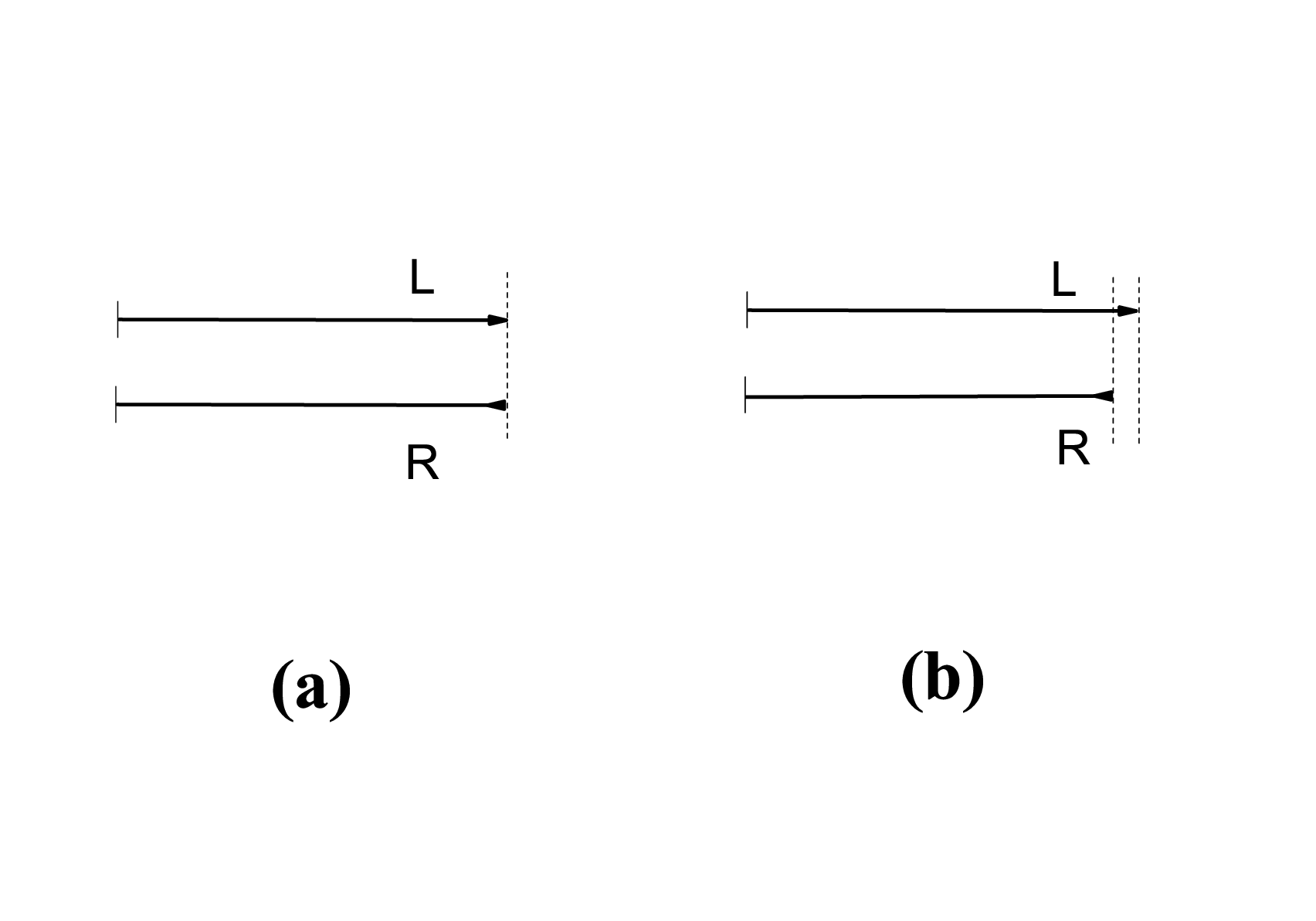}\caption{(Color online)
(a) An illustration of 1D variant with chiral symmetry; (b) An illustration of
1D variant without chiral symmetry}%
\end{figure}

To characterize the chirality of variants, we regroup the composite variant
$V_{\mathrm{\tilde{G}},d}^{\mathrm{chiral}}$ into two new sub-variant --
\emph{global sub-variant }$V_{g,\mathrm{\tilde{G}},d}^{\mathrm{sub}}%
[\Delta \phi_{g}^{\mu},\Delta x^{\mu},k_{g,0}^{\mu}]$ and \emph{relative
sub-variant }$V_{r,\mathrm{\tilde{G}},d}^{\mathrm{sub}}[\Delta \phi_{r}^{\mu
},\Delta x^{\mu},k_{r,0}^{\mu}]$.

On the other hand, the global sub-variant $V_{g,\mathrm{\tilde{G}}%
,d}^{\mathrm{sub}}[\Delta \phi_{g}^{\mu},\Delta x^{\mu},k_{g,0}^{\mu}]$
characterizes the synchronous changings of left-hand sub-variant and
right-hand sub-variant. Therefore, it is really a variant of non-compact Lie
group $\mathrm{\tilde{G}\otimes \tilde{U}(1).}$ Now, the corresponding
generator of global sub-variant along given direction becomes $T_{L}^{\mu
}\otimes \tau_{z}=T_{R}^{\mu}\otimes \tau_{z}$. The changing rate is denoted by
$k_{g,0}^{\mu}$.

On the other hand, the relative sub-variant $V_{r,\mathrm{\tilde{G}}%
,d}^{\mathrm{sub}}[\Delta \phi_{r}^{\mu},\Delta x^{\mu},k_{r,0}^{\mu}]$
characterizes the relative changings of left-hand sub-variant and right-hand
sub-variant. It is a left/right-hand variant of non-compact Lie group
$\mathrm{\tilde{G}.}$ Now, the corresponding generator of relative sub-variant
along given direction becomes $T_{L}^{\mu}\ $or $T_{R}^{\mu}$. Further
discussion is needed to determine whether it is $T_{L}^{\mu}\ $or $T_{R}^{\mu
}.$ See below discussion. The changing rate is $(k_{L,0}^{\mu}+k_{R,0}^{\mu})$.

Hence, for the composite variant with symmetric chirality, we have
$k_{r,0}^{\mu}=0.$ There doesn't exist the relative sub-variant; for the
variant with asymmetric chirality, we have $k_{r,0}^{\mu}\neq0.$ The variant
becomes a chiral type.

We take 1D chiral variant $V_{\mathrm{\tilde{G}},d}^{\mathrm{chiral}}$ of
non-compact $\mathrm{\tilde{U}(1)}$ Lie group as an example to show the
concept of chiral variant.

Now, the generators of left/right-hand sub-variants are same, i.e., $T_{L}%
^{x}=T_{R}^{x}=1$. When $k_{L,0}\neq-k_{R,0}$, the variant becomes a chiral
one without symmetric chirality. Consequently, the global sub-variant
$V_{g,\mathrm{\tilde{U}(1)},1}^{\mathrm{sub}}[\Delta \phi_{g},\Delta
x,k_{g,0}]$ is a variant of non-compact Lie group $\mathrm{\tilde{U}(1)}$
(really an Abelian sub-group of an \textrm{S}$\mathrm{\tilde{U}(2)}$ group).
The corresponding generator of global sub-variant becomes $\tau_{z}$. The
relative sub-variant $V_{r,\mathrm{\tilde{U}(1)},1}^{\mathrm{sub}}[\Delta
\phi_{r},\Delta x,k_{r,0}]$ is a left-hand (or right hand) variant of
non-compact Lie group $\mathrm{\tilde{U}(1)}$, of which the generator along
different direction is $1\mathrm{.}$ The changing rate is fixed to be
$(k_{L,0}^{\mu}+k_{R,0}^{\mu})$.

Another example of chiral variant is $V_{\mathrm{\tilde{S}\tilde{O}(3)}%
,d}^{\mathrm{chiral}}$. The generators of left/right-hand sub-variant are
$T_{L}^{\mu}=T_{R}^{\mu}=\sigma^{\mu}$. $k_{L/R,0}^{\mu}$\ characterizes the
changing rate along $\mu$-th spatial direction ($\mu=x,y,z$). When the
chirality of $V_{\mathrm{\tilde{S}\tilde{O}(3)},d}^{\mathrm{chiral}}$ becomes
asymmetry, we have $k_{r,0}^{\mu}=k_{L,0}^{\mu}+k_{R,0}^{\mu}\neq0.$ The
global sub-variant $V_{g,\mathrm{\tilde{S}\tilde{O}(3)},d}^{\mathrm{sub}%
}[\Delta \phi_{g}^{\mu},\Delta x^{\mu},k_{g,0}^{\mu}]$ becomes a variant of
non-compact Lie group $\mathrm{\tilde{S}\tilde{O}(4).}$ Now, the corresponding
generator of global sub-variant along given direction becomes $\sigma^{\mu
}\otimes \tau_{z}=\Gamma^{\mu}$. The relative sub-variant $V_{r,\mathrm{\tilde
{U}(1)},3}^{\mathrm{sub}}[\Delta \phi_{r}^{\mu},\Delta x^{\mu},k_{r,0}^{\mu}]$
is a left/right-hand variant of non-compact Lie group, of which the generator
along different direction is $\sigma^{\mu}\mathrm{.}$ The changing rate is
$k_{r,0}^{\mu}=(k_{L,0}^{\mu}+k_{R,0}^{\mu})$.

\subsubsection{Classification}

In this part, we classify the chiral variant $V_{\mathrm{\tilde{G}}%
,d}^{\mathrm{chiral}}$ of non-compact Lie group \textrm{\~{G}}.

Different chiral variants are classified by five values -- the non-compact Lie
group \textrm{\~{G} }that determines the whole structure (for example, our
universe is relevant to $\mathrm{\tilde{S}\tilde{O}}$\textrm{(d)} variant),
the dimension number $d$ of Cartesian space $\mathrm{C}_{d}$ (for example, in
our universe, $d=3$), the changing rate $k_{g,0}^{\mu}$ of global sub-variant,
and the changing rate $k_{r,0}^{\mu}$ of relevant sub-variant or the changing
rate $k_{L,0}^{\mu}$ of left-hand sub-variant, and the changing rate
$k_{R,0}^{\mu}$ of right-hand sub-variant. Addition values is about the
generation of relevant sub-variant (or the relationship between $k_{g/r,0}%
^{\mu}$ and $k_{L/R,0}^{\mu}$).

To classify a chiral variant $V_{\mathrm{\tilde{G}},d}^{\mathrm{chiral}}$, the
key point is the relationship between the two sub-variants of
$\{V_{g,\mathrm{\tilde{G}},d}^{\mathrm{sub}},V_{r,\mathrm{\tilde{G}}%
,d}^{\mathrm{sub}}\}$\ and tho other two $\{V_{L,\mathrm{\tilde{G}}^{L}%
,d}^{\mathrm{sub}},V_{R,\mathrm{\tilde{G}}^{R},d}^{\mathrm{sub}}\}$.

Because changing rate\ for the relative sub-variant is fixed to $k_{r,0}^{\mu
}=(k_{L,0}^{\mu}+k_{R,0}^{\mu}),$ we focus on the changing rate $k_{g,0}^{\mu
}$ for the global sub-variant of $V_{g,\mathrm{\tilde{G}},d}^{\mathrm{sub}}$.
To determine $k_{g,0}^{\mu},$ one must know the exact way to \emph{generate
}the relative sub-variant. If the relative sub-variant is generated by a
combination of both extra left-hand sub-variant and extra right-hand
sub-variant, we have $V_{r,\mathrm{\tilde{G}},d}^{\mathrm{sub}}=\alpha
V_{L,\mathrm{\tilde{G}}^{L},d}^{\mathrm{sub}}+\beta V_{R,\mathrm{\tilde{G}%
}^{R},d}^{\mathrm{sub}}.$ As a result, the changing rate of the relative
sub-variant becomes
\begin{equation}
k_{r,0}^{\mu}=\alpha k_{L,0}^{\mu}+\beta k_{R,0}^{\mu}.
\end{equation}
\begin{figure}[ptb]
\includegraphics[clip,width=0.92\textwidth]{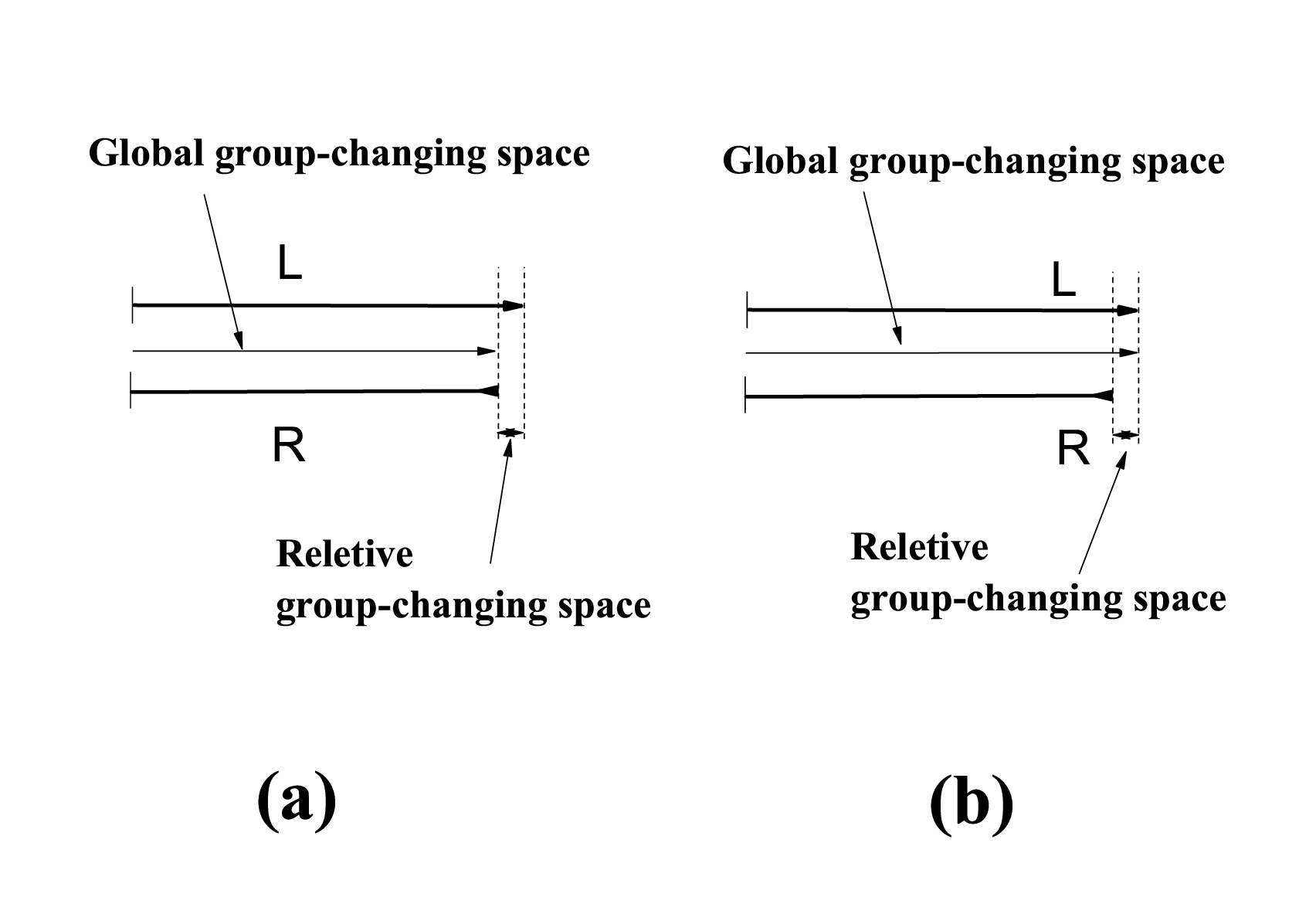}\caption{(Color online)
(a) A chiral variant with relative sub-variant fully induced by extra
left-hand sub-variant ($\alpha \neq0$ and $\beta=0$). The changing rate of the
global sub-variant is $k_{g,0}^{\mu}=k_{L,0}^{\mu}$ and the changing rate of
the relative sub-variant becomes $k_{r,0}^{\mu}=\alpha k_{R,0}^{\mu}$; (b) A
chiral variant with relative sub-variant fully induced by extra right-hand
sub-variant ($\alpha=0$ and $\beta \neq0$). The changing rate of the global
sub-variant is $k_{g,0}^{\mu}=k_{L,0}^{\mu}$ and the changing rate of the
relative sub-variant becomes $k_{r,0}^{\mu}=\beta k_{R,0}^{\mu}$}%
\end{figure}

For example, for a relative sub-variant fully induced by extra left-hand
sub-variant with $\alpha>0$, we have $\beta=0$ and $\alpha \neq0.$ The relative
sub-variant is given by $V_{r,\mathrm{\tilde{G}},d}^{\mathrm{sub}}=\alpha
V_{L,\mathrm{\tilde{G}}^{L},d}^{\mathrm{sub}}.$ The changing rate of the
global sub-variant becomes $k_{g,0}^{\mu}=k_{R,0}^{\mu}$ and the changing rate
of the relative sub-variant becomes $k_{r,0}^{\mu}=\alpha k_{L,0}^{\mu}.$ See
the illustration in Fig.38(a).

For a chiral variant with relative sub-variant fully induced by extra
right-hand sub-variant, we have $\alpha=0$ and $\beta \neq0.$ The relative
sub-variant is given by $V_{r,\mathrm{\tilde{G}},d}^{\mathrm{sub}}=\alpha
V_{R,\mathrm{\tilde{G}}^{R},d}^{\mathrm{sub}}.$ The changing rate of the
global sub-variant becomes $k_{g,0}^{\mu}=k_{L,0}^{\mu}$ and the changing rate
of the relative sub-variant becomes $k_{r,0}^{\mu}=\beta k_{R,0}^{\mu}.$ See
the illustration in Fig.38(b).

For the general cases, we have $\alpha \neq0$ and $\beta \neq0.$ The relative
sub-variant is given by $V_{r,\mathrm{\tilde{G}},d}^{\mathrm{sub}}=\alpha
V_{L,\mathrm{\tilde{G}}^{L},d}^{\mathrm{sub}}+\beta V_{R,\mathrm{\tilde{G}%
}^{R},d}^{\mathrm{sub}}.$ The changing rate of the global sub-variant becomes
$k_{g,0}^{\mu}=(1-\alpha)k_{L,0}^{\mu}=(1-\beta)k_{R,0}^{\mu}$ and the
changing rate of the relative sub-variant becomes $k_{r,0}^{\mu}=\alpha
k_{L,0}^{\mu}+\beta k_{R,0}^{\mu}.$ So, to classify a chiral variant, we can
just classify the two values, $\alpha$ and $\beta$, or $(\alpha,\beta).$

In particular, for our universe, the variant is a chiral one, of which the
relative sub-variant is fully induced by extra left-hand sub-variant. Now, we
have $\beta=0$ and $\alpha>0$.

\subsubsection{The changings of chiral variants}

In this part, we will classify the changings of chiral variants.

There are two types of changings of a chiral variant $V_{\mathrm{\tilde{G},}%
d}^{\mathrm{chiral}}$: one is topological, the other is non-topological. For
topological changings, the group-changing space is globally expand or
contract\emph{ }on Cartesian space $\mathrm{C}_{d}$. For non-topological
changings, there are global shift, local expand/contract and shape changings.
The size of the group-changing space doesn't change. Let give more detailed discussion.

There are five types of changings\ of the $d$-dimensional variant
$V_{\mathrm{\tilde{G},}d}^{\mathrm{chiral}}$:

1) Globally \emph{shifting} $\mathrm{C}_{L/R,\mathrm{\tilde{G}},d}$ without
changing its size on Cartesian space $\mathrm{C}_{d}$: Under globally shifts,
the U-variant is invariant. Therefore, this is a symmetric operation on a
U-variant, such as the global phase symmetry. Due to the existence left-hand
and right sub-variants, there are two kinds of globally shifting -- one for
left-hand, the other for right-hand. These two kinds of globally shifting can
also be transformed into another two -- one for global sub-variant and and
that for relative sub-variant;

2) Globally \emph{rotating} $\mathrm{C}_{L/R,\mathrm{\tilde{G}},d}$ from
$a$-direction to $b$-direction on Cartesian space $\mathrm{C}_{d}$: The
operation rotates the generator along one direction $T^{a}$ to another $T^{b}%
$. The operation of global rotation obeys a compact Lie group and thus will
not change the U-variant. Therefore, this is a also symmetric operation on an
U-variant. Under global rotation, the left-hand sub-variant and right-hand
sub-variant change synchronously. If we consider the global sub-variant and
the relative sub-variant, they also change synchronously;

3) Globally \emph{expanding} or \emph{contracting} $\mathrm{C}%
_{L/R,\mathrm{\tilde{G}},d}$\ with changing its corresponding size on
Cartesian space $\mathrm{C}_{d}$: The operation of contraction/expansion on
group-changing space changes topological number (or particle number). We point
out that globally expand/contract of group-changing space in a variant
corresponds to the generation/annihilate of particles in quantum mechanics.
Due to the existence left-hand and right sub-variants, there are two kinds of
globally expanding or contracting -- one for left-hand, the other for
right-hand. On the other hand, there are globally expanding or contracting for
global sub-variant or for relative sub-variant;

4) Locally rotating\emph{ }on Cartesian space $\mathrm{C}_{d}$: Locally
rotating of $\mathrm{C}_{L/R,\mathrm{\tilde{G}},d}$ on Cartesian space
$\mathrm{C}_{d}$\ ($d>1$) leads to the global sub-variant and the relative
sub-variant (or left-hand and right-hand) locally changing
synchronously.\emph{ }This type of changings of a variant leads to a curving spacetime;

5) Locally \emph{expanding} or \emph{contracting} $\mathrm{C}%
_{L/R,\mathrm{\tilde{G}},d}$\ without changing its corresponding size of
group-changing spaces on Cartesian space $\mathrm{C}_{d}$: There are two types
of local operations of contraction/expansion on group-changing spaces. One
type corresponds to the motion of elementary particles in quantum mechanics
with fixed particles' number and fixed changing rate $k_{L/R,0}^{\mu}$; the
other type corresponds to the motion of collective modes in quantum mechanics
without extra elementary particles and the changing rates become fluctuating,
i.e., $k_{g/r,0}^{\mu}\rightarrow k_{g/r,0}^{\mu}+\delta k_{g/r,0}^{\mu}(x).$

\subsubsection{Representations for chiral variants under the condition of
$k_{r,0}^{\mu}\ll k_{g,0}^{\mu}$}

It was known that a chiral variant $V_{\mathrm{\tilde{G}},d}^{\mathrm{chiral}%
}$ has two sub-variants -- the global sub-variant $V_{g,\mathrm{\tilde{G}}%
,d}^{\mathrm{subl}}[\Delta \phi_{g}^{\mu},\Delta x^{\mu},k_{g,0}^{\mu}]$ and
relative sub-variant $V_{r,\mathrm{\tilde{G}},d}^{\mathrm{sub}}[\Delta \phi
_{r}^{\mu},\Delta x^{\mu},k_{r,0}^{\mu}]$. In this part, we show the
representations for chiral variants $V_{\mathrm{\tilde{G}},d}^{\mathrm{chiral}%
}=\{V_{g,\mathrm{\tilde{G}},d}^{\mathrm{sub}},V_{r,\mathrm{\tilde{G}}%
,d}^{\mathrm{sub}}\}$ under the condition of $k_{r,0}^{\mu}\ll k_{g,0}^{\mu}$.
Now, the chiral symmetry for the chiral variant is weakly broken.

According to above discussion, there are two types of local operations of
contraction/expansion on group-changing spaces -- the motion of elementary
particles with fixed changing rate $k_{g/r,0}^{\mu}$ and the fluctuating of
changing rates without extra elementary particles. To characterize the two
types of local operations of contraction/expansion on group-changing spaces,
we use two different approaches to characterize the same chiral variants
$V_{\mathrm{\tilde{G}},d}^{\mathrm{chiral}}=\{V_{g,\mathrm{\tilde{G}}%
,d}^{\mathrm{sub}},V_{r,\mathrm{\tilde{G}},d}^{\mathrm{sub}}\}$ -- approach I
is to consider the chiral variant as 1-st order variant, approach II is to
consider the chiral variant as effective 2-nd order variant. Let us discuss
the representations based on the two approaches (I and II).

\paragraph{Representations based on approach I}

The approach I is to consider the chiral variant as 1-st order variant. With
help of this approach, we could characterize the motion of elementary
particles with fixed changing rate $k_{L/R,0}^{\mu}.$

Now, we have the usual definition of global/relevant sub-variant
$V_{g/r,\mathrm{\tilde{G}},d}^{\mathrm{sub}}[\Delta \phi_{g/r}^{\mu},\Delta
x^{\mu},k_{g/r,0}^{\mu}],$ that is denoted by\ a 1-st order mapping between a
d-dimensional group-changing space $\mathrm{C}_{g/r,\mathrm{\tilde{G},}d}%
$\textit{ }and Cartesian space\textit{ }$\mathrm{C}_{d}$, i.e.,%
\begin{align}
V_{g/r,\mathrm{\tilde{G}},d}^{\mathrm{sub}}[\Delta \phi_{g/r}^{\mu},\Delta
x^{\mu},k_{g/r,0}^{\mu}]  &  :\mathrm{C}_{g/r,\mathrm{\tilde{G},}d}=\{
\delta \phi_{g/r}^{\mu}\} \nonumber \\
&  \Longleftrightarrow^{g/r}\mathrm{C}_{d}=\{ \delta x^{\mu}\}
\end{align}
where $\Longleftrightarrow^{g}$\ denotes an ordered mapping under fixed
changing rate of integer multiple, $k_{g,0}^{\mu}$ and $\Longleftrightarrow
^{r}$\ denotes an ordered mapping under fixed changing rate of integer
multiple, $k_{r,0}^{\mu}<0.$ Here, we have the condition of $k_{r,0}^{\mu}\ll
k_{g,0}^{\mu}.$

Under the condition of $k_{r,0}^{\mu}\ll k_{g,0}^{\mu}$, the global
sub-variant $V_{g,\mathrm{\tilde{G}},d}^{\mathrm{sub}}[\Delta \phi_{g}^{\mu
},\Delta x^{\mu},k_{g,0}^{\mu}]$\ is assumed to be the background. Under
K-projection, the global sub-variant $V_{g,\mathrm{\tilde{G}},d}%
^{\mathrm{sub}}[\Delta \phi_{g}^{\mu},\Delta x^{\mu},k_{g,0}^{\mu}]$ is reduced
into a d-dimensional zero lattice. The d-dimensional zero lattice turns into a
curved space in continuum limit. So, we focus on the relevant sub-variant
$V_{r,\mathrm{\tilde{G}},d}^{\mathrm{sub}}[\Delta \phi_{r}^{\mu},\Delta x^{\mu
},k_{r,0}^{\mu}].$

Firstly, we take 1D variant $V_{\mathrm{\tilde{U}(1),}1}^{\mathrm{chiral}}$ of
non-compact $\mathrm{\tilde{U}(1)}$ Lie group as an example to show the
representation for relative sub-variant.

The relevant sub-variant $V_{r,\mathrm{\tilde{U}(1)},1}^{\mathrm{sub}}$ is
really a left/right-hand variant described by complex field $\mathrm{z}%
_{r}=e^{i\phi_{r}(x)}=\tilde{U}(\delta \phi_{r})\mathrm{z}_{0}$ where
$\tilde{U}(\delta \phi_{r})=\prod_{i}\tilde{U}(\delta \phi_{r,i}(x_{i}))$ denote
a series of group-changing operations\ with $\tilde{U}(\delta \phi_{r,i}%
(x_{i}))=e^{i((\delta \phi_{r,i})\cdot \hat{K}^{[1]})}$ and $\hat{K}%
^{[1]}=-i\frac{d}{d\phi_{r}}.$\ Here, the i-th group-changing operation
$\tilde{U}(\delta \phi_{r,i}(x))$ at $x$ generates a group-changing
element.\ For the case of single group-changing element $\delta \phi
_{r,i}(x_{i})$ on $\delta x_{i}$ at $x_{i},$ the function is given by
\begin{equation}
\phi_{r,i}(x)=\left \{
\begin{array}
[c]{c}%
-\frac{\delta \phi_{r,i}}{2},\text{ }x\in(-\infty,x_{i}]\\
-\frac{\delta_{r,i}}{2}+k_{r,0}x,\text{ }x\in(x_{i},x_{i}+\delta x_{i}]\\
\frac{\delta_{r,i}}{2},\text{ }x\in(x_{i}+\delta x_{i},\infty)
\end{array}
\right \}  .
\end{equation}
Therefore, the 1D uniform relevant sub-variant\ is described by the complex
field $\mathrm{z}_{r}(x)$ in Cartesian space as $\mathrm{z}_{r}(x)=\exp
(i\phi_{r}(x))$ where $\phi_{r}(x)=\phi_{0}+k_{r,0}x$.

For the 1D uniform relevant sub-variant with infinite size ($\Delta
x\rightarrow \infty$), we have the following relationship,%
\begin{equation}
\mathcal{T}_{r}(\delta x)\leftrightarrow \hat{U}(\delta \phi_{r})=e^{i\cdot
\delta \phi_{r}}%
\end{equation}
where $\mathcal{T}_{r}(\delta x)$ is the spatial translation operation and
$\hat{U}(\delta \phi_{r})$ is shifting operation on group-changing space, and
$\delta \phi_{r}=k_{r,0}\delta x$. As a result, for the uniform chiral variant,
its global sub-variant and$\ $reletive sub-variant both have 1-st order
variability, i.e.,
\begin{equation}
\mathcal{T}_{g}(\delta x)\leftrightarrow \hat{U}(\delta \phi_{g})=e^{i\cdot
\delta \phi_{g}\tau^{z}}%
\end{equation}
and
\begin{equation}
\mathcal{T}_{r}(\delta x)\leftrightarrow \hat{U}(\delta \phi_{r})=e^{i\cdot
\delta \phi_{r}}%
\end{equation}
where $\mathcal{T}_{g/r}(\delta x)$ is the spatial translation operation and
$\hat{U}(\delta \phi_{g/r})$ is shifting operation on group-changing space, and
$\delta \phi_{g}=k_{g,0}\delta x$ and $\delta \phi_{r}=k_{r,0}\delta x$.

Under geometric representation, we do the K-projection on the chiral variant
based on approach I.

Under K-projection, we have two zero lattices -- one for global sub-variant,
the other for relative sub-variant. Each zero corresponds to a solution of the
equation
\begin{equation}
\hat{P}_{\theta^{\lbrack1]}}[\mathrm{z}_{g/r}(x)]\equiv \xi_{g/r,\theta}(x)=0
\end{equation}
where $\xi_{g/r,\theta}(x)=\operatorname{Im}(\mathrm{z}_{g/r}(x))$. From the
zero-equation $\xi_{g,\theta}(x)=0$ or $\cos(k_{g,0}x-\theta)=0,$ we get the
zero-solutions of global sub-variant to be
\begin{equation}
x_{g}=l_{g}\cdot N_{g}/2+\frac{l_{g}}{2\pi}(\theta+\frac{\pi}{2})
\end{equation}
where $N_{g}$ is an integer number, and $l_{g}=2\pi/k_{g,0}$. From the
zero-equation $\xi_{r,\theta}(x)=0$ or $\cos(k_{r,0}x-\theta)=0,$ we get the
zero-solutions of global sub-variant to be
\begin{equation}
x_{r}=l_{r}\cdot N_{r}/2+\frac{l_{r}}{2\pi}(\theta+\frac{\pi}{2})
\end{equation}
where $N_{g}$ is an integer number, and $l_{r}=2\pi/k_{r,0}$. In this limit of
$k_{r,0}\ll k_{g,0}$, we have a very large unit cell, of which the length is
$l_{r}$. For a zero of relative sub-variant, there are $\frac{k_{g,0}}%
{k_{r,0}}$ zeroes of the global sub-variant.

For a higher-dimensional uniform chiral variant, there also exist two types of
zero lattices, one for global sub-variant, the other for relative sub-variant.
The situation is same to the case for a 1D uniform chiral variant by doing
knot-projection along its $\mu$-th spatial direction under "direction
projection" (D-projection). Then, under D-projection, we reduce the
higher-dimensional uniform chiral variant to $2\times d$ D-projected 1D
uniform chiral variant, $d$ for global sub-variant, and the other $d$ for
relevant sub-variant.

In the following parts, we will show that the zeroes of global sub-variant
correspond to Dirac type elementary particles (charged leptons\ and quarks),
and the zeroes for relevant sub-variant correspond to Weyl type of elementary
particles (neutrinos).

\paragraph{Representations based on approach II}

The approach II is to consider the chiral variant as effective 2-nd order
variant. With help of this approach, we could characterize the fluctuating of
changing rates without extra elementary particles.

To define 2-nd order variant, we split the group-changing space $\mathrm{C}%
_{g/r,\mathrm{\tilde{G}},d}^{\mathrm{sub}}(\Delta \phi_{g/r}^{\mu})$ of the
global/relevant sub-variant $V_{g/r,\mathrm{\tilde{G}},d}^{\mathrm{sub}%
}[\Delta \phi_{g/r}^{\mu},\Delta x^{\mu},k_{g/r,0}^{\mu}]$ into two kinds of
group-changing subspaces: one group-changing subspaces $\mathrm{C}%
_{g/r,\mathrm{\tilde{U}(1)\in \tilde{G}},1}^{\mathrm{sub}}(\Delta
\phi_{g/r,\mathrm{global}})$ is about global phase changing of the system
$\Delta \phi_{g/r,\mathrm{global}}=\sqrt{%
%TCIMACRO{\dsum \limits_{\mu}}%
%BeginExpansion
{\displaystyle \sum \limits_{\mu}}
%EndExpansion
(\Delta \phi_{g/r}^{\mu}(x))^{2}}$, the other is about internal relative angles.

Then, we introduce a globally-mapping between two group-changing
spaces\textit{ }$\mathrm{C}_{g,\mathrm{\tilde{U}(1)\in \tilde{G}},1}(\Delta
\phi_{g,\mathrm{global}})$ and\textit{ }$\mathrm{C}_{r,\mathrm{\tilde{U}%
(1)\in \tilde{G}},1}(\Delta \phi_{r,\mathrm{global}})$\textit{, i.e.,}%
\begin{equation}
\mathrm{C}_{g,\mathrm{\tilde{U}(1)\in \tilde{G}},1}(\Delta \phi
_{g,\mathrm{global}})\Longleftrightarrow \mathrm{C}_{r,\mathrm{\tilde{U}%
(1)\in \tilde{G}},1}(\Delta \phi_{r,\mathrm{global}}).
\end{equation}
In mathematic, it is defined by the mapping between global phase changings of
the two group-changing spaces\textit{,}
\begin{equation}
\delta \phi_{g,\mathrm{global}}=\lambda^{\lbrack gr]}\delta \phi
_{r,\mathrm{global}}%
\end{equation}
\textit{ }where $\lambda^{\lbrack gr]}$ is ratio between the changing rates of
the global phases for two group-changing spaces,\textit{ }$\delta
\phi_{g,\mathrm{global}}=\sqrt{%
%TCIMACRO{\dsum \limits_{\mu}}%
%BeginExpansion
{\displaystyle \sum \limits_{\mu}}
%EndExpansion
(\delta \phi_{g}^{\mu}(x))^{2}}$\textit{ }and\textit{ }$\delta \phi
_{r,\mathrm{global}}=\sqrt{%
%TCIMACRO{\dsum \limits_{\mu}}%
%BeginExpansion
{\displaystyle \sum \limits_{\mu}}
%EndExpansion
(\delta \phi_{r}^{\mu}(x))^{2}}$\textit{.} In this paper, we point out that
$\lambda^{\lbrack gr]}$ doesn't have to be an integer.

Next, we define 2-nd order variant by introducing higher-order mapping, i.e.,
a mapping between a group-changing space $\mathrm{C}_{r,\mathrm{\tilde{G},}%
d}^{[2]}$\ and another $\mathrm{C}_{g,\mathrm{\tilde{G},}d}^{[1]}$ that is
defined on $d$-dimensional Cartesian space $\mathrm{C}_{d}$. We point out that
$\mathrm{C}_{g,\mathrm{\tilde{G},}d}^{[1]}$ corresponds to a level-1
group-changing space and $\mathrm{C}_{r,\mathrm{\tilde{G},}d}^{[2]}$
corresponds to a level-2 group-changing space, respectively.

In general, the approach of 2-nd order variant\ is defined by%
\begin{equation}
V_{\mathrm{\tilde{G}},d}^{\mathrm{chiral}}:\mathrm{C}_{r,\mathrm{\tilde{G},}%
d}^{[2]}\Longleftrightarrow \mathrm{C}_{1,\mathrm{\tilde{G},}d}^{[1]}%
\Longleftrightarrow \mathrm{C}_{d},
\end{equation}
\textit{\ }of which one is the mapping between\textit{ }$\mathrm{C}%
_{g,\mathrm{\tilde{G},}d}^{[1]}\ $and Cartesian space $\mathrm{C}_{d}$\textit{
}with changing ratio $k_{g,0}^{\mu}$ i.e.,
\begin{equation}
\mathrm{C}_{g,\mathrm{\tilde{G},}d}^{[1]}\Longleftrightarrow \mathrm{C}_{d}.
\end{equation}
The other is between $\mathrm{C}_{r,\mathrm{\tilde{G},}d}^{[2]}$%
\textit{\ }with total size $(\Delta \phi_{r}^{\mu})$\textit{ }and
$\mathrm{C}_{g,\mathrm{\tilde{G},}d}^{[1]}$ with total size\ $(\Delta \phi
_{g}^{\mu}),$ i.e.,\textit{ }%
\begin{align}
\mathrm{C}_{r,\mathrm{\tilde{G},}d}^{[2]}(\Delta \phi_{r}^{\mu})  &
\Longleftrightarrow \mathrm{C}_{g\mathrm{\tilde{G},}d}^{[1]}(\Delta \phi
_{g}^{\mu})\\
&  \equiv \mathrm{C}_{r,\mathrm{\tilde{U}(1)\in \tilde{G}},1}(\Delta
\phi_{r,\mathrm{global}})\nonumber \\
&  \Longleftrightarrow \mathrm{C}_{g,\mathrm{\tilde{U}(1)\in \tilde{G}}%
,1}(\Delta \phi_{g,\mathrm{global}})\nonumber \\
&  \equiv \{ \delta \phi_{r,\mathrm{global}}\} \Leftrightarrow \{ \delta
\phi_{g,\mathrm{global}}\} \nonumber
\end{align}
where $\Longleftrightarrow$\ denotes an ordered mapping under fixed changing
rate along different directions. The elements of two subgroup-changing spaces
are $\delta \phi_{r,\mathrm{global}}^{[2]}=\sqrt{%
%TCIMACRO{\dsum \limits_{\mu}}%
%BeginExpansion
{\displaystyle \sum \limits_{\mu}}
%EndExpansion
(\delta \phi_{r}^{\mu})^{2}}$ and $\delta \phi_{g,\mathrm{global}}^{[1]}=\sqrt{%
%TCIMACRO{\dsum \limits_{\mu}}%
%BeginExpansion
{\displaystyle \sum \limits_{\mu}}
%EndExpansion
(\delta \phi_{g}^{\mu})^{2}},$ respectively. Therefore, the type of a 2-nd
order variant $V_{\mathrm{\tilde{G}},d}^{\mathrm{chiral}}$ is determined by
the ratio of the two changing rates,
\begin{equation}
\lambda^{\lbrack gr]}=\frac{\delta \phi_{g,\mathrm{global}}}{\delta
\phi_{r,\mathrm{global}}}.
\end{equation}

For 2-nd order U-variant $V_{\mathrm{\tilde{G}},d}$ with infinite size
($\Delta x^{\mu}\rightarrow \infty$), there exists 2-nd order variability. A
effective 2-nd order variability for 2-nd order U-variant $V_{\mathrm{\tilde
{G}},d}$ is described by the following two equations%
\begin{align}
\tilde{U}^{[1]}(\delta \phi_{g,\mathrm{global}}^{[1]})  &  \leftrightarrow
\hat{U}^{[2]}(\delta \phi_{r,\mathrm{global}}^{[2]})\nonumber \\
&  =\exp(i\lambda^{\lbrack gr]}\delta \phi_{g,\mathrm{global}}^{[1]}),
\end{align}
and
\begin{align}
\mathcal{T}(\delta x^{\mu})  &  \leftrightarrow \hat{U}^{[g]}(\delta \phi
_{g}^{\mu})\nonumber \\
&  =\exp(iT^{\mu}\delta \phi_{g}^{\mu})\\
&  =\exp(iT^{\mu}k_{g,0}^{\mu}\delta x^{\mu})),\nonumber
\end{align}
where $\delta \phi_{g,\mathrm{global}}^{[1]}=\sqrt{%
%TCIMACRO{\dsum \limits_{\mu}}%
%BeginExpansion
{\displaystyle \sum \limits_{\mu}}
%EndExpansion
(\delta \phi_{g}^{\mu})^{2}}$\textit{ }and\textit{ }$\delta \phi
_{r,\mathrm{global}}^{[1]}=\sqrt{%
%TCIMACRO{\dsum \limits_{\mu}}%
%BeginExpansion
{\displaystyle \sum \limits_{\mu}}
%EndExpansion
(\delta \phi_{r}^{\mu})^{2}}.$ $\mathcal{T}(\delta x^{\mu})$ is the spatial
translation operation along $x^{\mu}$-direction, $\hat{U}^{[g]}(\delta \phi
_{g}^{\mu})=\exp(iT^{\mu}\delta \phi_{g}^{\mu})$ and $\hat{U}^{[g]}(\delta
\phi_{g,\mathrm{global}}^{[1]})$\ are shift operations on $\mathrm{C}%
_{g,\mathrm{\tilde{G},}d}^{[1]}$, $\hat{U}^{[2]}(\delta \phi_{r,\mathrm{global}%
})$ is shift operation on $\mathrm{C}_{r,\mathrm{\tilde{G},}d}^{[2]}$. For
simplicity, we can denote a system with 2-nd order variability by the
following equation
\begin{equation}
\mathcal{T}\leftrightarrow \hat{U}^{[g]}\leftrightarrow \hat{U}^{[2]}%
\end{equation}
where $\hat{U}^{[2]}=\hat{U}^{[r]}$.

Under K-projection, we have two levels of composite zero lattices for a
higher-dimensional effective 2-nd order U-variant $V_{\mathrm{\tilde{G}}%
,d}^{\mathrm{chiral}}$: the group-changing space for $\mathrm{C}%
_{g,\mathrm{\tilde{G}}}^{[1]}$ is reduced into a $d$-dimensional level-1 zero
lattice on Cartesian space, of which each zero corresponds to $\lambda
^{\lbrack gr]}$ level-2 zeroes. The extra level-1 group-changing elements of
$\mathrm{C}_{g,\mathrm{\tilde{G}}}^{[1]}$ are effectively characterized by
local field of a compact \textrm{U(1)} group, and the extra level-2
group-changing elements of $\mathrm{C}_{r,\mathrm{\tilde{G}}}^{[2]}$ are
effectively characterized by local field of a compact \textrm{U(1)} group. As
a result, to characterize the effective 2-nd order variant, we introduce a
local \textrm{U(1)} gauge field. This is just the \textrm{U}$_{\mathrm{Y}}%
$\textrm{(1)} gauge field of hyperchange in weak interaction.

Because the relative sub-variant is generated by extra left/right hand
sub-variant $V_{r,\mathrm{\tilde{G}},d}^{\mathrm{sub}}=\alpha
V_{\mathrm{\tilde{G}}^{L},d}^{L}+\beta V_{\mathrm{\tilde{G}}^{R},d}^{R}$, the
level-2 group-changing space $\mathrm{C}_{r,\mathrm{\tilde{G},}d}^{[2]}$ has
similar properties to the group-changing space $\mathrm{C}_{L/R,\mathrm{\tilde
{G},}d}^{\mathrm{extra}}$ of extra left/right hand sub-variant. The changing
rate of extra left/right-hand sub-variant is same to that of the relative
sub-variant $k_{r,0}^{\mu}=\alpha k_{L,0}^{\mu}+\beta k_{R,0}^{\mu}$. This
particular equivalence leads to certain indistinguishability of an extra
left/right hand zero or a relative zero. As a result, there exists an
effective local \textrm{SU(2) }symmetry between an extra left/right hand zero
and a relative zero. This is just the \textrm{SU}$_{\mathrm{weak}}%
$\textrm{(2)} gauge field in weak interaction. In the following part, we will
map this effective 2-nd order variant to a usual 2-nd order variant with
$\lambda^{\lbrack12]}=2$ and use the effective local \textrm{SU(2) }symmetry
to characterize it.

In summary, a chiral variant can be effective considered as a 2-nd order
variant. The extra level-1 group-changing elements of $\mathrm{C}%
_{g,\mathrm{\tilde{G}}}^{[1]}$ are characterized by local field of a compact
\textrm{U(1)} group, and the extra level-2 group-changing elements of
$\mathrm{C}_{r,\mathrm{\tilde{G}}}^{[2]}$ are effectively characterized by
local field of a compact \textrm{U(1)} group and a usual \textrm{2}-component
field of compact \textrm{SU(2)} group. The \textrm{SU(2)} group comes from
certain indistinguishability between a zero of relevant sub-variant and that
of extra left/right hand sub-variant.\ As a result, by using both a usual
field of compact \textrm{U(1)} group and a usual \textrm{2}-component field of
compact \textrm{SU(2)} group, we fully describe the chiral variant
$V_{\mathrm{\tilde{G}},d}^{\mathrm{chiral}}$.

\subsubsection{Summary}

In this section, we develop a theory for chiral variants that are composite
variants with left-hand sub-variant and right-hand sub-variant. There are two
types of chiral variants -- those with chiral symmetry and those without. We
can use the global/reletive sub-variant to characterize the chirality of a
variant. In addition, we introduce two approaches (approach I and approach II)
to characterize different types of changings for a chiral variant. This
powerful mathematic theory will help us understand the non-local structure of
quantum field theory for the Standard Model.

\subsection{A new theoretical framework for the Standard Model -- "\textbf{All
from Changings}"}

In this section, based on higher-order chiral variant, we develop a new
framework on the foundation of the Standard Model -- an $\mathrm{SU}%
_{\mathrm{C}}\mathrm{(3)}\otimes \mathrm{SU}_{\mathrm{weak}}\mathrm{(2)}%
\otimes$\textrm{U}$_{\mathrm{Y}}$\textrm{(1)} gauge theory with Higgs
mechanism. Different physical laws emerge for the changings in different
levels. We call it "\emph{Tower of changings}". The base of the tower is 0-th
level physics structure that is the uniform physical variant named
"\emph{vacuum}" or "\emph{ground state}" in usual physics; above it is level-1
physics structure that is the expansion and contraction types of
"\emph{changings}" of the vacuum, named "\emph{matter}" or physical excited
states in usual physics; above it is level-2 physics structure that is the
time-dependent "\emph{changings}" of the expansion and contraction changings
of vacuum, named "\emph{motion}" of matter in usual physics.

\subsubsection{2--th order $\mathrm{\tilde{S}\tilde{O}}$\textrm{(d+1)}
physical chiral variants: concept and definition}

\emph{What's physical reality in a new theoretical framework of the Standard
Model?} In this paper, we point out that for the Standard Model, the physical
reality is ($d+1$) dimensional 2-nd order $\mathrm{\tilde{S}\tilde{O}}%
$\textrm{(d+1)} physical chiral variant with chiral asymmetry
$V_{\mathrm{\tilde{U}}^{[2]}\mathrm{(1)},\mathrm{\tilde{S}\tilde{O}}%
^{[1]}\mathrm{(d+1)},d+1}^{[2],\mathrm{chiral}}=\{V_{g,\mathrm{\tilde{U}%
}^{[2]}\mathrm{(1)},\mathrm{\tilde{S}\tilde{O}}^{[1]}\mathrm{(d+1)}%
,d+1}^{[2],\mathrm{sub}},V_{r,\mathrm{\tilde{S}\tilde{O}}^{[1]}\mathrm{(d)}%
,d}^{\mathrm{sub}}\}.$

Firstly, we give the detailed definition of ($d+1$)-dimensional 2-nd order
$\mathrm{\tilde{S}\tilde{O}}$\textrm{(d+1)} physical chiral variant
$V_{\mathrm{\tilde{U}}^{[2]}\mathrm{(1)},\mathrm{\tilde{S}\tilde{O}}%
^{[1]}\mathrm{(d+1)},d+1}^{[2],\mathrm{chiral}}$:

\textit{Definition -- }($d+1$)\textit{-dimensional} \textit{2-nd order
}$\mathrm{\tilde{S}\tilde{O}}$\textrm{(d+1)}\textit{ physical chiral variants
has two sub-variants -- the global sub-variant }$V_{g,\mathrm{\tilde{U}}%
^{[2]}\mathrm{(1)},\mathrm{\tilde{S}\tilde{O}}^{[1]}\mathrm{(d+1)}%
,d+1}^{[2],\mathrm{sub}}$\textit{ and the relative sub-variant }%
$V_{r,\mathrm{\tilde{S}\tilde{O}}^{[1]}\mathrm{(d)},d}^{\mathrm{sub}}%
.$\textit{ Here, the global sub-variant }$V_{g,\mathrm{\tilde{U}}%
^{[2]}\mathrm{(1)},\mathrm{\tilde{S}\tilde{O}}^{[1]}\mathrm{(d+1)}%
,d+1}^{[2],\mathrm{sub}}$\textit{ is a higher-order mapping between
}$\mathrm{C}_{\mathrm{\tilde{U}}^{[2]}\mathrm{(1)}}^{[2]}$\textit{,
$\mathrm{\tilde{S}\tilde{O}}$\textrm{(d+1)} Clifford group-changing space
}$\mathrm{C}_{g,\mathrm{\tilde{S}\tilde{O}(d+1)},d+1}^{[1]}$\textit{\ and a
rigid spacetime }$\mathrm{C}_{d+1},$\textit{ i.e.,}%
\begin{align}
V_{g,\mathrm{\tilde{U}}^{[2]}\mathrm{(1)},\mathrm{\tilde{S}\tilde{O}}%
^{[1]}\mathrm{(d+1)},d+1}^{[2],\mathrm{sub}}  &  :\mathrm{C}_{\mathrm{\tilde
{U}}^{[2]}\mathrm{(1)}}^{[2]}\nonumber \\
&  \Longleftrightarrow \mathrm{C}_{g,\mathrm{\tilde{S}\tilde{O}}^{[1]}%
\mathrm{(d+1)},d+1}^{[1]}\nonumber \\
&  \Longleftrightarrow \mathrm{C}_{d+1}%
\end{align}
\textit{ where }$\Leftrightarrow$\textit{\ between }$\mathrm{C}%
_{\mathrm{\tilde{U}}^{[2]}\mathrm{(1)}}^{[2]}$ \textit{and }$\mathrm{C}%
_{g,\mathrm{\tilde{S}\tilde{O}}^{[1]}\mathrm{(d+1)},d+1}^{[1]}$\textit{
denotes an ordered mapping under ratio of changing rates }$\lambda
^{\lbrack12]}=\left \vert \frac{\delta \phi_{\mathrm{global}}^{[2]}}{\delta
\phi_{g,\mathrm{global}}^{[1]}}\right \vert =\left \vert \frac{\delta
\phi^{\lbrack2]}}{\delta \phi_{g,\mathrm{global}}^{[1]}}\right \vert $\textit{,
}$\Leftrightarrow$\textit{\ between }$\mathrm{C}_{g,\mathrm{\tilde{S}\tilde
{O}}^{[1]}\mathrm{(d+1)},d+1}^{[1]}$\textit{ and} $\mathrm{C}_{d+1}$\textit{
denotes an ordered mapping under fixed changing rate of integer multiple
}$k_{0}$ or $\omega_{0}^{[1]},$\textit{\ and }$\mu$\textit{ labels the spatial
direction}.\textit{ The relevant sub-variant }$V_{r,\mathrm{\tilde{S}\tilde
{O}}^{[1]}\mathrm{(d)},d}$\textit{ is induced by extra left-hand sub-variant,
i.e.,}%
\begin{equation}
V_{r,\mathrm{\tilde{S}\tilde{O}}^{[1]}\mathrm{(d)},d}=\alpha
V_{L,\mathrm{\tilde{S}\tilde{O}}^{[1]}\mathrm{(d)},d}^{\mathrm{sub}}%
\end{equation}
\textit{where }$\alpha^{-1}=\lambda^{\lbrack gr]}+1$\textit{, and }%
$\lambda^{\lbrack gr]}$\textit{ is the ratio between the changing rate of
global sub-variant and that of relative sub-variant.}

Based on this physical chiral variant $V_{\mathrm{\tilde{U}}^{[2]}%
\mathrm{(1)},\mathrm{\tilde{S}\tilde{O}}^{[1]}\mathrm{(d+1)},d+1}%
^{[2],\mathrm{chiral}}$ as 0-th level physics structure, we develop a new,
complete theoretical framework for quantum gauge theory step by step.\textit{
}In particular, for our universe, we have
\begin{equation}
d=3,\text{ }\lambda^{\lbrack12]}=3,\  \lambda^{\lbrack gr]}=3\pi.
\end{equation}

\subsubsection{Higher-order Variability and its physical consequences}

Uniform 2-nd order $\mathrm{\tilde{S}\tilde{O}}$\textrm{(d+1)}\textit{
}physical\ chiral variant $V_{0,\mathrm{\tilde{U}}^{[2]}\mathrm{(1)}%
,\mathrm{\tilde{S}\tilde{O}}^{[1]}\mathrm{(d+1)},d+1}^{[2],\mathrm{chiral}%
}=\{V_{0,g,\mathrm{\tilde{U}}^{[2]}\mathrm{(1)},\mathrm{\tilde{S}\tilde{O}%
}^{[1]}\mathrm{(d+1)},d+1}^{[2],\mathrm{sub}},V_{0,r,\mathrm{\tilde{S}%
\tilde{O}}^{[1]}\mathrm{(d)},d}^{\mathrm{sub}}\}$ is a complex uniform
changing structure on Cartesian space. To characterize it, we consider its
higher-order variability.

Firstly, we discuss level-1 variability of level-1 group-changing space for
global sub-variant $V_{0,g,\mathrm{\tilde{U}}^{[2]}\mathrm{(1)},\mathrm{\tilde
{S}\tilde{O}}^{[1]}\mathrm{(d+1)},d+1}^{[2],\mathrm{sub}}$.

\emph{Level-1 spatial variability}: There exists variability along an
arbitrary spatial direction of the level-1 group-changing space $\mathrm{C}%
_{g,\mathrm{\tilde{S}\tilde{O}}^{[1]}\mathrm{(d+1)},d+1}^{[1]}$, i.e.,
\begin{align}
\mathcal{T}(\delta x^{i})  &  \rightarrow \hat{U}^{[1]}((\delta \phi_{g}%
^{i})^{[1]})=e^{i\cdot(\delta \phi_{g}^{i})^{[1]}\Gamma^{i}},\text{
}\nonumber \\
i  &  =x_{1},x_{2},\text{...},x_{d},
\end{align}
where $(\delta \phi_{g}^{i})^{[1]}=k_{g,0}\delta x^{i}$ and $\Gamma^{i}%
=\tau_{z}\otimes \sigma^{i}$ are the Gamma matrices obeying Clifford algebra
$\{ \Gamma^{i},\Gamma^{i}\}=2\delta^{ij}.$ The variability along an arbitrary
spatial direction of the group-changing space $\mathrm{C}_{r,\mathrm{\tilde
{S}\tilde{O}(d)},d}$, i.e.,
\begin{align}
\mathcal{T}(\delta x^{i})  &  \rightarrow \hat{U}^{[1]}(\delta \phi_{r}%
^{i})=e^{i\cdot(\delta \phi_{r}^{i})\sigma^{i}},\text{ }\nonumber \\
i  &  =x_{1},x_{2},\text{...},x_{d},
\end{align}
where $(\delta \phi_{r}^{i})^{[1]}=k_{r,0}\delta x^{i}$ and $\sigma^{i}$ are
the Pauli matrices.\ This level-1 spatial variability is relevant to the
Lorentz invariance of quantum physics;

\emph{Level-1 tempo variability}: There exists a variability along time
direction of the level-1 group-changing space for the global sub-variant
$\mathrm{C}_{g,\mathrm{\tilde{S}\tilde{O}}^{[1]}\mathrm{(d+1)},d+1}^{[1]}$, i.e.,%

\begin{equation}
\mathcal{T}(\delta t)\rightarrow \hat{U}^{[1]}((\delta \phi_{g}^{i}%
)^{[1]})=e^{i\cdot(\delta \phi_{g}^{i})^{[1]}\Gamma^{t}},
\end{equation}
where $(\delta \phi_{g}^{i})^{[1]}=\omega_{0}^{[g]}\delta t$ and $\Gamma^{t}$
is another Gamma matrix anticommuting with $\Gamma^{i},$ $\{ \Gamma^{i}%
,\Gamma^{t}\}=2\delta^{it}$. In particular, the system with 1-st order
variability along time direction also indicates a regular motion of the
level-1 group-changing space for the global sub-variant $\mathrm{C}%
_{g,\mathrm{\tilde{S}\tilde{O}}^{[1]}\mathrm{(d+1)},d+1}^{[1]}$ along
$\Gamma^{t}$ direction. In addition, we have $\omega_{0}^{[g]}=k_{g,0}+\delta$
where $\delta \ll k_{g,0}.$ This level-1 tempo variability is relevant to both
the quantization of quantum physics and Higgs mechanism of particle's masses.
The variability along tempo direction of the group-changing space of relative
sub-variant $\mathrm{C}_{r,\mathrm{\tilde{S}\tilde{O}(d)},d}$, i.e.,%
\begin{equation}
\mathcal{T}(\delta t)\rightarrow \hat{U}^{[1]}(\delta \phi_{r}^{i}%
)=e^{i\delta \phi_{r}^{i}},
\end{equation}
where $\delta \phi_{r}^{i}=\omega_{0}^{[r]}\delta t$ and $\omega_{0}%
^{[r]}=k_{0}^{[r]}$.

\emph{Level-1 rotation variability}: There exists a rotation variability of
both global sub-variant $V_{0,g,\mathrm{\tilde{U}}^{[2]}\mathrm{(1)}%
,\mathrm{\tilde{S}\tilde{O}}^{[1]}\mathrm{(d+1)},d+1}^{[2],\mathrm{sub}}$ and
relative sub-variant $V_{0,r,\mathrm{\tilde{S}\tilde{O}}^{[1]}\mathrm{(d)}%
,d}^{\mathrm{sub}}$ that is defined by
\begin{equation}
\hat{U}^{\mathrm{R}}\rightarrow \hat{R}_{\mathrm{space}}%
\end{equation}
where $\hat{U}^{\mathrm{R}}$ is \textrm{SO(d)} rotation operator on Clifford
group-changing space $\hat{U}^{\mathrm{R}}\Gamma^{I}\mathbf{(}\hat
{U}^{\mathrm{R}})^{-1}=\Gamma^{I^{\prime}},$ and $\hat{R}_{\mathrm{space}}$ is
\textrm{SO(d)} rotation operator on Cartesian space, $\hat{R}_{\mathrm{space}%
}x^{I}\hat{R}_{\mathrm{space}}^{-1}=x^{I^{\prime}}$. This level-1 rotation
variability is relevant to the spin dynamics and curved space in modern physics.

Next, we discuss level-2 variability of level-2 group-changing space
$\mathrm{C}_{\mathrm{U(1)}^{[2]},1}^{[2]}.$

\emph{Level-2 "spatial" variability}: There exists a "spatial" variability of
the level-2 group-changing space $\mathrm{C}_{\mathrm{U(1)}^{[2]},1}^{[1]}$,
i.e.,%
\begin{equation}
\hat{U}^{[1]}(\delta \phi_{g,\mathrm{global}}^{[1]})\leftrightarrow \hat
{U}^{[2]}(\delta \phi_{\mathrm{global}}^{[2]})=\exp(i\lambda^{\lbrack12]}%
\delta \phi_{g,\mathrm{global}}^{[1]}),
\end{equation}
where $\delta \phi_{g,\mathrm{global}}^{[1]}=\sqrt{%
%TCIMACRO{\dsum \limits_{\mu}}%
%BeginExpansion
{\displaystyle \sum \limits_{\mu}}
%EndExpansion
((\delta \phi_{g}^{\mu})^{[1]})^{2}}$. This level-2 "spatial" variability is
relevant to the gauge invariance of quantum physics.

\emph{Level-2 tempo variability}: There exists a tempo variability along time
direction of the level-2 group-changing space $\mathrm{C}_{\mathrm{U(1)}%
^{[2]},1}^{[2]}$, i.e.,%

\begin{equation}
\mathcal{T}(\delta t)\rightarrow \hat{U}^{[2]}(\delta \phi_{g}^{[2]}%
)=e^{i\delta \phi_{g}^{[2]}},
\end{equation}
where $\delta \phi_{g}^{[2]}=\omega_{0}^{[2]}\delta t$. This level-2 "tempo"
variability is relevant to the coupling constants of gauge fields.

\emph{Effective level-2 variability}: In addition, we have an effective 2-nd
order variability as%
\begin{equation}
\hat{U}^{[1]}(\delta \phi_{g,\mathrm{global}}^{[1]})\leftrightarrow \hat
{U}^{[2]}(\delta \phi_{r,\mathrm{global}}^{[2]})=\exp(i\lambda^{\lbrack
gr]}\delta \phi_{g,\mathrm{global}}^{[1]}),
\end{equation}
where $\delta \phi_{r,\mathrm{global}}^{[2]}=\sqrt{%
%TCIMACRO{\dsum \limits_{\mu}}%
%BeginExpansion
{\displaystyle \sum \limits_{\mu}}
%EndExpansion
((\delta \phi_{r}^{\mu})^{[2]})^{2}}.$ This effective level-2 "spatial"
variability is relevant to an effective gauge invariance.

\subsubsection{Classification of matter}

Next, we develop a theory about level-1 physics structure by classifying the
different types of matter.

In variant theory, matter correspond to different types of globally expand or
contract the three group-changing spaces, including $\mathrm{C}%
_{g,\mathrm{\tilde{S}\tilde{O}}^{[1]}\mathrm{(d+1)},d+1}^{[1]}$ or
$\mathrm{C}_{r,\mathrm{\tilde{S}\tilde{O}(d)},d}$, or $\mathrm{C}%
_{\mathrm{U(1)}^{[2]},1}^{[2]}$ by changing their corresponding sizes. In
general, different types of matter are quantized -- the units of
$\mathrm{C}_{g,\mathrm{\tilde{S}\tilde{O}}^{[1]}\mathrm{(d+1)},d+1}^{[1]}$ and
$\mathrm{C}_{r,\mathrm{\tilde{S}\tilde{O}(d)},d}$ are all fermionic elementary
particles and the unit of $\mathrm{C}_{\mathrm{U(1)}^{[2]},1}^{[2]}$ is color charge.

Firstly, we discuss the matter of level-1 group-changing space $\mathrm{C}%
_{g,\mathrm{\tilde{S}\tilde{O}}^{[1]}\mathrm{(d+1)},d+1}^{[1]}.$

Under D-projection, we have a translation symmetry protected knot/link along
spatial-tempo direction. Under K-projection along different directions on
spacetime, we have a (3+1)D uniform zero lattice. We point out that each zero
corresponds to an elementary particle and becomes the changing unit (or
information unit) for of level-1 group-changing space $\mathrm{C}%
_{g,\mathrm{\tilde{S}\tilde{O}}^{[1]}\mathrm{(d+1)},d+1}^{[1]}$. For the case
of a system with extra $N_{F}$ zero, the total phase changings of all
phase-changing elements of global sub-variant $\delta \phi_{g,\mathrm{global,}%
i}^{[1]}(x_{i})$ are equal to $\pm N_{F}\pi,$ i.e., $%
%TCIMACRO{\dsum \nolimits_{i}}%
%BeginExpansion
{\displaystyle \sum \nolimits_{i}}
%EndExpansion
\delta \phi_{g,\mathrm{global,}i}^{[1]}(x_{i})=\pm n\pi$. In physics, these
elementary particles are Dirac type of elementary particles, including charged
leptons or quarks.

Secondly, we discuss the matter of level-1 group-changing space of relative
sub-variant $\mathrm{C}_{r,\mathrm{\tilde{S}\tilde{O}(d)},d}.$ Under
D-projection and K-projection along different directions on spacetime, we also
have a (3+1)D uniform zero lattice. Each zero corresponds to a neutrino and
becomes the changing unit (or information unit) for of the level-1
group-changing space of relative sub-variant $\mathrm{C}_{r,\mathrm{\tilde
{S}\tilde{O}(d)},d}$.

Next, we discuss the matter of level-2 group-changing space $\mathrm{C}%
_{\mathrm{\tilde{U}(1)}^{[2]},1}^{[2]}.$

Under K-projection for $\mathrm{C}_{\mathrm{\tilde{U}}^{[2]}\mathrm{(1)}%
}^{[2]}$ on a level-1 zero of $\mathrm{C}_{\mathrm{\tilde{S}\tilde{O}}%
^{[1]}\mathrm{(d+1)},d+1}^{[1]}$, we have a level-2 zero lattice, of which the
lattice number is $\lambda^{\lbrack12]}$. For the case of a system with extra
$n^{[2]}$ level-2 zeroes, the total phase changings of phase-changing elements
$\delta \phi_{i}^{[2]}(x_{i})$ are equal to $\pm n^{[2]}\pi,$ i.e., $%
%TCIMACRO{\dsum \nolimits_{i}}%
%BeginExpansion
{\displaystyle \sum \nolimits_{i}}
%EndExpansion
\delta \phi_{i}^{[2]}(x_{i})=\pm n^{[2]}\pi$. Each level-2 zero corresponds to
a color charge of local field of \textrm{SU(}$\lambda^{\lbrack12]}$\textrm{)}
gauge symmetry.

We point out that the numbers $n^{[2]}$ and $\lambda^{\lbrack12]}$ fully
determine the types of matter for a 2-nd order $\mathrm{\tilde{S}\tilde{O}}%
$\textrm{(d+1)}\textit{ }physical variant $V_{0,g,\mathrm{\tilde{U}}%
^{[2]}\mathrm{(1)},\mathrm{\tilde{S}\tilde{O}}^{[1]}\mathrm{(d+1)}%
,d+1}^{[2],\mathrm{sub}}$. Due to the level-2 "spatial" variability, the
changings of the two group-changing spaces lock, i.e.,
\begin{equation}
\lambda^{\lbrack12]}\delta \phi_{g,\mathrm{global}}^{[1]}\equiv \delta
\phi_{\mathrm{global}}^{[2]}=\delta \phi^{\lbrack2]}.
\end{equation}
Under compactification,\ we have a constraint on the changings on the compact
phase angles, i.e.,
\begin{equation}
\delta \varphi_{g,\mathrm{global}}^{[1]}\equiv \lambda^{\lbrack12]}\delta
\varphi^{\lbrack2]}%
\end{equation}
where $\delta \varphi_{g,\mathrm{global}}^{[1]}\in(0,2\pi]$, $\delta
\varphi^{\lbrack2]}\in(0,2\pi]$.

Therefore, for an elementary particle (an level-1 zero) with $n^{[2]}$ level-2
zeroes ($\lambda^{\lbrack12]}>n^{[2]}>0$), if the phase changing of the
level-1 zero (elementary particle) is $\delta \phi_{g,\mathrm{global}}^{[1]},$
the phase changing of each level-2 zero is $\delta \phi_{g,\mathrm{global}%
}^{[1]}$ and the total phase changing of the level-2 zeroes is $\delta
\phi_{\mathrm{global}}^{[2]}=n^{[2]}\delta \phi_{g,\mathrm{global}}^{[1]}.$ For
the case of $n^{[2]}=\lambda^{\lbrack12]},$ we have $\delta \phi
_{\mathrm{global}}^{[2]}=\lambda^{\lbrack12]}\delta \phi_{\mathrm{global}%
}^{[1]}$. We set an electric charge for the case of $n^{[2]}=\lambda
^{\lbrack12]}$ to be unit. A level-2 zero has $1/\lambda^{\lbrack12]}$
electric charge and\ an elementary particle (an level-1 zero) with $n^{[2]}$
level-2 zeroes ($\lambda^{\lbrack12]}>n^{[2]}>0$) has $\frac{n^{[2]}}%
{\lambda^{\lbrack12]}}$ electric charge.

In summary, there are three types of matter of 2-nd order $\mathrm{\tilde
{S}\tilde{O}}$\textrm{(d+1)}\textit{ }physical variant $V_{\mathrm{\tilde{U}%
}^{[2]}\mathrm{(1)},\mathrm{\tilde{S}\tilde{O}}^{[1]}\mathrm{(d+1)}%
,d+1}^{[2],\mathrm{chiral}}=\{V_{g,\mathrm{\tilde{U}}^{[2]}\mathrm{(1)}%
,\mathrm{\tilde{S}\tilde{O}}^{[1]}\mathrm{(d+1)},d+1}^{[2],\mathrm{sub}%
},V_{r,\mathrm{\tilde{S}\tilde{O}}^{[1]}\mathrm{(d)},d}^{\mathrm{sub}}\}$ --
Dirac types of fermionic elementary particles corresponding to level-1 zeroes
for global sub-variant, Weyl types of fermionic elementary particles
corresponding to level-1 zeroes for relative sub-variant and color charges
corresponding to level-2 zeroes, i.e.,
\begin{align}
\text{Level-1 zero for global sub-variant}  &  \rightarrow \text{Quark or
charge lepton,}\\
\text{Level-1 zero for relative sub-variant}  &  \rightarrow \text{Neutrino,}%
\nonumber \\
\text{Level-2 zero }  &  \rightarrow \text{A color charge with }\nonumber \\
&  \frac{1}{\lambda^{\lbrack12]}}\text{ electric charge.}\nonumber
\end{align}

\subsubsection{Classification of motions}

In this section, we develop theory about level-2 physics structure by
classifying the types of motion that corresponds to different types of
time-dependent changings of 2-nd order physical variants $V_{\mathrm{\tilde
{U}}^{[2]}\mathrm{(1)},\mathrm{\tilde{S}\tilde{O}}^{[1]}\mathrm{(d+1)}%
,d+1}^{[2],\mathrm{chiral}}$ without changings the size of the group-changing
spaces. This comes from the changings of mappings between different
group-changing spaces $\mathrm{C}_{g,\mathrm{\tilde{S}\tilde{O}}%
^{[1]}\mathrm{(d+1)},d+1}^{[1]}$ or $\mathrm{C}_{r,\mathrm{\tilde{S}\tilde
{O}(d)},d}$, or $\mathrm{C}_{\mathrm{U(1)}^{[2]},1}^{[2]}$. In particular, for
different types of motions, the size of the three group-changing spaces
$\mathrm{C}_{g,\mathrm{\tilde{S}\tilde{O}}^{[1]}\mathrm{(d+1)},d+1}^{[1]}$ or
$\mathrm{C}_{r,\mathrm{\tilde{S}\tilde{O}(d)},d}$, or $\mathrm{C}%
_{\mathrm{U(1)}^{[2]},1}^{[2]}$ will never change.

In above section, we pointed out that globally expand/contract of the two
group-changing spaces $\mathrm{C}_{g,\mathrm{\tilde{S}\tilde{O}}%
^{[1]}\mathrm{(d+1)},d+1}^{[1]}$ or $\mathrm{C}_{r,\mathrm{\tilde{S}\tilde
{O}(d)},d}$ corresponds to the generation/annihilate of elementary particles
in quantum physics. In this part, we point out that there are two kinds of
motions, one is the motion of elementary particles (or locally expand/contract
of three group-changing spaces corresponds to the motion of particles in
quantum mechanics with fixed particles' number), the other is the collective
motion of the three zero lattices without extra zeroes. Let us show the results.

Firstly, we discuss the motions of level-1 zeroes that corresponds to the
locally expand/contract of level-1 group-changing spaces of global sub-variant
$\mathrm{C}_{g,\mathrm{\tilde{S}\tilde{O}}^{[1]}\mathrm{(d+1)},d+1}^{[1]}$ and
that of relative sub-variant $\mathrm{C}_{r,\mathrm{\tilde{S}\tilde{O}(d)},d}%
$. In Cartesian space\textit{ }$\mathrm{C}_{d+1},$ an elementary particle can
be divided into a group of group-changing elements. The evolution of
distribution of ordered level-1 group-changing elements of an elementary
particle in Cartesian space $\mathrm{C}_{d+1}$ are quantum motion in quantum
mechanics. Different distribution of group-changing elements of the elementary
particle are different states of quantum motion of particles. It is
wave-function that describes the different distribution of group-changing
elements of the elementary particle and characterize the changing of mapping
between level-1 group-changing spaces $\mathrm{(\mathrm{C}_{g,\mathrm{\tilde
{S}\tilde{O}}^{[1]}\mathrm{(d+1)},d+1}^{[1]}}$ or $\mathrm{\mathrm{C}%
_{r,\mathrm{\tilde{S}\tilde{O}(d)},d})}$ and Cartesian space\textit{
}$\mathrm{C}_{d+1}$.

The collective motion of the two level-1 zero lattices without extra zeroes is
gravitational wave that changes the shape of the physical variant. In this
paper, we will focus on another types of collective motion of two level-1 zero
lattices without extra zeroes comes from the fluctuations of the changing
rates, i.e., $k_{g,0}\rightarrow k_{g,0}+\delta k_{g}(x,t)$ and $k_{r,0}%
\rightarrow k_{r,0}+\delta k_{r}(x,t).$ In the following parts, we point out
that the changings of changing rate $\delta k_{g}(x,t)$ of global sub-variant
$V_{g,\mathrm{\tilde{U}}^{[2]}\mathrm{(1)},\mathrm{\tilde{S}\tilde{O}}%
^{[1]}\mathrm{(d+1)},d+1}^{[2],\mathrm{sub}}$ play the role of Higgs field,
and the changings of changing rate $\delta k_{r}(x,t)$ of relative sub-variant
$V_{r,\mathrm{\tilde{S}\tilde{O}}^{[1]}\mathrm{(d)},d}^{\mathrm{sub}}$ play
the role of gauge fields for weak interaction.

Next, we discuss the motions of level-2 zeroes that corresponds to the locally
expand/contract of level-2 group-changing space $\mathrm{C}_{\mathrm{\tilde
{U}}^{[2]}\mathrm{(1)}}^{[2]}$. There are two kinds of motions, one is the
motion of extra level-2 zeroes (or color charges), the other is the collective
motions of the level-2 zero lattice without defects.

On the one hand, there exists the motion of extra level-2 zeroes (or color
charges) inside an elementary particle (a level-1 zero). The motion of level-2
zeroes is then described by "quantum mechanics" inside an elementary particle.
This is an effective model on a 1D lattice with $\lambda^{\lbrack12]}$ sites.
Then, the "wave-function" of level-2 zeroes describes the different
distributions of level-2 group-changing elements inside an elementary particle
and characterize the changing of mapping between the two group-changing spaces
$\mathrm{C}_{\mathrm{\tilde{U}}^{[2]}\mathrm{(1)}}^{[2]}$ and $\mathrm{C}%
_{g,\mathrm{\tilde{S}\tilde{O}}^{[1]}\mathrm{(d+1)},d+1}^{[1]}$.

On the other hand, there are two types of the collective motions of the
level-2 zero lattice, one is about the global motion of the level-2 zeroes
inside a level-1 zero, the other is about relative motion of the level-2
zeroes inside a level-1 zero. The first type and second type of collective
modes are similar to the acoustic phonons and optical phonons in composite
crystals, respectively. In particular, the collective motion of level-2
group-changing space is described by the gauge fields on a rigid
spacetime\textit{ }$\mathrm{C}_{d+1}:$ The global motion of the level-2 zeroes
corresponds to the fluctuations of quantum fields of \textrm{U}$_{\mathrm{em}%
}$\textrm{(1)} gauge symmetry, and the relative motion of the level-2 zeroes
corresponds to the quantum fields of \textrm{SU(}$\lambda^{\lbrack12]}%
$\textrm{).}

In addition, under the approach II, the group-changing space of relative
sub-variant can be regarded as an effective level-2 group-changing space.
Under K-projection on level-1 zero lattice, we have an effective zero lattice.
There are two types of the collective motions of the effective level-2 zero
lattice, one is about the global motion of the level-2 zeroes inside a level-1
zero, the other is about relative motion of the level-2 zeroes from left-hand
sub-variant to relative sub-variant or vice versa inside a level-1 zero. The
global motion of the level-2 zeroes is really the fluctuations of the quantum
\textrm{U}$_{\mathrm{Y}}$\textrm{(1)} gauge field. The relative motion of the
level-2 zeroes corresponds to the quantum fields of \textrm{SU}%
$_{\mathrm{weak}}$\textrm{(}2\textrm{)} gauge fields for weak interaction.

\subsubsection{Invariance from variability}

According to above discussion, there are three levels of changings in physics.
We then discuss the invariances/symmetries for each levels of changings one by one.

\paragraph{Level-0 Invariant: The fixity of $6$ physical constants}

For the level-0 physics, we have a uniform 2-nd order physical chiral variant
$V_{0,\mathrm{\tilde{U}}^{[2]}\mathrm{(1)},\mathrm{\tilde{S}\tilde{O}}%
^{[1]}\mathrm{(d+1)},d+1}^{[2],\mathrm{chiral}}$. The \emph{fixity} of
changing rates of different group-changing spaces is known to be level-0
invariant. The \emph{fixity} indicates an invariant of physical laws (Lorentz
invariant, and quantization condition, gauge interaction, Schr\"{o}dinger
equation). The specific manifestation of invariance is the fixity of the
physical constants. There are totally $6$ physical constants, including light
speed $c$, Planck constant $\hslash$, the coupling constant for
electromagnetic interaction $e$, the number of generations, the coupling for
weak interaction, the value of Higgs condensation ... In other words, all
these physical constants don't change with time or place. So, we say that such
an invariance (or fixity) is protected by the 2-nd order variability.

Firstly, we consider light speed $c$ and the Lorentz invariant from spatial
variability of level-1 group-changing space for both global sub-variant
$\mathrm{C}_{g,\mathrm{\tilde{S}\tilde{O}}^{[1]}\mathrm{(d+1)},d+1}^{[1]}$ and
that of relative sub-variant $\mathrm{C}_{r,\mathrm{\tilde{S}\tilde{O}}%
^{[1]}\mathrm{(d)},d}^{[1]}$. After linearizing the dispersion of two
sub-variants at $k_{g}=k_{g/r,0}$, we have $\omega_{g/r}^{[1]}=\omega
_{g/r,0}^{[1]}+c(k_{g/r}-k_{g/r,0})$. Consequently, an effective "light"
velocity is $c=\frac{\partial \omega_{g}^{[1]}}{\partial k_{g}}\mid
_{k_{g}=k_{g,0}}=\frac{\partial \omega_{r}^{[1]}}{\partial k_{r}}\mid
_{k_{r}=k_{r,0}}$;

Secondly, we consider Planck constant $\hslash$ from tempo variability (or a
uniform motion) of level-1 group-changing space for global sub-variant
$\mathrm{C}_{g,\mathrm{\tilde{S}\tilde{O}}^{[1]}\mathrm{(d+1)},d+1}^{[1]}$ and
that for relative sub-variant $\mathrm{C}_{r,\mathrm{\tilde{S}\tilde{O}}%
^{[1]}\mathrm{(d)},d}^{[1],}$. Because $\rho_{E}(\omega_{g,0}^{[1]})$ is a
smooth function of $\omega_{0}$, we have $\frac{\delta \rho_{E}}{\delta
\omega_{g}^{[1]}}\mid_{\omega_{g}^{[1]}=\omega_{g,0}^{[1]}}=\frac{\delta
\rho_{E}}{\delta \omega_{r}^{[1]}}\mid_{\omega_{r}^{[1]}=\omega_{r,0}^{[1]}%
}=\rho_{J}^{E}$ where $\rho_{J}^{E}$ is the density of (effective) "angular
momentum". The "angular momentum" $\rho_{J}$ for an elementary particle is
just Planck constant $\hbar$.

Thirdly, we consider the value of Higgs condensation.

If the changing rates along spatial direction and that along tempo direction
for the global sub-variant are different, i.e., $\delta=\omega_{g,0}%
^{[1]}-k_{g,0}\neq0$, the elementary particles have finite mass. In other
words, the ratio $\gamma^{\lbrack st]}=\omega_{0}^{[g]}/k_{g,0}$ between the
changing rate along spatial directions $k_{g,0}$ and that along tempo
direction $\omega_{0}^{[g]}$ determines the particle's mass. The masses of all
fermionic elementary particles are proportional to $\left \vert \gamma^{\lbrack
st]}-1\right \vert $ that plays the role of Higgs condensation in the SM. If
the ratio $\gamma^{\lbrack st]}$ is exact $1$, the masses of elementary
particles are zero; if $\gamma^{\lbrack st]}\neq1,$ the masses of different
particles become finite.

Fourthly, we consider the total number of generations.

For a 2-nd order physical variant, an important value is the ratio
$\gamma^{\lbrack12]}=\frac{k_{0}^{[2]}}{k_{g,0}^{[1]}}$ between the level-2
group-changing space and level-1 group-changing space. In our universe, it is
$3$. On the one hand, this leads to the existence of $\mathrm{SU(3)\otimes
U}_{\mathrm{em}}\mathrm{(1)}$ gauge structure. On the other hand, the value
$3$ leads to three generations of different elementary particles.

Fifthly, we consider the coupling constant of weak interaction.

For the physical chiral variant with chiral asymmetry, we have the finite
changing rate $k_{r,0}$ of relative sub-variant. The ratio $\gamma^{\lbrack
gr]}=\frac{k_{g,0}}{k_{r,0}}$ between changing rate $k_{g,0}$ of global
sub-variant and the changing rate $k_{r,0}$ of relative sub-variant is an very
important value. In the following parts, it is $3\pi.$ On the one hand,
$\gamma^{\lbrack gr]}$ determines the mass spectra of all elementary
particles, on the other hand, $\gamma^{\lbrack gr]}$ determines the coupling
constant of weak interaction.

Sixthly, we consider the coupling constant of electromagnetic interaction.

It was known that $\mathrm{e}$ is really proportional to the effective Planck
constant $\hbar^{\lbrack2]}$ for a level-2 zero of the group-changing space
$\mathrm{C}_{\mathrm{\tilde{U}}^{[2]}\mathrm{(1)}}^{[2]}$. Therefore, the
electric charge is obtained as $\mathrm{e}=\hbar \lambda c,$ where $\lambda$ is
a dimensionless parameter as the ratio of Planck constants for the zeroes of
different levels, i.e., $\lambda=\frac{\hbar^{\lbrack2]}}{\hbar}$.

In summary, for 2-nd order physical chiral variant $V_{0,\mathrm{\tilde{U}%
}^{[2]}\mathrm{(1)},\mathrm{\tilde{S}\tilde{O}}^{[1]}\mathrm{(d+1)},d+1}%
^{[2]},$ there are six tunable parameters that determine six physical
constants in the SM. In next section, we will show how these six tunable
parameters determine mass spectra.

\paragraph{Level-1 Invariant: Topology stationarity}

Next, we discuss the invariant of level-1 physics structure for matter.

From point view of 2-nd order physical variant $V_{\mathrm{\tilde{U}}%
^{[2]}\mathrm{(1)},\mathrm{\tilde{S}\tilde{O}}^{[1]}\mathrm{(d+1)}%
,d+1}^{[2],\mathrm{chiral}}$, the total sizes of the three group-changing
spaces $\mathrm{C}_{g,\mathrm{\tilde{S}\tilde{O}}^{[1]}\mathrm{(d+1)}%
,d+1}^{[1]\mathrm{l}}$ or $\mathrm{C}_{r,\mathrm{\tilde{S}\tilde{O}(d)},d}$,
or $\mathrm{C}_{\mathrm{U(1)}^{[2]},1}^{[2]}$ are all topological invariables.
Because the motions in quantum physics are changing the mapping between
$\mathrm{C}_{g,\mathrm{\tilde{S}\tilde{O}}^{[1]}\mathrm{(d+1)},d+1}%
^{[1]\mathrm{l}}$ or $\mathrm{C}_{r,\mathrm{\tilde{S}\tilde{O}(d)},d},$ or
$\mathrm{C}_{\mathrm{U(1)}^{[2]},1}^{[2]}$, and $\mathrm{C}_{d+1},$ during the
processes of motion, the total sizes of the three group-changing spaces don't
change obviously. Therefore, we have topology stationarity for matter.

\paragraph{Level-2 Invariant: The symmetry of motion}

Finally, we discuss the invariant of level-2 physics structure for motions.

It was known that motion corresponds to locally expand or contract of the
three group-changing spaces, $\mathrm{C}_{g,\mathrm{\tilde{S}\tilde{O}}%
^{[1]}\mathrm{(d+1)},d+1}^{[1]}$ or $\mathrm{C}_{r,\mathrm{\tilde{S}\tilde
{O}(d)},d}$, or $\mathrm{C}_{\mathrm{U(1)}^{[2]},1}^{[2]}$ without changing
their corresponding sizes. Different states of motions correspond to different
mappings between $\mathrm{C}_{g,\mathrm{\tilde{S}\tilde{O}}^{[1]}%
\mathrm{(d+1)},d+1}^{[1]}$ or $\mathrm{C}_{r,\mathrm{\tilde{S}\tilde{O}(d)}%
,d}$, or $\mathrm{C}_{\mathrm{U(1)}^{[2]},1}^{[2]},$ and $\mathrm{C}_{d+1}$.
If two states (or different mappings between $\mathrm{C}_{g,\mathrm{\tilde
{S}\tilde{O}}^{[1]}\mathrm{(d+1)},d+1}^{[1]}$ or $\mathrm{C}_{r,\mathrm{\tilde
{S}\tilde{O}(d)},d}$, or $\mathrm{C}_{\mathrm{U(1)}^{[2]},1}^{[2]}$, and
$\mathrm{C}_{d+1}$) have \emph{same energy}, we call such a invariance to be
\emph{symmetry} of motions.

Firstly, we discuss the symmetry for a level-1 group-changing space.

According to the level-1 variability, under a globally shifting of the global
sub-variant and relative sub-variant, we have
\begin{align}
\mathcal{T}(\delta x^{\mu})  &  \leftrightarrow \hat{U}^{[1]}((\delta \phi
_{g}^{\mu})^{[1]})=\exp(i(T^{\mu})^{[1]}(\delta \phi_{g}^{\mu})^{[1]})\\
&  =\exp(i(T^{\mu})^{[1]}(k_{g,0}^{\mu}\delta x^{\mu})),\nonumber
\end{align}
and
\begin{align}
\mathcal{T}(\delta x^{\mu})  &  \leftrightarrow \hat{U}^{[1]}((\delta \phi
_{r}^{\mu})^{[1]})=\exp(i(T^{\mu})^{[1]}(\delta \phi_{r}^{\mu})^{[1]})\\
&  =\exp(i(T^{\mu})^{[1]}(k_{r,0}^{\mu}\delta x^{\mu})).\nonumber
\end{align}
For simplify, we denote it by $\mathcal{T}\leftrightarrow \hat{U}_{g}^{[1]},$
and $\mathcal{T}\leftrightarrow \hat{U}_{r}^{[1]}$. As a result, the energy of
whole system doesn't change.

We then do compactification on the system.

On the one hand, under compactification, the continuous translation operation
$\mathcal{T}(\delta x^{\mu})$ of the U-variant is reduced into the operation
with discrete translation symmetry $T(\delta x^{\mu})$ on two level-1 zero
lattices, i.e.,%
\begin{align}
\mathcal{T}(\delta x^{\mu})  &  \leftrightarrow \tilde{T}(\delta N_{g}%
^{\mu,[1]}),\\
\mathcal{T}(\delta x^{\mu})  &  \leftrightarrow \tilde{T}(\delta N_{r}%
^{\mu,[1]}).\nonumber
\end{align}
For level-1 zero lattices, one lattice site is equivalence to another. Then,
for the uniform level-1 zero lattices, we have a reduced translation symmetry.
However, the lattice constant of zero lattice of the global sub-variant is
much smaller than that of relative sub-variant, the true reduced translation
symmetry of the whole is determined by the zero lattice of relative
sub-variant. Hence, we denote the reduced translation symmetry by the
following equation
\begin{equation}
\tilde{T}(\delta N_{r}^{\mu})\leftrightarrow1.
\end{equation}
This fact indicates a big unit cell with a lot of level-1 zeroes of the global sub-variant.

On the other hand, under compactification, the operation $\hat{U}%
^{[1]}((\delta \phi_{g}^{\mu})^{[1]})$ of non-compact $\mathrm{\tilde{S}%
\tilde{O}}^{[1]}\mathrm{(d+1)}$ group is reduced to that of compact
$\mathrm{U}_{\mathrm{global}}^{[1]}\mathrm{(1)\otimes SO}^{[1]}$\textrm{(d+1)}
group. On each lattice site of level-1 zero lattice, we have an invariant
under global phase operation $\hat{U}_{\mathrm{U}_{\mathrm{global}}%
^{[1]}\mathrm{(1)}}^{[1]}(1)$ and global compact \textrm{SO}$^{[1]}%
$\textrm{(d+1)} operation $\hat{U}_{\mathrm{SO}^{[1]}\mathrm{(d+1)}}^{[1]}$,
i.e.,
\begin{equation}
\hat{U}_{g}^{\mu,[1]}\rightarrow \hat{U}_{\mathrm{U}_{\mathrm{global}}%
^{[1]}\mathrm{(1)}}^{[1]}(1)\times \hat{U}_{\mathrm{SO}^{[1]}\mathrm{(d+1)}%
}^{[1]}.
\end{equation}
For simplicity, we can denote it by the following equations
\begin{equation}
U_{\mathrm{U}_{\mathrm{global}}^{[1]}\mathrm{(1)}}^{[1]}(1)\leftrightarrow
1,\text{ }\hat{U}_{\mathrm{SO}^{[1]}\mathrm{(d+1)}}^{[1]}\leftrightarrow1.
\end{equation}

Therefore, $\tilde{T}(\delta N_{r}^{\mu,[1]})\leftrightarrow1$ indicates that
the system is invariance under a global translation on level-1 zero lattice;
global phase symmetry $\hat{U}_{\mathrm{U}_{\mathrm{global}}^{[1]}%
\mathrm{(1)}}^{[1]}(\delta \phi_{g/r,\mathrm{global}}^{[1]})\leftrightarrow1$
indicates that the system is invariance under the global phase rotation on all
level-1 zero lattice; global $\mathrm{SO}^{[1]}\mathrm{(d+1)}$ rotation
symmetry $\hat{U}_{\mathrm{SO}^{[1]}\mathrm{(d+1)}}^{[1]}\leftrightarrow1$
indicates that the system is invariance under a global spacetime rotation.

As a result, we unify the global translation symmetry, global phase symmetry
and global rotation symmetry into level-1 variability.

Next, we discuss the symmetry for a level-2 group-changing space of a 2-nd
order physical U-variant.

First of all, this symmetry for a level-2 group-changing space of a 2-nd order
physical U-variant is a local symmetry. Here, the word "\emph{local}" means
the symmetry comes from the invariance for an operation on a level-1 zero,
rather a global operation on the whole system.

Under level-1 compactification $\hat{U}^{[1]}(\delta \phi_{g,\mathrm{global}%
}^{[1]})\rightarrow \hat{U}_{I_{g}}^{[1]}(\delta \varphi_{I_{g},\mathrm{global}%
}^{[1]})$, we consider the projected level-2 variability on a level-1 zero of
global sub-variant,
\begin{equation}
\hat{U}_{I_{g}}^{[1]}(\delta \varphi_{I_{g},\mathrm{global}}^{[1]}%
)\leftrightarrow \hat{U}_{I_{g}}^{[2]}(\delta \phi_{I_{g}}^{[2]})
\end{equation}
where the index $I_{g}$ denotes $I_{g}$-th level-1 zero for global sub-variant
and $\varphi_{I_{g},\mathrm{global}}^{[1]}\in(0,2\pi]$. For simplify, we
denote it by $\hat{U}_{I_{g}}^{[1]}\leftrightarrow \hat{U}_{I_{g}}^{[2]}.$

According to the level-2 variability $\hat{U}_{I_{g}}^{[1]}\leftrightarrow
\hat{U}_{I_{g}}^{[2]}$, we do compactification and get a level-2 zero lattice
with a compact \textrm{U(1)} field.

On the one hand, under compactification, the continuous translation operation
$\hat{U}_{I_{g}}^{[1]}(\delta \varphi_{I_{g},\mathrm{global}}^{[1]})$ is
reduced into a discrete translation operation $\hat{U}_{I_{g}}^{[1]}(\delta
N_{I_{g}}^{[2]})$ on the level-2 zero lattice, i.e.,%
\begin{equation}
\hat{U}_{I_{g}}^{[1]}(\delta \varphi_{I_{g},\mathrm{global}}^{[1]}%
)\leftrightarrow \hat{U}_{I_{g}}^{[1]}(\delta N_{I_{g}}^{[2]}).
\end{equation}
For level-2 zero lattice, one lattice site is equivalence to another. Then, we
have a reduced discrete $\mathrm{Z}_{N}$ ($N=\lambda^{\lbrack1]}$) symmetry
denoted by the following equation
\begin{equation}
\hat{U}_{I_{g}}^{[1]}(\delta N_{I_{g}}^{[2]})\leftrightarrow1.
\end{equation}
This will lead to local \textrm{SU(N)} gauge symmetry.

On the other hand, under compactification, the operation $\hat{U}_{I_{g}%
}^{[2]}(\delta \phi_{I_{g}}^{[2]})$ of non-compact $\mathrm{\tilde{U}(1)}$
group is reduced to the operation $\hat{U}_{I_{g}}^{[2]}(\delta \varphi_{I_{g}%
}^{[2]})$ for a global compact \textrm{U}$^{[2]}$\textrm{(1)} group. On each
lattice site of level-2 zero lattice, we have an invariant under the global
compact \textrm{U}$^{[2]}$\textrm{(1)} group,
\begin{equation}
\hat{U}_{I_{g}}^{[2]}(\delta \varphi^{\lbrack2]})\leftrightarrow1.
\end{equation}

However, due to $\hat{U}_{I_{g}}^{[1]}\leftrightarrow \hat{U}_{I_{g}}^{[2]},$
the operation $\hat{U}_{I_{g}}^{[1]}(\delta \varphi_{I_{g},\mathrm{global}%
}^{[1]})$ and the operation $\hat{U}_{I_{g}}^{[2]}(\delta \varphi_{I_{g}}%
^{[2]})$ are not independent each other. Therefore, the symmetry from global
compact \textrm{U}$_{\mathrm{global}}^{[1]}$\textrm{(1)} group and that from
global compact \textrm{U}$^{[2]}$\textrm{(1)} group couple unify into a new
local \textrm{U(1)} gauge symmetry, i.e., $\hat{U}_{I_{g}}^{[2]}(\delta
\varphi_{I_{g}}^{[2]})\leftrightarrow \hat{U}_{I_{g}}^{[1]}(\delta
\varphi_{I_{g},\mathrm{global}}^{[1]})$ This is just the \textrm{U}%
$_{\mathrm{em}}$\textrm{(1)} gauge symmetry for electromagnetic field, i.e.,
\begin{equation}
\hat{U}_{I_{g}}^{[2]}(\delta \varphi^{\lbrack2]})=\hat{U}_{I_{g}}^{[1]}%
(\delta \varphi_{\mathrm{global}}^{[1]})\leftrightarrow1.
\end{equation}

As a result, the local \textrm{U}$_{\mathrm{em}}$\textrm{(1)} gauge symmetry
and the local \textrm{SU(N)} gauge symmetry are unified into level-2 variability.

Next, we discuss the symmetry of effective level-2 group-changing space of
relative sub-variant.

Under level-1 compactification $\hat{U}^{[1]}(\delta \phi_{g,\mathrm{global}%
}^{[1]})\rightarrow \hat{U}_{I_{g}}^{[1]}(\delta \varphi_{I_{g},\mathrm{global}%
}^{[1]})$, we consider the projected level-2 variability on a level-1 zero of
global sub-variant,
\begin{equation}
\hat{U}_{I}^{[1]}(\delta \varphi_{I_{g},\mathrm{global}}^{[1]})\leftrightarrow
\hat{U}_{r,\mathrm{global,}I_{g}}^{[2]}(\delta \phi_{r,I_{g}}^{[2]})
\end{equation}
where the index $I_{g}$ denotes $I_{g}$-th level-1 zero for global sub-variant
and $\varphi_{I_{g},\mathrm{global}}^{[1]}\in(0,2\pi]$. For simplify, we
denote it by $\hat{U}_{I_{g}}^{[1]}\leftrightarrow \hat{U}_{r,\mathrm{global}%
,I_{g}}^{[2]}.$ According to the effective level-2 variability $\hat
{U}_{\mathrm{global,}I_{g}}^{[1]}\leftrightarrow \hat{U}_{r,\mathrm{global}%
,I_{g}}^{[2]}$, we do compactification and get a level-2 zero with a compact
\textrm{U(1)} field. Under compactification, for level-2 zero lattice, because
$\lambda^{\lbrack gr]}$ is much smaller than 1, the situation is quite
different from the case of $\lambda^{\lbrack12]}>1.$ So, the level-1 zero
lattice of global sub-variant must be enlarged to that as same size of
relative variant. Hence, in a unit cell, there exists an effective level-2
zero (or the zero of relative sub-variant) on an enlarged level-1 zero lattice
for global sub-variant. The position of the effective level-2 zero on an
enlarged level-1 zero leads to the a local \textrm{U(1)} symmetry that is just
the local \textrm{U}$_{\mathrm{Y}}$\textrm{(1)} gauge symmetry.

On the one hand, there is an equivalence between the relevant sub-variant
$V_{r,\mathrm{\tilde{S}\tilde{O}}^{[1]}\mathrm{(d)},d}^{\mathrm{sub}}$ and
extra left-hand $V_{r,\mathrm{\tilde{S}\tilde{O}}^{[1]}\mathrm{(d)},d}=\alpha
V_{L,\mathrm{\tilde{S}\tilde{O}}^{[1]}\mathrm{(d)},d}^{\mathrm{sub}}$. Due to
$\hat{U}_{\mathrm{global,}I_{g}}^{L}=\hat{U}_{r,\mathrm{global}I_{g}}^{[2]},$
the operation of left-hand sub-variant $\hat{U}_{I_{g}}^{[1]}(\delta
\varphi_{I_{g},\mathrm{global}}^{L})$ and that of relative sub-variant
$\hat{U}_{I_{g}}^{[2]}(\delta \varphi_{r,\mathrm{global,}I_{g}}^{[2]})$ may
have same property. This leads to an effective \textrm{SU(2)} gauge symmetry
that is just the local \textrm{SU}$_{\mathrm{weak}}$\textrm{(2)} gauge
symmetry for weak interaction.

Finally, we give a brief summary.

The \emph{level-1} variability is reduced to different kinds of \emph{global}
symmetries: one is about spatial translation symmetry, the other is about the
rotating symmetry of compact \textrm{U}$_{\mathrm{global}}^{[1]}$\textrm{(1)}
group and $\mathrm{SO}^{[1]}$\textrm{(d+1)} (or $\mathrm{SO}^{[1]}%
$\textrm{(d)}) group. Different conserved quantities are physical consequences
of the level-1 variability of the original (uniform) 2-nd order variant: the
energy/momentum $E/p$ becomes a conserved quantity; the particle number $N$
becomes a conserved quantity; the angular momentum becomes a conserved quantity.

The \emph{level-2} variability for level-2 group-changing space is reduced to
different kinds of \emph{gauge} symmetries: one is about local \textrm{U}%
$_{\mathrm{em}}$\textrm{(1)} gauge symmetry, the other is about local
\textrm{SU}$_{\mathrm{C}}$\textrm{(N)} gauge symmetry. The \emph{level-2}
variability for effective level-2 group-changing space from relative variant
is also reduced to different kinds of \emph{gauge} symmetries: one is about
local \textrm{U}$_{\mathrm{Y}}$\textrm{(1)} gauge symmetry, the other is about
local \textrm{SU}$_{\mathrm{weak}}$\textrm{(2)} gauge symmetry. With
considering local \textrm{U}$_{\mathrm{em}}$\textrm{(1)} gauge symmetry, we
have the theory for electromagnetic interaction; with considering the local
\textrm{SU}$_{\mathrm{C}}$\textrm{(N)} gauge symmetry, we have the theory for
strong interaction, with considering local \textrm{SU}$_{\mathrm{weak}}%
$\textrm{(2)}$\times$\textrm{U}$_{\mathrm{Y}}$\textrm{(1)} gauge symmetry, we
have the theory for electro-weak interaction.\

In summary, different gauge theories are physical consequences of the level-2
variability of the original 2-nd order variant. Invariance/symmetry is really
the \emph{shadow} of variability

\subsection{Sector of fermionic elementary particles}

In particle physics, there exist different types fermionic elementary
particles: neutrinos, charged leptons and quarks. It was known that there are
three families:
\begin{align}
\text{Leptons}  &  \text{:}\quad%
\begin{pmatrix}
\nu_{e}\\
e
\end{pmatrix}
,%
\begin{pmatrix}
\nu_{\mu}\\
\mu
\end{pmatrix}
,%
\begin{pmatrix}
\nu_{\tau}\\
\tau
\end{pmatrix}
,\\
\text{Quarks}  &  \text{:}\quad%
\begin{pmatrix}
u\\
d
\end{pmatrix}
,%
\begin{pmatrix}
c\\
s
\end{pmatrix}
,%
\begin{pmatrix}
t\\
b
\end{pmatrix}
.\nonumber
\end{align}
The matter in our universe comes from the first family\ of particles. The lift
times of second- and third-generation particles are very tiny. However, except
for masses, the elementary particles behave identically. It looks like that
the laws of nature were designed in triplicate. \textquotedblleft \textit{Who
ordered that?}\textquotedblright \ Rabi asked. Why three of them? To explain
the existence of different types of elementary particles, people try to go
beyond SM. Pati and Salam \cite{Pati} had proposed preons to be the
fundamental constituent particles. Then, other types of models were developed,
for example, the Rishon Model proposed simultaneously by Harari and Shupe
\cite{Harari,Shupe}, the helon model by S. O. Bilson-Thompson\cite{helon}.
However, they all failed.\

In this part, a systematical theory for different types of elementary
particles is developed, based on which the hidden topological structure in
mass spectra is explored. As a result, we answer above questions and show why
there are three generations.

\subsubsection{Classifying types of fermionic elementary particles}

Our starting point is $3+1$ dimensional 2-nd order $\mathrm{\tilde{S}\tilde
{O}}$\textrm{(3+1)} physical chiral variant without chiral symmetry
$V_{\mathrm{\tilde{U}}^{[2]}\mathrm{(1)},\mathrm{\tilde{S}\tilde{O}}%
^{[1]}\mathrm{(3+1)},3+1}^{[2],\mathrm{chiral}}=\{V_{g,\mathrm{\tilde{U}%
}^{[2]}\mathrm{(1)},\mathrm{\tilde{S}\tilde{O}}^{[1]}\mathrm{(3+1)}%
,3+1}^{[2],\mathrm{sub}},V_{r,\mathrm{\tilde{S}\tilde{O}}^{[1]}\mathrm{(3)}%
,3}^{\mathrm{sub}}\}.$ that exactly describes our universe, of which there are
three group-changing spaces, $\mathrm{C}_{g,\mathrm{\tilde{S}\tilde{O}}%
^{[1]}\mathrm{(3+1)},3+1}^{[1]}$ or $\mathrm{C}_{r,\mathrm{\tilde{S}\tilde
{O}(3)},3}$, or $\mathrm{C}_{\mathrm{U(1)}^{[2]},1}^{[2]}.$\textit{ }The ratio
of the changing rates of $\mathrm{C}_{g,\mathrm{\tilde{S}\tilde{O}}%
^{[1]}\mathrm{(3+1)},3+1}^{[1]}$ and $\mathrm{C}_{\mathrm{U(1)}^{[2]},1}%
^{[2]}$ is $\lambda^{\lbrack12]}$\ that is an integer number $3$. The ratio of
the changing rates of $\mathrm{C}_{g,\mathrm{\tilde{S}\tilde{O}}%
^{[1]}\mathrm{(3)},3}^{[1]}$ and $\mathrm{C}_{r,\mathrm{\tilde{S}\tilde{O}%
(3)},3}$ is $\lambda^{\lbrack gr]}$ that is $3\pi$.

Elementary particles are basic block of spacetime and the spacetime is really
a multi-particle system and made of matter. When there exists an additional
zero corresponding to an elementary particle, the periodic boundary condition
of systems along arbitrary direction is changed into anti-periodic boundary
condition. That means the elementary particle obeys fermionic statistics.

In this part, we classfy the types of elementary particles.

Firstly, there are two types of fermionic elementary particles, one is zero of
$V_{g,\mathrm{\tilde{U}}^{[2]}\mathrm{(1)},\mathrm{\tilde{S}\tilde{O}}%
^{[1]}\mathrm{(3+1)},3+1}^{[2],\mathrm{sub}},$ the other is
$V_{r,\mathrm{\tilde{S}\tilde{O}}^{[1]}\mathrm{(3)},3}^{\mathrm{sub}}$. A zero
of global sub-variant $V_{g,\mathrm{\tilde{U}}^{[2]}\mathrm{(1)}%
,\mathrm{\tilde{S}\tilde{O}}^{[1]}\mathrm{(3+1)},3+1}^{[2],\mathrm{sub}}$
becomes a Dirac fermionic elementary particle (such as Quark, or charged
lepton), while a zero of relative sub-variant $V_{r,\mathrm{\tilde{S}\tilde
{O}}^{[1]}\mathrm{(3)},3}^{\mathrm{sub}}$ becomes a Weyl fermionic elementary
particle (neutrino).

Secondly, we classify different types of Dirac fermionic elementary particle.

By trapping different types of internal level-2 zeros, there exist different
types of elementary particles corresponding to the zero of
$V_{g,\mathrm{\tilde{U}}^{[2]}\mathrm{(1)},\mathrm{\tilde{S}\tilde{O}}%
^{[1]}\mathrm{(3+1)},3+1}^{[2],\mathrm{sub}}$. We then classify the types of
elementary particles by the number of level-2 zeroes $n^{[2]}$. For an
elementary particle, $n^{[2]}$ determines both color charge and electric
charge. Each level-2 zero has $1/\lambda^{\lbrack12]}$ electric charge. For an
elementary particle with $n^{[2]}$ level-2 zeroes, its electric charge is
$\mathrm{e}=n^{[2]}/\lambda^{\lbrack12]}.$ As a result, there are
$\lambda^{\lbrack12]}$ types of elementary particles that are labeled by
$n^{[2]}.$ Therefore, we have%
\begin{align}
n^{[2]}  &  =\lambda^{\lbrack12]}:\text{ Electron with }\\
&  1\text{ electric charge,}\nonumber \\
n^{[2]}  &  =\lambda^{\lbrack12]}-1:\text{ Quark with }\nonumber \\
&  \frac{\lambda^{\lbrack12]}-1}{\lambda^{\lbrack12]}}\text{ electric
charge,}\nonumber \\
n^{[2]}  &  =\lambda^{\lbrack12]}-2:\text{ Quark with }\nonumber \\
&  \frac{\lambda^{\lbrack12]}-2}{\lambda^{\lbrack12]}}\text{ electric
charge,}\nonumber \\
&  ...\nonumber \\
n^{[2]}  &  =1:\text{ Quark with }\nonumber \\
&  \frac{1}{\lambda^{\lbrack12]}}\text{ electric charge.}\nonumber
\end{align}
For the case of $\lambda^{\lbrack12]}=3$, we have three types of elementary
particles:%
\begin{align*}
\{n^{[1]}  &  =1,n^{[2]}=3\}:\text{ Electron with }\\
&  1\text{ electric charge,}\\
\{n^{[1]}  &  =1,n^{[2]}=2\}:\text{ u-Quark with }\\
&  \frac{2}{3}\text{ electric charge,}\\
\{n^{[1]}  &  =1,n^{[2]}=1\}:\text{ d-Quark with }\\
&  \frac{1}{3}\text{ electric charge.}%
\end{align*}

For our universe, we have $\lambda^{\lbrack12]}=3$. As a result, there are
four different types of elementary particles, neutrinos with zero electric
charge, charged leptons with $\pm \mathrm{e}$ electric charge, U-quarks with
$\pm \frac{1}{3}\mathrm{e}$ electric charge and D-quarks with $\pm \frac{2}%
{3}\mathrm{e}$ electric charge.

\paragraph{Effective model for elementary particles}

In this part, we derive an effective model for elementary particles.

At first step, we do compactification on the group-changing space of global
sub-variant. After compactification, we have a topological lattice, that is
characterized by $\phi_{g}^{\mu}(x)=2\pi N_{g}^{\mu}(x)+\varphi_{g}^{\mu}(x)$.
We then relabel position in phase space by phase angles of compact group
$\varphi^{\mu}(x)$ and the winding numbers $N_{g}^{\mu}(x).$ The lattice
distance is $l_{0}$ on Cartesian space $\mathrm{C}_{3+1}$. On each lattice
site of the topological lattice, there exist compact \textrm{U}%
$_{\mathrm{global}}$\textrm{(1)} group and compact \textrm{SO(d+1)} group.

The generation operator of elementary particle that corresponds to zero of
global sub-variant to be $c_{i}^{\dagger}\left \vert 0\right \rangle =\left \vert
i\right \rangle $. We then write down the hopping Hamiltonian. The hopping term
between two nearest neighbor sites $i$ and $j$ on topological lattice becomes
\begin{equation}
\mathcal{H}_{\left \{  i,j\right \}  }=Jc_{i}^{\dagger}(t)\mathbf{T}_{\left \{
i,j\right \}  }c_{j}(t)
\end{equation}
where $\mathbf{T}_{\left \{  i,j\right \}  }$ is the transfer matrix between two
nearest neighbor sites $i$ and $j$ and $c_{i}(t)$ is the annihilation operator
of elementary particle at the site $i$. $J=\frac{c}{2l_{p}}$ is an effective
coupling constant between two nearest-neighbor sites. $l_{p}=l_{0}/2$ is
Planck length and $c$ is light speed. We may set $c$ to be unit, i.e., $c=1$.

According to hybrid symmetry, $\left \vert i\right \rangle =e^{il_{p}(\hat
{k}^{\mu}\cdot \Gamma^{\mu})}\left \vert j\right \rangle ,$ the transfer matrix
$\mathbf{T}_{\left \{  i,j\right \}  }$ between $\left \vert i\right \rangle $ and
$\left \vert j\right \rangle $ is defined by $\mathbf{T}_{\left \{  i,j\right \}
}=e^{il_{p}(\hat{k}^{\mu}\cdot \Gamma^{\mu})}.$ After considering the
contribution of the terms from all sites, the effective Hamiltonian is
obtained as%
\begin{equation}
\mathcal{H}=%
%TCIMACRO{\dsum \limits_{\{i,j\}}}%
%BeginExpansion
{\displaystyle \sum \limits_{\{i,j\}}}
%EndExpansion
\mathcal{H}_{\left \{  i,j\right \}  }=J%
%TCIMACRO{\dsum \limits_{\{i,j\}}}%
%BeginExpansion
{\displaystyle \sum \limits_{\{i,j\}}}
%EndExpansion
c_{i}^{\dagger}\mathbf{T}_{\left \{  i,j\right \}  }c_{j}+h.c..
\end{equation}

In continuum limit, we have%
\begin{align}
\mathcal{H}  &  =J%
%TCIMACRO{\dsum \limits_{\{i,j\}}}%
%BeginExpansion
{\displaystyle \sum \limits_{\{i,j\}}}
%EndExpansion
c_{i}^{\dagger}(e^{il_{p}(\hat{k}^{\mu}\cdot \Gamma^{\mu})})c_{i+1}+h.c.\\
&  =2l_{p}J%
%TCIMACRO{\dsum \limits_{\mu}}%
%BeginExpansion
{\displaystyle \sum \limits_{\mu}}
%EndExpansion%
%TCIMACRO{\dsum \limits_{k^{\mu}}}%
%BeginExpansion
{\displaystyle \sum \limits_{k^{\mu}}}
%EndExpansion
c_{k^{\mu}}^{\dagger}[\cos(k^{\mu}\cdot \Gamma^{\mu})]c_{k^{\mu}}%
\end{align}
where%
\begin{align}
\Gamma^{t}  &  =\tau^{x}\otimes \vec{1}\mathbf{,}\text{ }\Gamma^{x}=\tau
^{z}\otimes \sigma^{x},\\
\Gamma^{y}  &  =\tau^{z}\otimes \sigma^{y},\text{ }\Gamma^{z}=\tau^{z}%
\otimes \sigma^{z}.\nonumber
\end{align}
The dispersion in long wave length limit around $\omega=\omega_{0}=\frac{\pi
}{2}\frac{1}{l_{p}}c$ and $\vec{k}=\vec{k}_{0}$ is
\begin{equation}
E_{k}\simeq \pm c\sqrt{[(\vec{k}-\vec{k}_{0})\cdot \vec{\Gamma}]^{2}%
+((\omega-\omega_{0}^{\ast})\cdot \Gamma^{t})^{2}},
\end{equation}
where $\vec{k}_{0}=\frac{1}{l_{p}}(\frac{\pi}{2},\frac{\pi}{2},\frac{\pi}%
{2}).$ The mass of the Dirac particles is $m=\omega_{0}^{\ast}-\omega_{0}$.

After doing continuation, the discrete numbers $N(x)$ is replaced by continuum
coordinate $x=2l_{p}\cdot N(x)=l_{0}N(x).$ We then re-write the effective
Hamiltonian to be
\begin{equation}
\mathcal{H}=\int(\Psi^{\dagger}\hat{H}\Psi)d^{3}x
\end{equation}
where
\begin{equation}
\hat{H}=\vec{\Gamma}\cdot \Delta \vec{p}+m\Gamma^{t}%
\end{equation}
$\vec{p}=\hbar \Delta \vec{k}$ is the momentum operator. This is a massive Dirac
model. The mass becomes the coupling between the left and right handed
elementary particles,
\[
m\Psi_{L}^{\ast}\Psi_{R}+m\Psi_{R}^{\ast}\Psi_{L}.
\]

Finally, the Lagrangian $L$\ of for different Dirac type elementary particles
becomes
\begin{equation}
L=\bar{\Psi}(i\gamma^{\mu}\hat{\partial}_{\mu}-m)\Psi
\end{equation}
where $\gamma^{\mu}$ are the Gamma matrices defined as $\gamma^{1}=\gamma
^{0}\Gamma^{x}$, $\gamma^{2}=\gamma^{0}\Gamma^{y},$ $\gamma^{3}=\gamma
^{0}\Gamma^{z}$, $\gamma^{0}=\Gamma^{t}.$ The Gamma matrices $\Gamma^{I}$
($I=x,y,z$) and $\Gamma^{t}$ obey Clifford algebra, i.e., $\{ \Gamma
^{I},\Gamma^{t}\}=0$, and $\{ \Gamma^{I},\Gamma^{J}\}=0.$

However, because $\lambda^{\lbrack gr]}\gg1,$ the size of unit cell of the
relative sub-variant is larger than that of global sub-variant.
Therefore,\ the true discrete translation symmetry of zero lattices is
determined by that of relative sub-variant rather than global sub-variant. As
a result, the true lattice constant between two nearest neighbor zeroes
increases from $l_{p}$ to $(3\pi)^{1/3}l_{p}$.

Now, the effective Hamiltonian for elementary particles turns into
\begin{align}
\hat{H}  &  =\vec{\Gamma}\cdot \Delta \vec{p}+m\Gamma^{t}\nonumber \\
&  =\hat{H}_{\mathrm{low}}+\hat{H}_{\mathrm{high}}%
\end{align}
where $\hat{H}_{\mathrm{low}}$ is low energy effective model based on the unit
cell of relative sub-variant%
\begin{equation}
\hat{H}_{\mathrm{low}}=\vec{\Gamma}\cdot \Delta \vec{p}+m\Gamma^{t}%
\end{equation}
and $\hat{H}_{\mathrm{high}}\ $desribes the high energy effective model inside
a unit cell of relative sub-variant. We then focus on the low energy physics.
The effective Hamiltonian is reduced into%
\begin{equation}
\hat{H}\simeq \hat{H}_{\mathrm{low}}=\vec{\Gamma}\cdot \Delta \vec{p}+m\Gamma
^{t}.
\end{equation}

Finally, we show the effective Hamiltonian for the Weyl type of elementary particles.

Without considering the coupling between the left and right handed elementary
particles, Weyl type of elementary particle (a relative zero) is equivalent to
left-hand elementary particles. So, we could use the Hamiltonian of left-hand
elementary particles to describe the Weyl type of elementary particles, i.e.,
\begin{equation}
\mathcal{H}=\int(\Psi^{\dagger}\hat{H}\Psi)d^{3}x
\end{equation}
where
\begin{equation}
\hat{H}=\vec{\sigma}\cdot \Delta \vec{p}.
\end{equation}
The effective Lagrangian becomes%
\begin{equation}
L=\Psi^{\dagger}i\hat{\partial}_{t}\Psi-\Psi^{\dagger}i\vec{\sigma}\cdot
\vec{\nabla}\Psi.
\end{equation}

\subsubsection{Generation as state-degeneracy from multimap}

In this part, we show why there are three generations in the SM.

To answer the question about generations, we focus on the mappings between
these three group-changing spaces, $\mathrm{C}_{g,\mathrm{\tilde{S}\tilde{O}%
}^{[1]}\mathrm{(3+1)},3+1}^{[1]}$ or $\mathrm{C}_{r,\mathrm{\tilde{S}\tilde
{O}(3)},3}$, or $\mathrm{C}_{\mathrm{U(1)}^{[2]},1}^{[2]},$
\begin{equation}
\mathrm{C}_{\mathrm{\tilde{U}}^{[2]}\mathrm{(1)}}^{[2]}\Longleftrightarrow
\mathrm{C}_{g,\mathrm{\tilde{S}\tilde{O}}^{[1]}\mathrm{(3)},3}^{[1]}%
\Longleftrightarrow \mathrm{C}_{r,\mathrm{\tilde{S}\tilde{O}(3)},3}.
\end{equation}
\textit{ }The mapping between $\mathrm{C}_{\mathrm{\tilde{U}}^{[2]}%
\mathrm{(1)}}^{[2]}\ $and $\mathrm{C}_{g,\mathrm{\tilde{S}\tilde{O}}%
^{[1]}\mathrm{(3)},3}^{[1]}$ is determined by the ratio $\lambda^{\lbrack
12]}=\left \vert \frac{\delta \phi_{\mathrm{global}}^{[2]}}{\delta
\phi_{g,\mathrm{global}}^{[1]}}\right \vert =\left \vert \frac{\delta
\phi^{\lbrack2]}}{\delta \phi_{g,\mathrm{global}}^{[1]}}\right \vert =3;$ The
mapping between $\mathrm{C}_{g,\mathrm{\tilde{S}\tilde{O}}^{[1]}%
\mathrm{(3)},3}^{[1]}\ $and $\mathrm{C}_{r,\mathrm{\tilde{S}\tilde{O}(3)},3}$
is determined by the ratio $\lambda^{\lbrack gr]}=3\pi.$ That means a zero of
$\mathrm{C}_{g,\mathrm{\tilde{S}\tilde{O}}^{[1]}\mathrm{(3)},3}^{[1]}$
corresponds to $3$ zeroes of $\mathrm{C}_{\mathrm{\tilde{U}}^{[2]}%
\mathrm{(1)}}^{[2]}$ and a zero of $\mathrm{C}_{r,\mathrm{\tilde{S}\tilde
{O}(3)},3}$ corresponds to $3\pi$ zeroes of $\mathrm{C}_{g,\mathrm{\tilde
{S}\tilde{O}}^{[1]}\mathrm{(3)},3}^{[1]}$ or $3\times3\pi$ zeroes of
$\mathrm{C}_{\mathrm{\tilde{U}}^{[2]}\mathrm{(1)}}^{[2]}.$ In particular, it
is multimap that leads to $\lambda^{\lbrack12]}$-fold degenerate states for
the case of $\lambda^{\lbrack gr]}\gg1$. The $\lambda^{\lbrack12]}$-fold
degenerate states indicate the existence of $N$ generations, i.e.,
\begin{equation}
N\equiv \lambda^{\lbrack12]}.
\end{equation}

We then have a zero lattice for $\mathrm{C}_{r,\mathrm{\tilde{S}\tilde{O}%
(3)},3}$ and a composite zero lattice for $\mathrm{C}_{g,\mathrm{\tilde
{S}\tilde{O}}^{[1]}\mathrm{(3)},3}^{[1]}$. According to above discussion, the
lattice constant of the zero lattice for $\mathrm{C}_{r,\mathrm{\tilde
{S}\tilde{O}(3)},3}$ is biggest. The zero of $\mathrm{C}_{r,\mathrm{\tilde
{S}\tilde{O}(3)},3}$ with $3\pi$ zeroes of $\mathrm{C}_{g,\mathrm{\tilde
{S}\tilde{O}}^{[1]}\mathrm{(3)},3}^{[1]}$ becomes true unit cell of the
universe. A trouble about incommensurability arise, i.e., $3\pi$ is not an
integer number.

Next, we introduce "\emph{commensurating}". We consider each zero of
$\mathrm{C}_{r,\mathrm{\tilde{S}\tilde{O}(3)},3}$ becomes a 1D crystal with
$10$ composite lattice site, each of which corresponds to 3 internal level-2
zeroes of $\mathrm{C}_{\mathrm{\tilde{U}}^{[2]}\mathrm{(1)}}^{[2]}.$ The
residue $3\pi-10$ zero is regarded as extra phase factor for the global
level-1 zero hopping on the composite lattice. As a result, each lattice
corresponds to $1-3\pi/10$ phase changing. The situation is similar to Peierls
lattice, on which the particle obtains extra phase factor during hopping.

As a result, the system has $\lambda^{\lbrack12]}$-fold degenerate states for
the case of $\lambda^{\lbrack gr]}\gg1$. The situation is very similar to 1D
CDW with $\lambda^{\lbrack12]}$-fold degenerate states.

For example, for the case $\lambda^{\lbrack12]}=2$, there are two degenerate
states. If we denote the position of level-2 zeroes on a global level-1 zero
to be \textrm{a} and \textrm{b}. One degenerate states $\left \vert
1\right \rangle $ becomes
\begin{equation}
...(\mathrm{ab})(\mathrm{ab})(\mathrm{ab})...;
\end{equation}
the other $\left \vert 2\right \rangle $ becomes
\begin{equation}
...(\mathrm{ba})(\mathrm{ba})(\mathrm{ba})...
\end{equation}
Here, $(..)$ denotes a global level-1 zero. When an extra level-2 zero
appears, the positions of level-2 zeroes globally shift one site. Hence, the
two degenerate states can be changed by generating or annihilating level-2
zeroes of $\mathrm{C}_{\mathrm{\tilde{U}}^{[2]}\mathrm{(1)}}^{[2]}$.

For the case $\lambda^{\lbrack12]}=3$, there are three degenerate states: one
degenerate state $\left \vert 1\right \rangle $ is denoted by
\begin{equation}
...(\mathrm{abc})(\mathrm{abc})(\mathrm{abc})...;
\end{equation}
the second $\left \vert 2\right \rangle $ is denoted by
\begin{equation}
...(\mathrm{bca})(\mathrm{bca})(\mathrm{bca})...;
\end{equation}
the third $\left \vert 3\right \rangle $ is denoted by%
\begin{equation}
...(\mathrm{cab})(\mathrm{cab})(\mathrm{cab})...
\end{equation}
Here, $(...)$ denotes a global level-1 zero.

When an extra level-2 zero appears, the positions of level-2 zeroes globally
shift one site; when two extra level-2 zeroes appears, the positions of
level-2 zeroes globally shift two sites. Hence, the three degenerate ground
states can be changed by generating or annihilating one or two level-2 zeroes
of $\mathrm{C}_{\mathrm{\tilde{U}}^{[2]}\mathrm{(1)}}^{[2]}$.

For the case with higher $\lambda^{\lbrack12]},$ there are $\lambda
^{\lbrack12]}$ degenerate ground states, i.e.,
\begin{align}
\left \vert 1\right \rangle  &  :\text{ }...(\mathrm{abc}...\mathrm{f}%
...)(\mathrm{abc}...\mathrm{f}...)(\mathrm{abc}...\mathrm{f}...)...;\\
\left \vert 2\right \rangle  &  :\text{ }...(\mathrm{bcd}...\mathrm{g}%
...)(\mathrm{bcd}...\mathrm{g}...)(\mathrm{bcd}...\mathrm{g}%
...)...;\nonumber \\
\left \vert 3\right \rangle  &  :\text{ }...(\mathrm{cde}...\mathrm{h}%
...)(\mathrm{cde}...\mathrm{h}...)(\mathrm{cde}...\mathrm{h}...);\nonumber \\
&  ...\nonumber
\end{align}
One can also change the different ground states by adding (or removing)
level-2 zeroes.

In summary, a ($d+1$) dimensional 2-nd order $\mathrm{\tilde{S}\tilde{O}}%
$\textrm{(d+1)} physical chiral variant with chiral asymmetry is described by
$\lambda^{\lbrack12]}$-generation fermionic elementary particles. According to
above discussion, there are $N=\lambda^{\lbrack12]}$ generations for each
types of elementary particles. Hence, there are $\lambda^{\lbrack12]}%
\times(\lambda^{\lbrack12]}+1)$ different fermionic elementary particles,
i.e., $\lambda^{\lbrack12]}$ different types of neutrinos, $\lambda
^{\lbrack12]}$ different types of charged leptons, and $\lambda^{\lbrack
12]}\times(\lambda^{\lbrack12]}-1)$ different types of quarks. For example, in
Fig.39, we show the elementary particles of $\lambda^{\lbrack12]}=2.$ There
are three different types of elementary particles -- neutrinos, charged
leptons, and quarks. For each type of elementary particle, there are two generations.

\begin{figure}[ptb]
\includegraphics[clip,width=0.92\textwidth]{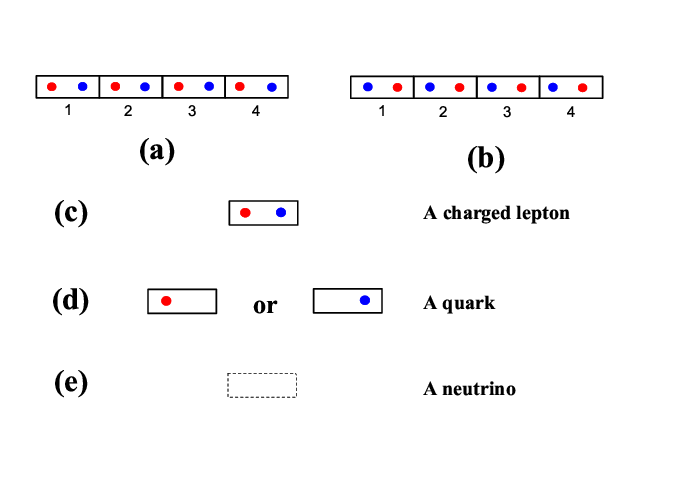}\caption{(Color online)
The illustration of elementary particles of $\lambda^{\lbrack12]}=2.$ There
are three different types of elementary particles -- neutrino, charged lepton,
and quark. For each type of elementary particle, there are two generations.}%
\end{figure}\

For our universe, we have $N=\lambda^{\lbrack12]}=3$. There are $3$\ different
types of neutrinos, $3$ different types of charged leptons, and $6$ different
types of quarks ($3$ for U-quarks and $3$ for D-quarks).

Finally, we can answer the question, "\textit{Why three of them}?" The answer
is \emph{multimap} under the condition of $\lambda^{\lbrack12]}=3$ and
$\lambda^{\lbrack gr]}\gg1.$ In other word, the "$3$" in non-Abelian
\textrm{SU(3)} gauge symmetry for strong interaction and the "$3$" generations
have same physical mechanism and are tightly locked each other.

\subsubsection{Quarks as topological defects in relative group-changing space}

According to above discussion, different elementary particles are classified
by the different number of their internal level-2 zeroes. In addition, there
are $\lambda^{\lbrack12]}$ generations for each type of elementary particle.
In our world, we have $\lambda^{\lbrack12]}=3$. Thus, there are 4 types of
elementary particles, each of which has three generations. In this part, we
show that quarks are really topological defects (or domain walls) for the
three generations.

Firstly, we consider the case of $\lambda^{\lbrack12]}=2.$

Now, we have three types of elementary particles -- chargeless neutrino with
$n^{[12]}=0$, the quark with half electric charge (or $n^{[12]}=1$), the
charged lepton with unit electric charge (or $n^{[12]}=2$). For each type of
elementary particle, there are two degenerate states (or two generations) --
$\left \vert 1\right \rangle $ ($...(\mathrm{ab})(\mathrm{ab})(\mathrm{ab})...$)
or $\left \vert 2\right \rangle $ ($...(\mathrm{ba})(\mathrm{ba})(\mathrm{ba}%
)...$). Because a quark is an object with an extra level-2 zero (extra
\textrm{a} or \textrm{b}), it becomes a topological defect that separates the
two degenerate states $\left \vert 1\right \rangle $ ($...(\mathrm{ab}%
)(\mathrm{ab})(\mathrm{ab})...$) and $\left \vert 2\right \rangle $
($...(\mathrm{ba})(\mathrm{ba})(\mathrm{ba})...$).

On the other hand, we may consider a domain wall between the two degenerate
states $\left \vert 1\right \rangle $ ($...(\mathrm{ab})(\mathrm{ab}%
)(\mathrm{ab})...$) and $\left \vert 2\right \rangle $ ($...(\mathrm{ba}%
)(\mathrm{ba})(\mathrm{ba})...$). According to the relationship between phase
changing and electric charge $\delta Q=-\frac{\delta \phi}{\pi}\left \vert
\mathrm{e}\right \vert $, it has $\pm \frac{1}{2}\left \vert \mathrm{e}%
\right \vert $ induced fractional number of charged leptons for a quark with
$\delta \phi=\frac{\pi}{2}$. This also indicates an extra level-2 zero (extra
\textrm{a} or \textrm{b}). The situation is similar to that in 1D dimerized
CDW, where the relationship between phase changing and electric charge $\delta
Q=-\frac{\delta \phi}{2\pi}\left \vert \mathrm{e}\right \vert .$

Secondly, we consider the case of $\lambda^{\lbrack12]}=3.$ \

Now, we have four types of elementary particles: chargeless neutrino, the U
quark with 1/3 electric charge, D quark with 2/3 electric charge, the charged
leptons with unit electric charge. For each type of elementary particles,
there are three degenerate states (or three generation) -- $\left \vert
1\right \rangle $ ($...(\mathrm{abc})(\mathrm{abc})(\mathrm{abc})...$),
$\left \vert 2\right \rangle $ ($...(\mathrm{bca})(\mathrm{bca})(\mathrm{bca}%
)...$), and $\left \vert 3\right \rangle $ ($...(\mathrm{cab})(\mathrm{cab}%
)(\mathrm{cab})...$). We map the system to 1D trimerized CDW. There are two
types of domain walls -- type-I is between the two nearest neighbor degenerate
states, for example, $\left \vert 1\right \rangle $ ($...(\mathrm{abc}%
)(\mathrm{abc})(\mathrm{abc})...$) and $\left \vert 2\right \rangle $
($...(\mathrm{bca})(\mathrm{bca})(\mathrm{bca})...$), the other is type-II
that is between the two next nearest neighbor degenerate states, for example,
$\left \vert 1\right \rangle $ ($...(\mathrm{abc})(\mathrm{abc})(\mathrm{abc}%
)...$), and $\left \vert 3\right \rangle $ ($...(\mathrm{cab})(\mathrm{cab}%
)(\mathrm{cab})...$). See the illustration\ in Fig.40.\

\begin{figure}[ptb]
\includegraphics[clip,width=0.7\textwidth]{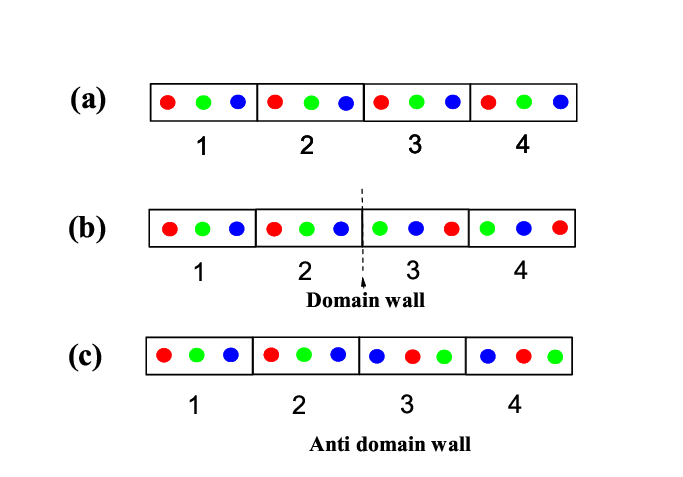}\caption{An illustration
of dommain wall and anti-domain wall for three generations of the ground
states (neutrinos)}%
\end{figure}

According to the relationship between phase changing and electric charge
$\delta Q=-\frac{\delta \phi}{\pi}\left \vert \mathrm{e}\right \vert $, a type-I
domain wall with two extra level-2 zeroes has $\frac{2}{3}\left \vert
\mathrm{e}\right \vert ,$ and a type-II domain wall with two extra level-2
zeroes has $-\frac{2}{3}\left \vert \mathrm{e}\right \vert $. So, we may
consider a type-I domain wall to be U-quark and a type-II domain wall to be an
anti-U-quark. The D-quark with $-\frac{1}{3}\left \vert \mathrm{e}\right \vert $
electric charge becomes a composite object of a U-quark with $\frac{2}%
{3}\left \vert \mathrm{e}\right \vert $ electric charge and a charged lepton
with $-\left \vert \mathrm{e}\right \vert $ electric charge, and a neutrino,
i.e.,
\begin{equation}
\text{D-quark = U-quark + Charged lepton + Neutrino.}%
\end{equation}
So, vice versa, one can consider a U-quark with $\frac{2}{3}\left \vert
\mathrm{e}\right \vert $ electric charge becomes a composite object of a
D-quark with $-\frac{1}{3}\left \vert \mathrm{e}\right \vert $ electric charge
and a charged lepton with $-\left \vert \mathrm{e}\right \vert $ electric
charge, and a neutrino.

It was known that a proton is cluster state of two u-quark and a d-quark and a
neutron is cluster state of two d-quark and a u-quark. We may consider a
proton as a composite with three domain wall together with a charged lepton
occupying one of three energy levels induced by domain walls. Therefore, a
neutron is regarded as a composite with three domain wall together with two
charged lepton occupying two of three energy levels induced by domain walls.

Thirdly, we consider the case of $\lambda^{\lbrack12]}=N>3.$ The situation is
similar to that in 1D CDW with $N$-aggregation.

Now, we have $N+1$ types of elementary particles, one is chargeless neutrino,
the $n^{[2]}$ type of quarks with $\frac{N-n^{[2]}}{N}$ electric charge
($n^{[2]}=1,2...N-1$), the electron with unit electric charge. For each type
of elementary particles, there are $N$ degenerate states (or $N$ generation),
i.e.,
\begin{align}
\left \vert 1\right \rangle  &  :\text{ }...(\mathrm{abc}...\mathrm{f}%
...)(\mathrm{abc}...\mathrm{f}...)(\mathrm{abc}...\mathrm{f}...)...;\\
\left \vert 2\right \rangle  &  :\text{ }...(\mathrm{bcd}...\mathrm{g}%
...)(\mathrm{bcd}...\mathrm{g}...)(\mathrm{bcd}...\mathrm{g}%
...)...;\nonumber \\
\left \vert 3\right \rangle  &  :\text{ }...(\mathrm{cde}...\mathrm{h}%
...)(\mathrm{cde}...\mathrm{h}...)(\mathrm{cde}...\mathrm{h}...);\nonumber \\
&  ...\nonumber
\end{align}

We consider a domain wall between the two nearest neighbor degenerate states,
for example, $\left \vert 1\right \rangle $ ($...(\mathrm{abc}...\mathrm{f}%
...)(\mathrm{abc}...\mathrm{f}...)(\mathrm{abc}...\mathrm{f}...)...$) and
$\left \vert 2\right \rangle $ ($...(\mathrm{bcd}...\mathrm{g}...)(\mathrm{bcd}%
...\mathrm{g}...)(\mathrm{bcd}...\mathrm{g}...)...$). The domain wall with
($N-1$) extra level-2 zeroes has $\frac{N-1}{N}\left \vert \mathrm{e}%
\right \vert $. So, we may consider a this domain wall to be U-quark. A D-quark
with $-\frac{1}{N}\left \vert \mathrm{e}\right \vert $ electric charge becomes a
composite object of a U-quark with $\frac{N-1}{N}\left \vert \mathrm{e}%
\right \vert $ electric charge and a charged lepton with $-\left \vert
\mathrm{e}\right \vert $ electric charge together with a neutrino, i.e.,
\begin{equation}
\text{D-quark = U-quark + Charged lepton + Neutrino.}%
\end{equation}
In a relative zero, the neutrino plays the role of "vacuum".

We then define a \emph{generalized proton} and a \emph{generalized neutron}
for the case of $\lambda^{\lbrack12]}=N>3.$ A generalized proton becomes
cluster state of $N-1$ U-quark and a D-quark and a neutron is cluster state of
$N-1$ D-quark and a U-quark. We may consider a generalized proton as a
composite with $N$ domain wall together with a neutrino and a charged lepton
occupying one of $N$ energy levels induced by domain walls. Therefore, a
generalized neutron is regarded as a composite with $N$ domain wall together
with $N-1$ charged lepton occupying $N-1$ of $N$ energy levels induced by
domain walls. So, a generalized proton has $N(N-2)\left \vert \mathrm{e}%
\right \vert $ electric charge, while neutron is still chargeless.

\subsubsection{Chiral para-statistics}

Before calculating mass spectra, we explore the hidden topological structure
for elementary particles of different generations. The non-trivial topological
changings come from a new type of quantum statistics. We call it \emph{chiral
para-statistics}. In this part, we focus on the chiral para-statistics for
elementary particles.

\paragraph{Review on usual non-Abelian statistics}

Firstly, we review usual non-Abelian statistics.

In general, for a system with non-Abelian statistics\cite{rea}, there are $N$
degenerate states denoted by $\psi_{\alpha}$ $,\alpha=1\dots N$. When we
exchange two particles $i$ and $j$, the corresponding operation becomes matrix
on the space of these degenerate states, i.e.,
\begin{equation}
\psi=(\psi_{1},\psi_{2},\dots,\psi_{N})^{T}\rightarrow \psi^{\prime}=U\psi
\end{equation}
where $U\equiv U_{\alpha \beta}$ is an $N\times N$ unitary matrix. If we
exchange other particles, for example, $k$ and $l$, we have
\begin{equation}
\psi \rightarrow \psi^{\prime}=V\psi
\end{equation}
with $V$ is another unitary $N\times N$ matrix. If $U$ and $V$ do not commute
$UV\neq VU$, there exists non-Abelian statistics for the space of $N$
degenerate states.

To characterize a non-Abelian statistics, we may use the theory about
\emph{unitary modular tensor categories}. Now, we have $N$ objects (or types
of particles). The fusion rules (rules for fusing two constituents into one
and also for splitting a particle into two constituents) and braiding rules
(rules for exchanging the positions of the two particles) determine the
structure of unitary modular tensor category. In addition to the braiding
rule, the fusion rule is also non-trivial, i.e.,
\begin{equation}
a\times b=\sum_{c}N_{ab}^{c}c
\end{equation}
where $a,b,c$ denote different kinds of particles. For the case of $N_{ab}%
^{c}>1$, $c$ is obtained in $N_{ab}^{c}$ ways. Therefore, the fusing of two
objects (or anyons) could lead to different types of particles with different probabilities.

An example for non-Abelian statistics is Majorana zero mode (MZM).

A single MZM can be described by a real fermionic field $\gamma=\int dr\left(
u_{0}\psi^{\ast}+u_{0}^{\ast}\psi \right)  $ obtained as a solution of the BdG
equations, where $\psi(r)$ is the wave function of the MZM and $\gamma
^{\dagger}=\gamma$. We can label two MZMs by complex fermions as $\gamma
_{1}=c_{1}+c_{1}^{\dagger}$, $\gamma_{2}=-i(c_{2}-c_{2}^{\dagger})$, and then,
the two MZMs can be used to represent the basis states $\left \vert
0\right \rangle _{\text{M}},$ $\left \vert 1\right \rangle _{\text{M}}.$
$\left \vert 0\right \rangle _{\text{M}}$ is a fermion-empty state, $\left \vert
1\right \rangle _{\text{M}}=C_{\text{M}}^{\dagger}\left \vert 0\right \rangle
_{\text{M}}$ is a fermion-occupied state. Here $C_{\text{M}}^{\dagger}$ is
composite fermionic operator, i.e., $C_{\text{M}}^{\dagger}=\gamma_{1}%
+i\gamma_{2}=(c_{1}+c_{1}^{\dagger})+(c_{2}-c_{2}^{\dagger})$.

On the one hand, the fusion rule of MZMs is given by $\sigma \times
\sigma=\mathbf{1}+\psi$, $\psi \times \psi=\mathbf{1}$ and $\psi \times
\sigma=\sigma$, where $\mathbf{1}$ is a vacuum sector, $\psi$ is the (complex)
fermion sector, and $\sigma$ is the MZM sector. Two $\sigma$-particles (MZMs)
may either annihilate to the vacuum or fuse into a $\psi$-particle; On the
other hand, if we exchange two MZMs ($\gamma_{1},$ $\gamma_{2}$), the
resulting exchange operation (the braiding operation) $\mathcal{R}_{\text{M}}$
is defined by $\gamma_{1}\rightarrow-\gamma_{2}$, $\gamma_{2}\rightarrow
\gamma_{1}$ and can be described by
\begin{equation}
\mathcal{R}_{\text{M}}=e^{i\frac{\pi}{4}\gamma_{1}\gamma_{2}}.
\end{equation}
We may call $\mathcal{R}_{\text{M}}$ to be Ivanov's braiding
operator\cite{Ivanov}. During the braiding process, the Berry phases for
${\left \vert 0\right \rangle }_{\text{M}}$ and $\left \vert 1\right \rangle
_{\text{M}}$ are $0$ and $\pi/2$, respectively.\ For the qubit $(\left \vert
0\right \rangle _{\text{M}},\left \vert 1\right \rangle _{\text{M}})^{T}$, the
braiding operator is obtained as $\mathrm{diag}\{1,i\}=e^{i\pi/4}%
\mathrm{diag}\{e^{-i\pi/4},e^{i\pi/4}\}$ which is known to be Ivanov's
braiding operator $\mathcal{R}_{\text{M}}$.

\paragraph{Chiral para-statistics}

In above part, we reviewed the usual non-Aeblian statistics. In particular,
the new quantum statistics is a \emph{(non-Abelian) para-statistics with
chiral asymmetry}, i.e., the "para-statistics" for left-hand elementary
particles and that for right hand become difference. So, in this part, we
introduce the concept of chiral para-statistics that includes global
para-statistics statistics and relative para-statistics (or left-hand
para-statistics and right-hand para-statistics).

Firstly, we discuss the chiral para-statistics for an arbitrary case of
$\lambda^{\lbrack12]}$. We assume that the parameter $\lambda^{\lbrack gr]}$
is large number, i.e., $\lambda^{\lbrack gr]}\gg1$. According to above
discussion, we have $N+1$ types of elementary particles, each of which has
$n^{[12]}$ level-2 zeroes ($n^{[12]}=0,1,2...N$), respectively. Each type of
elementary particle has $N$ generations (or flavors).

We already know that each level-2 zero traps $\frac{n^{[12]}}{\lambda
^{\lbrack12]}}$ electric charge and different types of elementary particles
are classified by different numbers of level-2 zeros (or $n^{[12]}$). Hence,
the chiral para-statistics characterizes the generation-transition for
elementary particles with fixed $n^{[12]}$. We may consider a given type of
elementary particle with fixed $n^{[12]}\ $to be a system with $N$ degenerate
states represented by $\psi_{\alpha}$ $,\alpha=1\dots N$. Because neutrinos
are elementary particles without internal level-1 zero, we could regarded
\emph{neutrinos as vacuum} for the system with $\lambda^{\lbrack gr]}$ lattice sites.\

In particular, for chiral para-statistics, instead of exchange two particles
on usual real space, we braid two elementary particles (\textrm{A} or
\textrm{B}) on group-changing space of a zero for relative sub-variant, of
which the lattice number is $\lambda^{\lbrack gr]}$. In addition, for chiral
para-statistics, to characterize the braiding rules between two elementary
particles (\textrm{A} or \textrm{B}), there are four matrices, $U_{\mathrm{AB}%
}^{g},$ $U_{\mathrm{AB}}^{r}$, $U_{\mathrm{BA}}^{g}$,\ $U_{\mathrm{BA}}^{r}$ :
$U_{\mathrm{AB}}^{g}$ and $U_{\mathrm{AB}}^{r}$ denote the changings of global
group-changing space and relative group-changing space for \textrm{A}
elementary particle, respectively; $U_{\mathrm{BA}}^{g}$ and $U_{\mathrm{BA}%
}^{r}$ denote the changings of global group-changing space and relative
group-changing space for \textrm{B} elementary particle, respectively. In
Fig.41, we show the difference between usual non-Abelian statistics and chiral para-statistics.

\begin{figure}[ptb]
\includegraphics[clip,width=0.7\textwidth]{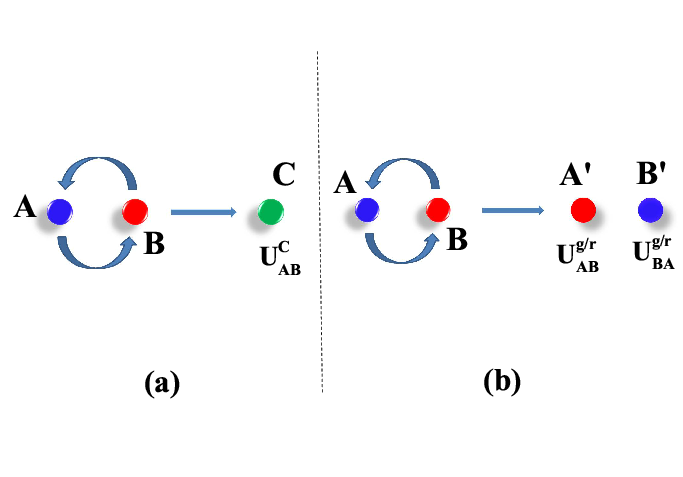}\caption{The difference
between usual non-Abelian statistics and chiral para-statistics}%
\end{figure}

When we do two braiding processes successively, we have two matrices,
$U_{1,\mathrm{AB}}^{g/r}$ for the first braiding process and $U_{2,\mathrm{AC}%
}^{g/r}$ for the second one. Here, $\mathrm{C}$ and $\mathrm{B}$ may be not
the same type of elementary particles. If, $U_{1,\mathrm{AB}}^{g/r}$ and
$U_{2,\mathrm{AC}}^{g/r}$ do not commute $U_{1,\mathrm{AB}}^{g/r}%
U_{2,\mathrm{AC}}^{g/r}\neq U_{2,\mathrm{AC}}^{g/r}U_{1,\mathrm{AB}}^{g/r}$,
the statistics becomes \emph{non-Abelian}.

Then, to obtain $U_{\mathrm{AB}}^{g}$ and $U_{\mathrm{AB}}^{r}$, we consider a
braiding process by moving an extra test object (or \textrm{B} elementary
particle) passing through \textrm{A} elementary particle. The corresponding
operations become two matrices on the space of degenerate states for
\textrm{A} elementary particle, i.e.,
\begin{align}
\psi_{g,\mathrm{A}}  &  =(\psi_{g,\mathrm{A,}1}\psi_{g,\mathrm{A}2}\dots
\psi_{g,\mathrm{A,}N})^{T}\rightarrow \psi_{g,\mathrm{A}}^{\prime
}=U_{\mathrm{AB}}^{g}\psi_{g,\mathrm{A}}\\
\psi_{r,\mathrm{A}}  &  =(\psi_{r,\mathrm{A,}1}\psi_{r,\mathrm{A}2}\dots
\psi_{r,\mathrm{A,}N})^{T}\rightarrow \psi_{r,\mathrm{A}}^{\prime
}=U_{\mathrm{AB}}^{r}\psi_{r,\mathrm{A}}\nonumber
\end{align}
where $U_{\mathrm{AB}}^{g}$ and $U_{\mathrm{AB}}^{r}$\ are two $N\times N$
unitary matrices for $\mathrm{A}$ elementary particles. Here the subscript
"$g$" or "$r$" denotes global or relative degrees of freedom on a zero of
relative group-changing space.

Next, to obtain $U_{\mathrm{BA}}^{g}$ and $U_{\mathrm{BA}}^{r}$, we consider a
braiding process by moving an extra test object (or \textrm{A} elementary
particle) passing through \textrm{B} elementary particle. The corresponding
operations become two matrices on the space of degenerate states for
\textrm{A} elementary particle, i.e.,
\begin{align}
\psi_{g,\mathrm{B}}  &  =(\psi_{g,\mathrm{B,}1}\psi_{g,\mathrm{B}2}\dots
\psi_{g,\mathrm{B,}N})^{T}\rightarrow \psi_{g,\mathrm{B}}^{\prime
}=U_{\mathrm{BA}}^{g}\psi_{g,\mathrm{B}}\\
\psi_{r,\mathrm{B}}  &  =(\psi_{r,\mathrm{B,}1}\psi_{r,\mathrm{B}2}\dots
\psi_{r,\mathrm{B,}N})^{T}\rightarrow \psi_{r,\mathrm{B}}^{\prime
}=U_{\mathrm{BA}}^{r}\psi_{r,\mathrm{B}}\nonumber
\end{align}
where $U_{\mathrm{AB}}^{g}$ and $U_{\mathrm{AB}}^{r}$\ are two $N\times N$
unitary matrices for $\mathrm{A}$ elementary particles. It is obvious that
$U_{\mathrm{BA}}^{g/r}$ and $U_{\mathrm{AB}}^{g/r}$ act on different spaces of
degenerate states, $U_{\mathrm{BA}}^{g/r}$ is not inverse matrix of
$U_{\mathrm{AB}}^{g/r}$.

During the braiding processes, both extra global phase factors $\delta
\phi_{g,\mathrm{A}}$ or $\delta \phi_{g,\mathrm{B}}$ and relative phase factors
$\delta \phi_{r,\mathrm{A}}$ or $\delta \phi_{r,\mathrm{B}}$ may be induced. It
was known the global phase factor is same for the two objects, $\delta \phi
_{g}=\delta \phi_{g,\mathrm{A}}=\delta \phi_{g,\mathrm{B}}$. The global phase
factor is obtained as
\begin{equation}
\delta \phi_{g}=\mathrm{e}_{\mathrm{A}}\cdot \mathrm{e}_{\mathrm{B}}\pi
=\frac{n_{\mathrm{A}}^{[12]}}{\lambda^{\lbrack12]}}\cdot \frac{n_{\mathrm{B}%
}^{[12]}}{\lambda^{\lbrack12]}}\cdot \pi.
\end{equation}
For example, for a D-quark and an U quark, we have $\delta \phi_{g}=(-\frac
{1}{3})\cdot \frac{2}{3}\cdot \pi=-\frac{2\pi}{9}.$ However, the relative phase
factors may be different for the two objects under braiding. According to the
relation from ratio between changings rates $\delta \phi_{g}=\lambda^{\lbrack
gr]}\delta \phi_{r}$, we have the following constraints,%
\begin{equation}
\delta \phi_{g}=\delta \phi_{g,\mathrm{A}}=\delta \phi_{g,\mathrm{B}}%
=\lambda^{\lbrack gr]}\delta \phi_{r}%
\end{equation}
where
\begin{equation}
\delta \phi_{r}=\delta \phi_{r,\mathrm{A}}+\delta \phi_{r,\mathrm{B}}%
\end{equation}
That means we must divide the total relative phase changing $\delta \phi_{r}$
into two parts, $\delta \phi_{r,\mathrm{A}}$ and $\delta \phi_{r,\mathrm{B}}$
under the following rule, i.e., for two quarks we have an inverse
relationship,
\begin{equation}
\frac{\delta \phi_{r,\mathrm{A}}}{\delta \phi_{r,\mathrm{B}}}=\frac
{\mathrm{e}_{\mathrm{B}}}{\mathrm{e}_{\mathrm{A}}}=\frac{n_{\mathrm{B}}%
^{[12]}}{n_{\mathrm{A}}^{[12]}};
\end{equation}
for a quark and a charged lepton, we have a zero relative phase changing for
quarks and the relative phase changing for charged lepton becomes $\delta
\phi_{r}=\delta \phi_{g}/\lambda^{\lbrack gr]}$.

In particular, we emphasize that when there exists relative phase changing,
the relative group-changing space shifts. As a result, one ground state may be
transformed into another. Without the changings of relative phase factor, the
ground states don't change any more.

We can use another four matrices, $U_{\mathrm{AB}}^{L},$ $U_{\mathrm{AB}}^{R}%
$, $U_{\mathrm{BA}}^{L}$,\ $U_{\mathrm{BA}}^{R}$ to characterize the
non-Abelian statistics. Here, $U_{\mathrm{AB}}^{L}$ and $U_{\mathrm{AB}}^{R}$
denote the changings of left-hand group-changing space and right-hand
group-changing space for \textrm{A} elementary particle, respectively;
$U_{\mathrm{BA}}^{L}$ and $U_{\mathrm{BA}}^{R}$ denote the changings of
left-hand group-changing space and right-hand group-changing space for
\textrm{B} elementary particle, respectively. For physical chiral variant of
our universe, we have
\begin{equation}
U_{\mathrm{AB}}^{g}=U_{\mathrm{AB}}^{R},\text{ }U_{\mathrm{BA}}^{g}%
=U_{\mathrm{BA}}^{R}%
\end{equation}
and
\begin{equation}
U_{\mathrm{AB}}^{L}=U_{\mathrm{AB}}^{g}U_{\mathrm{AB}}^{r},\text{
}U_{\mathrm{BA}}^{L}=U_{\mathrm{BA}}^{g}U_{\mathrm{BA}}^{r}.
\end{equation}
\ If $U_{\mathrm{AB}}^{g}=e^{i\delta \phi_{g,\mathrm{B}}}M_{\mathrm{AB}}$ and
$U_{\mathrm{AB}}^{r}=e^{i\delta \phi_{r,\mathrm{B}}}M_{\mathrm{AB}}$, (or
$U_{\mathrm{BA}}^{g}=e^{i\delta \phi_{g,\mathrm{A}}}M_{\mathrm{BA}}$ and
$U_{\mathrm{BA}}^{r}=e^{i\delta \phi_{r,\mathrm{A}}}M_{\mathrm{BA}}$), we have
\begin{equation}
U_{\mathrm{AB}}^{L}=e^{i\delta \phi_{g,\mathrm{B}}+i\delta \phi_{r,\mathrm{B}}%
}M_{\mathrm{AB}}%
\end{equation}
and
\begin{equation}
U_{\mathrm{BA}}^{L}=e^{i\delta \phi_{g,\mathrm{A}}+i\delta \phi_{r,\mathrm{A}}%
}M_{\mathrm{BA}}.
\end{equation}

In addition, we discuss the fusion rules for different elementary particles
with different internal level-2 zeroes.

We denote the sector of the elementary particle by $\sigma^{\lbrack n^{[12]}%
]}$. The fusion rule obeying usual associative law, i.e.,%
\begin{equation}
\sigma^{n_{1}^{[12]}}\times \sigma^{n_{2}^{[12]}}\times...\sigma^{n_{I}^{[12]}%
}=\sigma^{n_{1^{\prime}}^{[12]}}\times \sigma^{n_{2^{\prime}}^{[12]}}%
\times...\sigma^{n_{I^{\prime}}^{[12]}}.
\end{equation}
Here, "$\times$" denotes the particle fusion. There are two constraints -- one
is from conservation of the global level-1 zeroes (the number of fermionic
elementary particles), the other is from conservation of the level-2 zeroes
(the number of domain walls). As a result, we have
\begin{equation}
(%
%TCIMACRO{\dsum \limits_{i=1}^{I}}%
%BeginExpansion
{\displaystyle \sum \limits_{i=1}^{I}}
%EndExpansion
n_{i}^{[12]})\operatorname{mod}\lambda^{\lbrack12]}=(%
%TCIMACRO{\dsum \limits_{i^{\prime}=1}^{I^{\prime}}}%
%BeginExpansion
{\displaystyle \sum \limits_{i^{\prime}=1}^{I^{\prime}}}
%EndExpansion
n_{i^{\prime}}^{[12]})\operatorname{mod}\lambda^{\lbrack12]}.
\end{equation}

For example, for the fact of D-quark as a composite object of a U-quark, a
charged lepton, and a neutrino, we have fusion rule
\begin{align*}
\text{A D-quark (}n^{[12]}  &  =1\text{) = A U-quark (}n^{[12]}=2\text{)}\\
\text{+ A charged lepton (}n^{[12]}  &  =3\text{) + A neutrino (}%
n^{[12]}=0\text{).}%
\end{align*}
Now, we have
\begin{equation}
-1=(2-3+0)\operatorname{mod}3.
\end{equation}
This is just the decay process of elementary particles. In a relative zero,
the neutrino plays the role of "vacuum".

In summary, chiral para-statistics is a new type of quantum statistics that is
quite different from the usual one and will become the key point to calculate
mass spectra.

\paragraph{Examples}

In this part, we calculate $U_{\mathrm{AB}}^{g},$ $U_{\mathrm{AB}}^{r},$
$U_{\mathrm{BA}}^{r}$ and $U_{\mathrm{BA}}^{r}$ for different braiding processes.

Firstly, to characterize the $\lambda^{\lbrack12]}$-fold quasi-degenerate
ground states, we introduce a 1D effective lattice model with $\lambda
^{\lbrack12]}$ lattice sites under periodic boundary condition. Each of
lattice site corresponds to a degenerate state of the given elementary
particle. The processes of hopping between different lattice sites correspond
to the transition processes from one ground state to another.

In physics, because the quarks are domain walls separating different
degenerate states, when they pass through all lattice sites (the lattice
number is $\lambda^{\lbrack gr]}$), the states change from one to another. See
the illustration in Fig.42. According to above discussion, the processes of
changing ground states can be also considered by generating or eliminating the
level-2 zeroes of $\mathrm{C}_{\mathrm{\tilde{U}}^{[2]}\mathrm{(1)}}^{[2]}.$

\begin{figure}[ptb]
\includegraphics[clip,width=0.7\textwidth]{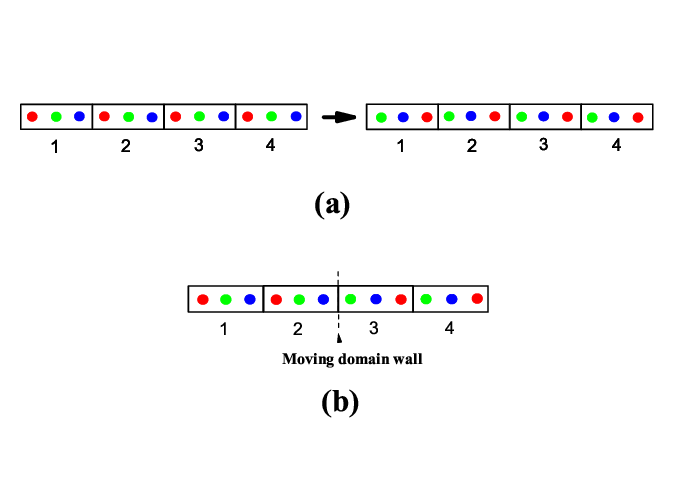}\caption{An illustration
of a domain wall and the processes of generation-transition}%
\end{figure}

Then, we show the rules for generation transitions and provide detailed calculations.

1) For the case of $\lambda^{\lbrack12]}=1,$ we have only two types of
elementary particles, one is chargeless neutrino with $n^{[12]}=0$, the other
is electron with unit electric charge (or $n^{[12]}=1$). It is obvious that
the statistics is Abelian one. We don't discuss this case.

2) For the case of $\lambda^{\lbrack12]}=2,$ we have three types of elementary
particles -- chargeless neutrino with $n^{[12]}=0$, the quark with half
electric charge (or $n^{[12]}=1$), the charged lepton with unit electric
charge (or $n^{[12]}=2$). For each type of elementary particle, there are two
degenerate states (or two generation) -- $\left \vert 1\right \rangle $
($...(\mathrm{ab})(\mathrm{ab})(\mathrm{ab})...$) or $\left \vert
2\right \rangle $ ($...(\mathrm{ba})(\mathrm{ba})(\mathrm{ba})...$). When a
quark (or a domain wall) passes through the system, one state is changed into
the other, $\left \vert 1\right \rangle \rightarrow \left \vert 2\right \rangle $
or $\left \vert 2\right \rangle \rightarrow \left \vert 1\right \rangle $. The
situation is similar to that in 1D dimerized CDW.

Let us calculate the matrices $U_{\mathrm{AB}}^{g},$ $U_{\mathrm{AB}}^{r},$
$U_{\mathrm{BA}}^{r}$ and $U_{\mathrm{BA}}^{r}$ for braiding rules:

\begin{enumerate}
\item Braiding rules for a quark (\textrm{q}) and a neutrino (\textrm{n}):
Neutrino is an elementary particle without internal level-1 zero and can be
regarded as "vacuum" on a relative zero. After a quark passing through the
neutrino, the type of flavors is changed. However, there doesn't exist extra
phase factor. As a result, the matrices $U_{\mathrm{nq}}^{g}$ and
$U_{\mathrm{nq}}^{r}$ for this braiding process are%
\begin{equation}
U_{\mathrm{nq}}^{g}=\left(
\begin{array}
[c]{cc}%
0 & 1\\
1 & 0
\end{array}
\right)  ,\text{ }U_{\mathrm{nq}}^{r}=\left(
\begin{array}
[c]{cc}%
0 & 1\\
1 & 0
\end{array}
\right)  .
\end{equation}
On the other hand, we calculate $U_{\mathrm{qn}}^{g}$ and $U_{\mathrm{qn}}%
^{r}$. After a neutrino passing through a quark, the braiding process is
trivial. Without changing the degenerate states and generating extra phase
factors, we have
\begin{equation}
U_{\mathrm{qn}}^{g}=\left(
\begin{array}
[c]{cc}%
1 & 0\\
0 & 1
\end{array}
\right)  ,\text{ }U_{\mathrm{qn}}^{r}=\left(
\begin{array}
[c]{cc}%
1 & 0\\
0 & 1
\end{array}
\right)  ;
\end{equation}

\item Braiding rules for a quark (\textrm{q}) and a charged lepton
(\textrm{l}): For a quark passing through a charged lepton, the type of
flavors for charged lepton is changed together with extra phase factors.
Because the quark is an object with $\pi/2$ phase changing, a charged lepton
feels $\frac{1}{2}\pi$ extra, global phase factor $\delta \phi_{g}$. So, we
have
\begin{equation}
U_{\mathrm{lq}}^{g}=\left(
\begin{array}
[c]{cc}%
0 & e^{i\zeta \frac{\pi}{2}}\\
e^{i\zeta \frac{\pi}{2}} & 0
\end{array}
\right)  .
\end{equation}
Here, $\zeta$ denotes a sign "$\pm$" that is the product of the signs of
\textrm{A} and \textrm{B.} For example, for the case of quark, we have
$\zeta=1$, and for the case of anti-quark, we have $\zeta=-1$. On the other
hand, for a charged lepton passing through a quark, the flavors of quark don't
change. However, there also exist extra global phase factors $\delta \phi
_{g}=\frac{1}{2}\pi$. So, we have
\begin{equation}
U_{\mathrm{ql}}^{g}=\left(
\begin{array}
[c]{cc}%
e^{i\zeta \frac{\pi}{2}} & 0\\
0 & e^{i\zeta \frac{\pi}{2}}%
\end{array}
\right)  .
\end{equation}
Due to the constraint relation from $\delta \phi_{g}=\lambda^{\lbrack
gr]}\delta \phi_{r}$, we have $\delta \phi_{r,\mathrm{l}}$ (the total relative
phase changing) to be $\frac{\pi}{2\lambda^{\lbrack gr]}}$. Because the
flavors of quark don't change during a charged lepton passing through a quark,
we have $\delta \phi_{r,\mathrm{q}}=0.$ So, the changings of relative phase
factor for quark is zero. Because the flavors of charged lepton change during
a quark passing through a charged lepton, we have $\delta \phi_{r,\mathrm{q}%
}\neq0.$ As a result, we divide the total relative phase factor $\delta
\phi_{r}$ into two parts $\delta \phi_{r,\mathrm{q}}$ and $\delta
\phi_{r,\mathrm{l}}$ for this braiding process,
\begin{equation}
\delta \phi_{r,\mathrm{q}}=0,\text{ and }\delta \phi_{r,\mathrm{l}}=\delta
\phi_{g}/\lambda^{\lbrack gr]}.
\end{equation}
Finally, we have%
\begin{equation}
U_{\mathrm{lq}}^{r}=\left(
\begin{array}
[c]{cc}%
0 & e^{i\frac{\zeta \pi}{2\lambda^{\lbrack gr]}}}\\
e^{i\frac{\zeta \pi}{2\lambda^{\lbrack gr]}}} & 0
\end{array}
\right)  .
\end{equation}
and
\begin{equation}
U_{\mathrm{ql}}^{r}=\left(
\begin{array}
[c]{cc}%
1 & 0\\
0 & 1
\end{array}
\right)  ;
\end{equation}

\item Braiding rules for a quark (\textrm{q}) and another quark (\textrm{q'}):
For a quark passing \textrm{A} through another quark \textrm{B}, the type of
flavors of quark \textrm{B} is changed together with extra total phase factors
$\frac{1}{2}\times \frac{1}{2}\pi=\frac{\pi}{4}$. We have
\begin{equation}
U_{\mathrm{qq}^{\prime}}^{g}=\left(
\begin{array}
[c]{cc}%
0 & e^{i\frac{\zeta \pi}{4}}\\
e^{i\frac{\zeta \pi}{4}} & 0
\end{array}
\right)  .
\end{equation}
Due to the constraint relation $\delta \phi_{g}=\lambda^{\lbrack gr]}\delta
\phi_{r}$, the total relative phase changing becomes $\frac{\pi}%
{4\lambda^{\lbrack gr]}}$. As a result, due to the symmetry for \textrm{A} (a
quark) and \textrm{B} (another quark), we divide the total relative phase
factor $\delta \phi_{r}$ into two equal parts $\delta \phi_{r,\mathrm{q}}$ and
$\delta \phi_{r,\mathrm{q}^{\prime}}$,
\begin{align}
\delta \phi_{r,\mathrm{q}}  &  =\frac{\delta \phi_{r}}{2}=\frac{\pi}%
{8\lambda^{\lbrack gr]}},\text{ }\\
\text{and }\delta \phi_{r,\mathrm{q}^{\prime}}  &  =\frac{\delta \phi_{r}}%
{2}=\frac{\pi}{8\lambda^{\lbrack gr]}}.\nonumber
\end{align}
Finally, the braiding rule for quark \textrm{q} and another quark \textrm{q'}
is described by
\begin{equation}
U_{\mathrm{qq}^{\prime}}^{r}=\left(
\begin{array}
[c]{cc}%
0 & e^{i\frac{\zeta \pi}{8\lambda^{\lbrack gr]}}}\\
e^{i\frac{\zeta \pi}{8\lambda^{\lbrack gr]}}} & 0
\end{array}
\right)  .
\end{equation}

\item Braiding rules for a neutrino (\textrm{n}) and charged lepton
(\textrm{l}): This is a trivial process without flavor changing and extra
phase factors. So, we have
\begin{equation}
U_{\mathrm{nl}}^{g}=U_{\mathrm{\ln}}^{g}=U_{\mathrm{nl}}^{r}=U_{\mathrm{\ln}%
}^{r}=\left(
\begin{array}
[c]{cc}%
1 & 0\\
0 & 1
\end{array}
\right)  ;
\end{equation}

\item Braiding rules for a neutrino (\textrm{n}) and another neutrino
(\textrm{n'}): This is trivial process without flavor changing and extra phase
factors. So, we have
\begin{equation}
U_{\mathrm{nn}^{\prime}}^{g}=U_{\mathrm{nn}^{\prime}}^{r}=\left(
\begin{array}
[c]{cc}%
1 & 0\\
0 & 1
\end{array}
\right)  ;
\end{equation}

\item Braiding rules for a charged lepton (\textrm{l}) and another charged
lepton (\textrm{l'}):\textrm{ }This is trivial process without flavor changing
and but with extra phase factors $\pi$. There doesn't exist relative phase
changing due to the result of global phase factor $\operatorname{mod}\pi$.
This is because the distance between two zeroes of the level-1 zero lattice of
a relative zero is $\pi$. So, we have%
\begin{equation}
U_{\mathrm{ll}^{\prime}}^{g}=\left(
\begin{array}
[c]{cc}%
-1 & 0\\
0 & -1
\end{array}
\right)  ,
\end{equation}
and
\begin{equation}
U_{\mathrm{ll}^{\prime}}^{r}=\left(
\begin{array}
[c]{cc}%
1 & 0\\
0 & 1
\end{array}
\right)  .
\end{equation}

\end{enumerate}

3) For the case of $\lambda^{\lbrack12]}=3,$ we have four types of elementary
particles: chargeless neutrino, U quark with 1/3 electric charge, D quark with
2/3 electric charge, and charged leptons with unit electric charge. For each
type of elementary particles, there are three degenerate states (or three
generations) -- $\left \vert 1\right \rangle $ ($...(\mathrm{abc})(\mathrm{abc}%
)(\mathrm{abc})...$), $\left \vert 2\right \rangle $ ($...(\mathrm{bca}%
)(\mathrm{bca})(\mathrm{bca})...$), and $\left \vert 3\right \rangle $
($...(\mathrm{cab})(\mathrm{cab})(\mathrm{cab})...$).

It was known that quarks play the role of domain walls that could change the
types of the degenerate states. The situation is similar to that in 1D
trimerized CDW. When a U-quark passes through another elementary particle, the
corresponding ground states change, i.e.,%
\begin{equation}
\left \vert 1\right \rangle \rightarrow \left \vert 2\right \rangle ,\text{
}\left \vert 2\right \rangle \rightarrow \left \vert 3\right \rangle ,\text{
}\left \vert 3\right \rangle \rightarrow \left \vert 1\right \rangle ;
\end{equation}
When an anti-U-quark passes through another elementary particle, the
corresponding ground states change, i.e.
\begin{equation}
\left \vert 1\right \rangle \rightarrow \left \vert 3\right \rangle ,\text{
}\left \vert 3\right \rangle \rightarrow \left \vert 2\right \rangle ,\text{
}\left \vert 2\right \rangle \rightarrow \left \vert 1\right \rangle .
\end{equation}

Let us calculate the braiding rules:

\begin{enumerate}
\item Braiding rules for a quark (\textrm{q}) and a neutrino (\textrm{n}):
Neutrino is an elementary particle without internal level-1 zero and can be
regarded as "vacuum" on a relative zero. After a quark passing through the
neutrino, the type of flavors is changed. However, there doesn't exist extra
phase factor. As a result, the matrices $U_{\mathrm{nq}}^{g}$ and
$U_{\mathrm{nq}}^{r}$ for this braiding process between a quark and a neutrino
are%
\begin{equation}
U_{\mathrm{nq}}^{g}=\left(
\begin{array}
[c]{ccc}%
0 & 1 & 0\\
0 & 0 & 1\\
1 & 0 & 0
\end{array}
\right)  ,\text{ }U_{\mathrm{nq}}^{r}=\left(
\begin{array}
[c]{ccc}%
0 & 1 & 0\\
0 & 0 & 1\\
1 & 0 & 0
\end{array}
\right)  .
\end{equation}
The matrices $U_{\mathrm{nq}}^{g}$ and $U_{\mathrm{nq}}^{r}$ for this braiding
process between an anti-quark and a neutrino are%
\begin{equation}
U_{\mathrm{nq}}^{g}=\left(
\begin{array}
[c]{ccc}%
0 & 0 & 1\\
1 & 0 & 0\\
0 & 1 & 0
\end{array}
\right)  ,\text{ }U_{\mathrm{nq}}^{r}=\left(
\begin{array}
[c]{ccc}%
0 & 0 & 1\\
1 & 0 & 0\\
0 & 1 & 0
\end{array}
\right)  .
\end{equation}
On the other hand, we calculate $U_{\mathrm{qn}}^{g}$ and $U_{\mathrm{qn}}%
^{r}$. After a neutrino passing through a quark, the braiding process is
trivial. Without changing the degenerate states and generating extra phase
factors, we have
\begin{equation}
U_{\mathrm{qn}}^{g}=\left(
\begin{array}
[c]{ccc}%
1 & 0 & 0\\
0 & 1 & 0\\
0 & 0 & 1
\end{array}
\right)  ,\text{ }U_{\mathrm{qn}}^{r}=\left(
\begin{array}
[c]{ccc}%
1 & 0 & 0\\
0 & 1 & 0\\
0 & 0 & 1
\end{array}
\right)  ;
\end{equation}

\item Braiding rules for a U-quark (\textrm{U}) and a charged lepton
(\textrm{l}): After a U quark passing through the electron, the type of
flavors is changed with extra phase factors. The U-quark is an object with
$-\frac{2}{3}\pi \zeta$ phase changing. As a result, the charged lepton feels
$-\frac{2}{3}\pi \zeta$ the global phase factor. So, we have%
\begin{equation}
U_{\mathrm{lU}}^{g}=\left(
\begin{array}
[c]{ccc}%
0 & e^{-i\frac{2\pi}{3}\zeta} & 0\\
0 & 0 & e^{-i\frac{2\pi}{3}\zeta}\\
e^{-i\frac{2\pi}{3}\zeta} & 0 & 0
\end{array}
\right)  .
\end{equation}
Here, $\zeta$\ is sign. The matrix for this braiding process between an
anti-quark and a charged lepton becomes $(U_{\mathrm{IU}}^{g})^{\dagger},$
i.e.,%
\begin{equation}
(U_{\mathrm{IU}}^{g})^{\dagger}=\left(
\begin{array}
[c]{ccc}%
0 & 0 & e^{i\frac{2\pi}{3}\zeta}\\
e^{i\frac{2\pi}{3}\zeta} & 0 & 0\\
0 & e^{i\frac{2\pi}{3}\zeta} & 0
\end{array}
\right)  .
\end{equation}
For a charged lepton passing through a U-quark, the type of flavors don't
change. However, the global phase factor changes $-\frac{2}{3}\zeta \pi$. So,
we have%
\begin{equation}
U_{\mathrm{Ul}}^{g}=\left(
\begin{array}
[c]{ccc}%
e^{-i\zeta \frac{2\pi}{3}} & 0 & 0\\
0 & e^{-i\zeta \frac{2\pi}{3}} & 0\\
0 & 0 & e^{-i\zeta \frac{2\pi}{3}}%
\end{array}
\right)
\end{equation}
where $\zeta$ is sign. From the constraint relation $\delta \phi_{g}%
=\lambda^{\lbrack gr]}\delta \phi_{r}$, we have the total relative phase
changing to be $-\frac{2\pi \zeta}{3\lambda^{\lbrack gr]}}$. Because the
flavors of quark don't change during a charged lepton passing through a quark,
we have $\delta \phi_{r,\mathrm{U}}=0.$ So, the changings of relative phase
factor for quark is zero. Because the flavors of charged lepton change during
a quark passing through a charged lepton, we have $\delta \phi_{r,\mathrm{U}%
}\neq0.$ As a result, we divide the total relative phase factor $\delta
\phi_{r}$ into two parts $\delta \phi_{r,\mathrm{U}}$ and $\delta
\phi_{r,\mathrm{l}}$ for this braiding process,
\begin{equation}
\delta \phi_{r,\mathrm{U}}=0,\text{ and }\delta \phi_{r,\mathrm{l}}=\delta
\phi_{g}/\lambda^{\lbrack gr]}.
\end{equation}
As a result, for U quark, we have
\begin{equation}
U_{\mathrm{IU}}^{r}=\left(
\begin{array}
[c]{ccc}%
0 & e^{-i\frac{2\pi}{3\lambda^{\lbrack gr]}}\zeta} & 0\\
0 & 0 & e^{-i\frac{2\pi}{3\lambda^{\lbrack gr]}}\zeta}\\
e^{-i\frac{2\pi}{3\lambda^{\lbrack gr]}}\zeta} & 0 & 0
\end{array}
\right)
\end{equation}
and for anti-U quark, we have%
\begin{equation}
(U_{\mathrm{lU}}^{r})^{\dagger}=\left(
\begin{array}
[c]{ccc}%
0 & 0 & e^{i\frac{2\pi}{3\lambda^{\lbrack gr]}}\zeta}\\
e^{i\frac{2\pi}{3\lambda^{\lbrack gr]}}\zeta} & 0 & 0\\
0 & e^{i\frac{2\pi}{3\lambda^{\lbrack gr]}}\zeta} & 0
\end{array}
\right)  .
\end{equation}
However, when a charge lepton passes through a U-quark, the ground state of
quark doesn't change. So, we have
\begin{equation}
U_{\mathrm{Ul}}^{r}=\left(
\begin{array}
[c]{ccc}%
1 & 0 & 0\\
0 & 1 & 0\\
0 & 0 & 1
\end{array}
\right)  ;
\end{equation}

\item Braiding rules for a U-quark (\textrm{U}) and another U-quark
(\textrm{U'}): After a U quark passing through the U-quark, the type of
flavors is changed together extra phase factor $\frac{2}{3}\times \frac{2\pi
}{3}\zeta=\frac{4\pi}{9}\zeta$. We have
\begin{equation}
U_{\mathrm{UU}^{\prime}}^{g}=\left(
\begin{array}
[c]{ccc}%
0 & e^{i\frac{4\pi}{9}\zeta} & 0\\
0 & 0 & e^{i\frac{4\pi}{9}\zeta}\\
e^{i\frac{4\pi}{9}\zeta} & 0 & 0
\end{array}
\right)  .
\end{equation}
or%
\begin{equation}
(U_{\mathrm{UU}^{\prime}}^{g})^{\dagger}=\left(
\begin{array}
[c]{ccc}%
0 & 0 & e^{-i\frac{4\pi}{9}\zeta}\\
e^{-i\frac{4\pi}{9}\zeta} & 0 & 0\\
0 & e^{-i\frac{4\pi}{9}\zeta} & 0
\end{array}
\right)  .
\end{equation}
Due to the constraint relation $\delta \phi_{g}=\lambda^{\lbrack gr]}\delta
\phi_{r}$, the total relative phase changing becomes $\frac{4\pi}%
{9\lambda^{\lbrack gr]}}$. As a result, due to the symmetry for \textrm{A} (a
quark) and \textrm{B} (another quark), we divide the total relative phase
factor $\delta \phi_{r}$ into two equal parts $\delta \phi_{r,\mathrm{U}}$ and
$\delta \phi_{r,\mathrm{U}^{\prime}}$ for this braiding process,
\begin{align}
\delta \phi_{r,\mathrm{U}}  &  =\frac{\delta \phi_{r}}{2}=\frac{1}{2}\frac{4\pi
}{9\lambda^{\lbrack gr]}}\zeta,\text{ }\\
\text{and }\delta \phi_{r,\mathrm{U}^{\prime}}  &  =\frac{\delta \phi_{r}}%
{2}=\frac{1}{2}\frac{4\pi}{9\lambda^{\lbrack gr]}}\zeta.\nonumber
\end{align}
Finally, we have
\begin{equation}
U_{\mathrm{UU}^{\prime}}^{r}=\left(
\begin{array}
[c]{ccc}%
0 & e^{i\frac{2\pi}{9\lambda^{\lbrack gr]}}\zeta} & 0\\
0 & 0 & e^{i\frac{2\pi}{9\lambda^{\lbrack gr]}}\zeta}\\
e^{i\frac{2\pi}{9\lambda^{\lbrack gr]}}\zeta} & 0 & 0
\end{array}
\right)  .
\end{equation}
The matrix $U_{\mathrm{UU}^{\prime}}^{r}$ for this braiding process between an
anti-U-quark and a U-quark are%
\begin{equation}
(U_{\mathrm{UU}^{\prime}}^{r})^{\dagger}=\left(
\begin{array}
[c]{ccc}%
0 & 0 & e^{-i\frac{2\pi}{9\lambda^{\lbrack gr]}}\zeta}\\
e^{-i\frac{2\pi}{9\lambda^{\lbrack gr]}}\zeta} & 0 & 0\\
0 & e^{-i\frac{2\pi}{9\lambda^{\lbrack gr]}}\zeta} & 0
\end{array}
\right)  ;
\end{equation}

\item Braiding rules for a U-quark (\textrm{U}) and D-quark (\textrm{D}):
According to above discussion, D-quark can be a composite object of a U-quark
and a charged lepton. Hence, the result is summation of that from two U-quarks
and that from a U-quark and a charged lepton. After one particle passing
through the other, the type of flavors is changed together extra phase factor
$-\frac{1}{3}\times \frac{2\pi}{3}\zeta=-\frac{2\pi}{9}\zeta$. We have
\begin{equation}
U_{\mathrm{UD}}^{g}=\left(
\begin{array}
[c]{ccc}%
0 & e^{-i\frac{2\pi}{9}\zeta} & 0\\
0 & 0 & e^{-i\frac{2\pi}{9}\zeta}\\
e^{-i\frac{2\pi}{9}\zeta} & 0 & 0
\end{array}
\right)
\end{equation}
or%
\begin{equation}
(U_{\mathrm{UD}}^{g})^{\dagger}=\left(
\begin{array}
[c]{ccc}%
0 & 0 & e^{i\frac{2\pi}{9}\zeta}\\
e^{i\frac{2\pi}{9}\zeta} & 0 & 0\\
0 & e^{i\frac{2\pi}{9}\zeta} & 0
\end{array}
\right)  .
\end{equation}
Due to the constraint relation $\delta \phi_{g}=\lambda^{\lbrack gr]}\delta
\phi_{r}$, the total relative phase changing becomes $-\frac{2\pi \zeta
}{9\lambda^{\lbrack gr]}}$.\ How to divide the total relative phase changing
$-\frac{2\pi}{9\lambda^{\lbrack gr]}}?$ Because the D-quark is a composite
object of a U-quark and a charged lepton,\ we firstly split the total relative
phase changing $-\frac{2\pi \zeta}{9\lambda^{\lbrack gr]}}$ into two parts --
one from braiding process of the charged lepton of the D-quark and U-quark,
the other from braiding process of U-quark of D-quark and another U-quark. For
the earlier process, another U-quark will not obtain the extra relative phase
changing, while the charged lepton has $-\frac{6\pi \zeta}{9\lambda^{\lbrack
gr]}};$ for the latter process, the U-quark of D-quark and another U-quark has
equal value $\frac{2\pi \zeta}{9\lambda^{\lbrack gr]}}$. So, we have
$\delta \phi_{r,\mathrm{D}}=-\frac{6\pi \zeta}{9\lambda^{\lbrack gr]}}%
+\frac{2\pi \zeta}{9\lambda^{\lbrack gr]}}=-\frac{4\pi \zeta}{9\lambda^{\lbrack
gr]}}$ and $\delta \phi_{r,\mathrm{U}}=\frac{2\pi \zeta}{9\lambda^{\lbrack gr]}%
}$. This\quad is consistent to the inverse relationship,
\begin{equation}
\frac{\delta \phi_{r,\mathrm{U}}}{\delta \phi_{r,\mathrm{D}}}=\frac
{\mathrm{e}_{\mathrm{D}}}{\mathrm{e}_{\mathrm{U}}}=\frac{-\frac{1}{3}}%
{\frac{2}{3}}.
\end{equation}
As a result, for the matrices of the D-quark, we have
\begin{equation}
U_{\mathrm{UD}}^{r}=\left(
\begin{array}
[c]{ccc}%
0 & e^{-i\frac{4\pi \zeta}{9\lambda^{\lbrack gr]}}} & 0\\
0 & 0 & e^{-i\frac{4\pi \zeta}{9\lambda^{\lbrack gr]}}}\\
e^{-i\frac{4\pi \zeta}{9\lambda^{\lbrack gr]}}} & 0 & 0
\end{array}
\right)
\end{equation}
or
\begin{equation}
(U_{\mathrm{UD}}^{r})^{\dagger}=\left(
\begin{array}
[c]{ccc}%
0 & 0 & e^{i\frac{4\pi \zeta}{9\lambda^{\lbrack gr]}}}\\
e^{i\frac{4\pi \zeta}{9\lambda^{\lbrack gr]}}} & 0 & 0\\
0 & e^{i\frac{4\pi \zeta}{9\lambda^{\lbrack gr]}}} & 0
\end{array}
\right)  .
\end{equation}
In addition, for the matrices of the U-quark, we have
\begin{equation}
U_{\mathrm{DU}}^{r}=\left(
\begin{array}
[c]{ccc}%
0 & e^{i\frac{2\pi \zeta}{9\lambda^{\lbrack gr]}}} & 0\\
0 & 0 & e^{i\frac{2\pi \zeta}{9\lambda^{\lbrack gr]}}}\\
e^{i\frac{2\pi \zeta}{9\lambda^{\lbrack gr]}}} & 0 & 0
\end{array}
\right)
\end{equation}
or
\begin{equation}
(U_{\mathrm{DU}}^{r})^{\dagger}=\left(
\begin{array}
[c]{ccc}%
0 & 0 & e^{-i\frac{2\pi \zeta}{9\lambda^{\lbrack gr]}}}\\
e^{-i\frac{2\pi \zeta}{9\lambda^{\lbrack gr]}}} & 0 & 0\\
0 & e^{-i\frac{2\pi \zeta}{9\lambda^{\lbrack gr]}}} & 0
\end{array}
\right)  .
\end{equation}

\item Braiding rules for a D-quark (\textrm{D}) and another D-quark
(\textrm{D'}): After a D quark passing through the D-quark, the type of
flavors is changed together extra phase factor $\frac{1}{3}\times \frac{\pi}%
{3}=\frac{\pi}{9}$. We have
\begin{equation}
U_{\mathrm{DD}^{\prime}}^{g}=\left(
\begin{array}
[c]{ccc}%
0 & 0 & e^{i\frac{\pi}{9}\zeta}\\
e^{i\frac{\pi}{9}\zeta} & 0 & 0\\
0 & e^{i\frac{\pi}{9}\zeta} & 0
\end{array}
\right)
\end{equation}
or%
\begin{equation}
(U_{\mathrm{DD}^{\prime}}^{g})^{\dagger}=\left(
\begin{array}
[c]{ccc}%
0 & e^{-i\frac{\pi}{9}\zeta} & 0\\
0 & 0 & e^{-i\frac{\pi}{9}\zeta}\\
e^{-i\frac{\pi}{9}\zeta} & 0 & 0
\end{array}
\right)  .
\end{equation}
Due to the constraint relation $\delta \phi_{g}=\lambda^{\lbrack gr]}\delta
\phi_{r}$, the total relative phase changing becomes $\frac{\pi}%
{9\lambda^{\lbrack gr]}}$. As a result, due to the symmetry for \textrm{A} (a
quark) and \textrm{B} (another quark), we divide the total relative phase
factor $\delta \phi_{r}$ into two equal parts $\delta \phi_{r,\mathrm{D}}$ and
$\delta \phi_{r,\mathrm{D}^{\prime}}$ for this braiding process,
\begin{align}
\delta \phi_{r,\mathrm{D}}  &  =\frac{\delta \phi_{r}}{2}=\frac{1}{2}\frac{\pi
}{9\lambda^{\lbrack gr]}},\text{ }\nonumber \\
\text{and }\delta \phi_{r,\mathrm{D}}  &  =\frac{\delta \phi_{r}}{2}=\frac{1}%
{2}\frac{\pi}{9\lambda^{\lbrack gr]}}.
\end{align}
Finally, we have
\begin{equation}
U_{\mathrm{DD}^{\prime}}^{r}=\left(
\begin{array}
[c]{ccc}%
0 & e^{i\frac{\pi}{18\lambda^{\lbrack gr]}}\zeta} & 0\\
0 & 0 & e^{i\frac{\pi}{18\lambda^{\lbrack gr]}}\zeta}\\
e^{i\frac{\pi}{18\lambda^{\lbrack gr]}}\zeta} & 0 & 0
\end{array}
\right)  .
\end{equation}
The matrix $U_{\mathrm{DD}^{\prime}}^{r}$ for this braiding process between an
anti-D-quark and a D-quark are%
\begin{equation}
(U_{\mathrm{DD}^{\prime}}^{r})^{\dagger}=\left(
\begin{array}
[c]{ccc}%
0 & 0 & e^{-i\frac{\pi}{18\lambda^{\lbrack gr]}}\zeta}\\
e^{-i\frac{\pi}{18\lambda^{\lbrack gr]}}\zeta} & 0 & 0\\
0 & e^{-i\frac{\pi}{18\lambda^{\lbrack gr]}}\zeta} & 0
\end{array}
\right)  ;
\end{equation}

\item Braiding rules for a D-quark (\textrm{D}) and a charged lepton
(\textrm{l}): Because the D-quark is a composite object of a U-quark and a
charged lepton,\ we firstly split the total global phase $\frac{\pi}{3}\zeta$
changing into two parts -- one from braiding process of the charged lepton of
the D-quark and another charged lepton, the other from braiding process of
U-quark of D-quark and another charged lepton. After a D-quark passing through
a charged lepton, the flavors of charged lepton are changed together with
extra phase factors. So, we have%
\begin{equation}
U_{\mathrm{lD}}^{g}=\left(
\begin{array}
[c]{ccc}%
0 & e^{i\frac{\pi}{3}\zeta} & 0\\
0 & 0 & e^{i\frac{\pi}{3}\zeta}\\
e^{i\frac{\pi}{3}\zeta} & 0 & 0
\end{array}
\right)
\end{equation}
or%
\begin{equation}
(U_{\mathrm{lD}}^{g})^{\dagger}=\left(
\begin{array}
[c]{ccc}%
0 & 0 & e^{-i\frac{\pi}{3}\zeta}\\
e^{-i\frac{\pi}{3}\zeta} & 0 & 0\\
0 & e^{-i\frac{\pi}{3}\zeta} & 0
\end{array}
\right)  .
\end{equation}
For a charged lepton passing through a D-quark, the type of flavors don't
change. So, we have%
\begin{equation}
U_{\mathrm{Dl}}^{g}=\left(
\begin{array}
[c]{ccc}%
e^{i\zeta \frac{\pi}{3}} & 0 & 0\\
0 & e^{i\zeta \frac{\pi}{3}} & 0\\
0 & 0 & e^{i\zeta \frac{\pi}{3}}%
\end{array}
\right)  .
\end{equation}
Using similar approach, we can derive the total relative phase changing as
$\frac{\pi}{3\lambda^{\lbrack gr]}}$. Because the charged lepton will not
change the flavors of the D-quark, we have $\delta \phi_{r,\mathrm{D}}=0$ and
$\delta \phi_{r,\mathrm{l}}=\frac{\pi}{3\lambda^{\lbrack gr]}}$. So, we have
\begin{equation}
U_{\mathrm{ID}}^{r}=\left(
\begin{array}
[c]{ccc}%
0 & e^{i\frac{\pi}{3\lambda^{\lbrack gr]}}\zeta} & 0\\
0 & 0 & e^{i\frac{\pi}{3\lambda^{\lbrack gr]}}\zeta}\\
e^{i\frac{\pi}{3\lambda^{\lbrack gr]}}\zeta} & 0 & 0
\end{array}
\right)
\end{equation}
or
\begin{equation}
(U_{\mathrm{lD}}^{r})^{\dagger}=\left(
\begin{array}
[c]{ccc}%
0 & 0 & e^{-i\frac{\pi}{3\lambda^{\lbrack gr]}}\zeta}\\
e^{-i\frac{\pi}{3\lambda^{\lbrack gr]}}\zeta} & 0 & 0\\
0 & e^{-i\frac{\pi}{3\lambda^{\lbrack gr]}}\zeta} & 0
\end{array}
\right)  .
\end{equation}
However, when a charge lepton passes through a D-quark, the ground state of
quark doesn't change. So, we have
\begin{equation}
U_{\mathrm{Dl}}^{r}=\left(
\begin{array}
[c]{ccc}%
1 & 0 & 0\\
0 & 1 & 0\\
0 & 0 & 1
\end{array}
\right)  ;
\end{equation}

\item Braiding rules for a neutrino (\textrm{n}) and charged lepton
(\textrm{l}): This is a trivial process without flavor changing and extra
phase factors. So, we have
\begin{equation}
U_{\mathrm{nl}}^{g}=U_{\text{\textrm{ln}}}^{g}=U_{\mathrm{nl}}^{r}%
=U_{\text{\textrm{ln}}}^{r}=\left(
\begin{array}
[c]{ccc}%
1 & 0 & 0\\
0 & 1 & 0\\
0 & 0 & 1
\end{array}
\right)  ;
\end{equation}

\item Braiding rules for a neutrino (\textrm{n}) and another neutrino
(\textrm{n'}): This is trivial process without flavor changing and extra phase
factors. So, we have
\begin{equation}
U_{\mathrm{nn}^{\prime}}^{g}=U_{\mathrm{nn}^{\prime}}^{r}=\left(
\begin{array}
[c]{ccc}%
1 & 0 & 0\\
0 & 1 & 0\\
0 & 0 & 1
\end{array}
\right)  ;
\end{equation}

\item Braiding rules for a charged lepton (\textrm{l}) and another charged
lepton (\textrm{l'}):\textrm{ }This is trivial process without flavor changing
and but with extra phase factors $\pi$. There doesn't exist relative phase
changing. So, we have%
\begin{equation}
U_{\mathrm{ll}^{\prime}}^{g}=\left(
\begin{array}
[c]{ccc}%
-1 & 0 & 0\\
0 & -1 & 0\\
0 & 0 & -1
\end{array}
\right)  ,
\end{equation}
and
\begin{equation}
U_{\mathrm{ll}^{\prime}}^{r}=\left(
\begin{array}
[c]{ccc}%
1 & 0 & 0\\
0 & 1 & 0\\
0 & 0 & 1
\end{array}
\right)  .
\end{equation}

\end{enumerate}

4) For a general case $\lambda^{\lbrack12]}=N$, we have $N+1$ types of
elementary particles: chargeless neutrino without electric charge, the quarks
with $\frac{n^{[2]}}{N}$ electric charge ($n^{[2]}=1,2...N-1$), the electron
with unit electric charge. For each type of elementary particles, there are
$N$ degenerate states (or $N$ generation). When a U-quark ($n^{[2]}=N-1$)
passes through another elementary particle, the corresponding ground states
change, i.e.,%
\begin{equation}
\left \vert 1\right \rangle \rightarrow \left \vert 2\right \rangle ,\text{
}\left \vert 2\right \rangle \rightarrow \left \vert 2\right \rangle ,\text{ ...
}\left \vert N\right \rangle \rightarrow \left \vert 1\right \rangle .
\end{equation}
A D-quark is composite object of a U-quark and a charged lepton ans shows same
behavior as U-quark. The situation is similar to that in 1D CDW with $N$-aggregation.

We can use similar approach to calculate the braiding processes. For example,
for two U-quarks, the changing of global phase factor is
\begin{equation}
\frac{1}{N-1}\times \frac{\pi \zeta}{N-1}=\frac{\pi \zeta}{(N-1)^{2}}%
\end{equation}
and the total changing of relative phase factor is
\begin{equation}
\frac{\pi \zeta}{(N-1)^{2}}\times \frac{\zeta}{\lambda^{\lbrack gr]}}.
\end{equation}
As a result, the changings of relative phase factors for each U-quark is
\begin{equation}
\frac{1}{2}\times \frac{\pi \zeta}{(N-1)^{2}}\times \frac{\zeta}{\lambda^{\lbrack
gr]}}=\frac{\pi \zeta}{2(N-1)^{2}\lambda^{\lbrack gr]}}.
\end{equation}
The corresponding matrices $U_{\mathrm{UU}^{\prime}}^{r}$ are obtained as
\begin{equation}
U_{\mathrm{UU}^{\prime}}^{r}=e^{i\frac{\pi \zeta}{2(N-1)^{2}\lambda^{\lbrack
gr]}}}\left(
\begin{array}
[c]{ccccc}%
0 & 1 & 0 & 0 & 0\\
0 & 0 & 1 & 0 & 0\\
0 & 0 & 0 & ... & 0\\
0 & 0 & ... & ... & 1\\
1 & 0 & 0 & 0 & 0
\end{array}
\right)
\end{equation}
or
\begin{equation}
(U_{\mathrm{UU}^{\prime}}^{r})^{\dagger}=e^{-i\frac{\pi \zeta}{2(N-1)^{2}%
\lambda^{\lbrack gr]}}}\left(
\begin{array}
[c]{ccccc}%
0 & 0 & 0 & 0 & 1\\
1 & 0 & 0 & 0 & 0\\
0 & 1 & 0 & ... & 0\\
0 & 0 & ... & ... & 0\\
0 & 0 & 0 & 1 & 0
\end{array}
\right)  .
\end{equation}

\subsubsection{Generalized ${{\mathbb{Z}}_{N}}$ clock model for
generation-transition}

In above part, we obtained the effective model for either Dirac type of
elementary particles or Weyl type of elementary particles. However, their
masses are still unknown. In this part, with help of chiral para-statistics,
we will calculate their masses from a generalized ${{\mathbb{Z}}_{N}}$ clock
model for generation-transition. Now, we have a free parameter -- the ratio
$\gamma^{\lbrack gr]}=3\pi$ between the changing rates of global sub-variant
and relative sub-variant.

The key point is to obtain the operator for mass spectra from
generation-transition. To obtain such operator, the following three conditions
are assumed to be satisfied:

1) \emph{Equivalence condition for different degenerate states}: This
condition leads to translation symmetry for the ${{\mathbb{Z}}_{N}}$ clock
model, $\hat{T}_{j}\hat{M}\hat{T}_{j}^{-1}=\hat{M}$ ($j\in \lbrack0,N-1]$). As
a result, the eigenstates are classified by effective "wave vectors $k$";

2)\emph{ Equivalence condition for different operators of
generation-transition}: We introduce a coherent operator for
generation-transition $R_{\mathrm{A}}$ that is the combination for the all
operators of generation-transition\ with equal weight;

3) \emph{The assumption of the mass psectra by a pair of coherent operator for
generation-transition }$R_{\mathrm{A}}$: The situation is similar to the
transition process induced by a pair of Majorana particles $\gamma_{k}%
\gamma_{-k}$ where $\gamma_{k}$ denotes Majorana fermion. We assume that
$\hat{M}\sim(R_{\mathrm{A}})^{2}.$

\paragraph{Generalized clock state and its operators}

A clock state is characterized by the operator $\sigma$ and $\tau$,
\begin{align}
\sigma^{N}  &  =1\ ,\  \tau^{N}=1,\  \\
\sigma^{\dagger}  &  =\sigma^{N-1},\  \tau^{\dagger}=\tau^{N-1}\ ,\text{
}\sigma \tau=\omega \, \tau \sigma,\nonumber
\end{align}
where $\omega \equiv e^{2\pi i/N}$. After diagonalizing $\sigma$, we have
\begin{align}
\sigma &  =%
\begin{pmatrix}
1 & 0 & 0 & \  \cdots \  & 0\\
0 & \omega & 0 & \  \cdots \  & 0\\
0 & 0 & \omega^{2} & \  & 0\\
\vdots & \vdots & 0 & 0 & \vdots \\
0 & 0 & 0 & \  \cdots \  & \omega^{N-1}%
\end{pmatrix}
,\quad \quad \nonumber \\
\tau &  =%
\begin{pmatrix}
0 & 0 & 0\  \cdots \  & 0 & 1\\
1 & 0 & 0\  \cdots \  & 0 & 0\\
0 & 1 & 0\  \cdots & 0 & 0\\
\vdots &  &  & \vdots & \vdots \\
0 & 0 & 0\  \cdots \  & 1 & 0
\end{pmatrix}
.
\end{align}

According to above discussion. for an elementary particle, we have $N$
degenerate states $\left \vert i\right \rangle $ that correspond to the $N$
generation. Due to $\left \vert i+N\right \rangle =\left \vert i\right \rangle
,$\ we have a generalized clock state.

On one hand, the diagonalizing operator $\sigma$ is
\begin{equation}
\sigma^{N}=1,\  \sigma^{\dagger}=\sigma^{N-1},\  \sigma \tau=\omega \, \tau \sigma.
\end{equation}
So, we have
\begin{equation}
\sigma=%
\begin{pmatrix}
1 & 0 & 0 & \  \cdots \  & 0\\
0 & \omega & 0 & \  \cdots \  & 0\\
0 & 0 & \omega^{2} & \  & 0\\
\vdots & \vdots & 0 & 0 & \vdots \\
0 & 0 & 0 & \  \cdots \  & \omega^{N-1}%
\end{pmatrix}
.
\end{equation}

Then, we use the U-quark to be a test object to do generation-transition.
Hence, the phase factor $\varphi_{r,\mathrm{A}}$ is determined by the matrix
$U_{\mathrm{AB}}^{r}$\ where $\mathrm{A}$ is object to be detected and
$\mathrm{B}$ must be U-quark (or domain wall with smallest mass in the SM).

For the case of $N=2,$ we have $\tau \rightarrow(%
\begin{array}
[c]{cc}%
0 & 1\\
0 & 0
\end{array}
)$\ and $\tau^{\dagger}\rightarrow(%
\begin{array}
[c]{cc}%
0 & 0\\
1 & 0
\end{array}
)$. For the case of $N>2$, we have the corresponding operator $\tau$ to be
\begin{equation}
\tau=U_{\mathrm{AB}}^{r}=e^{i\varphi_{r,\mathrm{A}}}%
\begin{pmatrix}
0 & 0 & 0\  \cdots \  & 0 & 1\\
1 & 0 & 0\  \cdots \  & 0 & 0\\
0 & 1 & 0\  \cdots & 0 & 0\\
\vdots & 0 & 0 & \vdots & \vdots \\
0 & 0 & 0\  \cdots \  & 1 & 0
\end{pmatrix}
\end{equation}
or
\begin{equation}
\tau^{\dagger}=(U_{\mathrm{AB}}^{r})^{\dagger}=e^{-i\varphi_{r,\mathrm{A}}}%
\begin{pmatrix}
0 & 1 & 0\  \cdots \  & 0 & 0\\
0 & 0 & 1\  \cdots \  & 0 & 0\\
0 & 0 & 0\  \cdots & 0 & 0\\
\vdots & 0 & 0 & \vdots & \vdots \\
1 & 0 & 0\  \cdots \  & 0 & 0
\end{pmatrix}
.
\end{equation}
Here, $e^{i\varphi_{r,\mathrm{A}}}$\ is phase factor of relative phase
changing of \textrm{A} elementary particle during the generation-transition.
The corresponding operators $\sigma$ and $\tau$ acting on the generalized
clock state satisfy
\begin{align}
\sigma^{N}  &  =1\ ,\  \tau^{N}=e^{iN\varphi_{r,\mathrm{A}}},\  \\
\sigma^{\dagger}  &  =\sigma^{N-1},\  \tau^{\dagger}=e^{iN\varphi
_{r,\mathrm{A}}}\tau^{N-1}\ ,\text{ }\sigma \tau=\omega \, \tau \sigma,\nonumber
\end{align}
where $\omega \equiv e^{2\pi i/N}$. Now, in the representation, $\sigma$
denotes the value of a clock variable, and $\tau$ denotes shifting the value.

\paragraph{Coherent operator for generation-transition}

According to equivalence condition for different operators of
generation-transition, we define a coherent operator for generation-transition
$R_{\mathrm{A}}$ that is the combination for the all operators of
generation-transition\ with equal weight.

\textit{Definition: Coherent operator for generation-transition }%
$R_{\mathrm{A}}$\textit{ is the combination for the following operators }%
$1,$\textit{ }$\tau$\textit{\ and }$\tau^{\dagger}$\textit{\ with equal
weight, i.e., }%
\begin{equation}
R_{\mathrm{A}}=%
%TCIMACRO{\dsum \limits_{j}}%
%BeginExpansion
{\displaystyle \sum \limits_{j}}
%EndExpansion
\tau^{j}%
\end{equation}
\textit{where }$j$\textit{ label the index of operators. Here, }$1$\textit{
denotes freezing, }$\tau$\textit{\ denotes shifting the state upward and
}$\tau^{\dagger}$ denotes \textit{shifting the state downward. }Coherent
operator $R_{\mathrm{A}}$ for generation-transition is a Hermitian operator.

For the case of $N=2,$ we have
\begin{equation}
R_{\mathrm{A}}=1+\tau+\tau^{\dagger}.
\end{equation}
Under $R_{\mathrm{A}},$ we have $\left \vert 1\right \rangle \rightarrow
\left \vert 1\right \rangle +\left \vert 2\right \rangle \ $or $\left \vert
2\right \rangle \rightarrow \left \vert 1\right \rangle +\left \vert 2\right \rangle
$.

For the case of $N=3,$ we have
\begin{equation}
R_{\mathrm{A}}=1+\tau+\tau^{\dagger}.\nonumber
\end{equation}
Under $R_{\mathrm{A}},$ we have $\left \vert i\right \rangle \rightarrow
\left \vert i\right \rangle +(e^{i\varphi_{r,\mathrm{A}}}\left \vert
\alpha \right \vert \left \vert i-1\right \rangle +e^{-i\varphi_{r,\mathrm{A}}%
}\left \vert \alpha \right \vert \left \vert i+1\right \rangle ).$

For the case of $N>3,$ $N$ degenerate states are $\left \vert 1\right \rangle ,$
$\left \vert 2\right \rangle $,\ ... and $\left \vert N\right \rangle $. The
generation-transition is characterized by
\begin{equation}
R_{\mathrm{A}}=1+U_{\mathrm{AB}}^{r}+(U_{\mathrm{AB}}^{r})^{\dagger
}+(U_{\mathrm{AB}}^{r})^{2}+((U_{\mathrm{AB}}^{r})^{2})^{\dagger}+...
\end{equation}
Under $R_{\mathrm{A}},$ we have $\left \vert i\right \rangle \rightarrow
\left \vert i\right \rangle +(e^{i\varphi_{r,\mathrm{A}}}\left \vert
\alpha \right \vert \left \vert i-1\right \rangle +e^{-i\varphi_{r,\mathrm{A}}%
}\left \vert \alpha \right \vert \left \vert i+1\right \rangle )+(e^{2i\varphi
_{r,\mathrm{A}}}\left \vert \alpha \right \vert ^{2}\left \vert i-1\right \rangle
+e^{-2i\varphi_{r,\mathrm{A}}}\left \vert \alpha \right \vert ^{2}\left \vert
i+1\right \rangle )+...$

Finally, we assume that the mass spectra is proportional to $(R_{\mathrm{A}%
})^{2}$.

\paragraph{$\left \vert \alpha \right \vert $ and Branch ratio}

Then, we focus on the effect of generation-transition and transform it to
usual hopping operator $a^{+}$ or $a$ of the ${{\mathbb{Z}}_{N}}$ clock model,
$a^{+}=\tau=U_{\mathrm{AB}}^{r}$ or $a=\tau^{\dagger}=(U_{\mathrm{AB}}%
^{r})^{\dagger}$. So, the hopping term for nearest neighbor hopping is written
as a linear combination between $a^{+}$ and $a,$ i.e.,
\begin{equation}
\alpha a^{+}+\beta a.
\end{equation}
Because the true physical processes are Hermitian, we have $\beta=\alpha
^{\ast}.$ So, the nearest neighbor hopping term must be written as $\alpha
a^{+}+\alpha^{\ast}a.$ Here, $\alpha=\left \vert \alpha \right \vert e^{i\varphi
}$ is a complex number, of which the phase factor $\varphi$ is just
$\varphi_{r,\mathrm{A}}$ that can be obtained by using the results from chiral
para-statistics. The true process must be a quantum state under equal
weighting superposition, $\left \vert \alpha \right \vert =\left \vert
\alpha^{\ast}\right \vert $.

What's its amplitude $\left \vert \alpha \right \vert $?

We assume that \emph{during the processes of generation-transition, the
initial state and final state with domain walls are all pure states of
different states with equivalent weight}. So, for the different degenerate
states have same weight, the corresponding normalization factors for an
initial state and a final state become $\frac{1}{\sqrt{N_{I}}}$\ and $\frac
{1}{\sqrt{N_{F}}}$. Here, $N_{I}$ is the number of initial degenerate states
and $N_{F}$ is the number the final degenerate states, respectively. So, we
have
\begin{equation}
\left \vert \alpha \right \vert =\sqrt{\frac{N_{I}}{N_{F}}}%
\end{equation}
We call $\frac{N_{I}}{N_{F}}$ to be \emph{branch ratio}. $\left \vert
\alpha \right \vert $ is the square root of branch ratio.

\begin{figure}[ptb]
\includegraphics[clip,width=0.92\textwidth]{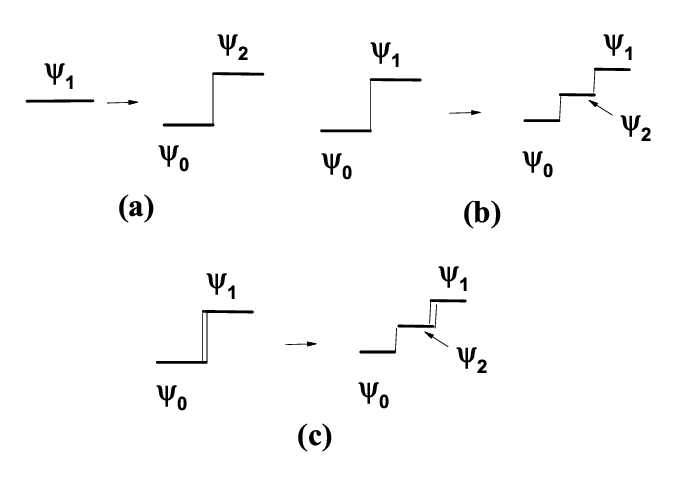}\caption{(Color
online) The illustration of branch ratio: (a) A processes of
generation-transition for a charged lepton; (b) A processes of
generation-transition for a D-quark; (c) A processes of generation-transition
for a U-quark. The double lines denote two internal level-2 zeroes.}%
\end{figure}

Let us give an example to show above result. See the illustration in Fig.43.

The first example is about processes of generation-transition for a charged lepton.

We consider the processes of generation-transition from a given lattice site
(for example, $i=s$).\ The initial state is $\left \vert s\right \rangle $. So,
the number of degenerate states $N_{I}$ is $1$. The normalization factor is
$1$. During generation-transition by $\alpha a^{+}+\alpha^{\ast}a$, the
quantum state on site $s$ turns into two quantum states $\left \vert
s+1\right \rangle =a^{+}\left \vert s\right \rangle $ and $\left \vert
s-1\right \rangle =a\left \vert s\right \rangle $ for the internal level-2
zeroes. Then, the original quantum state $\left \vert s\right \rangle $ becomes
a coherent quantum state $\left \vert \mathrm{coherent}\right \rangle =(\alpha
a^{+}+\alpha^{\ast}a)\left \vert s\right \rangle $ of the basis. If the two
degenerate quantum states $\left \vert s+1\right \rangle $ and $\left \vert
s-1\right \rangle $ are orthogonal,
\begin{align}
\left \langle s+1\mid s-1\right \rangle  &  =0,\text{ }\left \langle s\mid
s\right \rangle =1,\nonumber \\
\left \langle s+1\mid s+1\right \rangle  &  =1,\text{ }\left \langle s-1\mid
s-1\right \rangle =1
\end{align}
So, we could consider $\left \vert s+1\right \rangle $ and $\left \vert
s-1\right \rangle $ as a basis of quantum states. Now, with two degenerate
states $\left \vert s+1\right \rangle $ and $\left \vert s-1\right \rangle $, the
number of degenerate states $N_{F}$ is $2$. See the illustration in Fig.43.
For each state $\left \vert s+1\right \rangle $ or $\left \vert s-1\right \rangle
,$ the normalization factor is $1/\sqrt{2}$. On the other hand, the
normalization condition of the quantum coherent state is equal to
$\left \langle \mathrm{coherent}\mid \mathrm{coherent}\right \rangle =2\left \vert
\alpha \right \vert ^{2}$. Due to normalization condition $\left \langle
\mathrm{coherent}\mid \mathrm{coherent}\right \rangle =1,$ we derive
\begin{equation}
\left \vert \alpha \right \vert =\sqrt{\frac{N_{I}}{N_{F}}}=\frac{1}{\sqrt{2}}.
\end{equation}
This is just the square root of branch ratio $\left \vert \alpha \right \vert
=\sqrt{\frac{N_{I}}{N_{F}}}=\frac{1}{\sqrt{2}}.$ Under $R_{\mathrm{A}},$ we
have $\left \vert 2\right \rangle \rightarrow \left \vert 2\right \rangle +\frac
{1}{\sqrt{2}}(e^{i\varphi_{r,\mathrm{A}}}\left \vert 1\right \rangle
+e^{i\varphi_{r,\mathrm{A}}}\left \vert 3\right \rangle ).$

The second example is about processes of generation-transition for a D-quark.

On the one hand, a D-quark is an elementary particle with a internal level-2
zero. On the other hand,\ a D-quark is a domain wall between two degenerate
states $\left \vert s+1\right \rangle $ and $\left \vert s\right \rangle .$ For
the two degenerate states of a D-quark $\left \vert s+1\right \rangle $ and
$\left \vert s\right \rangle $, the number of degenerate states $N_{I}$ is $2$.
For each state $\left \vert s+1\right \rangle $ or $\left \vert s\right \rangle ,$
the normalization factor is $1/\sqrt{2}$. After the process of
generation-transition, there exist three degenerate states $\left \vert
s+1\right \rangle ,$ $\left \vert s-1\right \rangle $, and $\left \vert
s\right \rangle $ for final state. See the illustration in Fig.43. Now, the
number of degenerate states $N_{F}$ becomes $3$. For each state, the
normalization factor is $1/\sqrt{3}$. As a result, we have
\begin{equation}
\left \vert \alpha \right \vert =\sqrt{\frac{N_{I}}{N_{F}}}=\sqrt{\frac{2}{3}}.
\end{equation}
Under $R_{\mathrm{A}},$ we have $\left \vert 2\right \rangle \rightarrow
\left \vert 2\right \rangle +\sqrt{\frac{2}{3}}(e^{i\varphi_{r,\mathrm{A}}%
}\left \vert 1\right \rangle +e^{i\varphi_{r,\mathrm{A}}}\left \vert
3\right \rangle ).$

The third example is about processes of generation-transition for a U-quark.

It looks like we have same results as that for a D-quark. However, the
situation is different. For a U-quark, there are two internal level-2 zeroes.
We may assume the two internal level-2 zeroes move independently. As a result,
the number of degenerate initial states $N_{I}$ is $3$. For each state, the
normalization factor becomes $1/\sqrt{3}$. After the process of
generation-transition, the number of degenerate states becomes $3+1.$ See the
illustration in Fig.43. For each state, the normalization factor is
$1/\sqrt{4}$. As a result, we have
\begin{equation}
\left \vert \alpha \right \vert =\sqrt{\frac{N_{I}}{N_{F}}}=\sqrt{\frac{3}{4}}.
\end{equation}
Under $R_{\mathrm{A}},$ we have $\left \vert 2\right \rangle \rightarrow
\left \vert 2\right \rangle +\sqrt{\frac{3}{4}}(e^{i\varphi_{r,\mathrm{A}}%
}\left \vert 1\right \rangle +e^{i\varphi_{r,\mathrm{A}}}\left \vert
3\right \rangle ).$

\paragraph{${{\mathbb{Z}}_{N}}$ clock model for generation-transition}

To derive mass spectra, we introduce a ${{\mathbb{Z}}_{N}}$ clock\emph{ }model
for generation-transition. From above three conditions, particle's mass
operator is assumed to be
\begin{align}
\hat{M}_{\mathrm{A}}  &  =\left \vert b_{\mathrm{A}}\right \vert (%
%TCIMACRO{\dsum \limits_{i}}%
%BeginExpansion
{\displaystyle \sum \limits_{i}}
%EndExpansion
R_{\mathrm{A}}^{i})^{2}\\
&  =\left \vert b_{\mathrm{A}}\right \vert (%
%TCIMACRO{\dsum \limits_{i}}%
%BeginExpansion
{\displaystyle \sum \limits_{i}}
%EndExpansion
(%
%TCIMACRO{\dsum \limits_{j}}%
%BeginExpansion
{\displaystyle \sum \limits_{j}}
%EndExpansion
\tau^{j}))^{2}%
\end{align}
where $R_{\mathrm{A}}^{i}$ is the coherent operator for state $\left \vert
i\right \rangle $ and $b_{\mathrm{A}}$ is a constant value for a given type of
elementary particle.\

Now, we have $N$ generations. To characterize the processes of
generation-transition, we map the states of $N$ generations for each type of
elementary particles to a generalized clock model that is a 1D lattice model
with $N$ lattice sites under periodic boundary condition. Fig.44 is an
illustration of a generalized ${{\mathbb{Z}}_{N}}$ clock model, of which each
lattice site corresponds to a generation for a given elementary particle. For
the generalized ${{\mathbb{Z}}_{N}}$ clock model, we have periodic boundary
condition, $\left \vert s\right \rangle =\left \vert s+N\right \rangle $ where
$N=\lambda^{\lbrack12]}$. The hopping processes in the clock model correspond
to the transition processes from generation $i$ to $i+1$\ (or $i-1$) generated
by the operator $\tau=U_{\mathrm{AB}}^{r}\ $or $\tau^{\dagger}=(U_{\mathrm{AB}%
}^{r})^{\dagger}$.

\begin{figure}[ptb]
\includegraphics[clip,width=0.92\textwidth]{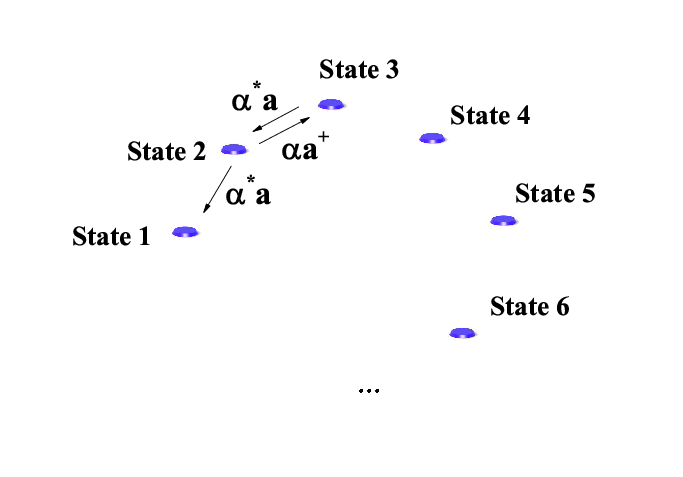}\caption{(Color online)
An illustration of a ${{\mathbb{Z}}_{N}}$ clock model for
generation-transition, of which each lattice site corresponds to a generation
for a given elementary particle}%
\end{figure}

Finally, for an arbitrary case, we obtain the operator for
generation-transition,%
\begin{align}
\hat{M}_{\mathrm{A}}  &  =\left \vert b_{\mathrm{A}}\right \vert (%
%TCIMACRO{\dsum \limits_{i}}%
%BeginExpansion
{\displaystyle \sum \limits_{i}}
%EndExpansion
R_{\mathrm{A}}^{i})^{2}\\
&  =\left \vert b_{\mathrm{A}}\right \vert (1+\left \vert \alpha \right \vert
e^{i\varphi_{r,\mathrm{A}}}%
\begin{pmatrix}
0 & 0 & 0\  \cdots \  & 0 & 1\\
1 & 0 & 0\  \cdots \  & 0 & 0\\
0 & 1 & 0\  \cdots & 0 & 0\\
\vdots &  &  & \vdots & \vdots \\
0 & 0 & 0\  \cdots \  & 1 & 0
\end{pmatrix}
\nonumber \\
&  +\left \vert \alpha \right \vert e^{-i\varphi_{r,\mathrm{A}}}%
\begin{pmatrix}
0 & 1 & 0\  \cdots \  & 0 & 0\\
0 & 0 & 1\  \cdots \  & 0 & 0\\
0 & 0 & 0\  \cdots & 0 & 0\\
\vdots &  &  & \vdots & \vdots \\
1 & 0 & 0\  \cdots \  & 0 & 0
\end{pmatrix}
)^{2}.\nonumber
\end{align}
where $\varphi_{r,\mathrm{A}}$ is determined by chiral para-statistics. After
replacing $a_{i}$ by $e^{i\varphi_{r,\mathrm{A}}}\left \vert i\right \rangle
\left \langle i+1\right \vert $ and $a_{i}^{+}$ by $e^{-i\varphi_{r,\mathrm{A}}%
}\left \vert i+1\right \rangle \left \langle i\right \vert $, ..., we have
\[
\hat{M}_{\mathrm{A}}=\left \vert b_{\mathrm{A}}\right \vert (%
%TCIMACRO{\dsum \limits_{i}}%
%BeginExpansion
{\displaystyle \sum \limits_{i}}
%EndExpansion
(1+\left \vert \alpha \right \vert e^{i\varphi_{r,\mathrm{A}}}\left \vert
i\right \rangle \left \langle i+1\right \vert +\left \vert \alpha \right \vert
e^{-i\varphi_{r,\mathrm{A}}}\left \vert i+1\right \rangle \left \langle
i\right \vert ))^{2}.
\]
Here, $i$ denotes a given generation that corresponds to a lattice site. Due
to periodic boundary condition, $\left \vert s\right \rangle =\left \vert
s+N\right \rangle ,$ we have $(a^{+})^{N}=e^{iN\varphi_{r,\mathrm{A}}}$ and
$(a)^{N}=e^{-iN\varphi_{r,\mathrm{A}}}$.

In the end, after Fourier transformation to "momentum space", we derive the
mass spectra for $\mathrm{A}$ type elementary particle that is eigenvalue of
$\hat{M}$,
\begin{equation}
M_{\mathrm{A}}=\left \vert b_{\mathrm{A}}\right \vert [1+(\left \vert
\alpha \right \vert e^{i\frac{2\pi j}{N}+i\varphi_{r,\mathrm{A}}}+\left \vert
\alpha \right \vert e^{-i\frac{2\pi j}{N}-i\varphi_{r,\mathrm{A}}})]^{2},\text{
}j=1,2,..N.
\end{equation}
In particular, for our universe, $\lambda^{\lbrack12]}=N=3,$ we have%
\begin{align}
M_{\mathrm{A}}  &  =\left \vert b_{\mathrm{A}}\right \vert [1+(\left \vert
\alpha \right \vert e^{i\frac{2\pi j}{3}+i\varphi_{r,\mathrm{A}}}+\left \vert
\alpha \right \vert e^{-i\frac{2\pi j}{3}-i\varphi_{r,\mathrm{A}}})]^{2}\\
&  =\left \vert b_{\mathrm{A}}\right \vert \left(  1+2\left \vert \alpha
\right \vert \, \cos{\left(  \frac{2\pi j}{3}+\varphi_{r,\mathrm{A}}\right)
}\right)  ^{2},\text{ }j=1,2,3.\nonumber
\end{align}

\subsubsection{Mass spectra for the Standard Model}

In above part, we have developed a theory of ${{\mathbb{Z}}_{N}}$ clock model
for generation-transition. In this part, we derive the mass spectra for
different elementary particles.

\paragraph{Mass spectra for charged leptons}

Firstly, we calculate the mass spectra of charged leptons. Now, we consider
the braiding processes by considering U-quark (the domain wall with smallest
mass) as an extra test object. So, we have $\mathrm{A}\rightarrow$ charged
lepton and $\mathrm{B}\rightarrow$ U-quark.

For charged leptons, we may consider the original quantum state $\left \vert
s\right \rangle $ to be one of the three degenerate states, $\left \vert
1\right \rangle $ ($...(\mathrm{abc})(\mathrm{abc})(\mathrm{abc})...$), or
$\left \vert 2\right \rangle $ ($...(\mathrm{bca})(\mathrm{bca})(\mathrm{bca}%
)...$), or $\left \vert 3\right \rangle $ ($...(\mathrm{cab})(\mathrm{cab}%
)(\mathrm{cab})...$). From $\left \vert \alpha \right \vert =\sqrt{\frac{N_{I}%
}{N_{F}}}=\frac{1}{\sqrt{2}}$ and $\varphi_{r,\mathrm{l}}=-\frac{2}{3}\pi
\frac{1}{3\pi}=-\frac{2}{9}$, we have%
\begin{align}
\hat{M}  &  =\left \vert b_{\mathrm{l}}\right \vert [%
%TCIMACRO{\dsum \limits_{i}}%
%BeginExpansion
{\displaystyle \sum \limits_{i}}
%EndExpansion
(1_{i}+\sqrt{2}e^{i\varphi_{r,\mathrm{l}}}\left \vert i\right \rangle
\left \langle i+1\right \vert +.h.c.)]^{2}\\
&  =2\left \vert b_{\mathrm{l}}\right \vert
%TCIMACRO{\dsum \limits_{i}}%
%BeginExpansion
{\displaystyle \sum \limits_{i}}
%EndExpansion
(1_{i}+\frac{1}{\sqrt{2}}e^{i\varphi_{r,\mathrm{l}}}\left \vert i\right \rangle
\left \langle i+1\right \vert \nonumber \\
&  +\frac{1}{2!}(\frac{1}{\sqrt{2}}e^{i\varphi_{r,\mathrm{l}}})(\frac{1}%
{\sqrt{2}}e^{i\varphi_{r,\mathrm{l}}})\left \vert i\right \rangle \left \langle
i+2\right \vert +.h.c.).\nonumber
\end{align}
After Fourier transformation to "momentum space", we derive the mass spectra
for charged leptons,
\begin{equation}
M_{\mathrm{l}}=\left \vert b_{\mathrm{l}}\right \vert \left(  1+\sqrt{2}%
\cos{\left(  \frac{2\pi j}{3}+\varphi_{r,\mathrm{l}}\right)  }\right)  ^{2}%
\end{equation}
where $j=1,2,3.$

If we set $\left \vert b_{\mathrm{A}}\right \vert =313.8~\mathrm{MeV,}$ the $e$,
$\mu,$ $\tau$\ masses are correct,
\begin{align}
m_{e}  &  =0.51\mathrm{MeV},\text{ }j=2,\\
m_{\mu}  &  =105.7\mathrm{MeV},\text{ }j=1,\nonumber \\
m_{\tau}  &  =1777~\mathrm{MeV,}\text{ }j=3.
\end{align}
This is consistent to the conjectural mass spectra from Koide
relation\cite{koide}
\begin{equation}
\frac{m_{e}+m_{\mu}+m_{\tau}}{(\sqrt{m_{e}}+\sqrt{m_{\mu}}+\sqrt{m_{\tau}%
})^{2}}=\frac{2}{3}.
\end{equation}

For the case of $\lambda^{\lbrack12]}>3,$ the mass spectra for charged leptons
are predicted by the following equation%
\begin{align}
M_{\mathrm{l}}  &  =\left \vert b_{\mathrm{l}}\right \vert [1+(\left \vert
\alpha \right \vert e^{i\frac{2\pi j}{N}+i\varphi_{r,\mathrm{l}}}+\left \vert
\alpha \right \vert e^{-i\frac{2\pi j}{N}-i\varphi_{r,\mathrm{l}}})\nonumber \\
&  +(\left \vert \alpha \right \vert ^{2}e^{i2\frac{2\pi j}{N}+2i\varphi
_{r,\mathrm{l}}}+\left \vert \alpha \right \vert ^{2}e^{-i2\frac{2\pi j}%
{N}-2i\varphi_{r,\mathrm{l}}})+...]^{2},\\
\text{ }j  &  =1,2,..\lambda^{\lbrack12]}\nonumber
\end{align}
where $\varphi_{r,\mathrm{l}}=-\frac{\lambda^{\lbrack12]}-1}{\lambda
^{\lbrack12]}}\pi \frac{1}{\lambda^{\lbrack gr]}}$ and $\left \vert
\alpha \right \vert =\sqrt{\frac{N_{I}}{N_{F}}}=\frac{1}{\sqrt{2}}$.

\paragraph{Mass spectra of neutrinos}

Secondly, we calculate the mass spectra of neutrino by consider the braiding
processes by considering a virtual U-quark as an extra test object.

According to above discussion, neutrino is considered as "vacuum" without
extra level-1 zero on its 1D lattice model of the level-1 relative zero. The
lattice number is $\lambda^{\lbrack gr]}+1+1$. An exra "$1$" comes from
rounding, the other extra "$1$" comes from a charged lepton induced by quantum
fluctuations. We will show its existence in next paragraph.

For neutrinos, we may consider the original quantum state $\left \vert
s\right \rangle $ to be one of the three degenerate states, $\left \vert
1\right \rangle $ ($...(\mathrm{abc})(\mathrm{abc})(\mathrm{abc})...$), or
$\left \vert 2\right \rangle $ ($...(\mathrm{bca})(\mathrm{bca})(\mathrm{bca}%
)...$), or $\left \vert 3\right \rangle $ ($...(\mathrm{cab})(\mathrm{cab}%
)(\mathrm{cab})...$).\ However, to calculate the transition from one
"\emph{vacuum}" to another, we must take into account the contribution from
"\emph{quantum tunneling effect}" by considering a \emph{virtual} U-quark
rather than a test U-quark.

In quantum physics, due to quantum tunneling effect, a ground state could turn
into another. We take quantum tunneling effect of topological degenerate
ground states for topological order as an example. There exist degenerate
ground states on a topological order under periodic boundary condition. When
an anyon moves across the whole system, one ground state turns into another.
The amplitude for the quantum tunneling effect is estimated to exponential
small on the size $L$ of the system, i.e., $\Delta \sim \exp(-L/\xi)$ where
$\xi$ is about correlation length of the system.

Using similar argument, the transition amplitude for the neutrino is about
$\exp(-\lambda^{\lbrack gr]}\kappa)$ where $\lambda^{\lbrack gr]}$ is the size
of the system and $\kappa$ is constant number. Let us calculate the value of
$\kappa$.

We firstly prepare a virtual U-quark at one end of the system and then
consider its motion to the other. Because there are two internal level-2
zeroes for a U-quark, we may assume that each zero could move freely in the
relative zero.

We then map the hopping system to a 1D SSH model, of which the lattice number
is $\lambda^{\lbrack gr]}+1+1$. See the illustration in Fig.45. The hopping
parameter between different unit cells is $1$ and the hopping parameters in
the unit cells is $1/3.$ The "$3$" comes from the three level-2 zeroes inside
a global level-1 zero. The moving level-2 zero can only choose $1$. The
probability is $1/3$ plays the role of hopping parameter. So, by mapping to
SSH model, the intra-cell and inter-cell hopping strengths are $t_{1}%
\rightarrow1/3$ and $t_{2}\rightarrow1$, respectively. In SSH, for a virtual
particle moving from one end to the other, the tunneling amplitude is
estimated about $\pm(\frac{t_{1}}{t_{2}})^{L}$ where $L$ is number of unit cells.

\begin{figure}[ptb]
\includegraphics[clip,width=0.92\textwidth]{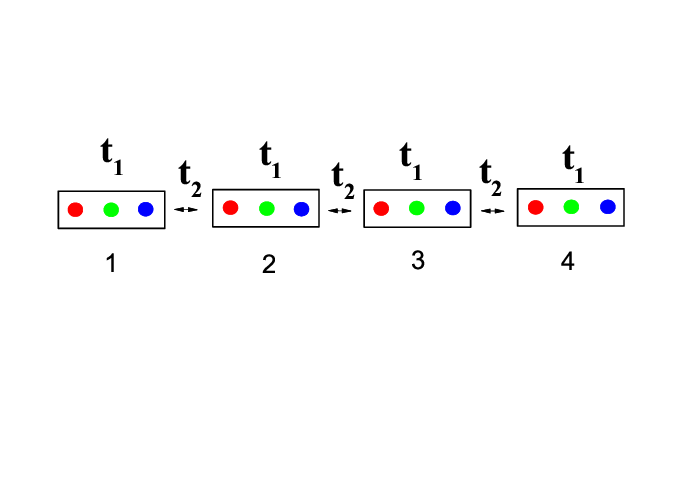}\caption{(Color
online) An illustration of effective 1D SSH model for the quantum tunneling
effect for neutrino}%
\end{figure}

Using the approach for quantum tunneling effect in SSH model, we get the
tunneling amplitude to be
\begin{equation}
3^{-(\lambda^{\lbrack gr]}-\operatorname{mod}\lambda^{\lbrack gr]}+1+1)}%
\end{equation}
for a 1 level-2 zero. As a result, for the two holes, the amplitude of quantum
tunneling effect is
\begin{equation}
3^{-2(\lambda^{\lbrack gr]}-\operatorname{mod}\lambda^{\lbrack gr]}+1+1)}.
\end{equation}

In addition, one can derive the result from an effective model for quantum
random walk -- when a level-2 zero moves a step, there are three
possibilities. After moving $L$ steps, the amplitude is reduced to $\frac
{1}{3^{L}}$ original value. Because we have two level-2 zeroes, the final
result is $\frac{1}{3^{L}}\times \frac{1}{3^{L}}=\frac{1}{3^{2L}}$.

Next, we consider the phase factor of the transition processes from quantum
tunneling effect. According to chiral para-statistics, the phase factor for
the transition process is $0$. However, when we consider quantum fluctuations,
the neutrino could turn into a charged lepton with $-\left \vert \mathrm{e}%
\right \vert $ electric charge by emitting a virtual $W^{\pm}$ particle. See
the illustration In Fig.46. This contributes an additional one lattice site.
Then, the generation-transition generates a phase factor from an extra charged
lepton $e^{-i\frac{2}{9}}$. Together with the changing of relative phase
factor $\frac{-\pi}{\lambda^{\lbrack gr]}+1}$ of the additional lattice from
the extra charged lepton, we have
\begin{equation}
\varphi_{r}=-\frac{2}{9}-\frac{\pi}{\lambda^{\lbrack gr]}+1}.
\end{equation}
In addition, we have $\left \vert \alpha \right \vert =\sqrt{\frac{N_{I}}{N_{F}}%
}=\frac{1}{\sqrt{2}}.$

\begin{figure}[ptb]
\includegraphics[clip,width=0.92\textwidth]{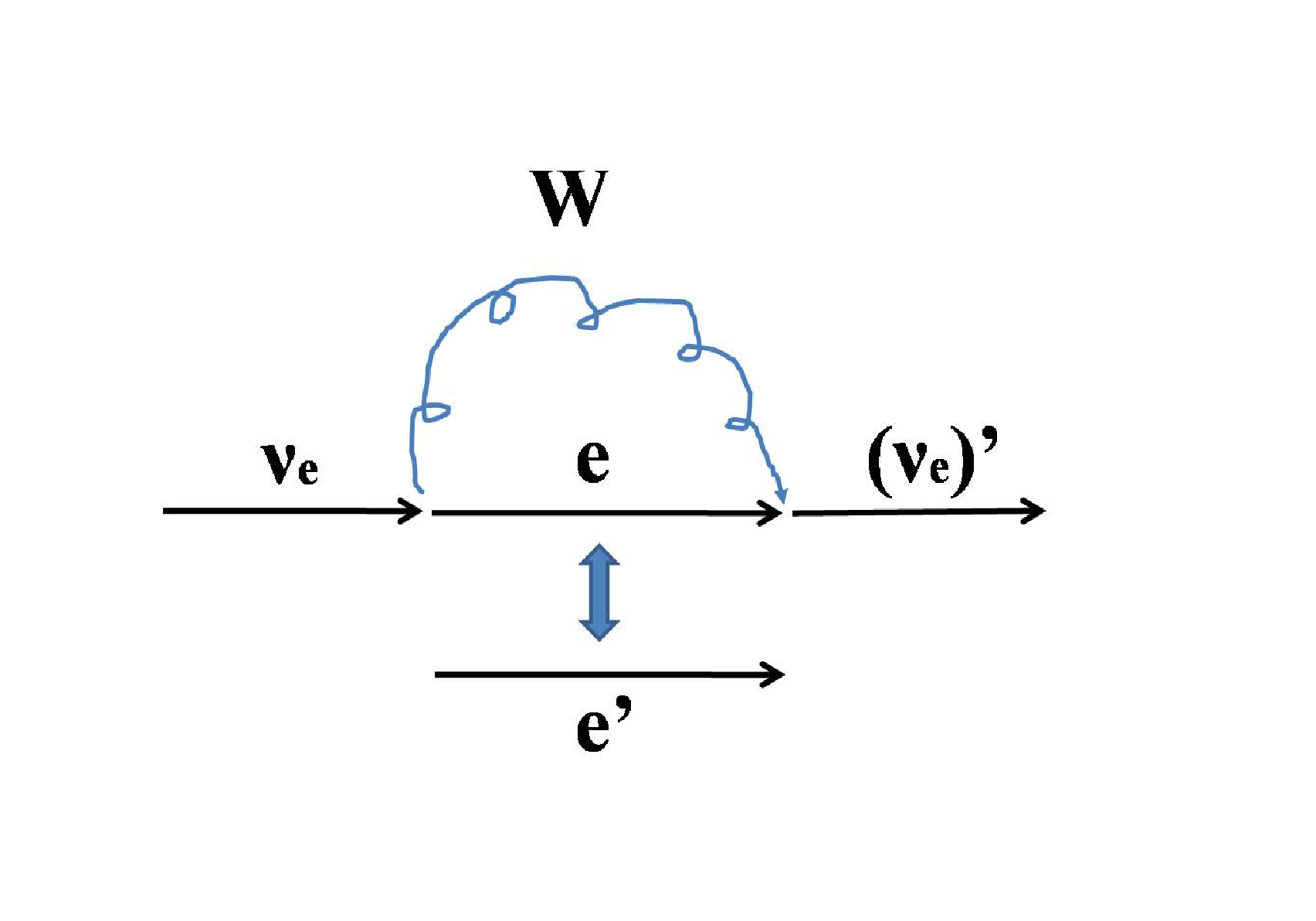}\caption{(Color online)
An illustration of effect of quantum fluctuations -- a neutrino $\nu$ turns
into an electron $\nu_{e}$ with$-\left \vert \mathrm{e}\right \vert $ electric
charge by emitting a virtual $W^{\pm}$ particle. After the process of
generation-transformation, it absorbs the virtual $W^{\pm}$ particle and
returns neutrino $\nu_{e}^{\prime}$ . }%
\end{figure}

As a result, the mass spectra for neutrinos are obtained
as\cite{Brannen:2010zz}
\begin{equation}
M_{\mathrm{n}}=\left \vert b_{\mathrm{n}}\right \vert \left(  1+\sqrt{2}%
\cos{\left(  \frac{2\pi j}{3}+\varphi_{r}\right)  }\right)  ^{2},\text{
}j=1,2,3
\end{equation}
where $\left \vert b_{\mathrm{n}}\right \vert =\frac{\left \vert b_{\mathrm{l}%
}\right \vert }{3^{2\lambda^{\lbrack gr]}-2\operatorname{mod}\lambda^{\lbrack
gr]}+4}}$ and $\varphi_{r}=-\frac{2}{9}-\frac{\pi}{\lambda^{\lbrack gr]}+1}$.
The masses of different neutrinos are%
\begin{align}
m_{\nu_{e}}  &  =0.01\mathrm{eV},\text{ }j=2,\\
m_{\nu_{\mu}}  &  =0.0005\mathrm{eV},\text{ }j=1,\nonumber \\
m_{\nu_{\tau}}  &  =0.05\mathrm{eV,}\text{ }j=3.
\end{align}
The neutrino masses are consistent to those from neutrino oscillation
measurements. In the limit of $L=(\lambda^{\lbrack gr]}-\operatorname{mod}%
\lambda^{\lbrack gr]}+1+1)\gg1$, the tunneling amplitude turns to zero,
$M_{\mathrm{n}}\rightarrow0$. This is the reason about small masses of neutrinos.

For the case of $\lambda^{\lbrack12]}>3,$ the mass spectra for neutrinos are
predicted by the following equation
\begin{align}
M_{\mathrm{n}}  &  =\left \vert b_{\mathrm{n}}\right \vert [1+(\left \vert
\alpha \right \vert e^{i\frac{2\pi j}{N}+i\varphi_{r,\mathrm{l}}}+\left \vert
\alpha \right \vert e^{-i\frac{2\pi j}{N}-i\varphi_{r,\mathrm{l}}})\nonumber \\
&  +(\left \vert \alpha \right \vert ^{2}e^{i2\frac{2\pi j}{N}+2i\varphi
_{r,\mathrm{l}}}+\left \vert \alpha \right \vert ^{2}e^{-i2\frac{2\pi j}%
{N}-2i\varphi_{r,\mathrm{l}}})+...]^{2},\text{ }\\
j  &  =1,2,..\lambda^{\lbrack12]}\nonumber
\end{align}
where $\left \vert b_{\mathrm{n}}\right \vert =\frac{\left \vert b_{\mathrm{l}%
}\right \vert }{(\lambda^{\lbrack12]})^{(2\lambda^{\lbrack gr]}%
-2\operatorname{mod}\lambda^{\lbrack gr]}+4)}},$ $\varphi_{r,\mathrm{l}%
}=-\frac{\lambda^{\lbrack12]}-1}{\lambda^{\lbrack12]}}\pi \frac{1}%
{\lambda^{\lbrack gr]}}-\frac{\pi}{\lambda^{\lbrack gr]}+1}$ and $\left \vert
\alpha \right \vert =\sqrt{\frac{N_{I}}{N_{F}}}=\frac{1}{\sqrt{2}}$.

\paragraph{Mass spectra for quarks}

Thirdly, we calculate the mass spectra of quarks.

Firstly, we consider mass spectra for U-quarks.

For a U-quark, we have $\left \vert \alpha \right \vert =\sqrt{\frac{N_{I}}%
{N_{F}}}=\sqrt{\frac{3}{4}}$ and $\varphi_{r,\mathrm{U}}=\frac{1}{3}\frac
{2}{9}=\frac{2}{27}.$ Thus, the mass spectra is obtained as
\begin{align}
M_{\mathrm{U}}  &  =\left \vert b_{\mathrm{U}}\right \vert \left(
1+2\sqrt{\frac{3}{4}}\cos{\left(  \frac{2\pi j}{3}+\varphi_{r,\mathrm{U}%
}\right)  }\right)  ^{2},\text{ }\\
j  &  =1,2,3.\nonumber
\end{align}
If we set $\left \vert b_{\mathrm{U}}\right \vert =22900\mathrm{MeV,}$
the\ masses of $u$, $c,$ $t$ quarks become
\begin{align}
m_{u}  &  =14.7\mathrm{MeV},\text{ }j=1\nonumber \\
m_{c}  &  =1.4\times10^{3}\mathrm{MeV},\text{ }j=2\nonumber \\
m_{t}  &  =1.7\times10^{5}\text{ }\mathrm{MeV,}\text{ }j=3.
\end{align}
In addition, we point out that $\left \vert \alpha \right \vert =\sqrt{\frac
{3}{4}}$ and $\varphi_{r,\mathrm{U}}=\frac{2}{27}$ are consistent to the
prediction of Koide formula\cite{z3}.

For the case of $\lambda^{\lbrack12]}>3,$ the mass spectra for U-quark with
$\lambda^{\lbrack12]}-1$ level-2 zeroes are predicted by the following
equation
\begin{align}
M_{\mathrm{U}}  &  =\left \vert b_{\mathrm{U}}\right \vert [1+(\left \vert
\alpha \right \vert e^{i\frac{2\pi j}{N}+i\varphi_{r,\mathrm{U}}}+\left \vert
\alpha \right \vert e^{-i\frac{2\pi j}{N}-i\varphi_{r,\mathrm{U}}})\nonumber \\
&  +(\left \vert \alpha \right \vert ^{2}e^{i2\frac{2\pi j}{N}+2i\varphi
_{r,\mathrm{l}}}+\left \vert \alpha \right \vert ^{2}e^{-i2\frac{2\pi j}%
{N}-2i\varphi_{r,\mathrm{l}}})+...]^{2},\text{ }\\
j  &  =1,2,..\lambda^{\lbrack12]}\nonumber
\end{align}
where $\varphi_{r,\mathrm{U}}=\frac{1}{2}(\frac{\lambda^{\lbrack12]}%
-1}{\lambda^{\lbrack12]}})^{2}\frac{\pi}{\lambda^{\lbrack gr]}}$ and
$\left \vert \alpha \right \vert =\sqrt{\frac{N_{I}}{N_{F}}}=\sqrt{\frac
{\lambda^{\lbrack12]}}{1+\lambda^{\lbrack12]}}}.$

Next, we consider mass spectra for D-quarks.

For a D-quark, we have $\left \vert \alpha \right \vert =\sqrt{\frac{N_{I}}%
{N_{F}}}=\sqrt{\frac{2}{3}}$ and $\varphi_{r,\mathrm{D}}=-\frac{2}{3}\frac
{2}{9}=-\frac{4}{27}.$ Thus, the mass spectra is obtained as
\begin{align}
M_{\mathrm{D}}  &  =\left \vert b_{\mathrm{D}}\right \vert \left(
1+2\sqrt{\frac{2}{3}}\cos{\left(  \frac{2\pi j}{3}+\varphi_{r,\mathrm{D}%
}\right)  }\right)  ^{2},\text{ }\nonumber \\
j  &  =1,2,3.
\end{align}
If we set $\left \vert b_{\mathrm{U}}\right \vert =627\mathrm{MeV,}$ the\ masses
of $d$, $s,$ $b$ quarks become
\begin{align}
m_{d}  &  =0.17\mathrm{MeV},\text{ }j=2,\\
m_{s}  &  =101\mathrm{MeV},\text{ }j=1,\nonumber \\
m_{b}  &  =4.3\times10^{3}\text{ }\mathrm{MeV,}\text{ }j=3.
\end{align}
In addition, we point out that $\left \vert \alpha \right \vert =\sqrt{\frac
{2}{3}}$ and $\varphi_{r,\mathrm{D}}=-\frac{4}{27}$ are consistent to the
prediction of Koide formula\cite{z3}.

We point out that the masses for $d,$ $u$ quarks are not so correct. This is
because that the mass of top quark is very large. Hence, one cannot reach too
many significant figures. In addition, the quark confinement may prevent
people from making precise measurements.

For the case of $\lambda^{\lbrack12]}>3,$ the mass spectra for D-quark with
$1$ level-2 zeroes are predicted by the following equation
\begin{align}
M_{\mathrm{D}}  &  =\left \vert b_{\mathrm{D}}\right \vert [1+(\left \vert
\alpha \right \vert e^{i\frac{2\pi j}{N}+i\varphi_{r,\mathrm{D}}}+\left \vert
\alpha \right \vert e^{-i\frac{2\pi j}{N}-i\varphi_{r,\mathrm{D}}})\nonumber \\
&  +(\left \vert \alpha \right \vert ^{2}e^{i2\frac{2\pi j}{N}+2i\varphi
_{r,\mathrm{l}}}+\left \vert \alpha \right \vert ^{2}e^{-i2\frac{2\pi j}%
{N}-2i\varphi_{r,\mathrm{l}}})+...]^{2},\text{ }\\
j  &  =1,2,..\lambda^{\lbrack12]}\nonumber
\end{align}
where $\varphi_{r,\mathrm{D}}=-\frac{\lambda^{\lbrack12]}-1}{\lambda
^{\lbrack12]}}\pi \frac{1}{\lambda^{\lbrack gr]}}+\frac{1}{2}(\frac
{\lambda^{\lbrack12]}-1}{\lambda^{\lbrack12]}})^{2}\frac{\pi}{\lambda^{\lbrack
gr]}}$ and $\left \vert \alpha \right \vert =\sqrt{\frac{2}{3}}.$

\subsubsection{Summary}

In this section, the physical mechanism of three generations in the SM is
explored. There exist very complex topological structure for elementary
particle -- \emph{chiral para-statistics}. The mass spectra of different
elementary particles are determined by the chiral para-statistics. According
to above discussion, with the help of three free parameters $\left \vert
b_{\mathrm{l}}\right \vert $, $\left \vert b_{\mathrm{U}}\right \vert ,$ and
$\left \vert b_{\mathrm{D}}\right \vert ,$ we derive the entire mass spectra. In
this end of this section, we point out that there exist additional
relationships between three parameters $\left \vert b_{\mathrm{l}}\right \vert
$, $\left \vert b_{\mathrm{U}}\right \vert $, $\left \vert b_{\mathrm{D}%
}\right \vert .$

An relationship is about $\left \vert b_{\mathrm{U}}\right \vert $ and
$\left \vert b_{\mathrm{D}}\right \vert $.

We consider an anomalous generation-transition between $c$, $b$, $t$ quarks.
For these three quarks, the mass spectra can be predicted by the following
equation:%
\begin{equation}
M=\left \vert b\right \vert \left(  1+2\sqrt{\frac{1}{2}}\cos{\left(  \frac{2\pi
j}{3}+\varphi_{r,\mathrm{U}}\right)  }\right)  ^{2},\text{ }j=1,2,3.
\end{equation}
where $\varphi_{r,\mathrm{U}}=\frac{2}{27}$ and $\left \vert b\right \vert
=29740\mathrm{MeV}$. Why?

We then give an explanation. According to chiral para-statistics, we cannot
explain this transition process mixing elementary particles with different
electric charge. However, when we consider quantum fluctuations, the b-quark
could turn into a virtual U-quark with $\frac{2}{3}$ electric charge by
emitting a virtual $W^{\pm}$ particle. Then, the generation-transition between
$c$, $b$, $t$ quarks, is no more forbidden. After the process of
generation-transition between them finishes, the virtual U-quark absorbs a
$W^{\pm}$ particle and returns the original type of elementary particle.
Therefore, we have relative phase factor $\varphi_{r,\mathrm{U}}=\frac{2}%
{27}.$ $\sqrt{\frac{1}{2}}$\ indicates the ratio between the number of initial
degenerate states $N_{I}$ and the final degenerate states $N_{F}$ is $1/2.$ As
a result, there are $6$ final states by considering the differences between
the true U-quarks and the virtual U-quark.

With the help of above relationship, when $\left \vert b_{\mathrm{U}%
}\right \vert $ is known, $\left \vert b_{\mathrm{D}}\right \vert $ could be
obtained simultaneously.

In addition, according to above discussion, we found an interesting
relationship between $\left \vert b_{\mathrm{l}}\right \vert $ and $\left \vert
b_{\mathrm{D}}\right \vert ,$ i.e.,%
\begin{equation}
\left \vert b_{\mathrm{D}}\right \vert =627\mathrm{MeV}\simeq2\left \vert
b_{\mathrm{l}}\right \vert =2\times313.8\mathrm{MeV}.
\end{equation}
Why? Although we don't understand the physical mechanism for this equation
($\left \vert b_{\mathrm{D}}\right \vert \simeq2\left \vert b_{\mathrm{l}%
}\right \vert $), we believe that $\left \vert b_{\mathrm{D}}\right \vert $ is
twice of $2\left \vert b_{\mathrm{l}}\right \vert ,$ i.e., $\left \vert
b_{\mathrm{D}}\right \vert \equiv2\left \vert b_{\mathrm{l}}\right \vert .$ With
the help of this equation ($\left \vert b_{\mathrm{D}}\right \vert
\equiv2\left \vert b_{\mathrm{l}}\right \vert $), when $\left \vert
b_{\mathrm{D}}\right \vert $ is known, $\left \vert b_{\mathrm{l}}\right \vert $
could be obtained simultaneously.

In the end, we draw the conclusion. For mass spectra, there is only two free
parameters -- one is the global coefficient for the whole mass spectra (or the
condensation of Higgs field), the other is the ratio $\lambda^{\lbrack gr]}$
between the changing rate of global sub-variant and that of relative
sub-variant. In addition, in the following part, we will show that the global
coefficient for the whole mass spectra (or the condensation of Higgs field) is
not a free parameter. Instead, it is determined by $\lambda^{\lbrack gr]}$. In
summary, in the Standard model, $\lambda^{\lbrack gr]}$ is the \emph{unique}
free parameter. In other words, if we set $\lambda^{\lbrack gr]}=3\pi,$ the
whole structure including all parameters are fixed!\begin{figure}[ptb]
\includegraphics[clip,width=0.7\textwidth]{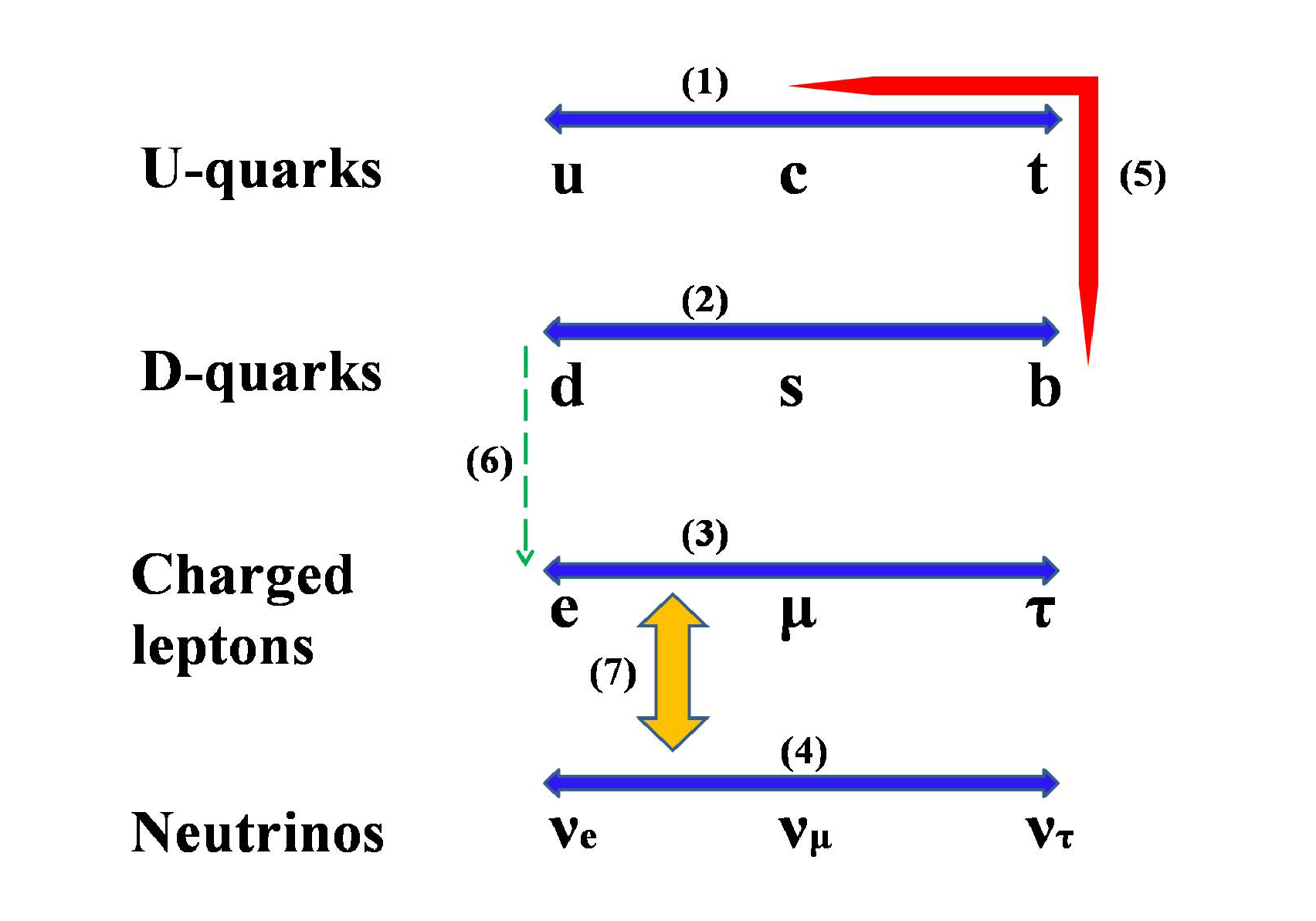}\caption{The approach to
obtain the entire mass spectra: Blue double arrow (1) denotes a generalized
Koide relation determines mass spectra of U-quarks; blue double arrow (2)
denotes a generalized Koide relation determines mass spectra of D-quarks; blue
double arrow (3) denotes a Koide relation determines mass spectra of charged
leptons; blue double arrow (4) denotes a generalized Koide relation determines
mass spectra of neutrinos; blue double arrow (5) denotes a generalized Koide
relation for mixed types of elementary particles determines the relationship
between $|b_{D}|$ and $|b_{U}|$; the green dotted line determines the
relationship between $|b_{D}|$ and $\left \vert b_{\mathrm{l}}\right \vert $
($\left \vert b_{\mathrm{D}}\right \vert \equiv2\left \vert b_{\mathrm{l}%
}\right \vert $); the orange double arrow determines the relationship between
$\left \vert b_{\mathrm{n}}\right \vert $ and $\left \vert b_{\mathrm{l}%
}\right \vert $ ($\left \vert b_{\mathrm{n}}\right \vert =\frac{\left \vert
b_{\mathrm{l}}\right \vert }{3^{2\lambda^{\lbrack gr]}-2\operatorname{mod}%
\lambda^{\lbrack gr]}+4}}$).}%
\end{figure}

In Fig.47, we show the approach to obtain the entire mass spectra: Blue double
arrow (1) denotes a generalized Koide relation determines mass spectra of
U-quarks; blue double arrow (2) denotes a generalized Koide relation
determines mass spectra of D-quarks; blue double arrow (3) denotes a Koide
relation determines mass spectra of charged leptons; blue double arrow (4)
denotes a generalized Koide relation determines mass spectra of neutrinos;
blue double arrow (5) denotes a generalized Koide relation for mixed types of
elementary particles determines the relationship between $|b_{D}|$ and
$|b_{U}|$; the green dotted line determines the relationship between $|b_{D}|$
and $\left \vert b_{\mathrm{l}}\right \vert $ ($\left \vert b_{\mathrm{D}%
}\right \vert \equiv2\left \vert b_{\mathrm{l}}\right \vert $); the orange double
arrow determines the relationship between $\left \vert b_{\mathrm{n}%
}\right \vert $ and $\left \vert b_{\mathrm{l}}\right \vert $ ($\left \vert
b_{\mathrm{n}}\right \vert =\frac{\left \vert b_{\mathrm{l}}\right \vert
}{3^{2\lambda^{\lbrack gr]}-2\operatorname{mod}\lambda^{\lbrack gr]}+4}}$).

\subsection{Sector of QED$\mathrm{\times}$QCD}

In this section, we discuss the sector of QED$\mathrm{\times}$QCD.

Now, we have a ($d+1$)-dimensional 2-nd order\textit{ }$\mathrm{\tilde
{S}\tilde{O}}$\textrm{(d+1) }physical chiral variant $V_{\mathrm{\tilde{U}%
}^{[2]}\mathrm{(1)},\mathrm{\tilde{S}\tilde{O}}^{[1]}\mathrm{(3+1)}%
,3+1}^{[2],\mathrm{chiral}}=\{V_{g,\mathrm{\tilde{U}}^{[2]}\mathrm{(1)}%
,\mathrm{\tilde{S}\tilde{O}}^{[1]}\mathrm{(3+1)},3+1}^{[2],\mathrm{sub}%
},V_{r,\mathrm{\tilde{S}\tilde{O}}^{[1]}\mathrm{(3)},3}^{\mathrm{sub}}\}.$ In
this part, we focus on $V_{g,\mathrm{\tilde{U}}^{[2]}\mathrm{(1)}%
,\mathrm{\tilde{S}\tilde{O}}^{[1]}\mathrm{(3+1)},3+1}^{[2],\mathrm{sub}}$ that
is a higher-order mapping between\textit{ }$\mathrm{C}_{\mathrm{\tilde{U}%
}^{[2]}\mathrm{(1)}}^{[2]}$\textit{, }$\mathrm{\tilde{S}\tilde{O}}%
$\textrm{(d+1)} Clifford group-changing space\textit{ }$\mathrm{C}%
_{g,\mathrm{\tilde{S}\tilde{O}(d+1)},d+1}^{[1]}$\textit{\ }and a rigid
spacetime\textit{ }$\mathrm{C}_{d+1}$\textit{.} The 2-nd order variability is%
\begin{equation}
\mathcal{T}(\delta x^{\mu})\leftrightarrow \hat{U}^{[1]}(\delta \phi_{g}%
^{[1]\mu})=\exp(i(T^{\mu}\delta \phi_{g}^{[1]\mu}))
\end{equation}
and%
\begin{equation}
\hat{U}^{[1]}(\delta \phi_{g,\mathrm{global}}^{[1]})\leftrightarrow \hat
{U}^{[2]}(\delta \phi^{\lbrack2]})=\exp(i\lambda^{\lbrack12]}\delta
\phi_{g,\mathrm{global}}^{[1]}).
\end{equation}
In particular, the changing rate $\lambda^{\lbrack12]}=\left \vert \frac
{\delta \phi^{\lbrack2]}}{\delta \phi_{g,\mathrm{global}}^{[1]}}\right \vert $ is
3. The corresponding 2-nd order variability is reduced into \textrm{U}%
$_{\mathrm{em}}$\textrm{(1)} gauge symmetry and $\mathrm{SU(3)}$ non-Abelian
gauge symmetry. The effective model is just QED$\mathrm{\times}$QCD.

\subsubsection{Matter}

We discuss the matter for global sub-variant $V_{g,\mathrm{\tilde{U}}%
^{[2]}\mathrm{(1)},\mathrm{\tilde{S}\tilde{O}}^{[1]}\mathrm{(3+1)}%
,3+1}^{[2],\mathrm{sub}}$ with changing rate $\lambda^{\lbrack12]}=\left \vert
\frac{\delta \phi^{\lbrack2]}}{\delta \phi_{g,\mathrm{global}}^{[1]}}\right \vert
.$ Matter corresponds to globally expand or contract the two group-changing
spaces $\mathrm{C}_{g,\mathrm{\tilde{S}\tilde{O}}^{[1]}\mathrm{(d+1)}%
,d+1}^{[1]}$ or $\mathrm{C}_{\mathrm{U(1)}^{[2]},1}^{[2]}$ with changing their
corresponding sizes. Therefore, an object is classified by two integer numbers
$\{n_{g}^{[1]},n^{[2]}\}$, the number of level-1 zeroes $n_{g}^{[1]}$ of
$\mathrm{C}_{g,\mathrm{\tilde{S}\tilde{O}}^{[1]}\mathrm{(d+1)},d+1}^{[1]}$ and
that of level-2 zeroes $n^{[2]}$, respectively. $n^{[2]}$ determines types of
elementary particles. For the case $\lambda^{\lbrack12]}=3$, there are four
types of elementary particles. After considering the three generations, there
are $3\times4=12$ Dirac types of elementary particles.

\subsubsection{Motion}

In this section, we discuss the motion. There are two types of motions -- one
is about the motions for elementary particles, the other is about the
collective motions of the two levels of zero lattices.

\paragraph{Local gauge symmetries}

Firstly, we discuss the quantum states for the level-2 group-changing elements
of a global level-1 zero.

Under compactification, the level-2 group-changing space turns into level-2
zero lattice. On each site of the level-2 zero lattice (we denote it by
$I^{[2]}$), there is a local field of compact U(1) symmetry. This\ compact
U(1) symmetry leads to the \textrm{U}$_{\mathrm{em}}$\textrm{(1)} gauge
symmetry for global motion of level-2 zeroes of a level-1 zero.

The motion of level-2 group-changing space comes from its local expansion and
contraction on different global level-1 zeroes. Because there are total
$\lambda^{\lbrack12]}$ lattice sites for level-2 zeroes of a global level-1
zero, we have $\lambda^{\lbrack12]}$\ level-2 phase factors $\delta
\varphi_{I^{[2]},I^{[1]}}^{[2]}$ for a level-2 group-changing element of a
global level-1 zero. We define a global level-2 phase factor and
$\lambda^{\lbrack12]}-1$ relative phase factors. The number $\lambda
^{\lbrack12]}-1$ corresponds to the number of Carton-Weyl base for
\textrm{SU(}$\lambda^{\lbrack12]}$\textrm{)} group. We denote its quantum
states by a matrix, i.e.,
\begin{equation}
\left(
\begin{array}
[c]{c}%
\left \vert \psi_{1^{[2]},I_{g}^{[1]}}^{[2]}\right \rangle \\
\left \vert \psi_{2^{[2]},I_{g}^{[1]}}^{[2]}\right \rangle \\
...\\
\left \vert \psi_{(\lambda^{\lbrack12]})^{[2]},I_{g}^{[1]}}^{[2]}\right \rangle
\end{array}
\right)  .
\end{equation}
Here, $\left \vert \psi_{I^{[2]},I_{g}^{[1]}}^{[2]}\right \rangle $ denotes the
state of the level-2 zero on the $I^{[2]}$-th lattice site of level-2 zero
lattice inside $I_{g}^{[1]}$-th global level-1 zero. Because there are
$\lambda^{\lbrack12]}$ lattice sites, the quantum states of $\left \vert
\psi_{I^{[2]},I_{g}^{[1]}}^{[2]}\right \rangle $ have $\lambda^{\lbrack12]}$ components.

For QCD, we have $\lambda^{\lbrack12]}=3$. There are three sites of level-2
zero lattice for an elementary particle that is labeled by $1$, $2$, $3.$ For
the case of $n^{[2]}=1$, there exists an level-2 zero inside the elementary
particle (or $d$-quark). Now, D-quark has three internal quantum states
described by, $\left(
\begin{array}
[c]{c}%
d_{1}\left \vert \mathrm{vac}\right \rangle \\
d_{2}\left \vert \mathrm{vac}\right \rangle \\
d_{3}\left \vert \mathrm{vac}\right \rangle
\end{array}
\right)  .$ For the case of $n^{[2]}=2$, there exists two level-2 zeroes
inside the elementary particle (or U-quark). Now, there exist a "hole" of the
level-2 zero. $u$-quark also has three internal quantum states described by
$\left(
\begin{array}
[c]{c}%
u_{1}\left \vert \mathrm{vac}\right \rangle \\
u_{2}\left \vert \mathrm{vac}\right \rangle \\
u_{3}\left \vert \mathrm{vac}\right \rangle
\end{array}
\right)  .$

In addition, the quantum states of the level-2 zero on different sites of the
level-2 zero-lattice are orthogonal, i.e.,%
\begin{equation}
\langle \psi_{J^{[2]},I_{g}^{[1]}}^{[2]}\left \vert \psi_{I^{[2]},I_{g}^{[1]}%
}^{[2]}\right \rangle =\delta_{J^{[2]}I_{g}^{[2]}}.
\end{equation}
Hence, $\left \vert \psi_{1^{[2]},I_{g}^{[1]}}^{[2]}\right \rangle ,$
$\left \vert \psi_{2^{[2]},I_{g}^{[1]}}^{[2]}\right \rangle ,$ ..., $\left \vert
\psi_{\lambda^{\lbrack12]},I_{g}^{[1]}}^{[2]}\right \rangle $ make up a
complete basis. The symmetry of different basis leads to non-Abelian,
$\mathrm{SU(\lambda^{\lbrack12]})}$ gauge symmetry for relative motion of
level-2 zeroes of a level-1 zero.

As a result, we have two types of local gauge symmetries, one is Abelian,
\textrm{U}$_{\mathrm{em}}$\textrm{(1)} gauge symmetry for global motion of
level-2 zeroes of a global level-1 zero, the other is non-Abelian,
$\mathrm{SU(\lambda^{\lbrack12]})}$ gauge symmetry for relative motion of
level-2 zeroes of a global level-1 zero. This leads to the underlying
mechanism of the \textrm{SU}$_{\mathrm{C}}$\textrm{(N)}$\times$\textrm{U}%
$_{\mathrm{em}}$\textrm{(1)} gauge symmetry from indistinguishable internal
state inside the elementary particles.

\paragraph{\textrm{U}$_{\mathrm{em}}$\textrm{(1)}$\times$\textrm{SU(N)} gauge
field}

Firstly, we discuss the \textrm{U}$_{\mathrm{em}}$\textrm{(1)} gauge field.

\textrm{U}$_{\mathrm{em}}$\textrm{(1)} gauge field characterizes the global
collective motion of level-2 zeroes. As a result, we introduce the vector
field $\mathrm{e}A_{I,I^{\prime}}=\varphi_{I}^{[2]}-\varphi_{I^{\prime}}%
^{[2]}$ that play the role of \textrm{U}$_{\mathrm{em}}$\textrm{(1)} gauge
field and the loop currents that plays the role of strength of gauge fields.

A simple approach is to consider the current around minimum loops that consist
of four links for four nearest neighbor lattice sites $\Phi_{\left \langle
IJKL\right \rangle }^{[2]}$ on the plaquette of $IJKL$ lattice site, i.e.,
$\Phi_{\left \langle IJKL\right \rangle }^{[2]}=\mathrm{e}A_{IJ}-\mathrm{e}%
A_{KL}=-\varphi_{I}^{[2]}+\varphi_{J}^{[2]}+\varphi_{K}^{[2]}-\varphi
_{L}^{[2]}.$ The quantum states for level-2 group-changing space on Cartesian
space are defined by $\{ \Phi_{\left \langle IJKL\right \rangle }^{[2]},$
$\left \langle IJKL\right \rangle \in \mathrm{All}\}.$ In continuum limit, we
have $\hat{U}_{I,\mathrm{U}_{\mathrm{em}}\mathrm{(1)}}(t)\rightarrow \hat
{U}_{\mathrm{U}_{\mathrm{em}}\mathrm{(1)}}(x,t)$, $A_{I,I^{\prime}}\rightarrow
A_{\mu}(x).$ $\Phi_{\left \langle IJKL\right \rangle }^{[2]}(I)$ is reduced to
the strength of gauge field $F_{\mu \nu},$%
\begin{equation}
\Phi_{\left \langle IJKL\right \rangle }^{[2]}\sim F_{\mu \nu}.
\end{equation}

Next, we discuss the \textrm{SU}$_{\mathrm{C}}$\textrm{(N)} gauge field.

The \textrm{SU}$_{\mathrm{C}}$\textrm{(N)} gauge field comes from the relative
collective motion of level-2 zero lattice. We then introduce the vector field
$\mathcal{A}_{I,I^{\prime}}=%
%TCIMACRO{\dsum \limits_{a}}%
%BeginExpansion
{\displaystyle \sum \limits_{a}}
%EndExpansion
A_{I,I^{\prime}}T^{a}$ of $\lambda^{\lbrack12]}\times \lambda^{\lbrack12]}$
matrix that plays the role of \textrm{SU}$_{\mathrm{C}}$\textrm{(N)} gauge
field. Here, $T^{a}$ is generate of \textrm{SU}$_{\mathrm{C}}$\textrm{(N)}
group along a-th direction. To characterize the relative collective motion of
level-2 group-changing space, we introduce colored loop currents
$\Phi_{\left \langle IJKL\right \rangle }^{[2]}$ on level-1 zero lattice.

A simple approach is to consider the colored loop currents along minimum loops
for four nearest neighbor lattice sites. We denote them by colored flux
$\Phi_{\left \langle IJKL\right \rangle }^{[2]}$ on the plaquette of $IJKL$
lattice site, i.e.,%
\begin{align}
\Phi_{\left \langle IJKL\right \rangle }^{[2]}  &  =%
%TCIMACRO{\dsum \limits_{a}}%
%BeginExpansion
{\displaystyle \sum \limits_{a}}
%EndExpansion
(\Phi_{\left \langle IJKL\right \rangle }^{a,[2]}T^{a})=g\mathcal{A}%
_{IJ}-g\mathcal{A}_{KL}\nonumber \\
&  =%
%TCIMACRO{\dsum \limits_{a}}%
%BeginExpansion
{\displaystyle \sum \limits_{a}}
%EndExpansion
(-\Delta \varphi_{I}^{a,[2]}+\Delta \varphi_{J}^{a,[2]}\nonumber \\
&  +\Delta \varphi_{K}^{a,[2]}-\Delta \varphi_{L}^{a,[2]})T^{a}%
\end{align}
In long wave limit, we have $\hat{U}_{J,\mathrm{SU_{\mathrm{C}}(N)}%
}(t)\rightarrow U_{\mathrm{SU_{\mathrm{C}}(N)}}(\vec{x},t),$ $\mathcal{A}%
_{I,I^{\prime}}\rightarrow \mathcal{A}_{\mu}(x)\ $and $\mathcal{\Phi
}_{\left \langle IJKL\right \rangle }^{[2]}\rightarrow \mathcal{G}_{\mu \nu}(x)=%
%TCIMACRO{\dsum \limits_{a}}%
%BeginExpansion
{\displaystyle \sum \limits_{a}}
%EndExpansion
(T^{a}F_{\mu \nu}^{a}).$

\paragraph{Effective model}

In this section, we review the effective model for the quantum gauge theory
with local \textrm{U}$_{\mathrm{em}}$\textrm{(1)}$\times$\textrm{SU}%
$_{\mathrm{C}}$\textrm{(}$\mathrm{N}$\textrm{)} symmetry.

In long wave limit, the effective Lagrangian $\mathcal{L}_{\mathrm{EM}}$ for
QED is written as
\begin{align}
\mathcal{L}_{\mathrm{EM}}  &  ={\bar{\Psi}}i\gamma^{\mu}\partial_{\mu}%
\Psi+m{\bar{\Psi}}\Psi \nonumber \\
&  +{A}_{\mu}{j}_{(em)}^{\mu}-\frac{1}{4}F_{\mu \nu}F^{\mu \nu}%
\end{align}
where ${j}_{(em)}^{\mu}=i\mathrm{e}{\bar{\Psi}}\gamma^{\mu}\Psi$.

For the elementary particles with extra level-2 zeroes, they may have
\emph{fractional }electric charge. The electric charge of the elementary
particle changes from $1$ to\ $\frac{n^{[2]}}{\lambda^{\lbrack12]}}$. For
example, for the case of $\lambda^{\lbrack12]}=3,$ and $n^{[2]}=1,$\ with only
one level-2 zero, the elementary particle has $1/3$ electric charge.

The Lagrangian of Non-Abelian gauge field is obtained as
\begin{equation}
\mathcal{L}_{\mathrm{YM}}(\mathrm{SU_{\mathrm{C}}(N)})=-\frac{1}{2}%
\mathrm{Tr}\left(  \mathcal{G}_{\mu \nu}\mathcal{G}^{\mu \nu}\right)
+\mathrm{Tr}\left(  J_{\mathrm{YM}}^{\mu}\mathcal{A}_{\mu}\right)
\end{equation}
where $J_{\mathrm{YM}}^{\mu}=ig\bar{\Psi}\gamma^{\mu}\Psi.$ The gauge strength
is defined by $\mathcal{G}_{\mu \nu}$ as%
\begin{equation}
\mathcal{G}_{\mu \nu}=\partial_{\mu}\mathcal{A}_{\nu}-\partial_{\nu}%
\mathcal{A}_{\mu}-ig\left[  \mathcal{A}_{\mu},\mathcal{A}_{\nu}\right]
\end{equation}
or
\begin{align}
\mathcal{G}_{\mu \nu}  &  =G_{\mu \nu}^{a}t^{a}\text{, }\nonumber \\
G_{\mu \nu}^{a}  &  =\partial_{\mu}A_{\nu}^{a}-\partial_{\nu}A_{\mu}%
^{a}+gf^{abc}A_{\mu}^{b}A_{\nu}^{c}.
\end{align}

An extra level-2 zero plays the role of source of $\mathrm{SU_{\mathrm{C}}%
(N)}$ gauge field and then carries color degree of freedom. There are
$n^{[2]}$ level-2 zeroes that determine the different quantum internal states
of an elementary particle. For the case of $\mathrm{SU_{\mathrm{C}}(3)}$,
there are three colors called red, blue and green that correspond to the three
positions $1$, $2$, $3$, of level-2 zero lattice. The collective modes of
$\mathrm{SU_{\mathrm{C}}(3)}$ gauge field are always called gluons that
correspond to the fluctuations of the level-2 zeroes. Therefore, two colored
particles interact by shaking their level-2 zero lattices.

\subsubsection{Summary}

In this section, we found that a 2-nd order physical variant is projected to
two coupled zero-lattices and naturally gives rise to gauge bosons (such as
photons and gluons) and Dirac fermions (such as electrons and quarks). Gauge
bosons are vibrations of level-2 zero-lattice, while Dirac fermions are extra
level-1 zeroes. The effective Lagrangian for QED$\mathrm{\times}$QCD is
obtained as
\begin{equation}
\mathcal{L}=\mathcal{L}_{\mathrm{fermion}}+\mathcal{L}_{\mathrm{EM}%
}(\mathrm{U_{\mathrm{em}}(1)})+\mathcal{L}_{\mathrm{YM}}%
(\mathrm{SU_{\mathrm{C}}(N)}).
\end{equation}

The 2-nd order variability is reduced into \textrm{U}$_{\mathrm{em}}%
$\textrm{(1)} local gauge symmetry and $\mathrm{SU_{\mathrm{C}}(N)}$
non-Abelian gauge symmetry. There are two types of the collective motions of
the level-2 zero lattice, one is about the global motion of the level-2 zeroes
inside a level-1 zero, the other is about relative motion of the level-2
zeroes inside a level-1 zero. The global motion of the level-2 zeroes
corresponds to the fluctuations of quantum fields of \textrm{U}$_{\mathrm{em}%
}$\textrm{(1)} gauge symmetry, and the relative motion of the level-2 zeroes
corresponds to the quantum fields of \textrm{SU}$_{\mathrm{C}}$\textrm{(}%
$\mathrm{N}$\textrm{). }

For the case of $n=3$, we have an effective $\mathrm{SU_{\mathrm{C}}%
(3)}\otimes$\textrm{U}$_{\mathrm{em}}$\textrm{(1)} gauge theory, of which the
the Lagrangian is
\begin{align}
\mathcal{L}  &  ={\bar{e}}(x)i\gamma^{\mu}\partial_{\mu}e(x)+{\bar{d}%
}(x)i\gamma^{\mu}\partial_{\mu}d(x)+{\bar{u}}(x)i\gamma^{\mu}\partial_{\mu
}u(x)\\
&  +m_{e}{\bar{e}}(x)e(x)+m_{d}{\bar{d}}(x)d(x)+m_{u}{\bar{u}}%
(x)u(x)\nonumber \\
&  -\frac{1}{4}F_{\mu \nu}F^{\mu \nu}+\mathrm{e}{A}_{\mu}(x){j}_{(em)}^{\mu
}(x)\nonumber \\
&  -\frac{1}{2}\mathrm{Tr}\mathcal{G}_{\mu \nu}\mathcal{G}^{\mu \nu}%
+\mathrm{Tr}J_{\mathrm{YM}}^{\mu}\mathcal{A}_{\mu}\nonumber
\end{align}
where $m_{e/d/u}$ are masses for u-quark, d-quark and electron, respectively.
The electric current is
\begin{align}
{j}_{(em)}^{\mu}(x)  &  =-i{\bar{e}}(x)\gamma^{\mu}e(x)+i\frac{2}{3}{\bar{u}%
}(x)\gamma^{\mu}u(x)\\
&  -i\frac{1}{3}{\bar{d}}(x)\gamma^{\mu}d(x).\nonumber
\end{align}
The $\mathrm{SU_{\mathrm{C}}(3)}$ color current for strong interaction is
\begin{equation}
J_{\mathrm{YM}}^{a,\mu}={\bar{u}}(x)i\gamma^{\mu}T^{a}u(x)+{\bar{d}}%
(x)i\gamma^{\mu}T^{a}d(x).
\end{equation}

In above paragraphs, we had considered a 2-nd order physical variant with
symmetric chirality. So, there doesn't exist neutrino and we have only one
generation. On the contrary, when we consider a 2-nd order physical variant
$V_{\mathrm{\tilde{U}}^{[2]}\mathrm{(1)},\mathrm{\tilde{S}\tilde{O}}%
^{[1]}\mathrm{(d+1)},d+1}^{[2],\mathrm{chiral}}$ without symmetric chirality
($\lambda^{\lbrack12]}=3,$\ $\lambda^{\lbrack gr]}=3\pi$), there exists
neutrino and we have three generations. Now, the we have an effective
$\mathrm{SU_{\mathrm{C}}(3)}\otimes$\textrm{U}$_{\mathrm{em}}$\textrm{(1)}
gauge theory, of which the the Lagrangian is
\begin{align}
\mathcal{L}  &  =%
%TCIMACRO{\dsum \limits_{i=1}^{3}}%
%BeginExpansion
{\displaystyle \sum \limits_{i=1}^{3}}
%EndExpansion
{\bar{l}}_{i}(x)i\gamma^{\mu}\partial_{\mu}l_{i}(x)+%
%TCIMACRO{\dsum \limits_{i=1}^{3}}%
%BeginExpansion
{\displaystyle \sum \limits_{i=1}^{3}}
%EndExpansion
{\bar{D}}_{i}(x)i\gamma^{\mu}\partial_{\mu}D_{i}(x)\nonumber \\
&  +%
%TCIMACRO{\dsum \limits_{i=1}^{3}}%
%BeginExpansion
{\displaystyle \sum \limits_{i=1}^{3}}
%EndExpansion
{\bar{U}}_{i}(x)i\gamma^{\mu}\partial_{\mu}U_{i}(x)+%
%TCIMACRO{\dsum \limits_{i=1}^{3}}%
%BeginExpansion
{\displaystyle \sum \limits_{i=1}^{3}}
%EndExpansion
{\bar{n}}_{i}(x)i\sigma^{\mu}\partial_{\mu}n_{i}(x)\\
&  +%
%TCIMACRO{\dsum \limits_{i=1}^{3}}%
%BeginExpansion
{\displaystyle \sum \limits_{i=1}^{3}}
%EndExpansion
m_{i}^{l}{\bar{l}}_{i}(x)l_{i}(x)+%
%TCIMACRO{\dsum \limits_{i=1}^{3}}%
%BeginExpansion
{\displaystyle \sum \limits_{i=1}^{3}}
%EndExpansion
m_{i}^{D}{\bar{D}}_{i}(x)D_{i}(x)+%
%TCIMACRO{\dsum \limits_{i=1}^{3}}%
%BeginExpansion
{\displaystyle \sum \limits_{i=1}^{3}}
%EndExpansion
m_{i}^{U}{\bar{U}}_{i}(x)U_{i}(x)\nonumber \\
&  -\frac{1}{4}F_{\mu \nu}F^{\mu \nu}+\mathrm{e}{A}_{\mu}(x){j}_{(em)}^{\mu
}(x)\nonumber \\
&  -\frac{1}{2}\mathrm{Tr}\mathcal{G}_{\mu \nu}\mathcal{G}^{\mu \nu}%
+\mathrm{Tr}J_{\mathrm{YM}}^{\mu}\mathcal{A}_{\mu}\nonumber
\end{align}
where $m_{i}^{U/D/l}$ are masses for U-quarks, D-quarks and charged leptons,
respectively. The electric current is
\begin{align}
{j}_{(em)}^{\mu}(x)  &  =-%
%TCIMACRO{\dsum \limits_{i=1}^{3}}%
%BeginExpansion
{\displaystyle \sum \limits_{i=1}^{3}}
%EndExpansion
{\bar{l}}_{i}(x)i\gamma^{\mu}l_{i}(x)-\frac{1}{3}%
%TCIMACRO{\dsum \limits_{i=1}^{3}}%
%BeginExpansion
{\displaystyle \sum \limits_{i=1}^{3}}
%EndExpansion
{\bar{D}}_{i}(x)i\gamma^{\mu}D_{i}(x)\\
&  +\frac{2}{3}%
%TCIMACRO{\dsum \limits_{i=1}^{3}}%
%BeginExpansion
{\displaystyle \sum \limits_{i=1}^{3}}
%EndExpansion
{\bar{U}}_{i}(x)i\gamma^{\mu}U_{i}(x).\nonumber
\end{align}
The $\mathrm{SU_{\mathrm{C}}(3)}$ color current for strong interaction is
\begin{equation}
J_{\mathrm{YM}}^{a,\mu}=%
%TCIMACRO{\dsum \limits_{i=1}^{3}}%
%BeginExpansion
{\displaystyle \sum \limits_{i=1}^{3}}
%EndExpansion
{\bar{U}}_{i}(x)i\gamma^{\mu}T^{a}U_{i}(x)+%
%TCIMACRO{\dsum \limits_{i=1}^{3}}%
%BeginExpansion
{\displaystyle \sum \limits_{i=1}^{3}}
%EndExpansion
{\bar{D}}_{i}(x)i\gamma^{\mu}T^{a}D_{i}(x).
\end{equation}
Because neutrino has no level-2 zero, it doesn't couple the
$\mathrm{SU_{\mathrm{C}}(3)}\otimes$\textrm{U}$_{\mathrm{em}}$\textrm{(1)}
gauge fields. This model gives the sector of QED and QCD.

\subsection{Sector of electro-weak interaction}

In this section, we discuss sector of electro-weak interaction.

\subsubsection{Equivalence for left-hand fermions}

For the Dirac type elementary particles $\psi_{D}=\left(
\begin{array}
[c]{c}%
l\\
D\\
U
\end{array}
\right)  $, there are left-hand and right-hand components:
\begin{equation}
\psi_{D,L/R}=[(1\mp \gamma_{5})/2]\psi,\quad \bar{\psi}_{D,L/R}=\bar{\psi}%
[(1\pm \gamma_{5})/2].
\end{equation}
For Weyl type fermionic elementary particles $\psi_{\nu,L}=n_{L}$, there are
only left-hand component.\ The effective Lagrangian of free elementary
particles becomes%
\begin{equation}
\mathcal{L}_{\mathrm{fermion}}=\bar{\psi}_{\nu,L}i\gamma^{\mu}\partial_{\mu
}\psi_{\nu,L}+\bar{\psi}_{D}i\gamma^{\mu}\partial_{\mu}\psi_{D}+\bar{\psi
}_{D,L}M\psi_{D,R}.
\end{equation}
There is no mass term for the left-hand Weyl type fermionic elementary
particles $\psi_{\nu,L}.$

For the physical chiral variant, the relevant sub-variant $V_{r,\mathrm{\tilde
{S}\tilde{O}}^{[1]}\mathrm{(d)},d}^{\mathrm{sub}}$ is induced by extra
left-hand sub-variant, i.e., $V_{r,\mathrm{\tilde{S}\tilde{O}}^{[1]}%
\mathrm{(d)},d}^{\mathrm{sub}}=\alpha V_{L,\mathrm{\tilde{S}\tilde{O}}%
^{[1]}\mathrm{(d)},d}^{\mathrm{sub}}.$ This means the expansion and
contraction of the relevant sub-variant $V_{r,\mathrm{\tilde{S}\tilde{O}%
}^{[1]}\mathrm{(d)},d}^{\mathrm{sub}}$ has similar properties to that of extra
left-hand sub-variant. In other words, left-hand Dirac type fermionic
elementary particles $\psi_{D,L}$ and left-hand Weyl type fermionic elementary
particles $\psi_{\nu,L}$\ are indistinguishable. This indicates that left-hand
Dirac type fermionic elementary particles $\psi_{D,L}$ and left-hand Weyl type
fermionic elementary particles $\psi_{\nu,L}$ become $\mathrm{SU}%
_{\mathrm{weak}}\mathrm{(2)}$ spinor,
\begin{equation}
\psi_{L}=\left(
\begin{array}
[c]{c}%
\psi_{D,L}\\
\psi_{\nu,L}%
\end{array}
\right)  .
\end{equation}
These considerations lead us to assign the left-handed components of the
elementary particles to $\mathrm{SU}_{\mathrm{weak}}\mathrm{(2)}$ doublets
\begin{equation}
\psi_{\mathrm{L}}=\frac{1}{2}(1+\gamma_{5})\left(
\begin{array}
[c]{c}%
\psi_{D}\\
\psi_{\nu}%
\end{array}
\right)  . \label{leptrepr1}%
\end{equation}
The right-handed components are assigned to $\mathrm{SU}_{\mathrm{weak}%
}\mathrm{(2)}$ singlets
\begin{equation}
\psi_{\mathrm{R}}=\frac{1}{2}(1-\gamma_{5})\left(
\begin{array}
[c]{c}%
\psi_{D}\\
\psi_{\nu}%
\end{array}
\right)  .
\end{equation}

Due to locally mixing $\mathrm{SU}_{\mathrm{weak}}\mathrm{(2)}$ doublets for
left-hand elementary particles, the $\mathrm{SU}_{\mathrm{weak}}\mathrm{(2)}$
symmetry is a local one,
\begin{align}
\psi_{\mathrm{L}}  &  \rightarrow e^{i\vec{\tau}\vec{\theta}(x,t)}%
\psi_{\mathrm{L}},\text{ }\nonumber \\
\psi_{\mathrm{R}}  &  \rightarrow \psi_{\mathrm{R}}%
\end{align}
with $\vec{\tau}$ the three Pauli matrices. To guarantee the local
$\mathrm{SU}_{\mathrm{weak}}\mathrm{(2)}$ gauge symmetry, we introduce the
corresponding gauge fields $W_{\mu}$, of which the corresponding field
strengths are $W_{\mu \nu}.$ The effective Lagrangian of elementary particles
turns into,
\begin{equation}
\mathcal{L}_{\mathrm{fermion}}=\mathrm{Tr}\bar{\psi}_{L}i\gamma^{\mu}%
(\partial_{\mu}-ig_{w}W_{\mu})\psi_{L}+\bar{\psi}_{R}i\gamma^{\mu}%
\partial_{\mu}\psi_{R}.
\end{equation}
The coupling constant is $g_{w}$.

\subsubsection{Electro-weak structure}

According to above discussion, left-hand Dirac type fermionic elementary
particles $\psi_{D,L}$ and left-hand Weyl type fermionic elementary particles
$\psi_{\nu,L}$\ are indistinguishable. However, left-hand Dirac type fermionic
elementary particles $\psi_{D,L}$ and left-hand Weyl type fermionic elementary
particles $\psi_{\nu,L}$ look very difference. For example, for the case of
$\lambda^{\lbrack12]}=3,$ a neutrino has zero electric charge (or zero level-2
zero) while a charged lepton has unit charge (or three missing level-2
zeroes); a D-quark has $-\frac{1}{3}$ electric charge (or one missing level-1
zero) while a U-quark has $\frac{2}{3}$ electric charge (or two extra level-2
zeroes). So, to characterize the equivalence, we introduce the
electro-weak\ structure -- weak charge and hypercharge.

\paragraph{Electro-weak structure for charged leptons and neutrinos}

Firstly, we consider the electro-weak structure for charged leptons and neutrinos.

According to above discussion, without considering the charge degree of
freedom, we cannot distinguish a left-hand electron from an e-neutrino. To
demonstrate the equivalence between a left-hand electron from an e-neutrino,
we consider them as composite object and split them into
\emph{opposite-component} and \emph{identical-component}. Their
opposite-component becomes the charge of weak $\mathrm{SU}_{\mathrm{weak}%
}\mathrm{(2)}$ gauge fields, their identical-component becomes hypercharge of
$\mathrm{U}_{\mathrm{Y}}\mathrm{(1)}$ gauge field.

We then consider a left-hand charged lepton $l$ ($l=e,$ $\mu,$ $\tau$) with
three level-2 zeroes to be a composite object with both weak charge and
hypercharge, i.e.,
\begin{equation}
n_{l}=\left(  n_{l}\right)  _{\mathrm{SU}_{\mathrm{weak}}\mathrm{(2)}}+\left(
n_{l}\right)  _{\mathrm{U}_{\mathrm{Y}}\mathrm{(1)}}=3
\end{equation}
where $\left(  n_{l}\right)  _{\mathrm{SU}_{\mathrm{weak}}\mathrm{(2)}}%
=\frac{3}{2}$ and $\left(  n_{l}\right)  _{\mathrm{U}_{\mathrm{Y}}%
\mathrm{(1)}}=\frac{3}{2}$ and a neutrino $\nu_{l}$ also to be a composite
object with weak charge and hypercharge
\begin{equation}
n_{\nu_{l}}=\left(  n_{\nu_{l}}\right)  _{\mathrm{SU}_{\mathrm{weak}%
}\mathrm{(2)}}+\left(  n_{\nu_{l}}\right)  _{\mathrm{U}_{\mathrm{Y}%
}\mathrm{(1)}}=0
\end{equation}
where $\left(  n_{\nu_{l}}\right)  _{\mathrm{SU}_{\mathrm{weak}}\mathrm{(2)}%
}=-\frac{3}{2}$ and $\left(  n_{\nu_{l}}\right)  _{\mathrm{U}_{\mathrm{Y}%
}\mathrm{(1)}}=\frac{3}{2}$. Then, we have
\begin{equation}
\left(  n_{l}\right)  _{\mathrm{SU}_{\mathrm{weak}}\mathrm{(2)}}=-\left(
n_{\nu_{l}}\right)  _{\mathrm{SU}_{\mathrm{weak}}\mathrm{(2)}}=\frac{3}{2}.
\end{equation}

Therefore, a left-hand neutrino $\nu_{l}$ and a left-hand charged lepton $l$
make up a lepton $\mathrm{SU}_{\mathrm{weak}}\mathrm{(2)}$ spinor
\begin{equation}
\psi_{\mathrm{Lepton}}=\left(
\begin{array}
[c]{c}%
\nu_{l}\\
l
\end{array}
\right)
\end{equation}
where
\begin{equation}
\psi_{\mathrm{Lepton,}L}=\frac{1}{2}(1+\gamma_{5})\left(
\begin{array}
[c]{c}%
\nu_{l}\\
l
\end{array}
\right)
\end{equation}
and
\begin{equation}
\psi_{\mathrm{Lepton,}R}={l}_{R}=\frac{1}{2}(1-\gamma_{5}){l}.
\end{equation}
As a result, we have
\begin{align}
\psi_{\mathrm{Lepton,}L}  &  \rightarrow e^{i\vec{\tau}\vec{\theta}(\vec{X}%
)}\psi_{\mathrm{Lepton,}L},\text{ }\nonumber \\
\psi_{\mathrm{Lepton,}R}  &  \rightarrow \psi_{\mathrm{Lepton,}R}%
\end{align}
with $\vec{\tau}$ the three Pauli matrices.

On the other hand, to characterize the property of identical-component for
leptons$\ $%
\begin{equation}
\left(  n_{l}\right)  _{\mathrm{U}_{\mathrm{Y}}\mathrm{(1)}}=\left(
n_{\nu_{l}}\right)  _{\mathrm{U}_{\mathrm{Y}}\mathrm{(1)}}=\frac{3}{2},
\end{equation}
a left-hand neutrino $\nu_{l}$ and a left-hand charged lepton $l$ have an
additional charge degree of freedom from $\mathrm{U}_{\mathrm{Y}}\mathrm{(1)}$
gauge symmetry. Such charge degree of freedom is called the \emph{hypercharge}
degrees of freedom. The hypercharge $\mathcal{Y}$ of left-hand charged lepton
and left-hand neutrino with $\left(  n_{l}\right)  _{\mathrm{U}_{\mathrm{Y}%
}\mathrm{(1)}}=\left(  n_{\nu_{l}}\right)  _{\mathrm{U}_{\mathrm{Y}%
}\mathrm{(1)}}=\frac{3}{2}$ is
\begin{equation}
\mathcal{Y}(\psi_{\mathrm{Lepton,}L})=-1.
\end{equation}
An important fact is that \emph{a missing level-2 zero has hypercharge }%
$Y$\emph{ to be }$-\frac{2}{3}$\emph{.}

For a right-hand charged lepton, there are $3$ level-2 zeroes. The
corresponding hypercharge $\mathcal{Y}$ is $3\times \left(  -\frac{2}%
{3}\right)  =-2$, i.e.,
\begin{equation}
\mathcal{Y}(\psi_{\mathrm{Lepton,}R})=-2.
\end{equation}

\paragraph{Electro-weak structure for quarks}

Next, we consider the electro-weak structure for quarks.

A D-quark has 1 missing level-2 zero $n_{\mathrm{D}}=1$ and an U-quark has 2
extra level-2 zero $n_{\mathrm{U}}=-2$.

We consider a left-hand D-quark to be a composite object as
\begin{equation}
n_{\mathrm{D}}=\left(  n_{\mathrm{D}}\right)  _{\mathrm{SU}_{\mathrm{weak}%
}\mathrm{(2)}}+\left(  n_{\mathrm{D}}\right)  _{\mathrm{U}_{\mathrm{Y}%
}\mathrm{(1)}}%
\end{equation}
where $\left(  n_{\mathrm{D}}\right)  _{\mathrm{SU}_{\mathrm{weak}%
}\mathrm{(2)}}=\frac{3}{2}$ and $\left(  n_{\mathrm{D}}\right)  _{\mathrm{U}%
_{\mathrm{Y}}\mathrm{(1)}}=-\frac{1}{2}.$ On the other hand, a left-hand
U-quark is also considered to be a object,
\begin{equation}
n_{\mathrm{U-quark}}=\left(  n_{\mathrm{U-quark}}\right)  _{\mathrm{SU}%
_{\mathrm{weak}}\mathrm{(2)}}+\left(  n_{\mathrm{U-quark}}\right)
_{\mathrm{U}_{\mathrm{Y}}\mathrm{(1)}}%
\end{equation}
where $\left(  n_{\mathrm{U}}\right)  _{\mathrm{SU}_{\mathrm{weak}%
}\mathrm{(2)}}=-\frac{3}{2}$ and $\left(  n_{\mathrm{U}}\right)
_{\mathrm{U}_{\mathrm{Y}}\mathrm{(1)}}=-\frac{1}{2}$.

To characterize the property of opposite-component for $U$ and $D$ as
\begin{equation}
\left(  n_{\mathrm{D}}\right)  _{\mathrm{SU}_{\mathrm{weak}}\mathrm{(2)}%
}=-\left(  n_{\mathrm{U}}\right)  _{\mathrm{SU}_{\mathrm{weak}}\mathrm{(2)}%
}=\frac{3}{2},
\end{equation}
$\bar{u}$ and $d$ make up a quark $\mathrm{SU}_{\mathrm{weak}}\mathrm{(2)}$
spinor%
\begin{equation}
\psi_{\mathrm{quark},L}=\frac{1}{2}(1+\gamma_{5})\left(
\begin{array}
[c]{c}%
U\\
D
\end{array}
\right)  .
\end{equation}

On the other hand, to characterize the property of identical-component for $U$
and $D$ as$\ $%
\begin{equation}
\left(  n_{\mathrm{D}}\right)  _{\mathrm{SU}_{\mathrm{weak}}\mathrm{(2)}%
}=\left(  n_{\mathrm{U}}\right)  _{\mathrm{SU}_{\mathrm{weak}}\mathrm{(2)}%
}=-\frac{1}{2},
\end{equation}
a left-hand D-quark and a left-hand U-quark have the hypercharge $\mathcal{Y}$
to be
\begin{equation}
\mathcal{Y}(\psi_{\mathrm{quark,}L})=-\frac{1}{2}\times \frac{2}{3}=\frac{1}%
{3}.
\end{equation}

Since a level-2 zero has hypercharge $\mathcal{Y}$ to be $-\frac{2}{3}$, we
find that the super charge for quarks $U_{R},$ $D_{R}$ to be
\begin{equation}
\mathcal{Y}(U_{R})=\frac{4}{3},\text{ }\mathcal{Y}(D_{R})=-\frac{2}{3}.
\end{equation}

\paragraph{Summary}

Finally, we write down the effective Lagrangian of fermionic elementary
particles for its electro-weak structure,
\begin{align}
&  \mathrm{Tr}\bar{\psi}_{L}i\gamma^{\mu}(\partial_{\mu}-ig_{w}W_{\mu}%
+i\frac{g^{\prime}}{2}B_{\mu})\psi_{L}\\
&  +\bar{\psi}_{R}i\gamma^{\mu}(\partial_{\mu}+ig^{\prime}B_{\mu})\psi
_{R}\nonumber
\end{align}
where $W_{\mu}$ and $B_{\mu}$ denote the gauge fields associated to weak
$\mathrm{SU}_{\mathrm{weak}}\mathrm{(2)}$ and hypercharge $\mathrm{U}%
_{\mathrm{Y}}\mathrm{(1)}$ respectively, of which the corresponding field
strengths are $W_{\mu \nu}$ and $B_{\mu \nu}$. The two coupling constants
$g_{w}$ and $g^{\prime}$ correspond to the groups $\mathrm{SU}_{\mathrm{weak}%
}\mathrm{(2)}$ and $\mathrm{U}_{\mathrm{Y}}\mathrm{(1),}$ respectively.
Because neutrino has only left-hand degrees of freedom, the charged $W$'s
couple only to the left-handed components of the lepton fields.

\subsubsection{Mathematical foundation for electro-weak gauge fields}

In the end of this section, we develop a theory for mathematical foundation of
electro-weak gauge fields.

Now, we consider $V_{r,\mathrm{\tilde{S}\tilde{O}}^{[1]}\mathrm{(d)}%
,d}^{\mathrm{sub}}$ to be an effective level-2 variant with the following
variability,
\begin{equation}
\hat{U}^{[1]}(\delta \phi_{g,\mathrm{global}}^{[1]})\leftrightarrow \hat
{U}^{[2]}(\delta \phi_{r,\mathrm{global}}^{[2]})=\exp(i\lambda^{\lbrack
gr]}\delta \phi_{g,\mathrm{global}}^{[1]}).
\end{equation}
It was known that the relative sub-variant is induced by extra left-hand
sub-variant,
\begin{equation}
V_{r,\mathrm{\tilde{S}\tilde{O}}^{[1]}\mathrm{(d)},d}^{\mathrm{sub}}=\alpha
V_{L,\mathrm{\tilde{S}\tilde{O}}^{[1]}\mathrm{(d)},d}^{\mathrm{sub}}.
\end{equation}
So, the group-changing space of extra left-hand sub-variant has same
variability. The equivalence between $V_{r,\mathrm{\tilde{S}\tilde{O}}%
^{[1]}\mathrm{(d)},d}^{\mathrm{sub}}\ $and $\alpha V_{L,\mathrm{\tilde
{S}\tilde{O}}^{[1]}\mathrm{(d)},d}^{\mathrm{sub}}$ leads to the emergence of
\textrm{SU}$_{\mathrm{weak}}$\textrm{(2)}$\times$\textrm{U}$_{\mathrm{Y}}%
$\textrm{(1)} gauge structure.

As a result, an extra level-2 zero of $V_{r,\mathrm{\tilde{S}\tilde{O}}%
^{[1]}\mathrm{(d)},d}^{\mathrm{sub}}$ and that of $\alpha V_{L,\mathrm{\tilde
{S}\tilde{O}}^{[1]}\mathrm{(d)},d}^{\mathrm{sub}}$ have same properties. So,
we map an zero of $V_{r,\mathrm{\tilde{S}\tilde{O}}^{[1]}\mathrm{(d)}%
,d}^{\mathrm{sub}}$ and that of $\alpha V_{\mathrm{\tilde{S}\tilde{O}}%
^{[1]}\mathrm{(d)},d}^{L}$ to those of an effective level-2 zero of 2-nd order
physical variant with $\lambda^{\lbrack12]}=2$. The following is the
correspondence,%
\begin{align}
&  \text{ Level-2 zero of }V_{r,\mathrm{\tilde{S}\tilde{O}}^{[1]}%
\mathrm{(d)},d}^{\mathrm{sub}}\nonumber \\
&  \rightarrow \text{State of a level-2 zero on position 1,}%
\end{align}
and%
\begin{align}
&  \text{ Level-2 zero of }\alpha V_{L,\mathrm{\tilde{S}\tilde{O}}%
^{[1]}\mathrm{(d)},d}^{\mathrm{sub}}\nonumber \\
&  \rightarrow \text{State of a level-2 zero on position 2.}%
\end{align}

Then, we study the effective level-2 zero of 2-nd order physical variant with
$\lambda^{\lbrack12]}=2.$

Matter corresponds to globally expand or contract of the different levels of
group-changing spaces. Therefore, elementary particle is classified by the
number of effective level-2 zeroes $n^{[2]}$. Here, we set $n^{[2]}=1$.

Under compactification, the effective level-2 group-changing space turns into
an effective level-2 zero lattice with two lattices. On each site of the
level-2 zero lattice (we denote it by $I^{[2]}$), there is a local field of
compact \textrm{U(1)} symmetry. This\ compact \textrm{U(1)} symmetry leads to
the \textrm{U}$_{\mathrm{Y}}$\textrm{(1)} gauge symmetry for global motion of
the effective level-2 zero of a level-1 zero.

On the other hand, there exists relative motion of the effective level-2 zero
of a level-1 zero.

Because there are $2$ lattice sites for the effective level-2 zeroes of a
level-1 zero, we have $2$\ level-2 phase factors $\delta \varphi_{I^{[2]}%
,I^{[1]}}^{[2]}$. As a result, the effective level-2 zero has two internal
states,
\begin{equation}
\left(
\begin{array}
[c]{c}%
\left \vert \psi_{1^{[2]},I_{g}^{[1]}}^{[2]}\right \rangle \\
\left \vert \psi_{2^{[2]},I_{g}^{[1]}}^{[2]}\right \rangle
\end{array}
\right)  .
\end{equation}
Here, $\left \vert \psi_{I^{[2]},I_{g}^{[1]}}^{[2]}\right \rangle $ denotes the
state of the effective level-2 zero on the $I^{[2]}$-th lattice site of
effective level-2 zero lattice inside $I_{g}^{[1]}$-th global level-1 zero.
Because there are $2$ lattice sites, the quantum states of $\left \vert
\psi_{I^{[2]},I_{g}^{[1]}}^{[2]}\right \rangle $ have $2$ components. The
quantum states of the effective level-2 zero on different sites of effective
the level-2 zero-lattice are orthogonal, i.e.,%
\begin{equation}
\langle \psi_{J^{[2]},I_{g}^{[1]}}^{[2]}\left \vert \psi_{I^{[2]},I_{g}^{[1]}%
}^{[2]}\right \rangle =\delta_{J^{[2]}I_{g}^{[2]}}.
\end{equation}
Hence, $\left \vert \psi_{1^{[2]},I_{g}^{[1]}}^{[2]}\right \rangle ,$
$\psi_{2^{[2]},I_{g}^{[1]}}^{[2]}$ make up a complete basis. The symmetry of
different basis leads to non-Abelian $\mathrm{SU(2)}$ gauge symmetry for
relative motion of effective level-2 zeroes of a level-1 zero.
This\ non-Abelian $\mathrm{SU(2)}$ gauge symmetry leads to the
$\mathrm{SU_{weak}(2)}$ non-Abelian gauge symmetry for weak interaction.

As a result, we have two types of local gauge symmetries, one is Abelian,
\textrm{U}$_{\mathrm{Y}}$\textrm{(1)} gauge symmetry for global motion of
effective level-2 zeroes of a level-1 zero, the other is non-Abelian,
$\mathrm{SU_{weak}(2)}$ gauge symmetry for relative motion of effective
level-2 zeroes of a level-1 zero. This leads to the underlying mechanism of
the $\mathrm{SU_{weak}(2)}\times$\textrm{U}$_{\mathrm{Y}}$\textrm{(1)} gauge
symmetry for electro-weak interaction. The effective model is an
\textrm{SU}$_{\mathrm{weak}}$\textrm{(2)}$\times$\textrm{U}$_{\mathrm{Y}}%
$\textrm{(1) }gauge field theory.

In long wave limit, the effective Lagrangian for electro-weak interaction is
written as%
\begin{align}
&  \mathcal{L}_{\mathrm{fermion}}+\mathcal{L}_{\mathrm{Y}}(\mathrm{U}%
_{\mathrm{Y}}\mathrm{(1)})+\mathcal{L}_{\mathrm{weak}}(\mathrm{SU}%
_{\mathrm{weak}}\mathrm{(2)}))\\
&  =\mathrm{Tr}\bar{\psi}_{L}i\gamma^{\mu}(\partial_{\mu}-ig_{w}W_{\mu}%
+i\frac{g^{\prime}}{2}B_{\mu})\psi_{L}\nonumber \\
&  +\bar{\psi}_{R}i\gamma^{\mu}(\partial_{\mu}+ig^{\prime}B_{\mu})\psi
_{R}\nonumber \\
&  -\frac{1}{4}B_{\mu \nu}B^{\mu \nu}-\mathrm{Tr}\frac{1}{2}W_{\mu \nu}W^{\mu \nu
}.\nonumber
\end{align}

\subsubsection{Coupling constant $g^{\prime}$}

In this part, we discuss the coupling constant of hypercharge $g^{\prime}$.
The coupling constant $g^{\prime}$ for global motion of an effective level-2
zero (a zero of relative sub-variant or a zero of extra left-hand sub-variant)
of a level-1 zero (or a level-1 zero of global sub-variant) is defined by the
ratio between the effective Planck constant $\hbar^{\lbrack2]}$ and Planck
constant $\hbar$.

It was known that there are $\lambda^{\lbrack gr]}$ level-1 zeroes for an
effective level-2 zero. We consider the time dependent phase factor
$e^{i\omega_{n}t}$ of wave function of the system, the level-1 zero has a
global motion with angular momentum $\omega_{0}^{[2]}=\frac{2\pi c}%
{\lambda^{\lbrack gr]}}$ in the group-changing space of the effective level-2
zero. Therefore, after considering the changing of energy, the density of
angular momentum is obtained as
\begin{equation}
\rho_{\omega_{0}^{[2]}}=\frac{\delta E_{n}}{\delta \omega}\mid_{\omega
=\omega_{0}^{[2]}}=\hbar.
\end{equation}
For each level-1 zero, we have effective Planck constant to be
\begin{equation}
\hbar^{\lbrack2]}=\hbar/\lambda^{\lbrack gr]}%
\end{equation}
that is an effective Planck constant by scaling the original one by a ratio
$\frac{1}{\lambda^{\lbrack gr]}}$. Finally, the unit of coupling constant
$g^{\prime}$ is derived as
\begin{equation}
g^{\prime}=c\hbar^{\lbrack2]}=\frac{1}{\lambda^{\lbrack gr]}}\hbar c.
\end{equation}

\subsection{Sector of Higgs fields}

\subsubsection{Higgs mechanism}

In this part we discuss the Higgs mechanism\cite{higg1,higg2,higg4}. For
simplicity, we take the case of $\lambda^{\lbrack12]}=0$ as example and don't
consider the effect of level-2 zeroes. Now, we focus on the effect of Higgs
field on electro-weak gauge fields and elementary particles of electron $e$
and neutrino $v_{e}$.

According to the theory of variant, the masses of elementary particles come
from the difference between the changing rate along spatial direction and that
along tempo direction, i.e., $\gamma^{\lbrack st]}=\omega_{0}^{[g]}%
/k_{g,0}\neq1$. If we consider the ratio $\gamma^{\lbrack st]}$ difference
from $1$, it plays the role of Higgs field $\Phi(x,t)$, i.e.,
\begin{equation}
(\gamma^{\lbrack st]}(x,t)-1)\rightarrow \Phi(x,t).
\end{equation}
The Higgs field $\Phi(x,t)$ couples left-hand elementary particles and
right-hand elementary particles. Or, the effect of $\Phi(x,t)$ is to change
$\psi_{\mathrm{L}}(x,t)$ to $\psi_{\mathrm{R}}(x,t).$ So, there appears an
extra term in Hamiltonian as
\begin{equation}
\psi_{R}^{\dagger}\Phi(x,t)\psi_{L}.
\end{equation}

On the other hand, due to
\begin{equation}
\psi_{\mathrm{L}}(x)\rightarrow e^{i\vec{\tau}\vec{\theta}(\vec{X})}%
\psi_{\mathrm{L}}(x),\text{ }\psi_{\mathrm{R}}(x)\rightarrow \psi_{\mathrm{R}%
}(x),
\end{equation}
$\Phi(x,t)$ must be an $\mathrm{SU}_{\mathrm{weak}}\mathrm{(2)}$ complex
doublet as $\Phi(x,t)=\left(
\begin{array}
[c]{c}%
\phi^{+}\\
\phi^{0}%
\end{array}
\right)  $ obeys $\Phi(\vec{X},t)\rightarrow e^{i\vec{\tau}\vec{\theta}%
(X)}\Phi(x,t).$

Next, we write down an effective Lagrangian of the Higgs field $\Phi(x,t).$

Because the $\Phi(x,t)$ is an $\mathrm{SU}_{\mathrm{weak}}\mathrm{(2)}$
complex doublet, we get its kinetic term as
\begin{equation}
|(\partial_{\mu}-ig_{w}\frac{\vec{\tau}}{2}\cdot \vec{W}_{\mu})\Phi(x,t)|^{2}.
\end{equation}
To obtain the finite expected value, we add a phenomenological potential term
$V(\Phi(x,t)).$ Finally, we have%
\begin{align}
\mathcal{L}_{\mathrm{Higgs}}  &  =|(\partial_{\mu}-ig_{w}\frac{\vec{\tau}}%
{2}\cdot \vec{W}_{\mu})\Phi(x,t)|^{2}-V(\Phi(x,t))\nonumber \\
&  +\bar{\psi}\Phi(x,t)\psi+h.c..
\end{align}

A finite value is given by minimizing $\Phi(x,t),$ of which the expected value
is $\phi_{0}$, i.e.,%
\begin{equation}
\left \langle \Phi(x,t)\right \rangle =\left(
\begin{array}
[c]{c}%
0\\
\phi_{0}%
\end{array}
\right)  +\delta \Phi(x,t).
\end{equation}
Then, the weak gauge symmetry is spontaneously broken. A finite value of Higgs
field creates a mass term for the Dirac type elementary particles, but leaving
the Weyl type elementary particle is also massless,
\begin{equation}
m_{e}=\phi_{0},\text{ }m_{v_{e}}=0.
\end{equation}

On the other hand, when $\phi_{0}\neq0$, there exists Higgs mechanism. Now,
the original gauge symmetry may be broken. The $\mathrm{SU}_{\mathrm{weak}%
}\mathrm{(2)}$ gauge fields obtain masses from the following terms
\begin{equation}
\frac{1}{2}g_{w}^{2}\phi_{0}^{2}(W_{\mu}W^{\mu}).
\end{equation}
The mass for the charged vector bosons $W_{\mu}\ $is $m_{W}=\phi_{0}g_{w}/2.$\

After considering the Higgs condensation $\phi_{0}\neq0,$ the low energy
effective Lagrangian becomes
\begin{align}
\mathcal{L}_{\mathrm{SM}}  &  =\mathrm{Tr}\bar{\psi}_{L}i\gamma^{\mu}%
(\partial_{\mu}-ig_{w}\frac{\vec{\tau}}{2}\cdot \vec{W}_{\mu})\psi
_{L}\nonumber \\
&  +\bar{\psi}_{R}i\gamma^{\mu}\partial_{\mu}\psi_{R}+m_{T}\bar{\psi}_{T}%
\psi_{T}\\
&  --\mathrm{Tr}\frac{1}{2}W_{\mu \nu}W^{\mu \nu}+\frac{1}{2}\Phi^{2}g_{w}%
^{2}W_{\mu}W^{\mu}+{g}_{w}W_{\mu}{j}_{w}^{\mu}\nonumber \\
&  +|\partial_{\mu}\Phi|^{2}+m_{\mathrm{Higgs}}\left \vert \Phi \right \vert
^{2}+...\nonumber
\end{align}
where the weak current is%
\begin{equation}
{j}_{w\text{ }-}^{\mu}=i\bar{e}\gamma_{\mu}\nu_{e},\text{ }{j}_{w\,+}^{\mu
}=i\bar{\nu}_{e}\gamma_{\mu}e.
\end{equation}

\subsubsection{Higgs mechanism in the Standard model}

In this part, we discuss the Higgs mechanism in the Standard model.

It is known that the difference between the changing rate along spatial
direction and that along tempo direction $(\gamma^{\lbrack st]}(x,t)-1)$ plays
the role of Higgs field $\Phi(x,t)$ in standard model. Let us firstly
calculate the hypercharge of $\Phi(x,t).$

The Higgs field changes $\psi_{L}(x)$ to $\psi_{R}(x),$ or, vice versa.
Because of $\mathcal{Y}(\psi_{\mathrm{Lepton,}L})=-1,$ $\mathcal{Y}%
(\psi_{\mathrm{Lepton,}R})=0,$ the $\mathrm{U}_{\mathrm{Y}}\mathrm{(1)}$
hypercharge of $\Phi(x,t)$ is unit, i.e.,
\begin{equation}
\mathcal{Y}(\omega(x,t))=\mathcal{Y}(\psi_{\mathrm{Lepton,}R})-\mathcal{Y}%
(\psi_{\mathrm{Lepton,}L})=1.
\end{equation}
On the other hand, due to $\psi_{L}(x)\rightarrow e^{i\vec{\tau}\vec{\theta
}(x)}\psi_{L}(x),$ $\psi_{R}(x)\rightarrow \psi_{R}(x),$ $\Phi(x,t)$ must be an
$\mathrm{SU}_{\mathrm{weak}}\mathrm{(2)}$ complex doublet $\Phi(x,t)=\left(
\begin{array}
[c]{c}%
\phi^{+}\\
\phi^{0}%
\end{array}
\right)  $ obeying $\Phi(x,t)\rightarrow e^{i\vec{\tau}\vec{\theta}(x)}%
\Phi(x,t).$ As a result, we get the kinetic term of Higgs field as
\begin{equation}
|(\partial_{\mu}-ig_{w}\frac{\vec{\tau}}{2}\cdot \vec{W}_{\mu}-i\frac
{g^{\prime}}{2}B_{\mu})\Phi|^{2}.
\end{equation}
By adding Yukawa coupling between the Higgs field and fermionic elementary
particles, the full Lagrangian about $\Phi(x,t)$ is given by%
\begin{align}
&  |(\partial_{\mu}-ig_{w}\frac{\vec{\tau}}{2}\cdot \vec{W}_{\mu}%
-i\frac{g^{\prime}}{2}B_{\mu})\Phi(x,t)|^{2}\nonumber \\
&  -V(\Phi(x,t))+\bar{\psi}_{\mathrm{Lepton,}L}G_{\mathrm{Lepton}}%
\Phi(x,t)\bar{\psi}_{\mathrm{Lepton,}R}\nonumber \\
&  +\bar{\psi}_{\mathrm{quark},L}G_{\mathrm{quark}}\Phi(x,t)\psi
_{\mathrm{quark},R}+h.c.
\end{align}
where $G_{\mathrm{Lepton}}$ and $G_{\mathrm{quark}}$ are coupling constants
for leptons and quarks with Higgs field. To derive the exact values of
$G_{\mathrm{Lepton}}$ and $G_{\mathrm{quark}},$ we use the approach to
calculate mass spectra. Here, we just consider $G_{\mathrm{Lepton}}$ and
$G_{\mathrm{quark}}$\ to be known parameters.

Next, we discuss the Higgs mechanism.

We assume that a finite value of $\Phi(x,t)$ is $\phi_{0}$, i.e.,
$\left \langle \Phi(x,t)\right \rangle =\frac{1}{\sqrt{2}}\left(
\begin{array}
[c]{c}%
0\\
\phi_{0}%
\end{array}
\right)  +\delta \Phi(x,t).$ A finite value of Higgs field creates a mass term
for the charged leptons, $m_{\mathrm{Lepton}}=\frac{1}{\sqrt{2}}%
G_{\mathrm{Lepton}}\phi.$ For the system with finite value of Higgs field, we
produce masses for the quarks given by $\frac{1}{\sqrt{2}}G_{\mathrm{quark}%
}^{\ast}\Phi.$ Because there is no right-hand neutrino, the neutrinos don't
couple the Higgs field. The masses of neutrinos come from the quantum
tunneling effect between different flavors. In addition, the Higgs field also
has mass, i.e., $m_{\mathrm{Higgs}}=\sqrt{2}\mu \neq0.$

On the other hand, the original $\mathrm{SU_{weak}(2)}\times$\textrm{U}%
$_{\mathrm{Y}}$\textrm{(1)} gauge symmetry is broken. the $\mathrm{SU}%
_{\mathrm{weak}}\mathrm{(2)}$ gauge fields obtain masses from the following
terms\cite{wein}
\begin{equation}
\frac{1}{8}(\phi^{0})^{2}[g_{w}^{2}(W_{\mu}^{1}W^{1\mu}+W_{\mu}^{2}W^{2\mu
})+(g^{\prime}B_{\mu}-g_{w}W_{\mu}^{3})^{2}].
\end{equation}
The mass for the charged vector bosons $W_{\mu}^{\pm}=(W_{\mu}^{1}\mp iW_{\mu
}^{2})/\sqrt{2}\ $becomes $m_{W}=\frac{\phi_{0}g_{w}}{2}.$ After
diagonalisation, the gauge fields $B_{\mu}$ and $W_{\mu}^{3}$ are transformed
into gauge fields $Z_{\mu}$ and $A_{\mu}$ from the following relations%
\begin{align}
Z_{\mu}  &  =\cos \theta_{W}B_{\mu}-\sin \theta_{W}W_{\mu}^{3},\\
A_{\mu}  &  =\cos \theta_{W}B_{\mu}+\sin \theta_{W}W_{\mu}^{3},\nonumber
\end{align}
with $\tan \theta_{W}=g^{\prime}/g_{w}$, of which the masses are
\begin{align}
m_{Z}  &  =\frac{\phi_{0}(g_{w}^{2}+{g^{\prime}}^{2})^{1}/2}{2}=\frac{m_{W}%
}{\cos \theta_{W}},\\
m_{A}  &  =0.\nonumber
\end{align}
The \textquotedblleft Weinberg angle\textquotedblright \ $\theta_{W}$ becomes
the angle between the original $\mathrm{U}_{\mathrm{Y}}\mathrm{(1)}$ and the
one left unbroken one $\mathrm{U}_{\mathrm{em}}\mathrm{(1)}$. \

The neutral gauge bosons $A_{\mu}$ are massless and will be identified with
the photons. Now, the gauge symmetry $\mathrm{U}_{\mathrm{em}}\mathrm{(1)}$
accompanying $A_{\mu}$ will never be broken. That is
\begin{align}
\psi(\vec{X})  &  =\left(
\begin{array}
[c]{c}%
\psi_{L}(x)\\
\psi_{R}(x)
\end{array}
\right)  \rightarrow \left(
\begin{array}
[c]{c}%
\psi_{L}(x)e^{-i\mathrm{e}\phi(x)}\\
\psi_{R}(x)e^{-i\mathrm{e}\phi(x)}%
\end{array}
\right) \\
&  \rightarrow e^{-i\mathrm{e}\phi(x)}\psi(x).\nonumber
\end{align}
Then, the electric charge operator $\mathrm{e}$ will be a linear combination
of $T_{3}$ and $\mathcal{Y}$ as
\begin{equation}
\mathrm{e}=T_{3}+\frac{\mathcal{Y}}{2}. \label{charge}%
\end{equation}

In the end, we address the values of different coupling constants $g_{w}$,
$g^{\prime}$ and $\mathrm{e}$.

According to above discussion, from higher-order variability, we had
determined the value of $\mathrm{e}$ and $g^{\prime}.$ Therefore, the
\textquotedblleft Weinberg angle\textquotedblright \ $\theta_{W}$ together with
the coupling constant of weak interaction $g_{w}$ becomes predictable.

\subsubsection{Property of Higgs field}

In this part, we study the property of Higgs field $\Phi(x,t)=(\gamma^{\lbrack
st]}(x,t)-1)$ where $\gamma^{\lbrack st]}(x,t)=\omega_{0}^{[g]}/k_{g}(x,t)$.
It was known that Higgs field is a scalar field with
spin-0\cite{higg1,higg2,higg4}. Its mass is about $125$\textrm{GeV}.

Firstly, we obtain the potential $V(\Phi(x,t))$ of Higgs field $\Phi(x,t).$

We consider the finite value of changing for changing rate of global
sub-variant $\delta k_{g}(x,t)$. Then, the ratio $\lambda^{\lbrack gr]}%
=\frac{\delta \phi_{g}}{\delta \phi_{r}}=\frac{k_{g}}{k_{r}}$ changes, i.e.,
$\lambda^{\lbrack gr]}\rightarrow \lambda^{\lbrack gr]}(x,t)=\frac{k_{g,0}%
}{k_{r}}+\frac{\delta k_{g}(x,t)}{k_{r}}$. Consequently, the masses of
neutrinos change, i.e., $M_{\mathrm{n}}=\left \vert b_{\mathrm{n}}\right \vert
\left(  1+\sqrt{2}\cos{\left(  \frac{2\pi j}{3}+\varphi_{r,\mathrm{l}}\right)
}\right)  ^{2}\rightarrow M_{\mathrm{n}}(\delta k_{g}(x,t)).$ Here,
$\left \vert b_{\mathrm{n}}\right \vert =\frac{\left \vert b_{\mathrm{l}%
}\right \vert }{3^{2\lambda^{\lbrack gr]}-2\operatorname{mod}\lambda^{\lbrack
gr]}+4}}$ and $\varphi_{r,\mathrm{l}}=-\frac{2}{9}-\frac{\pi}{\lambda^{\lbrack
gr]}+1}$ are both determined by $\lambda^{\lbrack gr]}$.\ It was known that
the neutrino masses play the role of the energy of vacuum. Therefore, the
masses of neutrinos determine the potential of the Higgs field!

We then consider small fluctuations $\delta \varphi$ on the phase factor
$\varphi_{r,\mathrm{l}}$, i.e., $\varphi_{r,\mathrm{l}}\rightarrow
\varphi_{r,\mathrm{l}}+\delta \varphi$ and have
\begin{align}
\delta \varphi &  =-\frac{\pi \lambda^{\lbrack gr]}}{(\lambda^{\lbrack
gr]}+1)^{2}}\Phi \simeq-\frac{\pi}{(\lambda^{\lbrack gr]}+1)}\Phi \\
&  \simeq-0.27245\Phi.\nonumber
\end{align}
The potential energy for Higgs field is obtained as
\begin{equation}
V(\delta \varphi)=\Lambda^{3}\frac{\left \vert b_{\mathrm{l}}\right \vert
}{3^{2\lambda^{\lbrack gr]}-2\operatorname{mod}\lambda^{\lbrack gr]}+4}%
}[1+\sqrt{2}\cos(\varphi_{0}+\delta \varphi)]^{2}%
\end{equation}
or
\begin{equation}
V(\Phi)=\Lambda^{3}\frac{\left \vert b_{\mathrm{l}}\right \vert }{3^{2\lambda
^{\lbrack gr]}-2\operatorname{mod}\lambda^{\lbrack gr]}+4}}[1+\sqrt{2}%
\cos(\varphi_{0}-\frac{\pi}{(\lambda^{\lbrack gr]}+1)}\Phi)]^{2}.
\end{equation}
Here, $\Lambda$ is cutoff that is estimated as reduced Planck energy,
$\Lambda=\frac{E_{p}}{\lambda^{\lbrack gr]}}$. The potential for Higgs field
indicates a surprising fact -- \emph{usual physical picture about spontaneous
symmetry breaking for Higgs mechanism may be wrong}.

Next, we estimate the mass of Higgs particle.

From the potential, we have%
\begin{align}
V(\delta \varphi)/\Lambda^{3}  &  =\frac{\left \vert b_{\mathrm{l}}\right \vert
}{3^{2\lambda^{\lbrack gr]}-2\operatorname{mod}\lambda^{\lbrack gr]}+4}%
}[1+\sqrt{2}\cos(\varphi_{0}+\delta \varphi)]^{2}\\
&  \simeq M_{0,\mathrm{n}}+(0.0025\mathrm{eV})(\delta \varphi)^{2}+...\nonumber
\end{align}
or
\begin{align}
V(\Phi)/\Lambda^{3}  &  \simeq M_{0,\mathrm{n}}+(0.0025\mathrm{eV}%
)(0.27245\Phi)^{2}+...\\
&  =M_{0,\mathrm{n}}+(0.00018557\mathrm{eV})\Phi^{2}+...\nonumber
\end{align}

Finally, after considering kinetic term for Higgs field $\kappa \Lambda
^{2}(\partial_{\mu}\Phi)^{2}$, we derive the the mass of Higgs field
\begin{equation}
\kappa^{-1/2}\sqrt{0.00018557\times \Lambda}\mathrm{eV}\sim \kappa^{-1/2}%
\times500\mathrm{GeV}%
\end{equation}
where $\kappa$\ is a dimensionless parameter, $\kappa \sim O(1)$.

According to this result, we understand why the energy scale of Higgs field
$10^{2}$\textrm{GeV} is much smaller than Planck energy $10^{19}$\textrm{GeV}.
A small energy scale for Higgs field comes from mixing the Planck energy scale
and masses of neutrinos,
\begin{equation}
E_{\mathrm{Higgs}}^{2}\sim E_{\mathrm{Planck}}\times E_{\mathrm{neutrino}}.
\end{equation}
The masses of neutrinos is exponentially suppressed by the value of
$\lambda^{\lbrack gr]}$ and become very small. Therefore, the energy scale of
Higgs field $E_{\mathrm{Higgs}}$ is between $E_{\mathrm{Planck}}$ and
$E_{\mathrm{neutrino}}$. So, we guess that the condensation of Higgs field is
not free parameter. Instead, it may be only determined by $\lambda^{\lbrack
gr]}$.

\subsection{"Dark" sector}

In universe, most of the matter is in the form of \textquotedblleft dark
matter,\textquotedblright \ a new type of particle beyond the SM, and most of
the energy is in the form of \textquotedblleft dark energy,\textquotedblright%
\ energy associated with vacuum. The word{}\textquotedblleft
dark\textquotedblright \ means that people can't "see" it. Unlike visible
matter (matter we can see), it is invisible and doesn't interact with other
matter except through gravity. However, the "dark" sector dominates the
structure and evolution of our universe, i.e., dark matter and dark energy
constitute about $95\%$ of the energy density of the universe. Therefore, both
dark matter and dark energy require extensions to our current understanding of
particle physics or point toward a breakdown of general relativity on
cosmological scales.

In this paper, based on the theory of variant, we resolve the puzzles about
dark matter and dark energy.

\subsubsection{Dark energy}

In general, dark energy is believed to be the energy of vacuum that has a
negative pressure and could drive the accelerated expansion of the
universe\cite{dark energy}. In our universe, the dark energy density
$V_{cosm}$ is estimated to be $(2.3\mathrm{meV})^{4}$. In usual quantum field
theory, the vacuum energy density $V_{cosm}$\ can be obtained by summing the
zero-point energy of all fields. This is a very large value, i.e., the vacuum
energy density contribution (Planck scale) exceeds $(2.3\mathrm{meV})^{4}$ by
120 orders of magnitude. As a result, the puzzle of dark energy is an open
fundamental challenge in modern physics.

In this part, we discuss the physical mechanism of dark energy.

According to above discussion, we had derived the masses of neutrino, i.e.,
\begin{equation}
M_{\mathrm{n}}=\left \vert b_{\mathrm{n}}\right \vert \left(  1+\sqrt{2}%
\cos{\left(  \frac{2\pi j}{3}+\varphi_{r,\mathrm{l}}\right)  }\right)  ^{2}%
\end{equation}
where $\left \vert b_{\mathrm{n}}\right \vert =\frac{\left \vert b_{\mathrm{l}%
}\right \vert }{3^{2\lambda^{\lbrack gr]}-2\operatorname{mod}\lambda^{\lbrack
gr]}+4}}$ and $\varphi_{r,\mathrm{l}}=-\frac{2}{9}-\frac{\pi}{\lambda^{\lbrack
gr]}+1}$. The masses of different neutrino are $m_{\nu_{e}}=0.01\mathrm{eV},$
$m_{\nu_{\mu}}=0.0005\mathrm{eV},$ $m_{\nu_{\tau}}=0.05\mathrm{eV}.$

In particular, the neutrino can be regarded as "vacuum" on relative
group-changing space. So, the masses of neutrino become the shift of the
energies of ground states, that determine the vacuum energy.\ As a result, the
range of density of vacuum energy is from $(0.0005$\textrm{eV}$)^{4}$ to
$(0.05$\textrm{eV}$)^{4}.$ The vacuum energy density $V_{cosm}$ observed in
cosmology is $(0.0023\mathrm{eV})^{4}$ that is indeed in the region between
$(0.0005$\textrm{eV}$)^{4}$ and $(0.05$\textrm{eV}$)^{4}.$

\subsubsection{Mechanism for abnormal galaxy rotation curve}

Firstly, we discuss the abnormal galaxy rotation curve.

It was known that dark energy could drive the accelerated expansion of the
universe. When there exists a tiny vacuum energy, the cosmology becomes expand
exponentially $a_{0}=c^{2}\Lambda_{c}/3$ where $\Lambda_{c}$ is the
cosmological constant. On a background of orthogonal intrinsic acceleration
$a_{0}=c^{2}\Lambda_{c}/3,$ the true acceleration turns into
\begin{equation}
\sqrt{a_{0}^{2}+a^{2}}-a_{0}=a\times \mathrm{f}(a/a_{0})
\end{equation}
where $\mathrm{f}(x)=[(1+4x^{2})^{1/2}-1]/2x.$ $a=\frac{dv}{dt}$ is
acceleration of classical object. The two limits of true acceleration are
\begin{equation}
\left \{
\begin{array}
[c]{l@{\quad:\quad}l}%
a & a\gg a_{0}\\
\frac{a^{2}}{a_{0}} & a\ll a_{0}%
\end{array}
\right.  . \label{nuj}%
\end{equation}
As a result, the rule of\ inertia is modified\cite{nar,des}. See the
illustration in Fig.48. Because the modified inertia is a rule for
acceleration on a background of orthogonal intrinsic acceleration $a_{0}%
\neq0,$ it becomes \emph{another class of higher-order variant}.

\begin{figure}[ptb]
\includegraphics[clip,width=0.92\textwidth]{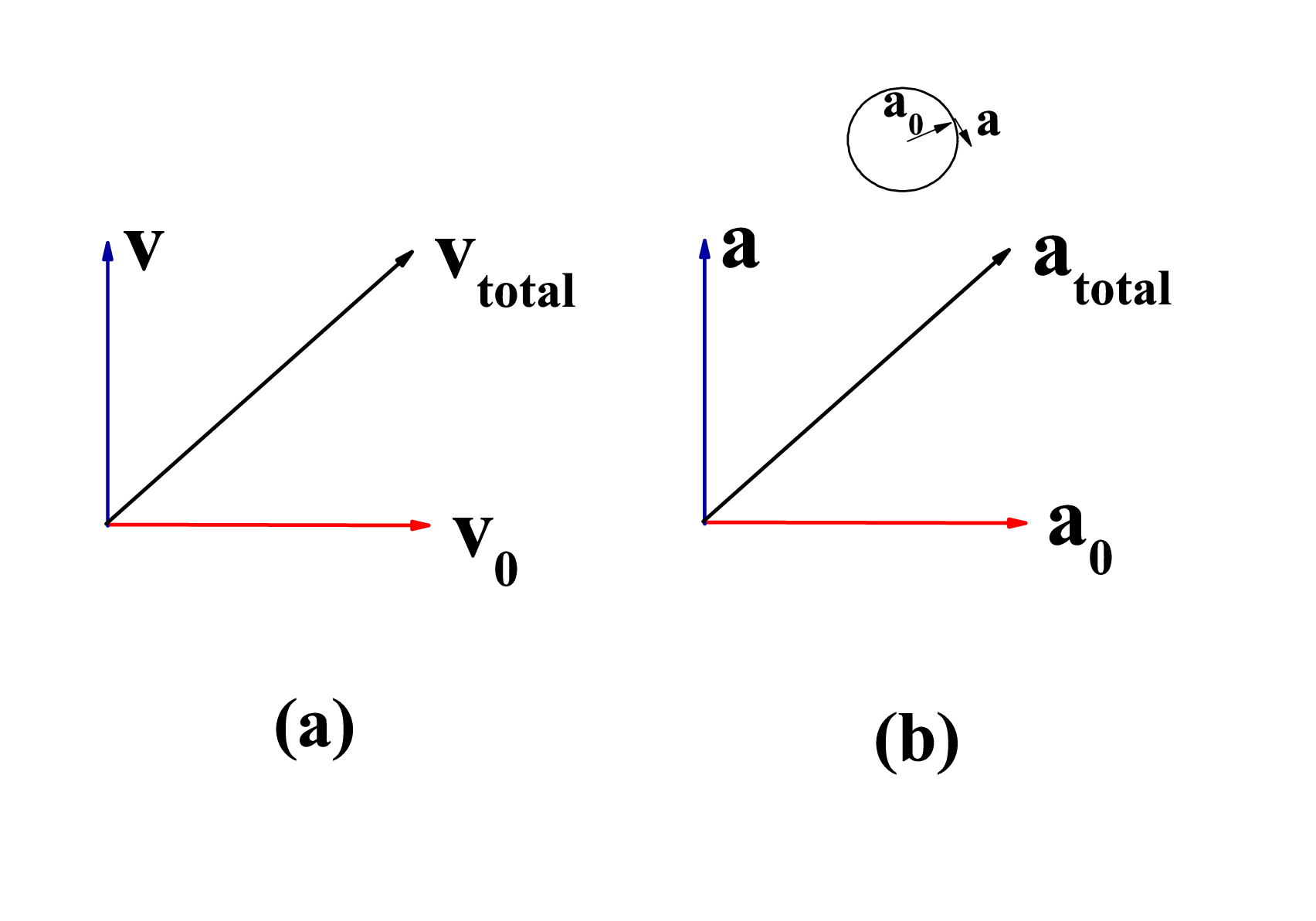}\caption{(Color
online) In this figure, we show why the rule of\ inertia is modified. (a)
modified rule for velocity; (b) modified for acceleration.}%
\end{figure}

In particular, above equation is just the formula of modification of Newtonian
dynamics (MOND)\cite{mond}. The result is firstly pointed by M. Milgrom at
1999\cite{mil}. Consequently, by using the approach of MOND, the abnormal
galaxy rotation curve is naturally explained!

\subsubsection{Possible candidate for dark matter}

Dark matter is a form of matter that does not emit, absorb, or reflect light,
making it invisible and detectable only through its gravitational
effects\cite{dark matter1, dark matter2, dark matter3}. It constitutes about
30.1\% of the matter-energy composition of the universe and plays a crucial
role in the formation and structure of galaxies. Despite being undetectable by
conventional means, scientists infer its existence from the gravitational
influence it exerts on visible matter, such as stars and galaxies. There are a
lot of candidates for dark matter beyond the Standard Model, such as
"supersymmetric particles", right-hand neutrinos and axions. However, till now
the searching for these dark matter candidates fails.

In this part, we show that a possible candidate for "dark matter" comes from
\emph{inhomogeneous cosmological constant} $\Lambda_{c}$ (or background of
orthogonal intrinsic acceleration $a_{0}=c^{2}\Lambda_{c}/3$). Now, we have
$\Lambda_{c}\rightarrow \Lambda_{c}(x,t)$ or $a_{0}\rightarrow a_{0}(x,t)$.

Based on this fundamental assumption, we discuss the effect of "dark matter"
in galaxy clusters.

For galaxy clusters, due to the effect of dark matter, the vacuum energy is
changed correspondingly together with $a_{0}$, i.e., $a_{0}\rightarrow
a_{0}(x,t)$. So, we may have two new physical quantities -- one is
"\emph{jerk}" that is the time rate of change of acceleration $\frac
{da_{0}(x,t)}{dt}$, the other is gradient field of acceleration, $\vec{\nabla
}a_{0}(x,t)$. This\quad gives additional contribution on the the rule
of\ inertia and thus provides a possible explanation on the Bullet galaxy 1E
0657-56\cite{bul}. Because the dark matter can be regarded as density
fluctuations of neutrinos in vacuum, people cannot use usual method to detect
its effect. This is why it looks "dark".

\subsection{Conclusions}

In the end of this part, we draw the conclusion.

The\emph{ starting point} of this theory is very simple -- a ($d+1$)
dimensional 2-nd order $\mathrm{\tilde{S}\tilde{O}}$\textrm{(d+1)} physical
chiral variant $V_{\mathrm{\tilde{U}}^{[2]}\mathrm{(1)},\mathrm{\tilde
{S}\tilde{O}}^{[1]}\mathrm{(3+1)},d+1}^{[2],\mathrm{chiral}}$ with 2-nd order
variability,%
\begin{equation}
\mathcal{T}(\delta x^{\mu})\leftrightarrow \hat{U}^{[1]}(\delta \phi_{g}%
^{[1]})=e^{i\cdot k_{g,0}\delta x^{\mu}\Gamma^{\mu}}.
\end{equation}
and
\begin{equation}
\hat{U}^{[1]}(\delta \phi_{g,\mathrm{global}}^{[1]})\leftrightarrow \hat
{U}^{[2]}(\delta \phi^{\lbrack2]})=\exp(i\lambda^{\lbrack12]}\delta
\phi_{g,\mathrm{global}}^{[1]}),
\end{equation}
where $\delta \phi_{g,\mathrm{global}}^{[1]}=\sqrt{%
%TCIMACRO{\dsum \limits_{\mu}}%
%BeginExpansion
{\displaystyle \sum \limits_{\mu}}
%EndExpansion
((\delta \phi_{g}^{\mu})^{[1]})^{2}}.$ The relevant sub-variant
$V_{r,\mathrm{\tilde{S}\tilde{O}}^{[1]}\mathrm{(3)},3}^{\mathrm{sub}}$ is
induced by extra left-hand sub-variant, $V_{r,\mathrm{\tilde{S}\tilde{O}%
}^{[1]}\mathrm{(3)},3}^{\mathrm{sub}}=\alpha V_{L,\mathrm{\tilde{S}\tilde{O}%
}^{[1]}\mathrm{(d)},d}^{\mathrm{sub}}$. The corresponding relation for 2-nd
order variability is given by%
\begin{equation}
\hat{U}^{[1]}(\delta \phi_{g,\mathrm{global}}^{[1]})\leftrightarrow \hat
{U}^{[2]}(\delta \phi_{r,\mathrm{global}}^{[2]})=\exp(i\lambda^{\lbrack
gr]}\delta \phi_{g,\mathrm{global}}^{[1]}),
\end{equation}
where $\delta \phi_{r,\mathrm{global}}^{[2]}=\sqrt{%
%TCIMACRO{\dsum \limits_{\mu}}%
%BeginExpansion
{\displaystyle \sum \limits_{\mu}}
%EndExpansion
((\delta \phi_{r}^{\mu})^{[2]})^{2}}.$ In particular, for our universe, we have
$d=3,$ $\lambda^{\lbrack12]}=3,$\ $\alpha^{-1}=\lambda^{\lbrack gr]}=3\pi$.
Based on the simple starting point, we develop a theory for the SM.

The level-1 zeroes for global sub-variant become Dirac elementary particles
(lepton or quark); the level-1 zeroes for relative sub-variant become
neutrinos. In addition, due to the existence of the relative sub-variant, for
each type of elementary particle, there are three generations, between which
there exists chiral para-statistics. From the chiral para-statistics, the
hidden structure of mass spectra for different elementary particles is explored.

The 2-nd order variability is reduced into \textrm{U}$_{\mathrm{em}}%
$\textrm{(1)} local gauge symmetry and $\mathrm{SU_{\mathrm{C}}(N)}$
non-Abelian gauge symmetry. The corresponding theory becomes
QED$\mathrm{\times}$QCD. The level-2 zero becomes color charge with $\frac
{1}{\lambda^{\lbrack12]}}$ electric charge. In addition, due to equivalence
between the relevant sub-variant\emph{\emph{ }}and the extra left hand
sub-variant, the effective 2-nd order variability is reduced into
\textrm{U}$_{\mathrm{Y}}$\textrm{(1)} local gauge symmetry and $\mathrm{SU}%
_{\mathrm{weak}}\mathrm{(2)}$ non-Abelian gauge symmetry. The corresponding
theory becomes electro-weak theory. The low energy effective theory of the
2-nd order $\mathrm{\tilde{S}\tilde{O}}$\textrm{(d+1)} physical chiral variant
with chiral asymmetry $V_{\mathrm{\tilde{U}}^{[2]}\mathrm{(1)},\mathrm{\tilde
{S}\tilde{O}}^{[1]}\mathrm{(3+1)},d+1}^{[2],\mathrm{chiral}}$ becomes the
Standard model -- an $\mathrm{SU_{\mathrm{C}}(3)}\otimes \mathrm{SU}%
_{\mathrm{weak}}\mathrm{(2)}\otimes$\textrm{U}$_{\mathrm{Y}}$\textrm{(1)}
gauge theory. The low energy effective Lagrangian density is obtained as
\begin{equation}
\mathcal{L}_{\mathrm{SM}}=\mathcal{L}_{\mathrm{fermion}}+\mathcal{L}%
_{\mathrm{EW}}+\mathcal{L}_{\mathrm{C}}(\mathrm{SU_{\mathrm{C}}(3)}%
)+\mathcal{L}_{\mathrm{Higgs}}.
\end{equation}
where $\mathcal{L}_{\mathrm{EW}}=\mathcal{L}_{\mathrm{Y}}(\mathrm{U}%
_{\mathrm{Y}}\mathrm{(1)})+\mathcal{L}_{\mathrm{weak}}(\mathrm{SU}%
_{\mathrm{weak}}\mathrm{(2)})$. See the logical structure of this part in
Fig.49. Now, the eight key points for the SM are fully understood.

\begin{figure}[ptb]
\includegraphics[clip,width=0.92\textwidth]{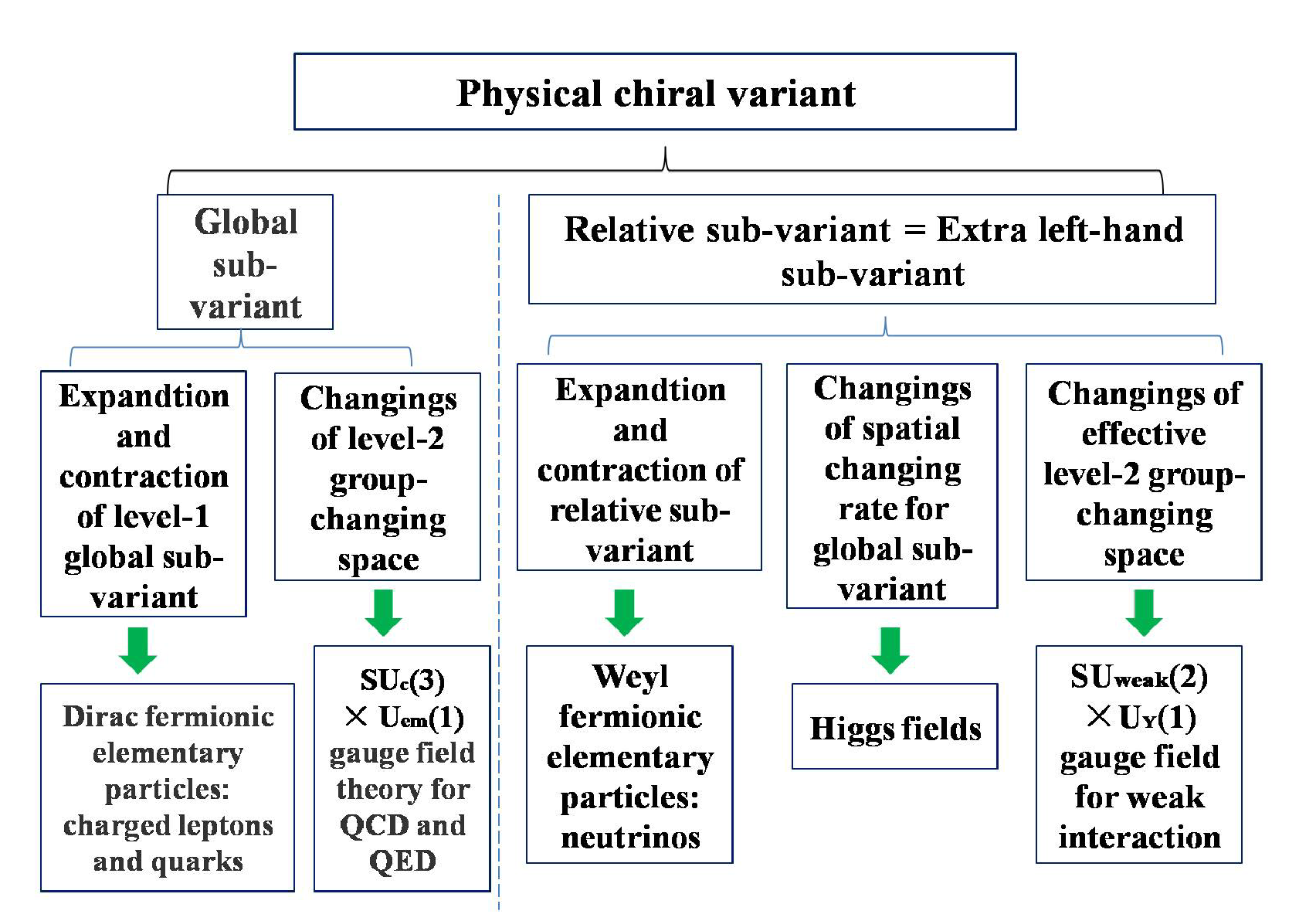}\caption{(Color online)
Logical scheme of this paper}%
\end{figure}

The value of $\lambda^{\lbrack gr]}$ determines different physical parameters.
Now, we have three energy scales : the energy scale of Planck energy
$E_{\mathrm{Planck}}$, the energy scale of Higgs field $E_{\mathrm{Higgs}},$
and the energy scale of masses of neutrinos $E_{\mathrm{neutrino}}$ (or the
energy scale of cosmological constant $\Lambda_{c}$). There exists a
relationship between them, $E_{\mathrm{Higgs}}^{2}\sim E_{\mathrm{Planck}%
}\times E_{\mathrm{neutrino}}.$

In the end of this paper, we answer most questions at beginning and show how
the troubles about quantum gauge fields disappear:

\emph{1. Problems for mass spectra}:

In this part, a hidden topological structure for elementary particles is
explored. The existence of non-trivial topological structure indicates a new
type of quantum statistics -- \emph{chiral para-statistics}. The mass spectra
of different elementary particles are determined by topological invariables
from chiral para-statistics. To obtain the entire mass spectra, we only need
to use only one free parameter $\lambda^{\lbrack gr]}$. The mass spectra for
charged leptons are predicted by the following equation,
\begin{equation}
M_{\mathrm{l}}=\left \vert b_{\mathrm{l}}\right \vert \left(  1+\left \vert
\alpha \right \vert \cos{\left(  \frac{2\pi j}{3}+\varphi_{r,\mathrm{l}}\right)
}\right)  ^{2}%
\end{equation}
where $j=1,2,3.$where $\varphi_{r,\mathrm{l}}=-\frac{\lambda^{\lbrack12]}%
-1}{\lambda^{\lbrack12]}}\pi \frac{1}{\lambda^{\lbrack gr]}}=-\frac{2}{3}%
\pi \frac{1}{3\pi}=-\frac{2}{9}$ and $\left \vert \alpha \right \vert =\sqrt
{\frac{N_{I}}{N_{F}}}=\frac{1}{\sqrt{2}}$. The mass spectra for U-quark with
$\lambda^{\lbrack12]}-1$ level-2 zeroes are predicted by the following
equation
\begin{align}
M_{\mathrm{U}}  &  =\left \vert b_{\mathrm{U}}\right \vert \left(
1+2\sqrt{\frac{3}{4}}\cos{\left(  \frac{2\pi j}{3}+\varphi_{r,\mathrm{U}%
}\right)  }\right)  ^{2},\text{ }\\
j  &  =1,2,3.\nonumber
\end{align}
where $\varphi_{r,\mathrm{U}}=\frac{1}{2}(\frac{\lambda^{\lbrack12]}%
-1}{\lambda^{\lbrack12]}})^{2}\frac{\pi}{\lambda^{\lbrack gr]}}=\frac{1}%
{3}\frac{2}{9}$ and $\left \vert \alpha \right \vert =\sqrt{\frac{N_{I}}{N_{F}}%
}=\sqrt{\frac{\lambda^{\lbrack12]}}{1+\lambda^{\lbrack12]}}}=\sqrt{\frac{3}%
{4}}.$ The mass spectra for D-quark with $1$ level-2 zeroes are predicted by
the following equation
\begin{align}
M_{\mathrm{D}}  &  =\left \vert b_{\mathrm{D}}\right \vert \left(
1+2\sqrt{\frac{2}{3}}\cos{\left(  \frac{2\pi j}{3}+\varphi_{r,\mathrm{D}%
}\right)  }\right)  ^{2},\text{ }\\
j  &  =1,2,3.\nonumber
\end{align}
where $\varphi_{r,\mathrm{D}}=-\frac{\lambda^{\lbrack12]}-1}{\lambda
^{\lbrack12]}}\pi \frac{1}{\lambda^{\lbrack gr]}}+\frac{1}{2}(\frac
{\lambda^{\lbrack12]}-1}{\lambda^{\lbrack12]}})^{2}\frac{\pi}{\lambda^{\lbrack
gr]}}=-\frac{2}{3}\frac{2}{9}$ and $\left \vert \alpha \right \vert =\sqrt
{\frac{2}{3}}.$ In addition, we have $\left \vert b_{\mathrm{D}}\right \vert
=2\left \vert b_{\mathrm{l}}\right \vert $.

\emph{2. Problems for neutrino oscillation}:

According to above discussion, neutrino oscillation becomes quantum tunneling
phenomenon under quantum fluctuations. As a result, neutrinos change from one
type to another. The mass spectra for neutrinos are obtained as
\begin{align}
M_{\mathrm{n}}  &  =\left \vert b_{\mathrm{n}}\right \vert \left(  1+\left \vert
\alpha \right \vert \cos{\left(  \frac{2\pi j}{3}+\varphi_{r}\right)  }\right)
^{2},\text{ }\\
j  &  =1,2,3\nonumber
\end{align}
where $\left \vert b_{\mathrm{n}}\right \vert =\frac{\left \vert b_{\mathrm{l}%
}\right \vert }{3^{2\lambda^{\lbrack gr]}-2\operatorname{mod}\lambda^{\lbrack
gr]}+4}}$ and $\varphi_{r}=-\frac{\lambda^{\lbrack12]}-1}{\lambda^{\lbrack
12]}}\pi \frac{1}{\lambda^{\lbrack gr]}}-\frac{\pi}{\lambda^{\lbrack gr]}%
+1}=-\frac{2}{9}-\frac{\pi}{\lambda^{\lbrack gr]}+1},$ and $\left \vert
\alpha \right \vert =\sqrt{\frac{N_{I}}{N_{F}}}=\frac{1}{\sqrt{2}}$. Because the
masses of neutrinos is exponentially suppressed by the value of $\lambda
^{\lbrack gr]},$ they are very small. This result indicates that the seesaw
mechanism is wrong and heavy right-hand neutrino doesn't exist.

\emph{3. Problems for Higgs field}:

In this part, we found that the difference between the changing rate along
spatial direction and that along tempo direction plays the role of Higgs field
$\Phi(x,t)$ in the Standard model. We obtain the potential of Higgs field
\begin{align}
V(\Phi)  &  =\Lambda^{3}\frac{\left \vert b_{\mathrm{l}}\right \vert
}{3^{2\lambda^{\lbrack gr]}-2\operatorname{mod}\lambda^{\lbrack gr]}+4}}[1+\\
&  \sqrt{2}\cos(\varphi_{0}-\frac{\pi}{(\lambda^{\lbrack gr]}+1)}\Phi
)]^{2}.\nonumber
\end{align}
According to this result, we understand why the energy scale of Higgs field
$10^{2}$\textrm{GeV} is much smaller than Planck energy $10^{19}$\textrm{GeV}.
A small energy scale for Higgs field comes from mixing the Planck energy scale
and masses of neutrinos, $E_{\mathrm{Higgs}}^{2}\sim E_{\mathrm{Planck}}\times
E_{\mathrm{neutrino}}.$ Therefore, the energy scale of Higgs field
$E_{\mathrm{Higgs}}$ is between $E_{\mathrm{Planck}}$ and
$E_{\mathrm{neutrino}}$.

\emph{4. Problems for Unification:} In this part, we found that different
types of interactions, including strong interaction, weak interaction and
electromagnetic interaction are indeed unified into a new framework -- ($d+1$)
dimensional 2-nd order $\mathrm{\tilde{S}\tilde{O}}$\textrm{(d+1)} physical
chiral variant $V_{\mathrm{\tilde{U}}^{[2]}\mathrm{(1)},\mathrm{\tilde
{S}\tilde{O}}^{[1]}\mathrm{(3+1)},d+1}^{[2],\mathrm{chiral}}.$

\emph{5. Problem for dark matter and dark energy. }In this part, the range of
vacuum energy is estimated. The masses of neutrinos become the shift of the
energies of ground states, that determine the value of the vacuum energy.\ A
possible candidate for dark matter comes from the fluctuations of
inhomogeneous cosmological constant $\Lambda_{c}(x,t)$.

In the end of the last section, we provide prospects on the application of our
theory for particle physics.

In this paper, to calculate mass spectra, we had assume that the
generation-transition is induced by a pair of Coherent operator $R_{\mathrm{A}%
}$ for the ${{\mathbb{Z}}_{N}}$ clock\emph{ }model. This assumption requires a
more solid physical foundation. During the calculation of the mass spectra, we
had consider the transition processes mixing different types of elementary
particles with considering quantum fluctuations. So, a systematic, dynamical
theory for mass spectra must be developed. In addition, we found an
interesting relationship between $\left \vert b_{\mathrm{l}}\right \vert $ and
$\left \vert b_{\mathrm{D}}\right \vert ,$ i.e., $\left \vert b_{\mathrm{D}%
}\right \vert =627\mathrm{MeV}\simeq2\left \vert b_{\mathrm{l}}\right \vert
=2\times313.8\mathrm{MeV}.$ Why? Although we don't understand the physical
mechanism for this equation ($\left \vert b_{\mathrm{D}}\right \vert
\simeq2\left \vert b_{\mathrm{l}}\right \vert $), we believe that $\left \vert
b_{\mathrm{D}}\right \vert $ is twice of $2\left \vert b_{\mathrm{l}}\right \vert
,$ i.e., $\left \vert b_{\mathrm{D}}\right \vert \equiv2\left \vert
b_{\mathrm{l}}\right \vert .$ In the future, we will study this problem and
develop a complete theory for mass spectra.

\newpage

\section{Conclusion}

In the end of this paper, we draw the conclusion.

In this paper, we pointed out that the truth about the physical reality are
relevant to the "\emph{changings}" (we call it variant) and the world becomes
a projection of the structure of "changings" (or variant). Our classical world
can be regarded as \emph{"non-changing"} structure that is described by usual
classical "field" on Cartesian space, i.e.,
\[
\text{"Field" on space: Non-changing structure;}%
\]
In the seven papers, we generalize usual classical "field" to "variant". We
call the new mathematic structure to be \emph{variant theory} that
characterizes a system with "\emph{changing}" or "\emph{operating}" structure,
i.e.,
\[
\text{"Space" on space: Changing structure.}%
\]
In particular, the "Space" denotes a space obeying noncommutative geometry
that is called group-changing spaces in the seven papers. Therefore, a variant
is theory describing the space dynamics rather than field dynamics on
Cartesian space.

\begin{figure}[ptb]
\includegraphics[clip,width=0.7\textwidth]{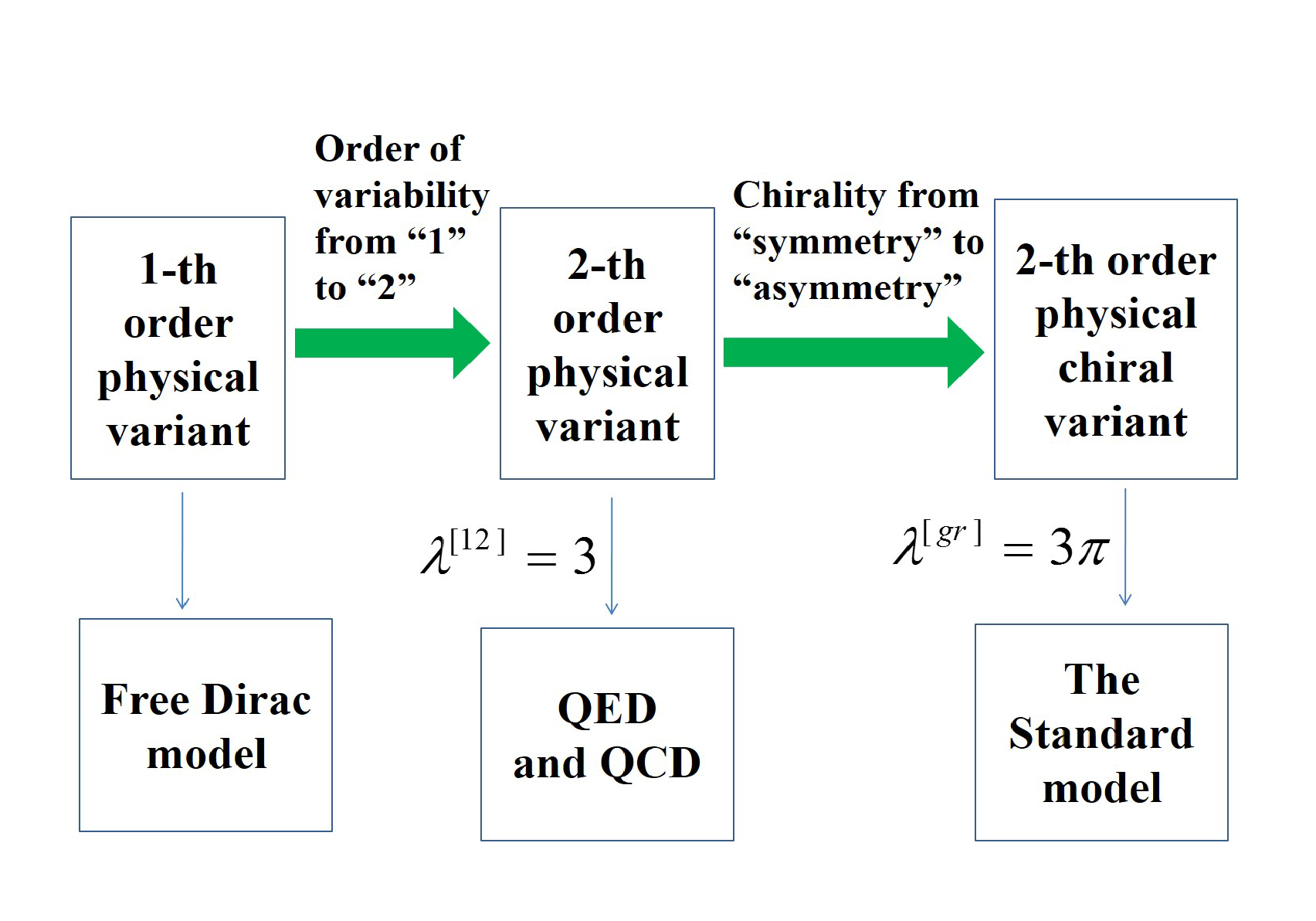}\caption{The logic
structure of the paper}%
\end{figure}

The physical reality in our universe is a ($d+1$)-dimensional $\mathrm{\tilde
{S}\tilde{O}}$\textrm{(d+1)} physical variant\textit{ }$V_{\mathrm{\tilde
{S}\tilde{O}(d+1)},d+1}$ that is a mapping between\textrm{ }$\mathrm{\tilde
{S}\tilde{O}}$\textrm{(d+1)} Clifford group-changing space\textit{
}$\mathrm{C}_{\mathrm{\tilde{S}\tilde{O}(d+1)},d+1}$\textit{\ }and a rigid
spacetime $\mathrm{C}_{d+1}.$\textit{ }Here, $\mathrm{C}_{\mathrm{\tilde
{S}\tilde{O}(d+1)},d+1}$\textit{ }is really a special spacetime obeying
noncommutative geometry. Therefore, our world is really a \emph{uniform},
\emph{holistic} changing structure. This uniform $\mathrm{\tilde{S}\tilde{O}}%
$\textrm{(d+1)}\textit{ }physical variant with a uniform changing structure in
Cartesian space becomes the starting point of the new theory. Now, physical
laws emerge from different changes of regular changes on spacetime. From this
spectacular scene about "changings", we say that "\emph{All from Changings}".
See the logic structure in Fig.50. In modern physics and modern mathematics,
"symmetry" or "invariant" is an important concept. So, to characterize
"changings", we then generalize "symmetry/invariant" of usual field to
(higher-order) variability. After projection, higher-order variability is
reduced into symmetry/invariant. So, we say that symmetry/invariant is the
shadow of variability and the theory for variant is just GUT!

Then we answer all the three fundamental questions at beginning:

1) How to give an exact definition of "\emph{classical object}" and how to
give an exact definition of "\emph{quantum object}"? And "how to unify the two
types of objects into a single framework"? A true theory beyond quantum
mechanics and classical mechanics must provide a complete understanding of the
physical reality (quantum object or classical object) rather than merely
describe its motion. In quantum mechanics, measurement is quite different from
that in classical mechanics. In quantum measurement processes,
\emph{randomness} appears. Why?

\textbf{Answer:}

The uniform $\mathrm{\tilde{S}\tilde{O}}$\textrm{(d+1)}\textit{ }physical
variant that a uniform changing structure in Cartesian space becomes the
starting point of the new theory. Now, physical laws emerge from different
changes of regular changes on spacetime that is characterized by 1-st order
spatial/tempo variability, i.e., $\mathcal{T}(\delta x^{\mu})\leftrightarrow
\hat{U}(\delta \phi^{\mu})=e^{i\cdot k_{0}\delta x^{\mu}\Gamma^{\mu}}.$ Such a
spatial-tempo variability indicates a uniform, holistic universe. Then, both
quantum mechanics and classical mechanics become \emph{phenomenological}
theory and are interpreted by using the concepts of the microscopic properties
of a single physical framework.

Matter and motion is about expanding or contracting $\mathrm{C}%
_{\mathrm{\tilde{S}\tilde{O}(d+1)},d+1}$ group-changing space on rigid space
$\mathrm{C}_{d+1}$. On the one hand, classical object is a \textquotedblleft
non-changing" object with disordered group-changing elements. Classical motion
describes globally motion of a quantum/classical object with
ordered/disordered group-changing elements. On the other hand, quantum object
is a \textquotedblleft changing" object with ordered group-changing elements.
Quantum motion describes the ordered relative motion between group-changing
elements of the elementary particles. As a result, classical motion describes
global motion on a rigid spacetime; quantum motion describes locally expanding
or contracting group-changing space.

In addition, we discuss the problem of quantum measurement. The most amazing
thing is the reversal of deterministic and stochastic characters! People used
to think that classical objects mean determinacy, and quantum objects mean
randomness. However, in this papers we point out that this point of view is
completely wrong -- classical objects mean randomness, and quantum objects
mean determinacy. As a result, the probability in quantum mechanics comes from
the surveyors or instruments during quantum measurement. During quantum
measurement, a quantum object (an ordered object) turn into a classical one
(or disordered one).

2) A satisfied GUT must provide a fully understanding many unsolved mysteries,
including quark confinement problem, Yang-Mill gap problem, Strong CP puzzle
and the existence of axion, Landau's pole problem. To solve above mysteries
satisfactorily, a complete, new theory beyond usual quantum gauge field theory
must be developed, rather than providing certain non-perturbative theories.

\textbf{Answer:}

In this paper, we provide a new foundation for quantum theory of gauge fields.
The key point of the theory is higher-order variability\emph{ }that
characterizes how the uniform changing on another changes. Therefore, in the
paper, we point out that the physical reality of quantum gauge fields is
really a 2-nd order physical variant. Within the new theory, usual quantum
gauge field theory become phenomenological theory and are interpreted by using
the concepts of the microscopic properties of a new physical framework. Now,
quantum gauge fields become one "\emph{changing}" structure on another
"\emph{changing}" structure.

Now, the principle of "symmetry induce interaction" is replaced by the
principle of "variability induce interaction". That means one can define a
physical system\ with usual gauge symmetry by using the concept of
higher-order variability directly, completely.

Under certain projection, 2-nd order variability will be reduced into usual
symmetry/invariant that characterizes the system indirectly, incompletely and
shows the extrinsic property of the system. All physical processes of quantum
gauge fields be intrinsically described by the processes of the changings of
"group-changing space" inside elementary particles. The 2-nd order variability
is reduced into \textrm{U}$^{\mathrm{em}}$\textrm{(1)} local gauge symmetry
and $\mathrm{SU(N)}$ non-Abelian gauge symmetry. The global motion of the
internal "group-changing space" corresponds to the fluctuations of quantum
fields of \textrm{U}$^{\mathrm{em}}$\textrm{(1)} gauge symmetry, and its
relative motion corresponds to the quantum fields of \textrm{SU(}$\mathrm{N}%
$\textrm{). }Therefore, the quantum gauge fields of \textrm{SU(3)} non-Abelian
gauge symmetry in QCD and \textrm{U}$^{\mathrm{em}}$\textrm{(1)} Abelian gauge
symmetry in QED are not separate each other. Instead, they are unified into
single physics structure -- 2--th order variant. QED and QCD describe the
global motion and relative motion of a single higher-order variant, respectively.

In addition, by using the new theory, we give possible solutions for the
troubles about quantum gauge fields, including quark confinement problem,
Strong CP Problem and the Landau's pole problem.

We point out that quark confinement is a problem of spacetime, rather than
just a field problem. This indicates the incompleteness of quantum field
theory. Now, the objects with fractional particle number induced by quarks or
gluons trap fractional number of magnetic monopole of spacetime. As a result,
a quark with fractional fermion number or fractional magnetic charge must have
infinite energy and thus cannot exist. This explains the quark confinement.

The solution about Strong CP Problem is to identify the correct type of
physical variants for our world, $V_{\mathrm{\tilde{U}}^{[2]}\mathrm{(1)}%
,\mathrm{\tilde{S}\tilde{O}}^{[1]}\mathrm{(d+1)},d+1}^{[2]}$ or
$V_{\mathrm{\tilde{U}}^{[2]}\mathrm{(1)},\mathrm{\tilde{S}\tilde{O}}%
^{[1]}\mathrm{(d+2)},d+2}^{[2]}$. Therefore, the Strong CP Problem is also
relevant to the dynamics of spacetime rather than a simple trouble of quantum
field theory. However, our world is the physical variant $V_{\mathrm{\tilde
{U}}^{[2]}\mathrm{(1)},\mathrm{\tilde{S}\tilde{O}}^{[1]}\mathrm{(d+1)}%
,d+1}^{[2]}$ with only one tempo dimension. There doesn't exist theta terms
and axion field at all.

3) The SM is more like a phenomenological model. There exists a lot of unsolve
problems of the SM, such as the trouble is about \emph{mass spectra},
\emph{neutrino oscillation}\cite{neu1,neu2,neu3}, the naturalness problem of
Higgs field, the trouble is about \emph{dark matter and dark energy. }A
satisfied GUT must provide a fully understanding these unsolved mysteries.

\textbf{Answer:}

In this paper, a new approach beyond the SM is proposed based on the theory of
chiral variant. The\emph{ }starting point of this theory is a ($d+1$)
dimensional 2-nd order $\mathrm{\tilde{S}\tilde{O}}$\textrm{(d+1)} physical
chiral variant that is defined by its global sub-variant $V_{g,\mathrm{\tilde
{U}}^{[2]}\mathrm{(1)},\mathrm{\tilde{S}\tilde{O}}^{[1]}\mathrm{(d+1)}%
,d+1}^{[2],\mathrm{sub}}$ and relative sub-variant $V_{r,\mathrm{\tilde
{S}\tilde{O}}^{[1]}\mathrm{(d)},d}^{\mathrm{sub}}.$ Here, the global
sub-variant $V_{g,\mathrm{\tilde{U}}^{[2]}\mathrm{(1)},\mathrm{\tilde{S}%
\tilde{O}}^{[1]}\mathrm{(d+1)},d+1}^{[2],\mathrm{sub}}$ is a higher-order
mapping between\textit{ }$\mathrm{C}_{\mathrm{\tilde{U}}^{[2]}\mathrm{(1)}%
}^{[2]}$,\textit{ $\mathrm{\tilde{S}\tilde{O}}$\textrm{(d+1)}} Clifford
group-changing space $\mathrm{C}_{g,\mathrm{\tilde{S}\tilde{O}(d+1)}%
,d+1}^{[1]}$\ and a rigid spacetime\textit{ }$\mathrm{C}_{d+1}.$\textit{ }The
ratio between the changing rate of global sub-variant and that of relative
sub-variant is $\lambda^{\lbrack gr]}=3\pi$.

The 2-nd order variability is reduced into \textrm{U}$_{\mathrm{em}}%
$\textrm{(1)} local gauge symmetry and $\mathrm{SU_{\mathrm{C}}(N)}$
non-Abelian gauge symmetry. The corresponding theory becomes
QED$\mathrm{\times}$QCD. The level-2 zero becomes color charge with $\frac
{1}{\lambda^{\lbrack12]}}$ electric charge. In addition, due to equivalence
between the relevant sub-variant and the extra left hand sub-variant, the
effective 2-nd order variability is reduced into \textrm{U}$_{\mathrm{Y}}%
$\textrm{(1)} local gauge symmetry and $\mathrm{SU}_{\mathrm{weak}%
}\mathrm{(2)}$ non-Abelian gauge symmetry. Now, the physical chiral variant
becomes fundamental physical object, of which elementary excitations are gauge
fields, fermionic particles and Higgs fields. The low energy effective theory
is just the Standard model -- an $\mathrm{SU}_{\mathrm{Strong}}\mathrm{(3)}%
\otimes \mathrm{SU}_{\mathrm{weak}}\mathrm{(2)}\otimes$\textrm{U}$_{\mathrm{Y}%
}$\textrm{(1)} gauge theory.

An important progress is about hidden topological structure for elementary
particles of different generations -- \emph{chiral para-statistics}. The mass
spectra of different elementary particles are determined by topological
invariables from chiral para-statistics. To obtain the entire mass spectra, we
only need to use one free parameter -- $\lambda^{\lbrack gr]}=3\pi.$

According to this result, we understand why the energy scale of Higgs field
$10^{2}$\textrm{GeV} is much smaller than Planck energy $10^{19}$\textrm{GeV}.
A small energy scale for Higgs field comes from mixing the Planck energy scale
and masses of neutrinos, $E_{\mathrm{Higgs}}^{2}\sim E_{\mathrm{Planck}}\times
E_{\mathrm{neutrino}}.$ Hence, a possible solution about the naturalness
problem is proposed.

In addition, in this paper, the vacuum energy is estimated that is just the
masses of neutrinos.\ Dark matter comes from the fluctuations of the density
of vacuum energy.

In the end of this paper, I pay respect to Einstein. At the beginning of this
paper, we have cited Einstein words, \textquotedblleft \textit{there is no
doubt that quantum mechanics has grasped the wonderful corner of truth... But
I don't believe that quantum mechanics is the starting point for finding basic
principles, just as people can't start from thermodynamics (or statistical
mechanics) to find the foundation of mechanics}.\textquotedblright \ In the
end, I marvel at his profound insight and believe that the new theory of
"space" dynamics for quantum physics will definitely satisfy him. Let us give
a short explanation. There are three key words here, "\emph{geometric}",
"\emph{unification}", and "\emph{nonlocality}": The new theory about "space"
dynamics is fully geometric. Einstein had guessed our world may be a geometric
one; The new theory naturally unifies quantum mechanics and general relativity
into a unique framework, of which quantum dynamics is dual to spacetime
curving. The unification of quantum mechanics and general relativity was
Einstein's dream; The new theory\ shows nonlocality, that is pursued by
Einstein for a long period.


\newpage

\begin{thebibliography}{99}                                                                                               %


\bibitem {we}H. Weyl, I. Z. Phys. \textbf{56}: 330 (1929).

\bibitem {yang}C. N. Yang and R. L. Mills, Phys. Rev. 96, 191 (1954).

\bibitem {1}L. O'Raifeartaigh and N. Straumann, Rev. Mod. Phys. 72 1 (2000).

\bibitem {stand}C. Quigg, \textit{Gauge Theories of the Strong, Weak, and
Electromagnetc Interactions,} Addison--Wesley Pub. Co., Menlo Park, (1983).

\bibitem {w1}S. Weinberg, \textit{The Quantum Theory of Fields. Vol. 1:
Foundations.} Cambridge University Press, 6, (2005). S. Weinberg, \textit{The
Quantum Theory of Fields. Vol. 2: Modern applications.} Cambridge University
Press, Cambridge University Press, 8, 2013.

\bibitem {bur}C. P. Burgess and G. D. Moore, \textit{The standard model: A
primer.} Cambridge University Press, 12, 2006.

\bibitem {tong}D. Tong, \textquotedblleft \textit{The Standard Model,}%
\textquotedblright \ https://www.damtp.cam.ac.uk/user/tong/sm/standardmodel.pdf.

\bibitem {so}H. Georgi and S. Glashow, Phys. Rev. Letts. 32. 438, (1974).

\bibitem {GSW}M.B. Green, J.H. Schwarz, and E. Witten, \textit{Superstring
Theory}, in 2 vols., Cambridge Univ. Press, 1987.

\bibitem {JPbook}J. Polchinski, \textit{String Theory}, in 2 vols., Cambridge
Univ. Press, 1998.

\bibitem {wen}{X.-G. Wen}, \textit{Quantum Field Theory of Many-Body Systems},
(Oxford Univ. Press, Oxford, 2004).

\bibitem {confinement}O. W. Greenberg, Phys. Rev. Lett. 13, 598 (1964). H. D.
Politzer, Phys. Rev. Lett. 30, 1346 (1973). D. J. Gross and F. Wilczek, Phys.
Rev. D 8, 3633 (1973).

\bibitem {axion}R. D. Peccei and H. R. Quinn, Phys. Rev. Lett. 38.25: 1440
(1977). S. Weinberg, Phys. Rev. Lett. 40.4: 223 (1978); F. Wilczek, Phys. Rev.
Lett. 40.5: 279 (1978).

\bibitem {landau}L.D. Landau, A.A. Abrikasov and I.M. Khalatnikov, Dokl. Akad.
Nauk. USSR 95 497; 773; 1173 (1954); Nuovo Cimento Suppl. X 3, 80 (1956).

\bibitem {neu1}B. Pontecorvo, \textquotedblleft Mesonium and
anti-mesonium,\textquotedblright \ Sov. Phys. JETP 6. 429 (1957).

\bibitem {neu2}Super-Kamiokande Collaboration, Y. Fukuda et al., Phys. Rev.
Lett. 81 1562 (1998).

\bibitem {neu3}L. A. Mikaelyan and V. V. Sinev, Phys. Atom. Nucl. 63. 1002 (2000).

\bibitem {higg1}P. W. Higgs, Phys. Lett. 12. 132 (1964). P. W. Higgs, Phys.
Rev. Lett. 13. 508 (1964). P. W. Higgs, Phys. Rev. 145. 1156 (1966).

\bibitem {higg2}F. Englert and R. Brout, Phys. Rev. Lett. 13. 321 (1964).

\bibitem {higg4}G. S. Guralnik, C. R. Hagen, and T. W. B. Kibble, Phys. Rev.
Lett. 13. 585 (1964). T. W. B. Kibble, Phys. Rev. 155. 1554 (1967).

\bibitem {dark matter1}F. Zwicky, Helv. Phys. Acta 6. 110 (1933). F. Zwicky,
Astrophys. J. 86. 217 (1937).

\bibitem {dark matter2}K. C. Freeman, Astrophys. J. 160. 811 (1970).

\bibitem {dark matter3}V. C. Rubin and W. K. Ford, Jr., Astrophys. J. 159. 379 (1970).

\bibitem {dark energy}D. Huterer and M. S. Turner, Phys. Rev. D 60. 081301 (1999).

\bibitem {ein}A. Einstein. Sitzungsber. Preuss. Akad. Wiss. Berlin (Math
Physics), 844 (1915).

\bibitem {jammer}M. Jammer\thinspace, \textit{The philosophy of quantum
mechanics} (Wiley, New York, 1974).

\bibitem {cabello}A. Cabello, \textit{Interpretations of Quantum Theory: A Map
of Madness,} In: \textit{What is Quantum Information}?, O. Lombardi, S.
Fortin, F. Holik, and C. Lopez (Eds.), Cambridge University Press, (2017).

\bibitem {de brogile}De Brogile, L., The Current interpretation of Wave
Mechanis. A Critical Study (Elsevier, Amsterdam, 1964).

\bibitem {Bohm1}D. Bohm, Phys. Rev. 85, 166 (1952); Phys. Rev. 85, 180 (1952).

\bibitem {many}H. Everett, Rev. Mod. Phys. 29, 454 (1957).

\bibitem {nelson}E. Nelson, Phys. Rev. 150, 1079 (1966).

\bibitem {con}A. Connes, \textit{Noncommutative Geometry}, Academic Press, 1994.

\bibitem {epr}A. Einstein, B. Podolsky, and N. Rosen, Phys. Rev. 47, 777, (1935).

\bibitem {dou}See a review, W. H. Zurek, Rev. Mod. Phys. 75, 715 (2003) and
references there.

\bibitem {fey}R. Feynman, R. B. Leighton, M. Sands, \textit{The Feynman
lectures on physics: Quantum mechanics Vol. 3, Chapter 1} (Addison-Wesley
Publishing Com. Inc., Reading, M.A., 1965).

\bibitem {sch3}E. Schrodinger, Naturwissenschaften. 23 (48): 807, (1935).

\bibitem {WH1}W. Heisenberg, Z. Phys\emph{.} \textbf{33}, 879 93 (1925). W.
Heisenberg, Z. Phys\emph{.} \textbf{43}, 172 98 (1927).

\bibitem {Como}N. Bohr, Nature \textbf{121} (suppl.), 580 90 (1928).

\bibitem {rov}C. Rovelli, Int. J. of Theor. Phys. 35 1637 (1996)

\bibitem {bohm}D. Bohm, \textit{Wholeness and the Implicate Order},
(Routledge, London, 1980).

\bibitem {confinement1}K. G. Wilson, Phys. Rev. D 10, 2445 (1974). M. Creutz,
Phys. Rev. D 21, 2308 (1980).

\bibitem {kou1}S.P. Kou, arXiv:1907.13436.

\bibitem {Kaluza}T. Kaluza, Sitzungsber. Preuss. Akad. Wiss. Berlin. (Math.
Phys.): 966 (1921). O. Klein, Zeitschrift f\"{u}r Physik A 37. 895 (1926). O.
Klein, Nature 118. 516 (1926).

\bibitem {ka}M. Kaku,\textit{ Introduction to Superstring and M-Theory 2nd
edition.} New York, Springer-Verlag (1999).

\bibitem {NiemiSemenoff-1986}A. J. Niemi and G. W. Semenoff, Phys. Rep.
\textbf{135}, 99 (1986).

\bibitem {ly}T. D. Lee and C. N. Yang, Phys. Rev. 104, 254 (1956).

\bibitem {wip1}C. J. Copi, D. N. Schramm, and M. S. Turner, Science 267. 192 (1995).

\bibitem {wip2}J. de Swart, G. Bertone, and J. van Dongen, Nature Astron. 1.
0059 (2017).

\bibitem {Pati}J. C. Pati, A. Salam, Phys. Rev. D10, 275 (1974).

\bibitem {Harari}H. Harari, Phys. Lett. B86, 83 (1979).

\bibitem {Shupe}M. Shupe, Phys. Lett. B86, 87 (1979).

\bibitem {helon}S. Bilson-Thompson,. arXiv:hep-ph/0503213 (2005).

\bibitem {rea}G. W. Moore and N. Read, Nucl. Phys. B 360, 362 (1991). N. Read
and E. Rezayi, Phys. Rev. B 59, 8084 (1999).

\bibitem {Ivanov}D. A. Ivanov, Phys. Rev. Lett. 86, 268 (2001).

\bibitem {koide}Y.~Koide, Phys.\ Rev.\ Lett.\  \textbf{47} (1981) 1241. Y.
Koide, Lett. Nuovo Cim. 34, 201 (1982). Y. Koide, Phys Lett. B 120, 161
(1983). Y.~Koide, Phys.\ Rev.\ D \textbf{28} (1983) 252. Y. Koide, hep-ph/0005137.

\bibitem {z3}P. \.{Z}enczykowski, Phys. Rev. D86 (2012) 117303. P.
\.{Z}enczykowski, Phys. Rev. D 87, 077302 (2013).

\bibitem {Brannen:2010zz}C.~A.~Brannen, AIP Conf.\ Proc.\  \textbf{1246}, 182
(2010). C.~A.~Brannen, Found.\ Phys.\  \textbf{40}, 1681 (2010).

\bibitem {scb}A. Rivero, arXiv:1111.7232.

\bibitem {Rosen}G. Rosen, Mod. Phys. Lett. A22 (2007) 283.

\bibitem {wein}S. Weinberg, Phys. Rev. Lett. 19, 1264 (1967).

\bibitem {nar}Narnhofer, Peter, \& Thirring 1996, Deser \& Levin 1997.

\bibitem {des}Deser, S. \& Levin, O. 1997, Class. Quant. Grav., 14, L163.

\bibitem {mond}M. Milgrom, ApJ, 270, 365 (1983).

\bibitem {mil}M. Milgrom, Phys. Lett. A, 253, 273 (1999).

\bibitem {bul}Tucker, W., Blanco, P., Rappoport, S., et al., ApJ, 496, L5+. (1998).
\end{thebibliography}
\end{document}